\documentclass[aps,secnumarabic,nobalancelastpage,amsmath,amssymb,nofootinbib]{revtex4}
\usepackage[a4paper, margin=1in]{geometry}
\usepackage{hyperref,graphicx}% Include figure files
\usepackage{dcolumn}% Align table columns on decimal point
\usepackage{bm}% bold math
\usepackage{bbm}
\usepackage{tikz}
\usepackage{physics}
\usepackage{enumitem}
\usepackage{amssymb,amsfonts}
\usepackage{epsfig}
\usepackage{epstopdf}
\usepackage[all]{xy}
\usepackage{amsthm}
\usepackage{dcolumn}
\usepackage{hyperref}
\usepackage{url}
\usepackage{dsfont}
\usepackage{slashed}
\usepackage{mathrsfs}
\usepackage{soul}
\usepackage{float}
\DeclareMathAlphabet{\mathpzc}{OT1}{pzc}{m}{it}

\newcommand{\be}{\begin{equation}}
\newcommand{\ee}{\end{equation}}
\newcommand{\bea}{\begin{eqnarray}}

\newcommand{\eea}{\end{eqnarray}}

\newcommand*\interior[1]{\mathring{#1}}
\newcommand{\red}[1]{\textcolor{red}{#1}}

\def\inbar{\,\vrule height1.5ex width.4pt depth0pt}
\def\IR{\relax{\rm I\kern-.18em R}}
\def\IC{\relax\hbox{$\inbar\kern-.3em{\rm C}$}}

\def\ud{\mathrm{d}}

\def\Z{\mathbb{Z}}

\def\N{\mathbb{N}}

\def\ii{\mathrm{i}}
\def\bu{\mathbbm{1}}
\usepackage{lipsum}  % For generating dummy text
\usepackage{geometry}  % To set page dimensions
\usepackage[a4paper, margin=1in]{geometry}

 \usepackage{hyperref}
\hypersetup{
    colorlinks=true,
    linkcolor=blue,
    citecolor=blue,
    urlcolor=blue
}

\begin{document}
$$\mbox{\red{\textbf{\Large{Important Note:}}}}$$ 
This draft marks the early stages of the book's 1st edition, officially published by Springer Nature in 2022, \emph{\textbf{albeit with a slightly modified title}}. Subsequently, in 2024, a 2nd edition emerged:
\begin{itemize}
\item{M. Enayati, J.P. Gazeau, H. Pejhan, and A. Wang, \textbf{\emph{The de Sitter (dS) Group and its Representations; An Introduction to Elementary Systems and Modeling the Dark Energy Universe}} (2nd edition), Springer Nature (2024). DOI: \href{https://doi.org/10.1007/978-3-031-56552-6}{\textcolor{blue}{10.1007/978-3-031-56552-6}}.}
\end{itemize}

\begin{figure}[H]
\begin{center}
\includegraphics[height=.8\textheight]{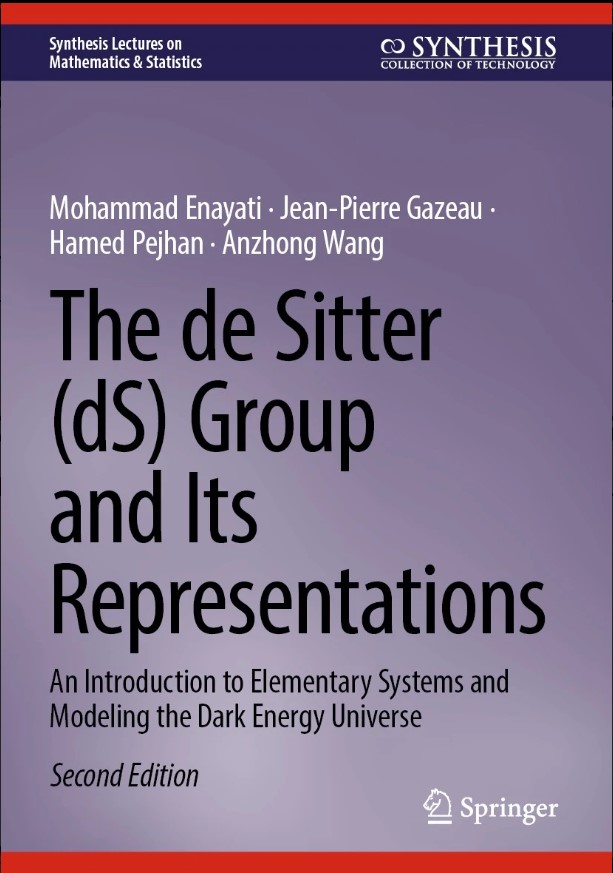}
\end{center}
\end{figure}

\newpage

\title{The de Sitter group and its representations:\\
a window on the notion of de Sitterian elementary systems}

\author{Mohammad Enayati$^1$}

\author{Jean-Pierre Gazeau$^{2}$\footnote{gazeau@apc.in2p3.fr}}

\author{Hamed Pejhan$^{3}$\footnote{pejhan@math.bas.bg}}

\author{Anzhong Wang$^{3,4}$\footnote{anzhong\_wang@baylor.edu}}

\affiliation{$^1$Department of Physics, Razi University, Kermanshah 6741414971, Iran}

\affiliation{$^2$Universit\'e de Paris, CNRS, Astroparticule et Cosmologie, F-75013 Paris, France}

\affiliation{$^3$Institute of Mathematics and Informatics, Bulgarian Academy of Sciences, Acad. G. Bonchev Str. Bl. 8, 1113, Sofia, Bulgaria}

\affiliation{$^4$GCAP-CASPER, Physics Department, Baylor University, Waco, Texas 76798-7316, USA}

\date{\today}

\begin{abstract}
We review the construction of (``free") elementary systems in de Sitter (dS) spacetime, in the Wigner sense, as associated with unitary irreducible representations (UIR's) of the dS (relativity) group. This study emphasizes the conceptual issues arising in the formulation of such systems and discusses known results in a mathematically rigorous way. Particular attention is paid to: ``smooth" transition from classical to quantum theory; physical content under vanishing curvature, from the point of view of a local (``tangent") Minkowskian observer; and thermal interpretation (on the quantum level), in the sense of the Gibbons-Hawking temperature. We review three decompositions of the dS group physically relevant for the description of dS spacetime and classical phase spaces of elementary systems living on it. We review the construction of (projective) dS UIR's issued from these group decompositions. (Projective) Hilbert spaces carrying the UIR's (in some restricted sense) identify quantum (``one-particle") states spaces of dS elementary systems. Adopting a well-established Fock procedure, based on the Wightman-G\"{a}rding axioms and on analyticity requirements in the complexified Riemannian manifold, we proceed with a consistent quantum field theory (QFT) formulation of elementary systems in dS spacetime. This dS QFT formulation closely parallels the corresponding Minkowskian one, while the usual spectral condition is replaced by a certain geometric Kubo-Martin-Schwinger (KMS) condition equivalent to a precise thermal manifestation of the associated ``vacuum" states. We end our study by reviewing a consistent and univocal definition of mass in dS relativity. This definition, presented in terms of invariant parameters characterizing the dS UIR's, accurately gives sense to terms like ``massive" and ``massless" fields in dS relativity according to their Minkowskian counterparts, yielded by the group contraction procedures.
\end{abstract}

\maketitle
\tableofcontents

\setcounter{equation}{0} \section{Introduction}
Quantum elementary systems are associated with (projective) unitary irreducible representations (UIR's)  of the (possibly extended) relativity group (or one of its covering). This seminal point of view was first put forward in the context of Einstein-Poincar\'{e} relativity by Wigner in his famous paper in $1939$ \cite{Wigner1939} (see also Ref. \cite{Newton/Wigner}), where the rest mass $m$ and the spin $s$ of an (Einsteinian) elementary system are shown to be the two invariants that characterize the associated UIR of the Poincar\'{e} group (the group of motions of flat Minkowski spacetime). He was followed by In\"{o}n\"{u} \cite{Wigner1952}, L\'{e}vy-Leblond \cite{Levy-Leblond}, and Voisin \cite{Voisin} who applied the Wigner ideas to Galilean systems, and by G\"{u}rsey \cite{Gursey1963} and Fronsdal \cite{Fronsdal 1,Fronsdal 2} who extended them to dS and anti-dS (AdS) systems, respectively.

In this paper, following the path initiated by Wigner and others in Refs. \cite{Wigner1939,Newton/Wigner,Wigner1952,Levy-Leblond,Voisin,Gursey1963,Fronsdal 1,Fronsdal 2}, we review the construction of (free) elementary systems in dS spacetime. To keep this review comprehensive, the level of exposition varies in its different parts. Hence, both experts and beginners can find something interesting and useful in this study.

\subsection{Motivations}
In the context of modern theories of elementary systems (both field theory and the phenomenological treatment), the formulation of a physical theory, and the interpretation in particular, rests upon the notions of energy, momentum, mass, and spin, whose existence literally stems from the principle of invariance under the Poincar\'{e} group \cite{Wigner1939,Newton/Wigner}. Physicists, however, are well aware that modern theories of elementary systems cannot in the end be based on the Poincar\'{e} group. What is needed is a theory of elementary systems, or at least a consistent framework, that respects the full general covariance of Einstein's view of spacetime as a Riemannian manifold. But, once one departs from flat Minkowski spacetime, due to the absence of nontrivial groups of motion in more general Riemannian spaces, a formidable obstacle to the extension of physical models appears; generally, no literal or unique extension of the aforementioned physical notions exists (?!).\footnote{Here, we put aside the suggestion that the important differential equations (Klein-Gordon and Dirac) may easily be generalized to forms that possess general covariance. In the above sense, this suggestion is almost totally irrelevant. Frankly speaking, the modern theories of elementary systems are not primarily studies in differential equations \cite{Fronsdal 1}.} Of course, there is a specific class of Riemannian spaces in which the road to generalizations is well marked, in the sense given by Fronsdal in $1965$ \cite{Fronsdal 1}: ``\emph{A physical theory that treats spacetime as Minkowskian flat must be obtainable as a well-defined limit of a more general physical theory, for which the assumption of flatness is not essential.}" Poincar\'{e} relativity indeed can be considered as the idealistic null-curvature limit of two possible curved-spacetime relativities of maximal symmetry. Technically, a four-dimensional Riemannian space may admit a continuous group of symmetry, preserving the metric $g^{}_{\mu\nu}$, with up to ten essential parameters. The maximum number (which is the same number as flat Minkowski spacetime) is merely realized for a space of constant curvature $1/R$ ($R$ being the radius of curvature, $0<R<\infty$).

Those spacetimes, which meet flat Minkowski spacetime as the curvature goes to zero ($R\rightarrow \infty$), are the ordinary dS and AdS spacetimes, the maximally symmetric solutions to the vacuum Einstein's equations with, respectively, positive and negative cosmological constant $\Lambda$ ($R=\sqrt{3/|\Lambda|}$) \cite{DeSitter1917}. The former, dS spacetime, admits SO$_0(1,4)$ (or its universal covering Sp$(2,2)$) as a group of motions. It is essentially finite in extension \cite{Wigner1950}; considering any point $p$ and any timelike direction in that point, the geodesics through $p$, perpendicular to the chosen timelike direction, are finite. AdS spacetime, on the other side, is infinite in extension; analogous geodesics possess infinite lengths and are completely spacelike. The AdS group of motions is SO$_0(2,3)$ (or its double covering Sp$(4,\mathbb{R})$, or even its universal covering $\widetilde{\mathrm{SO}_0(2,3)}$). Interestingly, as Minkowski spacetime is the limit $R\rightarrow \infty$ of the ordinary dS and AdS spacetimes, the Poincar\'{e} group can be obtained as a contraction of either SO$_0(1,4)$ or SO$_0(2,3)$ (or any of their coverings); UIR's of the dS and AdS groups, analogous to their shared Poincar\'{e} contraction limit, are characterized by two invariant parameters of the spin and energy scales (note that, in the AdS case, the latter should be read as the rest energy). These remarkable features, as already pointed out, allow the Wigner definition of elementary systems to be extended to dS and AdS relativities.\footnote{We note in passing that the (A)dS group-theoretical structures serve a wider variety of practical applications in modern physics than  what we have mentioned above. The study of the Hydrogen atom, for instance, well illustrates several aspects of the application of such structures in quantum mechanics (see, for instance, Refs. \cite{H-atom Perelomov,H-atom I,H-atom II,H-atom Musto,H-atom Pratt,H-atom Gazeau}).}

In the present paper, we are particularly interested in the dS case. Besides the above conceptual worries, it is motivated in part by the critical role that the dS metric plays in the inflationary cosmological scenarii (based upon which our Universe underwent a dS phase in the very early epochs of its life \cite{Linde}), and in part by the desire to construct possible models for late-time cosmology (since a small positive cosmological constant seems to be required by recent data \cite{Perlmutter}).

\subsection{Content at a glance}
As pointed out above, admitting a more general relativity group instead of the Poincar\'{e} one, to describe elementary systems, lies at the heart of the content of this review. In this sense, to make our discussions straight, let us first elaborate the notion and the role of relativity group in a general context. Technically, to study a physical system ${P}$, one needs a \emph{frame}, that is, a correspondence between ${P}$ and a mathematical structure ${M}$ describing the set of states of ${P}$ measured with respect to this frame. In this context, the relativity group $G$ is the group of frame transformations. Then, the rule ``physical laws are independent of the frame" turns into ``the structure of ${M}$ is invariant under $G$". This structure is a symplectic manifold (called phase space) on the classical level and a (projective) Hilbert space on the quantum level. This system is called an elementary system \cite{Wigner1939,Newton/Wigner}, when one does not deal with internal variables. Therefore, the different states, which appear, are merely due to a change of frames and nothing else. This implies that the action of the group $G$ on $M$ is transitive, i.e., is a co-adjoint representation or a (projective) UIR on the classical and the quantum levels, respectively.

In this paper, we study elementary systems in the above sense with the dS group Sp$(2,2)$. We start from scratch to be able to present the foundations step-by-step in a mathematically rigorous way. We employ three types of decomposition of the Sp$(2,2)$ group. The first one, called space-time-Lorentz decomposition, is nonstandard and yields a global, but nonunique, decomposition of the group, while the other two are well known in semi-simple group theory \cite{Knapp} and, respectively, called Cartan and Iwasawa decompositions. These group decompositions provide the basic mathematical ingredients of the discussion, namely, (related) families of group cosets Sp$(2,2)/\underline{\cal{S}}$, where $\underline{\cal{S}}$'s stand for (closed) subgroups of Sp$(2,2)$ yielded by these decompositions. As a matter of fact, each phase space of dS elementary systems, more accurately, each transitive manifold under the action of the Sp$(2,2)$ co-adjoint representations (say Sp$(2,2)$ co-adjoint orbit), being symplectic manifold and carrying a natural Sp$(2,2)$-invariant (Liouville) measure, is a homogeneous space homeomorphic to an even-dimensional group coset Sp$(2,2)/\underline{\cal{S}}$, where $\underline{\cal{S}}$ plays the role of stabilizer subgroup of some orbit point \cite{Kirillov,Kirillov1976}.

The Sp$(2,2)$ co-adjoint orbits, naively speaking, the group cosets Sp$(2,2)/\underline{\cal{S}}$, also possess very rich analytic structures, which underlie (projective) Hilbert spaces that carry UIR's of the Sp$(2,2)$ group. According to the physical point of view adopted in this paper, this remarkable feature in a well-established process allows for a ``smooth" transition from classical to quantum formulation of dS elementary systems; the phase spaces of dS elementary systems quantize into (projective) dS UIR's. These UIR's consist of three distinguished series, respectively, called principal, complementary, and discrete series \cite{Thomas1941,Newton1950,Dixmier,Takahashi',Martin1974}. The UIR's corresponding to the principal series contract to the massive UIR's of the Poincar\'{e} group \cite{Mickelsson,Garidi zero curvature limit}. Hence, they are called dS massive representations. The situation for the dS massless cases, however, is more subtle; the dS group has no UIR analogous to the so-called massless infinite-spin UIR's of the Poincar\'{e} group. Massless representations of the dS group then are naturally distinguished as those with a unique extension to the UIR's of the conformal group SO$_0(2,4)$, while that extension is equivalent to the conformal extension of the Poincar\'{e} massless UIR's \cite{Barut,Mack1977}. It follows that the dS massless scalar case coincides with a specific UIR of the complementary series, while the dS massless higher-spin cases correspond to the UIR's lying at the lower end of the discrete series. All other dS representations either have nonphysical Poincar\'{e} contraction limit or do not have Poincar\'{e} contraction limit at all.

Once dS massive and massless elementary systems are recognized with respect to group representation theory, one can deal with the corresponding covariant QFT's along the lines proposed by Wightman and G\"{a}rding in their seminal paper \cite{Wightman}. The very problem, that naturally arises here, is the absence of a true spectral condition, which plagues QFT in dS spacetime \cite{Streater,Nachtmann1967}. Actually, no matter what machinery of QFT is employed to quantize a field in dS spacetime, while it is rather straightforward to formalize the requirements of locality (microcausality) and covariance, it is impossible to formulate any condition on the spectrum of the ``energy" operator (even worse, it is impossible to define such a global object at all). Due to this ambiguity, for any single dS field model, many inequivalent QFT's appear (the phenomenon of nonuniqueness of the vacuum state), often each being relevant to a particular choice of time coordinate, which induces the corresponding frequence splitting. Therefore, to achieve a consistent QFT reading of dS elementary systems, besides the dS group representation theory (in the sense given by Wigner) and the Wightman and G\"{a}rding axioms, we still need a supplementary criterion to replace the usual spectral condition.

In Refs. \cite{GazeauPRL,Bros 2point func}, Bros et al. have argued that a suitable adaptation of some familiar notions of complex Minkowski spacetime to its (complex) dS counterpart can provide such a criterion to lift all ambiguities for dS QFT's and to select preferred vacuum states which, despite their thermal properties (in the sense given by Gibbons-Hawking \cite{Gibbons,Kay}), coincide with their corresponding Minkowski vacuum representations under vanishing curvature. Their original approach keeps from the Minkowskian case the idea that the analytic continuation properties of the QFT in the complexified spacetime are directly related to the energy content (in particular to the spectral condition) of the model considered. Technically, in order to apply this appealing idea to dS QFT's, they have put forward a genuine, global dS-Fourier type calculus, realized by the introduction of (coordinate-independent) \emph{dS plane waves} in their tube domains. [Such waves are the dS counterparts of the standard plane waves in Minkowski spacetime. They are well adapted to the dS group representations and also allow to control in a very suggestive way the null-curvature limit of dS QFT to its Minkowskian counterpart.] On this basis, they have shown that, for instance, in the simplest cases, i.e., linear dS QFT's which are of interest in the present study, the spectral condition is substituted by a certain geometric KMS condition \cite{Kubo,Martin}, equivalent to a precise thermal manifestation of the associated vacuum states (known in the literature under the name of Euclidean \cite{Gibbons} or Bunch-Davies \cite{Bunch} vacuum states).\footnote{For this, except the references cited above, see also Refs. \cite{Massive/Massless 1/2,Massless 1,Massive 1,BehrooziTakook,Massive 2,Massive 3/2}.}

Accordingly, in this paper, employing the dS group representation theory and its Wigner interpretation, on one hand and on the other hand, the Wightman and G\"{a}rding axioms equipped with analyticity requirements in the complexified dS manifold (in the sense given by Bros et al.), we encounter the QFT formulation of elementary systems in dS spacetime.

At the end, the question of finding a universal substitute to the notion of mass in dS relativity comes to fore. This demand leads us to adopt a consistent and univocal definition of mass in dS spacetime proposed by Garidi in $2003$ \cite{Garidimass}. The Garidi definition, presented in terms of the invariant parameters characterizing the dS UIR's, remarkably gives sense to terms like ``massive" and ``massless" fields in dS relativity with respect to their Minkowskian counterparts, yielded by the group contraction procedures. It also enjoys the advantage to encompass all mass formulas introduced within the dS context.

\subsection{Reading guide and conventions}
This review is divided into four parts. In part \ref{Part I}, to set the stage for better understanding the mathematical materials, we discuss $1+1$-dimensional dS relativity, which, despite its mathematical transparency, interestingly contains all essential ingredients of the realistic case, $1+3$-dimensional dS relativity. In part \ref{Part II}, the latter case is discussed on the group/algebra and representation levels, or in the sense given above, let us say the classical and quantum mechanics levels, respectively. In part \ref{Part plane waves}, we proceed with the corresponding QFT formulation. Finally, part \ref{Part mass} is devoted to the notion of mass in dS and, for the sake of comparison, AdS relativities.

The main conventions of our notations are:
\begin{itemize}
\item{Throughout this paper (unless noted otherwise), for the sake of simplicity, we consider the units $c = 1 = \hbar$, where $c$ and $\hbar$ are respectively the speed of light and the Planck constant.}
\item{We distinguish between $1+1$-dimensional dS spacetime and its $1+3$-dimensional counterpart by adding the relevant subscripts `$2$' (dS$_2$) and `$4$' (dS$_4$), respectively. Moreover, in order to distinguish between their relevant entities, in particular the ones that do not manifestly admit spacetime indices, we draw a line below those that are relevant to dS$_4$ relativity.}
\item{We use the letters $a,b,c,...$ for the indices $0,1,2$, the letters $\mu,\nu,\rho, ...$ for $0,1,2,3$, the letters $A,B,C,...$ for $0,1,2,3,4$, the letters $A^\prime, B^\prime, C^\prime, ...$ for $0,1,2,3,5$ (the number $4$ is left apart!), the letters $\mathpzc{A},\mathpzc{B},\mathpzc{C}, ...$ for $0,1,2,3,4,5$, the letters $\texttt{A},\texttt{B},\texttt{C},...$ for $1,2,3,4$, and finally the letters $i,j,k,...$ for $1,2,3$.}
\end{itemize}

%%%%%%%%%%%%%%%%%%%%%%%%%%%%%%%%%%%%%%%%%%%%%%%%%%%%%%%%%%%%%%%%%%%%%%%%%%%%%%%%%%%%%%%%%%%%%%%%%%%%%%%%%%%%%%%%%%%%%%%%
%%%%%%%%%%%%%%%%%%%%%%%%%%%%%%%%%%%%%%%%%%%%%%%%%%%%%%%%%%%%%%%%%%%%%%%%%%%%%%%%%%%%%%%%%%%%%%%%%%%%%%%%%%%%%%%%%%%%%%%%
%%%%%%%%%%%%%%%%%%%%%%%%%%%%%%%%%%%%%%%%%%%%%%%%%%%%%%%%%%%%%%%%%%%%%%%%%%%%%%%%%%%%%%%%%%%%%%%%%%%%%%%%%%%%%%%%%%%%%%%%
%%%%%%%%%%%%%%%%%%%%%%%%%%%%%%%%%%%%%%%%%%%%%%%%%%%%%%%%%%%%%%%%%%%%%%%%%%%%%%%%%%%%%%%%%%%%%%%%%%%%%%%%%%%%%%%%%%%%%%%%
%%%%%%%%%%%%%%%%%%%%%%%%%%%%%%%%%%%%%%%%%%%%%%%%%%%%%%%%%%%%%%%%%%%%%%%%%%%%%%%%%%%%%%%%%%%%%%%%%%%%%%%%%%%%%%%%%%%%%%%%
%%%%%%%%%%%%%%%%%%%%%%%%%%%%%%%%%%%%%%%%%%%%%%%%%%%%%%%%%%%%%%%%%%%%%%%%%%%%%%%%%%%%%%%%%%%%%%%%%%%%%%%%%%%%%%%%%%%%%%%%
%%%%%%%%%%%%%%%%%%%%%%%%%%%%%%%%%%%%%%%%%%%%%%%%%%%%%%%%%%%%%%%%%%%%%%%%%%%%%%%%%%%%%%%%%%%%%%%%%%%%%%%%%%%%%%%%%%%%%%%%

\part{As a preliminary: $1+1$-dimensional dS (dS$_2$) geometry and relativity}\label{Part I}

\setcounter{equation}{0} \section{DS$_2$ manifold and its symmetry group}
DS$_{2}$ spacetime is a globally hyperbolic spacetime with the topology of $\mathbb{R}^1\times\mathbb{S}^1$ ($\mathbb{R}^1$ being a timelike direction). This spacetime, by its embedding in a $1+2$-dimensional Minkowski spacetime $\mathbb{R}^{1+2}$ (by abuse of notation, let us say $\mathbb{R}^3$), can be conveniently described as a one-sheeted hyperboloid $M_R$ of constant radius $R$:
\begin{eqnarray}\label{1+1dS-M_R}
M_R \equiv \Big\{x = (x^0,x^1,x^2) \in\mathbb{R}^3 \;;\; (x)^2 \equiv x\cdot x = \eta^{}_{ab}x^a x^b =  -R^2 \Big\}\,, \;\;\;\;\;\;\; a,b = 0,1,2\,,
\end{eqnarray}
where $x^a$'s are the Cartesian coordinates in $\mathbb{R}^3$ and $\eta^{}_{ab} = \mbox{diag}(1,-1,-1)$ is the natural metric of $\mathbb{R}^3$. The dS$_{2}$ metric then is defined by inducing $\eta^{}_{ab}$ on $M_R$.

The dS$_2$ (relativity) group is SO$_0(1,2)$ (that is, the connected subgroup of O$(1,2)$) or its double-covering group SU$(1,1)$. The associated Lie algebra can be realized by the linear span of the following (three) Killing vectors (see appendix \ref{Killing}):
\begin{eqnarray}\label{Killing dS2}
K_{ab} = x_a \partial_b - x_b \partial_a\,, \;\;\;\;\;\;\; K_{ab} = - K_{ba}\,.
\end{eqnarray}

\subsection{Precision on SO$_0(1,2)$: a geometric viewpoint}\label{Subsec 2.1}
Let us begin with the group O$(1,2)$. By definition, it is the group of all linear transformations in $\mathbb{R}^3$ carrying $x$ into $x^\prime$ ($x,x^\prime \in \mathbb{R}^3$) such that the indefinite quadratic form $(x)^2= \eta^{}_{ab}x^a x^b$ remains unchanged. In this paper, among all such transformations, we are particularly interested in those that preserve the orientation of space. They form the subgroup SO$(1,2)$ consisting of the transformations with determinant $1$ (note that the transformations with determinant $-1$ are reflections). Among the transformations belonging to SO$(1,2)$, we also would like to restrict our attention to those carrying $x$ into $x^\prime$ in such a way that if $x^0 >0$ or $x^0<0$, then we get $x^{\prime 0} >0$ or $x^{\prime 0} <0$, respectively. [From the physical point of view, these transformations are important, because they take every positive timelike vector (characterizing actual motion) into another such vector.] Such transformations are connected to the identity. They form the so-called connected dS$_{2}$ group, denoted here by SO$_0(1,2)$. As already pointed out, we refer to the latter as the dS$_2$ (relativity) group.

In view of the approach adopted in this paper, it is convenient to take a closer look at the action of SO$_0(1,2)$ on $\mathbb{R}^3$. The fact that the quadratic form $(x)^2= \eta^{}_{ab}x^a x^b$ is not positive definite has a significant impact. Technically, the form $(x)^2=\eta^{}_{ab}x^a x^b = 0$ determines a cone, with vertex at the origin, in $\mathbb{R}^3$. It then follows that, under the action of SO$_0(1,2)$, $\mathbb{R}^3$ is divided into three domains: the cone itself $(x)^2=0$, the interior of the cone $(x)^2>0$, and the exterior of the cone $(x)^2<0$. More accurately, considering $(x)^2=r$ ($r$ being a real number), we have three types of \emph{orbit}\footnote{Let $G$ be a Lie group that acts on a manifold $M$. Given a point $p\in M$, the action of $G$ on $p$, symbolized here by $G \diamond p$, takes $p$ to various points in $M$. By definition, the orbit of $p$ under the action of $G$ is the subset of $M$ defined by $O(p) = \big\{ g \diamond p \; ; \; g\in G \big\}$. Of course, the orbit is independent of the choice of $p$ in the sense that, whenever $p^\prime \in O(p)$, we have $O(p) = O({p^\prime})$.}, apart from the trivial one $x^0 = x^1 = x^2 = 0$, in $\mathbb{R}^3$:
\begin{itemize}
\item{The upper and lower sheets of the cone, with $r=0$ and $x^0\gtrless0$, respectively.}
\item{The upper and lower sheets of the two-sheeted hyperboloids, with $r>0$ and $x^0\gtrless0$, respectively.}
\item{The one-sheeted hyperboloid, with $r<0$ (note that the dS$_2$ hyperboloid (\ref{1+1dS-M_R}) belongs to this type, with $r=-R^2$).}
\end{itemize}

\subsection{Homomorphism between SO$_0(1,2)$ and SU$(1,1)$ \label{Subsubsec 2.1.1}}
The dS$_2$ group SO$_0(1,2)$ is homomorphic to the SU$(1,1)$ group. The latter is the group of all $2\times 2$-matrices $g$ verifying the following conditions:
\begin{itemize}
\item{The unimodular condition; $\mbox{det}(g) = 1$.}
\item{The pseudo-unitary condition; $g^\dagger \sigma_3 g = \sigma_3$, where $g^\dagger$ stands for the Hermitian adjoint of $g$ and $\sigma_3 = \begin{pmatrix} 1 & 0 \\ 0 & -1 \end{pmatrix}$ is the third Pauli matrix.}
\end{itemize}
Then, the SU$(1,1)$ group can be viewed as:
\begin{eqnarray}\label{ginSU1,1}
\mathrm{SU}(1,1) = \Bigg\{ g = \begin{pmatrix}
{\alpha} & {\beta} \\
\beta^\ast & \alpha^\ast
\end{pmatrix}
\;;\; \alpha,\beta \in \mathbb{C}\,, \; \mbox{det}(g) = |\alpha|^2 - |\beta|^2 = 1 \Bigg\}\,,
\end{eqnarray}
where $\alpha^\ast$ and $\beta^\ast$ stand for the complex conjugate of $\alpha$ and $\beta$, respectively.

To make the homomorphism between SO$_0(1,2)$ and SU$(1,1)$ apparent, we associate with each point $x=(x^0,x^1,x^2)$ in $\mathbb{R}^3$ a Hermitian matrix:
\begin{eqnarray}\label{epoR3aHm}
h = \begin{pmatrix}
x^0 & x^1+ \mathrm{i} x^2 \\
x^1- \mathrm{i} x^2 & x^0
\end{pmatrix}\,,
\end{eqnarray}
where $\mbox{det}(h) = (x^0)^2 - (x^1)^2 - (x^2)^2 = (x)^2$. For each $g \in \mathrm{SU}(1,1)$ of the form given in Eq. (\ref{ginSU1,1}), one can define an action of SU$(1,1)$ on $\mathbb{R}^3$, symbolized here by $x\mapsto x^\prime = g \diamond x$ ($x,x^\prime \in \mathbb{R}^3$), as:
\begin{eqnarray}\label{ifginSUdaa}
ghg^\dagger =
\begin{pmatrix}
x^{\prime 0} & x^{\prime 1} + \mathrm{i} x^{\prime 2} \\
x^{\prime 1} - \mathrm{i} x^{\prime 2} & x^{\prime 0}
\end{pmatrix}\,.
\end{eqnarray}
Note that: (i) The transformed matrix $ghg^\dagger$ is also Hermitian and clearly corresponds to a point ${x^\prime} = (x^{\prime 0}, x^{\prime 1}, x^{\prime 2})$ in $\mathbb{R}^3$. (ii) This transformation is linear, because elements of the matrix $ghg^\dagger$ are linearly expressed in terms of the elements of $h$. (iii) $\mbox{det}(ghg^\dagger) = \mbox{det}(h) = (x)^2$, which means that this linear transformation preserves the quadratic form $(x)^2$. (iv) If $x^0>0$ or $x^0<0$, then $x^{\prime 0}>0$ or $x^{\prime 0}<0$, respectively. Consequently, the linear transformation $x^\prime = g \diamond x$, realized by the action (\ref{ifginSUdaa}), belongs to SO$_0(1,2)$, as well. More precisely, for every transformation in SO$_0(1,2)$, there are two elements $\pm g\in \mathrm{SU}(1,1)$, since, according to the action (\ref{ifginSUdaa}), we have $g \diamond x = (-g) \diamond x$. This means that SU$(1,1)$ is two-to-one homomorphic to SO$_0(1,2)$ (or in other words, the SU$(1,1)$ group is a two-valued representation or double covering of SO$_0(1,2)$).\footnote{The point to be noticed here is that, on the quantum level, we are not interested only in representations in the narrowest sense of the word, i.e., one-valued representations, but also in multi-valued representations. In this sense, it is not significant, from the point of view of representations, to distinguish between SO$_0(1,2)$, SU$(1,1)$, or their covering group. It is, however, significant to know how many times SO$_0(1,2)$ is covered by its covering group.} The \emph{kernel of this homomorphism}\footnote{Let $G$ and $H$ be groups and let $f$ be a group homomorphism from $G$ to $H$. If $e^{}_H$ is the identity element of $H$, then the kernel of $f$ (denoted by $\mbox{ker}(f)$) is: $\mbox{ker}(f) = \big\{ g\in G \;;\; f(g)=e^{}_H \big\} $.} $\{\pm \mathbbm{1}_2\}$, where $\mathbbm{1}_2$ stands for the $2\times 2$-unit matrix, is isomorphic to $\mathbb{Z}^2$, thus SU$(1,1)/\mathbb{Z}^2 \sim \mathrm{SO}_0(1,2)$.\footnote{In our notation, the symbol `$\sim$' stands for isomorphism between two groups. In the sequel, by abuse of notation, we also employ the same symbol to denote the homomorphism/homeomorphism between two groups/topological spaces.}

We also would like to remark that SU$(1,1)$ is isomorphic to the special linear group of order two with real entities, i.e., SL$(2,\mathbb{R})$. This isomorphism can be easily seen by considering a unitary $2\times 2$-matrix, for instance, $u = \frac{1}{\sqrt{2}} \begin{pmatrix} 1 & - \mathrm{i} \\ 1 & \mathrm{i} \end{pmatrix}$, based upon which one can associate with each $\bar{g}\in \mathrm{SL}(2,\mathbb{R})$ a matrix $g = u \bar{g} u^\dagger \in \mathrm{SU}(1,1)$.

Here, for more detailed discussions on the above topics, readers are referred to Ref. \cite{Vilenkin}.

\setcounter{equation}{0} \section{Relativistic meaning of the dS$_2$ group: group decomposition}\label{Sec dS2 decomposition}
In this section, we review three decomposition types of the dS$_2$ group SU$(1,1)$, respectively, called space-time-Lorentz, Cartan, and Iwasawa decompositions. The former is nonstandard whereas the other two are well known in the context of semi-simple group theory \cite{Knapp}. As we will show below, these group decompositions are physically relevant for the description of dS$_2$ spacetime, phase spaces of dS$_2$ elementary systems, and dS$_2$ timelike infinity or phase-spaces infinities.

\subsection{Space-time-Lorentz decomposition}\label{sec space-time-Lorentz}
Any element $g \in \mathrm{SU}(1,1)$, with respect to the group involution $\mathfrak{i}(g) : g\mapsto g^{\texttt{t}}$, where the superscript `$\texttt{t}$' denotes transposition, can be decomposed into \cite{Gazeaut1992}:
\begin{eqnarray}\label{space-time-Lorentz}
g= \begin{pmatrix}
\alpha & \beta \\
\beta^\ast & \alpha^\ast
\end{pmatrix} = j\;l\,,
\end{eqnarray}
where $ l \mathfrak{i}(l)= l l^\texttt{t} =\mathbbm{1}_2$ (that is, $l$ is orthogonal; $l^\texttt{t} = l^{-1}$). The set of all matrices $l$ forms the noncompact subgroup $\cal{L}$ of SU$(1,1)$:
\begin{eqnarray}\label{Lorentz}
{\cal{L}} = \Bigg\{ l = \begin{pmatrix} \cosh{\frac{\varphi}{2}} & \mathrm{i} \sinh{\frac{\varphi}{2}} \\ - \mathrm{i} \sinh{\frac{\varphi}{2}} & \cosh{\frac{\varphi}{2}} \end{pmatrix} \;;\; \varphi \in \mathbb{R} \Bigg\}\,,
\end{eqnarray}
which is isomorphic to $\mathrm{SO}_0(1,1)$. On the other hand, the element $j$, appeared in Eq. (\ref{space-time-Lorentz}), can be determined through the following equation:
\begin{eqnarray}\label{jjt..}
jj^{\texttt{t}} = gg^{\texttt{t}} = \begin{pmatrix} \alpha^2 + \beta^2 & 2\mbox{Re} (\alpha \beta^\ast) \\ 2\mbox{Re} (\alpha \beta^\ast) &  \alpha^{\ast 2} + \beta^{\ast 2} \end{pmatrix} \equiv \begin{pmatrix} e^{\mathrm{i} \theta}\cosh{\psi} & \sinh{\psi} \\ \sinh{\psi} & e^{- \mathrm{i} \theta}\cosh{\psi} \end{pmatrix}\,,
\end{eqnarray}
where $0 \leqslant \theta = \mbox{arg} (\alpha^2 + \beta^2) < 2\pi$ and $\psi = \sinh^{-1} (\alpha \beta^\ast + \alpha^\ast \beta)\in \mathbb{R}$. A possible solution to the above equation reads:
\begin{eqnarray}
j =  k(\theta) \; a(\psi)\,,
\end{eqnarray}
where:
\begin{eqnarray}
k(\theta) &=& \varrho \; \tilde{k}(\theta)\,,\;\;\;\;\;\;\; \mbox{with}\;\;\; \varrho = \pm \mathbbm{1}_2 \;\;\; \mbox{and} \;\;\; \tilde{k}(\theta) = \begin{pmatrix} e^{\mathrm{i} \theta/2} & 0 \\ 0 & e^{- \mathrm{i} \theta/2} \end{pmatrix}\,,\\
a(\psi) &=& \begin{pmatrix} \cosh{\frac{\psi}{2}} & \sinh{\frac{\psi}{2}} \\ \sinh{\frac{\psi}{2}} & \cosh{\frac{\psi}{2}} \end{pmatrix}\,.
\end{eqnarray}
Given $\theta$ and $\psi$, the parameter $\varphi$ is specified through the identity $l(\varphi) = j^{-1} g$:
\begin{eqnarray}
\varphi = 2\tanh^{-1} \Bigg( - \mathrm{i} \; \frac{ \beta - \alpha^\ast e^{\mathrm{i} \theta} \tanh \frac{\psi}{2} }{ \alpha - \beta^\ast e^{\mathrm{i} \theta} \tanh \frac{\psi}{2} } \Bigg)\,. \nonumber
\end{eqnarray}

The above identities provide a global, but nonunique, decomposition of the dS$_2$ group SU$(1,1)$:
\begin{eqnarray}\label{space-time-Lorentz 2}
\mathrm{SU}(1,1) \ni g = k(\theta) \; a(\psi) \; l(\varphi) = \varrho \exp (\theta Y_s) \; \exp (\psi Y_t) \; \exp (\varphi Y_l)\,,
\end{eqnarray}
where $Y_s$, $Y_t$, and $Y_l$ are the corresponding infinitesimal generators:
\begin{eqnarray}\label{dS2gene}
Y_s &=& \frac{\mathrm{d} \tilde{k}(\theta)}{\mathrm{d}\theta}\Big|_{\theta=0} = \frac{\mathrm{i}}{2} \begin{pmatrix} 1 & 0\\ 0 & -1 \end{pmatrix} = \frac{\mathrm{i}}{2} \sigma_3\,,\nonumber\\
Y_t &=& \frac{\mathrm{d} a(\psi)}{\mathrm{d}\psi}\Big|_{\psi=0} = \frac{1}{2} \begin{pmatrix} 0 & 1\\ 1 & 0 \end{pmatrix} = \frac{1}{2}\sigma_1\,,\nonumber\\
Y_l &=& \frac{\mathrm{d} l(\varphi)}{\mathrm{d}\varphi}\Big|_{\varphi=0} = \frac{1}{2} \begin{pmatrix} 0 & \mathrm{i} \\ - \mathrm{i} & 0 \end{pmatrix} = -\frac{1}{2}\sigma_2\,,
\end{eqnarray}
while $\sigma_k$'s ($k=1,2,3$) are the Pauli matrices. One can easily check that these generators obey the following commutation rules:
\begin{eqnarray}\label{dS2gencom''}
[Y_s, Y_l] = - Y_t\,, \;\;\;\;\;\;\; [Y_s, Y_t] = Y_l\,, \;\;\;\;\;\;\; [Y_l,Y_t]= Y_s\,,
\end{eqnarray}
which represent the $\mathfrak{su}(1,1)$ Lie algebra.

Now, we make clear in what sense we call the group decomposition (\ref{space-time-Lorentz 2}) the space-time-Lorentz decomposition of the dS$_2$ group SU$(1,1)$. In the above parametrization, the factor $j\equiv j(\theta,\psi)$ is indeed a kind of ``spacetime" square root, which provides a system of global coordinates for dS$_2$ spacetime. To see the point, we define the three coordinates in ${\mathbb{R}}^3$ as:
\begin{eqnarray}\label{global coordinates}
x^0 = R \sinh{\psi}\,,\;\;\;\;\;\;\; x^1 = R\cos{\theta}\cosh{\psi}\,,\;\;\;\;\;\;\; x^2 = R\sin{\theta}\cosh{\psi}\,,
\end{eqnarray}
or equivalently, as:
\begin{eqnarray}\label{jjt}
R jj^{\texttt{t}} \begin{pmatrix} 0 & 1 \\ 1 & 0 \end{pmatrix} = \begin{pmatrix} R\sinh{\psi} & Re^{\mathrm{i} \theta}\cosh{\psi} \\ Re^{- \mathrm{i} \theta}\cosh{\psi} & R\sinh{\psi} \end{pmatrix}
\equiv \begin{pmatrix} x^0 & x^1 + \mathrm{i} x^2 \\ x^1 - \mathrm{i} x^2 & x^0 \end{pmatrix} = \Upsilon (x)\,,
\end{eqnarray}
where, like before, $0<R<\infty$. The SU$(1,1)$ group acts on the $\Upsilon(x)$'s set by the left action on the set of matrices $j$:
\begin{eqnarray}\label{w}
\mathrm{SU}(1,1) \ni g \;:\; j \; \mapsto \; j^\prime \equiv g \diamond j\,, \;\;\;\;\;\;\; gj = j^\prime l^\prime\,,
\end{eqnarray}
and therefore:
\begin{eqnarray}\label{Upsilonx}
\Upsilon(x^\prime) &=& R\; j^\prime j^{\prime \texttt{t}} \;{\begin{pmatrix} 0 & 1 \\ 1 & 0 \end{pmatrix}}\nonumber\\
&=& R\; g\; j \; j^{\texttt{t}} g^{\texttt{t}} \;{\begin{pmatrix} 0 & 1 \\ 1 & 0 \end{pmatrix}}\nonumber\\
&=& g\; \bigg[ {R\; j\;j^{\texttt{t}} \begin{pmatrix} 0 & 1 \\ 1 & 0 \end{pmatrix}}\bigg] \; \bigg[{\begin{pmatrix} 0 & 1 \\ 1 & 0 \end{pmatrix} g^{\texttt{t}} \begin{pmatrix} 0 & 1 \\ 1 & 0 \end{pmatrix}}\bigg] \nonumber\\
&=& g\; \Upsilon(x)\; g^\dagger\,.
\end{eqnarray}
This action is linear and, as is obvious from Eq. (\ref{Upsilonx}), is determinant-preserving:
\begin{eqnarray}\label{wiprodet}
\mbox{det} \big(\Upsilon(x^\prime)\big) = \mbox{det} \big(\Upsilon(x)\big) = (x^0)^2 - (x^1)^2 - (x^2)^2 = -R^2\,.
\end{eqnarray}
The above identity clearly defines an embedded hyperboloid (of constant radius $R$) in ${\mathbb{R}}^3$ of the $1+1$ dS$_2$ spacetime, while each point of it is in one-to-one correspondence with each class of the left coset $\mathrm{SU}(1, 1)/\cal{L}$, namely, $\mathrm{SU}(1, 1) = \mbox{dS}_2 \times \cal{L}$.

In this realization, the subgroup $\cal{L}$ represents the \emph{stabilizer}\footnote{Let $G$ be a Lie group that acts on a manifold $M$. By definition, the stabilizer of a given point $p \in M$ is a subgroup of $G$ defined by ${\cal{S}}(p) = \big\{ g\in G \;;\; g\diamond p = p \big\}$. ${\cal{S}}(p)$ is also called \emph{little} or \emph{isotropy group} of $p$. Note that if $p^\prime$ is another point in $M$, for which there exists an element $g \in G$ such that $g \diamond p = p^\prime$ (or in other words, if $p^\prime$ belongs to the orbit of $p$ under the action of $G$; $p^\prime \in O(p)$), then the stabilizer of $p^\prime$ would be ${\cal{S}}({p^\prime}) = g {\cal{S}}(p) g^{-1}$, which is isomorphic to ${\cal{S}}(p)$.} of the point ${x}^{}_\odot = (0, R, 0)$, chosen as the origin of the dS$_2$ hyperboloid $M_R$, while the set of matrices $j$ maps this origin to any point $x=(x^0,x^1,x^2)$ in $M_R$:
\begin{eqnarray}\label{spacetime}
j\;\begin{pmatrix} 0 & R \\ R & 0 \end{pmatrix} j^\dagger = \begin{pmatrix} R\sinh{\psi} & Re^{\mathrm{i} \theta}\cosh{\psi} \\ Re^{- \mathrm{i} \theta}\cosh{\psi} & R\sinh{\psi} \end{pmatrix}
\equiv {\begin{pmatrix} x^0 & x^1 + \mathrm{i} x^2 \\ x^1 - \mathrm{i} x^2 & x^0 \end{pmatrix}}\,.
\end{eqnarray}
[The family $(\psi,\theta)$ actually provides a kind of global coordinates for the dS$_2$ hyperboloid $M_R$.] This reveals that SU$(1,1)$ acts \emph{transitively} on $M_R$; its action on $M_R$ has only one orbit, that is, the entire of $M_R$. In this sense, the dS$_2$ hyperboloid is called a SU$(1,1)$ \emph{homogeneous} space.

Moreover, Eq. (\ref{spacetime}) makes clear the interpretation of the transformations generated by $Y_s$ and $Y_t$ (given in Eq. (\ref{space-time-Lorentz 2})). They respectively generate the subgroups of the ``translations in space" isomorphic to $\mathrm{U}(1)$, as the maximal compact subgroup of SU$(1,1)$, and the ``translations in time" isomorphic to $\mathrm{SO}_0(1,1)$. On the other hand, $Y_l$ is interpreted as the generator of the subgroup of the ``Lorentz transformations", since the associated subgroup $\cal{L}$ (isomorphic to the other $\mathrm{SO}_0(1,1)$), which leaves a given point of dS$_2$ spacetime invariant, must be isomorphic to the ordinary $1+1$-dimensional homogeneous Lorentz group. The last point ensures that the neighborhood of any point of dS$_2$ spacetime behaves like flat Minkowski spacetime of special relativity. In this regard, one can also easily check the fact that the dS$_2$ spacetime, realized by the identity (\ref{wiprodet}), is locally Minkowskian by considering the left-invariant metric $\mathrm{d} s^2$, in the global coordinates:
\begin{eqnarray} \label{min coor}
t^{}_{\circ} = R \psi\,, \;\;\;\;\;\;\; \mbox{and} \;\;\;\;\;\;\; x^{}_{\circ} = R\theta\,,
\end{eqnarray}
which yield:
\begin{eqnarray}\label{locally Min}
\mathrm{d} s^2 = \big( \mathrm{d} t_{\circ} \big)^2 - \cosh^2 (R^{-1}t^{}_{\circ}) \; \big( \mathrm{d} x_{\circ} \big)^2\,.
\end{eqnarray}
Note that the coordinates (\ref{min coor}) are obtained by letting $\theta$ and $\psi$ tend to zero in the coordinates yielded by Eq. (\ref{spacetime}) (which exactly coincide with (\ref{global coordinates})).

\subsection{Cartan decomposition}\label{Subsec Cartan dS2}
The Cartan decomposition of SU$(1,1)$, denoted here by $\mathrm{SU}(1,1)=\cal{PK}$, implies that any element $g\in \mathrm{SU}(1,1)$ can be decomposed into two parts \cite{Knapp} (see also Refs. \cite{Gazeaut1992,GazeauBook,Gazeau/del Olmo}):
\begin{eqnarray}\label{Crtan fac.}
g = \begin{pmatrix}
\alpha & \beta \\
\beta^\ast & \alpha^\ast
\end{pmatrix} = p \; k\,,
\end{eqnarray}
with $p \in \cal{P}$ and $k \in \cal{K}$. This decomposition is technically carried out with respect to the Cartan involution $\mathfrak{i}(g) \;:\; g\mapsto (g^\dagger)^{-1}$, in the sense that $\cal{P}$ is made of all elements $p \in \mathrm{SU}(1,1)$ such that $\mathfrak{i}(p)= p^{-1}$, that is, $p \; \big(= p^\dagger \big)$ is Hermitian, while $\cal{K}$ is made of all elements $k \in \mathrm{SU}(1,1)$ unchanged under the involution, that is, $k^\dagger = k^{-1}$, which means that $k$ is unitary. Accordingly, besides the elements $k$ given by:
\begin{eqnarray}
k \equiv k(\theta) = \begin{pmatrix} e^{\mathrm{i} \theta/2}  & 0 \\ 0 & e^{- \mathrm{i} \theta/2} \end{pmatrix}\,, \;\;\;\;\;\;\;  0\leqslant \theta = 2 \;\mbox{arg}\; \alpha < 4\pi\,,
\end{eqnarray}
which were already encountered in the space-time-Lorentz factorization (see subsection \ref{sec space-time-Lorentz}$; {\cal{K}}\sim \mathrm{U}(1)$ is the maximal compact subgroup of SU$(1,1)$), the elements $p$ are determined by:
\begin{eqnarray}\label{facsu11p}
p \equiv p(z) = \begin{pmatrix} \delta & \delta z\\ \delta z^\ast & \delta \end{pmatrix}\,,\;\;\;\;\;\;\; z=\beta/\alpha^\ast, \;\;\; \delta = |\alpha| = (1-|z|^2)^{-1/2}\,,
\end{eqnarray}
with $|z| < 1$.

\begin{figure}[H]
\begin{center}
\includegraphics[height=.4\textheight]{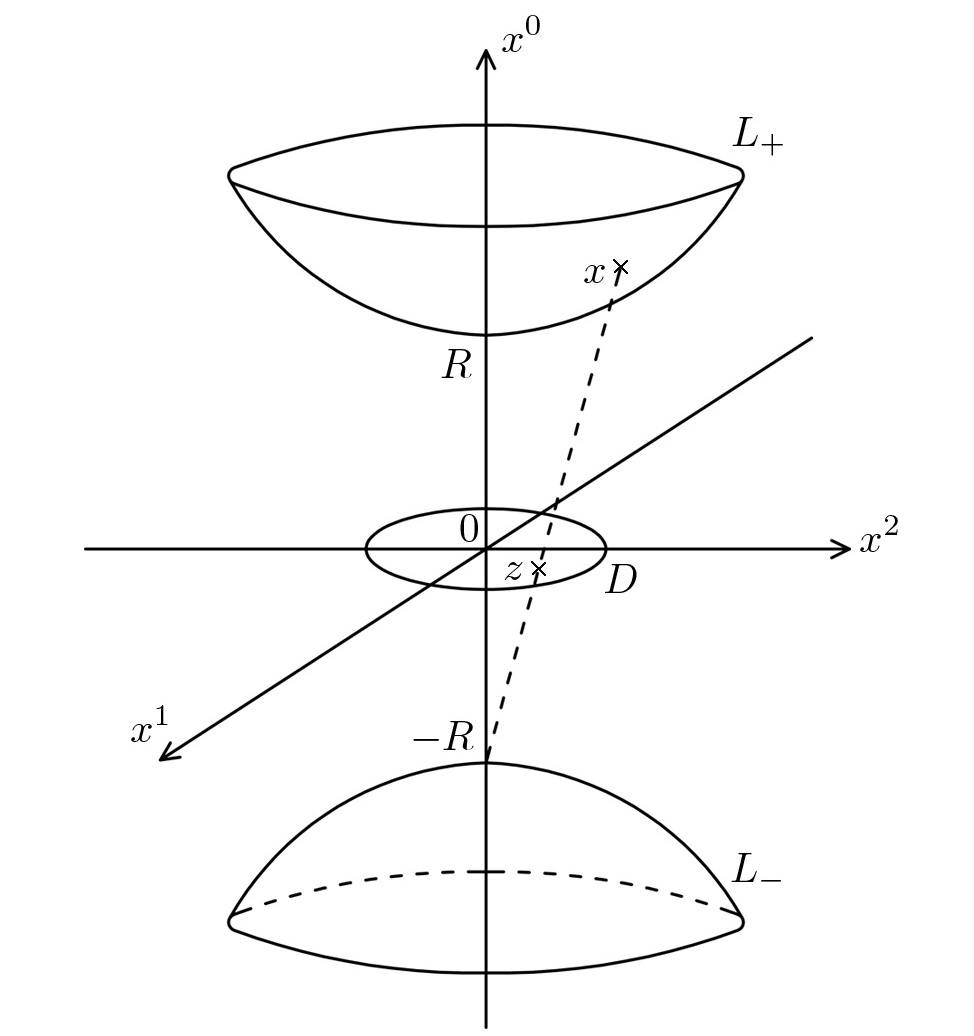}
\end{center}
\caption{$z\in D$ as the stereographic projection of a point $x=(x^0, x^1, x^2) \in L_+$.}
\label{FIG. stereographic}
\end{figure}

The subset of Hermitian matrices $\cal{P}$ is in one-to-one correspondence with the symmetric homogeneous space $\mathrm{SU}(1,1)/{\cal{K}}$. This coset space is in turn homeomorphic to the open unit-disk $D = \big\{ z\in \mathbb{C} \ ;\ |z|<1 \big\}$, which admits the coordinates $\varphi$ and $\varpi$:
\begin{eqnarray}\label{unidisk coor}
z =  e^{\mathrm{i} \varpi} \tanh \textstyle\frac{\varphi}{2}\,, \;\;\;\;\;\;\; \mbox{with} \;\; -\infty < \varphi <\infty \;\; \mbox{and} \;\; 0\leqslant \varpi < 2\pi\,.
\end{eqnarray}
To make this homeomorphism apparent, we first define:
\begin{eqnarray} \label{upper sheet}
\rho pp^\dagger \; \Big(= \rho p^2\Big) \; = \frac{\rho}{1-|z|^2} \begin{pmatrix} 1+|z|^2 & 2z\\ 2z^\ast & 1+|z|^2 \end{pmatrix} \equiv
\begin{pmatrix} x^0 & x^1 + \mathrm{i} x^2\\ x^1 - \mathrm{i} x^2 & x^0 \end{pmatrix}\,,
\end{eqnarray}
where $0<\rho<\infty$ and, again, $x^a$'s ($a = 0,1,2$) are the three Cartesian coordinates in $\mathbb{R}^3$. The action of SU$(1,1)$ on the set of matrices $\rho p^2$, representing the coset space $\mathrm{SU}(1,1)/{\cal{K}}$, can be found from its left action on the matrices $p$:
\begin{eqnarray}\label{1404}
\mathrm{SU}(1,1) \ni g \;:\; p(z) \; \mapsto \; p(z^\prime)\equiv p(g\diamond z)\,, \;\;\;\;\;\;\; g\; p(z) = p(z^\prime)\; k(\theta^\prime)\,,
\end{eqnarray}
from which, we have:
\begin{eqnarray}
\rho p(z^\prime) p^\dagger(z^{\prime}) \; \Big( = \rho p^2(z^\prime) \Big) \; &=& \rho \; \left( g \; p(z) \; k^{-1}(\theta^\prime) \right) \; \left( k(\theta^\prime) \; p(z) \; g^\dagger \right) \nonumber\\
&=& g \; \big( \rho p^2(z) \big) \; g^\dagger\,.
\end{eqnarray}
This action is clearly linear and determinant-preserving:
\begin{eqnarray}\label{2sheet norm}
\mbox{det} \big(\rho p^2(z^\prime)\big) = \mbox{det} \big(\rho p^2(z)\big) = (x^0)^2 - (x^1)^2 - (x^2)^2 = \rho^2\,.
\end{eqnarray}
The above identity shows that each element of the set of matrices $\rho p^2$ is in one-to-one correspondence with each point of the upper sheet $L_+$ of the two-sheeted hyperboloids $(x)^2 = \rho^2$ in $\mathbb{R}^3$:\footnote{Note that each element of the set of matrices $\rho p^2$ is simultaneously in one-to-one correspondence with each point of the lower sheet $L_- \equiv \big\{ x=(x^0, x^1, x^2) \in \mathbb{R}^3 \;;\; (x^0)^2 - (x^1)^2 - (x^2)^2 = \rho^2\,, \; x^0 \leqslant \rho \big\}$. To check this point it is sufficient to allocate a negative sign to $\rho$, namely, $\rho \mapsto -\rho$.}
\begin{eqnarray}\label{upper sheet 1}
L_+ \equiv \Big\{ x=(x^0, x^1, x^2) \in \mathbb{R}^3 \;;\; (x^0)^2 - (x^1)^2 - (x^2)^2 = \rho^2\,, \; x^0 \geqslant \rho \Big\}\,.\;\;\;\;\;
\end{eqnarray}
[The subgroup ${\cal{K}}$ is the stabilizer of the point ${x}^{}_\odot=(\rho,0,0)$ chosen as the origin of the upper sheet.] On the other hand, one can check that the coordinates $(\varphi, \varpi)$, given in Eq. (\ref{unidisk coor}) for the open unit-disk $D$, represent the (pseudo-)angular coordinates for the upper sheet as well:
\begin{eqnarray}\label{coordinate upper sheet}
x^0 &=& \rho\;\frac{1+|z|^2}{1-|z|^2} = \rho\cosh\varphi\,,\nonumber\\
x^1 &=& \rho\;\frac{z + z^\ast}{1-|z|^2}= \rho\sinh\varphi \cos\varpi\,, \nonumber\\
x^2 &=& \mathrm{i} \rho\;\frac{z^\ast - z}{1-|z|^2} = \rho\sinh\varphi \sin\varpi\,.
\end{eqnarray}
This explicitly reveals the well-known correspondence between $L_+$ (say the coset space $\mathrm{SU}(1,1)/{\cal{K}}$) and the open unit-disk $D$. The latter is actually the stereographic projection of the upper sheet (see FIG. \ref{FIG. stereographic}). This projection explicitly reads:
\begin{eqnarray}
L_+ \ni (x^0,x^1,x^2) \; \mapsto \; z = \frac{x^1 + \mathrm{i} x^2}{x^0 + \rho} = \sqrt{\frac{x^0 - \rho}{x^0 + \rho}} \; e^{\mathrm{i}({\mbox{arg}}\; z)} \in D\,.
\end{eqnarray}

Here, it is worthwhile noting that the relation (\ref{1404}) also defines the action of SU$(1,1)$ on the open unit-disk $D$ through the map $z^\prime \equiv g\diamond z$ as:
\begin{eqnarray}\label{transformation disk}
D \ni z \; \mapsto \; z^\prime = (\alpha z + \beta)(\beta^\ast z + \alpha^\ast)^{-1} \in D\,,
\end{eqnarray}
while:
\begin{eqnarray}
k(\theta^\prime) = \begin{pmatrix} \frac{\beta z^\ast + \alpha}{|\beta z^\ast + \alpha|} & 0 \\ 0 & \frac{\beta^\ast z + \alpha^\ast}{|\beta^\ast z + \alpha^\ast|} \end{pmatrix}\,.
\end{eqnarray}
The action of SU$(1,1)$ therefore leaves the set of matrices $p(z)$, where $z \in D$, invariant under the \emph{homographic} (or M\"{o}bius) type transformations (\ref{transformation disk}).

\subsection{Iwasawa decomposition}\label{Subsec Iwasawa dS2}
According to the so-called Iwasawa decomposition \cite{Knapp} (see also Refs. \cite{Gazeaut1992,GazeauBook}), any element $g \in \mathrm{SU}(1,1)$ admits a unique factorization in the following form:
\begin{eqnarray}\label{exsSUuni}
g = \begin{pmatrix}
\alpha & \beta \\
\beta^\ast & \alpha^\ast
\end{pmatrix} = k(\theta)\; a(\psi)\; n(\lambda)\,,
\end{eqnarray}
where:
\begin{eqnarray}
k(\theta) &=& \tilde{k}(\theta) \; \varrho = \begin{pmatrix} e^{\mathrm{i} \theta/2} & 0 \\ 0 & e^{- \mathrm{i} \theta/2} \end{pmatrix} \times \big(\pm \mathbbm{1}_2\big)\,,\;\;\;\;\;\;\; \mbox{elliptic class}\,, \\
a(\psi) &=& \begin{pmatrix} \cosh{\frac{\psi}{2}} & \sinh{\frac{\psi}{2}} \\ \sinh{\frac{\psi}{2}} & \cosh{\frac{\psi}{2}} \end{pmatrix}\,, \;\;\;\;\;\;\; \mbox{hyperbolic class}\,,\\
n(\lambda) &=& \begin{pmatrix} 1+ \mathrm{i} \lambda/2 & - \mathrm{i} \lambda/2 \\ \mathrm{i} \lambda/2 & 1- \mathrm{i} \lambda/2 \end{pmatrix}\,, \;\;\;\;\;\;\; \mbox{parabolic class}\,,
\end{eqnarray}
with:
\begin{eqnarray}
\theta &=& 2 \; \mbox{arg} (\alpha + \beta)\,, \;\;\; \varrho = \mathbbm{1}_2\,, \;\;\; \mbox{for} \;\;\; 0\leqslant \mbox{arg} (\alpha + \beta) <\pi\,,\nonumber\\
\theta &=& 2 \; \mbox{arg} (\alpha + \beta) - 2\pi\,, \;\;\; \varrho = - \mathbbm{1}_2\,, \;\;\; \mbox{for} \;\;\; \pi\leqslant \mbox{arg} (\alpha + \beta) <2\pi\,,\nonumber\\
\psi &=& 2 \; \ln |\alpha + \beta|\in \mathbb{R}\,, \;\;\; \mbox{and} \;\;\; \lambda = 2 \; \frac{\mbox{Im} (\alpha\beta^\ast)}{|\alpha + \beta|^2}\in\mathbb{R}\,.
\end{eqnarray}
The first two factors, $k(\theta)$ and $a(\psi)$, already appeared in the space-time-Lorentz factorization (see subsection \ref{sec space-time-Lorentz}). They respectively belong to the maximal compact subgroup, ${\cal{K}} \sim \mathrm{U}(1)$, and the Cartan maximal Abelian subgroup, ${\cal{A}}\sim \mathrm{SO}_0(1,1)$. The factor $n(\lambda)$, on the other hand, belongs to the nilpotent subgroup ${\cal{N}} \sim \mathbb{R}$. Note that the subgroups ${\cal{A}}$ and ${\cal{N}}$ are of noncompact type and \emph{simply connected}\footnote{A Lie group $G$ is said to be simply connected if every loop in its underlying manifold $M_G$ can be shrunk continuously to a point in $M_G$.}. The Iwasawa decomposition of SU$(1,1)$ then reads $\mathrm{SU}(1,1) = {\cal{KAN}}$.

In this context, we denote by ${\cal{M}} \sim \mathbb{Z}^2 \sim \{ \varrho = \pm \mathbbm{1}_2 \}$ the \emph{centralizer} of ${\cal{A}}$ in ${\cal{K}}$, that is, ${\cal{M}}= \big\{ k \in {\cal{K}} \;;\; ka = ak, \; \forall a \in {\cal{A}} \big\}$. Now, since both ${\cal{A}}$ and ${\cal{M}}$ \emph{normalize} ${\cal{N}}$,\footnote{This means that, for all $a\in {\cal{A}}$ and $m\in {\cal{M}}$, we have $a{\cal{N}}={\cal{N}}a$ and $m{\cal{N}}={\cal{N}}m$, respectively.} ${\cal{B}}= {\cal{MAN}}$ is a closed subgroup of SU$(1,1)$, called minimal parabolic subgroup (note that ${\cal{AN}}$ is a \emph{solvable}\footnote{A group $G$ is called solvable if there are subgroups $1=G_0<G_1<\;...\;<G_k=G$ such that $G_{j-1}$ normalizes $G_j$, and the quotient group $G_j/G_{j-1}$ is an abelian group, for all $j = 1, 2,\;...\;, k$.} connected subgroup). The coset space $ \mathrm{SU}(1,1)/{\cal{B}} \sim {\cal{K}} / {\cal{M}}$, yielded by the minimal parabolic subgroup, clearly characterizes the unit-circle $\mathbb{S}^1$ (the boundary of the open unit-disk $D$). With respect to the above group decomposition, the action of the SU$(1,1)$ group on this coset space can be found through the usual multiplication of the set of matrices $\tilde{k}(\theta)$ from the left:
\begin{eqnarray}
\mathrm{SU}(1,1) \ni g \; : \; \tilde{k}(\theta) \; \mapsto \; \tilde{k}(\theta^\prime)\equiv \tilde{k}(g\diamond \theta)\,, \;\;\;\;\;\;\; g\; \tilde{k}(\theta) = \tilde{k}(\theta^\prime)\; \varrho^\prime a^\prime n^\prime\,,
\end{eqnarray}
based upon which, we get:
\begin{eqnarray}\label{uni circle}
\mathbb{S}^1 \ni e^{\mathrm{i} \theta} \; \mapsto \; e^{\mathrm{i} \theta^\prime} = (\alpha e^{\mathrm{i} \theta} + \beta)(\beta^\ast e^{\mathrm{i} \theta} + \alpha^\ast)^{-1} \in \mathbb{S}^1\,.
\end{eqnarray}
Note that $\big({\alpha e^{\mathrm{i} \theta} + \beta}\big) \big/ \big({\beta^\ast e^{\mathrm{i} \theta} + \alpha^\ast}\big) = \big({\alpha e^{\mathrm{i} \frac{\theta}{2}} + \beta e^{\mathrm{-i} \frac{\theta}{2}}}\big) \big/ {\big(\alpha e^{\mathrm{i} \frac{\theta}{2}} + \beta e^{\mathrm{-i} \frac{\theta}{2}}\big)^\ast}$.

The map (\ref{uni circle}) explicitly recovers the action of SU$(1,1)$ on the open unit-disk $D$ extended to its boundary (see the relation (\ref{transformation disk})). Utilizing this map, one can also recover the action (\ref{Upsilonx}) of SU$(1,1)$ on the set of matrices $\Upsilon\big(x(\theta,\psi)\big) \equiv \begin{pmatrix} x^0 & x^1 + \mathrm{i} x^2 \\ x^1 - \mathrm{i} x^2 & x^0 \end{pmatrix}$ when, admitting the coordinates (\ref{global coordinates}), $\psi$ tends to infinity; symbolically:
\begin{eqnarray}\label{000000000}
\Upsilon \big(x(\theta^\prime,\psi^\prime)\big) = g \Big( \lim_{\psi \rightarrow \infty} \Upsilon\big(x(\theta,\psi)\big) \Big) g^\dagger\,.
\end{eqnarray}
Actually, for the elements $\Upsilon\big(x(\theta,\psi)\big)$, by letting $\psi$ tend to infinity, we get $x^0 / (x^1 - \mathrm{i} x^2) \approx e^{\mathrm{i} \theta}$, and consequently, $\mbox{det} \big(\Upsilon(x)\big) = (x^0)^2 - |x^1 - \mathrm{i} x^2|^2 \approx 0$. On the other hand, for the transformed elements $\Upsilon \big(x(\theta^\prime,\psi^\prime)\big)\equiv \begin{pmatrix} x^{\prime 0} & x^{\prime 1} + \mathrm{i} x^{\prime 2} \\ x^{\prime 1} - \mathrm{i} x^{\prime 2} & x^{\prime 0} \end{pmatrix}$, we have $x^{\prime 0} / (x^{\prime 1} - \mathrm{i} x^{\prime 2}) \approx (\alpha e^{\mathrm{i} \theta} + \beta)(\beta^\ast e^{\mathrm{i} \theta} + \alpha^\ast)^{-1} \equiv e^{\mathrm{i} \theta^\prime}$, and therefore, $\mbox{det} \big(\Upsilon(x^\prime)\big) \approx 0$. This result, as one can easily see, is clearly recovered by the map (\ref{uni circle}). Note that this realization of the map (\ref{uni circle}), yielded by the space-time-Lorentz decomposition, reveals an interesting manifestation of the unit-circle $\mathbb{S}^1$. As a matter of fact, taking the coordinates (\ref{min coor}) into account, the limit $\psi\rightarrow\infty$ implies that either $t_{\circ}$ tends to infinity, while $R$ is fixed, based upon which $\mathbb{S}^1$ is viewed as the dS$_2$ timelike infinity, or $R$ goes to zero, while $t_{\circ}$ is fixed, based upon which $\mathbb{S}^1$ is viewed as the projective null cone in $\mathbb{R}^3$. In the latter case, by proceeding as before, one can show that the nilpotent subgroup ${\cal{N}}$, appeared in the Iwasawa decomposition, is the stabilizer of the point $x^{}_\odot=(1,1,0)$ chosen as the origin of the upper sheet of the null cone in $\mathbb{R}^3$, and the set of matrices $k(\theta) a(\psi)$ maps this origin to any point $x=(x^0=e^\psi, x^1=e^\psi \cos\theta, x^2=e^\psi \sin\theta)$ on the upper sheet of the null cone:
\begin{eqnarray}
k(\theta) \; a(\psi) \begin{pmatrix} 1 & 1 \\ 1 & 1 \end{pmatrix} a^\dagger(\psi) \; k^\dagger(\theta) = e^\psi \begin{pmatrix} 1 & e^{\mathrm{i} \theta} \\ e^{- \mathrm{i} \theta} & 1 \end{pmatrix} \equiv \begin{pmatrix} x^0 & x^1 + \mathrm{i} x^2 \\ x^1 - \mathrm{i} x^2 & x^0 \end{pmatrix} = \Upsilon^c(x)\,, \;\;\;\;\;\;\; \mbox{det} \big(\Upsilon^c(x)\big) = 0\,.
\end{eqnarray}
Note that the same argument holds for the point $(-1,-1,0)$ considered as the origin of the lower sheet of the null cone in $\mathbb{R}^3$.

\setcounter{equation}{0} \section{DS$_2$ Lie algebra and classical phase spaces}\label{Sec dS2 phase}
In this section, having the above group-theoretical constructions in mind, we take a look at how to describe (free) dS$_2$ elementary systems on the classical level. We begin with a brief review of the notion of the orbits under the co-adjoint action of a Lie group $G$. Such orbits are indeed natural candidates for realizing phase spaces of classical systems \cite{Kirillov,Kirillov1976}.

Note that, since in the present paper we are concerned with the matrix realization of the dS$_2$ and, in the sequel, dS$_4$ groups and their respective algebras, we only review formulations of those statements relevant to our discussion. A more general consideration can be found in the book \cite{Kirillov1976} or in the survey \cite{Kirillov}.

\subsection{Co-adjoint orbits: a brief introduction}\label{Subsec coadjoint dS2-int}
A Lie group has natural actions on its Lie algebra and its dual, called adjoint and co-adjoint actions, respectively. To elaborate these notions, let $G$ denote a Lie group with Lie algebra $\mathfrak{g}$. For $g \in G$, one can define a differentiable map from $G$ to itself as $g^\prime\mapsto g\;g^\prime g^{-1}$. This map leaves the identity element $g^\prime=e$ invariant. The derivative of this map at $g^\prime=e$ defines the adjoint action, denoted here by $\mbox{Ad}_g$. This action is an invertible linear transformation of the Lie algebra $\mathfrak{g}$ onto itself. Technically, for $r \in (-\delta,\delta)$, with $\delta> 0$, such that $e^{rY}\in G$ and the infinitesimal generator $Y\in \mathfrak{g}$, the adjoint action reads:
\begin{eqnarray}\label{Ad_g}
\mbox{Ad}_g(Y) \equiv \frac{\mathrm{d}}{\mathrm{d} r}\Big[ g \; e^{rY} g^{-1} \Big] \Big|_{r=0}\,.
\end{eqnarray}
$\mbox{Ad}_g(Y)$ is actually a tangent vector in the tangent space at the identity, $\mbox{Ad}_g(Y) \in \mbox{T}_e G = \mathfrak{g}$. If $G$ is a matrix group, the adjoint action $\mbox{Ad}_g(Y)$ is simply matrix conjugation:
\begin{eqnarray}\label{Ad_g mat}
\mbox{Ad}_g(Y) = g \; Y g^{-1}\,.
\end{eqnarray}

The associated co-adjoint action of $g \in G$, denoted here by $\mbox{Ad}^\sharp_g$, is obtained by dualization; $\mbox{Ad}^\sharp_g$ acts on the \emph{dual linear space} to $\mathfrak{g}$, namely, ${\mathfrak{g}}_{}^{\circledast}$.\footnote{\label{foot}Strictly speaking, the dual space ${\mathfrak{g}}_{}^{\circledast}$ to the Lie algebra $\mathfrak{g}$ is the space of linear maps from $\mathfrak{g}$ to the base field $F$. [Here, for later use and by letting $\mathfrak{g}$ be a Lie algebra with a basis $\big\{ X_i \;;\; i=1,\;...\, , n \big\}$ and structure constants $c^k_{ij}$, it is also useful to point out that the space ${\mathfrak{g}}_{}^{\circledast}$, with coordinates given by the basis above, is naturally equipped with a Poisson structure, defined by the bivector $c=c^k_{ij} X_k {\partial^i} \wedge {\partial^j}$. This Poisson structure forms a Lie algebra isomorphic to $\mathfrak{g}$ \cite{Kirillov}.]} This action explicitly reads:
\begin{eqnarray}\label{Co Ad_g}
\langle \mbox{Ad}_g^\sharp({Y}_{}^{\circledast}) \;;\; Y\rangle = \langle {Y}_{}^{\circledast} \;;\; \mbox{Ad}_{g^{-1}}(Y) \rangle\,,\;\;\;\;\;\;\; {Y}_{}^{\circledast} \in {\mathfrak{g}}_{}^{\circledast}\,,\;\;\;\;\;\;\; Y\in \mathfrak{g}\,,
\end{eqnarray}
where $\langle \cdot \;;\; \cdot \rangle \equiv \langle \cdot \;;\; \cdot \rangle^{}_{{\mathfrak{g}}_{}^{\circledast},\mathfrak{g}}$ stands for the pairing between $\mathfrak{g}$ and its dual ${\mathfrak{g}}_{}^{\circledast}$. Under the co-adjoint action, one can split the vector space ${\mathfrak{g}}_{}^{\circledast}$ into a union of disjoint co-adjoint orbits. We call the set:
\begin{eqnarray}\label{gcoadorbit}
O({Y}_{}^{\circledast}) \equiv \Big\{\mbox{Ad}^\sharp_g({Y}_{}^{\circledast}) \;;\; g \in G \Big\}\,,
\end{eqnarray}
the co-adjoint orbit of ${Y}_{}^{\circledast} \in {\mathfrak{g}}_{}^{\circledast}$. $O({Y}_{}^{\circledast})$ is a homogeneous space for the co-adjoint action of $G$. The point to be noticed here is that the adjoint and co-adjoint actions of a group $G$ are generally inequivalent. They are equivalent if and only if $\mathfrak{g}$ admits a nondegenerate bilinear form, which is the case, for instance, for semi-simple Lie groups \cite{Kirillov,Kirillov1976}.

Physically, co-adjoint orbits are of great significance, since, according to the Kirillov-Souriau-Kostant theory \cite{Kirillov1976}, each co-adjoint orbit carries a natural $G$-invariant symplectic structure. This particularly means that the orbit is of even dimension and carries a natural $G$-invariant (Liouville) measure. In this sense, a co-adjoint orbit $O({Y}_{}^{\circledast})$ is a natural candidate for the phase-space realization of a classical system.

We now turn to our case; $G=\mathrm{SU}(1,1)$. As a byproduct of the discussions given in subsection \ref{sec space-time-Lorentz}, the matrix realization of the dS$_2$ Lie algebra $\mathfrak{su}(1,1)$ can be obtained by the linear span of the three infinitesimal generators $Y_s=\textstyle\frac{\mathrm{i}}{2}\sigma_3$, $Y_t=\textstyle\frac{1}{2}\sigma_1$, and $Y_l=-\textstyle\frac{1}{2}\sigma_2$ ($\sigma_k$'s, with $k = 1, 2, 3$, being the Pauli matrices):
\begin{eqnarray}\label{sugenele}
\mathfrak{su}(1,1) = \Bigg\{ \xi_s Y_s + \xi_t Y_t + \xi_l Y_l =
\frac{1}{2}\begin{pmatrix}
\mathrm{i} \xi_s & \xi_t + \mathrm{i} \xi_l \\
\xi_t - \mathrm{i} \xi_l & - \mathrm{i} \xi_s
\end{pmatrix} \equiv
\begin{pmatrix}
\mathrm{i} u & \zeta \\
\zeta^\ast & - \mathrm{i} u
\end{pmatrix} \;;\;\;\; \xi_s,\xi_t,\xi_l\in\mathbb{R}\Bigg\}\,.
\end{eqnarray}
It then follows that $\mathfrak{su}(1,1)$ is specified by (three) free real parameters $\xi_s$, $\xi_t$, and $\xi_l$, and hence, there is a one-to-one correspondence between this algebra and $\mathbb{R}^3$ (by abuse of notation, let us say $\mathfrak{su}(1,1) \sim \mathbb{R}^3$).

The $\mathfrak{su}(1,1)$ Lie algebra is simple, and admits the following symmetric bilinear form:
\begin{eqnarray}\label{su11bifo}
\langle Y_1 ; Y_2 \rangle \equiv \mbox{tr}(Y_1Y_2) = 2 \Big(-u_1u_2 + \mbox{Re}(\zeta^{}_1)\mbox{Re}(\zeta^{}_2) + \mbox{Im}(\zeta^{}_1)\mbox{Im}(\zeta^{}_2)\Big)\,,
\end{eqnarray}
which, as already expected for a simple Lie algebra, is nondegenerate\footnote{This form is proportional to the so-called Killing form for Lie algebras $\mathfrak{g}$, namely $K(X,Y)\equiv \mathrm{tr}(\mathrm{ad}_X\mathrm{ad}_Y)$, where the adjoint action $\mathrm{ad}_X$ is defined as the linear action $\mathfrak{g}\ni Y\mapsto \mathrm{ad}_X (Y) \equiv [X,Y]$. [Note that it is common in the literature to denote the adjoint action of Lie algebras on themselves by the symbol `$\mathrm{ad}$' (e.g., $\mathrm{ad}_X$), while `$\mathrm{Ad}$' is kept for the adjoint action of groups (e.g., $\mathrm{Ad}_g$).] This action is precisely the derivative of the adjoint group action \eqref{Ad_g mat} on its Lie algebra (more details on this topic can be found, as a byproduct of our discussion concerning the dS$_4$ complex Lie algebra, in appendix \ref{App Lie algebra B2}). In the present case, we have $\langle Y_1 ; Y_2 \rangle = K(Y_1,Y_2)$.}. Therefore, as mentioned above, the classification of its co-adjoint orbits would be equivalent to the classification of its adjoint orbits:
\begin{eqnarray}\label{adacsu11}
\mbox{Ad}_g(Y) = gYg^{-1} =
\begin{pmatrix}
\mathrm{i} u^\prime & \zeta^\prime\\
\zeta^{\prime\ast} & - \mathrm{i} u^\prime
\end{pmatrix}\,,
\end{eqnarray}
where, borrowing the notations used in Eq. (\ref{ginSU1,1}), $g = \begin{pmatrix} \alpha & \beta \\ {\beta^\ast} & {\alpha^\ast} \end{pmatrix}$, with $\alpha,\beta \in \mathbb{C}$ and $|\alpha|^2-|\beta|^2=1$, and consequently, $u^\prime = u\big( |\alpha|^2+|\beta|^2\big) - 2\mbox{Im}(\alpha \beta^\ast \zeta)$ and $\zeta^\prime = -2 \mathrm{i} \alpha\beta u + \alpha^2\zeta - \beta^2 {\zeta^\ast}$. The action (\ref{adacsu11}) is linear and determinant-preserving:
\begin{eqnarray}
\mbox{det}(gYg^{-1}) = \mbox{det}(Y) = \xi_s^2 - \xi_t^2 - \xi_l^2 = r\,, \;\;\;\;\;\;\; r\in\mathbb{R}\,.
\end{eqnarray}
Accordingly, proceeding as before (see subsection \ref{Subsec 2.1}), one can show that in the vector space $\mathfrak{su}(1,1) \sim \mathbb{R}^3$, under the (co-)adjoint action (\ref{adacsu11}), three types of orbits (apart from the trivial one $\xi_s = \xi_t = \xi_l = 0$ (the origin)) appear: the one-sheeted hyperboloids, with $r < 0$; the upper and lower sheets of the two-sheeted hyperboloids, with $r > 0$ ($\xi_s \gtrless 0$, respectively); and the upper and lower sheets of the cone, with $r = 0$ ($\xi_s \gtrless 0$, respectively).

Below, we will focus on the first type of the (co-)adjoint orbits, i.e., the one-sheeted hyperboloids, and following the lines sketched in Ref. \cite{three quantization}, we will show that it can be interpreted as a phase space for the set of free motions on dS$_2$ spacetime, with fixed ``energy" at rest.

\subsection{DS$_2$ (co-)adjoint orbits as possible phase spaces for motions on dS$_2$ spacetime}\label{Subsec dS2 phase}
The first type of the (co-)adjoint orbits, introduced above, corresponds to the transport of the particular element $2\kappa Y_t$, with $0 <\kappa<\infty$, under the action (\ref{adacsu11}). With respect to this action, the subgroup stabilizing the element $2\kappa Y_t$ is the dS$_2$ time-translation subgroup. [Recall from subsection \ref{sec space-time-Lorentz} that this subgroup is of noncompact type (hyperbolic class) and isomorphic to $\mathrm{SO}_0(1,1)$.] Actually, this family of the $\mathfrak{su}(1,1)$ (co-)adjoint orbits admits a homogeneous space realization identified by the coset space $O(2\kappa Y_t) \sim \mathrm{SU}(1,1)/\mathrm{SO}_0(1,1)$. According to the space-time-Lorentz decomposition (\ref{space-time-Lorentz 2}), this realization is achieved by applying the Lorentz boosts and space translations to transport the element $2\kappa Y_t$ under the action (\ref{adacsu11}):
\begin{eqnarray}\label{acton2kYt}
\mbox{Ad}_g(2\kappa Y_t) &=& k(\theta)\; l(\varphi) \; \Big( 2\kappa Y_t \Big)\; l^{-1}(\varphi)\; k^{-1}(\theta)\nonumber\\
&=& \kappa
\begin{pmatrix}
\mathrm{i} \sinh\varphi & e^{\mathrm{i} \theta}\cosh\varphi \\
e^{- \mathrm{i} \theta}\cosh\varphi & - \mathrm{i} \sinh\varphi
\end{pmatrix} \equiv Y(\theta,\varphi)\,,
\end{eqnarray}
where $k^{-1}(\theta) = k(-\theta)$ and $l^{-1}(\varphi) = l(-\varphi)$. The above action is linear and determinant-preserving, and since $\mbox{det} \big( Y(\theta,\varphi) \big) = -\kappa^2 < 0$, as already mentioned, it sets up a family of (co-)adjoint orbits of hyperbolic type in the vector space $\mathfrak{su}(1,1) \sim \mathbb{R}^3$, which divides the exterior of the cone (in ${\mathbb{R}}^3$) into a union of mutually disjoint hyperboloids of different radii $\kappa$.

Interestingly, for a given $\kappa$, there is an biunivocal correspondence between the matrix $Y(\theta,\varphi)$, representing a point in the (co-)adjoint orbit $O(2\kappa Y_t)$, and a point in \emph{the phase space of a relativistic test particle in dS$_2$ spacetime}\footnote{Note that such a phase space is defined by identifying classical states, which differ one from another by a time translation.}. To make this correspondence apparent, we first adopt the following parametrization:
\begin{eqnarray}
p = \kappa\sinh\varphi\,,\;\;\;\;\;\;\; \theta = \varpi + \tan^{-1}\frac{\kappa}{p}\,,\nonumber
\end{eqnarray}
where $p\in\mathbb{R}$ and $0 \leqslant \varpi < 2\pi$. Besides $(\theta,\varphi)$, the new parameters $(p,\varpi)$ also define a system of global coordinates for the (co-)adjoint orbit $O(2\kappa Y_t)$. Accordingly, we can rewrite the orbit generic element (\ref{acton2kYt}) as:
\begin{eqnarray}\label{tvtopbe}
Y(\theta,\varphi) \;\rightarrow\; {\mathtt{Y}}(p,\varpi) =
\begin{pmatrix}
\mathrm{i} p & (p+ \mathrm{i} \kappa)e^{\mathrm{i} \varpi}\\
(p- \mathrm{i} \kappa)e^{- \mathrm{i} \varpi} & - \mathrm{i} p
\end{pmatrix}\,.
\end{eqnarray}
Here, concerning the latter parameters (i.e., $(p,\varpi)$), we note in passing that, letting $\kappa$ ($0 <\kappa<\infty$) be proportional to the particle ``mass"\footnote{We will revisit this ambiguous notion of mass in Part \ref{Part mass}.}, one can define the Minkowskian-like ``energy" identity as $p^{}_0 \equiv |(p+ \mathrm{i} \kappa)e^{\mathrm{i} \varpi}|= \sqrt{p^2 + \kappa^2}$. This interpretation is actually quite interesting, since it associates with each dS$_2$ ``massive" test particle, with a specific ``mass", a (co-)adjoint orbit $O(2\kappa Y_t)$ of a certain radius $\kappa$, while on this orbit the aforementioned ``energy" identity remains unchanged (the latter point stems from the fact that the ``energy" identity is consistent with $\mbox{det}\big( {\mathtt{Y}}(p,\varpi) \big)=-\kappa^2$). Moreover, the latter parameters, based on the notion of Poisson manifolds \cite{Kirillov}, make apparent the canonical symplectic structure of the orbit. To see the point, one needs to proceed with the following diffeomorphism:
\begin{eqnarray}\label{sysdiff}
J_0 &\equiv& p\,,\nonumber\\
J_1 &\equiv& p\cos\varpi - \kappa\sin\varpi\,,\nonumber\\
J_2 &\equiv&  p\sin\varpi + \kappa\cos\varpi\,.
\end{eqnarray}
Note that the triplet $(J_0,J_1,J_2)$ are indeed cartesian coordinates for the orbit $O(2\kappa Y_t)$:
\begin{eqnarray}\label{111111111}
{\mathtt{Y}}(p,\varpi) =
\begin{pmatrix}
\mathrm{i} p & (p+ \mathrm{i} \kappa)e^{\mathrm{i} \varpi}\\
(p- \mathrm{i} \kappa)e^{- \mathrm{i} \varpi} & - \mathrm{i} p
\end{pmatrix} =
\begin{pmatrix}
\mathrm{i} J_0 & J_1 + \mathrm{i} J_2\\
J_1 - \mathrm{i} J_2 & - \mathrm{i} J_0
\end{pmatrix}\,,
\end{eqnarray}
with:
\begin{eqnarray}\label{phshyp}
J_2^2 + J_1^2 - J_0^2 = \kappa^2\,.
\end{eqnarray}
Now, following the simple instruction given in Ref. \cite{Piechocki}, one can show that the coordinates $(p,\varpi)$ are canonical with respect to the two-form $\Omega = \mathrm{d} p\wedge \mathrm{d} \varpi$; by defining the Poisson bracket on \emph{classical dS$_2$ observables}\footnote{The term (classical) ``observable" is allocated to an infinitely differentiable function $f$ on a classical phase space, equipped with some measure, susceptible to being measured within the framework imposed by some experimental or observational protocol. In the specific case of a test particle, since it does not change spacetime, it is truly expected that the local symmetries of the corresponding phase space are reflected by the algebra of all Killing vector fields. In this sense, the symmetry generators would provide the basic classical observables of the system.} $f(p,\varpi)$:
\begin{eqnarray}\label{ds2pobr}
\big\{ f_1,f_2 \big\} \equiv \frac{\partial f_1}{\partial p}\frac{\partial f_2}{\partial \varpi} - \frac{\partial f_1}{\partial\varpi}\frac{\partial f_2}{\partial p}\,,
\end{eqnarray}
one can show the canonical relation $\big\{ p,\varpi \big\} = 1$ and the commutation rules of the $\mathfrak{su}(1,1)$ algebra:
\begin{eqnarray}\label{su11corps}
\big\{J_0, J_1\big\}= -J_2\,,\;\;\;\;\;\;\; \big\{J_0, J_2\big\} = J_1\,, \;\;\;\;\;\;\; \big\{J_1, J_2\big\} = J_0\,.
\end{eqnarray}

In the above sense, we identify the canonical coordinates $(p,\varpi)$ as the phase-space parameters. As a matter of fact, the point $(p,\varpi)$ belongs to the space $\mathbb{R}^1 \times \mathbb{S}^1 = \big\{ x\equiv (p, \varpi)\; ; \; p\in\mathbb{R},\;\; 0 \leqslant \varpi < 2\pi \big\}$, which defines the phase space of a test particle moving on the unit circle ($p$ and $\varpi$, respectively, stand for a momentum and a position). Following the instruction given in appendix \ref{App UIR's SU(2)}, one can simply show that the invariant measure on this phase space, in terms of the coordinates $(p,\varpi)$, is given by $\mathrm{d} \mu (p,\varpi) = \mathrm{d} p \mathrm{d} \varpi$. Here, we also would like to emphasize the fact that the introduced matrix realization of the classical phase space for a ``massive" test particle in dS$_2$ spacetime is very convenient for describing the action of the symmetry group, since we simply have ${\mathtt{Y}}(p^\prime, \varpi^\prime) = g {\mathtt{Y}}(p,\varpi)g^{-1}$, while $\mbox{det} \big({\mathtt{Y}}(p^\prime, \varpi^\prime)\big) = \mbox{det} \big({\mathtt{Y}}(p,\varpi)\big) = - \kappa^2$, which means that the relation $p^{}_0 = \sqrt{p^2 + \kappa^2} = \sqrt{J_1^2 + J_2^2}$, determining the Minkowskian-like ``energy", remains intact.

As a final remark and without going into the details, we would like to point out that the phase space for ``massless" particles in dS$_2$ spacetime, strictly speaking, the ``massless" (co-)adjoint orbit, can be realized by a limiting process from the hyperbolic (``massive") (co-)adjoint orbits, possessing the generic element (\ref{111111111}), by letting $\kappa$ (say ``mass") tend to zero. One can easily show that this limiting process yields the two sheets of the cone in $\mathfrak{su}(1,1) \sim \mathbb{R}^3$, for which the nilpotent subgroup ${\cal{N}}$ (appeared in the Iwasawa decomposition of the dS$_2$ group) plays the role of stabilizer subgroup. To see more details on the last topic, readers are referred to Ref. \cite{deBievre}.

\setcounter{equation}{0} \section{UIR's of the dS$_2$ group and quantum version of dS$_2$ motions}\label{Sec Quant version dS2}
Quantization is generally understood as the transition from classical to quantum mechanics, where, in the latter, a physical system is described by states which are vectors (up to a phase) in a Hilbert space (and more generally, by density operators). Technically, to articulate more clearly the notion of quantum processing of a classical system, let $\Gamma$ denote a canonical phase space, equipped with some measure $\mathrm{d}\mu$ (for instance, its canonical phase-space measure); as pointed out above, $\Gamma$ can be identified by a co-adjoint orbit. As a measure space, $\Gamma$ (strictly speaking, $(\Gamma, \mathrm{d}\mu)$) provides us with a statistical reading of the set of all measurable real- or complex-valued functions $f(x)$ on $\Gamma$. [It allows to calculate, for instance, mean values on subsets with bounded measure.] In quantum mechanics, we are interested in quadratic mean values, therefore, the natural framework of study seems to be the Hilbert space $L^2(\Gamma)$ (strictly speaking, $L^2(\Gamma,\mathrm{d}\mu)$) of all square-integrable functions $f(x)$ on $\Gamma$:
\begin{eqnarray}
\int_\Gamma |f(x)|^2 \; \mathrm{d}\mu(x) < \infty\,.
\end{eqnarray}
The functions $f(x)$ might be referred to as (pure) quantum states in quantum mechanics. But, of course, not all square-integrable functions are eligible as quantum states. In order to select the true (projective) Hilbert space of quantum states, denoted here by ${\cal{H}}$ (that is, a closed subspace of $L^2(\Gamma)$), one needs a continuous map $\Gamma \ni x \mapsto |x\rangle \in {\cal{H}}$ (in \emph{Dirac notations}\footnote{As is well known, in Dirac notations, any quantum state $|x\rangle$ can be written in terms of an orthonormal basis $\big\{ |n\rangle \big\}_{n\in\mathbb{N}}$ of $\cal{H}$, which is in one-to-one correspondence with an orthonormal set $\big\{ f_n(x) \big\}_{n\in\mathbb{N}}$, as members of $L^2(\Gamma)$.}), which defines a set of states $\big\{ |x\rangle \big\}_{x\in\Gamma}$ verifying the following requirements:
\begin{itemize}
\item{normalization:
     \begin{eqnarray}\label{normalization}
     \langle x | x \rangle = 1\,,
     \end{eqnarray}}
\item{resolution of the unity in ${\cal{H}}$:
     \begin{eqnarray}\label{unity}
     \int_\Gamma |x \rangle \langle x| \; \mathrm{d}\nu(x) = \mathbbm{1}_{\cal{H}}\,,
     \end{eqnarray}}
\end{itemize}
where $\mathrm{d}\nu(x)$ is another measure on $\Gamma$, usually absolutely continuous in terms of $\mathrm{d}\mu(x)$; this implies that there is a positive measurable function $h(x)$, for which, we have $\mathrm{d}\nu(x) = h(x) \mathrm{d}\mu(x)$. Note that, in the context of quantum mechanics, a physical system possessing the above requirements is called an elementary system.

The quantization of a classical observable $J(x)$ then is carried out by associating with $J(x)$ the operator:
\begin{eqnarray}
\hat{J} \equiv \int_\Gamma J(x) |x \rangle \langle x| \; \mathrm{d}\nu(x)\,,
\end{eqnarray}
provided that for the classical observable $J(x)$, in the context of the theory of operators in Hilbert spaces, the above expansion is mathematically justified. In this regard, we must underline that the correspondence $J \mapsto \hat{J}$ is linear, and that the function $J(x)=1$ goes to the identity operator.

Generally, to get such families of quantum states $|x\rangle$, different approaches are considered in the literature (see Refs. \cite{GazeauBook,three quantization,Englis} and references therein, for an overview of some of the better known quantization techniques found in the current literature and used both by physicists and mathematicians). In this paper, we particularly interested in the group-theoretical quantization method. As a matter of fact, when the global symmetry of a classical phase-space $\Gamma$ is characterized by a Lie group $G$ with its Lie algebra $\mathfrak{g}$ being isomorphic to the Lie algebra of a local symmetry of $\Gamma$, which is exactly the case in our study, application of the group-theoretical quantization scheme, in the following sense, is quite rational. It simply includes finding a UIR of the symmetry group $G$ on a Hilbert space ${\cal{H}}$. The representation space ${\cal{H}}$ (in some restricted sense) identifies the corresponding quantum states space, since, in this context, the map $\Gamma \ni x \mapsto |x\rangle \in {\cal{H}}$ is well established: on one hand, a specific state, say, $|x_0 \rangle$, is transported along the orbit $\big\{ |g \diamond x_0 \equiv x \rangle, \; g \in G \big\}$ by the action of the group $G$ based upon which $\Gamma$ is a homogeneous space and, on the other hand, the requirements of unitarity, irreducibility (Schur lemma), and square integrability of the representation in some restricted sense automatically yield the identities (\ref{normalization}) and (\ref{unity}). Eventually, applying the Stone theorem \cite{Stone} to the UIR of one-parameter subgroups of $G$ gives the associated self-adjoint operators representing quantum observables. [More technically, if $G$ is represented on a Hilbert space ${\cal H}$ by unitary operators $U(g)$ which are continuous in $g$ ($g$ being group elements of $G$), every one-parameter subgroup may be expressed, according to the Stone theorem, in the form $U_t = \exp (- \mathrm{i} t\hat{J})$, where $\hat{J}$ is an (essentially) self-adjoint operator on ${\cal H}$. On the other hand, since every one-parameter subgroup is generated by an infinitesimal transformation (an element of the Lie algebra $\mathfrak{g}$) of the group $G$, there is a correspondence between the operators $\hat{J}$ and the elements of $\mathfrak{g}$; here, it is worth recalling from the previous section that the symmetry generators (the elements of $\mathfrak{g}$) provide us with the basic classical observables of the system.]

Regarding our group-theoretical approach to the quantization of classical phase spaces of dS$_2$ elementary systems, here it would be convenient to point out that all UIR's of the dS$_2$ group $\mathrm{SU}(1,1)\sim \mathrm{SL}(2,\mathbb{R})$ have been constructed a long time ago by Bargmann in his seminal work \cite{Bargmann}. These UIR's fall basically into three distinguished categories/series: the principal, complementary, and discrete series. In a shortcut, while we have in mind the three types of (co-)adjoint orbits (phase spaces) that appear for (free) elementary systems in the context of dS$_2$ relativity (see section \ref{Sec dS2 phase}), we assert that:
\begin{itemize}
\item{The quantization of the one-sheeted hyperboloid (co-)adjoint orbits leads to the principal or complementary series representations.}
\item{The quantization of the two-sheeted hyperboloid (co-)adjoint orbits leads to the holomorphic and anti-holomorphic discrete series.}
\item{The quantization of the two sheets of the cone in $\mathbb{R}^3$ yields the conformally extendable UIR's in the discrete series.}
\end{itemize}
The first two cases will be discussed in the coming section. For the third case, readers are referred to Ref. \cite{deBievre}.

\setcounter{equation}{0} \section{UIR's of the dS$_2$ group: global realization}
In this section, we aim to find (essentially) self-adjoint representations of the $\mathfrak{su}(1,1)$ algebra, which are integrable to the UIR's of the SU$(1,1)$ group. Technically, to do this, one may consider either a global procedure, as has been done by Bargmann \cite{Bargmann}, or a Lie algebraic method, as has been performed by Biedenharn et al. \cite{Biedenharn}. Here, we stick to the former approach, but for the sake of comparison, we first point out the gist of the latter approach and its results.

In the context of algebraic approach, in summary, the three dS$_2$ Killing vectors $K_{ab}$ (see Eq. (\ref{Killing dS2})) are represented by three (essentially) self-adjoint operators in the Hilbert space of square-integrable functions, according to some invariant inner product of Klein-Gordon type, on $M_R$ (or on the (co-)adjoint orbits (phase spaces) given in section \ref{Sec dS2 phase}). In the first case, these representations read:
\begin{eqnarray}
K_{ab} \;\mapsto\; L_{ab} = M_{ab} = -\mathrm{i} (x_a \partial_b - x_b \partial_a)\,.
\end{eqnarray}
The associated second-order Casimir operator then reads:
\begin{eqnarray}
Q = - \frac{1}{2} M_{ab} M^{ab} = - t(t+1)\mathbbm{1}\,,
\end{eqnarray}
where its eigenvalues characterize the set of dS$_2$ UIR's such that the values $t = -\frac{1}{2} - \mathrm{i} v$, with $-\infty < v < \infty$, refer to the principal series, $-1 < t < 0$ to the complementary series, and $t = -1, -3/2, -2, ...$ to the (holomorphic) discrete series, respectively.

Below, as mentioned above, we will elaborate these results in a global realization.

\subsection{Group representations: a brief introduction}\label{Subsec rep,int}
A (linear) representation of a group $G$ is a continuous function $G \ni g \mapsto U(g)$, which admits values in the group of nonsingular continuous linear transformations of a vector space $V$ on $\mathbb{R}$ or on $\mathbb{C}$ and verifies the functional equations $U(g_1 g_2) = U(g_1) U(g_2)$, for all $g_1,g_2 \in G$, and $U(e) = \mathbbm{1}$, where $e$ and $\mathbbm{1}$ refer to the identity element of $G$ and the identity operator in the vector space $V$, respectively. It then follows that $U(g^{-1}) = U^{-1}(g)$. In this sense, the representation $U(g)$ is a homomorphism of $G$ into the group of nonsingular continuous linear transformations of $V$.

A representation is called \emph{unitary}, if the linear operators $U(g)$ are unitary according to an inner product $\langle \cdot , \cdot \rangle$ defined on $V$.\footnote{Note that all different types of inner (or scalar) products, which are appeared in this paper, are marked by the same symbol `$\langle \cdot , \cdot \rangle$'. Whenever, it is necessary to distinguish between two different types of inner products, proper subscripts will be considered.} This implies that $\langle U(g)\mathfrak{v}_1 , U(g)\mathfrak{v}_2 \rangle = \langle \mathfrak{v}_1 , \mathfrak{v}_2 \rangle$, for all vectors $\mathfrak{v}_1,\mathfrak{v}_2 \in V$. On the other hand, a representation is called \emph{irreducible}, if there is no nontrivial subspace $V_0 \subset V$, which is invariant under the operators $U(g)$. Technically, this means that there is no nontrivial subspace $V_0 \subset V$, such that for all vectors $\mathfrak{v}_0 \in V_0$ and all $g \in G$, we have $U(g)\mathfrak{v}_0 \in V_0$.

\subsubsection{Group representations by shift operators}\label{Subsub Shift operators}
Let $G$ be a \emph{transformation group of a set $S$},\footnote{By definition, a group $G$ is called a transformation group of a set $S$, if with each $g \in G$, one can associate a transformation $s \mapsto s^\prime = g \diamond s$ of $S$, while, for any two elements $g_1,g_2 \in G$ and $s \in S$, we have $(g_1g_2) \diamond s = g_1 \diamond (g_2 \diamond s)$.} and $V$ be a linear space of functions $f(s)$ for $s \in S$. Then, for each invariant subspace $V_0 \subset V$, a representation $U(g)$ of the group $G$ can be realized by the left-shift operator $U(g) f \equiv f^\prime $, defined in such a way that:
\begin{eqnarray}\label{shift}
f^\prime(s^\prime) = f(g^{-1} \diamond s^\prime) = f(s)\,.
\end{eqnarray}
Note that the introduction of the inverse $g^{-1}$ in Eq. (\ref{shift}) ensures that the action defines a group homomorphism:
\begin{eqnarray}\label{zzzzzzz}
\left(U(g^{}_1)\left( U(g^{}_2)f\right)\right)(s) = \left(U(g^{}_1) f\right)(g_2^{-1}\diamond s) = f\big(g_2^{-1}\diamond (g^{-1}_1\diamond s) \big) = f\big( (g^{}_1g^{}_2)^{-1}\diamond s \big) = \left(U(g_1g_2) f\right)(s)\,,
\end{eqnarray}
for all $g_1, g_2 \in G$ and $f \in V_0$.

\subsubsection{Induced representations}\label{Subsub Induced Rep.}
The regular representation is indeed a special case of induced representations, which were first given a firm footing by Mackey in the 1940's and early 1950's \cite{Mackey,Mackey'}. In summary, induced representations are constructed in the following way. Let $Q$ and $T(q)$ ($q\in Q$), respectively, denote a subgroup of a group $G$ and its representation in a vector space $V^{}_Q$. We also consider $V^{}_G$ as the set of vector functions $f(s)$ on $S=G$, taking values in $V^{}_Q$ and verifying the following conditions:
\begin{itemize}
\item{First, for any element $\mathfrak{q}$ of $V^{}_Q$, the scalar function $\langle f(s) , \mathfrak{q} \rangle$ on $S=G$ is measurable with respect to the left-invariant measure $\mathrm{d} \mu(s)$ on $S=G$, as well as $\langle f(s) , f(s) \rangle$, where:
    \begin{eqnarray}
    \int \langle f(s) , f(s) \rangle \; \mathrm{d} \mu(s) < \infty\,.
    \end{eqnarray}}
\item{Second, for any element $q$ of $Q$, we have:
    \begin{eqnarray}
    f(s\diamond q) = T(q) f(s)\,.
    \end{eqnarray}}
\end{itemize}
On this basis, one can easily show that the space $V^{}_G$ is invariant for the left-shift operator $\big( U(g) f \big)(s) = f(g^{-1}\diamond s)$, for all $g\in G$. Actually, the first condition is trivially fulfilled by $f(g^{-1}\diamond s)$, and the second condition by:
\begin{eqnarray}
\left(U(g) f\right)(s\diamond q) = f(g^{-1}\diamond s\diamond q) = T(q) f(g^{-1}\diamond s) = T(q) \left(U(g) f\right)(s)\,.
\end{eqnarray}
It follows from the invariance of $V^{}_G$ that the shift operator $U(g)$ is a representation of $G$, known as \emph{the representation induced by the representation $T(q)$ of the subgroup $Q$}.

Note that the regular representation of $G$, pointed out in subsubsection \ref{Subsub Shift operators}, is induced by the trivial representation of the identity subgroup $Q=\{e\}$, for which the second condition is automatically verified. Another interesting case appears when we consider the trivial representation of any subgroup $Q$, namely, $T(q)={\mathbbm{1}}$ for all $q \in Q$, which is realized in the space of constant functions on $Q$. In this case, the second condition turns into $f(s\diamond q) = f(s)$, based upon which the trivial representation $T(q)={\mathbbm{1}}$ induces the representation $\big(U(g) f \big)(s) = f(g^{-1}\diamond s)$, in the space of vector functions $f(s)$ on the homogeneous coset space $S = G/Q$ (for which, $Q$ being the stabilizer of the orbit points, we automatically have $f(s\diamond q) = f(s)$ for all $q \in Q$).

\subsubsection{Representations of groups with operator factors}\label{Subsubsec operator factor}
We again invoke the notations/mathematical materials introduced in subsubsection \ref{Subsub Shift operators}. A representation of the form of shift operators on the (infinite dimensional) function space $V_0$ (in the sense given in subsubsections \ref{Subsub Shift operators} and \ref{Subsub Induced Rep.}) typically includes a wide variety of important subrepresentations, including representations on subspaces of polynomials, continuous functions, smooth (differentiable) functions, analytic functions, normalizable ($L^2$) functions, and so on, however it is not yet quite general enough for our purposes. Accordingly, we consider a more general construction of representations for transformation groups, in which shifts are combined with multiplications by an operator-valued function $N(g,s)$, where $g\in G$ and $s\in S$. $N(g,s)$ is indeed an automorphic factor verifying:
\begin{equation}
\label{Ngs}
N(e,s)=\bu\, , \quad N(g_1 g_2,s) = N(g_1, s)N(g_2,g_1^{-1} \diamond s) \, \Rightarrow\, N^{-1}(g, s)=N(g^{-1},g^{-1}\diamond s)\,.
\end{equation}
Then, a \emph{multiplier representation} explicitly reads:
\begin{eqnarray}\label{multi rep}
\big( U(g) f\big)(s) = N(g,s) \; f(g^{-1}\diamond s)\,.
\end{eqnarray}
Proceeding as Eq. (\ref{zzzzzzz}), one can easily show that the above action defines a group homomorphism.

For more detailed discussions on the above topics, readers are referred to Refs. \cite{Vilenkin,Lipsman}.

\subsection{Principal series}
In this subsection, we discuss the construction of the principal series representations of SU$(1,1)$ with respect to the Mackey's method of induced representations \cite{Mackey,Mackey'} (see also Refs. \cite{Vilenkin,Lipsman,Takahashi}), which is based on the existence of a characteristic solvable connected subgroup. We employ the Iwasawa decomposition of SU$(1,1)$ (see subsection \ref{Subsec Iwasawa dS2}) to make the associated solvable connected subgroup explicit.

According to the Iwasawa decomposition of $\mathrm{SU}(1,1) = {\cal{KAN}}$, this group admits the minimal parabolic subgroup ${\cal{B}}= {\cal{MAN}}$. [We recall that ${\cal{K}}\sim \mathrm{U}(1)$ is the maximal compact subgroup, ${\cal{A}}\sim \mathrm{SO}_0(1,1)$ is the Cartan maximal abelian subgroup, ${\cal{M}}$ is the centralizer of ${\cal{A}}$ in ${\cal{K}}$, and ${\cal{N}}\sim\mathbb{R}$ is the nilpotent subgroup. Moreover, we point out again that ${\cal{AN}}$ is a solvable connected subgroup.] In the context of our current discussion, the existence of the subgroup ${\cal{B}}$ is of great significance, since the quotient manifold $\mathrm{SU}(1,1)/{\cal{B}} \sim {\cal{K}} / {\cal{M}}$, homeomorphic to the unit-circle $\mathbb{S}^1$, carries the principal series UIR's of SU$(1,1)$.\footnote{For the role of parabolic subgroups in the construction of unitary representations, one can refer to Refs. \cite{Wolf,Wolf'}.} The principal series is indeed induced by a continuous homomorphism of the minimal parabolic subgroup ${\cal{B}}$ on $\mathbb{S}^1$ (i.e. a character of ${\cal{B}}$), for which the Lie algebra of ${\cal{AN}}$ acts as a \emph{real polarization} for the differential of this character.\footnote{Let $\mathfrak{g}$ be an arbitrary Lie algebra (over $\mathbb{R}$) and ${\mathfrak{g}}_{}^{\circledast}$ be the dual vector space to $\mathfrak{g}$. If $\langle f;X\rangle$ denotes the pairing of ${\mathfrak{g}}_{}^{\circledast}\times \mathfrak{g}\rightarrow \mathbb{R}$, then we may define an alternating bilinear form $B_f$ on $\mathfrak{g}$ by:
\begin{eqnarray}
B_f(X,Y) = \langle f;[Y,X]\rangle\,, \;\;\;\; f\in {\mathfrak{g}}_{}^{\circledast} \;\mbox{and}\; X,Y\in\mathfrak{g}\,,\nonumber
\end{eqnarray}
where $[Y,X]$ is the bracket in $\mathfrak{g}$. By definition, a subalgebra $\mathfrak{h}\subset\mathfrak{g}$ is said to be a real polarization at $f$ if $\mathfrak{h}$ is a maximal totally isotropic subspace for $B_f$, i.e., $\langle f;[X,\mathfrak{h}]\rangle =0 \iff X\in\mathfrak{h}$. See more details in Ref. \cite{Lipsman}.} Note that the UIR of ${\cal{M}}$, denoted here by $\mathfrak{m}$, reduces to $\pm \mathbbm{1}$ (``even" or ``odd"), while $\mathfrak{a}$ as the UIR of ${\cal{A}}$ is a character of $\mathbb{R}$. The associated UIR's of the principal series then form the following set of representations:
\begin{eqnarray}
\label{USU11Ind}
U_{}^{\mbox{\small{ps}}} = \mbox{Ind}_{\cal{B}}^{\mathrm{SU}(1,1)} (\mathfrak{m} \times \mathfrak{a})\,.
\end{eqnarray}
Below, admitting the so-called ``compact" realization \cite{Lipsman}, we will describe these UIR's.

\subsubsection{Representation space}
We here follow Vilenkin in \cite{Vilenkin} for better understanding  the role of the unit circle in the description of Hilbert space carrying the principal and complementary series. Let $(\varepsilon,t)$ denote the set of pairs of numbers $\varepsilon$, taking the values $0$ and $1/2$, and complex numbers $t$. With each such family we associate a space $\mathfrak{D}^{}_{\varepsilon,t}$ of all complex-valued functions $\varphi(z)$ of the complex variable $z=x+\ii y$, such that:
\begin{itemize}
\item{The functions $\varphi(z)$ are infinitely differentiable in $x$ and $y$ at all points $z=x+\ii y\in\mathbb{C}$, with the exception of $z=0$.}
\item{The functions $\varphi(z)$ have given parity:
    \begin{eqnarray}
    \varphi(-z) = (-1)^{2\varepsilon}\varphi(z)\,,
    \end{eqnarray}
    where, for $\varepsilon=0$, they are even and, for $\varepsilon=1/2$, they are odd.}
\item{The functions $\varphi(z)$ are homogeneous, with the degree of homogeneity $2t$:
    \begin{eqnarray}
    \varphi(a z) = a^{2t}\varphi(z)\,,
    \end{eqnarray}
    for any positive number $a$.}
 \end{itemize}

Now, let $\daleth$ denote some curve on the complex plane $\mathbb{C}$, with one and only one intersection with any straight line passing through the origin $z=0$. Then, having the aforementioned parity and homogeneity properties in mind, each function $\varphi(z) \in \mathfrak{D}^{}_{\varepsilon,t}$ can be uniquely determined by its values on $\daleth$  through the homogeneity property:
\begin{equation}
\label{phizz0}
\varphi(z)= \left\vert\frac{z}{z_0}\right\vert^{2t}\left(\frac{z}{z_0}\right)^{2\varepsilon} \varphi(z_0)\,,
\end{equation}
where $z_0$  is the point of intersection of the curve $\daleth$ and  the line joining this point to the origin. Accordingly, $\mathfrak{D}^{}_{\varepsilon,t}$ can be viewed as a space of functions on $\daleth$. Of course, if $\daleth$ intersects the straight lines passing through the origin at several points, the space $\mathfrak{D}^{}_{\varepsilon,t}$ can be realized by the space of functions given on $\daleth$ which verifies certain additional criterions, arising from the parity and homogeneity of $\varphi(z)$ in $\mathfrak{D}^{}_{\varepsilon,t}$. For instance, $\daleth$ being the unit circle $\mathbb{S}^1$ (compact realization), the space $\mathfrak{D}^{}_{\varepsilon,t}$, for $\varepsilon=0$, can be realized by the space of \emph{even} infinitely differentiable (in $x$ and $y$) functions on $\mathbb{S}^1$ (they take equal values at diametrically opposite points) and, for $\varepsilon=1/2$, by the space of \emph{odd} infinitely differentiable (in $x$ and $y$) functions on $\mathbb{S}^1$ (they take values at diametrically opposite points which differ only in sign).

It is actually more convenient to merge the two cases $\varepsilon=0$ and $\varepsilon=1/2$ into a unique one by defining in each case a  function $f$ on $\mathbb{S}^1$ as:
\begin{align}
\label{fctfS1e}
\mathfrak{D}^{}_{0,t}&\ni \varphi \;\mapsto\; f\left(e^{\ii\varpi}\right) =
  \varphi\left(e^{\ii\frac{\varpi}{2}}\right)\,, \\
\label{fctfS1o}\mathfrak{D}^{}_{1/2,t}&\ni \varphi \;\mapsto\; f\left(e^{\ii\varpi}\right) = e^{\ii\frac{\varpi}{2}}   \varphi\left(e^{\ii\frac{\varpi}{2}}\right)\,.
\end{align}
Hence, for any  pair $(\varepsilon,t)$ the space $\mathfrak{D}^{}_{\varepsilon,t}$ can be realized as the space $\mathfrak{D}$ of infinitely differentiable (in $x$ and $y$) functions on $\mathbb{S}^1$. However there is a deep  topological meaning behind the existence of the two possibilities, even ($\varepsilon=0$) and  odd ($\varepsilon=1/2$). It is related to that notion of \emph{spin} making the difference between the unit circle (one complete round) and its double covering (two complete rounds). The even functions in $\mathfrak{D}^{}_{0,t}$, when they are expressed as $ \varphi\left(e^{\ii\frac{\varpi}{2}}\right)$, do not feel the difference and can be qualified of scalar fields while the odd functions in $\mathfrak{D}^{}_{1/2,t}$ feel it and recover their original value after two complete rounds, and hance can be qualified of spinorial fields.

\subsubsection{Representations}
Let us first introduce the following representation of SU$(1,1)$ in the space $\mathfrak{D}_{\varepsilon,t}$:
\begin{equation}
\label{repDet}
\mathrm{SU}(1,1) \ni g =
\begin{pmatrix}
    \alpha & \beta \\
   \beta^\ast & \alpha^\ast
\end{pmatrix}: \,\varphi(z) \;\mapsto\; \Big(T_{\varepsilon,t}(g)\varphi\Big)(z)= \varphi(\alpha^\ast z - \beta z^\ast)\,,
\end{equation}
One easily check that the parity $\varepsilon$ and the degree of  homogeneity $t$ are conserved and the representation property $T_{\varepsilon,t}(g_1)T_{\varepsilon,t}(g_2)=T_{\varepsilon,t}(g_1g_2)$ holds true. Let us now express this representation in terms of actions on the space of functions $f$ in $\mathfrak{D}$ as there introduced in Eqs. \eqref{fctfS1e} and \eqref{fctfS1o}. By using the homogeneity property and a minimum of calculus with complex numbers of unit modulus, one finds that the action \eqref{repDet} becomes the multiplier representation which we now denote by $U_{\varepsilon,t}$ in order to be consistent with \eqref{USU11Ind}:
\begin{equation}
\label{repD}
\begin{split}
\mathrm{SU}(1,1) \ni g =
\begin{pmatrix}
    \alpha & \beta \\
   \beta^\ast & \alpha^\ast
\end{pmatrix}: \,f\left(e^{\ii\varpi}\right) \;\mapsto\; \Big(U_{\varepsilon,t}(g)f\Big)\left(e^{\ii\varpi}\right)&= (-\beta^\ast e^{\ii\varpi} + \alpha)^{t+\varepsilon} (-\beta e^{-\ii\varpi} + \alpha^\ast)^{t-\varepsilon}\,f\left(\frac{\alpha^\ast e^{\ii\varpi} -\beta}{-\beta^\ast e^{\ii\varpi} + \alpha}\right)\\
&\equiv  N_{\varepsilon,t}\left(g, e^{\ii\varpi}\right)\,f\left(g^{-1}\diamond e^{\ii\varpi}\right)\,.
\end{split}
\end{equation}
Recall that according to the Iwasawa decomposition of SU$(1,1)$, we have :
\begin{eqnarray}
\mathbb{S}^1 \ni e^{\ii\varpi} \;\mapsto\; e^{\ii\varpi^\prime} = g^{-1}\diamond e^{\ii\varpi} = (\alpha^\ast e^{\ii\varpi} - \beta)(-\beta^\ast e^{\ii\varpi} + \alpha)^{-1} \in \mathbb{S}^1\,.
\end{eqnarray}

\subsubsection{Hilbert space and condition for being unitary}
In the sense of \eqref{USU11Ind} and the above material, the compact realization of the principal series UIR's of SU$(1,1)$ entails the action of the representation operators $U_{\varepsilon,t}$ in the Hilbert space  $L^2_\mathbb{C}(\mathbb{S}^1,\ud \varpi/2\pi)\equiv L^2_\mathbb{C}(\mathbb{S}^1) $ \emph{densely}\footnote{\label{dense} A subspace $M_d$ of a topological space $M$ is called dense in $M$, if every point in $M$ either belongs to $M_d$ or is a limit point of $M_d$, that is, the closure of $M_d$ constitutes the whole set $M$; for instance, the rational numbers are a dense subspace of the real numbers because every real number either is a rational number or has a rational number arbitrarily close to it. Note that, equivalently, $M_d$ is dense in $M$ if and only if the smallest closed subspace of $M$ containing $M_d$ is $M$ itself.}  generated by all those complex-valued  functions $f\left(e^{\ii\varpi}\right)$, $\varpi\in [0,2\pi)$ mod\,$2\pi$, in $\mathfrak{D}$, which are square integrable with respect to the scalar product:
\begin{eqnarray}\label{norm principal dS2}
\langle f_1 , f_2 \rangle = \frac{1}{2\pi} \int_{\mathbb{S}^1} \; f_1^\ast\left(e^{\ii\varpi}\right) f^{}_2\left(e^{\ii\varpi}\right) \; \mathrm{d} \varpi\,,
\end{eqnarray}
namely:
\begin{eqnarray}
{\| f \| }^2 \equiv \langle f , f \rangle = \frac{1}{2\pi} \int_{\mathbb{S}^1} \; | f\left(e^{\ii\varpi}\right) |^2 \; \mathrm{d} \varpi < \infty\,.
\end{eqnarray}
The representations (\ref{repD}) are unitary if we set $t= - \frac{1}{2} - \mathrm{i} v$, while $v \in \mathbb{R}$. The proof is based on the transformation of the differential $\ud z$ under the homographic action $z\mapsto z^{\prime}= (a z+ b)(cz+d)^{-1}$ in the complex plane:
\begin{equation}
\label{dzdzp}
\ud z^{\prime}= \ud \frac{a z + b}{c z + d}= \frac{ad -bc}{(cz+d)^2}\ud z\,,
\end{equation}
and in particular for $z= e^{\ii\varpi}$, $z^{\prime}= e^{\ii\varpi^{\prime}}$, and $g = \begin{pmatrix} \alpha & \beta \\ \beta^\ast & \alpha ^\ast \end{pmatrix}\in$ SU$(1,1)$:
\begin{equation}
\label{dthetdthetp}
\ud \varpi^{\prime}= \frac{\ud \varpi}{\left\vert\alpha e^{\ii \varpi}+ \beta\right\vert^2}\,.
\end{equation}
Hence, for $g\in$ SU$(1,1)$, and with $e^{\ii\varpi^{\prime}}= g^{-1}\diamond e^{\ii \varpi}$:
\begin{equation}\label{unitary PS2}
\begin{split}
\langle U_{\varepsilon,t} (g)f_1 , U_{\varepsilon,t} (g) f_2 \rangle &=\frac{1}{2\pi} \int_{\mathbb{S}^1} \; f_1^\ast\left(g^{-1}\diamond e^{\ii \varpi}\right) \; \left\vert N_{\varepsilon,t}(g,z)\right\vert^2 \; f_2\left(g^{-1}\diamond e^{\ii \varpi}\right) \; \mathrm{d} \varpi\\
&= \frac{1}{2\pi} \int_{\mathbb{S}^1} \; f_1^\ast\left( e^{\ii \varpi^{\prime}}\right) f_2\left( e^{\ii \varpi^{\prime}}\right) \, \left\vert\alpha^\ast e^{\ii \varpi} -\beta \right\vert^{4\mathrm{Re}(t)+2 }\,\mathrm{d}\varpi^{\prime}\\
&= \langle f_1 , f_2 \rangle\quad \mbox{if}\quad t=-\frac{1}{2} -\ii v\,,
\end{split}
\end{equation}
which implies that the representations $U_{\varepsilon,t=-\frac{1}{2}- \mathrm{i} v}\equiv U^{\mbox{\small{ps}}}_{\varepsilon,t=-\frac{1}{2}- \mathrm{i} v}$ are unitary. From the above, it is also obvious that the two families of the representations $U^{\mbox{\small{ps}}}_{\varepsilon,t=-\frac{1}{2}- \mathrm{i} v}$ and $U_{\varepsilon,-t-1=-\frac{1}{2}+ \mathrm{i} v}^{\mbox{\small{ps}}}$ are unitary equivalent.

Below, we will show that there is a correspondence between the parameters $v$ and $\kappa$ (the latter was already encountered in subsection \ref{Subsec dS2 phase}, and denotes the radius (``mass") of the SU$(1,1)$ (co-)adjoint orbits of hyperbolic (``massive") type), as $\kappa= \pm v$, for $v \gtrless 0$, respectively.

\subsubsection{Irreducibility and infinitesimal operators}
Now, we show that the unitary representations $U^{\mbox{\small{ps}}}_{\varepsilon,t=-\frac{1}{2}- \mathrm{i} v}$, with $\varepsilon=0,1/2$ and $v\in\mathbb{R}$, are irreducible. To do this, the first task is to find an expression for the infinitesimal operators of $U^{\mbox{\small{ps}}}_{\varepsilon,t}$. Considering the Stone theorem \cite{Stone}, while the one-parameter subgroups of SU$(1,1)$ appeared in Eq. (\ref{space-time-Lorentz 2}) are taken into account, below we encounter this matter.

We begin with the subgroup ${\cal{L}}$ of matrices of the form $l(\varphi) = \begin{pmatrix} \cosh{\frac{\varphi}{2}} & \mathrm{i} \sinh{\frac{\varphi}{2}} \\ - \mathrm{i} \sinh{\frac{\varphi}{2}} & \cosh{\frac{\varphi}{2}} \end{pmatrix}$, with $\varphi \in \mathbb{R}$, for which we have:
\begin{eqnarray}
l^{-1}(\varphi) = \begin{pmatrix} \cosh{\frac{\varphi}{2}} & - \mathrm{i} \sinh{\frac{\varphi}{2}} \\ \mathrm{i} \sinh{\frac{\varphi}{2}} & \cosh{\frac{\varphi}{2}} \end{pmatrix}\,.
\end{eqnarray}
From Eq. \eqref{repD}, we have:
\begin{eqnarray}
\Big(U^{\mbox{\small{ps}}}_{\varepsilon,t} \big(l(\varphi)\big) \; f\Big) (e^{\mathrm{i}\varpi}) = \Big(\mathrm{i} \sinh{\textstyle\frac{\varphi}{2}}e^{\mathrm{i}\varpi} + \cosh{\textstyle\frac{\varphi}{2}}\Big)^{t+\varepsilon} \Big(-\mathrm{i}\sinh{\textstyle\frac{\varphi}{2}} e^{-\mathrm{i}\varpi} + \cosh{\textstyle\frac{\varphi}{2}}\Big)^{t-\varepsilon} \; f\Bigg( \frac{\cosh{\frac{\varphi}{2}} e^{\mathrm{i}\varpi} - \mathrm{i}\sinh{\frac{\varphi}{2}}}{\mathrm{i}\sinh{\frac{\varphi}{2}} e^{\mathrm{i}\varpi} + \cosh{\frac{\varphi}{2}}} \Bigg)\,.
\end{eqnarray}
After adjoining the usual factor $\mathrm{i}$, the corresponding infinitesimal operator reads:
\begin{eqnarray} \label{hat J_1}
\hat{Y}_l\equiv\hat{J}_1 = \frac{\mathrm{i} \partial\; \Big(U^{\mbox{\small{ps}}}_{\varepsilon,t} \big(l(\varphi)\big)\Big)}{\partial\varphi}\Big|_{\varphi=0} = -(t+\varepsilon) \frac{e^{\mathrm{i}\varpi}}{2} + (t-\varepsilon) \frac{e^{-\mathrm{i}\varpi}}{2} - \mathrm{i}\cos\varpi \frac{\mathrm{d}}{\mathrm{d}\varpi}\,.
\end{eqnarray}

In the same way, to the one-parameter subgroups ${\cal{A}}$ and ${\cal{K}}$, respectively constituted by matrices of the forms $a(\psi) = \begin{pmatrix} \cosh{\frac{\psi}{2}} & \sinh{\frac{\psi}{2}} \\ \sinh{\frac{\psi}{2}} & \cosh{\frac{\psi}{2}} \end{pmatrix}$, with $\psi \in \mathbb{R}$, and $k(\theta) = \begin{pmatrix} e^{\mathrm{i}\theta/2} & 0 \\ 0 & e^{-\mathrm{i}\theta/2} \end{pmatrix}$, with $0\leqslant \theta <4\pi$, correspond the respective infinitesimal operators:
\begin{eqnarray} \label{hat J_2}
\hat{Y}_t\equiv\hat{J}_2 = \frac{\mathrm{i}\partial\; \Big(U^{\mbox{\small{ps}}}_{\varepsilon,t}\big(a(\psi)\big)\Big)}{\partial\psi}\Big|_{\psi=0} = {-\mathrm{i}}(t+\varepsilon) \frac{e^{\mathrm{i}\varpi}}{2} - {\mathrm{i}}(t-\varepsilon) \frac{e^{-\mathrm{i}\varpi}}{2} + \mathrm{i}\sin\varpi \frac{\mathrm{d}}{\mathrm{d}\varpi}\,,
\end{eqnarray}
and:
\begin{eqnarray} \label{hat J_0}
\hat{Y}_s\equiv\hat{J}_0 = \frac{\mathrm{i} \partial\; \Big(U^{\mbox{\small{ps}}}_{\varepsilon,t}\big(k(\theta)\big)\Big)}{\partial\theta}\Big|_{\theta=0} = -\varepsilon - \mathrm{i} \frac{\mathrm{d}}{\mathrm{d}\varpi}\,.
\end{eqnarray}
Considering the latter, the operators $\hat{J}_1$ and $\hat{J}_2$ can also be rewritten as:
\begin{eqnarray} \label{hat Js}
\hat{J}_1 &=& - \mathrm{i} t \sin\varpi + \cos\varpi \hat{J}_0\,,\nonumber\\
\hat{J}_2 &=& - \mathrm{i} t \cos\varpi - \sin\varpi \hat{J}_0\,.
\end{eqnarray}
These operators obey the following commutation relations:
\begin{eqnarray}
[\hat{J}_0,\hat{J}_1] = - \mathrm{i} \hat{J}_2\,, \;\;\;\;\;\;\; [\hat{J}_0,\hat{J}_2] = \mathrm{i} \hat{J}_1\,, \;\;\;\;\;\;\; [\hat{J}_1,\hat{J}_2] = \mathrm{i} \hat{J}_0\,,
\end{eqnarray}
which generate the $\mathfrak{su}(1,1)$ algebra.

Here, it must be underlined that the operators $\hat{J}_a$ ($a=0,1,2$) are unbounded, and hence, they cannot be defined on the whole Hilbert space $L^2_\mathbb{C}(\mathbb{S}^1)$. They are indeed essentially self-adjoint operators on a suitable common \emph{dense subspace} $\Delta \subset L^2_\mathbb{C}(\mathbb{S}^1)$ (see footnote \ref{dense}). In the following, we will show that such a $\Delta$ exists and can be spanned with respect to all finite linear combinations of elements of the orthonormal basis $\big\{ |n\rangle \big\} \equiv \big\{\exp(- \mathrm{i} n\varpi)\big\}$, with $n \in \mathbb{Z}$, in $L^2_\mathbb{C}(\mathbb{S}^1)$ such that the aforementioned infinitesimal operators establish the UIR's $U^{\mbox{\small{ps}}}_{\varepsilon,t}$ of the SU$(1,1)$ group on it. To prove the irreducibility of the representations $U^{\mbox{\small{ps}}}_{\varepsilon,t}$, then, it would be sufficient to show that there are no subspaces $\Delta_0 \subset \Delta$, other than $\{0\}$ and $\Delta$ itself, invariant under the action of all the operators $\hat{J}_0$, $\hat{J}_1$, and $\hat{J}_2$. Here, however, instead of $\hat{J}_0$, $\hat{J}_1$, and $\hat{J}_2$, it would be more convenient to use their linear combinations $\hat{J}^\prime_0 = -\hat{J}^{}_0$ and $\hat{J}_\pm = - \mathrm{i} \hat{J}_2 \mp \hat{J}_1$. Besides the former which is quite obvious, the other two, considering Eqs. (\ref{hat J_1}) and (\ref{hat J_2}), explicitly read:
\begin{eqnarray}\label{J pm}
\hat{J}_+  &=& \mathrm{i} e^{- \mathrm{i} \varpi} \frac{\mathrm{d}}{\mathrm{d}\varpi} - (t-\varepsilon)e^{- \mathrm{i} \varpi}\,,\nonumber\\
\hat{J}_-  &=& - \mathrm{i} e^{ \mathrm{i} \varpi} \frac{\mathrm{d}}{\mathrm{d}\varpi} - (t+\varepsilon)e^{\mathrm{i} \varpi}\,.
\end{eqnarray}
They verify $[\hat{J}^{}_+,\hat{J}^{}_-] = 2 \hat{J}^\prime_0$ and $[\hat{J}^\prime_0,\hat{J}^{}_\pm] = \pm \hat{J}^{}_\pm$. In the sense that this new set of operators is linear combinations of the former, to prove the irreducibility of the representations, it would be sufficient to show the absence of a nontrivial subspace invariant for $\hat{J}^\prime_0$, $\hat{J}^{}_+$, and $\hat{J}^{}_-$.

Technically, the aforementioned orthonormal basis $\big\{ |n\rangle \;;\; n \in \mathbb{Z} \big\}$, which is supposed to generate the common dense invariant subspace $\Delta \subset L^2_\mathbb{C}(\mathbb{S}^1)$, is constituted by the eigenfunctions of $\hat{J}^\prime_0$:
\begin{eqnarray}\label{J'0}
\hat{J}^\prime_0 \; |n\rangle = (\varepsilon + n) \; |n\rangle\,.
\end{eqnarray}
On the other hand, the operators $\hat{J}_+$ and $\hat{J}_-$ act on the functions of this basis, respectively, as raising and lowering operators:
\begin{eqnarray}
\hat{J}_+ \; |n\rangle &=& (\varepsilon -t + n) \;  |n+1\rangle\,,
\end{eqnarray}
\begin{eqnarray}\label{Jpm}
\hat{J}_- \; |n\rangle &=& - (\varepsilon + t + n) \; |n-1\rangle\,.
\end{eqnarray}
Therefore, in the allowed ranges of parameters, the operators $\hat{J}^\prime_0$ and $\hat{J}_\pm$ are well defined in the common dense $\Delta$. Moreover, $\Delta$ is clearly invariant, with no nontrivial subspace invariant, for the given infinitesimal operators. The irreducibility of the representations $U^{\mbox{\small{ps}}}_{\varepsilon,t}$ then is proved.

\subsubsection{Quantum Casimir operator}
The associated quadratic quantum Casimir operator reads:
\begin{eqnarray}\label{Casimir ps}
Q = \hat{J}_2^2 + \hat{J}_1^2 - \hat{J}_0^2 = \frac{-\hat{J}_+ \hat{J}_- - \hat{J}_- \hat{J}_+}{2} - \hat{J}_0^{\prime 2} = - t(t+1)\mathbbm{1}\,, \;\;\;\;\;\;\; t= -\frac{1}{2} - \mathrm{i} v\,, \;\;\; v\in\mathbb{R}\,.
\end{eqnarray}
By adjusting $\kappa= \pm v$, respectively, for $v \gtrless 0$, one can check the correspondence between the above result and the given classical Casimir form (\ref{phshyp}), associated with the phase space of a relativistic ``massive" test particle, with ``mass" $\kappa$, in dS$_2$ spacetime:
\begin{eqnarray}\label{555555555}
Q = \hat{J}_2^2 + \hat{J}_1^2 - \hat{J}_0^2 = (\kappa^2 + \frac{1}{4}) {\mathbbm{1}}\,.
\end{eqnarray}
Considering the above, one can also check that the functions $f(z) \in \Delta$, which characterize the (true) quantum states carrying the principal series representations, are indeed eigenfunctions of the quadratic Casimir operator $Q$ for the eigenvalues $(\kappa^2 + \frac{1}{4})$, namely, $\big( Q - (\kappa^2 + \frac{1}{4}) \big) f(z) = 0$. This delicate point (extended to $1+3$ dimension and of course to the whole three series of the UIR's) will be used later in part \ref{Part plane waves}, when the spacetime realization of the dS$_4$ representations is taken into account, to give the ``wave equations" of dS$_4$ elementary systems.

\subsection{Complementary series}
The complementary series of the SU$(1,1)$ UIR's, quite analogous to its principal counterpart (presented in the previous subsection), is realized by induction from the minimal parabolic subgroup UIR's. As a matter of fact, the formula (\ref{repD}) defines the unitary representations $U^{\mbox{\small{ps}}}_{\varepsilon,t=-\frac{1}{2}- \mathrm{i} v}$ (belonging to the principal series) in $L^2_{\mathbb{C}}(\mathbb{S}^1)$ only for $\mathrm{Re}(t)= -1/2$, while the complementary series of the SU$(1,1)$ unitary representations comes to fore by the construction of a Hilbert space, i.e., by the definition of a scalar product, in such a way that the formula (\ref{repD}) defines unitary representations in that space for a range of real values of $t$. More precisely, it is shown \cite{Bargmann} (see also Refs. \cite{Naimark,Takahashi}) that the complementary series UIR's, denoted here by $U^{\mbox{\small{cs}}}$, is specified by $-1 < t < 0$ and $\varepsilon = 0$. The corresponding Hilbert space $L^2_{\mathbb{C}}(\mathbb{S}^1 \times \mathbb{S}^1)$ is defined as the space of functions $f\left(e^{\ii\varpi}\right)$  on the unit-circle ${\mathbb{S}}^1$, which verify:
\begin{eqnarray}
\langle f_1 , f_2 \rangle = c \iint_{\mathbb{S}^1 \times \mathbb{S}^1} f_1^\ast \left(e^{\ii\varpi_1}\right) f_2\left(e^{\ii\varpi_2}\right) \; |\varpi_1 - \varpi_2|^{-2t - 2} \; \mathrm{d} \varpi_1 \mathrm{d} \varpi_2 <\infty\,,
\end{eqnarray}
where $c$ is an arbitrary constant. The operators $U^{\mbox{\small{cs}}}_{0,t}$ act in the Hilbert space as:
\begin{equation}
\begin{split}
\mathrm{SU}(1,1) \ni g =
\begin{pmatrix}
    \alpha & \beta \\
   \beta^\ast & \alpha^\ast
\end{pmatrix}: \,f\left(e^{\ii\varpi}\right) \;\mapsto\; \Big(U^{\mbox{\small{cs}}}_{0,t}(g)f\Big)\left(e^{\ii\varpi}\right)&= \left\vert-\beta^\ast e^{\ii\varpi} + \alpha\right\vert^{2t} \,f\left(\frac{\alpha^\ast e^{\ii\varpi} -\beta}{-\beta^\ast e^{\ii\varpi} + \alpha}\right)\\
&\equiv  N_{0,t}\left(g, e^{\ii\varpi}\right)\,f\left(g^{-1}\diamond e^{\ii\varpi}\right)\,.
\end{split}
\end{equation}
Note that the representations $U^{\mbox{\small{cs}}}_{0,t}$ and $U^{\mbox{\small{cs}}}_{0,-t-1}$ are unitary equivalent.

Considering the above action along with the one-parameter subgroups of SU$(1,1)$ appeared in Eq. (\ref{space-time-Lorentz 2}), the corresponding representatives of the $\mathfrak{su}(1,1)$ Lie algebra elements (\ref{dS2gene}), after adjoining the usual factor $\mathrm{i}$, read:
\begin{eqnarray}
Y_s \;\mapsto\; \hat{Y}_s\equiv\hat{J}_0 &=& - \mathrm{i} \frac{\mathrm{d}}{\mathrm{d}\varpi}\,,\\
Y_l \;\mapsto\; \hat{Y}_l\equiv\hat{J}_1 &=& - \mathrm{i} t \sin\varpi + \cos\varpi \hat{J}_0\,,\\
Y_t \;\mapsto\; \hat{Y}_t\equiv\hat{J}_2 &=& - \mathrm{i} t \cos\varpi -\sin\varpi \hat{J}_0\,,
\end{eqnarray}
with:
\begin{eqnarray}\label{comruk012}
[\hat{J}_0,\hat{J}_1] = - \mathrm{i} \hat{J}_2\,, \;\;\;\;\;\;\; [\hat{J}_0,\hat{J}_2] = \mathrm{i} \hat{J}_1\,, \;\;\;\;\;\;\; [\hat{J}_1,\hat{J}_2] = \mathrm{i} \hat{J}_0\,.
\end{eqnarray}
The associated quantum quadratic Casimir operator is:
\begin{eqnarray}
Q = \hat{J}_2^2 + \hat{J}_1^2 - \hat{J}_0^2 = -t(t+1) {\mathbbm{1}}\,, \;\;\;\;\;\;\; -1 < t < 0\,.
\end{eqnarray}

\subsection{Discrete series}
We come now to a brief description of the discrete series representations of SU$(1,1)$, which is issued from the Cartan decomposition of the latter (see subsection \ref{Subsec Cartan dS2}). [For more detailed discussions, one can refer to Refs. \cite{Lipsman,Perelomov,Gazeau/del Olmo}.] We denote here the representation operators by $U^{\mbox{\small{ds}}}$, which are characterized by the parameters $(\varepsilon,t)$ taking values $t=-1,-2,\;...\;$, for $\varepsilon=0$, and $t=-1/2,-3/2, \;...\;$, for $\varepsilon=1/2$. These operators act in the Fock-Bargmann Hilbert space $L^2_{\mathbb{C}}(D)$ of holomorphic (respectively, anti-holomorphic) functions $f(z)$ (respectively, $f(z^\ast)$), which are analytic inside the unit circle $D = \big\{ z\in \mathbb{C} \ ;\ |z|<1 \big\}$, and satisfy:
\begin{equation}\label{holoff(z)}
\| f \|^2 = \frac{-2t -1}{2\pi} \iint_D |f(z)|^2 \; (1-|z|^2)^{-2t -2} \mathrm{d}^2 z < \infty\,, \quad \mathrm{d}^2 z= \frac{\ii}{2}\mathrm{d} z \wedge \mathrm{d} z^\ast = \ud \big(\mathrm{Re}(z)\big)\, \ud\big(\mathrm{Im}(z)\big)\,,
\end{equation}
for $t\neq -1/2$,\footnote{Note that a limit procedure is needed in the case $t=-1/2$, based upon which we have to involve the universal covering of SU$(1,1)$ (in this regard, see Ref. \cite{GazeauBook}).}. The action of the representation operators $U^{\mbox{\small{ds}}}_{\varepsilon,t} (g)$ on the functions $f(z)$ is defined by:
\begin{equation}
\begin{split}
\mathrm{SU}(1,1) \ni g =
\begin{pmatrix}
    \alpha & \beta \\
   \beta^\ast & \alpha^\ast
\end{pmatrix}: \,f(z) \;\mapsto\; \Big(U^{\mbox{\small{ds}}}_{\varepsilon,t}(g) f\Big) (z) &= (-\beta^\ast z + \alpha)^{2t}\,f\left(\frac{\alpha^\ast z -\beta}{-\beta^\ast z + \alpha}\right)\\&\equiv N_{\varepsilon,t}(g,z) \; f(g^{-1}\diamond z)\,.
\end{split}
\end{equation}
Remind that according to the Cartan decomposition of SU$(1,1)$, we have :
\begin{eqnarray}
D \ni z \;\mapsto\; z^\prime = g^{-1}\diamond z = (\alpha^\ast z - \beta)(-\beta^\ast z + \alpha)^{-1} \in D\,.
\end{eqnarray}

Pursuing a similar procedure leading to the infinitesimal operators (\ref{hat J_1}), (\ref{hat J_2}), and (\ref{hat J_0}) in the context of the principal series representations, here we obtain the following representatives of the $\mathfrak{su}(1,1)$ Lie algebra elements (\ref{dS2gene}):
\begin{eqnarray}
Y_s \;\mapsto\; \hat{Y}_s\equiv\hat{J}_0 &=& z \frac{\mathrm{d}}{\mathrm{d} z} - t\,,\\
Y_l \;\mapsto\; \hat{Y}_l\equiv\hat{J}_1 &=& \frac{1 + z^2}{2}\frac{\mathrm{d}}{\mathrm{d} z} - t z\,,\\
Y_t \;\mapsto\; \hat{Y}_t\equiv\hat{J}_2 &=& - \mathrm{i} \Big(\frac{1 - z^2}{2}\frac{\mathrm{d}}{\mathrm{d} z} + t z\Big)\,,
\end{eqnarray}
which obey the commutation rules:
\begin{eqnarray}\label{comruk012}
[\hat{J}_0,\hat{J}_1] = - \mathrm{i} \hat{J}_2\,, \;\;\;\;\;\;\; [\hat{J}_0,\hat{J}_2] = \mathrm{i} \hat{J}_1\,, \;\;\;\;\;\;\; [\hat{J}_1,\hat{J}_2] = \mathrm{i} \hat{J}_0\,.
\end{eqnarray}
Finally, the quantum quadratic Casimir operator takes the form:
\begin{eqnarray}
Q = \hat{J}_2^2 + \hat{J}_1^2 - \hat{J}_0^2 = - t (t + 1) {\mathbbm{1}}\,, \;\;\;\;\;\;\; t= -1, -3/2, -2, ... \;.
\end{eqnarray}

%%%%%%%%%%%%%%%%%%%%%%%%%%%%%%%%%%%%%%%%%%%%%%%%%%%%%%%%%%%%%%%%%%%%%%%%%%%%%%%%%%%%%%%%%%%%%%%%%%%%%%%%%%%%%%%%%%%%%%%%
%%%%%%%%%%%%%%%%%%%%%%%%%%%%%%%%%%%%%%%%%%%%%%%%%%%%%%%%%%%%%%%%%%%%%%%%%%%%%%%%%%%%%%%%%%%%%%%%%%%%%%%%%%%%%%%%%%%%%%%%
%%%%%%%%%%%%%%%%%%%%%%%%%%%%%%%%%%%%%%%%%%%%%%%%%%%%%%%%%%%%%%%%%%%%%%%%%%%%%%%%%%%%%%%%%%%%%%%%%%%%%%%%%%%%%%%%%%%%%%%%
%%%%%%%%%%%%%%%%%%%%%%%%%%%%%%%%%%%%%%%%%%%%%%%%%%%%%%%%%%%%%%%%%%%%%%%%%%%%%%%%%%%%%%%%%%%%%%%%%%%%%%%%%%%%%%%%%%%%%%%%
%%%%%%%%%%%%%%%%%%%%%%%%%%%%%%%%%%%%%%%%%%%%%%%%%%%%%%%%%%%%%%%%%%%%%%%%%%%%%%%%%%%%%%%%%%%%%%%%%%%%%%%%%%%%%%%%%%%%%%%%
%%%%%%%%%%%%%%%%%%%%%%%%%%%%%%%%%%%%%%%%%%%%%%%%%%%%%%%%%%%%%%%%%%%%%%%%%%%%%%%%%%%%%%%%%%%%%%%%%%%%%%%%%%%%%%%%%%%%%%%%
%%%%%%%%%%%%%%%%%%%%%%%%%%%%%%%%%%%%%%%%%%%%%%%%%%%%%%%%%%%%%%%%%%%%%%%%%%%%%%%%%%%%%%%%%%%%%%%%%%%%%%%%%%%%%%%%%%%%%%%%

\part{$1+3$-dimensional dS (dS$_4$) geometry and relativity (classical and quantum mechanics)}\label{Part II}
In this part, we extend the above group-theoretical construction to dS$_4$ relativity. Of course, this part is not the exact parallel of the previous one; while, the arguments, that have counterparts in the previous part, are somewhat shortened, we supplement our discussions with some physical considerations that are naturally of more significance in this realistic case, such as, the causal structure of dS$_4$ spacetime and physical content of the theory under vanishing curvature. We begin with a brief description of dS$_4$ spacetime and its causal structure.

\setcounter{equation}{0} \section{DS$_4$ manifold and its causal structure}\label{Sec dS4 causal structure}
DS$_4$ spacetime is topologically $\mathbb{R}^1 \times \mathbb{S}^3$ ($\mathbb{R}^1$ being a timelike direction, the notion of ``time" having to be carefully defined in the dS context), and can be conveniently visualized as a one-sheeted hyperboloid embedded in a $1+4$-dimensional Minkowski spacetime $\mathbb{R}^{1+4}$ (by abuse of notation, let us say $\mathbb{R}^5$):
\begin{eqnarray}\label{1+3dS-M_R}
\underline{M}_R \equiv \Big\{x = (x^0, \;...\; , x^4) \in\mathbb{R}^5 \;;\; (x)^2 \equiv x\cdot x = \eta^{}_{AB}x^A x^B = -R^2 \Big\}\,, \;\;\;\;\;\;\; A,B = 0,1,2,3,4\,,
\end{eqnarray}
where $x^A$'s stand for the corresponding Cartesian coordinates and $\eta^{}_{AB} = \mbox{diag}(1,-1,-1,-1,-1)$ for the ambient Minkowski metric. From a cosmological viewpoint, the (constant) radius of curvature is given by $R = H^{-1}$, where $H$ is the Hubble constant (which fixes the rate of expansion of the dS$_4$ spatial sections).

The global causal ordering of $\underline{M}_R$ is induced by that of $\mathbb{R}^5$. Technically, to see the point, let:
\begin{eqnarray}\label{causal ordering dS2}
{\underline{V}}^+ \equiv \Big\{x \in\mathbb{R}^5 \;;\; (x)^2 = x\cdot x \geqslant 0,\; x^0 > 0 \Big\}\,.
\end{eqnarray}
[For future use, we also denote the interior of ${\underline{V}}^+$ by $\interior{\underline{V}}^+ \equiv \big\{x \in\mathbb{R}^5 \;;\; (x)^2 > 0,\; x^0 > 0 \big\}$.] Then, for two ``events" $x,x^\prime \in \underline{M}_R$, we say that $x^\prime$ is future connected to $x$ (symbolized here by $x^\prime\geqslant x$), if $x^\prime - x \in {{\underline{V}}^+}$, i.e., $(x^\prime - x)^2 \geqslant 0$ (or equivalently,\footnote{Note that $(x^\prime - x)^2 = -2(R^2 + x\cdot x^\prime)$, for $x,x^\prime \in \underline{M}_R$.} $x\cdot x^\prime \leqslant -R^2$), with $(x^{\prime 0} - x^0) > 0 $. Indeed, the future (respectively, past) cone of a given event $x \in \underline{M}_R$ is determined by $\underline{\Sigma}^+(x)$ (respectively, by $\underline{\Sigma}^-(x)$):
\begin{eqnarray}
\underline{\Sigma}^{+}(x) \; \big(\mbox{respectively,} \; \underline{\Sigma}^{-}(x)\big) = \Big\{ x^\prime \in \underline{M}_R \;;\; x^\prime \geqslant x \;\; \big(\mbox{respectively,} \; x^\prime \leqslant x\big)\Big\}\,.
\end{eqnarray}
With this definition, the ``light-cone" $\partial\underline{\Sigma}(x)$, as the boundary set of $\underline{\Sigma}^+(x) \bigcup \underline{\Sigma}^-(x)$, is the union of all linear generatrices of $\underline{M}_R$ containing the event $x$:
\begin{eqnarray}
\partial\underline{\Sigma}(x) = \Big\{ x^\prime \in \underline{M}_R \;;\; (x^\prime - x)^2 = 0 \;\; \big(\mbox{or equivalently,}\; x\cdot x^\prime = -R^2\big) \Big\}\,.
\end{eqnarray}
On the other hand, two events $x,x^\prime \in \underline{M}_R$ are said in ``acausal relation" or ``spacelike separated", if $x^\prime \notin \underline{\Sigma}^+(x) \bigcup \underline{\Sigma}^-(x)$, i.e., if $(x^\prime - x)^2 < 0$ (or equivalently, $x\cdot x^\prime > -R^2$).

Finally, let us point out that the (pseudo-)distance $d(x,x^\prime)$ on $\underline{M}_R$ is implicitly defined by \cite{Bros 2point func}:
\begin{eqnarray}
\cosh \left(\frac{d(x,x^\prime)}{R}\right) &=& - \frac{x\cdot x^\prime}{R^2}\,, \;\;\;\;\;\;\; \mbox{for $x$ and $x^\prime$ timelike separated ,} \nonumber\\
\cos \left(\frac{d(x,x^\prime)}{R}\right) &=& - \frac{x\cdot x^\prime}{R^2}\,, \;\;\;\;\;\;\; \mbox{for $x$ and $x^\prime$ spacelike separated such that $| x\cdot x^\prime | < R^2$ .}
\end{eqnarray}

\setcounter{equation}{0} \section{DS$_4$ relativity group SO$_0(1,4)$ and its covering Sp$(2,2)$}\label{Sec Sp(2,2) group}
The relativity group of dS$_4$ spacetime, as the Lorentz group of the ambient Minkowski spacetime $\mathbb{R}^5$, is SO$_0(1,4)$. It is the ten-parameter group of all linear transformations in $\mathbb{R}^5$ which leave invariant the quadratic form $(x)^2 = \eta^{}_{AB}x^A x^B$, have determinant $1$, and do not reverse the direction of the ``time" variable $x^0$. [Analogous to the $1+1$-dimensional case (see subsection \ref{Subsec 2.1}), this invariant quadratic form $(x)^2 = \eta^{}_{AB}x^A x^B$, besides the origin $x^A = 0$ (with $A=0,\;...\;,4$), determines three types of orbits in $\mathbb{R}^5$: the upper and lower sheets of the cone, with $(x)^2 = 0$ ($x^0 \gtrless 0$, respectively); the upper and lower sheets of the two-sheeted hyperboloids, with $(x)^2 > 0$ ($x^0 \gtrless 0$, respectively); and the one-sheeted hyperboloid, with $(x)^2 < 0$. Note that the dS$_4$ manifold $\underline{M}_R$ belongs to the latter case.] A familiar realization of the corresponding Lie algebra is achieved by the linear span of the ten Killing vectors:
\begin{eqnarray}\label{Killing dS4}
K_{AB} = x_A \partial_B - x_B \partial_A\,, \;\;\;\;\;\;\; K_{AB} = - K_{BA}\,.
\end{eqnarray}

The \emph{universal covering} of the dS$_4$ relativity group is the symplectic Sp$(2,2)$ group.\footnote{The phrase ``universal covering" refers to the fact that Sp$(2,2)$, as the covering group of SO$_0(1,4)$, is simply connected.} The latter comes to fore when dealing with half-integer spins (in the same way as SO$(3)$ is replaced by SU$(2)$, or SO$_0(1,3)$ is replaced by SL$(2,\mathbb{C})$). The Sp$(2,2)$ group is suitably described as the group of all $2 \times 2$-matrices $\underline{g}$, with \emph{quaternionic components}\footnote{Note that, in this paper, we merely consider the $2\times 2$-matrix representation of the set of quaternions, based upon which the quaternionic basis is given by $\big\{ \textbf{1} \equiv \mathbbm{1}_2, {\textbf{e}}^{}_k \equiv (-1)^{k+1} \mathrm{i} \sigma_k \;;\; k=1,2,3 \big\}$, where $\sigma_k$'s are the Pauli matrices. For more details, see appendix \ref{App quat}.}, which verify the unimodular and pseudo-unitary conditions:
\begin{eqnarray}\label{sp22somq}
\mathrm{Sp}(2,2) = \Bigg\{\underline{g} =
\begin{pmatrix}
\textbf{a} & \textbf{b}\\
\textbf{c} & \textbf{d}
\end{pmatrix}
\; ; \;\; \textbf{a},\textbf{b},\textbf{c},\textbf{d}\in\mathbb{H}, \;\; \mbox{det}(\underline{g}) = 1,\;\; \underline{g}^\dagger \gamma^0 \underline{g} = \gamma^0 \Bigg\}\,,
\end{eqnarray}
where $\underline{g}^\dagger = {\underline{g}^{\scriptscriptstyle\bigstar}}^{\texttt{t}}$, $\underline{g}^{\scriptscriptstyle\bigstar}$ being the \emph{quaternionic conjugate} of $\underline{g}$ (see more details in appendix \ref{App quat}) and ${\underline{g}}^{\texttt{t}}$ the transpose of $\underline{g}$, and $\gamma^0 = \begin{pmatrix} \textbf{1} & \textbf{0} \\ \textbf{0} & \textbf{-1} \end{pmatrix}$, with components $\textbf{1}$ and $\textbf{0}$ being, respectively, the unit and zero $2\times 2$ matrices.

It is worth noting that this matrix $\gamma^0$ is a part of the Clifford algebra determined by:
\begin{equation}\label{clifford}
\gamma^A \gamma^B + \gamma^B \gamma^A = 2\eta^{AB} \mathbbm{1}_4\,, \;\;\;\;\;\;\; {\gamma^A}^\dagger = \gamma^0 \gamma^A \gamma^0\,,
\end{equation}
where, with quaternionic components, the other four matrices read:
\begin{equation}\label{gamexp}
\gamma^k = (-1)^{k+1} \begin{pmatrix} \textbf{0} & \mathrm{i} \sigma_k \\ \mathrm{i} \sigma_k & \textbf{0} \end{pmatrix} = \begin{pmatrix} \textbf{0} & {\textbf{e}}^{}_k \\ {\textbf{e}}^{}_k & \textbf{0} \end{pmatrix}\,,\;\;\;\;\;\;\; \gamma^4 = \begin{pmatrix} \textbf{0} & \textbf{1}\\ -\textbf{1} & \textbf{0} \end{pmatrix}\,,
\end{equation}
with $k=1,2,3$. Note that, in the above, $\mathbbm{1}_4$ stands for the $4\times 4$-unit matrix. Among the set of the $\gamma^A$'s, $\gamma^0$ is the unique one to belong to Sp$(2,2)$. Due to \eqref{clifford}, one checks that ${\gamma^A}^\dagger\gamma^0 \gamma^A = -\gamma^0$ for all $A\neq0$.

Moreover, we should point out that the determinant of $\underline{g}$ is given by:
\begin{eqnarray}\label{det}
\mbox{det}(\underline{g}) = | \textbf{a} |^2 | \textbf{d} - \textbf{c}\textbf{a}^{-1}\textbf{b} |^2 = | \textbf{b} |^2 | \textbf{c} - \textbf{d}\textbf{b}^{-1}\textbf{a} |^2 = | \textbf{c} |^2 | \textbf{b} - \textbf{a}\textbf{c}^{-1}\textbf{d} |^2 = | \textbf{d} |^2 | \textbf{a} - \textbf{b}\textbf{d}^{-1}\textbf{c} |^2\,,
\end{eqnarray}
where $|\cdot|$ stands for the quaternion norm or modulus. These expressions are properly extended in case $\textbf{a}, \textbf{b}, \textbf{c}$, and $\textbf{d}$ are zero.

From the pseudo-unitary condition $\underline{g}^\dagger \gamma^0 \underline{g} = \gamma^0$, on one hand, we obtain the following auxiliary relations between the (quaternionic) matrix elements:
\begin{eqnarray}\label{rere1}
|\textbf{a}|^2 - |\textbf{c}|^2 = {1}\,, \;\;\;\;\;\;\; |\textbf{d}|^2 - |\textbf{b}|^2 = {1}\,, \;\;\;\;\;\;\; {\textbf{a}}^{\scriptscriptstyle\bigstar} \textbf{b} = {\textbf{c}}^{\scriptscriptstyle\bigstar} \textbf{d}\,,
\end{eqnarray}
and, on the other hand, since $\mbox{det} (\underline{g}) \neq 0$, we get:
\begin{eqnarray}\label{g-1qua}
\underline{g}^{-1} = \gamma^0 \underline{g}^\dagger \gamma^0 =
\begin{pmatrix}
{\textbf{a}}^{\scriptscriptstyle\bigstar} & -{\textbf{c}}^{\scriptscriptstyle\bigstar} \\
-{\textbf{b}}^{\scriptscriptstyle\bigstar} & {\textbf{d}}^{\scriptscriptstyle\bigstar}
\end{pmatrix}\,.
\end{eqnarray}
Now, considering the fact that $\underline{g}\underline{g}^{-1} = \underline{g}^{-1}\underline{g} = \begin{pmatrix} \textbf{1} & \textbf{0} \\ \textbf{0} & \textbf{1} \end{pmatrix}$, along with the identity (\ref{g-1qua}), we obtain:
\begin{eqnarray}\label{rere2}
|\textbf{a}|^2 - |\textbf{b}|^2 = {1}\,, \;\;\;\;\;\;\; |\textbf{d}|^2 - |\textbf{c}|^2 = {1}\,, \;\;\;\;\;\;\; \textbf{a} {\textbf{c}}^{\scriptscriptstyle\bigstar} = \textbf{b} {\textbf{d}}^{\scriptscriptstyle\bigstar}\,.
\end{eqnarray}
Finally, comparing the identities given in (\ref{rere1}) with those in (\ref{rere2}) results in:
\begin{eqnarray}\label{equality in norm}
|\textbf{a}| = |\textbf{d}|\,, \;\;\;\;\;\;\; |\textbf{b}| = |\textbf{c}|\,.
\end{eqnarray}
One can easily check that the matrix $\underline{g} = \begin{pmatrix} \textbf{a} & \textbf{b}\\ \textbf{c} & \textbf{d} \end{pmatrix}$, with generic quaternionic components, possessing the conditions (\ref{rere1}), (\ref{rere2}), and (\ref{equality in norm}), verifies $\mbox{det} (\underline{g}) = 1$, as well. Moreover, one can check that the constraints (\ref{rere1}) or, equivalently, (\ref{rere2}), reduce to $10$ the $16$ parameters of a generic $2\times 2$-quaternionic matrix.

\subsection{Homomorphism between SO$_0(1,4)$ and Sp$(2,2)$ and some discrete symmetries}\label{Subsec homo between dS4 groups}
In order to give an explicit realization of the homomorphism between the groups SO$_0(1,4)$ and Sp$(2,2)$, with any $x \in \mathbb{R}^5$ we associate the matrix $\slashed{x}$ defined by:
\begin{eqnarray}\label{slashx}
\slashed{x} = x^A \gamma_A =
\begin{pmatrix}
\textbf{1} x^0 & - \textbf{x} \\
{\textbf{x}}^{\scriptscriptstyle\bigstar} & - \textbf{1} x^0
\end{pmatrix}\,,\;\;\;\;\;\;\; \slashed{x}^2 = \big( (x^0)^2 - |\textbf{x}|^2 \big)\bu_4 = (x)^2\bu_4\,,
\end{eqnarray}
where $\textbf{x} = (x^4, \vec{x}) \in \mathbb{H}$ is a quaternion (again, in the $2\times 2$-matrix notations). On the other hand, a given matrix of the form $\slashed{x}$ uniquely determines a point $x \in \mathbb{R}^5$ in such a way that its components are:
\begin{eqnarray}\label{cogmuv}
x^A = \frac{1}{4} \mbox{tr}\big(\gamma^A \slashed{x}\big)\,.
\end{eqnarray}
Accordingly, this correspondence defines a one-to-one map between $\mathbb{R}^5$ and the set of $4\times 4$-matrices $\slashed{x}$.

Considering the above, the action of Sp$(2,2)$ on $\mathbb{R}^5$, for each element $\underline{g}\in \mathrm{Sp}(2,2)$, is given by:
\begin{eqnarray}\label{xxxxxxxxx}
\slashed{x}^\prime = \underline{g} \slashed{x} \underline{g}^{-1} =
\begin{pmatrix}
\textbf{1} x^{\prime 0} & -\textbf{x}^{\prime}\\
{\textbf{x}}^{\prime{\scriptscriptstyle\bigstar}} & - \textbf{1} x^{\prime 0}
\end{pmatrix}\,.
\end{eqnarray}
[A full justification of this action is given in the next section.] The transformed matrix $\slashed{x}^\prime$ represents a unique point $x^\prime $ in $\mathbb{R}^5$, with the components:
\begin{eqnarray}
x^{\prime A} &=& \frac{1}{4} \mbox{tr}\big(\gamma^A \slashed{x}^\prime\big) \nonumber\\
&=& \frac{1}{4} \mbox{tr} \big(\gamma^A \underline{g} \slashed{x} \underline{g}^{-1}\big)\nonumber\\
&=& \frac{1}{4} \mbox{tr} \big(\gamma^A \underline{g}\gamma_B \underline{g}^{-1}\big)x^B\,.
\end{eqnarray}
For the sake of simplicity, we symbolize the above action by $x^{\prime}= \underline{g} \diamond x$. This action is clearly linear and determinant-preserving; $\mbox{det} \big(\underline{g} \slashed{x} \underline{g}^{-1} \big) = \mbox{det} (\slashed{x}) = \left((x)^2\right)^2$. Moreover, this action does not change the signature of $x^0$, i.e., if $x^0>0$ then we get $x^{\prime 0} = \frac{1}{4}\mbox{tr} \big(\gamma^0 \underline{g} \gamma_B \underline{g}^{-1} \big)x^B >0$. To see the point, let us set $x = (1,0,0,0,0)$, then:
\begin{eqnarray}
x^{\prime 0} &=& \frac{1}{4}\mbox{tr}\big(\gamma^0 \underline{g} \gamma_0 \underline{g}^{-1}\big)\nonumber\\
&=& \frac{1}{4} \big(|\textbf{a}|^2 + |\textbf{b}|^2 + |\textbf{c}|^2 + |\textbf{d}|^2\big) > 0\,.
\end{eqnarray}
These facts show that the linear transformation $x^{\prime}= \underline{g} \diamond x$, which leaves invariant the quadratic form $(x)^2$, belongs to SO$_0(1,4)$, as well. Actually, for every transformation in SO$_0(1,4)$, there are two elements $\pm \underline{g}\in \mathrm{Sp}(2,2)$, since $\underline{g} \diamond x = (-\underline{g}) \diamond x $. In this sense, Sp$(2,2)$ is two-to-one homomorphic to SO$_0(1,4)$, with the kernel isomorphic to $\mathbb{Z}^2$:
\begin{eqnarray}
\mathrm{Sp}(2,2)/\mathbb{Z}^2 \sim \mathrm{SO}_0(1,4)\,.
\end{eqnarray}

Another interesting point to be highlighted here is that, as an element of Sp$(2,2)$, the group action of $\gamma^0$ corresponds to the discrete symmetry:
\begin{equation}\label{actgam0}
\slashed{x} \;\mapsto\; \slashed{x}^\prime = \gamma^0 \slashed{x} \left(\gamma^0\right)^{-1} = \gamma^0 \slashed{x} \gamma^0 =
\begin{pmatrix}
\textbf{1} x^0 & \textbf{x} \\
-{\textbf{x}}^{\scriptscriptstyle\bigstar} & - \textbf{1} x^0
\end{pmatrix}
= \slashed{x}^{\dag}\,, \;\;\;\;\;\;\; \mbox{i.e.,} \quad x=(x^0, \textbf{x}) \;\mapsto\; x^{\prime}=(x^0, -\textbf{x})\,.
\end{equation}
Whereas they are not elements of Sp$(2,2)$, the other elements (\ref{gamexp}) of the Clifford basis give rise to the following four discrete symmetries:\footnote{Note that, since $\gamma^A$'s, with $A\neq0$, are not elements of the Sp$(2,2)$ group, their respective inverses $\left(\gamma^A\right)^{-1}$'s do not verify the identity (\ref{g-1qua}), which merely holds true for the Sp$(2,2)$ elements.}
\begin{align}\label{discsym4-k}
\slashed{x} & \;\mapsto\; \slashed{x}^\prime = \gamma^k \slashed{x} \left(\gamma^k\right)^{-1}= -\gamma^k \slashed{x} \gamma^k=
\begin{pmatrix}
-\textbf{1} x^0 & -{\textbf{e}}^{}_k{\textbf{x}}^{\scriptscriptstyle\bigstar}{\textbf{e}}^{}_k \\
 {\textbf{e}}^{}_k\textbf{x}{\textbf{e}}^{}_k &  \textbf{1} x^0
\end{pmatrix}\,, \;\;\;\;\;\;\; \mbox{i.e.,} \quad x=(x^0, \textbf{x}) \;\mapsto\; x^{\prime}=(-x^0, {\textbf{e}}^{}_k\textbf{x}^{\scriptscriptstyle\bigstar}{\textbf{e}}^{}_k)\,,
\\ \nonumber \\ \label{discsym4-4}
\slashed{x} & \;\mapsto\; \slashed{x}^\prime = \gamma^4 \slashed{x} \left(\gamma^4\right)^{-1}= -\gamma^4\slashed{x} \gamma^4=
\begin{pmatrix}
-\textbf{1} x^0 & -{\textbf{x}}^{\scriptscriptstyle\bigstar} \\
\textbf{x}& \textbf{1} x^0
\end{pmatrix}\,, \;\;\;\;\;\;\; \mbox{i.e.,} \quad x=(x^0, \textbf{x}) \;\mapsto\; x^{\prime}=(-x^0, \textbf{x}^{\scriptscriptstyle\bigstar})\,.
\end{align}
It follows that the action of $\gamma^0 \gamma^k$ changes the sign of the components $x^0$ and $x^k$, $\gamma^4 \gamma^k$ changes the sign of the components $x^4$ and $x^k$, and finally $\gamma^0 \gamma^4$ changes the sign of the components $x^0$ and $x^4$. Note that while the so-called antipodal symmetry $x\mapsto -x $ (see, for instance, Ref. \cite{FolacciSanchez}) cannot be obtained through such actions, it can be yielded by combining the action of $\gamma^0 \gamma^4$ with quaternionic conjugation $\slashed{x} \mapsto \slashed{x}^{\scriptscriptstyle\bigstar}$:
\begin{equation} \label{antipodal}
\slashed{x} \;\mapsto\; \slashed{x}^\prime = \gamma^0 \gamma^4 \slashed{x}^{\scriptscriptstyle\bigstar} \left(\gamma^0 \gamma^4\right)^{-1} = \gamma^0 \gamma^4 \slashed{x}^{\scriptscriptstyle\bigstar} \gamma^0 \gamma^4 = - \slashed{x} \;\;\;\;\;\;\; \mbox{i.e.,} \quad x=(x^0, \textbf{x}) \;\mapsto\; -x\,.
\end{equation}

\setcounter{equation}{0} \section{Relativistic meaning of the dS$_4$ group: group decomposition}\label{Sec dS4 group decomposition}

\subsection{Space-time-Lorentz decomposition}\label{sec space-time-Lorentz dS4}
Any $\underline{g} \in \mathrm{Sp}(2,2)$, with respect to the group involution $\mathfrak{i}(\underline{g}) : \underline{g} \;\mapsto\; \gamma^0\gamma^4 \underline{g}^\dagger \gamma^0\gamma^4$, can be decomposed into:
\begin{eqnarray}\label{jl}
\underline{g} =
\begin{pmatrix}
\textbf{a} & \textbf{b} \\
\textbf{c} & \textbf{d}
\end{pmatrix}
=\underline{j} \; \underline{l}\,,
\end{eqnarray}
where the factor $\underline{l}$ is an element of the subgroup:
\begin{eqnarray} \label{involu equ}
{\underline{\cal{L}}} = \Big\{  {\underline{l}} \in \mathrm{Sp}(2,2) \; ; \; \mathfrak{i}({\underline{l}}) = {\underline{l}}^{-1} \Big\}\,.
\end{eqnarray}
Considering $\underline{l} = \begin{pmatrix} \textbf{a}_l & \textbf{b}_l \\ \textbf{c}_l & \textbf{d}_l \end{pmatrix}$, with generic quaternionic components, the definition (\ref{involu equ}) along with the identity (\ref{g-1qua}) result in:
\begin{eqnarray} \label{a_l b_l}
\textbf{a}_l = \textbf{d}_l\,,\;\;\;\;\;\;\; \textbf{b}_l = - \textbf{c}_l\,.
\end{eqnarray}
On the other hand, it is truly expected that the quaternionic components of $\underline{l} \; \big(\in \mathrm{Sp}(2,2)\big)$ verify the conditions (\ref{rere1}) and (\ref{rere2}), as well. This directly implies that:
\begin{eqnarray}\label{a_l b_l '}
\textbf{a}_l^{\scriptscriptstyle\bigstar} \textbf{b}^{}_l + \textbf{b}_l^{\scriptscriptstyle\bigstar} \textbf{a}^{}_l = \textbf{0}\,, \;\;\;\;\;\;\; \textbf{a}_l^{} \textbf{b}^{\scriptscriptstyle\bigstar}_l + \textbf{b}_l^{} \textbf{a}^{\scriptscriptstyle\bigstar}_l = \textbf{0}\,,
\end{eqnarray}
for which, by defining $\textbf{a}^{}_l \equiv (a_l^4, \vec{a}^{}_l)$ and $\textbf{b}^{}_l \equiv (b_l^4, \vec{b}^{}_l)$, we obtain the condition $a_l^4 b_l^4 + \vec{a}^{}_l \cdot \vec{b}^{}_l = 0$. Then, a possible solution to the set of Eqs. (\ref{a_l b_l '}) can be given by allocating to $\textbf{a}^{}_l$ and $\textbf{b}^{}_l$, respectively:
\begin{eqnarray}\label{a solu}
\textbf{a}^{}_l = \cosh\textstyle\frac{\varphi}{2} \big( 1, \vec{0} \big)\,,\;\;\;\;\;\;\; \textbf{b}^{}_l = \sinh\textstyle\frac{\varphi}{2} \big( 0, \vec{u} \big)\,,
\end{eqnarray}
where $\varphi \in \mathbb{R}$ and the \emph{pure vector quaternion}\footnote{Note that the term ``pure vector quaternion" is used with respect to the representation of a quaternion in the scalar-vector notations (see appendix \ref{App quat}).} $\vec{\textbf{u}} \equiv \big( 0, \vec{u} \big)$, with $\vec{u}=\big( u^1,u^2,u^3 \big)$, belongs to $\mathrm{SU}(2)$ (that is, $|\vec{\textbf{u}}| = |\vec{u}| = \sqrt{(u^1)^2 + (u^2)^2 + (u^3)^2}= 1$). The above solution can be simply generalized by readjusting ${\textbf{a}}^{}_l \mapsto \textbf{v} {\textbf{a}}^{}_l$ and ${\textbf{b}}^{}_l \mapsto \textbf{v} {\textbf{b}}^{}_l$, where $\textbf{v} \equiv \big(\cos\frac{\vartheta}{2}, \sin\frac{\vartheta}{2} \vec{v}\big) \in \mathrm{SU}(2)$, with $0\leqslant \vartheta < 2\pi$, $\vec{v}=\big( v^1,v^2,v^3 \big)$, and $|\textbf{v}|= |\vec{v}|= 1$. Accordingly, a generic form of the matrix $\underline{l}$ reads:
\begin{eqnarray}\label{solu lorentz}
\underline{l} &=&
\begin{pmatrix}
\textbf{v} & \textbf{0} \\
\textbf{0} & \textbf{v}
\end{pmatrix}
\begin{pmatrix}
\textbf{1}\cosh \frac{\varphi}{2} & \vec{\textbf{u}}\sinh \frac{\varphi}{2}\\
- \vec{\textbf{u}}\sinh\frac{\varphi}{2} & \textbf{1}\cosh\frac{\varphi}{2}
\end{pmatrix}\,.
\end{eqnarray}
Here, in a shortcut, we would like to point out that the subgroup ${\underline{\cal{L}}}$, with the above generic element, is actually the Lorentz subgroup of Sp$(2,2)$ and that the above factorization is called the Cartan factorization of this subgroup.

Now, we deal with the factor $\underline{j}$, appeared in Eq. (\ref{jl}). Having Eq. (\ref{involu equ}) in mind, we have:\footnote{Note that $\gamma^0 \gamma^4 = - \gamma^4 \gamma^0$ and $\gamma^0 \gamma^4 \gamma^0 \gamma^4 = \gamma^0 \gamma^0 = - \gamma^4 \gamma^4 = \mathbbm{1}_4$, and that the quaternion field, as a multiplicative group, is $\mathbb{H} \sim \mathbb{R}_+ \times \mathrm{SU}(2)$ (for the latter see appendix \ref{App quat}).}
\begin{eqnarray}\label{ji(j) dS4}
\underline{j} \; \mathfrak{i}(\underline{j}) = \underline{g} \; \mathfrak{i}(\underline{g}) =
\begin{pmatrix}
\textbf{a} {\textbf{d}}^{\scriptscriptstyle\bigstar} + \textbf{b}{\textbf{c}}^{\scriptscriptstyle\bigstar} & & \textbf{a} {\textbf{b}}^{\scriptscriptstyle\bigstar} + \textbf{b} {\textbf{a}}^{\scriptscriptstyle\bigstar}\\
\textbf{c} {\textbf{d}}^{\scriptscriptstyle\bigstar} + \textbf{d} {\textbf{c}}^{\scriptscriptstyle\bigstar} & & \textbf{c} {\textbf{b}}^{\scriptscriptstyle\bigstar} + \textbf{d} {\textbf{a}}^{\scriptscriptstyle\bigstar}
\end{pmatrix}
\equiv
\begin{pmatrix}
{\textbf{w}}^2 \cosh\psi & \textbf{1}\sinh\psi \\
\textbf{1}\sinh\psi & {\textbf{w}}^{{\scriptscriptstyle\bigstar} 2} \cosh\psi
\end{pmatrix}\,,
\end{eqnarray}
where $\psi\in\mathbb{R}$ and ${\textbf{w}} \equiv \big( \cos\frac{\theta}{2}, \sin\frac{\theta}{2} \vec{w} \big) \in \mathrm{SU}(2)$, with $0\leqslant \theta < 2\pi$, $\vec{w}=\big( w^1,w^2,w^3 \big)$, and $|\textbf{w}|= |\vec{w}|= 1$. A possible solution for the factor $\underline{j}$, verifying Eq. (\ref{ji(j) dS4}), is:
\begin{eqnarray} \label{j solu}
\underline{j} &=&
\begin{pmatrix}
{\textbf{w}} & \textbf{0} \\
\textbf{0} & {\textbf{w}}^{\scriptscriptstyle\bigstar}
\end{pmatrix}
\begin{pmatrix}
\textbf{1}\cosh \frac{\psi}{2} & \textbf{1}\sinh\frac{\psi}{2}\\
\textbf{1}\sinh\frac{\psi}{2} & \textbf{1}\cosh \frac{\psi}{2}
\end{pmatrix}\,.
\end{eqnarray}

In summary, the above construction presents a global, but nonunique, decomposition of Sp$(2,2)$ as:
\begin{eqnarray}\label{space-time-Lorentz dS4}
\mathrm{Sp}(2,2) \ni \underline{g} = \Big(\prod_{k}\exp(\theta_k\underline{X}_k)\Big)\exp(\psi \underline{X}_0)\Big(\prod_{k}\exp(\vartheta_k{\underline{Y}}_k)\Big) \Big(\prod_{k}\exp(\varphi_k\underline{Z}_k)\Big)\,,
\end{eqnarray}
where $\underline{X}_k, \underline{X}_0, \underline{Y}_k$ and $\underline{Z}_k$, with $k=1,2,3$, denote the corresponding infinitesimal generators:
\begin{eqnarray}\label{Xk}
\underline{X}_k = \frac{\mathrm{d}}{\mathrm{d}\theta_k}
\begin{pmatrix}
{\textbf{w}} & \textbf{0} \\
\textbf{0} & {\textbf{w}}^{\scriptscriptstyle\bigstar}
\end{pmatrix}\Big|_{\theta_k=0} = \frac{1}{2}
\begin{pmatrix}
{\textbf{e}}^{}_k & \textbf{0} \\
\textbf{0} & - {\textbf{e}}^{}_k
\end{pmatrix}\,,
\end{eqnarray}
\begin{eqnarray}\label{X0}
\underline{X}_0 = \frac{\mathrm{d}}{\mathrm{d}\psi}
\begin{pmatrix}
\textbf{1}\cosh \frac{\psi}{2} & \textbf{1}\sinh\frac{\psi}{2}\\
\textbf{1}\sinh\frac{\psi}{2} & \textbf{1}\cosh \frac{\psi}{2}
\end{pmatrix}\Big|_{\psi=0} = \frac{1}{2}
\begin{pmatrix}
\textbf{0} & \textbf{1} \\
\textbf{1} & \textbf{0}
\end{pmatrix}\,,
\end{eqnarray}
\begin{eqnarray}\label{Yk}
\underline{Y}_k = \frac{\mathrm{d}}{\mathrm{d}\vartheta_k}
\begin{pmatrix}
{\textbf{v}} & \textbf{0} \\
\textbf{0} & {\textbf{v}}
\end{pmatrix}\Big|_{\vartheta_k=0} = \frac{1}{2}
\begin{pmatrix}
{\textbf{e}}^{}_k & \textbf{0} \\
\textbf{0} & {\textbf{e}}^{}_k
\end{pmatrix}\,,
\end{eqnarray}
\begin{eqnarray}\label{Zk}
\underline{Z}_k = \frac{\mathrm{d}}{\mathrm{d}\varphi}_k
\begin{pmatrix}
\textbf{1}\cosh \frac{\varphi}{2} & \vec{\textbf{u}}\sinh \frac{\varphi}{2}\\
- \vec{\textbf{u}}\sinh\frac{\varphi}{2} & \textbf{1}\cosh\frac{\varphi}{2}
\end{pmatrix}\Big|_{\varphi_k=0} = \frac{1}{2}
\begin{pmatrix}
\textbf{0} & {\textbf{e}}^{}_k \\
- {\textbf{e}}^{}_k & \textbf{0}
\end{pmatrix}\,.
\end{eqnarray}
They obey the following commutation relations:
\begin{eqnarray}\label{commutation relations dS4}
\big[\underline{{Y}}_i,\underline{{Y}}_j\big] &=& {{\cal{E}}_{ij}}^{k} \;\underline{{Y}}_k\,,\nonumber\\
\big[\underline{{Y}}_i,\underline{{X}}_j\big] &=& {{\cal{E}}_{ij}}^{k} \;\underline{{X}}_k\,,\nonumber\\
\big[\underline{{X}}_i,\underline{{X}}_j\big] &=& {{\cal{E}}_{ij}}^{k} \;\underline{{Y}}_k\,,\nonumber\\
\big[\underline{{Y}}_i,\underline{{Z}}_j\big] &=& {{\cal{E}}_{ij}}^{k} \;\underline{{Z}}_k\,,\nonumber\\
\big[\underline{{X}}_i,\underline{{Z}}_j\big] &=& -\delta^{}_{ij} \;\underline{{X}}_0\,,\nonumber\\
\big[\underline{{Z}}_i,\underline{{Z}}_j\big] &=& -{{\cal{E}}_{ij}}^{k} \;\underline{{Y}}_k\,,\nonumber\\
\big[\underline{{X}}_0,\underline{{X}}_i\big] &=& -\underline{{Z}}_i\,,\nonumber\\
\big[\underline{{X}}_0,\underline{{Z}}_i\big] &=& -\underline{{X}}_i\,,\nonumber\\
\big[\underline{{X}}_0,\underline{{Y}}_i\big] &=& 0\,,
\end{eqnarray}
where $i,j,k = 1,2,3$ and ${{\cal{E}}_{ij}}^{k}$ is the three-dimensional totally antisymmetric Levi-Civita symbol. The above commutation relations can be brought into a form which explicitly exhibits the dS$_4$ Lie algebra $\mathfrak{sp}(2,2)$, by defining:
\begin{eqnarray}
K_{4k} \equiv \underline{X}_k\,,\;\;\;\;\;\;\; K_{04} \equiv \underline{X}_0\,,\;\;\;\;\;\;\; K_{ki} \equiv {{\cal{E}}_{ki}}^{j} \;\underline{Y}_j\,,\;\;\;\;\;\;\; K_{0k} \equiv \underline{Z}_k\,,
\end{eqnarray}
based upon which, we get:
\begin{eqnarray}
[K_{AB},K_{CD}] = - \big( \eta^{}_{AC} {K^{}_{BD}} + \eta^{}_{BD} {K^{}_{AC}} - \eta^{}_{AD} {K^{}_{BC}} - \eta^{}_{BC} {K^{}_{AD}} \big)\,, \;\;\;\;\;\;\; A,B=0,1,2,3,4\,.
\end{eqnarray}
Note that $K_{AB} = - K_{BA}$.

Here, we must underline that, quite analogous to the $1+1$-dimensional case, the factor $\underline{j}$ (see Eq. (\ref{j solu})) plays the role of ``spacetime" square root, which exemplifies the dS$_4$ topology $\mathbb{R}^1 \times \mathbb{S}^3$. To make this point apparent, following the instruction given in subsection \ref{sec space-time-Lorentz}, we define the coordinates $(x^0,\textbf{x})$ in $\mathbb{R}^5$ as:
\begin{eqnarray}\label{spacetime dS4}
\underline{\Upsilon}(x) = R \; \underline{j} \; \mathfrak{i}(\underline{j}) \; (-\gamma^4) = R
\begin{pmatrix}
\textbf{1}\sinh\psi & - {\textbf{w}}^2 \cosh\psi \\
{\textbf{w}}^{{\scriptscriptstyle\bigstar} 2} \cosh\psi & - \textbf{1}\sinh\psi
\end{pmatrix} \equiv
\begin{pmatrix}
\textbf{1} x^0 & - \textbf{x} \\
{\textbf{x}}^{\scriptscriptstyle\bigstar} & - \textbf{1} x^0
\end{pmatrix} = \slashed{x}\,,
\end{eqnarray}
where $0<R<\infty$. The action of Sp$(2,2)$ on the $\underline{\Upsilon}(x)$'s set is realized by its action on the set of matrices $\underline{j}$ from the left:
\begin{eqnarray}
\mathrm{Sp}(2,2) \ni \underline{g} \;:\; \underline{j} \;\mapsto\; \underline{j}^\prime \equiv \underline{g}\diamond \underline{j}\,, \;\;\;\;\;\;\; \underline{g}\underline{j} = \underline{j}^\prime \underline{l}^\prime\,,
\end{eqnarray}
from which, recalling that $\underline{l} \in \underline{\cal{L}}$ (that is, $\mathfrak{i}(\underline{l}) = \underline{l}^{-1}$), $\gamma^0 \gamma^4 = - \gamma^4 \gamma^0$, and $\gamma^0 \gamma^4 \gamma^0 \gamma^4 = \gamma^0 \gamma^0 = - \gamma^4 \gamma^4 = \mathbbm{1}_4$, we obtain:
\begin{eqnarray}\label{action spacetime}
\underline{\Upsilon}(x^\prime) &=& R \; \underline{j}^\prime \; \mathfrak{i}(\underline{j}^\prime) \; (-\gamma^4) \nonumber\\
&=& R \; \Big(\underline{g} \underline{j} \underline{l}^{\prime{-1}}\Big) \; \Big[\gamma^0 \gamma^4 \Big((\underline{l}^{\prime{-1}})^\dagger \underline{j}^\dagger \underline{g}^\dagger\Big) \gamma^0 \gamma^4\Big] \; (-\gamma^4) \nonumber\\
&=& R \; \underline{g} \underline{j} \underbrace{\Big[ \underline{l}^{\prime{-1}}\; \gamma^0 \gamma^4 \Big((\underline{l}^{\prime{-1}})^\dagger\Big) \gamma^0 \gamma^4 \Big]}_{=\mathbbm{1}_4} \Big[\gamma^0 \gamma^4 \Big(\underline{j}^\dagger \underline{g}^\dagger\Big) \gamma^0 \gamma^4 \Big] \; (-\gamma^4) \nonumber\\
&=& \underline{g} \underbrace{\Big[R \; \underline{j}\; \big(\gamma^0 \gamma^4 \underline{j}^\dagger \gamma^0 \gamma^4\big)\Big](-\gamma^4)}_{=\underline{\Upsilon}(x)} \; \underbrace{\gamma^4 \Big[\gamma^0 \gamma^4 \underline{g}^\dagger \gamma^0 \gamma^4 \Big](-\gamma^4)}_{= \gamma^0 \underline{g}^\dagger \gamma^0}\nonumber\\
&=& \underline{g} \underline{\Upsilon}(x) \underline{g}^{-1} = \slashed{x}^\prime\,.
\end{eqnarray}
This is the justification of the group action already given in (\ref{xxxxxxxxx}). Since this group action preserves the determinant:
\begin{eqnarray}
\mbox{det} \big(\underline{\Upsilon}(x^\prime)\big) = \mbox{det} \big(\underline{\Upsilon}(x)\big) = \left((x^0)^2 - |\textbf{x}|^2\right)^2 = R^4\,,
\end{eqnarray}
as well as the identity $(x^0)^2 - |\textbf{x}|^2 = -R^2$ (see Eq. (\ref{slashx})):
\begin{eqnarray}
(\slashed{x}^{\prime})^2 = g\slashed{x}g^{-1}g\slashed{x}g^{-1} = (\slashed{x})^2 = \left((x^0)^2 - |\textbf{x}|^2\right)\bu_4 = -R^2 \bu_4\,,
\end{eqnarray}
we understand from the factorisation (\ref{jl}) that each point of the corresponding dS$_4$ hyperboloid is in one-to-one correspondence with each class of the left coset $\mathrm{Sp}(2,2)/\underline{\cal{L}}$, i.e., $\mathrm{Sp}(2,2) = \mbox{dS}_4 \times \underline{\cal{L}}$, topologically.

Technically, in the above group decomposition, the factor $\underline{l} \in \underline{{\cal{L}}}$ leaves the point $x^{}_\odot = (0,0,0,0,R)$, chosen as the origin of the dS$_4$ hyperboloid $\underline{M}_R$, invariant.\footnote{Just remember that this choice is purely arbitrary due to the SO$_0(1,4)$ or Sp$(2,2)$ symmetry; any other point is physically equivalent. If we were to deal with the unit sphere $\mathbb{S}^4$ as representing a four-dimensional manifold in $\mathbb{R}^5$, we would not have any hesitation to acknowledge this property. Dealing with the representation of the dS$_4$ manifold embedded in $\mathbb{R}^5$ as a hyperboloid might be misleading in this regard because of its deformed shape.} The tangent space to $\underline{M}_R$ on the base point $x^{}_\odot$ (that is, the hyperplane $\{x\in\mathbb{R}^5 \; ;\; x^4 = R \}$) is considered as the $1+3$-dimensional Minkowski spacetime (with pseudo-metric $\mathrm{d} x_0^2 - \mathrm{d} x_1^2 - \mathrm{d} x_2^2 - \mathrm{d} x_3^2$) onto which dS$_4$ spacetime can be contracted at the zero-curvature limit. In this sense, the subgroup ${\cal{L}}$ (isomorphic to $\mathrm{SO}_0(1,3)$), which is the stabilizer of this base point, is interpreted as the Lorentz group of the tangent space. This makes clear the interpretation of the corresponding infinitesimal transformations generated by $\underline{Y}_k$ and $\underline{Z}_k$ (see Eqs. (\ref{Yk}) and (\ref{Zk})). They are indeed the ``space rotations" and ``boost transformations", respectively. Correspondingly, the parameters $\textbf{v}$, $\vec{\textbf{u}}$ and $\varphi$ are respectively presumed to carry the meaning of space rotation, boost velocity direction, and rapidity.

On the other hand, the set of matrices $\underline{j}$ maps the origin $x^{}_\odot = (0,0,0,0,R)$ to any point of $\underline{M}_R$:
\begin{eqnarray}\label{454545454545}
\underline{j}\begin{pmatrix}
\textbf{0} & - \textbf{1} R \\
\textbf{1} R & \textbf{0}
\end{pmatrix} \underline{j}^{-1} = R
\begin{pmatrix}
\textbf{1}\sinh\psi & - {\textbf{w}}^2 \cosh\psi \\
{\textbf{w}}^{{\scriptscriptstyle\bigstar} 2} \cosh\psi & - \textbf{1}\sinh\psi
\end{pmatrix} \equiv
\begin{pmatrix}
\textbf{1} x^0 & - \textbf{x} \\
{\textbf{x}}^{\scriptscriptstyle\bigstar} & - \textbf{1} x^0
\end{pmatrix} = \slashed{x}\,.
\end{eqnarray}
The family $(\psi, {\textbf{w}}^2)$ actually provides global coordinates for $\underline{M}_R$:
\begin{eqnarray}\label{222222222222}
x^0 = R \sinh\psi\,,\;\;\;\;\;\;\; \textbf{x} \; \Big(= (x^4, \vec{x}) \Big) = R{\textbf{w}}^2 \cosh\psi\,.
\end{eqnarray}
[Recall that ${\textbf{w}} \equiv \big( \cos\frac{\theta}{2}, \sin\frac{\theta}{2} \vec{w} \big) \in \mathrm{SU}(2)$, where $0\leqslant \theta <2\pi$ and $\vec{w}=\big( w^1,w^2,w^3 \big)$, with $(w^1)^2 + (w^2)^2 + (w^3)^2= 1$. One can also show that ${\textbf{w}}^2 = \big(\cos{\theta}, \sin{\theta} \vec{w}\big) \in \mathrm{SU}(2)$.] In this sense, the corresponding generators $\underline{X}_k$ and $\underline{X}_0$ (see Eqs. (\ref{Xk}) and (\ref{X0})) are interpreted as the ``space translations" and ``time translations", respectively.

Here, taking the coordinates (\ref{222222222222}) into account, the fact that dS$_4$ spacetime is locally Minkowskian can also be viewed by considering the left-invariant metric $\mathrm{d} s^2$ in the following global coordinates (yielded by letting $\psi,\theta \rightarrow 0$ in the coordinates (\ref{222222222222})):
\begin{eqnarray}\label{coor Min dS4}
t^{}_{\circ} = R \psi\,, \;\;\;\;\;\;\; \mbox{and} \;\;\;\;\;\;\; \vec{x}^{}_{\circ} \equiv (x_{\circ}^1,x_{\circ}^2,x_{\circ}^3) = R\theta\vec{w}\,,
\end{eqnarray}
from which, we get:
\begin{eqnarray}\label{coor Min dS4'}
\mathrm{d} s^2 = \big(\mathrm{d} t_{\circ}\big)^2 - \cosh^2 (R^{-1}t^{}_{\circ}) \; \big(\mathrm{d} \vec{x}_{\circ}\big)^2\,.
\end{eqnarray}

\subsection{Cartan decomposition}\label{Sec Cartan dS4}
The Cartan decomposition of $\mathrm{Sp}(2,2) = \underline{{\cal{P}}} \underline{{\cal{K}}}$ is carried out with respect to the Cartan involution $\mathfrak{i}(\underline{g})\; : \; \underline{g} \mapsto (\underline{g}^\dagger)^{-1}$, based upon which any $\underline{g} \in \mathrm{Sp}(2,2)$ can be decomposed into \cite{Lipsman}:
\begin{eqnarray}
\underline{g} =
\begin{pmatrix}
\textbf{a} & \textbf{b}\\
\textbf{c} & \textbf{d}
\end{pmatrix} =
\underline{p} \; \underline{k}\,,
\end{eqnarray}
where the elements $\underline{p}\in\underline{{\cal{P}}}$ and $\underline{k}\in\underline{{\cal{K}}}$ are respectively determined by the conditions $\mathfrak{i}(\underline{p}) = \underline{p}^{-1}$, which means that $\underline{p}$ is Hermitian ($\underline{p} = \underline{p}^{\dagger}$), and $\mathfrak{i}(\underline{k}) = \underline{k}$, which means that $\underline{k}$ is unitary ($\underline{k}^\dagger = \underline{k}^{-1}$). Utilizing the conditions (\ref{rere1}), (\ref{rere2}), and (\ref{equality in norm}) along with the identities given in appendix \ref{App quat}, one can show that:
\begin{eqnarray}\label{888888888888}
\underline{p} =
\begin{pmatrix}
|\textbf{a}| \textbf{1} & |\textbf{a}| \textbf{q}\\
|\textbf{a}| \textbf{q}^{\scriptscriptstyle\bigstar} & |\textbf{a}| \textbf{1}
\end{pmatrix} \in \underline{{\cal{P}}}\,, \;\;\;\;\;\;\; \mbox{and} \;\;\;\;\;\;\;
\underline{k} =
\begin{pmatrix}
\frac{\textbf{a}}{|\textbf{a}|} & \textbf{0} \\
\textbf{0} & \frac{\textbf{d}}{|\textbf{d}|}
\end{pmatrix} \in \underline{{\cal{K}}}\,,
\end{eqnarray}
where $\textbf{q} = \textbf{b} \textbf{d}^{-1}$, $\textbf{q}^{\scriptscriptstyle\bigstar} = (\textbf{b} \textbf{d}^{-1})^{\scriptscriptstyle\bigstar} = (\textbf{d}^{-1})^{\scriptscriptstyle\bigstar} \textbf{b}^{\scriptscriptstyle\bigstar} = {\textbf{d}\textbf{b}^{\scriptscriptstyle\bigstar}}/{|\textbf{d}|^2} = {\textbf{c}\textbf{a}^{\scriptscriptstyle\bigstar}}/{|\textbf{a}|^2} = \textbf{c} \textbf{a}^{-1}$, and $\frac{\textbf{a}}{|\textbf{a}|},\; \frac{\textbf{d}}{|\textbf{d}|} \in \mathrm{SU}(2)$, while $|\textbf{a}| = |\textbf{d}| = (1- |\textbf{q}|^2)^{-1/2}$, with $|\textbf{q}| < 1$. The subgroup ${\underline{{\cal{K}}}}$, as the maximal compact subgroup of Sp$(2,2)$, can be realized by the isomorphism ${\underline{{\cal{K}}}} \sim \mathrm{SU}(2) \times \mathrm{SU}(2)$.\footnote{\label{foot2}Recall that \emph{direct} product of two groups $G_1$ and $G_2$, denoted here by $G_1 \times G_2$, consists in glueing the two groups together, without interaction; $G = G_1 \times G_2$ is simply their Cartesian product, endowed with the group law:
\begin{equation}
(g^{}_1, g^{}_2)(g^\prime_1, g^\prime_2) = (g^{}_1 g^\prime_1, g^{}_2 g^\prime_2)\, , \;\;\;\;\;\;\; g^{}_1, g^\prime_1 \in G_1, \;\;\; g^{}_2, g^\prime_2 \in G_2\, . \nonumber
\end{equation}
In this context, neutral element is $(e_1, e_2)$. With the identifications $ g_1 \sim (g_1, e_2), \, g_2 \sim (e_1, g_2)$, both $G_1$ and $ G_2$ are invariant subgroups of $G_1 \times G_2$. In the matrix realization, if $G_1$ and $G_2$ are respectively considered to be groups of $n\times n$ and $m\times m$ matrices, the direct product $G_1 \times G_2$ represents the group $\Bigg\{ \begin{pmatrix} g_1 & 0 \\ 0 & g_2 \end{pmatrix} \; ; \; g_1\in G_1,\; g_2\in G_2 \Bigg\}$ of block diagonal $(n+m)\times (n+m)$ matrices, since $\begin{pmatrix} g_1 & 0 \\ 0 & g_2 \end{pmatrix}\begin{pmatrix} g^\prime_1 & 0 \\ 0 & g^\prime_2 \end{pmatrix} = \begin{pmatrix} g^{}_1g^\prime_1 & 0 \\ 0 & g^{}_2 g^\prime_2 \end{pmatrix}$.} Moreover, one can define:
\begin{eqnarray}\label{coor tan}
\textbf{q} = \textbf{b} \textbf{d}^{-1} \equiv \tanh{\frac{\varphi}{2}} \; \frac{\textbf{b}}{|\textbf{b}|} \frac{\textbf{d}^{\scriptscriptstyle\bigstar}}{|\textbf{d}|}\,,
\end{eqnarray}
where $\tanh\frac{\varphi}{2} = \frac{|\textbf{b}|}{|\textbf{d}|}$, with $\varphi\in\mathbb{R}$, and $\frac{\textbf{b}}{|\textbf{b}|},\; \frac{\textbf{d}^{\scriptscriptstyle\bigstar}}{|\textbf{d}|} \in \mathrm{SU}(2)$. It then follows that:
\begin{eqnarray}
|\textbf{a}| = |\textbf{d}| = \cosh{\frac{\varphi}{2}}\,,\;\;\;\;\;\;\; |\textbf{b}| = |\textbf{c}| = \sinh{\frac{\varphi}{2}}\,.
\end{eqnarray}

In this context, the subset of Hermitian matrices $\underline{{\cal{P}}}$ can be put in one-to-one correspondence with the symmetric homogeneous space $\mathrm{Sp}(2,2)/ \underline{{\cal{K}}}$. This coset space is in turn homeomorphic to the open unit-ball $B = \big\{ \textbf{q} \; ; \; |\textbf{q}| < 1\big\}$, well-described by the coordinates $\Big(\varphi,\frac{\textbf{b}}{|\textbf{b}|} \frac{\textbf{d}^{\scriptscriptstyle\bigstar}}{|\textbf{d}|}\equiv \textbf{v} \in \mbox{SU}(2)\Big)$ (see Eq. (\ref{coor tan})). To make this homeomorphism apparent, we define:
\begin{eqnarray}\label{coordinate ball}
\rho \underline{p} \underline{p}^\dagger \gamma^0 \; \Big(= \rho \underline{p}^2 \gamma^0 \Big) \; = \frac{\rho}{1 - |\textbf{q}|^2}
\begin{pmatrix}
 \textbf{1}(1 + |\textbf{q}|^2) & -2\textbf{q} \\
2\textbf{q}^{\scriptscriptstyle\bigstar} & - \textbf{1}(1 + |\textbf{q}|^2)
\end{pmatrix} \equiv
\begin{pmatrix}
\textbf{1} x^0 & - \textbf{x} \\
{\textbf{x}}^{\scriptscriptstyle\bigstar} & -\textbf{1} x^0
\end{pmatrix} = \slashed{x}\,,
\end{eqnarray}
where $0<\rho<\infty$ and, again, $x^A$'s $\big( = \frac{1}{4}\mbox{tr}(\gamma^A \slashed{x}) \big)$ are the Cartesian coordinates in $\mathbb{R}^5$. The action of Sp$(2,2)$ on the set of matrices $\rho \underline{p}^2 \gamma^0$, representing the coset space $\mathrm{Sp}(2,2)/\underline{{\cal{K}}}$, can be found from the usual multiplication of the matrices $\underline{p}(\textbf{q})$ from the left:
\begin{eqnarray}\label{1400}
\mathrm{Sp}(2,2) \ni \underline{g} \; : \; \underline{p}(\textbf{q}) \;\mapsto\; \underline{p}(\textbf{q}^\prime)\equiv \underline{p}(\underline{g} \diamond\textbf{q})\,, \;\;\;\;\;\;\; \underline{g}\; \underline{p}(\textbf{q}) = \underline{p}(\textbf{q}^\prime)\; \underline{k}^\prime\,,
\end{eqnarray}
from which, we get:
\begin{eqnarray}
\rho \;\underline{p}({\textbf{q}}^\prime) \;\underline{p}^\dagger({\textbf{q}}^\prime) \;\gamma^0 \;\; \Big(= \rho \;\underline{p}^2({\textbf{q}}^\prime) \;\gamma^0 \Big) \; &=& \rho \; \big( \underline{g} \; \underline{p}(\textbf{q}) \; \underline{k}^{\prime -1} \big) \big( \underline{k}^{\prime} \; \underline{p}(\textbf{q}) \; \underline{g}^\dagger \big) \gamma^0 \nonumber\\
&=& \underline{g} \; \big( \rho \underline{p}^2(\textbf{q}) \gamma^0 \big) \; \gamma^0 \underline{g}^\dagger \gamma^0 \nonumber\\
&=& \underline{g} \; \big( \rho \underline{p}^2(\textbf{q}) \gamma^0 \big) \; \underline{g}^{-1}\,.
\end{eqnarray}
The above action is linear and determinant-preserving:
\begin{eqnarray}
\mbox{det} \Big(\rho \underline{p}^2(\textbf{q}^\prime) \gamma^0\Big) = \mbox{det} \Big(\rho \underline{p}^2(\textbf{q}) \gamma^0\Big) = \left((x^0)^2 - |\textbf{x}|^2\right)^2 = \rho^4\,.
\end{eqnarray}
Moreover, it preserves $(x^0)^2 - |\textbf{x}|^2 = \rho^2$ (see Eq. (\ref{slashx})):
\begin{eqnarray}
(\slashed{x}^{\prime})^2 = g\slashed{x}g^{-1}g\slashed{x}g^{-1} = (\slashed{x})^2 = \left((x^0)^2 - |\textbf{x}|^2\right)\bu_4 = \rho^2 \bu_4\,.
\end{eqnarray}
These identities clearly reveal that each element of the $\rho \underline{p}^2(\textbf{q}) \gamma^0$'s set is in one-to-one correspondence with each point of the upper sheet $\underline{L}_+$ of the two-sheeted hyperboloids $(x)^2 = \rho^2$ in $\mathbb{R}^5$:\footnote{One can also check the one-to-one correspondence between each element of the $\rho \underline{p}^2(\textbf{q}) \gamma^0$'s set and each point of the lower sheet $\underline{L}_- \equiv \big\{ x=(x^0, \textbf{x}) \in \mathbb{R}^5 \;;\; (x^0)^2 - |\textbf{x}|^2 = \rho^2, \; x^0 \leqslant \rho \big\}$ by giving a negative sign to $\rho$, i.e., $\rho \mapsto -\rho$.}
\begin{eqnarray}
\underline{L}_+ \equiv \Big\{ x=(x^0, \textbf{x}) \in \mathbb{R}^5 \;;\; (x^0)^2 - |\textbf{x}|^2 = \rho^2, \; x^0 \geqslant \rho \Big\}\,.
\end{eqnarray}
Note that the subgroup $\underline{{\cal{K}}}$ stabilizes the point $x^{}_\odot=(\rho,\textbf{0})$, which is considered as the origin of upper sheet. Taking into account the Cartesian coordinates (\ref{coordinate ball}) along with Eq. (\ref{coor tan}), the upper sheet $\underline{L}_+$ can also be described by the coordinates $\Big(\varphi,\frac{\textbf{b}}{|\textbf{b}|} \frac{\textbf{d}^{\scriptscriptstyle\bigstar}}{|\textbf{d}|}\equiv \textbf{v} \in \mbox{SU}(2)\Big)$:
\begin{eqnarray}
x^0 &=& \rho\frac{1+|\textbf{q}|^2}{1-|\textbf{q}|^2} = \rho\cosh\varphi\,,\nonumber\\
\textbf{x} &=& \rho\frac{2\textbf{q}}{1-|\textbf{q}|^2} = \rho\sinh\varphi \; \frac{\textbf{b}}{|\textbf{b}|} \frac{\textbf{d}^{\scriptscriptstyle\bigstar}}{|\textbf{d}|}\,.
\end{eqnarray}
This makes apparent the correspondence between each point $(x^0,\textbf{x})$ of $\underline{L}_+$ and the points $\textbf{q}$ in the open unit-ball $B$. Actually, the latter is the stereographic projection of $\underline{L}_+$. This projection explicitly reads:
\begin{eqnarray}
\underline{L}_+ \ni (x^0,\textbf{x}) \;\mapsto\; \textbf{q} = \frac{\textbf{x}}{x^0 + \rho} = \sqrt{\frac{x^0 - \rho}{x^0 + \rho}} \; \frac{\textbf{b}}{|\textbf{b}|} \frac{\textbf{d}^{\scriptscriptstyle\bigstar}}{|\textbf{d}|} \in B\,.
\end{eqnarray}

We end this subsection by pointing out that Eq. (\ref{1400}) also defines the action of Sp$(2,2)$ on the open unit-ball $B$ through the map $\textbf{q}^\prime \equiv \underline{g} \diamond \textbf{q}$ as:
\begin{eqnarray} \label{map B}
B \ni \textbf{q} \;\mapsto\; \textbf{q}^\prime = (\textbf{a} \textbf{q} + \textbf{b})(\textbf{c} \textbf{q} + \textbf{d})^{-1} \in B\,,
\end{eqnarray}
and correspondingly $\textbf{q}^{\prime{\scriptscriptstyle\bigstar}} = (\textbf{d} \textbf{q}^{\scriptscriptstyle\bigstar} + \textbf{c})(\textbf{b} \textbf{q}^{\scriptscriptstyle\bigstar} + \textbf{a})^{-1}$, while:
\begin{eqnarray}
\underline{k}^\prime = \begin{pmatrix} \frac{\textbf{b} \textbf{q}^{\scriptscriptstyle\bigstar} + \textbf{a}}{|\textbf{b} \textbf{q}^{\scriptscriptstyle\bigstar} + \textbf{a}|} & \textbf{0} \\ \textbf{0} & \frac{\textbf{c} \textbf{q} + \textbf{d}}{|\textbf{c} \textbf{q} + \textbf{d}|} \end{pmatrix}\,.
\end{eqnarray}

\subsection{Iwasawa decomposition}\label{Sec Iwasawa dS4}
The Iwasawa decomposition of $\mathrm{Sp}(2,2) = \underline{\cal{K}} \underline{\cal{A}} \underline{\cal{N}}$ implies that any element $\underline{g} \in \mathrm{Sp}(2,2)$ can be factorized in a unique way as \cite{Takahashi'}:
\begin{eqnarray}\label{Iwasawa dS4}
\underline{g} =
\begin{pmatrix}
\textbf{a} & \textbf{b}\\
\textbf{c} & \textbf{d}
\end{pmatrix} =
\underline{k} \; \underline{a} \; \underline{n}\,,
\end{eqnarray}
with:\footnote{Again, the conditions (\ref{rere1}), (\ref{rere2}), and (\ref{equality in norm}) along with the identities given in appendix \ref{App quat} are used here.}
\begin{eqnarray}\label{k a}
\underline{k} &=& \begin{pmatrix} \textbf{v} & \textbf{0} \\ \textbf{0} & \textbf{w} \end{pmatrix} \in \underline{\cal{K}}\,, \\
\underline{a} &=& \begin{pmatrix} \textbf{1}\cosh{\frac{\psi}{2}} & \textbf{1}\sinh{\frac{\psi}{2}} \\ \textbf{1}\sinh{\frac{\psi}{2}} & \textbf{1}\cosh{\frac{\psi}{2}} \end{pmatrix} \in \underline{\cal{A}}\,,\\
\underline{n} &=& \begin{pmatrix} \textbf{1}+\vec{\textbf{y}} & - \vec{\textbf{y}} \\ \vec{\textbf{y}} & \textbf{1}-\vec{\textbf{y}} \end{pmatrix} \in \underline{\cal{N}}\,,
\end{eqnarray}
where:
\begin{eqnarray}\label{v w}
\textbf{v} = \frac{\textbf{a} + \textbf{b}}{|\textbf{a} + \textbf{b}|} \in \mathrm{SU}(2)\,, \;\;\;\;\;\;\; \textbf{w}  = \frac{\textbf{c} + \textbf{d}}{|\textbf{c} + \textbf{d}|} \in \mathrm{SU}(2)\,,
\end{eqnarray}
\begin{eqnarray}
e^{\psi/2} = |\textbf{a} + \textbf{b}| = |\textbf{c} + \textbf{d}|\,,
\end{eqnarray}
and the pure vector quaternion $\vec{\textbf{y}}$ is:
\begin{eqnarray}
\vec{\textbf{y}} = \frac{\textbf{b}^{\scriptscriptstyle\bigstar} \textbf{a} - \textbf{a}^{\scriptscriptstyle\bigstar} \textbf{b}}{2|\textbf{a} + \textbf{b}|^2} = \frac{\textbf{d}^{\scriptscriptstyle\bigstar} \textbf{c} - \textbf{c}^{\scriptscriptstyle\bigstar} \textbf{d}}{2|\textbf{c} + \textbf{d}|^2}\,.
\end{eqnarray}
As for the Cartan decomposition the subgroup $\underline{\cal{K}} \sim \mathrm{SU}(2) \times \mathrm{SU}(2)$ is the maximal compact subgroup, while $\underline{\cal{A}} \sim \mathbb{R}$ and $\underline{\cal{N}} \sim \mathbb{R}^3$ are, respectively, a one-parameter hyperbolic (Cartan) subgroup and a three-parameter abelian (nilpotent) subgroup of Sp$(2,2)$.

In the above group decomposition, the minimal parabolic subgroup is represented by $\underline{\cal{B}} = \underline{\cal{M}} \underline{\cal{A}} \underline{\cal{N}}$, where $\underline{\cal{M}} = \Bigg\{ \underline{\varrho}=\begin{pmatrix} \textbf{w} & 0 \\ 0 & \textbf{w} \end{pmatrix}\; ; \; \textbf{w}\in \mathrm{SU}(2) \Bigg\}$ is the centralizer of $\underline{\cal{A}}$ in $\underline{\cal{K}}$; both $\underline{\cal{A}}$ and $\underline{\cal{M}}$ normalize $\underline{\cal{N}}$. Considering:
\begin{eqnarray}
\underline{k} = \begin{pmatrix} \textbf{v} & \textbf{0} \\ \textbf{0} & \textbf{w} \end{pmatrix} = \begin{pmatrix} \textbf{v}\textbf{w}^{\scriptscriptstyle\bigstar} & \textbf{0} \\ \textbf{0} & \textbf{1} \end{pmatrix}
\begin{pmatrix} \textbf{w} & \textbf{0} \\ \textbf{0} & \textbf{w} \end{pmatrix} \equiv \underline{\tilde{k}}(\textbf{u}) \;\underline{\varrho}\,,
\end{eqnarray}
where $\textbf{u} \equiv \textbf{v}\textbf{w}^{\scriptscriptstyle\bigstar} = (\textbf{a} + \textbf{b}) (\textbf{c} + \textbf{d})^{-1} \in \mathrm{SU}(2)$ (see Eq. (\ref{v w})), one can easily check that $\mathrm{Sp}(2,2)/\underline{\cal{B}} \sim \underline{\cal{K}}/\underline{\cal{M}} \sim \mathrm{SU}(2)$. The latter is in turn homeomorphic to $\mathbb{S}^3$ (the boundary of the open unit-ball $B$).\footnote{For this homeomorphism, see appendix \ref{App UIR's SU(2)}.} The action of Sp$(2,2)$ on the unit-sphere $\mathbb{S}^3$ is found by the usual left action on the set of matrices $\underline{\tilde{k}}(\textbf{u})$:
\begin{eqnarray}
\mathrm{Sp}(2,2) \ni \underline{g} \; : \; \underline{\tilde{k}}(\textbf{u}) \;\mapsto\; \underline{\tilde{k}}(\textbf{u}^\prime) \equiv \underline{\tilde{k}}(\underline{g}\diamond\textbf{u}) \,, \;\;\;\;\;\;\; \underline{g}\; \underline{\tilde{k}}(\textbf{u}) = \underline{\tilde{k}}(\textbf{u}^\prime)\; \underline{\varrho}^\prime \underline{a}^\prime \underline{n}^\prime\,,
\end{eqnarray}
where $\underline{\varrho}^\prime \underline{a}^\prime \underline{n}^\prime \in \underline{\cal{B}}$ and $\textbf{u}^\prime \equiv \underline{g}\diamond \textbf{u}$ is given by:
\begin{eqnarray}\label{map S3}
\mathbb{S}^3 \ni \textbf{u} \;\mapsto\; \textbf{u}^\prime = (\textbf{a} \textbf{u} + \textbf{b})(\textbf{c} \textbf{u} + \textbf{d})^{-1} \in \mathbb{S}^3\,.
\end{eqnarray}

The map (\ref{map S3}) clearly recovers the action of Sp$(2,2)$ on the open unit-ball $B$ (see Eq. (\ref{map B})) extended to the boundary of $B$. It also recovers the action of Sp$(2,2)$ on the set of matrices $\underline{\Upsilon}\big(x(\psi, {\textbf{w}}^2)\big) \equiv \begin{pmatrix} \textbf{1} x^0 & - \textbf{x} \\ {\textbf{x}}^{\scriptscriptstyle\bigstar} & - \textbf{1} x^0 \end{pmatrix}$ (see Eqs. (\ref{spacetime dS4}) and (\ref{action spacetime})), when $\psi$ goes to infinity; symbolically:
\begin{eqnarray}
\underline{\Upsilon}\big(x(\psi^\prime, {\textbf{w}}^{\prime 2})\big) = \underline{g} \Big(\lim_{\psi \rightarrow \infty} \underline{\Upsilon}\big(x(\psi, {\textbf{w}}^2)\big)\Big) \underline{g}^{-1}\,.
\end{eqnarray}
To see the argument lying behind the latter case, one must notice, on one hand, that for the elements $\underline{\Upsilon}\big(x(\psi, {\textbf{w}}^2)\big)$, by letting $\psi \rightarrow \infty$, we obtain $(\textbf{1} x^0) ({\textbf{x}}^{\scriptscriptstyle\bigstar})^{-1} \approx {\textbf{w}}^2 \in \mathrm{SU}(2)$, which implies that $\mbox{det} \big(\underline{\Upsilon}(x)\big) = \left((x^0)^2 - |{\textbf{x}}|^2\right)^2 \approx 0$ and, on the other hand, that for the transformed elements $\underline{\Upsilon}(x^\prime) \equiv \begin{pmatrix} \textbf{1} x^{\prime 0} & - \textbf{x}^{\prime} \\ {\textbf{x}}^{\prime{\scriptscriptstyle\bigstar}} & - \textbf{1} x^{\prime 0} \end{pmatrix}$, we get $(\textbf{1} x^{\prime 0}) ({\textbf{x}}^{\prime{\scriptscriptstyle\bigstar}})^{-1} \approx (\textbf{a} {\textbf{w}}^2 + \textbf{b})(\textbf{c} {\textbf{w}}^2 + \textbf{d})^{-1} \equiv {\textbf{w}}^{\prime 2} \in \mathrm{SU}(2)$, and hence, $\mbox{det} \big(\underline{\Upsilon}(x^\prime)\big) \approx 0$. In this realization of the map (\ref{map S3}), admitting the coordinates (\ref{coor Min dS4}) interestingly entails two manifestations of the unit-sphere $\mathbb{S}^3$. Actually, the limit $\psi\rightarrow\infty$ means that either $t_{\circ}\rightarrow\infty$ and $R$ being fixed, based upon which $\mathbb{S}^3$ depicts the dS$_4$ timelike infinity, or $R\rightarrow 0$ and $t_{\circ}$ being fixed, based upon which $\mathbb{S}^3$ represents the projective null cone in $\mathbb{R}^5$. Regarding the latter case, for the upper (respectively, the lower) sheet of the projective null cone in $\mathbb{R}^5$, the subgroup $\underline{{\cal{N}}}$, present in the Iwasawa decomposition of Sp$(2,2)$, acts as the stabilizer of the base point $x^{}_\odot = (1,0,0,0,1)$ (respectively, $(-1,0,0,0,-1)$), while the other two subgroups involved in this group decomposition, i.e., $\underline{\cal{K}}$ and $\underline{\cal{A}}$, map this base point to any point of the upper (respectively, the lower) sheet of the null cone:
\begin{eqnarray}
\underline{k} \; \underline{a} \begin{pmatrix} \pm\textbf{1} & \mp\textbf{1} \\ \pm\textbf{1} & \mp\textbf{1} \end{pmatrix} \underline{a}^{-1} \underline{k}^{-1} = \pm e^\psi \begin{pmatrix} \textbf{1} & - \textbf{u} \\ \textbf{u}^{\scriptscriptstyle\bigstar} & -\textbf{1} \end{pmatrix}
\equiv \pm \begin{pmatrix} \textbf{1} x^0 & - \textbf{x} \\ {\textbf{x}}^{\scriptscriptstyle\bigstar} & - \textbf{1} x^0 \end{pmatrix} = \underline{\Upsilon}^c(x)\,, \;\;\;\;\;\;\; \mbox{det} \big(\underline{\Upsilon}^c(x)\big) = 0\,.
\end{eqnarray}
Recall that $\textbf{u} = \textbf{v}\textbf{w}^{\scriptscriptstyle\bigstar}$ and $\psi \in \mathbb{R}$.

\subsection{Four integration formulas on Sp$(2,2)$}
In this subsection, considering the above group decompositions, we point out three important integration formulas on Sp$(2,2)$. Let $\mathrm{d}\mu(\underline{g})$ be a Haar measure on Sp$(2,2)$. Then \cite{Takahashi'}:
\begin{itemize}
\item{From the space-time-Lorentz decomposition $\mathrm{Sp}(2,2) \ni \underline{g} = \underline{j}(\textbf{w},\psi) \; \underline{l}(\textbf{v},\vec{\textbf{u}},\varphi)$ (see subsection \ref{sec space-time-Lorentz dS4}), we have:
    \begin{eqnarray}
    \int_{\mathrm{Sp}(2,2)} f(\underline{g}) \; \mathrm{d}\mu(\underline{g}) = \int_{-\infty}^{+\infty} e^{3\psi} \; \mathrm{d}\psi \int_{-\infty}^{+\infty} e^{3\varphi} \; \mathrm{d}\varphi \int_{\mathbb{S}^3} \mathrm{d}\mu(\textbf{w}) \int_{\mathbb{S}^3} \mathrm{d}\mu(\textbf{v}) \int_{\mathbb{S}^3} f\big(\underline{g}(\textbf{w},\psi,\textbf{v},\vec{\textbf{u}},\varphi)\big) \; \mathrm{d}\mu(\vec{\textbf{u}})\,.
    \end{eqnarray}}
\item{From the Cartan decomposition $\mathrm{Sp}(2,2) \ni \underline{g} = \underline{p}(\textbf{q}) \; \underline{k}(\textbf{v},\textbf{w})$ (see subsection \ref{Sec Cartan dS4}), we have:
    \begin{eqnarray}
    \int_{\mathrm{Sp}(2,2)} f(\underline{g}) \; \mathrm{d}\mu(\underline{g}) = \int_{B} (1-|\textbf{q}|^2)^{-4} \; \mathrm{d}\mu(\textbf{q}) \int_{\mathbb{S}^3} \mathrm{d}\mu(\textbf{v}) \int_{\mathbb{S}^3} f\big(\underline{g}(\textbf{q},\textbf{v},\textbf{w})\big) \; \mathrm{d}\mu(\textbf{w})\,.
    \end{eqnarray}}
\item{From the Iwasawa decomposition $\mathrm{Sp}(2,2) \ni \underline{g} = \underline{k}(\textbf{v},\textbf{w}) \; \underline{a}(\psi) \; \underline{n}(\vec{\textbf{y}})$ (see subsection \ref{Sec Iwasawa dS4}), we have:
    \begin{eqnarray}
    \int_{\mathrm{Sp}(2,2)} f(\underline{g}) \; \mathrm{d}\mu(\underline{g}) = 2^3 \int_{-\infty}^{+\infty} e^{3\psi} \; \mathrm{d}\psi \int_{\mathbb{S}^3} \mathrm{d}\mu(\textbf{v}) \int_{\mathbb{S}^3} \mathrm{d}\mu(\textbf{w}) \int_{\mathbb{R}^3} f\big(\underline{g}(\textbf{v},\textbf{w},\psi,\vec{\textbf{y}})\big) \; \mathrm{d}^3\vec{\textbf{y}}\,.
    \end{eqnarray}}
\item{From the (nonunique) ``$KAK$" decomposition of Sp$(2,2)$, which implies that any $\underline{g} \in \mathrm{Sp}(2,2)$ can be decomposed as:
    \begin{eqnarray}
    \underline{g} = \begin{pmatrix} \textbf{a} & \textbf{b} \\ \textbf{c} & \textbf{d} \end{pmatrix} = \begin{pmatrix} \textbf{1} & \textbf{0} \\ \textbf{0} & \textbf{u}\end{pmatrix} \begin{pmatrix} \textbf{1}\cosh\frac{\psi}{2} & \textbf{1}\sinh\frac{\psi}{2} \\ \textbf{1}\sinh\frac{\psi}{2} & \textbf{1}\cosh\frac{\psi}{2} \end{pmatrix} \begin{pmatrix} \textbf{v} & \textbf{0} \\ \textbf{0} & \textbf{w} \end{pmatrix}\,,
    \end{eqnarray}
    where $\textbf{u} \equiv \frac{\textbf{c}}{|\textbf{c}|} \frac{\textbf{a}^{\scriptscriptstyle\bigstar}}{|\textbf{a}|} \in \mathrm{SU}(2)$, $\cosh\frac{\psi}{2} \equiv |\textbf{a}|$ and $\sinh\frac{\psi}{2} \equiv |\textbf{b}|$, with $\psi\geqslant 0$, and finally $\textbf{v} \equiv \frac{\textbf{a}}{|\textbf{a}|} \in \mathrm{SU}(2)$ and $\textbf{w} \equiv \frac{\textbf{b}}{|\textbf{b}|} \in \mathrm{SU}(2)$, we have the integration formula:
    \begin{eqnarray}
    \int_{\mathrm{Sp}(2,2)} f(\underline{g}) \; \mathrm{d}\mu(\underline{g}) = 2\pi^2 \int_{\mathbb{S}^3} \mathrm{d}\mu(\textbf{u}) \int_{\mathbb{S}^3} \mathrm{d}\mu(\textbf{v}) \int_{\mathbb{S}^3} \mathrm{d}\mu(\textbf{w}) \int_{0}^{+\infty} f\big(\underline{g}(\textbf{u},\psi,\textbf{v},\textbf{w})\big) \; (\sinh\psi)^3 \; \mathrm{d}\psi\,.
    \end{eqnarray}}
\end{itemize}

\setcounter{equation}{0} \section{Relativistic meaning of the dS$_4$ group: group (algebra) contraction}\label{Sec Group contraction dS4}

\subsection{Group (algebra) contraction: a brief introduction}\label{Subsec Group contraction, int}
At the origin of the concept of group contraction, initiated by Segal \cite{Segal contraction} and by In\"{o}n\"{u} and Wigner \cite{Wigner contraction} and then developed by Saletan \cite{Saletan}, stands a delicate question: what is the relevance between two symmetry groups when the theories respectively invariant under the action of them are obtained one from the other through a limiting procedure? For instance, in special relativity, it is well known that the transition to the limit $c\rightarrow\infty$ ($c$ being the speed of light) allows to pass from the description of the free relativistic particle to the free Galilean particle. Now, what can we say about their associated symmetry groups (respectively, called Poincar\'{e} and Galileo groups)? In other words, how to formulate this limit on the group level?

Technically, in the literature, authors employ different contractions of Lie algebras instead of the Lie groups. Besides differences, they all have a feature in common, arising from the following well-known facts. The Lie algebra of a given Lie group is indeed a linear space described by basis elements (generators of the algebra) along with the commutation relations between them. As shown by Cartan, whenever one changes the basis by a nonsingular transformation, the obtained basis will describe the generators of an algebra isomorphic to the former. However, if the transformation is singular, the situation would be different and one may get a new algebra, provided that the properties of the commutator to be the commutator of a Lie algebra are fulfilled. In practice, the contraction of a Lie algebra can be carried out with respect to a sequence of transformations (rather than one transformation) of basis elements, and correspondingly, of the associated commutation relations. Transformations in the sequence depend on one (or more) parameter $\lambda$ such that for all values of $\lambda$ except one (usually, $\lambda=0$) the corresponding transformation is regular and in that particular point the transformation becomes singular. The contraction of the algebra then is performed, if by letting $\lambda$ tend to that particular point (say $\lambda\rightarrow 0$), the obtained generators form a new Lie algebra \cite{Mickelsson}. Here, to articulate more clearly the mechanism of group (algebra) contraction, following the instruction given in Ref. \cite{Garidi Thesis}, we explain the mechanism in a way which is neither the most rigorous nor the most modern one, but which allows one to understand simply the procedure.

We consider a Lie algebra constituted by the basis $X_1,\;...\;,X_r$ with the commutation relations $[X_\sigma,X_\iota] = c_{\sigma\iota}^\upsilon X_\upsilon$, where $\sigma,\iota,\upsilon=1,\;...\;,r$ and $c_{\sigma\iota}^\upsilon$'s stand for the structure constants of the algebra. For a subset $X_1,\;...\;,X_s$ of this basis, with $s\leqslant r$, we define:
\begin{eqnarray}
Y_i \equiv \lambda X_i\,, \;\;\;\;\;\;\; i= 1\,,\;...\;,s\leqslant r\,,
\end{eqnarray}
and accordingly, we rewrite the commutation relations in terms of $Y_i$'s as:
\begin{eqnarray}
[Y_i,Y_j] = \lambda c_{ij}^k Y_k\,, \;\;\;\;\;\;\; [Y_i,X_m] = c_{im}^k Y_k + \lambda c_{im}^n X_n\,, \;\;\;\;\;\;\; [X_m,X_n] = \lambda^{-1} c_{mn}^k Y_k + c_{mn}^l X_l\,,
\end{eqnarray}
with $i,j,k \leqslant s$ and $s < m,n,l \leqslant r$. The contraction of the Lie algebra (when it is possible) is carried out by letting $\lambda$ tend to zero, provided that we specify under which conditions the elements $Y_1,\;...\;,Y_s,X_{s+1},\;...\;,X_r$ form a new Lie algebra designated as the contracted algebra. Considering the above, a necessary condition is:
\begin{eqnarray}
c_{mn}^k = 0\,, \;\;\;\;\; \mbox{for} \;\; k \leqslant s \;\; \mbox{and} \;\; s < m,n\leqslant r\,.
\end{eqnarray}
Note that, in this case, the generators $X_{s+1},\;...\;,X_r$, possessing the commutation relations $[X_m,X_n] = c_{mn}^l X_l$, form a Lie subalgebra, which remains intact during the contraction procedure. This subalgebra corresponds to the subgroup based upon which, it is said that, the contraction is carried out.

\subsection{DS$_4$ group (algebra) contractions}
DS$_4$ relativity involves the universal length $R$, denoting the radius of curvature of the dS$_4$ hyperboloid $\underline{M}_R$, and the universal speed of light $c$. As a matter of fact, as long as $c$ is not normalized to unity, the radius of curvature of $\underline{M}_R$ is given by $R=cH^{-1}$, where $H$ is like the Hubble constant (within a cosmological framework). Taking these fundamental physical constants into account, the physical analysis of dS$_4$ relativity, specially from the point of view of the question of mass (see part \ref{Part mass}), is irremediably relevant to its flat ($R\rightarrow\infty$) and nonrelativistic ($c\rightarrow\infty$) limits. Therefore, mathematically, we have to encounter the question of contraction limit on different levels of complexities, i.e., geometry, Lie algebra, and group representations.\footnote{Note that the dS$_4$ group, as a group of (pseudo-)rotations, has parameters which are (pseudo-)angles, i.e. pure numbers; no physical dimension is needed on that level. Physical dimensions necessarily appear with contractions, one per contraction parameter.} We are here solely concerned with the algebraic level; the geometrical level has been already discussed in subsection \ref{sec space-time-Lorentz dS4}, and the representation level will be discussed in subsection \ref{Subsec massive contraction}.

The set of all possibilities of contractions from the dS$_4$ (relativity) group algebra, with respect to symmetry principles and some physically reasonable assumptions, namely \emph{space is isotropic}, \emph{parity and time-reversal are group automorphisms}, \emph{boosts are non-compact}, has been precisely presented by Bacry and L\'{e}vy-Leblond \cite{BacryLevi}. [Note that in Ref. \cite{Levy-Nahas}, focusing on abstract groups, relevant results have been rigorously established in terms of \emph{inverse contraction}, i.e., \emph{deformation}\footnote{Just to get the gist, let us restrict ourselves to the algebraic level: roughly speaking, by contraction (as explained above) is meant a transformation of a Lie algebra into a ``more Abelian" one by making some structure constants vanish, and by deformation is meant a transformation of a Lie algebra into a ``less Abelian" one by producing some nonzero structure constants. Note that ``more Abelian'' is generally associated with the appearance of \emph{direct} or \emph{semi-direct} product structures in the contracted group. [We have already given a brief reminder on \emph{direct} product of two groups (see subsection \ref{Sec Cartan dS4}, footnote \ref{foot2}). Hence, let us remind here just \emph{semi-direct} product of two groups $G_1$ and $G_2$. Suppose given a homomorphism $\alpha$ from $G_2$ into the group ${\rm Aut} (G_1)$ of automorphisms of $G_1$. Then, we define the semi-direct product $G = G_1 \rtimes G_2$ as the Cartesian product, endowed with the group law:
\begin{equation}
\label{sdprod}
(g^{}_1, g^{}_2)(g_1^{\prime}, g_2^{\prime}) = (g^{}_1 \alpha_{g^{}_2}(g_1^{\prime}) , g^{}_2g_2^{\prime})\,, \nonumber
\end{equation}
Neutral element is $(e_1,e_2)$ and the inverse of $(g_1, g_2)$ is $(g_1, g_2)^{-1} = \left(\left[\alpha_{g_2^{-1}}(g_1)\right]^{-1}, g_2^{-1}\right)$.]}.] Here, as pointed out above, we are particularly interested in the flat and nonrelativistic contraction limits of the dS$_4$ group algebra, which respectively lead to the Poincar\'{e} and the so-called Newton (relativity) group algebras. Both Poincar\'{e} and Newton group algebras in turn contract towards the algebra of the Galileo group. Symbolically, we have:
\begin{eqnarray}
\begin{array}{ccccccc}
{\mbox{dS$_4$ group (algebra)}} & \longrightarrow & {\mbox{Poincar\'{e} group (algebra)}} & \\
\\
\big\downarrow &  & \big\downarrow &\\
\\
{\mbox{Newton group (algebra)}} & \longrightarrow & {\mbox{Galilei group (algebra)}} &
\end{array}
\end{eqnarray}
where the arrows represent group (algebra) contractions. Below, we elaborate these contraction limits.

\subsubsection{Contraction of the dS$_4$ group $\longrightarrow$ the Poincar\'{e} group $\longrightarrow$ the Galileo group}\label{Subsubsec 10.2.1}
Here, following the lines sketched in Refs. \cite{Levy-Nahas,BacryLevi}, we explicitly present the contraction of the dS$_4$ group algebra towards the algebra of the Poincar\'{e} group, and then, the contraction of the latter towards the Galileo group algebra. We first recall that the dS$_4$ Lie algebra $\mathfrak{sp}(2,2)$ is defined by:
\begin{eqnarray}
[K_{AB},K_{CD}] = - \big( \eta^{}_{AC} {K_{BD}} + \eta^{}_{BD} {K_{AC}} - \eta^{}_{AD} {K_{BC}} - \eta^{}_{BC} {K_{AD}} \big)\,,
\end{eqnarray}
where $A,B = 0,1,2,3,4$ and $\eta^{}_{AB} = \mbox{diag} (1,-1,-1,-1,-1)$. Moreover, we recall that the proper Poincar\'{e} group is the semi-direct product of the group of spacetime translations with the orthochronous Lorentz group $\mathrm{SO}_0(1,3)$ or its universal covering SL$(2,\mathbb{C})$. Now, we define $j_{\mu\nu} \equiv K_{\mu\nu}$ and $p_\mu \equiv R^{-1} K_{4\mu}$, where $\mu,\nu=0,1,2,3$, while $R$, interpreted as the radius of curvature of $\underline{M}_R$, plays the role of contraction parameter. Accordingly, at the null-curvature limit ($R\rightarrow \infty$), we get:
\begin{eqnarray}
[p_\mu,p_\nu] = R^{-2} [K_{4\mu},K_{4\nu}] = R^{-2} K^{}_{\mu\nu} = R^{-2} j^{}_{\mu\nu} \rightarrow 0\,,
\end{eqnarray}
\begin{eqnarray}
[p_\rho,j_{\mu\nu}] = R^{-1} [K_{4\rho},K_{\mu\nu}] = - R^{-1} \big( \eta^{}_{\rho\nu} {K_{4\mu}} - \eta^{}_{\rho\mu} {K_{4\nu}} \big) \rightarrow - \big( \eta^{}_{\rho\nu} {p_{\mu}} - \eta^{}_{\rho\mu} {p_{\nu}} \big)\,,
\end{eqnarray}
\begin{eqnarray}\label{Lorentz algebra}
[j_{\mu\nu},j_{\rho\sigma}] = [K_{\mu\nu},K_{\rho\sigma}] &=& - \big( \eta^{}_{\mu\rho} {K_{\nu\sigma}} + \eta^{}_{\nu\sigma} {K_{\mu\rho}} - \eta^{}_{\mu\sigma} {K_{\nu\rho}} - \eta^{}_{\nu\rho} {K_{\mu\sigma}} \big) \nonumber\\
&=& - \big( \eta^{}_{\mu\rho} {j_{\nu\sigma}} + \eta^{}_{\nu\sigma} {j_{\mu\rho}} - \eta^{}_{\mu\sigma} {j_{\nu\rho}} - \eta^{}_{\nu\rho} {j_{\mu\sigma}} \big)\,.
\end{eqnarray}
Note that all the indices, appeared above, only take the values $0,1,2,3$. The above commutation relations between the (ten) generators $p_\mu$ and $j_{\mu\nu}$ interestingly display the Lie algebra of the Poincar\'{e} group; the infinitesimal generators $p_\mu$ and $j_{\mu\nu}$ respectively represent the four basic elements of the Lie algebra of the spacetime-translations subgroup and the six basic elements of the Lie algebra of the Lorentz subgroup $\mathrm{SO}_0(1,3)$ (or SL$(2,\mathbb{C})$); note that $j_{\mu\nu}=-j_{\nu\mu}$. According to the terminology given in subsection \ref{Subsec Group contraction, int}, the contraction that we have just described is carried out with respect to the Lorentz subgroup $\mathrm{SO}_0(1,3)$ (or SL$(2,\mathbb{C})$), whose Lie algebra, given by Eq. (\ref{Lorentz algebra}), remains intact during the contraction procedure.

We now turn to the contraction of the Poincar\'{e} group towards the Galileo one. The latter is the relativity group of classical mechanics. This group is constituted by the spacetime translations (corresponding to the four generators $p_\mu$, with $\mu=0,1,2,3$), the space rotations (three generators denoted here by $r^{}_{ik}$, with $i,k = 1,2,3$ (note that $r^{}_{ik}=-r^{}_{ki}$)), and the inertial transformations (three generators denoted here by $b^{}_i$). Thus, the Galileo group also admits ten infinitesimal generators.

The contraction of the Poincar\'{e} group algebra towards the Galileo one is obtained by adjusting $r^{}_{ik}\equiv j^{}_{ik}$ and $b^{}_i \equiv c^{-1} j^{}_{0i}$. In this case, the parameter $c$, interpreted as the speed of light, represents the contraction parameter. When $c\rightarrow\infty$, the Lie algebra of the Poincar\'{e} group, given above, turns into:
\begin{eqnarray}
[r^{}_{il}, b^{}_k] = c^{-1} [j^{}_{il},j_{0k}^{}] = c^{-1} \big( \eta^{}_{lk} j^{}_{0i} - \eta^{}_{ik} j^{}_{0l} \big) \rightarrow \eta^{}_{lk} b^{}_{i} - \eta^{}_{ik} b^{}_{l}\,,
\end{eqnarray}
\begin{eqnarray}
[b^{}_{i}, b^{}_k] = c^{-2} [j^{}_{0i},j_{0k}^{}] = - c^{-2} j^{}_{ik} = - c^{-2} r^{}_{ik} \rightarrow 0\,,
\end{eqnarray}
\begin{eqnarray}
[b^{}_{i}, p_\mu] = c^{-1} [j^{}_{0i},p_\mu] = c^{-1} \big( \eta^{}_{\mu i} {p_{0}} - \eta^{}_{\mu 0} {p_{i}} \big) \rightarrow 0\,,
\end{eqnarray}
with the following unchanged commutation relations:
\begin{eqnarray}
[p_\mu,p_\nu] = 0\,, \;\;\;\;\;\;\; \mbox{and} \;\;\;\;\;\;\; [r^{}_{ik},r^{}_{lm}] = - \big( \eta^{}_{il} {r^{}_{km}} + \eta^{}_{km} {r^{}_{il}} - \eta^{}_{im} {r^{}_{kl}} - \eta^{}_{kl} {r^{}_{im}} \big)\,,
\end{eqnarray}
where the indices $i,k,l,m = 1,2,3$ and $\mu,\nu=0,1,2,3$. These commutation relations form the Lie algebra of the Galileo group. [Considering the latter set of commutation relations, it is useful here to make the link with the angular momentum components, for instance, by defining $\ell^{}_i \equiv r^{}_{jk}$, where $(i,j,k)$ is even permutation of $(1,2,3)$, so that, $[\ell^{}_i,\ell^{}_j]=\ell^{}_k$.]

\subsubsection{Contraction of the dS$_4$ group $\longrightarrow$ the Newton group $\longrightarrow$ the Galileo group}\label{Subsubsec 10.2.2}
The contraction of the dS$_4$ group algebra with respect to the infinitesimal generators of the time-translation and rotation subgroups leads to the algebra of the Newton group \cite{BacryLevi} (see also Refs. \cite{Tian2005,Banerjee}). Technically, this contraction is performed by adjusting $p^\prime_0\equiv cR^{-1} K_{40}$, $p^\prime_i\equiv R^{-1} K_{4i}$, $r^{}_{ik}\equiv K_{ik}$, and finally $b_i\equiv c^{-1} K_{0i}$, where $i,k=1,2,3$, while $c$ and $R$, again, refer to the speed of light and the radius of $\underline{M}_R$, respectively. In this case, $c$ and $R$ play the role of contraction parameters. Letting $c,R\rightarrow\infty$, while $\tau^{}_\mathfrak{a} = R/c$ remains unchanged,\footnote{Note that the reason for this naming convention, strictly speaking, for the subscript `$\mathfrak{a}$' in $\tau^{}_\mathfrak{a}$, will be clarified in the next section.} we get:
\begin{eqnarray}
[r^{}_{il}, b^{}_k] = c^{-1} [K^{}_{il},K_{0k}^{}] = c^{-1} \big( \eta^{}_{lk} K^{}_{0i} - \eta^{}_{ik} K^{}_{0l} \big) \rightarrow \eta^{}_{lk} b^{}_{i} - \eta^{}_{ik} b^{}_{l}\,,
\end{eqnarray}
\begin{eqnarray}
[p^\prime_0,p^\prime_i] = cR^{-2} [K_{40},K_{4i}] = cR^{-2} {K_{0i}} \rightarrow \tau^{-2}_\mathfrak{a} b_{i}\,,
\end{eqnarray}
\begin{eqnarray}
[p^\prime_0,b^{}_i] = R^{-1} [K_{40},K_{0i}] = R^{-1} K_{4i} \rightarrow p^\prime_i\,,
\end{eqnarray}
\begin{eqnarray}
[p^\prime_i,p^\prime_k] = R^{-2} [K_{4i},K_{4k}] = R^{-2} K_{ik} \rightarrow 0\,,
\end{eqnarray}
\begin{eqnarray}
[b_i,b_k] = c^{-2} [K_{0i},K_{0k}] = - c^{-2} K_{ik} \rightarrow 0\,,
\end{eqnarray}
\begin{eqnarray}
[p^\prime_i,b_k] = c^{-1}R^{-1} [K_{4i},K_{0k}] = - c^{-1}R^{-1} \eta^{}_{ik} K_{40} \rightarrow 0\,,
\end{eqnarray}
with the following unchanged commutation relations:
\begin{eqnarray}
[r^{}_{ik},r^{}_{lm}] = - \big( \eta^{}_{il} {r^{}_{km}} + \eta^{}_{km} {r^{}_{il}} - \eta^{}_{im} {r^{}_{kl}} - \eta^{}_{kl} {r^{}_{im}} \big)\,,
\end{eqnarray}
where the indices $i,k,l,m = 1,2,3$. The above commutation relations characterize the Newton group algebra. Finally, one can check that letting $\tau^{}_\mathfrak{a}\rightarrow \infty$ in the above commutation relations immediately yields those of the Galileo algebra.

\setcounter{equation}{0} \section{DS$_4$ Lie algebra and classical phase spaces}\label{Sec dS4 phase spaces}
Quite analogous to the $1+1$-dimensional case (see subsection \ref{Subsec coadjoint dS2-int}), dS$_4$ elementary systems on the classical level can be understood along the traditional phase-space approach, well established through the notion of the orbits of the Sp$(2,2)$ co-adjoint action \cite{Kirillov,Kirillov1976}; such orbits are symplectic manifolds, and each of them, carrying a natural Sp$(2,2)$-invariant (Liouville) measure, is a homogeneous space homeomorphic to an even-dimensional group coset $\mathrm{Sp}(2,2)/\underline{\cal{S}}$, where $\underline{\cal{S}}$ is the stabilizer subgroup of some orbit point. Of course, one must notice that, since Sp$(2,2)$ is a simple group, its adjoint action on the Lie algebra $\mathfrak{sp}(2,2)$ would be equivalent to its co-adjoint action on the dual (defined as a vector space through the nondegenerate Killing form) of $\mathfrak{sp}(2,2)$ \cite{Kirillov,Kirillov1976}.

Here, it is worthwhile noting that:
\begin{itemize}
\item{The dS$_4$ Lie algebra $\mathfrak{sp}(2,2)$ (in quaternionic notations, with the quaternionic basis $\{ \textbf{1} \equiv \mathbbm{1}_2, {\textbf{e}}^{}_k \equiv (-1)^{k+1} \mathrm{i} \sigma_k ; \; k=1,2,3 \}$, where $\sigma_k$'s refer to the Pauli matrices) can be realized by the linear span of the infinitesimal generators $\underline{X}_k,\; \underline{X}_0,\; \underline{Y}_k$, and $\underline{Z}_k$, respectively, given in Eqs. (\ref{Xk}), (\ref{X0}), (\ref{Yk}), and (\ref{Zk}):\footnote{\label{foot3}Here, for future use and by referring to the commutation relations (\ref{commutation relations dS4}), we would like to highlight the fact that the space-rotation generators $\underline{Y}_k$ ($k=1,2,3$) constitute a subalgebra of $\mathfrak{sp}(2,2)$ (more accurately, of the Lorentz subalgebra of $\mathfrak{sp}(2,2)$), which is isomorphic to $\mathfrak{su}(2)$, and which commutes with the time-translations generator $\underline{X}_0$. Moreover, referring to the subsubsection \ref{Subsubsec 10.2.1}, we would like to recall that the Lorentz subalgebra of $\mathfrak{sp}(2,2)$, containing the aforementioned $\mathfrak{su}(2)$, remains intact during the Poincar\'{e} contraction limit. [In other words, the Lie algebra of the subgroup $\mathrm{SO}_0(1,3)$ (or SL$(2,\mathbb{C})$) is recognized as the Lorentz subalgebra in both Poincar\'{e} and dS$_4$ relativities.] This directly implies that the mentioned $\mathfrak{su}(2)$ subalgebra exactly coincides with the $\mathfrak{su}(2)$ (rotations) subalgebra of the Lorentz one in the context of Poincar\'{e} relativity. Just in passing and as a final remark, we also would like to add that the latter $\mathfrak{su}(2)$ gives sense to the notion of spin in (quantum) Poincar\'{e} elementary systems. We will come back to these important facts in the sequel.}
\begin{eqnarray}\label{algebra dS4}
\mathfrak{sp}(2,2) = \Bigg\{ 2a^k \underline{X}_k + 2j^k \underline{Y}_k + 2d^0 \underline{X}_0 + 2d^k \underline{Z}_k &=&
\begin{pmatrix} (a^k+j^k) \textbf{e}^{}_k & d^0 \textbf{1} + d^k \textbf{e}^{}_k \\ d^0 \textbf{1} - d^k \textbf{e}^{}_k & (-a^k+j^k) \textbf{e}^{}_k \end{pmatrix} \nonumber\\
&\equiv& \begin{pmatrix} {\vec{\textbf{n}}}^{(l)} & \textbf{d} \\ {\textbf{d}}^{\scriptscriptstyle\bigstar} & {\vec{\textbf{n}}}^{(r)} \end{pmatrix} \;;\; a^k, j^k, d^0, d^k \in \mathbb{R} \Bigg\}\,,
\end{eqnarray}
where $\textbf{d}, {\vec{\textbf{n}}}^{(l)}, {\vec{\textbf{n}}}^{(r)} \in \mathbb{H}$ ($\vec{\textbf{n}}^{(l)}$ and $\vec{\textbf{n}}^{(r)}$ are pure vector quaternions).\footnote{Note that the reason for this naming convention, strictly speaking, for the superscripts `$(l)$' and `$(r)$' in $\vec{\textbf{n}}^{(l)}$ and $\vec{\textbf{n}}^{(r)}$, is to make a connection with the discussions that will be given in the next section.} It follows that the $\mathfrak{sp}(2,2)$ algebra, specified by the (ten) free real parameters $a^k, j^k, d^0,$ and $d^k$ ($k=1,2,3$), is in one-to-one correspondence with $\mathbb{R}^{10}$ (again, by abuse of notation, let us say $\mathfrak{sp}(2,2)\sim\mathbb{R}^{10}$). Taking $a^k, j^k, d^0,$ and $d^k$ ($k=1,2,3$) as cartesian coordinates on the dual of $\mathfrak{sp}(2,2)$ (again, the Lie algebra of a simple (generally, semi-simple) Lie group is isomorphic to its dual), their Poisson brackets (see subsection \ref{Subsec coadjoint dS2-int}, footnote \ref{foot}) are given directly by the commutation relations (\ref{commutation relations dS4}) among the corresponding Lie algebra generators:
\begin{eqnarray}\label{dddd}
\big\{j^m,j^n\big\} &=& {{\cal{E}}^{mn}}_{k} \; j^k\,,\nonumber\\
\big\{j^m,a^n\big\} &=& {{\cal{E}}^{mn}}_{k} \; a^k\,,\nonumber\\
\big\{a^m,a^n\big\} &=& {{\cal{E}}^{mn}}_{k} \; j^k\,,\nonumber\\
\big\{j^m,d^n\big\} &=& {{\cal{E}}^{mn}}_{k} \; d^k\,,\nonumber\\
\big\{a^m,d^n\big\} &=& -\delta^{mn} \; d^0\,,\nonumber\\
\big\{d^m,d^n\big\} &=& -{{\cal{E}}^{mn}}_{k} \; j^k\,,\nonumber\\
\big\{d^0,a^m\big\} &=& -d^m\,,\nonumber\\
\big\{d^0,d^m\big\} &=& -a^m\,,\nonumber\\
\big\{d^0,j^m\big\} &=& 0\,,
\end{eqnarray}
where $m,n,k = 1,2,3$ and again ${{\cal{E}}^{mn}}_{k}$ is the three-dimensional totally antisymmetric Levi-Civita symbol.}

\item{The adjoint action of the Sp$(2,2)$ group on its Lie algebra $\mathfrak{sp}(2,2)$ is defined by:
\begin{eqnarray}\label{Ad_g dS4}
\underline{g} \in \mathrm{Sp}(2,2), \; \underline{X} \in \mathfrak{sp}(2,2) \; ; \; \mbox{Ad}_{\underline{g}}(\underline{X}) = \underline{g} \underline{X} \underline{g}^{-1}\,.
\end{eqnarray}}
\end{itemize}

Below, we will briefly discuss four (related) families of (co-)adjoint orbits and their phase-space interpretations.

\subsection{Phase space for scalar ``massive"/``massless" elementary systems in dS$_4$ spacetime}\label{Subsec scalar massive co-adjoint}
We study here a particular family of (co-)adjoint orbits of the $\mathfrak{sp}(2,2)$ algebra, each being related to the transport of the element $2 \kappa \underline{X}_0 = \kappa\begin{pmatrix} \textbf{0} & \textbf{1} \\ \textbf{1} & \textbf{0} \end{pmatrix}$, with a given $0<\kappa<\infty$, under the (co-)adjoint action (\ref{Ad_g dS4}). In this case, the subgroup stabilizing the element $2 \kappa \underline{X}_0$ is made up with the space-rotations and time-translations subgroups appeared in the space-time-Lorentz decomposition of Sp$(2,2)$ (see subsection \ref{sec space-time-Lorentz dS4}):
\begin{eqnarray}
\underline{{\cal{S}}} = \Bigg\{ \underline{g} =
\begin{pmatrix}
{\textbf{v}} & \textbf{0} \\
\textbf{0} & {\textbf{v}}
\end{pmatrix}
\begin{pmatrix}
\textbf{1}\cosh \frac{\psi}{2} & \textbf{1}\sinh\frac{\psi}{2}\\
\textbf{1}\sinh\frac{\psi}{2} & \textbf{1}\cosh \frac{\psi}{2}
\end{pmatrix}\; ; \; {\textbf{v}} \in \mathrm{SU}(2),\; \psi \in \mathbb{R} \Bigg\} \sim \mathrm{SU}(2) \times \mathrm{SO}_0(1,1)\,.
\end{eqnarray}
This family of the (co-)adjoint orbits therefore can be identified by the group coset $O(2 \kappa \underline{X}_0) \sim \mathrm{Sp}(2,2)/\big(\mathrm{SU}(2) \times \mathrm{SO}_0(1,1)\big)$. The latter in turn, with respect to the space-time-Lorentz decomposition of Sp$(2,2)$, can be realized by applying the space-translations and Lorentz-boosts subgroups to transport the element $2 \kappa \underline{X}_0$ under the effective (co-)adjoint action (\ref{Ad_g dS4}). Let:
\begin{eqnarray}
\Big\{ \mbox{space translations} \times \mbox{Lorentz boosts}\Big\} &=& \Bigg\{
\begin{pmatrix}
{\textbf{w}} & \textbf{0} \\
\textbf{0} & {\textbf{w}}^{\scriptscriptstyle\bigstar}
\end{pmatrix}
\begin{pmatrix}
\textbf{1}\cosh \frac{\varphi}{2} & \vec{\textbf{u}}\sinh \frac{\varphi}{2}\\
- \vec{\textbf{u}}\sinh\frac{\varphi}{2} & \textbf{1}\cosh\frac{\varphi}{2}
\end{pmatrix} \;;\; {\textbf{w}},\vec{\textbf{u}} \in \mathrm{SU}(2),\; \varphi \in \mathbb{R}
\Bigg\} \nonumber\\
&\equiv& \Big\{ \underline{s}({\textbf{w}}) \; \underline{b}(\varphi,\vec{\textbf{u}}) \Big\}\,,
\end{eqnarray}
then, according to the (co-)adjoint action (\ref{Ad_g dS4}), we have:
\begin{eqnarray}\label{zzzzz}
\mbox{Ad}_{\underline{g}}(2 \kappa \underline{X}_0) &=& \underline{s}({\textbf{w}}) \; \underline{b}(\varphi,\vec{\textbf{u}}) \; \Big( 2 \kappa \underline{X}_0 \Big)\; \underline{b}^{-1}(\varphi,\vec{\textbf{u}}) \; \underline{s}^{-1}({\textbf{w}}) \nonumber\\
&=&
\begin{pmatrix}
\vec{\textbf{p}} & p_0 {\textbf{w}}^2 \\
p_0 {\textbf{w}}^{2{\scriptscriptstyle\bigstar}} & -{\textbf{w}}^{2{\scriptscriptstyle\bigstar}} \vec{\textbf{p}} {\textbf{w}}^2
\end{pmatrix} \equiv \begin{pmatrix}
\vec{\textbf{p}} & p_0 {\textbf{z}} \\
p_0 {\textbf{z}}^{{\scriptscriptstyle\bigstar}} & -{\textbf{z}}^{{\scriptscriptstyle\bigstar}} \vec{\textbf{p}} {\textbf{z}}
\end{pmatrix}\equiv\underline{X}({\textbf{z}},\vec{\textbf{p}}) \,,
\end{eqnarray}
where $\vec{\textbf{p}}\equiv \kappa {\textbf{w}} \vec{\textbf{u}} {\textbf{w}}^{\scriptscriptstyle\bigstar} \sinh{\varphi}$ is a pure vector quaternion and $p_0 = \kappa\cosh\varphi = \left(\kappa^2 + \big|\vec{\textbf{p}}\big|^2\right)^{1/2}$. This parametrization makes clear the topological nature ${\mathbb{S}}^3 \times {\mathbb{R}}^3 = \big\{ \underline{X}({\textbf{z}},\vec{\textbf{p}}) \; ; \; {\textbf{z}} \in \mathrm{SU}(2) \sim {\mathbb{S}}^3,\; \vec{\textbf{p}} \sim {\mathbb{R}}^3 \big\}$ of the ($6$-dimensional) (co-)adjoint orbits $O(2 \kappa \underline{X}_0)$. Note that the invariant measure on $O(2 \kappa \underline{X}_0)$'s, in terms of the coordinates $({\textbf{z}},\vec{\textbf{p}})$, reads \cite{Rabeie}:
\begin{eqnarray}
\mathrm{d}\mu ({\textbf{z}},\vec{\textbf{p}}) = \mathrm{d}\mu({\textbf{z}}) \; \mathrm{d}^3 \vec{\textbf{p}} \,,
\end{eqnarray}
where $\mathrm{d}\mu({\textbf{z}})$ is the $O(4)$-invariant measure on $\mathbb{S}^3$.

Now, we turn to the phase-space interpretation of the above construction. First of all, for the sake of reasoning, let us identify the generic element $\underline{X}({\textbf{z}},\vec{\textbf{p}})$ of the (co-)adjoint orbits $O(2 \kappa \underline{X}_0)$, as a member of the $\mathfrak{sp}(2,2)$ Lie algebra (see Eq. (\ref{algebra dS4})), by:
\begin{eqnarray}\label{ggggg}
\begin{pmatrix}
\vec{\textbf{p}} & p_0 {\textbf{z}} \\
p_0 {\textbf{z}}^{{\scriptscriptstyle\bigstar}} & -{\textbf{z}}^{{\scriptscriptstyle\bigstar}} \vec{\textbf{p}} {\textbf{z}}
\end{pmatrix}
\equiv \begin{pmatrix} \vec{\textbf{n}}^{(l)} = (0,\vec{a}+\vec{j}) & \;\;\textbf{d}=(d^0,\vec{d}) \\ {\textbf{d}}^{\scriptscriptstyle\bigstar}=(d^0,-\vec{d}) & \;\;\vec{\textbf{n}}^{(r)} = (0,-\vec{a}+\vec{j}) \end{pmatrix}\,.
\end{eqnarray}
[Above, we have used the scalar-vector representation of the quaternionic components introduced in Eq. (\ref{algebra dS4}).] With respect to the relations that hold between the components of the (co-)adjoint generic element $\underline{X}({\textbf{z}},\vec{\textbf{p}})$, it is quite straightforward to show that ${\textbf{d}}^{\scriptscriptstyle\bigstar}\vec{\textbf{n}}^{(l)}=-\vec{\textbf{n}}^{(r)}{\textbf{d}}^{\scriptscriptstyle\bigstar}$ and $\big|{\textbf{d}}\big|^2 - \big|\vec{\textbf{n}}^{(l)}\big|^2 = \kappa^2$. These identities respectively result in the following conditions:
\begin{eqnarray} \label{conservation laws}
\vec{j} &=& \frac{1}{d^0} \vec{d}\times\vec{a}\,, \;\;\;\;\;\;\; (\vec{j}\cdot\vec{d}=0=\vec{j}\cdot\vec{a}) \nonumber\\
\kappa^2 &=& (d^0)^2 + \vec{d}\cdot\vec{d} - \vec{a}\cdot\vec{a} - \vec{j}\cdot\vec{j}\,,
\end{eqnarray}
where the symbols `$\cdot$' and `$\times$' respectively refer to the Euclidean inner product and the cross product in $\mathbb{R}^3$. The very point to be noticed here is that, due to the conditions (\ref{conservation laws}) between $\vec{a},\vec{j},d^0$, and $\vec{d}$, the number of degrees of freedom in (the dual of) $\mathfrak{sp}(2,2)\sim\mathbb{R}^{10}$ reduces from $10$ to $6$, which is exactly the degrees of freedom on each $O(2 \kappa \underline{X}_0)$ (that is where we began from). Hence, although one may still find other relations between the components $\vec{\textbf{n}}^{(l)},\vec{\textbf{n}}^{(r)},\textbf{d}$, and ${\textbf{d}}^{\scriptscriptstyle\bigstar}$ of the (co-)adjoint generic element $\underline{X}({\textbf{z}},\vec{\textbf{p}})$, for instance, from $\mbox{det}\big( \underline{X}({\textbf{z}},\vec{\textbf{p}}) \big)=\kappa^4$ (the (co-)adjoint action (\ref{Ad_g dS4}) is determinant-preserving), they will ultimately result in the same conditions as (\ref{conservation laws}). In this sense, we argue that the conditions (\ref{conservation laws}) precisely characterize the aforementioned family of (co-)adjoint orbits $O(2 \kappa \underline{X}_0)$ in (the dual of) $\mathfrak{sp}(2,2)\sim\mathbb{R}^{10}$:
\begin{eqnarray}\label{massive orbit4}
O(2 \kappa \underline{X}_0) = \Big\{ (\vec{a},\vec{j},d^0,\vec{d}) \;;\; \vec{j} = \frac{1}{d^0} \vec{d}\times\vec{a}, \; \kappa^2 = (d^0)^2 + \vec{d}\cdot\vec{d} - \vec{a}\cdot\vec{a} - \vec{j}\cdot\vec{j} \Big\}\,.
\end{eqnarray}

Interestingly, for a given $\kappa$, recognizing the (co-)adjoint orbit $O(2 \kappa \underline{X}_0)$ as the phase space of a scalar dS$_4$ elementary system, the conditions (\ref{conservation laws}) can be interpreted as the conservation laws for the system. To make the point clear, we invoke the universal length $R$, as the radius of curvature of the dS$_4$ hyperboloid $\underline{M}_R$, the universal speed of light $c$, and a ``mass"\footnote{Readers should notice that the name ``mass" is purely formal here; for a comprehensive discussion on the notion of mass in dS$_4$ relativity, one can refer to part \ref{Part mass}.} $m$. Technically, they allow to associate with the variables $(\vec{a},d^0,\vec{d})$ proper physical dimensions:
\begin{eqnarray}\label{phys dimension 0}
\vec{a} = \kappa \; \frac{\vec{p}}{mc}\,, \;\;\;\;\;\;\; d^0 = \kappa \; \frac{E}{mc^2}\,, \;\;\;\;\;\;\; \vec{d} = \kappa \; \frac{\vec{q}}{R}\,,
\end{eqnarray}
with $\kappa=mc^2$. The conditions (\ref{conservation laws}) then read:
\begin{eqnarray}\label{conservation laws'}
\vec{j} &=& \kappa \; \frac{c}{E R} \vec{l}\,,\;\;\;\;\;\;\; \mbox{with} \;\;\;\; \vec{l}\equiv\vec{q}\times\vec{p}\,, \nonumber\\
0 &=& E^4 + E^2\Big( - m^2c^4 - c^2 (\vec{p}\cdot\vec{p}) + \frac{m^2c^4}{R^2} (\vec{q}\cdot\vec{q}) \Big) - \frac{m^2c^6}{R^2} (\vec{l}\cdot\vec{l})\,.
\end{eqnarray}
At the Poincar\'{e} contraction limit $R\rightarrow\infty$ (see subsubsection \ref{Subsubsec 10.2.1}), the above dS$_4$ construction coincides with the mass shell hyperboloid:
\begin{eqnarray}
E^2 - c^2 (\vec{p}\cdot\vec{p}) = m^2c^4\,,
\end{eqnarray}
which describes the co-adjoint orbits of massive scalar elementary systems in Poincar\'{e} relativity \cite{Cari1990}. On the other hand, at the Newton contraction limit $c,R\rightarrow\infty$, $\tau^{}_\mathfrak{a}=R/c$ being fixed (see section \ref{Subsubsec 10.2.2}), this dS$_4$ construction leads to:
\begin{eqnarray}
E = mc^2 + \Big(\frac{1}{2m} (\vec{p}\cdot\vec{p}) - \frac{m}{2\tau^2_\mathfrak{a}} (\vec{q}\cdot\vec{q})\Big) + \frac{1}{2mc^2} \Big(\frac{1}{2m} (\vec{p}\cdot\vec{p}) - \frac{m}{2\tau^2_\mathfrak{a}} (\vec{q}\cdot\vec{q})\Big)^2 + {\cal{O}}\Big(\frac{1}{c^2}\Big)\,,
\end{eqnarray}
which exhibits the energy of a system constituted by a relativistic free particle (with the rest energy $mc^2$) and an \emph{anti}-harmonic oscillator with time constant $\tau^{}_\mathfrak{a}$ arising from the dS$_4$ curvature.

We end our discussions in this subsection by pointing out that:
\begin{itemize}
\item{The phase space for dS$_4$ ``massless" scalar particles, quite similar to what we have mentioned in $1+1$ dimension (see subsection \ref{Subsec dS2 phase}), can be realized by the ``massless" limit ($\kappa\rightarrow 0$) of the ``massive" (co-)adjoint orbits (\ref{massive orbit4}).}
\item{There is a bit tricky realization of two other (equivalent) families of (co-)adjoint orbits in the $\mathfrak{sp}(2,2)$ algebra, which can also be extracted from the generic element $\underline{X}({\textbf{z}},\vec{\textbf{p}})$. To see the point, having in mind (respectively, the first and second columns of) the matrix exhibition of $\underline{X}({\textbf{z}},\vec{\textbf{p}})$ given in Eqs. (\ref{zzzzz}) and (\ref{ggggg}), let:
    \begin{eqnarray}
    \textbf{q} \equiv {\vec{\textbf{n}}^{(l)}}{\big(\textbf{d}^{\scriptscriptstyle\bigstar}\big)}^{-1} &=& {\vec{\textbf{p}}} \; \big({p_0^{} {\textbf{z}}^{\scriptscriptstyle\bigstar}}\big)^{-1} \nonumber\\
    &=& {\kappa {\textbf{w}} \vec{\textbf{u}} {\textbf{w}}^{\scriptscriptstyle\bigstar} \sinh{\varphi}} \; {\big(\kappa {\textbf{w}}^{2{\scriptscriptstyle\bigstar}} \cosh\varphi\big)^{-1}} \nonumber\\
    &=& {\textbf{w}} \vec{\textbf{u}} {\textbf{w}} \tanh{\varphi} \,,
    \end{eqnarray}
    and:
    \begin{eqnarray}
    \bar{\textbf{q}} \equiv {\textbf{d}} \; {\big(\vec{\textbf{n}}^{(r)}\big)}^{-1} &=& {p_0^{} {\textbf{z}}} \; \big( -\textbf{z}^{\scriptscriptstyle\bigstar} {\vec{\textbf{p}}} \textbf{z} \big)^{-1} \nonumber\\
    &=& \kappa {\textbf{w}}^{2} \cosh\varphi \; \big( -{\kappa {\textbf{w}}^{\scriptscriptstyle\bigstar} \vec{\textbf{u}} {\textbf{w}} \sinh{\varphi}} \big)^{-1} \nonumber\\
    &=& {\textbf{w}} \vec{\textbf{u}} {\textbf{w}} \coth{\varphi} \,.
    \end{eqnarray}
    Independent of the values of $\kappa$, these two identities respectively characterize the unit-ball $B$ and its exterior, as two families of (co-)adjoint orbits, in the $\mathfrak{sp}(2,2)$ algebra (since $\textbf{q}$ and $\bar{\textbf{q}}$, as two general quaternions, respectively verify $|\textbf{q}| < 1$ and $|\bar{\textbf{q}}| > 1$). Now, to get a more explicit realization of these orbits, let $2\kappa(\underline{Z}_3+\epsilon\underline{Y}_2) = \kappa\begin{pmatrix} \epsilon\textbf{e}_2 & \textbf{e}_3 \\ -\textbf{e}_3 & \epsilon \textbf{e}_2 \end{pmatrix} \in O(2 \kappa \underline{X}_0)$, with $\epsilon \approx 0$,\footnote{Here, considering the two identities (given above Eq. (\ref{conservation laws})) which characterize the generic element $\underline{X}({\textbf{z}},\vec{\textbf{p}}) \; \big( \in O(2 \kappa \underline{X}_0) \big)$, we must underline that the components of the chosen point $\kappa\begin{pmatrix} \epsilon\textbf{e}_2 & \textbf{e}_3 \\ -\textbf{e}_3 & \epsilon \textbf{e}_2 \end{pmatrix}$ literarily verify the first identity and, for an infinitesimally small $\epsilon$, approximately verify the second one. Therefore, the chosen point is not exactly located in the orbit $O(2 \kappa \underline{X}_0)$, but it is infinitely close to the orbit such that, by abuse of notation, we can yet say $\kappa\begin{pmatrix} \epsilon\textbf{e}_2 & \textbf{e}_3 \\ -\textbf{e}_3 & \epsilon \textbf{e}_2 \end{pmatrix} \in O(2 \kappa \underline{X}_0)$.} determine their origins; accordingly, we have indeed $\textbf{q}_\odot = \epsilon\textbf{e}_2 (-\textbf{e}_3)^{-1} \approx \textbf{0}$ as the origin of the unit-ball $B$ and $\bar{\textbf{q}}_\odot = \textbf{e}_3 (\epsilon\textbf{e}_2)^{-1} \approx \boldsymbol{\infty}$ as the origin of its exterior. One can check that the maximal compact subgroup of Sp$(2,2)$:
    \begin{eqnarray}
    \underline{{\cal{S}}} = \Bigg\{ \underline{g} =
    \begin{pmatrix}
    {\textbf{v}} & \textbf{0} \\ \textbf{0} & {\textbf{v}}
    \end{pmatrix}
    \begin{pmatrix}
    {\textbf{w}} & \textbf{0} \\ \textbf{0} & {\textbf{w}}^{\scriptscriptstyle\bigstar}
    \end{pmatrix}
    \; ; \; {\textbf{v}}, {\textbf{w}} \in \mathrm{SU}(2) \Bigg\} \sim {\mathrm{SU}(2) \times \mathrm{SU}(2)} \,,
    \end{eqnarray}
    stabilizes these origins \emph{simultaneously}, namely:
    \begin{eqnarray}
    \underline{{\cal{S}}}(\textbf{v},\textbf{w}) \; \Big( 2\kappa(\underline{Z}_3+\epsilon\underline{Y}_2) \Big) \; \underline{{\cal{S}}}^{-1}(\textbf{v},\textbf{w}) &=& \kappa
    \begin{pmatrix}
    \textbf{vw}^{} \epsilon\textbf{e}_2^{} \textbf{w}^{\scriptscriptstyle\bigstar}\textbf{v}^{\scriptscriptstyle\bigstar} & \textbf{vw}^{}\textbf{e}_3^{} \textbf{w}\textbf{v}^{\scriptscriptstyle\bigstar} \\ -\textbf{v} \textbf{w}^{\scriptscriptstyle\bigstar} \textbf{e}_3^{} \textbf{w}^{\scriptscriptstyle\bigstar} \textbf{v}^{\scriptscriptstyle\bigstar} & \textbf{v}\textbf{w}^{\scriptscriptstyle\bigstar} \epsilon\textbf{e}_2^{} \textbf{w}\textbf{v}^{\scriptscriptstyle\bigstar}
    \end{pmatrix}
    \equiv \begin{pmatrix} \vec{\textbf{n}}^{(l)}(\textbf{v},\textbf{w})  & \;\;\textbf{d}(\textbf{v},\textbf{w}) \\ {\textbf{d}}^{\scriptscriptstyle\bigstar}(\textbf{v},\textbf{w}) & \;\;\vec{\textbf{n}}^{(r)}(\textbf{v},\textbf{w}) \end{pmatrix} \,, \nonumber\\
    &\;\Rightarrow\;& \textbf{q}(\textbf{v},\textbf{w}) =  \vec{\textbf{n}}^{(l)}(\textbf{v},\textbf{w}) \; \big( {\textbf{d}}^{\scriptscriptstyle\bigstar}(\textbf{v},\textbf{w}) \big)^{-1} \approx \textbf{0} \,, \nonumber\\
    &\;\Rightarrow\;& \bar{\textbf{q}}(\textbf{v},\textbf{w}) = \textbf{d}(\textbf{v},\textbf{w}) \; \big( \vec{\textbf{n}}^{(r)}(\textbf{v},\textbf{w}) \big)^{-1} \approx \boldsymbol{\infty} \,.
    \end{eqnarray}
    In this sense, the unit-ball $B$ and its exterior can be referred to as two equivalent families of (co-)adjoint orbits in the $\mathfrak{sp}(2,2)$ algebra, since both are identified by the (\emph{same}) group coset $\mathrm{Sp}(2,2)/\big(\mathrm{SU}(2) \times \mathrm{SU}(2)\big)$. [Note that these two equivalent families of (co-)adjoint orbits come to fore when we get involved with the scalar discrete series representations of the dS$_4$ group in subsection \ref{Subsec discrete dS4}.] The invariant measure on the unit-ball $B$ is given by \cite{Takahashi'}:
    \begin{eqnarray}
    \big( 1-|\textbf{q}|^2 \big)^{-4} \mathrm{d}\mu({\textbf{q}}) \,.
    \end{eqnarray}}
\end{itemize}

Finally, readers who find it interesting to compare the above results with the AdS$_4$ case are referred to Refs. \cite{Debievre1992,Debievre1994}.

\subsection{Phase space for ``spin" ``massive" elementary systems in dS$_4$ spacetime}
In this subsection, taking steps parallel to those pointed out above, we go a bit further and study another family of (co-)adjoint orbits of the $\mathfrak{sp}(2,2)$ algebra, each being related to the transport of the element $ \sqrt{2} \kappa ( \underline{X}_0 + \underline{Y}_3 ) = {\sqrt{2}}\kappa/{2} \begin{pmatrix} \textbf{e}^{}_3 & \textbf{1} \\ \textbf{1} & \textbf{e}^{}_3 \end{pmatrix}$, again with a given $0<\kappa<\infty$, under the (co-)adjoint action (\ref{Ad_g dS4}). In this case, the stabilizer subgroup is:
\begin{eqnarray}
\underline{{\cal{S}}} = \Bigg\{ \underline{g} =
\begin{pmatrix}
\textbf{v}^{}_3 & \textbf{0} \\
\textbf{0} & \textbf{v}^{}_3
\end{pmatrix}
\begin{pmatrix}
\textbf{1}\cosh \frac{\psi}{2} & \textbf{1}\sinh\frac{\psi}{2}\\
\textbf{1}\sinh\frac{\psi}{2} & \textbf{1}\cosh \frac{\psi}{2}
\end{pmatrix}\; ; \; \textbf{v}^{}_3 \; \big(\equiv\underbrace{ v^{}_4 \textbf{1} + v^{}_3 \textbf{e}_3}_{\substack{v^{}_4,v^{}_3 \in \mathbb{R} \\ \\ (v^{}_4)^2 + (v^{}_3)^2 = 1}} \big) \in \mathrm{U}(1),\; \psi \in \mathbb{R} \Bigg\} \sim \mathrm{U}(1) \times \mathrm{SO}_0(1,1)\,,
\end{eqnarray}
and hence, this family of the (co-)adjoint orbits can be identified with the group coset $O\big(\sqrt{2} \kappa ( \underline{X}_0 + \underline{Y}_3 )\big) \sim \mathrm{Sp}(2,2)/\big(\mathrm{U}(1) \times \mathrm{SO}_0(1,1)\big)$. An explicit realization of the latter can be achieved by considering the transportation of the element $ \sqrt{2} \kappa ( \underline{X}_0 + \underline{Y}_3 ) $ under the action (\ref{Ad_g dS4}), when $\underline{g}$ involved in the action belongs to:
\begin{eqnarray}
\Big\{ \underline{s}({\textbf{w}}) \; \underline{b}(\varphi,\vec{\textbf{u}}) \; \underline{r}^\prime(\textbf{v}^{}_{\slashed{3}}) \Big\} &=& \Bigg\{
\begin{pmatrix}
{\textbf{w}} & \textbf{0} \\
\textbf{0} & {\textbf{w}}^{\scriptscriptstyle\bigstar}
\end{pmatrix}
\begin{pmatrix}
\textbf{1}\cosh \frac{\varphi}{2} & \vec{\textbf{u}}\sinh \frac{\varphi}{2}\\
- \vec{\textbf{u}}\sinh\frac{\varphi}{2} & \textbf{1}\cosh\frac{\varphi}{2}
\end{pmatrix}
\begin{pmatrix}
\textbf{v}^{}_{\slashed{3}} & \textbf{0} \\
\textbf{0} & \textbf{v}^{}_{\slashed{3}}
\end{pmatrix} \;;\; \nonumber\\
&& {\textbf{w}},\vec{\textbf{u}} \in \mathrm{SU}(2),\; \textbf{v}^{}_{\slashed{3}} \; \big(\equiv\underbrace{ v^{\prime}_4 \textbf{1} + v^{}_1 \textbf{e}_1 + v^{}_2 \textbf{e}_2}_{\substack{v^{\prime}_4,v^{}_1,v^{}_2 \in \mathbb{R} \\ \\ (v^{\prime}_4)^2 + (v^{}_1)^2 + (v^{}_2)^2 = 1}} \big) \in \mathrm{SU}(2)/\mathrm{U}(1) \sim \mathbb{S}^2,\; \varphi \in \mathbb{R}
\Bigg\} \,,
\end{eqnarray}
based upon which, we obtain:
\begin{eqnarray}\label{rather}
\mbox{Ad}_{\underline{g}}\big( \sqrt{2} \kappa ( \underline{X}_0 + \underline{Y}_3 ) \big) &=& \underline{s}({\textbf{w}}) \; \underline{b}(\varphi,\vec{\textbf{u}}) \; \underline{r}^\prime(\textbf{v}^{}_{\slashed{3}}) \; \Big( \sqrt{2} \kappa ( \underline{X}_0 + \underline{Y}_3 ) \Big) \; \underline{r}^{\prime -1}(\textbf{v}^{}_{\slashed{3}}) \; \underline{b}^{-1}(\varphi,\vec{\textbf{u}}) \; \underline{s}^{-1}({\textbf{w}}) \nonumber\\ \nonumber\\
&=& \frac{\sqrt{2}}{2}
\begin{pmatrix}
\vec{\textbf{p}}(\textbf{w},\vec{\textbf{u}},\varphi) + \vec{\textbf{p}}^\prime(\vec{\textbf{v}},\textbf{w},\vec{\textbf{u}},\varphi) & & \Big( p^{}_0(\varphi) + p^{\prime}_0(\vec{\textbf{v}},\textbf{w},\vec{\textbf{u}},\varphi) \Big)\textbf{w}^2 \\ \\
\textbf{w}^{2{\scriptscriptstyle\bigstar}} \Big( p^{}_0(\varphi) - p^{\prime}_0(\vec{\textbf{v}},\textbf{w},\vec{\textbf{u}},\varphi) \Big) & & {\textbf{w}}^{2{\scriptscriptstyle\bigstar}} \Big( - \vec{\textbf{p}}(\textbf{w},\vec{\textbf{u}},\varphi) + \vec{\textbf{p}}^\prime(\vec{\textbf{v}},\textbf{w},\vec{\textbf{u}},\varphi) \Big) {\textbf{w}}^2
\end{pmatrix} \nonumber\\ \nonumber\\
& \equiv & \frac{\sqrt{2}}{2}
\begin{pmatrix}
\vec{\textbf{p}}(\textbf{w},\vec{\textbf{u}},\varphi) + \vec{\textbf{p}}^\prime(\vec{\textbf{v}},\textbf{w},\vec{\textbf{u}},\varphi) & & \Big( p^{}_0(\varphi) + p^{\prime}_0(\vec{\textbf{v}},\textbf{w},\vec{\textbf{u}},\varphi) \Big)\textbf{z}(\textbf{w}) \\ \\
\textbf{z}^{{\scriptscriptstyle\bigstar}}(\textbf{w}) \Big( p^{}_0(\varphi) - p^{\prime}_0(\vec{\textbf{v}},\textbf{w},\vec{\textbf{u}},\varphi) \Big) & & {\textbf{z}}^{{\scriptscriptstyle\bigstar}}(\textbf{w}) \Big( - \vec{\textbf{p}}(\textbf{w},\vec{\textbf{u}},\varphi) + \vec{\textbf{p}}^\prime(\vec{\textbf{v}},\textbf{w},\vec{\textbf{u}},\varphi) \Big) {\textbf{z}}(\textbf{w})
\end{pmatrix} \nonumber\\ \nonumber\\
&\equiv& \underline{X}({\textbf{z}},\vec{\textbf{p}},\vec{\textbf{v}}) \,,
\end{eqnarray}
where: (i) like the previous subsection, $\vec{\textbf{p}}(\textbf{w},\vec{\textbf{u}},\varphi) \; \big(= \vec{\textbf{p}} \big) \equiv \kappa {\textbf{w}} \vec{\textbf{u}} {\textbf{w}}^{\scriptscriptstyle\bigstar} \sinh{\varphi}$ and $p_0(\varphi) \; \big(= p^{}_0 \big) \equiv \kappa\cosh\varphi$, (ii) $\vec{\textbf{p}}^\prime(\vec{\textbf{v}},\textbf{w},\vec{\textbf{u}},\varphi) \; \big(= \vec{\textbf{p}}^\prime \big) \equiv \kappa\textbf{w}\big( \vec{\textbf{v}} \cosh^2 \frac{\varphi}{2} + \vec{\textbf{u}} \vec{\textbf{v}} \vec{\textbf{u}} \sinh^2 \frac{\varphi}{2} \big) \textbf{w}^{\scriptscriptstyle\bigstar}$ is a pure vector quaternion, while $\vec{\textbf{v}} \equiv \textbf{v}^{}_{\slashed{3}} \textbf{e}^{}_{3} \textbf{v}^{{\scriptscriptstyle\bigstar}}_{\slashed{3}}$ as a pure vector quaternion belongs to SU$(2)$, strictly speaking, to the group coset $\mathrm{SU}(2)/\mathrm{U}(1)$ homeomorphic to $\mathbb{S}^2$, and finally (iii) $p^\prime_0(\vec{\textbf{v}},\textbf{w},\vec{\textbf{u}},\varphi) \; \big(= p_0^\prime \big) \equiv \kappa \textbf{w} ( \vec{\textbf{u}} \vec{\textbf{v}} - \vec{\textbf{v}} \vec{\textbf{u}}) \textbf{w}^{\scriptscriptstyle\bigstar} \sinh\frac{\varphi}{2}\cosh\frac{\varphi}{2}$. [Here, since the generic element (\ref{rather}) has a rather complicated form, it is perhaps worthwhile reviewing how we claim that the parameters $({\textbf{z}},\vec{\textbf{p}},\vec{\textbf{v}})$ cover the whole degrees of freedom of the orbit, or in other words, how they constitute a system of global coordinates for the orbit. At first glance, the generic element (\ref{rather}), possessing $8$ degrees of freedom, is characterized by the parameters $\textbf{w}$ ($3$ degrees of freedom), $\vec{\textbf{u}}$ ($2$ degrees), $\varphi$ ($1$ degree), and finally $\vec{\textbf{v}}$ ($2$ degrees). In the alternative exhibition $\underline{X}({\textbf{z}},\vec{\textbf{p}},\vec{\textbf{v}})$ instead, the coordinate $\textbf{z}$ covers the $3$ degrees of freedom of $\textbf{w}$, $\vec{\textbf{p}}$ subsequently gives the $2+1$ degrees of freedom carried by $\vec{\textbf{u}}$ and $\varphi$, and eventually the remaining $2$ degrees of freedom carried by $\vec{\textbf{v}}$ are covered by $\vec{\textbf{v}}$ itself.] According to this parametrization, the topological nature of the ($8$-dimensional) (co-)adjoint orbits $O\big( \sqrt{2} \kappa ( \underline{X}_0 + \underline{Y}_3 ) \big)$ is ${\mathbb{S}}^3 \times {\mathbb{R}}^3 \times \mathbb{S}^2 = \big\{ \underline{X}({\textbf{z}},\vec{\textbf{p}},\vec{\textbf{v}}) \; ; \; {\textbf{z}} \in \mathrm{SU}(2) \sim {\mathbb{S}}^3,\; \vec{\textbf{p}} \sim {\mathbb{R}}^3, \; \vec{\textbf{v}} \in \mathrm{SU}(2)/\mathrm{U}(1) \sim \mathbb{S}^2 \big\}$. Note that the invariant measure on $O\big( \sqrt{2} \kappa ( \underline{X}_0 + \underline{Y}_3 ) \big)$'s, in terms of the coordinates $({\textbf{z}},\vec{\textbf{p}},\vec{\textbf{v}})$, is given by:
\begin{eqnarray}
\mathrm{d}\mu ({\textbf{z}},\vec{\textbf{p}},\vec{\textbf{v}}) = \mathrm{d}\mu({\textbf{z}}) \; \mathrm{d}^3 \vec{\textbf{p}} \; \mathrm{d}\mu^\prime(\vec{\textbf{v}}) \,,
\end{eqnarray}
where $\mathrm{d}\mu({\textbf{z}})$ and $\mathrm{d}\mu^\prime(\vec{\textbf{v}})$ are the invariant measures on $\mathbb{S}^3$ and $\mathbb{S}^2$, respectively.

Now, let us identify the generic element (\ref{rather}) of the (co-)adjoint orbits $O\big( \sqrt{2} \kappa ( \underline{X}_0 + \underline{Y}_3 ) \big)$, as a member of the $\mathfrak{sp}(2,2)$ Lie algebra (see Eq. (\ref{algebra dS4})), by:
\begin{eqnarray}
\frac{\sqrt{2}}{2}\begin{pmatrix}
\vec{\textbf{p}} + \vec{\textbf{p}}^\prime & ( p^{}_0 + p^{\prime}_0 )\textbf{z} \\
\textbf{z}^{{\scriptscriptstyle\bigstar}} ( p^{}_0 - p^{\prime}_0 ) & {\textbf{z}}^{{\scriptscriptstyle\bigstar}} ( - \vec{\textbf{p}} + \vec{\textbf{p}}^\prime ) {\textbf{z}}
\end{pmatrix}
\equiv \begin{pmatrix} \vec{\textbf{n}}^{(l)}=(0,\vec{a}+\vec{j}) & \;\;\textbf{d}=(d^0,\vec{d}) \\ {\textbf{d}}^{\scriptscriptstyle\bigstar}=(d^0,-\vec{d}) & \;\;\vec{\textbf{n}}^{(r)}=(0,-\vec{a}+\vec{j}) \end{pmatrix}\,.
\end{eqnarray}
Trivially, there must be two independent conditions between $\vec{a},\vec{j},d^0$, and $\vec{d}$, to reduce the number of degrees of freedom in (the dual of) $\mathfrak{sp}(2,2)\sim\mathbb{R}^{10}$ from $10$ to $8$, i.e., to the degrees of freedom on each $O\big( \sqrt{2} \kappa ( \underline{X}_0 + \underline{Y}_3 ) \big)$. The first condition is issued from the relation $\big|\vec{\textbf{n}}^{(l)}\big|^2 + \big|\vec{\textbf{n}}^{(r)}\big|^2 = 2\big|{\textbf{d}}\big|^2$, that holds between the components of the (co-)adjoint generic element, and the other from the fact that the (co-)adjoint action (\ref{Ad_g dS4}) is determinant-preserving. These conditions respectively read:
\begin{eqnarray} \label{conservation laws new}
0 &=& (d^0)^2 + \vec{d}\cdot\vec{d} - \vec{a}\cdot\vec{a} - \vec{j}\cdot\vec{j}\,, \nonumber\\
\kappa^2 &=& |\textbf{d}|^2 \left| 1 - \left( \frac{1}{|\textbf{d}|^2} + \frac{\textbf{d}^2}{|\textbf{d}|^4}\right) \vec{\textbf{n}}^{(r)}\vec{\textbf{n}}^{(l)} + \frac{\textbf{d}}{|\textbf{d}|^4}\vec{\textbf{n}}^{(r)}\textbf{d}\;\vec{\textbf{n}}^{(l)} \right| \nonumber\\
&=& (d^0)^2 + \vec{d}\cdot\vec{d} - \vec{a}\cdot\vec{a} + \vec{j}\cdot\vec{j} + {\cal{O}}\left(\frac{1}{(d^0)^2 + \vec{d}\cdot\vec{d}}\right) \,.
\end{eqnarray}
Then, the aforementioned family of (co-)adjoint orbits $O\big( \sqrt{2} \kappa ( \underline{X}_0 + \underline{Y}_3 ) \big)$ in (the dual of) $\mathfrak{sp}(2,2)\sim\mathbb{R}^{10}$ is identified by:
\begin{eqnarray}
O\big( \sqrt{2} \kappa ( \underline{X}_0 + \underline{Y}_3 ) \big) = \Bigg\{ (\vec{a},\vec{j},d^0,\vec{d}) \;;\; 0 &=& (d^0)^2 + \vec{d}\cdot\vec{d} - \vec{a}\cdot\vec{a} - \vec{j}\cdot\vec{j}, \nonumber\\
\kappa^2 &=& (d^0)^2 + \vec{d}\cdot\vec{d} - \vec{a}\cdot\vec{a} + \vec{j}\cdot\vec{j} + {\cal{O}}\left(\frac{1}{(d^0)^2 + \vec{d}\cdot\vec{d}}\right) \Bigg\}\,.
\end{eqnarray}

Proceeding as before, by taking into account the universal length $R$, the universal speed of light $c$, and a mass $m$ as:
\begin{eqnarray}\label{phys dimension}
\vec{a} = \kappa \; \frac{\vec{p}}{mc}\,, \;\;\;\;\;\;\; \vec{j} &=& \kappa \; \frac{c}{E R} \vec{l}\,, \;\;\;\;\;\;\; d^0 = \kappa \; \frac{E}{mc^2}\,, \;\;\;\;\;\;\; \vec{d} = \kappa \; \frac{\vec{q}}{R}\,, \;\;\;\;\;\;\; \mbox{with}\;\; \kappa=\frac{mc^2}{R}\,,
\end{eqnarray}
the (co-)adjoint orbit $O\big( \sqrt{2} \kappa ( \underline{X}_0 + \underline{Y}_3 ) \big)$ can be interpreted as the phase space of a ``spin" ``massive" dS$_4$ elementary system. Subsequently, the conditions (\ref{conservation laws new}), reading as:
\begin{eqnarray}
0 &=& \frac{E^2}{R^2} - \frac{c^2}{R^2} (\vec{p}\cdot\vec{p}) + \frac{m^2c^4}{R^4} (\vec{q}\cdot\vec{q}) - \frac{m^2c^6}{E^2R^4} (\vec{l}\cdot\vec{l})\,, \nonumber\\
m^2c^4 &=& E^2 - c^2 (\vec{p}\cdot\vec{p}) + \frac{m^2c^4}{R^2} (\vec{q}\cdot\vec{q}) + \frac{m^2c^6}{E^2R^2} (\vec{l}\cdot\vec{l}) + {\cal{O}}\left(\frac{1}{E^2 + \frac{m^2c^4}{R^2} (\vec{q}\cdot\vec{q})}\right)\,,
\end{eqnarray}
can be interpreted as the conservation laws for the system.

The very point to be noticed here is that, quite contrary to the dS$_4$ scalar ``massive" (co-)adjoint orbits for which taking the ``massless" limit ($\kappa \rightarrow 0$) yields their ``massless" counterpart, for the ``spin" ``massive" cases such a procedure does not hold true; their ``massless" limit does not lead to the corresponding ``spin" ``massless" (co-)adjoint orbit. Finding the latter is our task in the coming subsection.

\subsection{Phase space for ``spin" (or helicity) ``massless" elementary systems in dS$_4$ spacetime}
Eventually, we come to a brief study of the last (related) family of (co-)adjoint orbits of the $\mathfrak{sp}(2,2)$ algebra, which is relevant to the transport of the element $(\underline{Y}_3 - \underline{X}_3) = \begin{pmatrix} \textbf{0} & \textbf{0} \\ \textbf{0} & \textbf{e}_3 \end{pmatrix}$ under the (co-)adjoint action (\ref{Ad_g dS4}); note that $\mbox{det}( \underline{Y}_3 - \underline{X}_3 ) = 0$. Having the Cartan decomposition of the Sp$(2,2)$ group in mind (see subsection \ref{Sec Cartan dS4}), the stabilizer subgroup of this element is:
\begin{eqnarray}
\underline{{\cal{S}}} = \Bigg\{ \underline{g} =
\begin{pmatrix}
{\textbf{u}} & \textbf{0} \\
\textbf{0} & {\textbf{v}}^{}_3
\end{pmatrix}
\; ; \; {\textbf{u}} \in \mathrm{SU}(2),\; {\textbf{v}}^{}_3 \; \big(\equiv\underbrace{ v^{}_4 \textbf{1} + v^{}_3 \textbf{e}_3}_{\substack{v^{}_4,v^{}_3 \in \mathbb{R} \\ \\ (v^{}_4)^2 + (v^{}_3)^2 = 1}} \big) \in \mathrm{U}(1) \Bigg\} \sim \mathrm{SU}(2) \times \mathrm{U}(1)\,,
\end{eqnarray}
and subsequently, the (co-)adjoint orbit $O(\underline{Y}_3 - \underline{X}_3)$, homeomorphic to the group coset $\mathrm{Sp}(2,2)/\big(\mathrm{SU}(2) \times \mathrm{U}(1)\big)$, can be realized by considering the transportation of the element $ (\underline{Y}_3 - \underline{X}_3) $ under the action (\ref{Ad_g dS4}), when $\underline{g}$ involved in this action belongs to:
\begin{eqnarray}
\Big\{ \underline{p}({\textbf{q}}) \; \underline{k}^\prime(\textbf{v}^{}_{\slashed{3}}) \Big\} = \Bigg\{
\big( 1-|\textbf{q}|^2 \big)^{-1/2}
\begin{pmatrix}
{\textbf{1}} & \textbf{q} \\
\textbf{q}^{\scriptscriptstyle\bigstar} & {\textbf{1}}
\end{pmatrix}
\begin{pmatrix}
{\textbf{1}} & \textbf{0} \\
\textbf{0} & \textbf{v}^{}_{\slashed{3}}
\end{pmatrix} \;&;&\; {\textbf{q}} \in B \; (\mbox{that is,} \;|\textbf{q}|<1), \nonumber\\
&& \textbf{v}^{}_{\slashed{3}} \; \big(\equiv\underbrace{ v^{\prime}_4 \textbf{1} + v^{}_1 \textbf{e}_1 + v^{}_2 \textbf{e}_2}_{\substack{v^{\prime}_4,v^{}_1,v^{}_2 \in \mathbb{R} \\ \\ (v^{\prime}_4)^2 + (v^{}_1)^2 + (v^{}_2)^2 = 1}} \big) \in \mathrm{SU}(2)/\mathrm{U}(1) \sim \mathbb{S}^2 \Bigg\} \,.\;\;\;
\end{eqnarray}
Accordingly, we have:
\begin{eqnarray}
\mbox{Ad}_{\underline{g}}(\underline{Y}_3 - \underline{X}_3) &=& \underline{p}({\textbf{q}}) \; \underline{k}^\prime(\textbf{v}^{}_{\slashed{3}}) \; \Big( \underline{Y}_3 - \underline{X}_3 \Big)\; \underline{k}^{\prime\;-1}(\textbf{v}^{}_{\slashed{3}}) \; \underline{p}^{-1}({\textbf{q}}) \nonumber\\
&=& \big( 1-|\textbf{q}|^2 \big)^{-1}
\begin{pmatrix}
-\textbf{q}\vec{\textbf{v}}\textbf{q}^{\scriptscriptstyle\bigstar} & {\textbf{q}}\vec{\textbf{v}} \\
-\vec{\textbf{v}}{\textbf{q}}^{{\scriptscriptstyle\bigstar}} & \vec{\textbf{v}}
\end{pmatrix}
\equiv \underline{X}(\textbf{q},\vec{\textbf{v}}) \,,
\end{eqnarray}
where, again, the pure vector quaternion $\vec{\textbf{v}} \equiv \textbf{v}^{}_{\slashed{3}} \textbf{e}^{}_{3} \textbf{v}^{{\scriptscriptstyle\bigstar}}_{\slashed{3}} \in \mathrm{SU}(2)/\mathrm{U}(1) \sim \mathbb{S}^2$. The topological nature of the ($6$-dimensional) (co-)adjoint orbit $O(\underline{Y}_3 - \underline{X}_3)$ then is $B\times{\mathbb{S}}^2 = \big\{ \underline{X}({\textbf{q}},\vec{\textbf{v}}) \; ; \; {\textbf{q}} \in B,\; \vec{\textbf{v}} \sim {\mathbb{S}}^2 \big\}$. The corresponding invariant measure, in terms of the coordinates $({\textbf{q}},\vec{\textbf{v}})$, reads:
\begin{eqnarray}
\mathrm{d}\mu ({\textbf{q}},\vec{\textbf{v}}) = \big( 1-|\textbf{q}|^2 \big)^{-4} \mathrm{d}\mu({\textbf{q}}) \; \mathrm{d} \mu^\prime(\vec{\textbf{v}}) \,,
\end{eqnarray}
where $\mathrm{d} \mu^\prime(\vec{\textbf{v}})$ is the invariant measure on $\mathbb{S}^2$.

Now, let us identify the generic element $\underline{X}({\textbf{q}},\vec{\textbf{v}})$ of the (co-)adjoint orbit $O(\underline{Y}_3 - \underline{X}_3)$, as a member of the $\mathfrak{sp}(2,2)$ Lie algebra (see Eq. (\ref{algebra dS4})), by:
\begin{eqnarray}
\big( 1-|\textbf{q}|^2 \big)^{-1}
\begin{pmatrix}
-\textbf{q}\vec{\textbf{v}}\textbf{q}^{\scriptscriptstyle\bigstar} & {\textbf{q}}\vec{\textbf{v}} \\
-\vec{\textbf{v}}{\textbf{q}}^{{\scriptscriptstyle\bigstar}} & \vec{\textbf{v}}
\end{pmatrix}
\equiv \begin{pmatrix} \vec{\textbf{n}}^{(l)}=(0,\vec{a}+\vec{j}) & \;\;\textbf{d}=(d^0,\vec{d}) \\ {\textbf{d}}^{\scriptscriptstyle\bigstar}=(d^0,-\vec{d}) & \;\;\vec{\textbf{n}}^{(r)}=(0,-\vec{a}+\vec{j}) \end{pmatrix}\,.
\end{eqnarray}
In this case, there are four independent conditions between $\vec{a},\vec{j},d^0$, and $\vec{d}$, which reduce the number of degrees of freedom in (the dual of) $\mathfrak{sp}(2,2)\sim\mathbb{R}^{10}$ from $10$ to $6$, i.e., to the degrees of freedom on $O(\underline{Y}_3 - \underline{X}_3)$. The first three conditions are issued from the relation $\vec{\textbf{n}}^{(r)} = {\textbf{d}}^{\scriptscriptstyle\bigstar} \big(\vec{\textbf{n}}^{(l)}\big)^{-1} {\textbf{d}}$, and fourth one from $\big| \vec{\textbf{n}}^{(l)} \big| \big| \vec{\textbf{n}}^{(r)} \big|= \big|{\textbf{d}}\big|^2$; other possible relations between the components of the (co-)adjoint generic element $\underline{X}({\textbf{q}},\vec{\textbf{v}})$, for instance, the one that obtains from $\mbox{det}\big( \underline{X}({\textbf{q}},\vec{\textbf{v}}) \big)= 0$, do not lead to any new condition (see the relevant argument in subsection \ref{Subsec scalar massive co-adjoint}). On this basis, we get the following conditions:
\begin{eqnarray} \label{conservation laws newwww}
\vec{a}-\vec{j} &=& \left(\vec{d}(\vec{d}\cdot) + (d^0)^2 - 2d^0(\vec{d}\times) + \vec{d}\times (\vec{d}\times) \right) \frac{(\vec{a}+\vec{j})}{(\vec{a}+\vec{j})\cdot(\vec{a}+\vec{j})}\,, \nonumber\\
0 &=& \big((d^0)^2 + \vec{d}\cdot\vec{d}\big)^2 - \big(\vec{a}\cdot\vec{a} + \vec{j}\cdot\vec{j}\big)^2 + 4(\vec{a}\cdot\vec{j})^2\,.
\end{eqnarray}
Then, in (the dual of) $\mathfrak{sp}(2,2)\sim\mathbb{R}^{10}$, the (co-)adjoint orbit $O(\underline{Y}_3 - \underline{X}_3)$ reads:
\begin{eqnarray}\label{massless orbit4}
O(\underline{Y}_3 - \underline{X}_3) = \Bigg\{ (\vec{a},\vec{j},d^0,\vec{d}) \;;\; \vec{a}-\vec{j} &=& \left(\vec{d}(\vec{d}\cdot) + (d^0)^2 - 2d^0(\vec{d}\times) + \vec{d}\times (\vec{d}\times) \right) \frac{(\vec{a}+\vec{j})}{(\vec{a}+\vec{j})\cdot(\vec{a}+\vec{j})} \,, \nonumber\\
0 &=& \big((d^0)^2 + \vec{d}\cdot\vec{d}\big)^2 - \big(\vec{a}\cdot\vec{a} + \vec{j}\cdot\vec{j}\big)^2 + 4(\vec{a}\cdot\vec{j})^2 \Bigg\}\,.
\end{eqnarray}

Again, considering the universal length $R$, the universal speed of light $c$, and a \emph{mass}\footnote{The phrase ``mass", that used here, may cause confusion. It is indeed an invariant parameter with ``mass" dimension. We will revisit this ambiguous notion in Part \ref{Part mass}.} $m$:
\begin{eqnarray}
\vec{a} = \; \frac{\vec{p}}{mc}\,, \;\;\;\;\;\;\; \vec{j} = \; \frac{c}{E R} \vec{l}\,, \;\;\;\;\;\;\; d^0 = \; \frac{E}{mc^2}\,, \;\;\;\;\;\;\; \vec{d} = \; \frac{\vec{q}}{R}\,,
\end{eqnarray}
the (co-)adjoint orbit $O(\underline{Y}_3 - \underline{X}_3)$ can be interpreted as the phase space of a ``spin" ``massless" dS$_4$ elementary system,\footnote{The point based upon which we refer to (\ref{massless orbit4}) as the dS$_4$ ``spin" ``massless" (co-)adjoint orbit will be clarified later, when we consider the notions given in subsections \ref{Subsec discrete dS4} and \ref{Subsec Massless UIR's}.} while the conditions (\ref{conservation laws newwww}), interpreted as the conservation laws for the system, take the forms:
\begin{eqnarray}
\frac{\vec{p}}{mc}-\frac{c}{E R} \vec{l} &=& \left(\frac{\vec{q}}{R}\left(\frac{\vec{q}}{R}\cdot\right) + \left(\frac{E}{mc^2}\right)^2 - 2\frac{E}{mc^2}\left(\frac{\vec{q}}{R}\times\right) + \frac{\vec{q}}{R}\times \left(\frac{\vec{q}}{R}\times\right) \right) \frac{\left(\frac{\vec{p}}{mc}+\frac{c}{E R} \vec{l}\right)}{\left(\frac{\vec{p}}{mc}+\frac{c}{E R} \vec{l}\right)\cdot\left(\frac{\vec{p}}{mc}+\frac{c}{E R} \vec{l}\right)} \,, \nonumber\\
0 &=& \left( \frac{E^2}{m^2c^4} + \frac{1}{R^2} (\vec{q}\cdot\vec{q}) \right)^2 - \left(\frac{1}{m^2c^2} (\vec{p}\cdot\vec{p}) + \frac{c^2}{E^2R^2} (\vec{l}\cdot\vec{l})\right)^2 + \frac{4}{E^2R^2m^2} (\vec{p}\cdot\vec{l})^2\,.
\end{eqnarray}

\setcounter{equation}{0} \section{UIR's of the dS$_4$ group and quantum version of dS$_4$ motions}\label{Sec Dixmier}
Quantum counterparts of the classical phase spaces of dS$_4$ elementary systems correspond in a biunivocal way to the UIR's of the dS$_4$ group Sp$(2,2)$ (see section \ref{Sec Quant version dS2}). Technically, on the quantum level, the associated (ten) dS$_4$ Killing vectors are represented by (essentially) self-adjoint operators in the Hilbert space of (spinor-)tensor valued functions, square integrable according to some invariant inner product of Klein-Gordon type (or else) on $\underline{M}_R$ (or on the phase spaces given in the previous section). In the first case, these representations read as:
\begin{eqnarray}\label{LLLLLLL}
K_{AB} \;\mapsto\; L_{AB} = M_{AB} + S_{AB}\,,\;\;\;\;\;\;\; A,B=0,1,2,3,4\,,
\end{eqnarray}
where the orbital part is given by $M_{AB} = - \mathrm{i} (x_A\partial_B - x_B\partial_A)$ and the spinorial part $S_{AB}$ acts on indices of the given (spinor-)tensor valued functions in a certain permutational way. These operators obey the usual commutation rules of the $\mathfrak{sp}(2,2)$ algebra:
\begin{eqnarray}\label{adjoint commutator}
[L^{}_{AB},L^{}_{CD}] = - \mathrm{i} \big( \eta^{}_{AC} {L^{}_{BD}} + \eta^{}_{BD} {L^{}_{AC}} - \eta^{}_{AD} {L^{}_{BC}} - \eta^{}_{BC} {L^{}_{AD}} \big)\,.
\end{eqnarray}
In this context, there are two Casimir operators:
\begin{eqnarray}\label{Casimir 2}
Q^{(1)} = - \frac{1}{2} L_{AB} L^{AB} \;\;\;\;\;\;\; \mbox{(quadratic)}\,,
\end{eqnarray}
\begin{eqnarray}\label{Casimir 4}
Q^{(2)} = - W_A W^A \;\;\;\;\;\;\;\;\;\;\;\;\;\; \mbox{(quartic)}\,,
\end{eqnarray}
where the $W_A$'s, as the dS$_4$ counterparts of the Pauli-Lubanski operators (see appendix \ref{App Com UIR's Poincare}), are given by:
\begin{eqnarray} \label{WAform}
W_A = - \frac{1}{8} {\cal{E}}_{\tiny{ABCDE}} L^{BC} L^{DE}\,.
\end{eqnarray}
Here, ${\cal{E}}_{\tiny{ABCDE}}$ is the five-dimensional totally antisymmetric Levi-Civita symbol. The operators $W_A$ transform like vectors:
\begin{eqnarray}
[L_{AB},W_{C}] = \mathrm{i} \big( \eta^{}_{BC} {W_{A}} - \eta^{}_{AC} {W_{B}} \big)\,.
\end{eqnarray}
Particularly, one can show that:
\begin{eqnarray}
{W_{\texttt{A}}} = \mathrm{i} [W_{0},L_{\texttt{A} 0}]\,,\;\;\;\;\;\;\; \texttt{A} = 1,2,3,4\,.
\end{eqnarray}
Moreover, they obey the following commutation rules:
\begin{eqnarray}
[W_{A},W_{B}] = - \mathrm{i} {\cal{E}}_{ABCDE}W^C L^{DE}\,.
\end{eqnarray}

Here, it must be reminded that the introduced Casimir operators commute with all generator representatives $L^{}_{AB}$; they act like constants on all states in a given dS$_4$ UIR. Accordingly, their eigenvalues are used in a well-defined way to categorize the dS$_4$ UIR's \cite{Dixmier}. Below, we give the gist of this procedure.

\begin{figure}[H]
\begin{center}
\includegraphics[height=0.3\textheight]{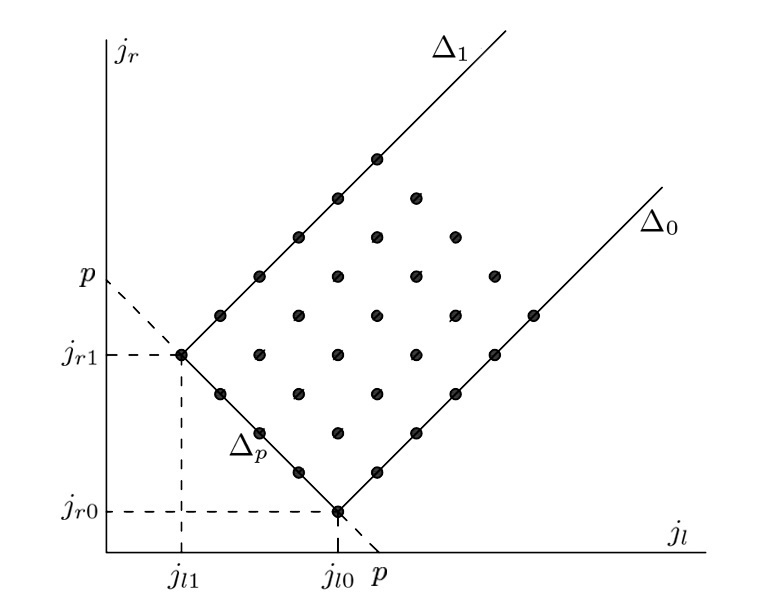}
\end{center}
\caption{Dixmier parameters $(p,q)$ for classifying the unitary discrete dual of Sp$(2,2)$ (a generic case for notations); the half-lines $\Delta_p, \Delta_0, \Delta_1$ are defined by $j_r+j_l = p$, $j_r-j_l = q_0$, and $j_r-j_l = q_1$, respectively.}
\label{FIG. Dix. 1}
\end{figure}

\begin{figure}[H]
\begin{center}
\includegraphics[height=0.2\textheight]{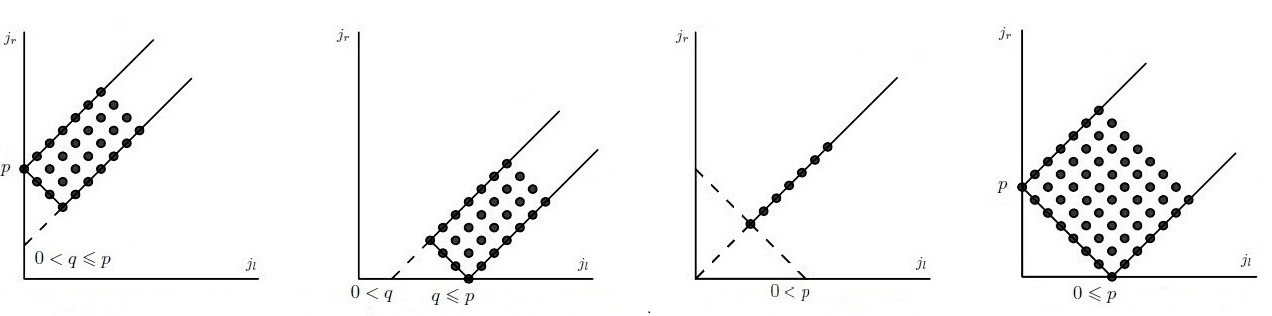}
\end{center}
\caption{Dixmier parameters $(p,q)$ for classifying the unitary dual of Sp$(2,2)$; from left to right, the first and second figures correspond, respectively, to the discrete series elements $\Pi_{p,q}^+$ and $\Pi_{p,q}^-$, the third figure to the remarkable (``degenerate") series $\Pi_{p,0}$, with $p=1,2,\;...\;$, lying at the bottom of the discrete series, and finally the last one to the principal and complementary series.}
\label{FIG. Dix. 2}
\end{figure}

We begin by pointing out that $W_0$ shows the difference of two commuting $\mathfrak{su}(2)$-Casimirs. This operator at first glance reads as:
\begin{eqnarray}\label{W0}
W_0 = - \big( L_{12}L_{34} + L_{23}L_{14} + L_{31}L_{24} \big) = - \textbf{J} \cdot \textbf{A}\,,
\end{eqnarray}
where $\textbf{J} \equiv (L_{23}, L_{31}, L_{12})^{\texttt{t}}$ and $\textbf{A} \equiv (L_{14}, L_{24}, L_{34})^{\texttt{t}}$ (again, the superscript `$\texttt{t}$' denotes transposition) constitute a basis for the maximal compact subalgebra $\mathfrak{so}(4)$ (see the first three commutation relations in (\ref{dddd})):
\begin{eqnarray}\label{rrrr}
[\textbf{J}_i,\textbf{J}_j] = \mathrm{i} {{\cal{E}}_{ij}}^{k} \; \textbf{J}_k\,,\;\;\;\;\;\;\; [\textbf{J}_i,\textbf{A}_j] = \mathrm{i} {{\cal{E}}_{ij}}^{k} \; \textbf{A}_k\,,\;\;\;\;\;\;\; [\textbf{A}_i,\textbf{A}_j] = \mathrm{i} {{\cal{E}}_{ij}}^{k} \;\textbf{J}_k\,,
\end{eqnarray}
where, again, ${{\cal{E}}_{ij}}^{k}$ ($i,j,k = 1,2,3$) is the three-dimensional totally antisymmetric Levi-Civita symbol. On the other hand, the two commuting families of generators of the $\mathfrak{su}(2)$ algebra are:
\begin{eqnarray}\label{ccccc}
\textbf{N}^{(L)} \equiv \frac{1}{2} (\textbf{A} + \textbf{J})\,,\;\;\;\;\;\;\; \textbf{N}^{(R)} \equiv \frac{1}{2} (\textbf{A} - \textbf{J})\,,
\end{eqnarray}
with:
\begin{eqnarray}\label{uuuuu}
\Big[ \textbf{N}_i^{(L)}, \textbf{N}_j^{(L)} \Big] = \mathrm{i} {{\cal{E}}_{ij}}^{k} \;\textbf{N}_k^{(L)}\,,\;\;\;\;\;\;\; \Big[ \textbf{N}_i^{(R)}, \textbf{N}_j^{(R)} \Big] = - \mathrm{i} {{\cal{E}}_{ij}}^{k} \;\textbf{N}_k^{(R)}\,.
\end{eqnarray}
Accordingly, as already mentioned, one can rewrite $W_0$ as the difference of two commuting $\mathfrak{su}(2)$-Casimirs:
\begin{eqnarray}
W_0 = - \textbf{J} \cdot \textbf{A} = - \textbf{A} \cdot \textbf{J} = \big(\textbf{N}^{(L)}\big)^2 - \big(\textbf{N}^{(R)}\big)^2\,.
\end{eqnarray}
The spectrum of $W_0$, as an operator on a direct sum of the $\mathrm{SU}(2)$ UIR's, is therefore made of the numbers $j_l(j_l + 1) - j_r(j_r + 1)$, with $j_l,j_r \in \mathbb{N}/2$. On this basis, a complete classification of the set of dS$_4$ UIR's \cite{Dixmier} is accomplished with respect to the following property. Let the map $\mathrm{Sp}(2,2) \ni \underline{g}\mapsto \underline{U}(\underline{g}) \in \mbox{Aut} ({\cal{H}})$ determine a UIR of the dS$_4$ group Sp$(2,2)$ acting in a Hilbert space ${\cal{H}}$ (the symbol `$\mbox{Aut}({\cal{H}})$' stands for the set of all automorphism of ${\cal{H}}$). The restriction to $\underline{{\cal{K}}} \sim \mathrm{SU}(2) \times \mathrm{SU}(2)$, as the maximal compact subgroup of Sp$(2,2)$, then would be quite reducible:
\begin{eqnarray}
{\cal{H}} = \oplus_{(j_l,j_r)\in \Gamma_{\underline{U}}} \;{\cal{H}}_{j_l,j_r}\,,\;\;\;\;\;\;\; {\cal{H}}_{j_l,j_r} \sim \mathbb{C}^{2j_l+1} \times \mathbb{C}^{2j_r+1}\,,
\end{eqnarray}
where $\Gamma_{\underline{U}} \subset \mathbb{N}/2 \times \mathbb{N}/2$ stands for the set of pairs $(j_l,j_r)$ in such a way that the UIR ${\cal{D}}^{j_l} \otimes {\cal{D}}^{j_r}$ of the maximal compact subgroup $\underline{{\cal{K}}}$ appears once and only once in the reduction of the restriction $\underline{U}\;\big|_{\underline{{\cal{K}}}}$. In Ref. \cite{Dixmier}, Dixmier defines $p\in \N/2$ as the \emph{infimum (greatest lower bound)} of the set of $j_l+j_r$:\footnote{Note that the parameter $p$ by construction is the natural candidate for carrying the notion of spin in dS$_4$ relativity. Below, we clarify this point in two steps:
\begin{itemize}
\item{First, we briefly recall the construction of the representations of $\textbf{J}_i$'s (with $i=1,2,3$), which constitute a basis for the shared $\mathfrak{su}(2)$ subalgebra (of the shared Lorentz subalgebra) in both dS$_4$ and Poincar\'{e} relativities (see section \ref{Sec dS4 phase spaces}, footnote \ref{foot3}). [Again, this $\mathfrak{su}(2)$ subalgebra gives sense to the notion of spin in Poincar\'{e} relativity.] Technically, following the approach familiar from the standard quantum mechanics texts, we define $\textbf{J}_\pm \equiv \textbf{J}_1 \pm \mathrm{i} \textbf{J}_2$ so that the first set of the commutation relations in (\ref{rrrr}) become $[\textbf{J}_3,\textbf{J}_\pm]= \pm \textbf{J}_\pm$ and $[\textbf{J}_+,\textbf{J}_-]=2\textbf{J}_3$. These generators are supposed to be represented by some linear transformations $\textbf{J}_3\rightarrow T(\textbf{J}_3)$ and $\textbf{J}_\pm\rightarrow T(\textbf{J}_\pm)$ (the so-called  \emph{raising/lowering operators}), which act on some vector space $V$. To construct this space, we begin from a \emph{highest weight vector} $|v_j\rangle$, and define the actions of $T(\textbf{J}_3)$ and the raising operator $T(\textbf{J}_+)$ on it respectively by $T(\textbf{J}_3)|v_j\rangle = j|v_j\rangle$ and $T(\textbf{J}_+)|v_j\rangle = 0 $, where $j \in \mathbb{N}/2$ is the corresponding \emph{highest weight} (or the spin parameter in the context of Poincar\'{e} relativity). Subsequently, by applying the lowering operator $T(\textbf{J}_-)$ on the highest weight vector $|v_j\rangle$ repeatedly ($T^{n} (\textbf{J}_-)|v_j\rangle \propto |v_{j-n}\rangle$, with $0 \leqslant n \leqslant 2j$), we will get all other (relevant) $2j+1$ weight vectors $|v_j\rangle,\; ...\;, |v_{-j}\rangle$ of $V$. [The latter actually consists of all linear combinations of the $|v\rangle$'s.]}
\item{Second, we draw attention to the definitions given in (\ref{ccccc}), based upon which we get $\textbf{J}_i = \textbf{N}^{(L)}_i - \textbf{N}^{(R)}_i$ (with $i=1,2,3$). Then, defining $\textbf{N}^{(L)}_{\pm}\equiv \textbf{N}^{(L)}_1 \pm \mathrm{i}\textbf{N}^{(L)}_2$ and $\textbf{N}^{(R)}_{\pm}\equiv \textbf{N}^{(R)}_1 \mp \mathrm{i}\textbf{N}^{(R)}_2$, we have $\textbf{J}^{}_{\pm} = \textbf{N}^{(L)}_{\pm} - \textbf{N}^{(R)}_{\mp}$. Interestingly, the linearity of the transformations $T$ allows us to write $T(\textbf{J})$'s in terms of $T\big(\textbf{N}^{(L)}\big)$'s and $T\big(\textbf{N}^{(R)}\big)$'s, i.e., $T(\textbf{J}) = T\big(\textbf{N}^{(L)}\big) - T\big(\textbf{N}^{(R)}\big)$. The point to be noticed here is that $T\big(\textbf{N}^{(L)}\big)$'s (like $T(\textbf{J})$'s) form the \emph{left-handed} representations of SU$(2)$, while $T\big(\textbf{N}^{(R)}\big)$'s form the \emph{right-handed} ones. Actually, the definitions for $\textbf{N}^{(L)}_{\pm}$ and $\textbf{N}^{(R)}_{\pm}$ are given above in such a way that is consistent with this very point; $T\big(\textbf{N}^{(L)}_\pm\big)$ and $T\big(\textbf{N}^{(R)}_\pm\big)$ are the raising/lowerng operators in the respective representations. We now turn back to an aforementioned identity, which in this new context reads as $T(\textbf{J}_+)|v_j\rangle = \big( \; T\big(\textbf{N}^{(L)}_+\big) - T\big(\textbf{N}^{(R)}_-\big) \; \big) |v_j\rangle = 0$. Since the actions of $T\big(\textbf{N}^{(L)}_+\big)$ and $T\big(\textbf{N}^{(R)}_-\big)$ on any $|v\rangle\in V$ generally result in two different weight vectors in $V$, the latter identity holds if and only if the highest weight vector $|v_j\rangle$ (of $T(\textbf{J})$'s) is considered respectively as a highest weight vector for the left-handed representations $T\big(\textbf{N}^{(L)}\big)$ and as a \emph{lowest weight vector} for the right handed ones $T\big(\textbf{N}^{(R)}\big)$, that is, $T\big(\textbf{N}^{(L)}_+\big)|v_j\rangle = 0 = T\big(\textbf{N}^{(R)}_-\big) |v_j\rangle$. Accordingly, by allocating $j_l,j_r \in \mathbb{N}/2$ as the respective highest weights to $T\big(\textbf{N}^{(L)}\big)$'s and $T\big(\textbf{N}^{(R)}\big)$'s, we have $T\big(\textbf{N}^{(L)}_3\big)|v_j\rangle = j_l |v_j\rangle$ and $T\big(\textbf{N}^{(R)}_3\big)|v_j\rangle = -j_r |v_j\rangle$. Now, considering the action $T(\textbf{J}_3)|v_j\rangle = \big( \; T\big(\textbf{N}^{(L)}_3\big) - T\big(\textbf{N}^{(R)}_3\big) \; \big) |v_j\rangle = j |v_j\rangle$, the highest weight $j$ can be clearly viewed as the infimum (greatest lower bound) of the set of $j_l+j_r$, i.e., $j = \inf{(j_l+j_r)} \equiv p$ (in the Dixmier notations).}
\end{itemize}
In conclusion, as far as we are concerned with `meaningful dS$_4$ representations from the Minkowskian point of view', $p \equiv \inf{(j_l+j_r)}$ can be referred to as the parameter which carries the notion of spin in dS$_4$ relativity. Supplementary information describing `meaningful dS$_4$ ...' will be given in section \ref{Sec contraction}.}
\begin{eqnarray} \label{defp}
p \equiv \inf_{(j_l,j_r)\in \Gamma_{\underline{U}}} (j_l+j_r)\,.
\end{eqnarray}
Moreover, he defines $q_0\in \Z/2$ and $q_1\in \Z/2$, respectively, as:
\begin{eqnarray} \label{defp}
q_0 \equiv \min_{\substack{(j_l,j_r) \in \Gamma_{\underline{U}}\\j_l+j_r = p}} (j_r - j_l)\,,\;\;\;\;\;\;\; q_1 \equiv \max_{\substack{(j_l,j_r) \in \Gamma_{\underline{U}}\\j_l+j_r = p}} (j_r -j_l)\,.
\end{eqnarray}
The generic case for pairs $(j_l,j_r)$ is shown in FIG. \ref{FIG. Dix. 1}, where the half-lines $\Delta_p, \Delta_0, \Delta_1$ delimit the allowed values of these pairs in the quadrant $\{j_r\geqslant 0,j_l\geqslant 0\}$, and $q_0\leqslant q\leqslant q_1$ in the case of the discrete series described below. Hence, the following exhaustive possibilities for the set of UIR's of the Sp$(2,2)$ group in the discrete series and its lowest limit (in the Dixmier notations \cite{Dixmier}) hold:
\begin{itemize}
\item{$q_1 = p$ and $0 < q_0 \equiv q \leqslant p$, characterizing elements of the discrete series, denoted by $\Pi^+_{p,q}$.}
\item{$q_0 = -p$ and $0 < -q_1 \equiv q \leqslant p$, characterizing elements of the discrete series, denoted by $\Pi^-_{p,q}$.}
\item{$q_0 = q_1 =0 \equiv q$, characterizing elements lying at the lower end of the discrete series, denoted by $\Pi_{p,0}$.}
\end{itemize}
These cases are shown in FIG. \ref{FIG. Dix. 2}. For the principal and complementary series denoted by $\Upsilon_{p,\sigma}$, $\sigma\equiv q(1-q)$, one still has $p \in \mathbb{N}/2$, but now the parameter $q$ becomes real or complex and is constrained by $-2<\sigma<-\infty$.

Below, following the seminal work by Dixmier \cite{Dixmier}, we will provide more details on these three series of the dS$_4$ UIR's.

\subsection{Discrete series}\label{Subsec Discrete/Dix}
As already pointed out, the dS$_4$ UIR's belonging to the discrete series (in the Dixmier notations) are symbolized by $\Pi^{\pm}_{p,q}$. In this case, the two Casimir operators are explicitly given by:
\begin{eqnarray} \label{Casimir dis}
Q^{(1)} = \Big(-p(p+1) - (q+1)(q-2)\Big) \mathbbm{1}\,,
\end{eqnarray}
\begin{eqnarray}
Q^{(2)} = \Big(-p(p+1)q(q-1)\Big) \mathbbm{1}\,.
\end{eqnarray}
In this series, two distinguished categories of the UIR's appear:
\begin{itemize}
\item{The nonsquare-integrable scalar cases $\Pi_{p,0}$, with $p=1,2,...\;$.}
\item{The spinorial cases $\Pi^{\pm}_{p,q}$, with $p = 1/2,1,3/2,...$ and $q = p,p-1,...,1$ or $1/2$ ($q>0$). Note that the representations determined by $q = 1/2$, i.e., $\Pi^{\pm}_{p,\frac{1}{2}}$, are not square integrable.}
\end{itemize}
Note that the symmetric cases $\Pi^{\pm}_{p,p}$, with $p>0$,  correspond to the so-called massless representations with helicity $s=p$. More details will be given in section \ref{Sec contraction}.

\subsection{Principal series}\label{Subsec Principal/Dix}
The dS$_4$ UIR's belonging to the principal series (in the Dixmier notations) are characterized by ${\underline{U}}^{\mbox{\small{ps}}}_{s,\nu} \equiv \Upsilon_{p=s,\sigma=q(1-q)}$, where $q = \frac{1}{2} \pm \mathrm{i} \nu$. In this context, the parameter $p$ possesses a spin meaning and the two Casimir operators explicitly read:
\begin{eqnarray}\label{Casiprise}
Q^{(1)} = \Big(- p(p+1) \;\;\; + \underbrace{(\sigma + 2)}_{-(q+1)(q-2)} \Big) \mathbbm{1} =  \left(- s(s+1) + \left(\frac{9}{4} + \nu^2\right) \right) \mathbbm{1}\,,
\end{eqnarray}
\begin{eqnarray}
Q^{(2)} = \Big( p(p+1) \sigma\Big) \mathbbm{1} = \left( s(s+1)\left(\frac{1}{4} + \nu^2\right)\right) \mathbbm{1}\,.
\end{eqnarray}
In this series, two distinguished categories of the UIR's appear:
\begin{itemize}
\item{The integer spin principal cases, with $\nu \in \mathbb{R}$ (i.e., $\sigma \geqslant \frac{1}{4}$) and $s = 0,1,2,...$ .}
\item{The half-integer spin principal cases, with $\nu \in \mathbb{R} - \{0\}$ (i.e., $\sigma > \frac{1}{4}$) and $s = 1/2,3/2,...$ .}
\end{itemize}
In both categories, the two sets of representations ${\underline{U}}^{\mbox{\small{ps}}}_{s,\nu}$ and ${\underline{U}}^{\mbox{\small{ps}}}_{s,-\nu}$ are equivalent. [Note that the parameter $\tau = -\frac{3}{2} - \mathrm{i} \nu$ will also be used in the sequel. The aforementioned equivalence then holds under $\tau \mapsto -3 - \tau$.] Moreover, we should point out that in the case $\nu = 0$ (or equivalently, $\tau = -3/2$), namely $q = 1/2$, and for a given $s = 1/2, 3/2, 5/2, ...$, the associated representation would be the direct sum of two UIR's in the discrete series:
\begin{eqnarray}\label{salam}
{\underline{U}}^{\mbox{\small{ps}}}_{s,0} = \Pi^+_{s,\frac{1}{2}} \oplus \Pi^-_{s,\frac{1}{2}}\,.
\end{eqnarray}

\subsection{Complementary series}\label{Subsec Complementary/Dix}
The UIR's belonging to the complementary series are symbolized by ${\underline{U}}^{\mbox{\small{cs}}}_{s,\nu} \equiv \Upsilon_{p=s,\sigma= q(1-q)}$, where $q = \frac{1}{2} \pm \nu$, for which the parameter $p$ has a spin meaning and the two Casimir operators explicitly read:
\begin{eqnarray}\label{Compleeee}
Q^{(1)} = \Big(- p(p+1) \;\;\; + \underbrace{(\sigma + 2)}_{-(q+1)(q-2)} \Big) \mathbbm{1} =  \left(- s(s+1) + \left(\frac{9}{4} - \nu^2\right) \right) \mathbbm{1}\,,
\end{eqnarray}
\begin{eqnarray}
Q^{(2)} = \Big(p(p+1)\sigma\Big) \mathbbm{1} = \left(s(s+1)\left(\frac{1}{4} - \nu^2\right)\right) \mathbbm{1}\,.
\end{eqnarray}
[Note that, here too, another parameter will also be used in the sequel: $\tau = - \frac{3}{2} - \nu $.] In this series, two distinguished categories of the UIR's appear:
\begin{itemize}
\item{The scalar case ${\underline{U}}^{\mbox{\small{cs}}}_{0,\nu}$, with $\nu \in \mathbb{R}$ and $0 < |\nu| < 3/2$ (namely, $-2 < \sigma < 1/4$).}
\item{The spinorial cases ${\underline{U}}^{\mbox{\small{cs}}}_{s,\nu}$, with $\nu \in \mathbb{R}$ and $0 < |\nu| < 1/2$ (namely, $0 < \sigma < 1/4$) and $s = 1,2,3,... $ .}
\end{itemize}
In both categories, the two sets of representations ${\underline{U}}^{\mbox{\small{cs}}}_{s,\nu}$ and ${\underline{U}}^{\mbox{\small{cs}}}_{s,-\nu}$ are equivalent (in other words, the equivalence holds under $\tau \mapsto - 3 - \tau$).

\subsection{Discussion: a natural fuzzyness of dS$_4$ spacetime}
Here, having the above mathematical materials in mind, we would like to highlight in passing another appealing characteristic of the dS$_4$ group structure relevant to the possibility of constructing a \emph{noncommutative spacetime}\footnote{For noncommutative spacetimes, as one of the approaches to Planck scale physics, see Ref. \cite{Steinacker} and references therein.}. Actually, considering $W_A$'s (the dS$_4$ counterparts of the Pauli-Lubanski operators introduced above), one can define a noncommutative structure for dS$_4$ spacetime by substituting the classical variables $x_{}^A$ of the ambient Minkowski spacetime $\mathbb{R}^5$ with the suitably normalized operators $W^A$, through the ``fuzzy" variables $\hat{x}_{}^A$:
\begin{eqnarray}
x_{}^A \; \mapsto \; \hat{x}_{}^A = l W^A\,,
\end{eqnarray}
where the (positive) real parameter $l$ is supposed to be of length dimension; in practice, for any given dS$_4$ UIR characterized by the parameters $p$ and $q$, a specific $l$ has to be introduced ($l \equiv l^{}_{p,q}$). Then, the following noncommutative relation holds:
\begin{eqnarray}
[\hat{x}_{}^A , \hat{x}_{}^B] = -i l {\cal{E}}^{ABCDE} \; \hat{x}^{}_C L_{DE}\,.
\end{eqnarray}
This identity tends to zero when $l$ does. On the other hand, the classical constraint $-x^{}_A x_{}^A = R^2$, describing the dS$_4$ hyperboloid of radius $R$ in $\mathbb{R}^5$, is replaced in the fuzzy case by:
\begin{eqnarray}
- \hat{x}^{}_A \hat{x}_{}^A = -l^2 p(p+1)q(q-1)\,,
\end{eqnarray}
which gives the identification:
\begin{eqnarray}
R = l \sqrt{- p(p+1)q(q-1)}\,.
\end{eqnarray}
Note that the parameter $l$, determining the dimensionality of the fuzzy variables, can be naturally interpreted as a Compton length of the associated particle (corresponding to the given UIR). Such an interpretation entails various interesting scenarios, the study of which is beyond the scope of this review. Hence, we end our brief introduction here while for further discussion in this regard, we refer readers to Ref. \cite{Gazeau fuzzyness}.

\setcounter{equation}{0} \section{UIR's of the dS$_4$ group: global realization}
In this section, generally following the lines sketched in the seminal paper \cite{Takahashi'} by Takahashi, we review the global realization of the Sp$(2,2)$ UIR's. Technically, to make the mathematical details easier to grasp, we begin our discussion with the scalar principal series UIR's of Sp$(2,2)$ (issued from the Iwasawa decomposition of the latter), which are quite analogous to the principal series UIR's of SU$(1,1)$ discussed in the previous part.

\subsection{Principal series: scalar case}\label{Subsec dS4 Principal scalar}

\subsubsection{Hilbert space and representations}
Let $\underline{U}^{\mbox{\small{ps}}}_{0,\tau}$ denote the scalar representation operators of the Sp$(2,2)$ group, associated with the principal series, where the subscript `$0$' refers to the scalar case (with $s=0$) and $\tau$ is a complex number. The operators $\underline{U}^{\mbox{\small{ps}}}_{0,\tau}$ act in the Hilbert space $L^2_{\mathbb{C}}(\mathbb{S}^3)$ as the space of all complex-valued functions $f(\textbf{z})$ on the unit-sphere $\mathbb{S}^3$ ($\mathbb{S}^3 \ni \textbf{z} \mapsto f(\textbf{z}) \in \mathbb{C}$),\footnote{Recall from the Iwasawa decomposition of Sp$(2,2)$ that $\mathrm{Sp}(2,2)/\underline{\cal{B}}\sim \underline{\cal{K}}/\underline{\cal{M}} \sim \mathbb{S}^3$ (see subsection {\ref{Sec Iwasawa dS4}}).} which are infinitely differentiable in terms of $\textbf{z}$, and are square integrable with respect to:
\begin{eqnarray}\label{norm scalar prin dS4}
\langle f_1 , f_2 \rangle = \int_{\mathbb{S}^3} f^{\ast}_1(\textbf{z})f_2(\textbf{z}) \; \mathrm{d}\mu(\textbf{z})\,,
\end{eqnarray}
where $\mathrm{d}\mu(\textbf{z})= \sin^2\psi\sin\theta \;\mathrm{d}\psi \mathrm{d}\theta \mathrm{d}\phi$, with $0 \leqslant \psi , \theta \leqslant \pi$ and $0 \leqslant \phi < 2\pi$, is the $O(4)$-invariant measure on $\mathbb{S}^3$. The action of $\underline{U}^{\mbox{\small{ps}}}_{0,\tau}$ on the functions $f(\textbf{z})$ is given by:
\begin{eqnarray}\label{U principal dS4}
\mathrm{Sp}(2,2) \ni \underline{g} = \begin{pmatrix} \textbf{a} & \textbf{b} \\ \textbf{c} & \textbf{d} \end{pmatrix} \;:\; f(\textbf{z}) \;\mapsto\; \Big(\underline{U}^{\mbox{\small{ps}}}_{0,\tau} (\underline{g}) f\Big) (\textbf{z}) &=& \big|-\textbf{b}^{\scriptscriptstyle\bigstar}\textbf{z} + \textbf{d}^{\scriptscriptstyle\bigstar}\big|^{2\tau} f\Big( \frac{\textbf{a}^{\scriptscriptstyle\bigstar}\textbf{z} - \textbf{c}^{\scriptscriptstyle\bigstar}}{-\textbf{b}^{\scriptscriptstyle\bigstar}\textbf{z} + \textbf{d}^{\scriptscriptstyle\bigstar}} \Big) \nonumber\\
&\equiv& N(\underline{g},\textbf{z}) \; f(\underline{g}^{-1}\diamond \textbf{z})\,,
\end{eqnarray}
while, according to the Iwasawa decomposition of the Sp$(2,2)$ group (see subsection \ref{Sec Iwasawa dS4}), the following map holds:
\begin{eqnarray}\label{g.z dS4}
\mathbb{S}^3 \ni \textbf{z} \;\mapsto\; \textbf{z}^\prime \equiv \underline{g}^{-1}\diamond \textbf{z} = (\textbf{a}^{\scriptscriptstyle\bigstar}\textbf{z} - \textbf{c}^{\scriptscriptstyle\bigstar})(-\textbf{b}^{\scriptscriptstyle\bigstar}\textbf{z} + \textbf{d}^{\scriptscriptstyle\bigstar})^{-1} \in \mathbb{S}^3\,.
\end{eqnarray}
[Again, the quaternionic components $\textbf{a}, \textbf{b}, \textbf{c}$, and $\textbf{d}$ are presented in terms of the basis $\big\{ \textbf{1} \equiv \mathbbm{1}_2, \; {\textbf{e}}^{}_k \equiv (-1)^{k+1} \mathrm{i} \sigma_k ; \; k=1,2,3 \big\}$, where $\sigma_k$'s are the Pauli matrices.]

\subsubsection{Condition for being unitary}
The above representations are unitary if and only if $\tau = -\frac{3}{2} - \mathrm{i} \nu$, with $\nu\in\mathbb{R}$. The proof, quite analogous to what we have pointed out in the previous part, is based on the transformation of the differential $\mathrm{d}\mu(\textbf{z})$ under the homographic action $\textbf{z} \;\mapsto\; \textbf{z}^\prime = (\textbf{a}\textbf{z} + \textbf{b})(\textbf{c}\textbf{z} + \textbf{d})^{-1}$, when $\textbf{z}$ and $\textbf{z}^\prime$ belong to $\mathbb{S}^3$ and $\underline{g} = \begin{pmatrix} \textbf{a} & \textbf{b} \\ \textbf{c} & \textbf{d} \end{pmatrix} \in \mathrm{Sp}(2,2)$, that is:
\begin{eqnarray}
\mathrm{d}\mu(\textbf{z}^\prime) = \frac{1}{|\textbf{c}\textbf{z} + \textbf{d}|^{6}} \; \mathrm{d}\mu (\textbf{z})\,.
\end{eqnarray}
Then, for $\underline{g} \in \mathrm{Sp}(2,2)$ and $\textbf{z}^\prime = \underline{g}^{-1}\diamond \textbf{z}$, we have:
\begin{eqnarray}\label{unitary PS4}
\langle \underline{U}^{\mbox{\small{ps}}}_{0,\tau}(\underline{g}) f^{}_1 , \underline{U}^{\mbox{\small{ps}}}_{0,\tau}(\underline{g}) f^{}_2 \rangle &=& \int_{\mathbb{S}^3} \; f_1^\ast(\underline{g}^{-1}\diamond \textbf{z}) \; |N(\underline{g},\textbf{z})|^2 \; f^{}_2(\underline{g}^{-1}\diamond \textbf{z}) \; \mathrm{d}\mu(\textbf{z}) \nonumber\\
&=& \int_{\mathbb{S}^3} \; f_1^\ast(\textbf{z}^\prime) \; \big|-\textbf{b}^{\scriptscriptstyle\bigstar}\textbf{z} + \textbf{d}^{\scriptscriptstyle\bigstar}\big|^{4\mathrm{Re}(\tau)+6} \; f^{}_2(\textbf{z}^\prime) \; \mathrm{d}\mu(\textbf{z}^\prime) \nonumber\\
&=& \langle f^{}_1 , f^{}_2 \rangle \quad \mbox{if} \quad \tau = -\frac{3}{2} - \mathrm{i} \nu\,,
\end{eqnarray}
which means that the representations $\underline{U}^{\mbox{\small{ps}}}_{0,\tau= -\frac{3}{2} - \mathrm{i} \nu}$ are unitary. Considering the above, it is also quite clear that two sets of representations $\underline{U}^{\mbox{\small{ps}}}_{0,\tau = -\frac{3}{2} - \mathrm{i} \nu}$ and $\underline{U}^{\mbox{\small{ps}}}_{0,-3-\tau = -\frac{3}{2} + \mathrm{i} \nu}$ are unitary equivalent.

\subsubsection{Irreducibility and infinitesimal operators}\label{Irre dS4}
In order to check the irreducibility of the representations $\underline{U}^{\mbox{\small{ps}}}_{0,\tau= -\frac{3}{2} - \mathrm{i} \nu}$ ($\nu\in\mathbb{R}$), an expression for the infinitesimal operators of $\underline{U}^{\mbox{\small{ps}}}_{0,\tau}$ is required. Technically, to fulfill this requirement, we employ the following system of global coordinates to describe any $\textbf{z} = ({z}^4, \vec{z}) \in \mathrm{SU}(2) \sim \mathbb{S}^3$:
\begin{eqnarray}\label{z set}
{z}^4 &=& \cos\psi\,,\nonumber\\
{z}^1 &=& \sin\psi \sin\theta \cos\phi\,,\nonumber\\
{z}^2 &=& \sin\psi \sin\theta \sin\phi\,,\nonumber\\
{z}^3 &=& \sin\psi \cos\theta\,,
\end{eqnarray}
where $0 \leqslant \psi , \theta \leqslant \pi$ and $0 \leqslant \phi < 2\pi$, for which $\mathrm{d}\mu(\textbf{z})= \sin^2\psi\sin\theta \;\mathrm{d}\psi \mathrm{d}\theta \mathrm{d}\phi$. Now, let $\underline{g}(t)$ denote a one-parameter subgroup (with a parameter $t$) of the Sp$(2,2)$ group. According to the Stone theorem \cite{Stone}, the associated infinitesimal operator $\underline{\hat{X}}$ is obtained by:
\begin{eqnarray}
\frac{\mathrm{i}\partial \Big( \underline{U}^{\mbox{\small{ps}}}_{0,\tau} \big(\underline{g}(t)\big) f(\psi,\theta,\phi) \Big)}{\partial t}\Bigg|_{t=0} &=& \mathrm{i} \Bigg[\frac{\partial N(\underline{g},\textbf{z}) }{\partial t} f(\psi^\prime,\theta^\prime,\phi^\prime) + \big(N(\underline{g},\textbf{z})\big) \frac{\partial f(\psi^\prime,\theta^\prime,\phi^\prime)}{\partial t}\Bigg]_{t=0}\nonumber\\
&=& \mathrm{i} \Bigg[\frac{\partial N(\underline{g},\textbf{z}) }{\partial t}\Big|_{t=0} + \frac{\partial\psi^\prime}{\partial t}\Big|_{t=0} \frac{\partial}{\partial\psi} + \frac{\partial\theta^\prime}{\partial t}\Big|_{t=0}\frac{\partial}{\partial\theta} + \frac{\partial\phi^\prime}{\partial t}\Big|_{t=0} \frac{\partial}{\partial\phi}\Bigg] f(\psi,\theta,\phi)\nonumber\\
&=& \underline{\hat{X}} \; f(\psi,\theta,\phi)\,,\nonumber
\end{eqnarray}
where $f(\textbf{z})\equiv f(\psi,\theta,\phi)$ and $f(\underline{g}^{-1}\diamond\textbf{z}) = f(\textbf{z}^\prime) \equiv f(\psi^\prime,\theta^\prime,\phi^\prime)$. Then, it follows that:
\begin{eqnarray}
\underline{\hat{X}} = \mathrm{i} \Bigg[\frac{\partial N(\underline{g},\textbf{z}) }{\partial t}\Big|_{t=0} + \frac{\partial\psi^\prime}{\partial t}\Big|_{t=0}\frac{\partial}{\partial\psi} + \frac{\partial\theta^\prime}{\partial t}\Big|_{t=0}\frac{\partial}{\partial\theta} + \frac{\partial\phi^\prime}{\partial t}\Big|_{t=0}\frac{\partial}{\partial\phi}\Bigg]\,.
\end{eqnarray}

Proceeding as above, while the one-parameter subgroups appeared in the space-time-Lorentz decomposition of the Sp$(2,2)$ group (see subsection \ref{sec space-time-Lorentz dS4}) are taken into account, we now present the infinitesimal operators of the representations $\underline{U}^{\mbox{\small{ps}}}_{0,\tau}$. We begin with the subgroup of space rotations, spanned by the infinitesimal generators $\underline{Y}_k$, with $k=1,2,3$ (see Eq. (\ref{Yk})); $\underline{g}_k^{}(t) = \begin{pmatrix} \textbf{a} & \textbf{b} \\ \textbf{c} & \textbf{d} \end{pmatrix} \equiv \exp(\underline{Y}^{}_k t) = \begin{pmatrix} \exp(\textstyle\frac{1}{2} {\textbf{e}}^{}_k t) & \textbf{0} \\ \textbf{0} & \exp(\textstyle\frac{1}{2} {\textbf{e}}^{}_k t) \end{pmatrix}$. For $k=1$, it yields:
\begin{eqnarray}
(\textbf{a}^{\scriptscriptstyle\bigstar}\textbf{z} - \textbf{c}^{\scriptscriptstyle\bigstar}) &=& \exp(\textstyle\frac{1}{2} {\textbf{e}}^{\scriptscriptstyle\bigstar}_1 t) \textbf{z}\,,\nonumber\\
(-\textbf{b}^{\scriptscriptstyle\bigstar}\textbf{z} + \textbf{d}^{\scriptscriptstyle\bigstar}) &=& \exp(\textstyle\frac{1}{2} {\textbf{e}}^{\scriptscriptstyle\bigstar}_1 t)\,,\nonumber\\
(-\textbf{b}^{\scriptscriptstyle\bigstar}\textbf{z} + \textbf{d}^{\scriptscriptstyle\bigstar})^{-1} &=& \frac{(-\textbf{b}^{\scriptscriptstyle\bigstar}\textbf{z} + \textbf{d}^{\scriptscriptstyle\bigstar})^{\scriptscriptstyle\bigstar}}{|-\textbf{b}^{\scriptscriptstyle\bigstar}\textbf{z} + \textbf{d}^{\scriptscriptstyle\bigstar}|^2} = \exp(\textstyle\frac{1}{2} {\textbf{e}}^{}_1 t)\,.\nonumber
\end{eqnarray}
Accordingly, with respect to the coordinates (\ref{z set}) and the identity (\ref{exp(quat)}), the components of $\textbf{z}^\prime = (\textbf{a}^{\scriptscriptstyle\bigstar}\textbf{z} - \textbf{c}^{\scriptscriptstyle\bigstar})(-\textbf{b}^{\scriptscriptstyle\bigstar}\textbf{z} + \textbf{d}^{\scriptscriptstyle\bigstar})^{-1}$ are:
\begin{eqnarray}
{z}^{\prime 4} = {z}^4\; &\Rightarrow& \; \cos\psi^\prime = \cos\psi\,,\nonumber\\
{z}^{\prime 1} = {z}^1 \; &\Rightarrow& \; \sin\psi^\prime \sin\theta^\prime \cos\phi^\prime = \sin\psi \sin\theta \cos\phi\,,\nonumber\\
{z}^{\prime 2} = {z}^3 \sin t + {z}^2 \cos t \; &\Rightarrow& \; \sin\psi^\prime \sin\theta^\prime \sin\phi^\prime = (\sin\psi \cos\theta) \sin t + (\sin\psi \sin\theta \sin\phi) \cos t \,,\nonumber\\
{z}^{\prime 3} = {z}^3 \cos t - {z}^2 \sin t \; &\Rightarrow& \; \sin\psi^\prime \cos\theta^\prime = (\sin\psi \cos\theta) \cos t - (\sin\psi \sin\theta \sin\phi) \sin t \,.\nonumber
\end{eqnarray}
These equations allow to calculate the derivatives of $\psi^\prime(t), \theta^\prime(t), \phi^\prime(t)$, and $N(\underline{g}_k,\textbf{z})$ with respect to $t$:
\begin{eqnarray}
\frac{\partial\psi^\prime}{\partial t}\Big|_{t=0} &=& 0\,,\nonumber\\
\frac{\partial\theta^\prime}{\partial t}\Big|_{t=0} &=& \sin\phi\,,\nonumber\\
\frac{\partial\phi^\prime}{\partial t}\Big|_{t=0} &=& \cot\theta \cos\phi\,,\nonumber\\
\frac{\partial N(\underline{g}_{k=1},\textbf{z}) }{\partial t}\Big|_{t=0} &=& 0\,.\nonumber
\end{eqnarray}
The associated infinitesimal operator then takes the form:
\begin{eqnarray}
\underline{\hat{Y}}_1 = \mathrm{i}\Big(\sin\phi \;\frac{\partial}{\partial\theta} + \cot\theta\cos\phi \;\frac{\partial}{\partial\phi} \Big)\,.
\end{eqnarray}
Similarly, for $k=2$ and $k=3$, we respectively obtain:
\begin{eqnarray}
\underline{\hat{Y}}_2 &=& \mathrm{i}\Big( -\cos\phi \;\frac{\partial}{\partial\theta} + \cot\theta\sin\phi \;\frac{\partial}{\partial\phi} \Big)\,,\\
\underline{\hat{Y}}_3 &=& \mathrm{i}\Big( -\frac{\partial}{\partial\phi} \Big)\,.
\end{eqnarray}

The other infinitesimal operators associated with the scalar representations $\underline{U}^{\mbox{\small{ps}}}_{0,\tau}$ are obtained in the same way. They respectively read \cite{Rabeie}:
\begin{itemize}
\item{The infinitesimal operator of time translations:}
\begin{eqnarray}
\underline{\hat{X}}_0 = \mathrm{i}\Big(-\tau\cos\psi + \sin\psi \;\frac{\partial}{\partial\psi}\Big)\,.
\end{eqnarray}
\item{The infinitesimal operators of space translations:}
\begin{eqnarray}
\underline{\hat{X}}_1 &=& \mathrm{i}\Big( -\sin\theta\cos\phi \;\frac{\partial}{\partial\psi} - \cot\psi\cos\theta\cos\phi \;\frac{\partial}{\partial\theta} + \frac{\cot\psi\sin\phi}{\sin\theta} \;\frac{\partial}{\partial\phi}\Big)\,,\\
\underline{\hat{X}}_2 &=& \mathrm{i}\Big( -\sin\theta\sin\phi \;\frac{\partial}{\partial\psi} - \cot\psi\cos\theta\sin\phi \;\frac{\partial}{\partial\theta} - \frac{\cot\psi\cos\phi}{\sin\theta} \;\frac{\partial}{\partial\phi}\Big)\,,\\
\underline{\hat{X}}_3 &=& \mathrm{i}\Big( -\cos\theta \;\frac{\partial}{\partial\psi} + \cot\psi\sin\theta \;\frac{\partial}{\partial\theta}\Big)\,.
\end{eqnarray}
\item{The infinitesimal operators of boosts:}
\begin{eqnarray}
\underline{\hat{Z}}_1 &=& \mathrm{i}\Big(-\tau\sin\psi\sin\theta\cos\phi - \cos\psi\sin\theta\cos\phi \;\frac{\partial}{\partial\psi} - \frac{\cos\theta\cos\phi}{\sin\psi} \;\frac{\partial}{\partial\theta} + \frac{\sin\phi}{\sin\psi\sin\theta} \;\frac{\partial}{\partial\phi}\Big)\,,\\
\underline{\hat{Z}}_2 &=& \mathrm{i}\Big(-\tau\sin\psi\sin\theta\sin\phi - \cos\psi\sin\theta\sin\phi \;\frac{\partial}{\partial\psi} - \frac{\cos\theta\sin\phi}{\sin\psi} \;\frac{\partial}{\partial\theta} - \frac{\cos\phi}{\sin\psi\sin\theta} \;\frac{\partial}{\partial\phi} \Big)\,,\\
\underline{\hat{Z}}_3 &=& \mathrm{i}\Big(-\tau\sin\psi\cos\theta - \cos\psi\cos\theta \;\frac{\partial}{\partial\psi} + \frac{\sin\theta}{\sin\psi} \;\frac{\partial}{\partial\theta} \Big)\,.
\end{eqnarray}
\end{itemize}

One can check that the given infinitesimal operators verify the following commutation relations:
\begin{eqnarray}\label{2000}
\big[\underline{\hat{Y}}_i,\underline{\hat{Y}}_j\big] &=& \mathrm{i} {{\cal{E}}_{ij}}^{k} \; \underline{\hat{Y}}_k\,,\nonumber\\
\big[\underline{\hat{Y}}_i,\underline{\hat{X}}_j\big] &=& \mathrm{i} {{\cal{E}}_{ij}}^{k} \; \underline{\hat{X}}_k\,,\nonumber\\
\big[\underline{\hat{X}}_i,\underline{\hat{X}}_j\big] &=& \mathrm{i} {{\cal{E}}_{ij}}^{k} \; \underline{\hat{Y}}_k\,,\nonumber\\
\big[\underline{\hat{Y}}_i,\underline{\hat{Z}}_j\big] &=& \mathrm{i} {{\cal{E}}_{ij}}^{k} \; \underline{\hat{Z}}_k\,,\nonumber\\
\big[\underline{\hat{X}}_i,\underline{\hat{Z}}_j\big] &=& - \mathrm{i} \delta^{}_{ij} \; \underline{\hat{X}}_0\,,\nonumber\\
\big[\underline{\hat{Z}}_i,\underline{\hat{Z}}_j\big] &=& - \mathrm{i} {{\cal{E}}_{ij}}^{k} \; \underline{\hat{Y}}_k\,,\nonumber\\
\big[\underline{\hat{X}}_0,\underline{\hat{X}}_i\big] &=& - \mathrm{i} \underline{\hat{Z}}_i \,,\nonumber\\
\big[\underline{\hat{X}}_0,\underline{\hat{Z}}_i\big] &=& - \mathrm{i} \underline{\hat{X}}_i \,,\nonumber\\
\big[\underline{\hat{Y}}_i,\underline{\hat{X}}_0\big] &=& 0\,,
\end{eqnarray}
where, again, $i,j,k = 1,2,3$ and ${{\cal{E}}_{ij}}^{k}$ is the three-dimensional totally antisymmetric Levi-Civita symbol. The above infinitesimal operators demonstrate the orbital part of the $\mathfrak{sp}(2,2)$ algebra. Actually, by defining:\footnote{An interesting point to be noticed here is that usually when we speak about the orbital part of the $\mathfrak{sp}(2,2)$ algebra, we are thinking about actions on the functions of the ambient space coordinates of the dS$_4$ hyperboloid, but now we present the orbital part in terms of functions on $\mathbb{S}^3$. The link between these two realizations, let us say, the spacetime and the $\mathbb{S}^3$ realizations, will be given in part \ref{Part plane waves}.}
\begin{eqnarray}\label{M_AB}
M_{4k} \equiv \underline{\hat{X}}_k\,,\;\;\;\;\;\;\; M_{04} \equiv \underline{\hat{X}}_0\,,\;\;\;\;\;\;\; M_{ki} \equiv {{\cal{E}}_{ki}}^{j} \; \underline{\hat{Y}}_j\,,\;\;\;\;\;\;\; M_{0k} \equiv \underline{\hat{Z}}_k\,,
\end{eqnarray}
we explicitly get:
\begin{eqnarray}
[M_{AB},M_{CD}] = - \mathrm{i} \big( \eta^{}_{AC} {M^{}_{BD}} + \eta^{}_{BD} {M^{}_{AC}} - \eta^{}_{AD} {M^{}_{BC}} - \eta^{}_{BC} {M^{}_{AD}} \big)\,, \;\;\;\;\;\;\; A,B=0,1,2,3,4\,.
\end{eqnarray}
Note that $M_{AB}=-M_{BA}$.

Quite analogous to the situation in the $1+1$-dimensional case, the above operators cannot be defined on the whole Hilbert space $L^2_{\mathbb{C}} (\mathbb{S}^3)$, since they are unbounded. Below, we will show that they are indeed essentially self-adjoint operators on the common dense invariant subspace $\underline{\Delta} \; \big(\subset L^2_{\mathbb{C}} (\mathbb{S}^3) \big)$ made of all finite linear combinations of elements of the orthonormal basis $\big\{ |{Llm}\rangle \big\} \equiv \big\{ {\cal{Y}}_{Llm}(\textbf{z}) \big\} $, where ${\cal{Y}}_{Llm}(\textbf{z})$'s, with $(L,l,m) \in \mathbb{N} \times \mathbb{N} \times \mathbb{Z}$, $0\leqslant l \leqslant L$ and $-l\leqslant m \leqslant l$, are the hyperspherical harmonics on $\mathbb{S}^3$ (for the explicit form of ${\cal{Y}}_{Llm}(\textbf{z})$'s, see appendix \ref{App some}). Then, to check the irreducibility of the corresponding representations $\underline{U}^{\mbox{\small{ps}}}_{0,\tau}$, it is sufficient to show that $\underline{\Delta}$ does not contain any nontrivial subspace invariant for the given infinitesimal operators. Considering the relations given in appendix \ref{App some}, the actions of the above infinitesimal operators on the basis $\big\{ |{Llm}\rangle \big\}$'s respectively read \cite{Rabeie}:
\begin{itemize}
\item{The actions of the space-rotation operators:
\begin{eqnarray}
\underline{\hat{Y}}_1 \; |{Llm}\rangle &=& \mathrm{i} \Big(\sin\phi \;\frac{\partial}{\partial\theta} + \cot\theta\cos\phi \;\frac{\partial}{\partial\phi} \Big) \; |{Llm}\rangle \nonumber\\
&=& -\frac{1}{2} \sqrt{(l+m)(l-m+1)} \; |{Ll,m-1}\rangle - \frac{1}{2} \sqrt{(l-m)(l+m+1)} \; |{Ll,m+1}\rangle\,,
\end{eqnarray}
\begin{eqnarray}
\underline{\hat{Y}}_2 \; |{Llm}\rangle &=& \mathrm{i}\Big( -\cos\phi \;\frac{\partial}{\partial\theta} + \cot\theta\sin\phi \;\frac{\partial}{\partial\phi} \Big) \; |{Llm}\rangle \nonumber\\
&=& -\frac{\mathrm{i}}{2} \sqrt{(l+m)(l-m+1)} \; |{Ll,m-1}\rangle + \frac{\mathrm{i}}{2} \sqrt{(l-m)(l+m+1)} \; |{Ll,m+1}\rangle\,,
\end{eqnarray}
\begin{eqnarray}
\underline{\hat{Y}}_3 \; |{Llm}\rangle = \mathrm{i}\Big( -\frac{\partial}{\partial\phi}\Big) \; |{Llm}\rangle = m \; |{Llm}\rangle\,.
\end{eqnarray}}
\item{The action of the time-translation operator:
\begin{eqnarray}
\underline{\hat{X}}_0 \; |{Llm}\rangle &=& \mathrm{i}\Big(-\tau\cos\psi + \sin\psi \;\frac{\partial}{\partial\psi}\Big) \; |{Llm}\rangle \nonumber\\
&=& \frac{\mathrm{i}}{2} \sqrt{\frac{(L-l+1)(L+l+2)}{(L+1)(L+2)}} \; (-\tau+L) \; |{L+1,lm}\rangle \nonumber\\
&& - \frac{\mathrm{i}}{2} \sqrt{\frac{(L+l+1)(L-l)}{L(L+1)}} \; (\tau+L+2) \; |{L-1,lm}\rangle\,.
\end{eqnarray}}
\item{The actions of the space-translation operators:
\begin{eqnarray}
\underline{\hat{X}}_3 \; |{Llm}\rangle &=& \mathrm{i}\Big( -\cos\theta \;\frac{\partial}{\partial\psi} + \cot\psi\sin\theta \;\frac{\partial}{\partial\theta}\Big) \; |{Llm}\rangle \nonumber\\
&=& \mathrm{i} \sqrt{\frac{(l-m+1)(L+l+2)(L-l)(l+m+1)}{(2l+1)(2l+3)}} \; |{L,l+1,m}\rangle \nonumber\\
&& - \mathrm{i} \sqrt{\frac{(l-m)(L+l+1)(L-l+1)(l+m)}{(2l+1)(2l-1)}} \; |{L,l-1,m}\rangle\,,
\end{eqnarray}
\begin{eqnarray}
\underline{\hat{X}}_2 \; |{Llm}\rangle = - \mathrm{i} [\underline{\hat{X}}_3, \underline{\hat{Y}}_1] \; |{Llm}\rangle &=& \frac{1}{2} \Bigg( \sqrt{\frac{(l-m+1)(l-m+2)(L+l+2)(L-l)}{(2l+1)(2l+3)}} \; |{L,l+1,m-1}\rangle \nonumber\\
&& + \sqrt{\frac{(l+m-1)(l+m)(L+l+1)(L-l+1)}{(2l+1)(2l-1)}} \; |{L,l-1,m-1}\rangle \nonumber\\
&& + \sqrt{\frac{(l+m+1)(l+m+2)(L+l+2)(L-l)}{(2l+1)(2l+3)}} \; |{L,l+1,m+1}\rangle \nonumber\\
&& + \sqrt{\frac{(l-m)(l-m-1)(L+l+1)(L-l+1)}{(2l+1)(2l-1)}} \; |{L,l-1,m+1}\rangle \Bigg)\,,\;\;\;\;\;
\end{eqnarray}
\begin{eqnarray}
\underline{\hat{X}}_1 \; |{Llm}\rangle = - \mathrm{i} [\underline{\hat{X}}_2, \underline{\hat{Y}}_3] \; |{Llm}\rangle &=& - \frac{\mathrm{i}}{2} \Bigg( \sqrt{\frac{(l-m+1)(l-m+2)(L+l+2)(L-l)}{(2l+1)(2l+3)}} \; |{L,l+1,m-1}\rangle \nonumber\\
&& + \sqrt{\frac{(l+m-1)(l+m)(L+l+1)(L-l+1)}{(2l+1)(2l-1)}} \; |{L,l-1,m-1}\rangle \nonumber\\
&& - \sqrt{\frac{(l+m+1)(l+m+2)(L+l+2)(L-l)}{(2l+1)(2l+3)}} \; |{L,l+1,m+1}\rangle \nonumber\\
&& - \sqrt{\frac{(l-m)(l-m-1)(L+l+1)(L-l+1)}{(2l+1)(2l-1)}} \; |{L,l-1,m+1}\rangle \Bigg)\,. \;\;\;\;\;
\end{eqnarray}}
\item{The actions of the boost operators:
\begin{eqnarray}
\underline{\hat{Z}}_3 \; |{Llm}\rangle = \mathrm{i} [\underline{\hat{X}}_0, \underline{\hat{X}}_3] \; |{Llm}\rangle &=& \frac{\mathrm{i}(-\tau+L)}{2} \Bigg(
- \sqrt{\frac{(l-m)(l+m)(L-l+1)(L-l+2)}{(L+1)(L+2)(2l+1)(2l-1)}} \; |{L+1,l-1,m}\rangle \nonumber\\
&& + \sqrt{\frac{(l-m+1)(l+m+1)(L+l+2)(L+l+3)}{(L+1)(L+2)(2l+1)(2l+3)}} \; |{L+1,l+1,m}\rangle \Bigg) \nonumber\\
&& - \frac{\mathrm{i}(\tau+L+2)}{2} \Bigg( \sqrt{\frac{(l-m)(l+m)(L+l+1)(L+l)}{L(L+1)(2l+1)(2l-1)}} \; |{L-1,l-1,m}\rangle \nonumber\\
&& - \sqrt{\frac{(l-m+1)(l+m+1)(L-l)(L-l-1)}{L(L+1)(2l+1)(2l+3)}} \; |{L-1,l+1,m}\rangle \Bigg)\,,
\end{eqnarray}
\begin{eqnarray}
\underline{\hat{Z}}_2 \; |{Llm}\rangle = \mathrm{i} [\underline{\hat{X}}_0, \underline{\hat{X}}_2] \; |{Llm}\rangle &=& \frac{-\tau+L}{4} \Bigg( \sqrt{\frac{(l+m-1)(l+m)(L-l+1)(L-l+2)}{(L+1)(L+2)(2l+1)(2l-1)}} \; |{L+1,l-1,m-1}\rangle \nonumber\\
&& + \sqrt{\frac{(l-m+1)(l-m+2)(L+l+2)(L+l+3)}{(L+1)(L+2)(2l+1)(2l+3)}} \; |{L+1,l+1,m-1}\rangle \nonumber\\
&& + \sqrt{\frac{(l+m+1)(l+m+2)(L+l+2)(L+l+3)}{(L+1)(L+2)(2l+1)(2l+3)}} \; |{L+1,l+1,m+1}\rangle \nonumber\\
&& + \sqrt{\frac{(l-m)(l-m-1)(L-l+1)(L-l+2)}{(L+1)(L+2)(2l+1)(2l-1)}} \; |{L+1,l-1,m+1}\rangle \Bigg) \nonumber\\
&& + \frac{\tau+L+2}{4} \Bigg( \sqrt{\frac{(l+m-1)(l+m)(L+l+1)(L+l)}{L(L+1)(2l+1)(2l-1)}} \; |{L-1,l-1,m-1}\rangle \nonumber\\
&& + \sqrt{\frac{(l-m)(l-m-1)(L+l+1)(L+l)}{L(L+1)(2l+1)(2l-1)}} \; |{L-1,l-1,m+1}\rangle \nonumber\\
&& + \sqrt{\frac{(l+m+1)(l+m+2)(L-l)(L-l-1)}{L(L+1)(2l+1)(2l+3)}} \; |{L-1,l+1,m+1}\rangle \nonumber\\
&& + \sqrt{\frac{(l-m+1)(l-m+2)(L-l)(L-l-1)}{L(L+1)(2l+1)(2l+3)}} \; |{L-1,l+1,m-1}\rangle \Bigg)\,,
\end{eqnarray}
\begin{eqnarray}
\underline{\hat{Z}}_1 \; |{Llm}\rangle = \mathrm{i} [\underline{\hat{X}}_0, \underline{\hat{X}}_1] \; |{Llm}\rangle &=& \frac{\mathrm{i}(-\tau+L)}{4} \Bigg( -\sqrt{\frac{(l+m-1)(l+m)(L-l+1)(L-l+2)}{(L+1)(L+2)(2l+1)(2l-1)}} \; |{L+1,l-1,m-1}\rangle \nonumber\\
&& - \sqrt{\frac{(l-m+1)(l-m+2)(L+l+2)(L+l+3)}{(L+1)(L+2)(2l+1)(2l+3)}} \; |{L+1,l+1,m-1}\rangle \nonumber\\
&& + \sqrt{\frac{(l+m+1)(l+m+2)(L+l+2)(L+l+3)}{(L+1)(L+2)(2l+1)(2l+3)}} \; |{L+1,l+1,m+1}\rangle \nonumber\\
&& + \sqrt{\frac{(l-m)(l-m-1)(L-l+1)(L-l+2)}{(L+1)(L+2)(2l+1)(2l-1)}} \; |{L+1,l-1,m+1}\rangle \Bigg) \nonumber\\
&& - \frac{\mathrm{i}(\tau+L+2)}{4} \Bigg( \sqrt{\frac{(l+m-1)(l+m)(L+l+1)(L+l)}{L(L+1)(2l+1)(2l-1)}} \; |{L-1,l-1,m-1}\rangle \nonumber\\
&& - \sqrt{\frac{(l-m)(l-m-1)(L+l+1)(L+l)}{L(L+1)(2l+1)(2l-1)}} \; |{L-1,l-1,m+1}\rangle \nonumber\\
&& - \sqrt{\frac{(l+m+1)(l+m+2)(L-l)(L-l-1)}{L(L+1)(2l+1)(2l+3)}} \; |{L-1,l+1,m+1}\rangle \nonumber\\
&& + \sqrt{\frac{(l-m+1)(l-m+2)(L-l)(L-l-1)}{L(L+1)(2l+1)(2l+3)}} \; |{L-1,l+1,m-1}\rangle \Bigg)\,.
\end{eqnarray}}
\end{itemize}
The above relations reveal that, on $\underline{\Delta} \subset L^2_{\mathbb{C}} (\mathbb{S}^3)$, the given infinitesimal operators of the scalar representations $\underline{U}^{\mbox{\small{ps}}}_{0,\tau}$ are well defined (in the allowed ranges of parameters), and also that the common dense $\underline{\Delta}$ is invariant, with no nontrivial subspace invariant, under the actions of the aforementioned operators. The irreducibility of the given representations then is proved.

\subsubsection{Quantum Casimir operator}
Having in mind the identities given in Eq. (\ref{M_AB}), the corresponding (scalar) quadratic Casimir operator reads:\footnote{Note that the quartic Casimir operator vanishes in this case.}
\begin{eqnarray}\label{6666666666}
Q_0^{(1)} = -\frac{1}{2} M_{AB}M^{AB} = \underline{\hat{X}}_0^2 + \sum_{i=1}^{3} \underline{\hat{Z}}_i^2 - \sum_{i=1}^{3} \underline{\hat{X}}_i^2 - \sum_{i=1}^{3} \underline{\hat{Y}}_i^2 = -\tau(\tau+3) \mathbbm{1}\,,
\end{eqnarray}
with $\tau=-\frac{3}{2}-\mathrm{i}\nu$ ($\nu\in\mathbb{R}$). Considering the above, and quite analogous to the discussion (subsequent to Eq. (\ref{555555555})) given in the previous part, we here would like to draw attention to the fact that the functions $f(\textbf{z}) \in \underline{\Delta}$, which identify the (true) quantum states carrying the dS$_4$ scalar principal representations, are indeed eigenfunctions of the quadratic Casimir operator $Q_0^{(1)}$ for the eigenvalues $(\frac{9}{4} + \nu^2)$, namely, $\big( Q_0^{(1)} - (\frac{9}{4} + \nu^2) \big) f(\textbf{z}) = 0$. Again, this point (extended to the whole three series of the dS$_4$ UIR's) will be employed in part \ref{Part plane waves}, when the spacetime realization of the dS$_4$ UIR's is considered, to present the ``wave equations" of dS$_4$ elementary systems.

\subsection{Principal series: general case}\label{Subsec Principal dS4 gen.}
In a general case possessing spin $s$, the Hilbert space carrying the principal series UIR's of Sp$(2,2)$ is characterized by $L^2_{{\mathbb{C}}^{2s+1}}(\mathbb{S}^3)$. The action of the representation operators $\underline{U}^{\mbox{\small{ps}}}_{s,\nu}$ on the functions $f(\textbf{z}) \in L^2_{{\mathbb{C}}^{2s+1}}(\mathbb{S}^3)$ reads:
\begin{eqnarray}\label{U principal dS4 gen}
\mathrm{Sp}(2,2) \ni \underline{g} = \begin{pmatrix} \textbf{a} & \textbf{b} \\ \textbf{c} & \textbf{d} \end{pmatrix} \;:\; f(\textbf{z}) \;\mapsto\; \Big(\underline{U}^{\mbox{\small{ps}}}_{s,\nu} (\underline{g}) f\Big) (\textbf{z}) = N(\underline{g},\textbf{z}) \; {\cal{D}}^s \Big( \frac{-\textbf{z}^{\scriptscriptstyle\bigstar}\textbf{b} + \textbf{d}}{|-\textbf{z}^{\scriptscriptstyle\bigstar}\textbf{b} + \textbf{d}|} \Big) \; f(\underline{g}^{-1}\diamond \textbf{z})\,,
\end{eqnarray}
where the definitions of $N(\underline{g},\textbf{z})$ and $\underline{g}^{-1}\diamond \textbf{z}$ are precisely the same as those given in Eq. (\ref{U principal dS4}), and ${\cal{D}}^s$'s stand for the $2s+1$-dimensional UIR's of $\mathrm{SU}(2)$ (to see more on the $\mathrm{SU}(2)$ UIR's, one can refer to appendix \ref{App UIR's SU(2)}). Clearly, by adjusting $s=0$, the UIR's (\ref{U principal dS4 gen}) coincide with the scalar ones (\ref{U principal dS4}).

Taking steps parallel to those pointed out in subsubsection \ref{Irre dS4}, one can check that the infinitesimal operators $L_{AB}$ of the UIR's $\underline{U}^{\mbox{\small{ps}}}_{s,\nu}$ are constituted by an orbital part $M_{AB}$, which is exactly the one already given in Eq. (\ref{M_AB}), and a spinorial part $S_{AB}$ ($L_{AB} = M_{AB} + S_{AB}$). These infinitesimal operators explicitly read:
\begin{eqnarray}
L_{4k} &=& \big(M_{4k}\big) \mathbbm{1}_{2s+1} + \hat{J}_k\,, \nonumber\\
L_{04} &=& \big(M_{04}\big) \mathbbm{1}_{2s+1} - \sum_{k=1}^3 z^k\hat{J}_k\,, \nonumber\\
L_{ki} &=& \big(M_{ki}\big) \mathbbm{1}_{2s+1} - {{\cal{E}}_{ki}}^{j} \; \hat{J}_j\,, \nonumber\\
L_{0k} &=& \big(M_{0k}\big) \mathbbm{1}_{2s+1} + z^4\hat{J}_k + {{\cal{E}}_{ki}}^{j} \; z^i \hat{J}_j\,,
\end{eqnarray}
where $i,j,k=1,2,3$ and $z^1,\; ... \;, z^4$ are given by the identities in Eq. (\ref{z set}), while the matrix elements of the $(2s+1)\times(2s+1)$ matrices $\hat{J}_k$ are given by:
\begin{eqnarray}\label{iiiiiiiii}
\left(\hat{J}_{k=1}\right)_{mm^\prime} &=& \frac{1}{2} \sqrt{(s+m)(s-m+1)}\; \delta^{}_{m,m^\prime+1} + \frac{1}{2} \sqrt{(s-m)(s+m+1)}\; \delta^{}_{m,m^\prime-1}\,, \nonumber\\
\left(\hat{J}_{k=2}\right)_{mm^\prime} &=& \frac{1}{2\mathrm{i}} \sqrt{(s+m)(s-m+1)}\; \delta^{}_{m,m^\prime+1} - \frac{1}{2\mathrm{i}} \sqrt{(s-m)(s+m+1)}\; \delta^{}_{m,m^\prime-1}\,, \nonumber\\
\left(\hat{J}_{k=3}\right)_{mm^\prime} &=& m \; \delta^{}_{m,m^\prime}\,,
\end{eqnarray}
where $m$ and $m^\prime$ are such that $-s \leqslant m,m^\prime \leqslant s$ and $s\pm m,\; s\pm m^\prime$ are integers. Note that the $\hat{J}_k$'s set exhibits the matrix realization of the spin $s$ representation of the $\mathfrak{su}(2)$ Lie algebra:
\begin{eqnarray}
[ \hat{J}_k, \hat{J}_i ] = - \mathrm{i} {{\cal{E}}_{ki}}^{j} \; \hat{J}_j\,.
\end{eqnarray}
Considering the above, one can also check that the infinitesimal operators $L_{AB}$ verify the commutation relations (\ref{adjoint commutator}), and that the two Casimir operators (\ref{Casimir 2}) and (\ref{Casimir 4}) respectively take the following forms:
\begin{eqnarray}
Q^{(1)} = \Big(- s(s+1) - \tau(\tau+3) \Big) \mathbbm{1}\,,
\end{eqnarray}
\begin{eqnarray}
Q^{(2)} = \Big( - s(s+1)(\tau+1)(\tau+2) \Big) \mathbbm{1}\,,
\end{eqnarray}
again, with $\tau=-3/2-\mathrm{i}\nu$ ($\nu\in\mathbb{R}$).

\subsection{Principal series: restriction to the maximal compact subgroup $\mathrm{SU}(2) \times \mathrm{SU}(2)$}
In this subsection, we explicitly present the given action of $\underline{U}^{\mbox{\small{ps}}}_{s,\nu}({\underline{g}})$'s in the Hilbert space $L^2_{{\mathbb{C}}^{2s+1}} (\mathbb{S}^3)$ (see Eq. (\ref{U principal dS4 gen})), when it is restricted to the elements ${\underline{g}}$ belonging to the maximal compact subgroup of Sp$(2,2)$, that is, $\underline{\cal{K}} \sim  \mathrm{SU}(2) \times \mathrm{SU}(2)$. Such elements are determined by ${\underline{g}} = \begin{pmatrix} \textbf{v} & \textbf{0} \\ \textbf{0} & \textbf{w} \end{pmatrix}$, with $\textbf{v},\textbf{w} \in \mathrm{SU}(2)$ (see subsection \ref{Sec Cartan dS4}). On this basis, the action (\ref{U principal dS4 gen}) takes the form:
\begin{eqnarray}
\Big(\underline{U}^{\mbox{\small{ps}}}_{s,\nu} (\underline{k}) f\Big) (\textbf{z}) = {\cal{D}}^s(\textbf{w}) f(\textbf{v}^{\scriptscriptstyle\bigstar}\textbf{z}\textbf{w})\,.
\end{eqnarray}
Now, let ${\textbf{D}}^{(j_l,j_r)}$ denote the $(2j_l+1)\times(2j_r+1)$-dimensional UIR of $\mathrm{SU}(2) \times \mathrm{SU}(2)$ on the vectors $f \in L^2_{{\mathbb{C}}^{2s+1}} (\mathbb{S}^3)$; note that, again, the subscripts `$l$' and `$r$' refer to the left and right regular UIR's of $\mathrm{SU}(2)$, respectively, and that ${\cal{D}}^s \sim {\textbf{D}}^{(0,s)}$. Then, the representations $\underline{U}^{\mbox{\small{ps}}}_{s,\nu} (\underline{g})$, with $\underline{g} \in \mathrm{SU}(2) \times \mathrm{SU}(2)$, decompose into the infinite direct sum:
\begin{eqnarray}
\bigoplus_{j\in \mathbb{N}/2} {\textbf{D}}^{(0,s)} \otimes {\textbf{D}}^{(j,j)} = \bigoplus_{j\in \mathbb{N}/2} \; \bigoplus_{|j-s|\leqslant i \leqslant j+s} {\textbf{D}}^{(j,i)}\,.
\end{eqnarray}
In the above decomposition, the UIR ${\textbf{D}}^{(j,i)}$ appears once and only once in the reduction of the restriction $\underline{U}^{\mbox{\small{ps}}}_{s,\nu}\;\big|^{}_{\mathrm{SU}(2) \times \mathrm{SU}(2)}$.

\subsection{Complementary series}\label{Subsec complementary dS4}
The complementary series of the Sp$(2,2)$ UIR's, like its principal counterpart, is issued from the Iwasawa factorization of Sp$(2,2)$. The carrier Hilbert space of the complementary series UIR's, in a general case with spin $s=0,1,2, \; ... \;$, is $L^2_{{\mathbb{C}}^{2s+1}} (\mathbb{S}^3 \times \mathbb{S}^3)$, while the measure can be found through a reproducing kernel (see the process pointed out in the dS$_2$ case). For instance, in the scalar case $s=0$, we have \cite{Takahashi'}:
\begin{eqnarray}
\langle f_1 , f_2 \rangle^{}_{-\tau-3} = \frac{\Gamma(-\tau-1) \; \Gamma(\tau+3)}{2\pi^2 \; \Gamma(-2\tau-3)} \iint_{\mathbb{S}^3 \times \mathbb{S}^3} f^{\ast}_1(\textbf{z}_1)f_2(\textbf{z}_2) \; |\textbf{z}_1 - \textbf{z}_2|^{-2\tau-6} \; \mathrm{d}\mu(\textbf{z}_1) \mathrm{d}\mu(\textbf{z}_2)\,,
\end{eqnarray}
where $\tau$ is real and bounded, $-3 < \tau < -3/2$, and again $\mathrm{d}\mu(\textbf{z})$ is the $O(4)$-invariant measure on $\mathbb{S}^3$. Note that by expanding the kernel in terms of the hyperspherical harmonics on $\mathbb{S}^3$ (see appendix \ref{App kernel}):\footnote{Note that the value $\tau = -3$, for which the kernel reduces to $1$, represents the critical value separating the scalar complementary series UIR's from the discrete ones. We will discuss this matter in the following subsection.}
\begin{eqnarray}\label{Kernel complementary}
\frac{\Gamma(-\tau-1) \; \Gamma(\tau+3)}{2\pi^2 \; \Gamma(-2\tau-3)} \; |\textbf{z} - \textbf{z}_2|^{-2\tau-6} = \sum_{L,l,m} \frac{\Gamma(L+\tau+3)}{\Gamma(L-\tau)} {\cal{Y}}_{Llm}(\textbf{z}_1) {\cal{Y}}^\ast_{Llm}(\textbf{z}_2)\,,
\end{eqnarray}
we achieve the following orthonormal basis of the corresponding Hilbert space $L^2_{\mathbb{C}} (\mathbb{S}^3 \times \mathbb{S}^3)$:
\begin{eqnarray}
\Bigg\{ \widetilde{{\cal{Y}}}_{Llm}^\tau (\textbf{z}) \equiv \sqrt{\frac{\Gamma(L-\tau)}{\Gamma(L+\tau+3)}} \; {\cal{Y}}_{Llm}(\textbf{z}),\;\; L\in\mathbb{N},\; 0\leqslant l \leqslant L,\; -l\leqslant m \leqslant l \Bigg\}\,.
\end{eqnarray}

The action of the associated representation operators $\underline{U}^{\mbox{\small{cs}}}_{0,\tau}$ on the functions $f(\textbf{z})\in L^2_{\mathbb{C}} (\mathbb{S}^3 \times \mathbb{S}^3)$ reads:
\begin{eqnarray}
\mathrm{Sp}(2,2) \ni \underline{g} = \begin{pmatrix} \textbf{a} & \textbf{b} \\ \textbf{c} & \textbf{d} \end{pmatrix} \;:\; f(\textbf{z}) \;\mapsto\; \Big(\underline{U}^{\mbox{\small{cs}}}_{0,\tau} (\underline{g}) f\Big) (\textbf{z}) &=& \big|-\textbf{b}^{\scriptscriptstyle\bigstar}\textbf{z} + \textbf{d}^{\scriptscriptstyle\bigstar}\big|^{2\tau} f\Big( \frac{\textbf{a}^{\scriptscriptstyle\bigstar}\textbf{z} - \textbf{c}^{\scriptscriptstyle\bigstar}}{-\textbf{b}^{\scriptscriptstyle\bigstar}\textbf{z} + \textbf{d}^{\scriptscriptstyle\bigstar}} \Big) \nonumber\\
&\equiv& N(\underline{g},\textbf{z}) \; f(\underline{g}^{-1}\diamond \textbf{z})\,.
\end{eqnarray}
In a spinorial case, with spin $s=1,2,3,\;...\;$, this action extends to:
\begin{eqnarray}
\mathrm{Sp}(2,2) \ni \underline{g} = \begin{pmatrix} \textbf{a} & \textbf{b} \\ \textbf{c} & \textbf{d} \end{pmatrix} \;:\; f(\textbf{z}) \;\mapsto\; \Big(\underline{U}^{\mbox{\small{cs}}}_{s,\tau} (\underline{g}) f\Big) (\textbf{z}) &=& \big|-\textbf{b}^{\scriptscriptstyle\bigstar}\textbf{z} + \textbf{d}^{\scriptscriptstyle\bigstar}\big|^{2\tau} \; {\cal{D}}^s \Big( \frac{-\textbf{z}^{\scriptscriptstyle\bigstar}\textbf{b} + \textbf{d}}{|-\textbf{z}^{\scriptscriptstyle\bigstar}\textbf{b} + \textbf{d}|} \Big) \; f\Big( \frac{\textbf{a}^{\scriptscriptstyle\bigstar}\textbf{z} - \textbf{c}^{\scriptscriptstyle\bigstar}}{-\textbf{b}^{\scriptscriptstyle\bigstar}\textbf{z} + \textbf{d}^{\scriptscriptstyle\bigstar}} \Big) \nonumber\\
&\equiv& N(\underline{g},\textbf{z}) \; {\cal{D}}^s \Big( \frac{-\textbf{z}^{\scriptscriptstyle\bigstar}\textbf{b} + \textbf{d}}{|-\textbf{z}^{\scriptscriptstyle\bigstar}\textbf{b} + \textbf{d}|} \Big) \; f(\underline{g}^{-1}\diamond \textbf{z})\,,
\end{eqnarray}
where, in this case, the functions $f(\textbf{z})\in L^2_{{\mathbb{C}}^{2s+1}} (\mathbb{S}^3 \times \mathbb{S}^3)$ and, again, ${\cal{D}}^s$'s are the $2s+1$-dimensional UIR's of $\mathrm{SU}(2)$, while $-2<\tau<-3/2$. Apart from the the values of $\tau$, the above actions are quite similar to those given in the principal case (see Eqs. (\ref{U principal dS4}) and (\ref{U principal dS4 gen}), respectively). Clearly, the infinitesimal operators and correspondingly the two Casimir operators in this series of the UIR's also take the same forms as those appeared in the principal one (again, with the exception of the values of $\tau$).

\subsection{Discrete series}\label{Subsec discrete dS4}
The discrete series of the Sp$(2,2)$ UIR's is issued from the Cartan decomposition of the group (see subsection \ref{Sec Cartan dS4}). The corresponding UIR's are specified by two parameters $p$ and $q$, namely, $\underline{U}^{\mbox{\small{ds}}} \equiv \Pi^\pm_{p,q}$, such that $p,q \in \mathbb{N}/2$, $1\leqslant q \leqslant p$, and $p-q \in \mathbb{N}$ \cite{Takahashi'}. In a general case with given $p$ and $q$ in their respective allowed ranges, these UIR's act in the Hilbert space $L^2_{\mathbb{C}^{2p+1}}(B)$ as the space of vector-valued functions $f(\textbf{z})$, analytic inside the open unit-ball $B$, with values in $\mathbb{C}^{2p+1}$; the latter determines the carrier space of either UIR's ${\cal{D}}^p\otimes \mathbbm{1}$ or $\mathbbm{1} \otimes {\cal{D}}^p$ of the maximal compact subgroup $\underline{{\cal{K}}} \sim \mathrm{SU}(2) \times \mathrm{SU}(2)$ of Sp$(2,2)$,\footnote{Note that ${\cal{D}}^p$'s refer to the $2p+1$-dimensional UIR's of $\mathrm{SU}(2)$ (see appendix \ref{App UIR's SU(2)}).} which are respectively referred to by the superscripts `$-$' and `$+$' in the representation operators (see Eq. (\ref{33333333333})). The functions $f(\textbf{z})$ are also supposed to be square integrable with respect to the scalar product \cite{Takahashi'}:
\begin{eqnarray}
\langle f_1 , f_2 \rangle = c_{p,q} \int_{B} \langle f_1(\textbf{z}) , f_2(\textbf{z}) \rangle^{}_{\mathbb{C}^{2p+1}} \; \big(1 - |\textbf{z}|^2 \big)^{2q-2} \mathrm{d}\mu(\textbf{z})\,,
\end{eqnarray}
where:
\begin{eqnarray}
c_{p,q} = \frac{(2q-1)(p+q)(p-q+1)}{\pi^2}\,,
\end{eqnarray}
while $\langle \cdot , \cdot \rangle^{}_{\mathbb{C}^{2p+1}}$ stands for the scalar product in $\mathbb{C}^{2p+1}$, and $\big(1 - |\textbf{z}|^2 \big)^{2q-2} \mathrm{d}\mu(\textbf{z})$ for the invariant measure. [Note that for the limit cases $(p,q=1/2)$ and $(p,q=0)$, a specific treatment is needed. We will study the case $(p,q=0)$ in the following subsubsection.] The action of the representation operators $\Pi^\pm_{p,q}$ in the Hilbert space $L^2_{\mathbb{C}^{2p+1}}(B)$ reads:
\begin{eqnarray}\label{33333333333}
\mathrm{Sp}(2,2) \ni \underline{g} = \begin{pmatrix} \textbf{a} & \textbf{b} \\ \textbf{c} & \textbf{d} \end{pmatrix} \;:\; f(\textbf{z}) \;\mapsto\; \Big(\Pi^\pm_{p,q} (\underline{g}) f\Big) (\textbf{z}) &=& \big|-\textbf{b}^{\scriptscriptstyle\bigstar}\textbf{z} + \textbf{d}^{\scriptscriptstyle\bigstar}\big|^{2\tau} \; {\cal{D}}^p \Big( \big(\eta_{}^\pm (\underline{g}, \textbf{z})\big)^{-1} \Big) \; f\Big( \frac{\textbf{a}^{\scriptscriptstyle\bigstar}\textbf{z} - \textbf{c}^{\scriptscriptstyle\bigstar}}{-\textbf{b}^{\scriptscriptstyle\bigstar}\textbf{z} + \textbf{d}^{\scriptscriptstyle\bigstar}} \Big) \nonumber\\
&\equiv& N(\underline{g},\textbf{z}) \; {\cal{D}}^p \Big( \big(\eta_{}^\pm (\underline{g}, \textbf{z})\big)^{-1} \Big) f(\underline{g}^{-1}\diamond \textbf{z})\,,
\end{eqnarray}
where $\tau= -q-1$ and:
\begin{eqnarray}
\big(\eta_{}^+ (\underline{g}, \textbf{z})\big)^{-1} = \frac{-\textbf{z}^{\scriptscriptstyle\bigstar} \textbf{b} + \textbf{d}}{|-\textbf{z}^{\scriptscriptstyle\bigstar} \textbf{b} + \textbf{d}|} \,, \;\;\;\;\;\;\; \big(\eta_{}^- (\underline{g}, \textbf{z})\big)^{-1} = \frac{-\textbf{z} \textbf{c} + \textbf{a}}{|-\textbf{z}^{\scriptscriptstyle\bigstar} \textbf{b} + \textbf{d}|} \,.
\end{eqnarray}
Note that $|-\textbf{z}^{\scriptscriptstyle\bigstar} \textbf{b} + \textbf{d}| = |-\textbf{z} \textbf{c} + \textbf{a}|$. Moreover, considering the Cartan decomposition of Sp$(2,2)$ (see subsection \ref{Sec Cartan dS4}), we have:
\begin{eqnarray}\label{g.z dS4 B}
B \ni \textbf{z} \;\mapsto\; \textbf{z}^\prime \equiv \underline{g}^{-1}\diamond \textbf{z} = (\textbf{a}^{\scriptscriptstyle\bigstar}\textbf{z} - \textbf{c}^{\scriptscriptstyle\bigstar})(-\textbf{b}^{\scriptscriptstyle\bigstar}\textbf{z} + \textbf{d}^{\scriptscriptstyle\bigstar})^{-1} \in B\,,
\end{eqnarray}

Proceeding as before, it is a matter of straightforward calculations to show that the infinitesimal operators $L_{AB}$ of the UIR's $\Pi^\pm_{p,q}$, constituted by an orbital part $M_{AB}$ and a spinorial part $S_{AB}$ ($L_{AB} = M_{AB} + S_{AB}$), are given respectively by:
\begin{eqnarray}
L_{4k} &=& \mathrm{i}\Big(z_k \frac{\partial}{\partial z^4} - z_4 \frac{\partial}{\partial z^k}\Big) \mathbbm{1}_{2s+1} \pm \hat{J}_k\,, \nonumber\\
L_{04} &=& \mathrm{i}\Big( -\tau z_4 + z_4 \sum_{\sigma=1}^4 z^\sigma \frac{\partial}{\partial z^\sigma} - \frac{|\textbf{z}|^2 + 1}{2} \frac{\partial}{\partial z^4} \Big) \mathbbm{1}_{2s+1} \mp \sum_{k=1}^3 \frac{z^k}{|\textbf{z}|}\hat{J}_k\,, \nonumber\\
L_{ki} &=& - \mathrm{i}\Big( z_k \frac{\partial}{\partial z^i} - z_i \frac{\partial}{\partial z^k} \Big) \mathbbm{1}_{2s+1} - {{\cal{E}}_{ki}}^{j} \; \hat{J}_j\,, \nonumber\\
L_{0k} &=& \mathrm{i} \Big( -\tau z_k + z_k \sum_{\sigma=1}^4 z^\sigma \frac{\partial}{\partial z^\sigma} - \frac{|\textbf{z}|^2 + 1}{2} \frac{\partial}{\partial z^k} \Big) \mathbbm{1}_{2s+1} \pm \frac{z^4}{|\textbf{z}|}\hat{J}_k + {{\cal{E}}_{ki}}^{j} \; \frac{z^i}{|\textbf{z}|} \hat{J}_j\,,
\end{eqnarray}
where, again, $i,j,k=1,2,3$ and the $(2s+1)\times(2s+1)$-matrices $\hat{J}_k$ stand for the matrix realization (\ref{iiiiiiiii}) of the spin $s$ representation of the $\mathfrak{su}(2)$ Lie algebra. One can easily show that the above infinitesimal operators obey the commutation relations (\ref{adjoint commutator}), and that the two Casimir operators (\ref{Casimir 2}) and (\ref{Casimir 4}) explicitly read:
\begin{eqnarray}
Q^{(1)} = \Big(-p(p+1) - \tau(\tau+3)\Big) \mathbbm{1}\,,
\end{eqnarray}
\begin{eqnarray}
Q^{(2)} = \Big(-p(p+1)(\tau+1)(\tau+2)\Big) \mathbbm{1}\,,
\end{eqnarray}
again, with $\tau= -q-1$.

Here, we underline again that the above representations form the square-integrable part of the discrete series UIR's of Sp$(2,2)$. Below, we will study the scalar discrete series representations, lying at the ``lowest limit" of this series, which are not square integrable, and as we will point out below, they rather deserve the appellation ``degenerate scalar complementary series".

\subsubsection{Discrete series: scalar case ($\Pi_{p\geqslant 1,0}$)}
Let ${\cal{H}}_{p-1} \equiv L^2_{\mathbb{C}} (\mathbb{S}^3)$ denote the carrier Hilbert space of $\Pi_{p,0}$ ($p=1,2,...$), that is, the closure of the linear span of all square-integrable functions $\mathbb{S}^3 \ni \textbf{z} \mapsto f(\textbf{z}) \in \mathbb{C}$, according to the inner product \cite{Takahashi'}:
\begin{eqnarray}
\langle f_1 , f_2 \rangle^{}_{p-1} = \frac{(-1)^{p+1}}{4\pi^2(2p-1)!} \iint_{\mathbb{S}^3 \times \mathbb{S}^3} f_1^\ast(\textbf{z}_1) f_2(\textbf{z}_2) \; |\textbf{z}_1 - \textbf{z}_2|^{2(p-1)} \log |\textbf{z}_1 - \textbf{z}_2|^{-2} \; \mathrm{d}\mu(\textbf{z}_1) \mathrm{d}\mu(\textbf{z}_2)\,,
\end{eqnarray}
where $\mathrm{d}\mu(\textbf{z})$ is the $O(4)$-invariant measure on $\mathbb{S}^3$. The functions $f(\textbf{z})$ are also supposed to verify the following orthogonality condition:
\begin{eqnarray}
\int_{\mathbb{S}^3} {\cal{Y}}_{Llm}^\ast(\textbf{z}) f(\textbf{z}) \; \mathrm{d}\mu(\textbf{z}) = 0\,,
\end{eqnarray}
for all triplets $(Llm)$, with $0\leqslant L \leqslant p-1$. Therefore, if the function $f(\textbf{z})$ belongs to $L^2_{\mathbb{C}} (\mathbb{S}^3)$, it lies in the subspace orthogonal to the finite-dimensional subspace $V_{p-1}$ with the orthonormal basis $\big\{ {\cal{Y}}_{Llm}\; ; \; 0\leqslant L \leqslant p-1, \; 0\leqslant l \leqslant L, \; -l\leqslant m \leqslant l \big\}$. The subspace $V_{p-1}$, considering the allowed ranges of $L$, $l$, and $m$, is of $p(p+1)(2p+1)/6$ dimension, and carries the irreducible (nonunitary!) dS$_4$ finite-dimensional representations, which, with respect to the notations given in appendix \ref{App Lie algebra B2}, are determined by $(n_1=0,n_2=p-1)$. These representations are \emph{Weyl equivalent}\footnote{If two representations are Weyl equivalent, then they share same Casimir eigenvalue.} to the UIR's $\Pi_{p,0}$. [We will revisit the above mathematical structure in detail later in subsection \ref{Subsec generating}, when the spacetime realization of the representations is taken into account.]

An orthonormal basis for the Hilbert space ${\cal{H}}_{p-1}$, with respect to the identities given in appendix \ref{App kernel}, is:
\begin{eqnarray}
\Bigg\{ \widetilde{{\cal{Y}}}_{Llm}^{\tau=-p-2}(\textbf{z}) \equiv \sqrt{\frac{(L+p+1)!}{(L-p)!}} \; {\cal{Y}}_{Llm}(\textbf{z}),\;\; L\geqslant p,\; 0\leqslant l \leqslant L,\; -l\leqslant m \leqslant l \Bigg\}\,.
\end{eqnarray}
The action of the UIR's $\Pi_{p,0}$ in ${\cal{H}}_{p-1}$ is realized by:
\begin{eqnarray}
\mathrm{Sp}(2,2) \ni \underline{g} = \begin{pmatrix} \textbf{a} & \textbf{b} \\ \textbf{c} & \textbf{d} \end{pmatrix} \;:\; f(\textbf{z}) \;\mapsto\; \Big(\Pi_{p,0} (\underline{g}) f\Big) (\textbf{z}) &=& \big|-\textbf{b}^{\scriptscriptstyle\bigstar}\textbf{z} + \textbf{d}^{\scriptscriptstyle\bigstar}\big|^{2\tau} f\Big( \frac{\textbf{a}^{\scriptscriptstyle\bigstar}\textbf{z} - \textbf{c}^{\scriptscriptstyle\bigstar}}{-\textbf{b}^{\scriptscriptstyle\bigstar}\textbf{z} + \textbf{d}^{\scriptscriptstyle\bigstar}} \Big) \nonumber\\
&\equiv& N(\underline{g},\textbf{z}) \; f(\underline{g}^{-1}\diamond \textbf{z})\,,
\end{eqnarray}
with $\tau=-p-2$. Note that the infinitesimal operators and correspondingly the two Casimir operators in the scalar discrete representations again take the same forms as those appeared in the scalar principal series, except the values of $\tau$ which here, as already mentioned, are given by $\tau=-p-2$.

The fact that the above representations are the UIR's of Sp$(2,2)$ corresponding to what we could call the ``complementary degenerate" series rests on proofs which necessitate expansion formulas of kernels of the type $|\textbf{z}_1 - \textbf{z}_2|^{2p-2} \log |\textbf{z}_1 - \textbf{z}_2|^{-2} $, given in appendix \ref{App kernel}.

\setcounter{equation}{0} \section{``Massive"/``massless" dS$_4$ UIR's and the Poincar\'{e} contraction}\label{Sec contraction}
At this stage, it is critical to understand the physical content of the dS$_4$ UIR's in terms of their Poincar\'{e} contraction limit ($R\rightarrow\infty$).\footnote{Here, it is perhaps worthwhile mentioning that the contraction of the dS UIR's to the Poincar\'{e} ones was first put forward in Ref. \cite{Phillips} in $1+1$ dimension and then in Ref. \cite{Mickelsson} for the representations of the dS group in $1+n$ dimension.} On this basis, naturally, three categories of the dS$_4$ UIR's come to fore: those which contract to the Poincar\'{e} massive UIR's; those which possess a massless content; and finally those which either have nonphysical Poincar\'{e} contraction limit or do not have Poincar\'{e} contraction limit at all. Below, we will briefly study the first two categories. But, before that, it would be convenient to take a look at the notion of group contraction on the representation level. In this regard, we again follow the lines sketched in Ref. \cite{Garidi Thesis}, and present such a notion in a way that is more suited to the needs of this paper.

Note that, in this section, the parameters $c$ (the speed of light) and $\hbar$ (the Planck constant) are no longer normalized to unity. Together with $R$, the radius of curvature of the dS$_4$ hyperboloid $\underline{M}_R$, they represent dimensionally independent quantities, which are employed to fix the natural unit of ``mass" $\hbar/cR$ in dS$_4$ relativity.

\subsection{Group contraction (the representation level): a brief introduction}
\begin{figure}[H]
\begin{center}
\includegraphics[height=.2\textheight]{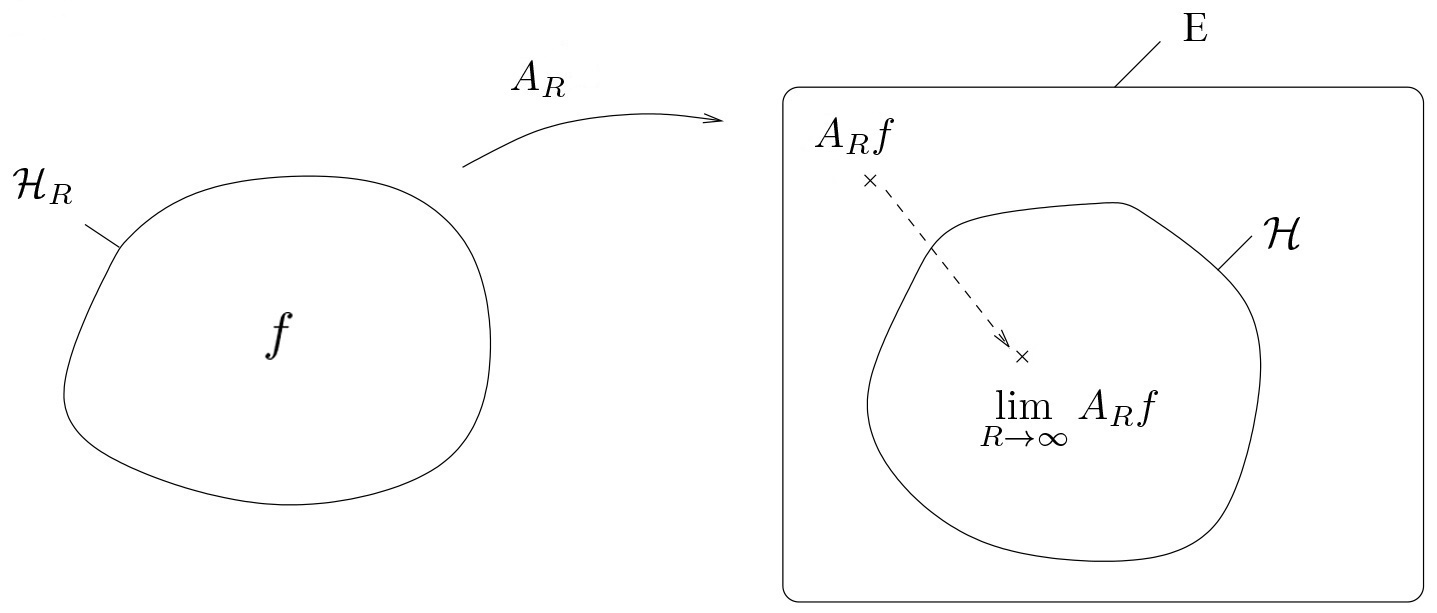}
\end{center}
\caption{Topological space $E$, in which the representations of $G$ and $G^\prime$ can be written \cite{Garidi Thesis}.}
\label{FIG. Contraction}
\end{figure}

Let $U^R$ and $U$ respectively denote a family of representations of a group $G$ acting in a Hilbert space ${\cal{H}}_R$ and a family of representations of a group $G^\prime$ acting in a Hilbert space ${\cal{H}}$. The two given groups $G$ and $G^\prime$ are supposed to be close enough to be put into bijection $i\; : \; G \rightarrow G^\prime$ (which is not a homomorphism). We also need a topological space $E$ (see FIG. \ref{FIG. Contraction}), in which we can write the representations of $G$ and $G^\prime$. This is achieved by having an injective map $A_R \;:\; {\cal{H}}_R \mapsto E \; \big(\supset {\cal{H}}\big)$ such that, for all $f \in {\cal{H}}_R$, we get:
\begin{eqnarray}
\lim_{R\rightarrow \infty} A_R f = h\,, \;\;\;\;\;\;\; h\in {\cal{H}}\,.
\end{eqnarray}
Considering the above, the representation $U$ is said to be the contraction of the representation $U^R$, symbolized here by $U^R \longrightarrow U$, if:
\begin{eqnarray}
\mbox{for all}\; f \in {\cal{H}}_R\,, \;\;\;\;\;\;\; \lim_{R\rightarrow \infty} A_R U^R(g) f = U(g^\prime) h = U(g^\prime) \lim_{R\rightarrow \infty} A_R f\,,
\end{eqnarray}
where $g^\prime \in G^\prime$ corresponds to $g \in G$ by the bijection $i$.

\subsection{DS$_4$ massive UIR's}\label{Subsec massive contraction}
We now turn to the aforementioned categories of the dS$_4$ UIR's, given in terms of their null-curvature limit. For the first category, i.e., those dS$_4$ UIR's which contract to the Poincar\'{e} massive ones, the dS$_4$ principal series UIR's $\underline{U}^{\mbox{\small{ps}}}_{s,\nu}$, with $s\in \mathbb{N}/2$ and $\nu\in\mathbb{R}$, are merely involved. In this sense, they are usually called dS$_4$ massive representations. Technically, for these representations, by introducing a relation between the dS$_4$ representation parameter $\nu$ and the Poincar\'{e}-Minkowski mass $m$ as:
\begin{eqnarray}\label{minandds}
m = \frac{\hbar\nu}{cR}\,,
\end{eqnarray}
the Poincar\'{e} contraction explicitly reads \cite{Mickelsson,Garidi zero curvature limit}:
\begin{eqnarray}\label{massive contraction}
\underline{U}^{\mbox{\small{ps}}}_{s,\nu} \;\; \underset{\underset{\hbar \nu/cR = m}{R \rightarrow \infty,\; \nu \rightarrow \infty}}{\longrightarrow} \;\; c_>{\cal{P}}^>_{s,m} \; \oplus \; c_<{\cal{P}}^<_{s,m} \,,
\end{eqnarray}
where ${\cal{P}}^\gtrless_{s,m}$ respectively stand for the positive/negative energy Wigner UIR's of the Poincar\'{e} group, possessing mass $m$ and spin $s$. Among the `coefficients' $c^{}_>$ and $c^{}_<$, one can be set $1$, while the other is zero. [We will go through the mathematical details of this contraction limit later in subsections \ref{Subsec flat limit} and \ref{Subsec 2point function}, when the spacetime realization of the representations is taken into account.]

Here, having in mind the lines sketched in the previous sections, we would like to once again underline that during the Poincar\'{e} contraction procedure, symbolized by Eq. (\ref{massive contraction}), the Lorentz subgroup remains intact. In other words, the contraction procedure is carried out with respect to the Lorentz subgroup in the dS$_4$ and Poincar\'{e} groups. This is actually the point that gives sense to the notion of spin in the context of dS$_4$ massive elementary systems (associated with the principal series UIR's), since it arises from the same $\mathrm{SU}(2)$ (that is, the rotations subgroup of the Lorentz group) that the notion of spin in Poincar\'{e} relativity does.

Another point to be noticed here is the possible breaking of the irreducibility of the dS$_4$ principal (massive) UIR's, through the Poincar\'{e} contraction limit, into a direct sum of two Poincar\'{e} massive UIR's with positive and negative energies. Actually, from this evidence of the possible breaking of irreducibility, the inquiry into the concept of ``rest energy" in dS$_4$ (generally, dS) relativity leads to ambiguity, when one follows the procedure of the Poincar\'{e} contraction. We will come back to this important point in subsection \ref{Subsec Energy operator}.

\subsection{DS$_4$ massless UIR's}\label{Subsec Massless UIR's}
In comparison with the dS$_4$ massive UIR's, the massless ones are more subtle. As a matter of fact, the dS$_4$ group has no UIR analogous to the so-called massless infinite-spin UIR's of the Poincar\'{e} group. Then, the dS$_4$ massless UIR's are distinguished as those UIR's with a unique extension to the UIR's of the conformal group $\mathrm{SO}_0(2,4)$, while that extension is equivalent to the conformal extension of the Poincar\'{e} massless UIR's \cite{Barut,Mack1977}. Accordingly, two different categories of the dS$_4$ massless UIR's come to fore (see appendix \ref{App massless UIRs}):
\begin{itemize}
\item{The massless scalar case, which involves the only physical representation (in the sense of contraction/extension to the Poincar\'{e}/conformal UIR's) of the dS$_4$ complementary series, labelled in our notations by $\underline{U}^{\mbox{\small{cs}}}_{s=0, \nu=\frac{1}{2}}$:
   \begin{eqnarray}
    \left. \begin{array}{ccccccc}
    & & {\cal{C}}^{>}_{1,0,0} & & {\cal{C}}^{>}_{1,0,0} & \hookleftarrow & {\cal{P}}^{>}_{0,0} \\
    \underline{U}^{\mbox{\small{cs}}}_{0,\frac{1}{2}} & \hookrightarrow & \oplus & \underset{R\rightarrow \infty}{\longrightarrow} & \oplus & &\oplus \\
    & & {\cal{C}}^{<}_{-1,0,0} & & {\cal{C}}^{<}_{-1,0,0} & \hookleftarrow & {\cal{P}}^{<}_{0,0} \,.
    \end{array} \right.
    \end{eqnarray}}
\item{The ``helicity $= \pm s$" cases, which involve all representations $\Pi^{\pm}_{p=s,q=s}$,\footnote{Here, the superscript `$\pm$' stands for the helicities $\pm s$.} with $s>0$, lying at the lower limit of the dS$_4$ discrete series:
    \begin{eqnarray}
    \left. \begin{array}{ccccccc}
    & & {\cal{C}}^{>}_{s+1,0,s} & & {\cal{C}}^{>}_{s+1,0,s} & \hookleftarrow & {\cal{P}}^{>}_{-s,0} \\
    \Pi^+_{s,s} & \hookrightarrow & \oplus & \underset{R\rightarrow \infty}{\longrightarrow} & \oplus & & \oplus \\
    & & {\cal{C}}^{<}_{-s-1,0,s} & & {\cal{C}}^{<}_{-s-1,0,s} & \hookleftarrow & {\cal{P}}^{<}_{-s,0} \,.
    \end{array} \right.
    \end{eqnarray}
    \begin{eqnarray}
    \left. \begin{array}{ccccccc}
    & & {\cal{C}}^{>}_{s+1,s,0} & & {\cal{C}}^{>}_{s+1,s,0} & \hookleftarrow & {\cal{P}}^{>}_{s,0} \\
    \Pi^-_{s,s} & \hookrightarrow & \oplus & \underset{R\rightarrow \infty}{\longrightarrow} & \oplus & & \oplus \\
    & & {\cal{C}}^{<}_{-s-1,s,0} & & {\cal{C}}^{<}_{-s-1,s,0} & \hookleftarrow & {\cal{P}}^{<}_{s,0}\,,
    \end{array} \right.
    \end{eqnarray}}
\end{itemize}
Note that: (i) Above, we have denoted by the arrows `$\hookrightarrow$' unique extension and by ${\mathcal P}^{\gtrless}_{s,0}$, respectively, the positive/negative energy Poincar\'{e} massless representations, possessing helicity $s$. (ii) Conformal invariance involves the discrete series representations (and their lower end) of the (universal covering of the) conformal group or its double covering group $\mathrm{SO}_0(2,4)$ or its fourth covering group $\mathrm{SU}(2,2)$. Above, the associated conformal UIR's are denoted by ${\mathcal C}^{\gtrless}_{E_0,j_l, j_r}$, where the parameters $(j_l,j_r) \in \mathbb{N}/2 \times \mathbb{N}/2$ label the UIR's of $\mathrm{SU}(2) \times \mathrm{SU}(2)$, while $E_0$ refers to the positive/negative conformal energy.

\subsection{Discussion: rehabilitating the dS$_4$ physics from the point of view of a local Minkowskian observer}\label{Subsec Energy operator}
We here draw attention to Eq. (\ref{massive contraction}), in particular, to the possible breaking of the irreducibility of the dS$_4$ principal UIR's, during the Poincar\'{e} contraction limit, into a direct sum of two Poincar\'{e} UIR's possessing positive and negative energies. This phenomenon can be discussed on two levels.

On one hand, it refers to the energy ambiguity in dS$_4$ (generally, dS) relativity, which actually originates from the existence of the discrete symmetry \eqref{actgam0} in the dS$_4$ group sending any point $(x^0,\textbf{x}) \in \underline{M}_R$ to its mirror image with respect to the $x^0$-axis, that is, $(x^0,-\textbf{x}) \in \underline{M}_R$.\footnote{To see the latter point, one can also consider Eq. (\ref{454545454545}), when $\psi=0$ and ${\textbf{w}}^2 \; \big( = (\cos{\theta}, \sin{\theta} \vec{w}) \big) = -1$ (that is, $\theta=\pi$).} Considering this discrete symmetry, the dS$_4$ infinitesimal generators $L_{\texttt{A} 0}$ (see Eq. (\ref{LLLLLLL})), with $\texttt{A}=1,2,3,4$, transform into their respective opposites with possibly different signs of the corresponding conserved charges, depending on the sign of $\textbf{x}$. This, for instance, implies that whether the generator $L_{40}$, which contracts to the Poincar\'{e} energy operator, moves us forwards or backwards in time (towards increasing or decreasing $x^0$) depends on the sign of $\textbf{x}$, and hence, cannot be precisely determined. In this sense, this is the best we can do: there is no positive conserved energy in dS$_4$ (generally, dS) spacetime.

On the other hand and besides the global considerations concerning the notion of energy in dS$_4$ (generally, dS) relativity, it is yet critical to understand the physical content of dS$_4$ relativity with respect to its null-curvature limit at a given point $x \in \underline{M}_R$, namely, from the point of view of a local (tangent) Minkowskian observer, for whom the fundamental physical conservation laws are understood from the principles of Einstein-Poincar\'{e} relativity. Considering Eq. (\ref{massive contraction}), it seems that, even at a specific point (say $x \in \underline{M}_R$), one cannot give a precise meaning to the dS$_4$ ``rest energy" in terms of the Poincar\'{e} contraction of the representations. In part \ref{Part plane waves}, we will come back to this significant point. We will show that there is a proper choice of dS$_4$ (global) modes, which, at the zero-curvature limit, tend to the usual plane waves with exclusively positive frequencies, as far as their analyticity domain has been chosen properly. Respecting the analyticity prerequisite of these modes, the Poincar\'{e} contraction of the dS$_4$ UIR's can be controlled in such a way that they contract merely to the Poincar\'{e} UIR's with positive energy.\footnote{Note that, besides the analyticity prerequisite which is of particular interest in this paper, there is also another possible way out to lift the ambiguity of the dS$_4$ ``rest energy" in terms of the Poincar\'{e} contraction of the representations which is based on a causality de Sitterian semi-group. For this approach, readers are referred to Ref. \cite{Mizony1984}.} This rehabilitates dS$_4$ (generally, dS) relativity from the point of view of the interpretation that can be made by considering the Poincar\'{e} contraction limit. Of course, this argument by no means implies that the energy concept can be defined globally in dS$_4$ (generally, dS) spacetime. As a matter of fact, under Bogoliubov transformations, the given modes at a point $x \in \underline{M}_R$ may turn into modes at some point $x^\prime \in \underline{M}_R$ which their flat limit is of negative energy. See the details in part \ref{Part plane waves}.

\subsection{For comparison: AdS$_4$ UIR's and the Poincar\'{e} contraction}
AdS$_4$ spacetime is most easily described as embedded in $\mathbb{R}^5$ provided with the metric $\eta^{}_{A^{\prime} B^{\prime}} = \mbox{diag}(1,-1,-1,-1,1)$, where the indices $A^{\prime}$ and $B^{\prime}$ take the values $0,1,2,3,5$. [Note that the missing number $4$ is left apart for a possible extension to conformal theories.] Points in $\mathbb{R}^5$, therefore, are denoted by $x=(x^0,x^1,x^2,x^3,x^5)$. In this context, AdS$_4$ spacetime can be visualized as (the covering space of) the connected hyperboloid $(x)^2=\eta^{}_{A^{\prime}B^{\prime}} x^{A^{\prime}} x^{B^{\prime}}= (x^0)^2 - (x^1)^2 - (x^2)^2 - (x^3)^2 + (x^5)^2 = R^2$, where $R$ is the radius of curvature.

The AdS$_4$ relativity group is SO$_0(2,3)$, or its double covering $\mathrm{Sp}(4,\mathbb{R})$, or even its universal covering $\widetilde{\mathrm{SO}_0(2,3)}$.\footnote{The time in this spacetime has a periodic nature, and is proportional to the rotation parameter relevant to the subgroup $SO(2)$. This periodicity of time can be circumvented by considering the covering space of group of motions $\widetilde{\mathrm{SO}_0(2,3)}$. Then, the time is not bounded.} A realization of the associated Lie algebra, quite similar to the dS$_4$ case, is achieved by the linear span of the following ten Killing vectors:
\begin{eqnarray}
K^{}_{A^{\prime}B^{\prime}} = x^{}_{A^{\prime}} \partial^{}_{B^{\prime}} - x^{}_{B^{\prime}} \partial^{}_{A^{\prime}}\,, \;\;\;\;\;\;\; K^{}_{A^{\prime}B^{\prime}} = - K^{}_{B^{\prime}A^{\prime}}\,.
\end{eqnarray}
Here, however, contrary to the dS$_4$ case, there exists one globally timelike Killing vector, i.e., $K_{50}$.

On the quantum level, the above Killing vectors are represented by (essentially) self-adjoint operators in the Hilbert space of (spinor-)tensor valued functions, square integrable according to some invariant inner product of Klein-Gordon type (or else) on the AdS$_4$ manifold (or on respective phase spaces). In the former case, these representations read as: $K_{A^{\prime}B^{\prime}} \mapsto L_{A^{\prime}B^{\prime}} = M_{A^{\prime}B^{\prime}} + S_{A^{\prime}B^{\prime}}$, where the definitions of $M_{A^{\prime}B^{\prime}}$ and $S_{A^{\prime}B^{\prime}}$ are the same as the dS$_4$ case (see Eq. (\ref{LLLLLLL}) and its subsequent discussions). In this context, and on the physical level, three classes of the AdS$_4$ UIR's (besides the trivial representations) appear:
\begin{itemize}
\item{The first class is constituted by those AdS$_4$ UIR's, which belong to the \emph{holomorphic} discrete series and its lower limit. In this class, the spectrum of the ``energy" operator $L_{50}$ has a lower (positive) bound, say $\varsigma>0$, while the spin operator $S_{12}$ admits eigenvalues $-s, \;...\; , s$, with $s \in \mathbb{N}/2$.}
\item{The second class, constituted by the AdS$_4$ \emph{anti-holomorphic} discrete series UIR's, possesses the same features as the first one, but with an upper bound for the spectrum of the ``energy" operator, i.e., $-\varsigma>0$.}
\item{The third class includes those AdS$_4$ UIR's with unbounded energy and spin.}
\end{itemize}

In the first case (i.e., the physically meaningful AdS$_4$ UIR's), the representations are usually denoted by $D_{s,\varsigma}$, with $s \in \mathbb{N}/2$ and $\varsigma \geqslant s+1$ (with the exception of a few of them),\footnote{To be more precise, the parameter $\varsigma$ takes the values $\varsigma \in \mathbb{N}/2$, for the UIR's in the strict sense of the $\mathrm{Sp}(4,\mathbb{R})$ discrete series, and the values $\varsigma \in \mathbb{R}$, while $\varsigma \geqslant s+1$, for the ``discrete" series UIR's of $\widetilde{\mathrm{SO}_0(2,3)}$.} where the parameters $s$ and $\varsigma$ carry the physical meanings of spin and the lowest ``energy", respectively. [Note that, to some extent, the parameter $\varsigma$ plays the role of the parameter $\nu$ in the classification of the dS$_4$ UIR's.] Among the AdS$_4$ UIR's $D_{s,\varsigma}$, one must distinguish between the UIR's with $\varsigma > s+1$ and the two significant limit cases:
\begin{itemize}
\item{The limit scalar cases $D_{0,1}$ and $D_{0,\frac{1}{2}}$ (the latter is named the ``Rac" \cite{Fronsdal}).}
\item{The limit spinorial or tensorial cases $D_{s,s+1}$ and $D_{\frac{1}{2},1}$ (the latter is named the ``Di" \cite{Fronsdal}).}
\end{itemize}

Here, quite similar to the dS$_4$ case, there exist two Casimir operators and their eigenvalues entirely characterize the AdS$_4$ UIR's. These Casimir operators explicitly read:
\begin{eqnarray}
Q^{(1)} &=& - \frac{1}{2} L_{A^{\prime}B^{\prime}} L^{A^{\prime}B^{\prime}}\,,\\
Q^{(2)} &=& - W_{A^\prime} W^{A^\prime}\,, \;\;\;\;\;\;\; W_{^\prime} = - \frac{1}{8} {\cal{E}}_{\tiny{A^\prime B^\prime C^\prime D^\prime E^\prime}} L^{B^\prime C^\prime} L^{D^\prime E^\prime}\,,
\end{eqnarray}
and, according to the parameters introduced above, their respective eigenvalues are:
\begin{eqnarray}\label{Caeigena}
\langle Q^{(1)} \rangle^{}_{\text{\tiny{AdS$_4$}}} &=& s(s+1) + \varsigma(\varsigma - 3)\,,\\
\langle Q^{(2)} \rangle^{}_{\text{\tiny{AdS$_4$}}} &=& -s(s+1)(\varsigma - 1)(\varsigma - 2)\,.
\end{eqnarray}

Proceeding as the previous subsections, we now point out those AdS$_4$ UIR's which contract to the Poincar\'{e} massive UIR's (the so-called AdS$_4$ massive representations) and those AdS$_4$ UIR's which have a massless content (the so-called AdS$_4$ massless representations). For the AdS$_4$ massive cases, the (holomorphic) AdS$_4$ discrete series representations $D_{s,\varsigma}$, with $s\in\mathbb{N}/2$ and $\varsigma >s+1$, are only involved \cite{Evans67}. For them, defining the Poincar\'{e}-Minkowski mass $m$ as:
\begin{eqnarray}\label{minandads}
m = \frac{\hbar \varsigma}{cR}\,,
\end{eqnarray}
the Poincar\'{e} contraction yields \cite{Evans67}:
\begin{eqnarray}
D_{s,\varsigma} \underset{{\underset{\hbar \varsigma/cR = m}{R \rightarrow \infty,\; \varsigma\rightarrow\infty}}}{\longrightarrow} {\cal P}^>_{s,m} \,.
\end{eqnarray}
Evidently, for the AdS$_4$ (generally, AdS) massive cases, quite contrary to the dS$_4$ (generally, dS) relativity (see Eq. (\ref{massive contraction})), the Poincar\'{e} contraction entails no energy ambiguity; the AdS$_4$ (generally, AdS) massive UIR's contract merely to the positive energy Poincar\'{e} massive UIR's (in their respective dimensions). It is also interesting to note that the Poincar\'{e} massive UIR's with negative energy can be obtained when we choose the \emph{anti-holomorphic} discrete series representations $D_{s,-\varsigma}$:
\begin{eqnarray}
D_{s,-\varsigma} \underset{R \rightarrow \infty,\; \varsigma\rightarrow\infty}{\longrightarrow} {\cal P}^<_{s,m} \,.
\end{eqnarray}

On the other hand, for the AdS$_4$ massless (conformal) representations, two different categories appear \cite{Barut,Angelopoulos}:
\begin{itemize}
\item{The massless scalar case, involving the UIR $D_{0, 1}$.}
\item{The spinorial or tensorial cases, involving all UIR's $D_{s, s + 1}$, with $s > 0$, lying at the lower end of the holomorphic AdS$_4$ discrete series.}
\end{itemize}
For the above massless representations, the following extensions hold \cite{Barut,Angelopoulos}:
\begin{eqnarray}
D_{0, 1}\;\; \hookrightarrow \;\;{\cal C}^>_{1,0,0} \;\;\underset{R \rightarrow \infty}{\longrightarrow}\;\; {\cal C}^>_{1,0,0} \;\;\hookleftarrow \;\;{\cal P}^>_{0,0} \,,
\end{eqnarray}
\begin{eqnarray}
    \left. \begin{array}{ccccccc}
    & & {\cal{C}}^{>}_{s+1,s,0} & & {\cal{C}}^{>}_{s+1,s,0} & \hookleftarrow & {\cal{P}}^{>}_{s,0} \\
    D_{s, s+1} & \hookrightarrow & \oplus & \underset{R \rightarrow \infty}{\longrightarrow} & \oplus & & \oplus \\
    & & {\cal{C}}^{>}_{s+1,0,s} & & {\cal{C}}^{>}_{s+1,0,s} & \hookleftarrow & {\cal{P}}^{>}_{-s,0} \,.
    \end{array} \right.
\end{eqnarray}
The arrows `$\hookrightarrow$', as before, denote unique extension. As is obvious from the above, for the AdS$_4$ massless cases, contrary to dS$_4$ relativity, the extensions entail no energy ambiguity. There is, however, an ambiguity concerning helicity. As a matter of fact, the notion of helicity is not defined in AdS$_4$ relativity at all. At the end, we must underline that all other AdS$_4$ representations have either nonphysical Poincar\'{e} contraction limit or do not have Poincar\'{e} contraction limit.

%%%%%%%%%%%%%%%%%%%%%%%%%%%%%%%%%%%%%%%%%%%%%%%%%%%%%%%%%%%%%%%%%%%%%%%%%%%%%%%%%%%%%%%%%%%%%%%%%%%%%%%%%%%%%%%%%%%%%%%%
%%%%%%%%%%%%%%%%%%%%%%%%%%%%%%%%%%%%%%%%%%%%%%%%%%%%%%%%%%%%%%%%%%%%%%%%%%%%%%%%%%%%%%%%%%%%%%%%%%%%%%%%%%%%%%%%%%%%%%%%
%%%%%%%%%%%%%%%%%%%%%%%%%%%%%%%%%%%%%%%%%%%%%%%%%%%%%%%%%%%%%%%%%%%%%%%%%%%%%%%%%%%%%%%%%%%%%%%%%%%%%%%%%%%%%%%%%%%%%%%%
%%%%%%%%%%%%%%%%%%%%%%%%%%%%%%%%%%%%%%%%%%%%%%%%%%%%%%%%%%%%%%%%%%%%%%%%%%%%%%%%%%%%%%%%%%%%%%%%%%%%%%%%%%%%%%%%%%%%%%%%
%%%%%%%%%%%%%%%%%%%%%%%%%%%%%%%%%%%%%%%%%%%%%%%%%%%%%%%%%%%%%%%%%%%%%%%%%%%%%%%%%%%%%%%%%%%%%%%%%%%%%%%%%%%%%%%%%%%%%%%%
%%%%%%%%%%%%%%%%%%%%%%%%%%%%%%%%%%%%%%%%%%%%%%%%%%%%%%%%%%%%%%%%%%%%%%%%%%%%%%%%%%%%%%%%%%%%%%%%%%%%%%%%%%%%%%%%%%%%%%%%
%%%%%%%%%%%%%%%%%%%%%%%%%%%%%%%%%%%%%%%%%%%%%%%%%%%%%%%%%%%%%%%%%%%%%%%%%%%%%%%%%%%%%%%%%%%%%%%%%%%%%%%%%%%%%%%%%%%%%%%%

\part{$1+3$-dimensional dS (dS$_4$) geometry and relativity (QFT)}\label{Part plane waves}
In the previous part, we have employed the dS$_4$ relativity group Sp$(2,2)$ and its UIR's (in the Wigner sense) to provide a robust mathematical structure describing (free) elementary systems in dS$_4$ spacetime on the classical and quantum mechanics levels. In the present part, utilizing the above mathematical materials, on one hand and on the other hand, adopting the Wightman-G\"{a}rding axioms along with analyticity requirements in the complexified pseudo-Riemanian manifold, we proceed with a consistent QFT description of dS$_4$ elementary systems. Technically, to manage the group representations, the whole process is performed in terms of coordinate-independent (global) dS$_4$ plane waves. The latter, defined in their relevant tube domains of the complex dS$_4$ manifold, are the formal analogue of the usual plane waves in Minkowski spacetime. As a matter of fact, letting the curvature go to zero, the dS$_4$ plane waves (for massive cases that such a limit exists; see section \ref{Sec contraction}) precisely reduce to their Minkowskian counterparts in such a way that, as far as the analyticity domain has been chosen properly, no negative frequency mode appears in this limiting process. These waves also provide us with a remarkable (global) dS$_4$ Fourier transform which turns into the ordinary Fourier transform at the null-curvature limit.\footnote{The dS$_4$ (generally, dS$_d$) plane waves are also of great significance in constructing possible models of dS$_4$/CFT$_3$ (generally, dS$_d$/CFT$_{d-1}$) correspondence. For this case, which is beyond the scope of this paper, readers are referred to Refs. \cite{CFT Hollands,CFT Tanhayi,CFT Wrochna}.}

This approach to QFT reading of elementary systems in dS$_4$ spacetime is justified by the fact that the resulting theory has all the properties which one might require from a (free) quantum field on a spacetime with high symmetry. Indeed, a quantum field in this context is, roughly speaking, a distribution $\hat{\Psi}$ on dS$_4$ spacetime $\underline{M}_R$, solution to the field equation, with values in a set of symmetric operators in some inner product (Fock) space ${\mathscr{H}}$, and fulfilling the following physical requirements:
\begin{itemize}
\item{Covariance, in the usual strong sense; there is a unitary representation $\underline{\mathscr{U}}$ of the dS$_4$ group Sp$(2,2)$ on the Fock space of states ${\mathscr{H}}$, such that $\underline{\mathscr{U}}(\underline{g}) \hat{\Psi}(x) \underline{\mathscr{U}}^{-1}(\underline{g}) = \hat{\Psi}(\underline{g}\diamond x)$ for any $\underline{g}\in \mathrm{Sp}(2,2)$ and $x\in \underline{M}_R$. [Borrowing the notations used in the previous part, let $\underline{U}$ be the natural representation of the dS$_4$ group in a Hilbert space ${\cal{H}}$. Then, $\underline{\mathscr{U}}$ denotes the extension of $\underline{U}$ to the Fock space ${\mathscr{H}}$ built on ${\cal{H}}$.]}
\item{Existence of a distinguished state $\Omega \in {\mathscr{H}}$, called vacuum, which is invariant under the representation $\underline{\mathscr{U}}(\underline{g})$, for all $\underline{g} \in \mathrm{Sp}(2,2)$.}
\item{Locality/(anti-)commutativity, with respect to the dS$_4$ causal structure (see section \ref{Sec dS4 causal structure}).}
\item{Positive definiteness of all physical states.}
\end{itemize}
Moreover, this construction allows to control the thermal properties and, as pointed out above, the zero-curvature limit of the fields (if exists!).

Below, mainly considering dS$_4$ scalar fields, we will make this QFT construction explicit. Technically, dS$_4$ scalar fields, in spite of their simplicity, provide us with a complete framework to discuss in detail all essential ingredients of this QFT formulation of dS$_4$ elementary systems. Moreover, it has been already shown that all other (spinor-)tensor fields in dS$_4$ spacetime can be given in terms of a copy of dS$_4$ scalar fields: for the massive and massless spin-$\frac{1}{2}$ fields, see Ref. \cite{Massive/Massless 1/2}, for the massive and massless spin-$1$ fields, respectively, Refs. \cite{Massive 1} and \cite{Massless 1}, for the massive and massless spin-$\frac{3}{2}$ fields, Ref. \cite{Massive 3/2}, for the massive spin-$2$ field, Ref. \cite{Massive 2}, and finally for the massless spin-$2$ field (the dS$_4$ linear quantum gravity), Refs. \cite{Massless 2,Massless 2',Massless 2'',Bamba 1,dS gravity 1,dS gravity 2,DehghaniTakook}. Then, in this sense also, the choice of dS$_4$ scalar fields seems quite natural for our discussions.

\setcounter{equation}{0} \section{DS$_4$ wave equations}\label{Sec wave Eqs.}
The whole QFT construction that is meant to be presented here, as pointed out above, is carried out in terms of dS$_4$ plane waves, defined globally on the dS$_4$ hyperboloid $\underline{M}_R$. For a detailed explanation of these global waves, we first need to introduce the corresponding ``wave equations". This is our task in this section.

Technically, the dS$_4$ plane waves are considered as \emph{eigendistributions}\footnote{See the next section.} of the quadratic Casimir operator $Q^{(1)}$ for the eigenvalues associated with the principal, complementary, and discrete series of the dS$_4$ UIR's.\footnote{Here, it is perhaps worthwhile to recall the arguments subsequent to Eq. (\ref{6666666666}).} Accordingly, with respect to the Dixmier notations (see section \ref{Sec Dixmier}), the wave equations read:
\begin{eqnarray}\label{Wave Eq. Gen}
\Big( Q^{(1)} + p(p+1) + (q+1)(q-2) \Big) \Psi(x) = 0\,, \;\;\;\;\;\;\; x\in \underline{M}_R\,,
\end{eqnarray}
with the specific allowed ranges of values assumed by the parameters $p$ and $q$ for the three series of the dS$_4$ UIR's (see subsections \ref{Subsec Discrete/Dix}, \ref{Subsec Principal/Dix}, and \ref{Subsec Complementary/Dix}).

For the scalar waves (denoted here by $\phi(x)$), which are of particular interest in our study, two different cases appear: those associated with the scalar principal and complementary series which are determined by $p=0$, and the ones associated with the scalar discrete series which are given by $q=0$ (see section \ref{Sec Dixmier}). In both cases, the associated plane waves are characterized by solutions to the following wave equations:
\begin{eqnarray}\label{Wave Eq. scalar}
\Big( Q^{(1)} + \tau(\tau+3) \Big) \phi(x) = 0\,, \;\;\;\;\;\;\; x\in \underline{M}_R\,,
\end{eqnarray}
where the unifying complex parameter $\tau$, for the scalar principal series, takes the values $\tau = -q-1 = -3/2- \mathrm{i} \nu$, with $\nu \in \mathbb{R}$, for the scalar complementary series, the values $\tau = -q-1 = -3/2-\nu$, with $\nu \in \mathbb{R}$ and $0 < |\nu| <3/2$, and for the scalar discrete series, the values $\tau = p-1$ or $ \tau= -p-2$, with $p=1,2,...$ . In a shortcut, we assert that the solutions to the wave equations (\ref{Wave Eq. scalar}) are well defined for all values of $\tau$ with $\mbox{Re}(\tau) < 0$ (see subsection \ref{Subsec generating}). Accordingly, for the scalar discrete series, the values $\tau = p-1$ need to be dropped. Indeed, this series begins with $\tau=-3$, which is precisely where the complementary series ends on its left.

\subsection{Computation in ambient notations}
In ambient space notations, dS$_4$ fields are represented by symmetric (spinor-)tensor fields on the hyperboloid:
\begin{eqnarray}
\underline{M}_R \ni x \;\mapsto\; \Psi(x) \equiv \Psi^{(r)}_{A_1 \;...\; A_n}(x)\,,
\end{eqnarray}
where $r=n+1/2$ ($n$ being the tensorial rank) and $A_1, \;...\;,\; A_n = 0,1,2,3,4$. Note that: (i) For the sake of simplicity, the spinorial index, characterizing the four spinor components, has been omitted. (ii) From now on, whenever is possible, we also omit the tensorial indices.

The (spinor-)tensor fields $\Psi^{(r)}(x)$ are supposed to be homogeneous functions in the $\mathbb{R}^5$-variables $x^A$ with some arbitrarily chosen homogeneity degree $\ell$:
\begin{eqnarray}\label{homoge}
x\cdot\partial \Psi^{(r)}(x) \; \Big( \equiv x^A \frac{\partial}{\partial x^A} \Big) = \ell \Psi^{(r)}(x)\,.
\end{eqnarray}
Trivially, every homogeneous (spinor-)tensor field $\Psi^{(r)}(x)$ of the $\mathbb{R}^5$ variables does not represent a dS$_4$ physical entity. Actually, to make sure that the field $\Psi^{(r)}(x)$ lies in the dS$_4$ tangent spacetime, it also must verify the transversality requirement for all indices $A_1, \;...\;,\; A_n$:
\begin{eqnarray}\label{transversality}
x^{A_i}_{} \Psi^{(r)}_{A_1 \;...\; A_i \;...\; A_n}(x) = (x\cdot \Psi)^{(r-1)}_{A_1 \;...\; \breve{A}_i \;...\; A_n}(x) = 0\,,
\end{eqnarray}
or more concisely $x \cdot \Psi^{(r)}(x) = 0$, where $\breve{A}_i$ means that this index is omitted.

We define here the symmetric, ``transverse projector" $\theta^{}_{AB} = \eta^{}_{AB} + R^{-2} x_A x_B$, verifying $\theta^{}_{AB} x^A = \theta^{}_{AB} x^B = 0$. [This projector is indeed the transverse form of the dS$_4$ metric in ambient space notations; this point will be clarified in the next subsection.] We employ $\theta^{}_{AB}$ to construct transverse entities, such as $\bar{\partial}_A = \theta^{}_{AB}\partial^B = \partial_A + R^{-2} x_A x\cdot\partial$, called transverse derivative, for which we have:
\begin{eqnarray}
\bar{\partial}_A x_B = \theta^{}_{AB}\,,\;\;\;\;\;\;\; \mbox{and} \;\;\;\;\;\;\; \bar{\partial}_A (x)^2=0\,.
\end{eqnarray}
The latter identity shows that the differential operator $\bar{\partial}$ commutates with $(x)^2$, which means that $\bar{\partial}$ is intrinsically defined on the dS$_4$ hyperboloid $(x)^2 = - R^{2}$. Considering the above, for a general (spinor-)tensor field $\Psi^{(r)}_{A_1 \;...\; A_n}(x)$, the operator $\mathfrak{T}$ with the following definition: \begin{eqnarray}
(\mathfrak{T} \Psi)^{(r)}_{A_1 \;...\; A_n}(x) = \Bigg( \prod_{i=1}^n \theta^{B_i}_{A_i} \Bigg) \Psi^{(r)}_{B_1 \;...\; B_n}(x)\,,
\end{eqnarray}
guarantees the transversality in each tensorial index; since the degree of homogeneity of $\theta^{}_{AB}$ is zero, the above definition does not change the degree of homogeneity of the given (spinor-)tensor field.

We now turn to the explicit description of the dS$_4$ quadratic Casimir operator in ambient space notations. As already mentioned (see section \ref{Sec Dixmier}), this operator is written in terms of the self-adjoint operators $L_{AB}$ associated with each of the ten Killing vectors (\ref{Killing dS4}). In the Hilbert space of symmetric (spinor-)tensors $\Psi^{(r)}_{A_1 ...}(x)$ on $\underline{M}_R$, square integrable according to some invariant inner (Klein-Gordon type) product, the generator representatives $L_{AB}^{(r)}$ are defined by \cite{Moylan}:
\begin{eqnarray}
L_{AB}^{(r)} = M_{AB} + S_{AB}^{(n)} + S^{(\frac{1}{2})}_{AB}\,,
\end{eqnarray}
where the orbital part is given by $M_{AB} = - \mathrm{i} (x_A \partial_B - x_B \partial_A ) = - \mathrm{i} (x_A \bar\partial_B - x_B \bar\partial_A )$, while the spinorial parts $S_{AB}^{(n)}$ and $S^{(\frac{1}{2})}_{AB}$ respectively act on the tensorial indices as:
\begin{eqnarray}
S_{AB}^{(n)}\Psi^{(r)}_{A_1 \;...\; A_n} = - \mathrm{i} \sum_{i=1}^{n}\Big(\eta^{}_{AA_i}\Psi^{(r)}_{A_1 \;...\; (A_i\rightarrow B) \;...\; A_n} - (A \rightleftharpoons B)\Big)\,,\nonumber
\end{eqnarray}
and on the spinorial indices by $S^{(\frac{1}{2})}_{AB} = -\frac{\mathrm{i}}{4}[\gamma_A,\gamma_B]$ (recall from section \ref{Sec Sp(2,2) group} that the five $4\times 4$-matrices $\gamma_A$ generate the Clifford algebra). The action of the quadratic Casimir operator on a given spinor-tensor field $\Psi^{(r)}_{A_1 \;...\; A_n}(x)$ then reads:
\begin{eqnarray}\label{777777777777}
Q_r^{(1)} \; \Psi^{(r)} (x) = - \frac{1}{2} L^{(r)}_{AB} L^{(r)AB} \; \Psi^{(r)} (x) &=& -\frac{1}{2} \Big( M_{AB} + S_{AB}^{(n)} + S^{(\frac{1}{2})}_{AB} \Big) \Big( M^{AB} +  S_{}^{(n)AB} + S_{}^{(\frac{1}{2})AB} \Big) \Psi^{(r)} (x) \nonumber\\
&=& -\frac{1}{2} \Big( M_{AB}M^{AB} + S_{AB}^{(n)}S^{(n)AB} + S^{(\frac{1}{2})}_{AB} S^{(\frac{1}{2})AB} \Big) \Psi^{(r)} (x)\nonumber\\
&& - \Big( M_{AB}  S_{}^{(n)AB} + M_{AB}  S_{}^{(\frac{1}{2})AB} + S^{(\frac{1}{2})}_{AB} S_{}^{(n)AB} \Big) \Psi^{(r)} (x)\,,
\end{eqnarray}
with:
\begin{eqnarray}\label{Q0}
\frac{1}{2} M_{AB}M^{AB} \; \Psi^{(r)} \; \Big( = - Q_0^{(1)} \; \Psi^{(r)} \Big) &=& R^{2} \bar\partial^2\,, \nonumber\\
\frac{1}{2} S_{AB}^{(n)}  S_{}^{(n)AB} \; \Psi^{(r)} &=& n(n+3) \Psi^{(r)} - 2\Sigma_2 \eta \Psi^{(r)\prime}\,, \nonumber\\
\frac{1}{2} S_{AB}^{(\frac{1}{2})}  S_{}^{(\frac{1}{2})AB} \; \Psi^{(r)} &=& \frac{5}{2} \Psi^{(r)}\,, \nonumber\\
M_{AB}  S_{}^{(n)AB} \; \Psi^{(r)} &=& 2\Sigma_1 \partial x\cdot \Psi^{(r)} - 2\Sigma_1 x \partial\cdot \Psi^{(r)} - 2n\Psi^{(r)}\,, \nonumber\\
M_{AB}  S_{}^{(\frac{1}{2})AB} \; \Psi^{(r)} &=& - \frac{\mathrm{i}}{2} \gamma_A \gamma_B M^{AB} \Psi^{(r)}\,, \nonumber\\
S^{(\frac{1}{2})}_{AB} S_{}^{(n)AB} \; \Psi^{(r)} &=& n \Psi^{(r)} - \Sigma_1 \gamma \big( \gamma \cdot \Psi^{(r)} \big)\,.
\end{eqnarray}
Note that: (i) The subscript `$r$' in $Q_r^{(1)}$ refers to the fact that the carrier space is constituted by rank $r$ spinor-tensors. (ii) $\Sigma_m$ designates the symmetrizer of the tensor product of two symmetric (spinor-)tensors $\Phi$ and $\Psi$ of, respectively, tensorial rank $m$ and $n-m$, where $m \leqslant \lfloor n/2 \rfloor$, based upon which the components of the symmetrized tensor product are given by:
\begin{eqnarray}
\big( \Sigma_m \Phi\Psi \big)^{}_{A_1 \;...\; A_n} = \sum_{i_1<i_2< \;...\; <i_m} \big(\Phi^{}_{A_{i_1}A_{i_2} \;...\; A_{i_m}}\big) \big(\Psi^{}_{A_1, \;...\; \breve{A}_{i_1} \;...\; \breve{A}_{i_2} \;...\; \breve{A}_{i_m} \;...\; A_n}\big)\,,
\end{eqnarray}
where $\breve{A}_{i_1},\breve{A}_{i_2}$, and $\breve{A}_{i_m}$ mean that these terms are omitted. (iii) The trace, associated with the tensorial part, of the spinor-tensor field $\Psi^{(r)}$ of tensorial rank $n$ is denoted by $\Psi^{(r)\prime}$, which is a symmetric spinor-tensor field of tensorial rank $n-2$, given by:
\begin{eqnarray}
\Psi^{(r)\prime}_{A_1 \;...\; A_{n-2}} = \eta^{A_{n-1}A_n}_{} \Psi^{(r)}_{A_1 \;...\; A_{n-2}A_{n-1}A_n}\,.
\end{eqnarray}
(iv) The operators $Q_r^{(1)}$ and $L^{(r)}_{AB}$ commutate with $(x)^2$. Therefore, they are intrinsically defined on the dS$_4$ hyperboloid $(x)^2 = -R^{2}$.

Finally, taking all the above identities into account, we can explicitly rewrite Eq. (\ref{777777777777}) as:
\begin{eqnarray}\label{Q1}
Q_r^{(1)} \; \Psi^{(r)} (x) &=& \Big( Q_0^{(1)} + \frac{\mathrm{i}}{2} \gamma_A \gamma_B M^{AB} - n(n+2) - \frac{5}{2} \Big) \Psi^{(r)} \nonumber\\
&& - 2\Sigma_1 \partial x\cdot \Psi^{(r)} + 2\Sigma_1 x \partial\cdot \Psi^{(r)} + \Sigma_1 \gamma \big( \gamma \cdot \Psi^{(r)} \big) + 2\Sigma_2 \eta \Psi^{(r)\prime}\,.
\end{eqnarray}

Now, substituting Eq. (\ref{Q1}) into (\ref{Wave Eq. Gen}), we get the explicit form of the wave equations (\ref{Wave Eq. Gen}) in terms of ambient notations. The very point that must be noticed here is that clearly due to the form of $Q_r^{(1)}$ in ambient notations (see Eq. (\ref{Q1})), for a given (spinor-)tensor field, the space of solutions to the relevant wave equation contains some invariant subspaces which must be eliminated if one wishes to be left with the space that solely carries the corresponding dS$_4$ UIR's. In this sense, the aforementioned list of requirements (the homogeneity and transversality) for a given (spinor-)tensor field $\Psi^{(r)}_{A_1 \;...\; A_n}(x)$ must be supplemented as follows:
\begin{eqnarray}\label{conditions UIR's}
x \cdot \partial \Psi^{(r)} &=& 0\,, \;\;\;\;\;\;\; \mbox{homogeneity ,} \nonumber\\
x \cdot \Psi^{(r)} &=& 0\,, \;\;\;\;\;\;\; \mbox{transversality ,} \nonumber\\
\partial \cdot \Psi^{(r)} &=& 0\,, \;\;\;\;\;\;\; \mbox{divergencelessness ,} \nonumber\\
\gamma \cdot \Psi^{(r)} &=& 0\,, \;\;\;\;\;\;\; \mbox{tracelessness (associated with the spinorial part) conditions}\,.
\end{eqnarray}
Note that: (i) Here, for the sake of simplicity, we set the degree of homogeneity $\ell =0$. Accordingly, for instance, regarding formulas that will be given in the next subsection, one can easily show that the d'Alembertian operator $\square_R \equiv \nabla_\mu \nabla^\mu$ on dS$_4$ spacetime ($\nabla_\mu$, with $\mu=0,1,2,3$, being the covariant derivative given in local (intrinsic) coordinates) coincides with its counterpart $\square_5 \equiv \partial^2$ on $\mathbb{R}^5$. (ii) The transversality and divergencelessness conditions together yield $\Psi^{(r)\prime} = 0$.

Considering the above conditions along with the relations leading to Eq. (\ref{Q1}), the dS$_4$ wave equations for spinor-tensor fields $\Psi^{(r=n+1/2)}_{A_1 \;...\; A_n}(x)$ (see Eq. (\ref{Wave Eq. Gen})), in ambient notations, take the form:
\begin{eqnarray}\label{Wave Eq. Gen ambient}
\Big( Q_0^{(1)} + \frac{\mathrm{i}}{2} \gamma_A \gamma_B M^{AB} - n(n+2) - \frac{5}{2} + [p(p+1) + (q+1)(q-2)] \Big) \Psi^{(r)}_{A_1 \;...\; A_n}(x) = 0\,,
\end{eqnarray}
for tensor fields $\Psi^{(r=n)}_{A_1 \;...\; A_n}(x)$:
\begin{eqnarray}\label{Wave Eq. Gen-ten ambient}
\Big( Q_0^{(1)} - n(n+1) + [p(p+1) + (q+1)(q-2)] \Big) \Psi^{(n)}_{A_1 \;...\; A_n}(x) = 0\,,
\end{eqnarray}
and particulary, for scalar fields $\Psi^{(r=0)}(x) \equiv \phi(x)$:
\begin{eqnarray}\label{Wave Eq. scalar ambient}
\Big( Q_0^{(1)} + \tau(\tau+3) \Big) \phi(x) = 0\,,
\end{eqnarray}
where the allowed ranges of $\tau$, corresponding to the three series of the dS$_4$ scalar UIR's, have been already listed below Eq. (\ref{Wave Eq. scalar}).

\subsection{Link to intrinsic coordinates}\label{Subsec Link to intrinsic coordinates}
Since, often in the literature concerning dS$_4$ (generally, dS) QFT, the fields are presented in terms of local (intrinsic) coordinates, it would be convenient here to demonstrate the link between the intrinsic and ambient coordinates.

The intrinsic field $\Psi^{(r)}_{{\mu^{}_1} \;...\; {\mu^{}_n}}(X)$ is locally characterized by $\Psi^{(r)}_{A_1 \;...\; A_n}(x)$ through the following identity:
\begin{eqnarray}\label{tt}
\Psi^{(r)}_{{\mu^{}_1} \;...\; {\mu^{}_n}}(X) = x^{A_1}_{\,\,,\,\mu_1} \;...\; x^{A_n}_{\,\,,\,\mu_n} \; \Psi^{(r)}_{A_1 \;...\; A_n}\big(x(X)\big)\,,
\end{eqnarray}
where $x^{A_i}_{\,\,,\,\mu_i} = \partial x^{A_i}/\partial X^{\mu_i}$ and $X^\mu$'s, with $\mu=0,1,2,3$, refer to the four local spacetime coordinates for the dS$_4$ hyperboloid $\underline{M}_R$. The corresponding dS$_4$ metric $g_{\mu\nu}$ is determined by inducing the natural metric of $\mathbb{R}^5$ on $\underline{M}_R$:
\begin{eqnarray}
\mathrm{d} s^2 = \eta^{}_{AB} \mathrm{d} x^A_{} \mathrm{d} x^B_{} \Big|_{(x)^2=-R^{2}} = g_{\mu\nu} \mathrm{d} X^\mu \mathrm{d} X^\nu\,.
\end{eqnarray}
Considering Eq. (\ref{tt}), it follows that $\theta^{}_{AB}$ is the only symmetric and transverse tensor which is connected to the dS$_4$ metric; $g_{\mu\nu}= x^A_{\,\,,\,\mu} x^B_{\,\,,\,\nu} \theta^{}_{AB}$. In this context, the covariant derivatives are transformed as:
\begin{eqnarray}
\nabla_\mu \nabla_\nu \cdot\cdot\cdot \nabla_\rho \; \Psi^{(r)}_{\lambda_1 \;...\; \lambda_n} = x^A_{\,\,,\,\mu} x^B_{\,\,,\,\nu} \;...\; x^C_{\,\,,\,\rho} \;\; x^{D_1}_{\,\,,\,\lambda_1} \;...\; x^{D_n}_{\,\,,\,\lambda_n} \; \mathfrak{T}\bar{\partial}^{}_A \mathfrak{T}\bar{\partial}^{}_B \;...\; \mathfrak{T}\bar{\partial}^{}_C \; \Psi^{(r)}_{D_1 \;...\; D_n}\,.
\end{eqnarray}

For scalar fields, which are of particular interest in this paper, the d'Alembertian operator reads:
\begin{eqnarray}
\square_R \phi = g^{\mu\nu} \nabla_\mu \nabla_\nu \phi &=& g^{\mu\nu} x^A_{\,\,,\,\mu} x^B_{\,\,,\,\nu} \Big( \bar\partial_A \bar\partial_B - R^{-2} x^{}_B \bar\partial_A \Big)\phi \nonumber\\
&=& \theta^{AB} \Big( \bar\partial_A \bar\partial_B - R^{-2} x^{}_B \bar\partial_A \Big)\phi = \bar\partial^2 \; \phi\,.
\end{eqnarray}
Considering this equation along with the first identity given in (\ref{Q0}), we get $Q^{(1)}_0 = - R^{2} \square_R$. Then, the scalar-wave equations (\ref{Wave Eq. scalar ambient}), by adjusting $-\tau(\tau+3)= R^2 m^2_R$ ($m_R^{}$ being a ``mass''\footnote{The name ``mass'' is purely formal here, being in general not related to at rest energy in the Poincar\'e symmetry sense.} term ($\hbar = c = 1$)), yield the ordinary dS$_4$ Klein-Gordon-like field equations $\big( \Box_R + m^2_R \big)\phi(x) = 0$.

\setcounter{equation}{0} \section{Plane-wave type solutions}\label{Sec plane wave}
From now on, as already mentioned, to avoid technical (but not conceptual) complications, we merely stick to dS$_4$ scalar fields. According to Refs. \cite{GazeauPRL,Bros 2point func}, there is a continuous family of simple solutions, called dS$_4$ plane waves, to the scalar-wave equations (\ref{Wave Eq. scalar ambient}) defined by:
\begin{eqnarray}\label{dSplwa}
\phi^{}_{\tau,\xi}(x) = \Big(\frac{x\cdot\xi}{R}\Big)^\tau,
\end{eqnarray}
where $x\in \underline{M}_R$ and $\xi$ lies in the null-cone $C$ in $\mathbb{R}^5$:
\begin{eqnarray}
C = \Big\{ \xi\in \mathbb{R}^5 \; ; \; (\xi)^2 =0 \Big\}\,.
\end{eqnarray}
The plane waves (\ref{dSplwa}), as functions of $\xi$ in the null-cone $C$, are homogeneous with degree of homogeneity $\tau$. In this sense, they can be completely characterized by specifying their values on a well-chosen curve (the orbital basis) $\daleth$ of $C$. [Here, it is worthwhile noting that: (i) On the dS$_4$ submanifold $\underline{M}_R$ ($\subset\mathbb{R}^5$) defined by $x\cdot x = -R^2$, with $R$ being constant, these waves are also homogeneous with degree of homogeneity $\tau$. (ii) As functions of $\mathbb{R}^5$, they are homogeneous, but with degree of homogeneity zero, since in this case $R$ should be viewed as a function of $x$: $R(x)= - \sqrt{- x\cdot x}$.]

Here, in a shortcut and without getting involved in mathematical detail, we would like to highlight two critical features of the dS$_4$ plane waves (\ref{dSplwa}). First, these waves, as \emph{functions} on the dS$_4$ manifold $\underline{M}_R$, are only locally defined on connected open subsets of $\underline{M}_R$, because they are singular on some specific lower dimensional subsets of the manifold, for instance, on the spatial boundaries given by $x^0 = \pm x^4 \Leftrightarrow (x^1)^2 + (x^2)^2 + (x^3)^2 = R^{2}$; to see the latter point, it is sufficient to associate with these boundaries, respectively, $\xi=(\xi^0=\pm\xi^4,0,0,0,\xi^4) \in C$, for which we get $x\cdot\xi = 0$ (recall that $\tau$ is generally a complex number, with $\mbox{Re}(\tau)<0$). Moreover, these waves, as \emph{functions} on $\underline{M}_R$, are multi-valued, since $x\cdot\xi$ can take negative values, as well. In order to get a single-valued, global definition of these waves, they need to be considered as \emph{distributions}, strictly speaking, as the boundary values of analytic continuations of the solutions (\ref{dSplwa}) to suitable domains in the complexified dS$_4$ manifold $\underline{M}_R^{(\mathbb{C})}$. It turns out that the minimal domains of analyticity, which lead to such a single-valued, global definition of the dS$_4$ plane waves, are the \emph{forward} and \emph{backward tubes} of $\underline{M}_R^{(\mathbb{C})}$.

Second, the plane waves (\ref{dSplwa}) are not square integrable with respect to the Klein-Gordon inner product. However, they can be considered as generating functions for physically meaningful dS$_4$ entities, like square-integrable eigenfunctions of the dS$_4$ quadratic Casimir operator. [Such square-integrable eigenfunctions, being constructed through continuous superpositions of these waves (by varying $\xi$ in $C$), generate (projective) Hilbert spaces carrying the dS$_4$ UIR's. They actually generate the spacetime realization of those Hilbert spaces that have been already presented in the previous part in the context of the $\mathbb{S}^3$ realization of the representations.] In this sense, in dS$_4$ (generally, dS) relativity, the above plane waves behave quite analogue to the usual ones in Minkowskian or Galilean quantum mechanics, where, by superposition of nonsquare-integrable plane waves, one can build up physical wave functions (wave packets).

In the following subsections, strongly inspired by Refs. \cite{GazeauPRL,Bros 2point func,sigma}, we will elaborate these points in details, respectively. Then, following Ref. \cite{Garidi zero curvature limit}, we will end our discussions in this section by investigating the behavior of the dS$_4$ plane waves in the flat (Minkowskian) limit. We will show that, as far as the analyticity domain has been chosen properly, no negative frequency plane wave appears in this limiting process, whatever the point around which the flat limit is calculated.

\subsection{A global definition: dS$_4$ plane waves in their tube domains}\label{Subsec tube}
Before getting involved with the definition of dS$_4$ plane waves in their tube domains, we need to elaborate some geometrical notions relevant to the dS$_4$ complex hyperboloid:
\begin{eqnarray}\label{complex dS}
\underline{M}_R^{(\mathbb{C})} \equiv \Big\{ z=x+ \mathrm{i} y \in {\mathbb{C}}^5 \;;\; (z)^2 = \eta^{}_{AB} z^A_{} z^B_{} = (z^0)^2 - (z^1)^2 - (z^2)^2 - (z^3)^2 - (z^4)^2 = -R^2 \Big\}\,,
\end{eqnarray}
where ${\mathbb{C}}^5$ stands for the ambient complex Minkowski spacetime. We begin with the dS$_4$ complex hyperboloid $\underline{M}_R^{(\mathbb{C})}$ itself, which equivalently can also be realized by the following set:
\begin{eqnarray}
\underline{M}_R^{(\mathbb{C})} \equiv \Big\{ (x,y) \in {\mathbb{R}}^5 \times {\mathbb{R}}^5 \;;\; (x)^2 - (y)^2 = -R^2, \; x\cdot y=0 \Big\}\,.
\end{eqnarray}
In order to provide a clear visualization of $\underline{M}_R^{(\mathbb{C})}$, we first point out that the identity $(x)^2 - (y)^2 = -R^2$ designates the following distinguished sets of points $(x,y)$ in ${\mathbb{R}}^5 \times {\mathbb{R}}^5$:
\begin{itemize}
\item {$(x)^2 < 0$, or equivalently $(y)^2 < R^2$,}
\item{$x=0$ (that is, $x=(0,0,0,0,0)$), or equivalently $(y)^2 = R^2$, and finally}
\item{$(x)^2 \geqslant 0$ ($x\neq0$), or equivalently $(y)^2 \geqslant R^2$.}
\end{itemize}
The latter case, however, has no intersection with $x\cdot y=0$, and therefore, must be dropped. To make the latter point apparent, utilizing the quaternionic notations introduced in part \ref{Part II}, we denote the points $x$ and $y$ belonging to the latter case by $(x^0,\textbf{x})$ and $(y^0,\textbf{y})$, respectively. On this basis, we have:\footnote{Note that the dot product of two quaternions is exactly their usual vector dot product.}
\begin{eqnarray}\label{xy1}
\mbox{for} \;\; x^0 y^0 > 0\;;\;\; x\cdot y = |x^0| |y^0| - \textbf{x}\cdot\textbf{y}\,, \;\;\;\;\;\;\; \mbox{and} \;\;\;\;\;\;\; \mbox{for} \;\; x^0 y^0 < 0 \;;\;\; x\cdot y = -|x^0| |y^0| - \textbf{x}\cdot\textbf{y}\,.
\end{eqnarray}
Then, according to the Cauchy-Schwarz inequality (which implies that $- |\textbf{x}| |\textbf{y}| \leqslant\textbf{x}\cdot\textbf{y}\leqslant |\textbf{x}| |\textbf{y}|$) and the facts that $(x)^2 \geqslant 0$ (i.e., $|x^0|\geqslant |\textbf{x}|$) and $(y)^2 \geqslant R^2>0$ (i.e., $|y^0| > |\textbf{y}|$), we have:
\begin{eqnarray}\label{xy2}
\mbox{for} \;\; x^0 y^0 > 0\;;\;\; x\cdot y \geqslant |x^0| |y^0| - |\textbf{x}| |\textbf{y}| > 0\,, \;\;\;\;\;\;\; \mbox{and} \;\;\;\;\;\;\; \mbox{for} \;\; x^0 y^0 < 0 \;;\;\; x\cdot y \leqslant - |x^0| |y^0| + |\textbf{x}| |\textbf{y}| < 0\,,
\end{eqnarray}
which both inequalities explicitly reveal that $x\cdot y\neq 0$. Accordingly, dropping the third case (with $(x)^2 \geqslant 0$, $x\neq0$), $\underline{M}_R^{(\mathbb{C})}$ can be visualized as the set of points $(x,y) \in {\mathbb{R}}^5 \times {\mathbb{R}}^5$ with $(x)^2 < 0$ or $x=0$ (equivalently $(y)^2 \leqslant R^2$).

We also need to remind the notions of forward and backward tubes in ${\mathbb{C}}^5$, respectively, denoted by $T^+$ and $T^-$. By definition, $T^\pm = \mathbb{R}^5 + \mathrm{i} \interior{\underline{V}}^\pm$, where the domains $\interior{\underline{V}}^\pm \equiv \big\{y \in\mathbb{R}^5 \;;\; (y)^2 > 0,\; y^0 \gtrless 0 \big\}$ stem from the causal structure on $\underline{M}_R$ (see section (\ref{Sec dS4 causal structure})). Indeed, $T^+$ and $T^-$ can be visualized as the set of points $z=x+ \mathrm{i} y \in \mathbb{C}^5$ ($(x,y) \in {\mathbb{R}}^5 \times {\mathbb{R}}^5$), such that $(y)^2 > 0$, with $y^0 > 0$ and $y^0 < 0$, respectively. [Here, in a shortcut, it is convenient to point out that the tubes $T^\pm$ are the (standard) analyticity domains of $1+4$-dimensional Minkowskian quantum fields, which verify the positivity of the spectrum of the energy operator (see section \ref{Sec QFT}).]

We finally define the following open subsets of $\underline{M}_R^{(\mathbb{C})}$:
\begin{eqnarray}\label{forward and back tube dS}
{\cal{T}}^+ = T^+ \cap \underline{M}_R^{(\mathbb{C})}\,, \;\;\;\;\;\;\; {\cal{T}}^- = T^- \cap \underline{M}_R^{(\mathbb{C})}\,,
\end{eqnarray}
where ${\cal{T}}^+$ and ${\cal{T}}^-$ are respectively called forward and backward tubes of $\underline{M}_R^{(\mathbb{C})}$. Considering the above, they can be viewed as the set of points $z=x+ \mathrm{i} y \in {\mathbb{C}}^5$, with $-R^2 < (x)^2 < 0$ or $x=0$ (equivalently $0 < (y)^2 \leqslant R^2$), while $y^0 > 0$ and $y^0 < 0$, respectively. Here, one must notice that the tubes ${\cal{T}}^\pm$ are indeed domains and \emph{tuboids}\footnote{A tuboid is a domain bordered by a set with real coordinates.} above $\underline{M}_R$ in $\underline{M}_R^{(\mathbb{C})}$, from which, in a well-defined way, one can take \emph{the boundary value on $\underline{M}_R$ of analytic functions in the distribution sense}\footnote{Let $\phi(z=x+ \mathrm{i} y)$ be an analytic function in a given local tube $\Theta = \Delta + \mathrm{i} \Xi$. Then, for all $f(x)$ in $\mathfrak{D}(\Delta)$ (the latter being the space of infinitely differentiable functions with \emph{compact support} on the $\mu$-measure space $\Delta$), the sequence of distributions $\big\{ \phi^{}_y \;;\; y\in\Xi \big\}$ defined in $\mathfrak{D}^\prime(\Delta)$ by:
\begin{eqnarray}
\phi_y(f) = \int_\Delta \phi(x+ \mathrm{i} y) f(x) \; \mathrm{d}\mu(x)\,, \nonumber
\end{eqnarray}
is weakly convergent when $y$ goes to zero in $\Xi$; this limit ($\phi_y |_{y\in \Xi, \; y\rightarrow 0}$) gives a distribution in $\mathfrak{D}^\prime(\Delta)$, called boundary value of $\phi(z)$ on $\Delta$ from the local tube $\Theta$. To see more detailed discussions on this topic, one can refer to Refs. \cite{Bros 2point func,Neeb}. Moreover, to get acquainted with the notion of \emph{distribution} and its relevant objects, readers are referred to Ref. \cite{Streater}. [As a final remark in this footnote, we would like to point out that, by definition, the support of a function $f$, denoted in the sequel by $\mbox{supp} (f)$, is the subset of the domain containing those elements which are not mapped to zero. If the domain of $f$ is a topological space, the support of $f$ is defined as the smallest closed set containing all points that are not mapped to zero. Functions with compact support on a topological space are those whose closed support is a compact subset of the space.]} \cite{Bros 2point func}.

Now, we are in a position to encounter our main task in this subsection, that is, presenting a single-valued, global definition of the dS$_4$ plane waves. By analytic continuation of the plane waves (\ref{dSplwa}) to the dS$_4$ complex hyperboloid $\underline{M}_R^{(\mathbb{C})}$, the obtained complexified waves $(z\cdot\xi/R)^\tau$ are defined globally and single valued, provided that $z$ varies in ${\cal{T}}^+$ or ${\cal{T}}^-$ and $\xi$ lies in the future null-cone $C^+ \equiv \big\{ \xi\in \mathbb{R}^5 \; ; \; (\xi)^2 =0, \; \xi^0 > 0 \big\}$, because then the imaginary part of $(z\cdot\xi/R)$ has a fixed sign and moreover $z\cdot\xi \neq 0$. To see the point, let $z = x+ \mathrm{i} y \in {\cal{T}}^+$ or ${\cal{T}}^-$ (that is, $y=(y^0,\textbf{y}) = \mbox{Im}(z)$ verify $0 < (y)^2 \leqslant R^2$, with $y^0>0$ and $y^0<0$, respectively) and, again using the quaternionic notations, $\xi = (\xi^0, \boldsymbol{\xi}) \in C^+$ (that is, $\xi^0 > 0$ and $|\boldsymbol{\xi}|=\xi^0$). Taking parallel steps to the ones that led to Eqs. (\ref{xy1}) and (\ref{xy2}), one can show that:
\begin{eqnarray}
\left \{ \begin{array}{rl} y \cdot \xi >0 \;\;\;\;\;\;\; \mbox{for} \;\;\; z\in {\cal{T}}^+ \,,\vspace{1mm}\\
\vspace{1mm} y \cdot \xi <0 \;\;\;\;\;\;\; \mbox{for} \;\;\; z\in {\cal{T}}^- \,.\end{array}\right.
\end{eqnarray}
Therefore, for all $z=x+ \mathrm{i} y$ in the domains ${\cal{T}}^\pm$ and $\xi\in C^+$, we have $z\cdot\xi\neq 0$; no matter $x\cdot\xi\neq0$ or not, the term $y\cdot\xi$ is always nonzero. On the other hand, for an arbitrary value of $\tau$ and a fixed sign of the imaginary part of $(z\cdot\xi/R)$, a single-valued determination of $(z\cdot\xi/R)^\tau$ is given by:
\begin{eqnarray}
\Big(\frac{z\cdot\xi}{R}\Big)^\tau = \exp \Big(\tau \Big[\mathrm{i} \; \mbox{arg}\Big(\frac{z\cdot\xi}{R}\Big) + \log\Big|\frac{z\cdot\xi}{R}\Big| \Big] \Big)\,, \;\;\;\;\;\;\; \mbox{arg} \Big(\frac{z\cdot\xi}{R}\Big) \in \; ]-\pi,\pi[\,.
\end{eqnarray}
Accordingly, one can define the single-valued, global dS$_4$ plane waves as the boundary values, in the sense of distributions, of analytic continuation to the forward (${\cal{T}}^+$) or backward (${\cal{T}}^-$) tube of the waves (\ref{dSplwa}):
\begin{eqnarray}\label{Fourier x.xi}
\phi^{\pm}_{\tau,\xi}(f) &=& \int_{\underline{M}_R} c \; \Big( \frac{(x+ \mathrm{i} y)\cdot\xi}{R} \Big)^\tau \Big|_{\xi \in C^+,\; y \in \interior{\underline{V}}^\pm, \; y \rightarrow 0} \; f(x) \; \mathrm{d}\mu(x) \nonumber\\
&=& \int_{\underline{M}_R} \underbrace{\Bigg( c \; \Big[ \vartheta\Big(\frac{x\cdot\xi}{R}\Big) + \vartheta\Big(-\frac{x\cdot\xi}{R}\Big) e^{\pm \mathrm{i} \pi\tau} \Big] \; \Big|\frac{x\cdot\xi}{R}\Big|^\tau \Bigg)}_{\equiv \phi^{\pm}_{\tau,\xi}(x)} \; f(x) \; \mathrm{d}\mu(x)\,,
\end{eqnarray}
where, in the above ``Fourier transform", $f(x) \in \mathfrak{D}(\underline{M}_R)$ ($\mathfrak{D}(\underline{M}_R)$ being the space of infinitely differentiable functions with compact support on $\underline{M}_R$), while $\mathrm{d}\mu(x)$ stands for the invariant measure on $\underline{M}_R$, $\vartheta$ for the Heaviside function, and $c$ for a real-valued constant (this constant, as we will discuss in section \ref{Sec QFT}, is fixed by applying the local Hadamard condition on the corresponding two-point function).

\subsection{Precision on orbital basis of the cone}\label{Subsec orbial basis}
We now elaborate on the notion of orbital basis $\daleth$ of the future null-cone $C^+$. Let $x_e^{}$ and $\underline{{\cal{S}}}(x_e^{})$ respectively denote a unit vector (in the sense $|(x_e)^{2}| =1$) in $\mathbb{R}^5$ and its stabilizer subgroup in Sp$(2,2)$. In the present context, two types of orbits are of interest:
\begin{itemize}
\item{If $x_e^{} \in \interior{\underline{V}}^+$, then $\daleth$ would be the section of $C^+$ characterized by a hyperplane of the form $x_e^{}\cdot\xi = a$ ($a>0$), i.e., an orbit (of spherical type) of the stabilizer subgroup $\underline{{\cal{S}}}(x_e^{}) \sim SO(4)$.\footnote{We recall that under the action of Sp$(2,2)$ in $\mathbb{R}^5$, the region $\interior{\underline{V}}^+$ is divided into a union of mutually disjoint orbits of different radii, for which, according to the Cartan decomposition of Sp$(2,2)$, the subgroup $SO(4) \sim \mathrm{SU}(2)\times \mathrm{SU}(2)$, isomorphic to the maximal compact subgroup of Sp$(2,2)$, plays the role of stabilizer subgroup (see subsection \ref{Sec Cartan dS4}).} To see the point, let us set $x_e^{}=(1,0,0,0,0)$. With this choice of the unit vector $x_e^{} \in \interior{\underline{V}}^+$, the orbit $\daleth \equiv \daleth_0$ reads:
    \begin{eqnarray}\label{aaaaaaa}
    \daleth_0 = \Big\{ \xi \in C^+ \;;\; \xi^0 = a \Big\} = \Big\{ \xi \in C^+ \;;\; (\xi^1)^2+(\xi^2)^2+(\xi^3)^2+(\xi^4)^2 = a^2 \Big\}\,.
    \end{eqnarray}}
\item{If $(x_e^{})^2=-1$, $\daleth$ represents the union of the sections of $C^+$ characterized by two hyperplanes of the forms $x_e^{}\cdot\xi = \pm b$ ($b>0$), i.e., the union of two hyperboloid sheets, which are orbits of the stabilizer subgroup $\underline{{\cal{S}}}(x_e^{}) \sim \mathrm{SO}_0(1,3)$.\footnote{We recall that under the action of Sp$(2,2)$ in $\mathbb{R}^5$, the exterior of the cone $(x)^2=0$, which here $x_e^{}$ belongs to, is divided into a union of mutually disjoint orbits of different radii, for which, according to the space-time-Lorentz decomposition of Sp$(2,2)$, the subgroup $\mathrm{SO}_0(1,3)$, isomorphic to the Lorentz subgroup, plays the role of stabilizer subgroup (see subsection \ref{sec space-time-Lorentz dS4}).} To see the point, we choose $x_e^{}=(0,0,0,0,1)$. With this choice, the orbit $\daleth\equiv\daleth_4$ takes the form:
    \begin{eqnarray}\label{orbit gamma4}
    \daleth^{}_4 = \daleth^+_4 \cup \daleth^-_4 &=& \Big\{ \xi \in C^+ \;;\; \xi^4 = +b \Big\} \bigcup \Big\{ \xi \in C^+ \;;\; \xi^4 = -b \Big\}\nonumber\\
    &=& \Big\{ \xi \in C^+ \;;\; (\xi^0)^2-(\xi^1)^2-(\xi^2)^2-(\xi^3)^2 = b^2 \Big\}\,.
    \end{eqnarray}}
\end{itemize}
Here, in a shortcut, we assert that the latter parametrization is the most proper choice when one deals with the zero-curvature limit of dS$_4$ fields. In this context, the null-vector $\xi \in C^+$ is considered in terms of the four-momentum $(k^0,\vec{k})$ of a Minkowskian particle with mass $m$ as follows:
\begin{eqnarray}\label{orbit gamma4'}
\xi^\pm \; \big( \in \daleth^{\pm}_4 \big) = \Bigg( \frac{k^0}{m} = \sqrt{\frac{\vec{k}\cdot\vec{k}}{m^2} +1}, \frac{\vec{k}}{m}, \pm 1 \Bigg)\,,
\end{eqnarray}
based upon which, we have $(k^0)^2 - \vec{k}\cdot\vec{k} = m^2$. We will come back to this important point later.

\subsection{DS$_4$ plane waves as generating functions for square-integrable eigenfunctions}\label{Subsec generating}
For the sake of reasoning, in this subsection, we invoke a system of bounded global coordinates appropriate for describing a bounded version of dS$_4$ spacetime, namely $\mathbb{S}^3\times (-\pi/2,\pi/2)$. This system of intrinsic coordinates, known as conformal coordinates, is given by:
\begin{eqnarray}\label{cbgicoo}
x = (x^0, \textbf{x}) = \big(R \tan\rho , R (\cos\rho)^{-1}\textbf{u} \big)\,,
\end{eqnarray}
where $-\pi /2< \rho <\pi /2$ and $\textbf{u} \in \mathbb{S}^3$ (for an explicit form of $\textbf{u}$, see Eq. (\ref{z set})). Note that the coordinate $\rho$ is actually timelike, and plays the role of a conformal time; the closure of the $\rho$-interval is taken into account, when one studies compactified spacetime under conformal action.

\subsubsection{DS$_4$ plane waves as generating functions}
Again, employing the quaternionic notations, let $\xi = (\xi^0,\boldsymbol{\xi}) \in C^+$ (that is, $\xi^0 > 0$ and $\xi^0 = |\boldsymbol{\xi}|$). According to the identities given in appendices \ref{App quat} and \ref{App UIR's SU(2)}, one can demonstrate $\boldsymbol{\xi} \in\mathbb{H}\sim\mathbb{R}^4$ as $\boldsymbol{\xi} = |\boldsymbol{\xi}|\textbf{v}$, with $\textbf{v}\in \mathrm{SU}(2) \sim \mathbb{S}^3$. The dot product $x\cdot\xi/R$ then can be given, in terms of the conformal coordinates (\ref{cbgicoo}), as follows:
\begin{eqnarray}\label{dopHx.xi}
\frac{x\cdot\xi}{R} = (\tan\rho)\xi^0 - \frac{1}{\cos\rho}\textbf{u}\cdot\boldsymbol{\xi} = \frac{\xi^0 e^{\mathrm{i} \rho}}{2 \mathrm{i} \cos\rho}\big(1+r^2-2r (\textbf{u}\cdot\textbf{v})\big)\,, \;\;\;\;\;\;\; r = \mathrm{i} e^{- \mathrm{i} \rho}\,.
\end{eqnarray}
Considering the above identity along with the generating function for the Gegenbauer polynomials $C_n^{-\tau} (x)$ (see Eq. (\ref{G1})), it is straightforward to show that:
\begin{eqnarray}\label{dsplwgege}
\Big(\frac{x\cdot\xi}{R}\Big)^\tau &=& \Bigg(\frac{\xi^0e^{\mathrm{i} \rho}}{2 \mathrm{i} \cos\rho}\Bigg)^\tau \big(1+r^2-2r (\textbf{u}\cdot\textbf{v})\big)^\tau \nonumber\\
&=& \Bigg(\frac{\xi^0e^{\mathrm{i} \rho}}{2 \mathrm{i} \cos\rho}\Bigg)^\tau \sum_{n=0}^\infty r^n \; C_n^{-\tau}(\textbf{u}\cdot\textbf{v}) \,, \;\;\;\;\;\;\; \mbox{Re}(\tau) < \frac{1}{2}\,.
\end{eqnarray}
Note that the above expansion is not valid in the function sense, because the generating function for the Gegenbauer polynomials (as is manifest above) is merely convergent for $|r|<1$, while, in our case, we have $|r|= |\mathrm{i} e^{- \mathrm{i} \rho}| =1$. This failure, however, is circumvented here by giving a negative imaginary part to the angle $\rho \rightarrow \rho - \mathrm{i} \epsilon$ ($\epsilon > 0$). This process, ensuring the convergence of the expansion, indeed amounts to extend the ambient coordinates to the backward tube ${\cal{T}}^-$ (see subsection \ref{Subsec tube}).

Utilizing Eqs. (\ref{G1}) and (\ref{G2}) along with (\ref{G3}), we get the following auxiliary relation:
\begin{eqnarray}\label{comG123}
\big( 1+r^2 - 2r (\textbf{u}\cdot\textbf{v}) \big)^{\tau} = 2\pi^2 \sum_{Llm} r^L \; {\cal P}_L^{-\tau} (r^2) \; {\cal Y}_{Llm}(\textbf{u}){\cal Y}_{Llm}^\ast(\textbf{v})\,,
\end{eqnarray}
where, again, ${\cal Y}_{Llm}$'s, with $(L,l,m) \in \mathbb{N} \times \mathbb{N} \times \mathbb{Z}$, $0\leqslant l \leqslant L$ and $-l\leqslant m \leqslant l$, are the hyperspherical harmonics on the unit-sphere $\mathbb{S}^3$ (see appendix \ref{App some}), while the functions ${\cal P}_L^{-\tau} (r^2)$ are defined in terms of the hypergeometric functions as:
\begin{eqnarray}\label{pllz}
{\cal P}_L^{-\tau} (r^2) = \frac{1}{(L+1)!}\frac{\Gamma(L-\tau)}{\Gamma(-\tau)} \; ^{}_2F^{}_1(L-\tau, -\tau -1 ; L +2 ; r^2)\,,
\end{eqnarray}
with the integral representation:
\begin{eqnarray}
r^L \; {\cal P}_L^{-\tau} (r^2) \; {\cal Y}_{Llm}(\textbf{u}) = \frac{1}{2\pi^2}\int_{\mathbb{S}^3} \big( 1+r^2 -2r (\textbf{u}\cdot\textbf{v}) \big)^{\tau} \; {\cal Y}_{Llm}(\textbf{v}) \; \mathrm{d}\mu (\textbf{v})\,.
\end{eqnarray}

Now, combining the above auxiliary relation with Eq. (\ref{dsplwgege}), and substituting $r = \mathrm{i} e^{- \mathrm{i} \rho}$, the expansion of the dS$_4$ plane waves takes the form:
\begin{eqnarray}\label{nexdspw}
\Big(\frac{x\cdot\xi}{R}\Big)^\tau = 2\pi^2 \sum_{Llm} \Phi_{Llm}^\tau(x) \; (\xi^0)^\tau \; {\cal Y}_{Llm}^\ast(\textbf{v})\,,
\end{eqnarray}
where we introduce three sets of functions $\big\{\Phi_{Llm}^\tau(x)\big\}$, in the allowed ranges of the unifying complex parameter $\tau$ corresponding to the three series of the dS$_4$ scalar representations, on $\underline{M}_R$ as:
\begin{eqnarray}\label{ufuods}
\Phi_{Llm}^\tau(x) &=& \frac{\mathrm{i}^{L-\tau} \; e^{- \mathrm{i} (L-\tau)\rho}}{(2\cos\rho)^\tau} \; {\cal P}_L^{-\tau}\big(-e^{-2 \mathrm{i} \rho}\big) \; {\cal Y}_{Llm}(\textbf{u})\nonumber\\
&=& \frac{\mathrm{i}^{L-\tau} \; e^{- \mathrm{i} (L-\tau)\rho}}{(2\cos\rho)^\tau} \; \frac{\Gamma(L-\tau)}{(L+1)! \; \Gamma(-\tau)}\; ^{}_2F^{}_1(L-\tau \;, -\tau -1 ; L +2 ; -e^{-2 \mathrm{i} \rho}) \; {\cal Y}_{Llm}(\textbf{u})\,.
\end{eqnarray}
By using the Euler's transformation (see Eq. (\ref{Euler's transformation})), one can also obtain the following alternative form of Eq. (\ref{ufuods}):
\begin{eqnarray}\label{ufuods alternative}
\Phi_{Llm}^\tau (x) = \mathrm{i}^{L-\tau} \; e^{- \mathrm{i} (L+\tau +3)\rho} \; (2\cos\rho)^{\tau +3} \; \frac{\Gamma(L-\tau)}{(L+1)! \; \Gamma(-\tau)} \; {}_2F^{}_1\big(\tau +2, L+\tau +3;L+2;-e^{-2 \mathrm{i} \rho}\big) \; {\cal Y}_{Llm}(\textbf{u})\,.
\end{eqnarray}
Here, we must underline that:
\begin{itemize}
\item{The introduced functions $\Phi_{Llm}^\tau (x)$ are well defined for all values of $\tau$ with $\mbox{Re}(\tau) < 0$,\footnote{Note that this restriction on $\tau$, namely, $\mbox{Re}(\tau) < 0$, is issued from the domain of the Gamma function $\Gamma(-\tau)$ appeared in the denominator of (\ref{ufuods}) (to see the properties of the Gamma functions, we refer readers to Ref. \cite{Magnus}).} and therefore, are well defined for all the dS$_4$ scalar UIR's.}
\item{In the conformal coordinates $x=x(\rho , \textbf{u})$, these functions are infinitely differentiable in their respective allowed ranges of parameters.}
\item{Due to the linear independence of ${\cal Y}_{Llm}$'s, the functions $\Phi_{Llm}^\tau(x)$ play the role of the usual solutions to the scalar-wave equations (\ref{Wave Eq. scalar ambient}) (again, in the respective allowed ranges of $\tau$), when one proceeds with the proper separation of variables. [We will clarify this point in subsubsection \ref{Subsubsec separation}.]}
\item{Considering the orthonormality of ${\cal Y}_{Llm}$'s on $\mathbb{S}^3$, the integral representation of $\Phi_{Llm}^\tau (x)$'s, say the ``Fourier transform" on $\mathbb{S}^3$, is given by:
    \begin{eqnarray}\label{Fourier principal}
    \Phi_{Llm}^\tau (x) = \frac{1}{2\pi^2\big(\xi^0\big)^\tau}\int_{\mathbb{S}^3}\Big(\frac{x\cdot\xi}{R}\Big)^\tau {\cal Y}_{Llm}(\textbf{v}) \; \mathrm{d}\mu (\textbf{v})\,.
    \end{eqnarray}
    [We will also come back to the meaning of this Fourier transform later in subsubsection \ref{Subsubsec versus}.]}
\item{At the limit $\rho \rightarrow \pm \pi/2$, namely, the infinite dS$_4$ ``future"/``past", the behavior of these functions, with respect to the form (\ref{ufuods}), is determined by (see Eq. (\ref{limit})):
    \begin{eqnarray}\label{LimitPhi 5.7}
    \Phi_{Llm}^\tau (x) \approx^{}_{\rho\rightarrow \pm \frac{\pi}{2}} \;\; \frac{\mathrm{i}^{L-\tau} \; e^{- \mathrm{i} (L-\tau)\rho}}{(2\cos\rho)^\tau} \; \frac{\Gamma(L-\tau) \; \Gamma(2\tau +3)}{\Gamma(-\tau) \; \Gamma(\tau +2) \; \Gamma(L+\tau+3)} \; {\cal Y}_{Llm} (\textbf{u})\,,
    \end{eqnarray}
    for $-\frac{3}{2} < \mbox{Re}(\tau) < 0$, and, with respect to the alternative form (\ref{ufuods alternative}), by (see Eq. (\ref{limit})):
    \begin{eqnarray}\label{LimitPhi 5.8}
    \Phi_{Llm}^\tau (x) \approx^{}_{\rho\rightarrow \pm \frac{\pi}{2}} \;\; \mathrm{i}^{L-\tau} \; e^{- \mathrm{i} (L+\tau +3)\rho} \; (2\cos\rho)^{\tau +3} \; \frac{\Gamma(-2\tau -3)}{\Gamma(-\tau) \; \Gamma(-\tau -1)} \; {\cal Y}_{Llm} (\textbf{u})\,,
    \end{eqnarray}
    for $\mbox{Re}(\tau) < -\frac{3}{2}$.\footnote{Note that for $\mbox{Re}(\tau) = -\frac{3}{2}$, which is the case for the scalar principal series $\tau=-3/2- \mathrm{i} \nu$ ($\nu\in\mathbb{R}$), we need to give a small imaginary part to $\nu$, then we can use the above asymptotic formulas (see Ref. \cite{Bros 2002}).} Considering the asymptotic relations (\ref{LimitPhi 5.7}) and (\ref{LimitPhi 5.8}) in their respective domains of the parameter $\tau$, it is clear that the functions $\Phi_{Llm}^\tau (x)$ are singular at the limit $\rho \rightarrow \pm \pi/2$ for all $\mbox{Re}(\tau) < -3$ corresponding to the relation (\ref{LimitPhi 5.8}), where the asymptotic behavior is dominated by the factor $(\cos\rho)^{\tau+3}$. Of course, one must notice that this singularity is nothing but a direct result of the choice of coordinates, that we have already made in order to express the dot product $x\cdot\xi$. Technically, this singularity occurs for the scalar discrete series ($\tau=-p-2$) with $p \geqslant 2$, which includes all the scalar discrete series UIR's $\Pi_{p,0}$, with the exception of $\Pi_{p=1,0}$ ($\tau=-3$) associated with the so-called minimally coupled scalar field (this point will be used later in subsection \ref{Subsec mc}).}
\end{itemize}

\subsubsection{Normalized eigenfunctions}\label{Subsubsec Normalized eigen}
Here, we proceed with the examination of the introduced three sets of functions $\big\{ \Phi_{Llm}^\tau(x) \big\}$, in the respective allowed ranges of $\tau$, as three sets of basis elements for the spacetime realization of the (respective) carrier Hilbert spaces of the dS$_4$ scalar (principal, complementary, and discrete series) UIR's. Such basis elements are supposed to be scalar-valued functions on $\underline{M}_R$, infinitely differentiable, and solutions to the scalar-wave equations (\ref{Wave Eq. scalar ambient}),\footnote{Again, for the latter point, it is convenient to recall the arguments subsequent to Eq. (\ref{6666666666}).} which, so far, all are well fulfilled by $\Phi_{Llm}^\tau(x)$'s. These elements are also supposed to be square integrable with respect to the so-called Klein-Gordon inner product. This is actually the only criterion left here that needs to be examined concerning the introduced sets of functions. Technically, for given solutions $\Phi_1$ and $\Phi_2$ to the scalar-wave equations (\ref{Wave Eq. scalar ambient}), the Klein-Gordon inner product is defined by:
\begin{eqnarray}
\langle \Phi_1(x) , \Phi_2(x) \rangle = \mathrm{i} \int_\Sigma \Phi_1^\ast(x)\big(\overset{\rightarrow}{\partial}_\nu - \overset{\leftarrow}{\partial}_\nu \big)\Phi_2(x) \; \mathrm{d} \sigma^\nu \equiv \mathrm{i} \int_\Sigma \Phi_1^\ast(x)\overset{\leftrightarrow}{\partial}_\nu \Phi_2(x) \; \mathrm{d} \sigma^\nu\,,
\end{eqnarray}
where $\Sigma$ and $\mathrm{d} \sigma^\nu$ respectively stand for a \emph{Cauchy surface}\footnote{That is a spacelike surface in such a way that the Cauchy data on it uniquely define a solution to the wave equations (\ref{Wave Eq. scalar ambient}).} and the area element vector on it. The Klein-Gordon inner product is dS$_4$ invariant and of course independent of the choice of $\Sigma$. Regarding the choice of global coordinates $x=x(\rho , \textbf{u})$, that we have already made, this product takes the form:
\begin{eqnarray}\label{Kl-Go inpro}
\langle \Phi_1(x), \Phi_2(x) \rangle = \mathrm{i} R^2 \int_{\rho =0} \Phi_1^\ast (\rho,\textbf{u}) \; \overset{\leftrightarrow}{\partial}_\rho \; \Phi_2(\rho,\textbf{u}) \; \mathrm{d} \mu(\textbf{u})\,,
\end{eqnarray}
where $\mathrm{d} \mu(\textbf{u})$ is the invariant measure on $\mathbb{S}^3$ (see appendix \ref{App UIR's SU(2)}).

Considering the orthogonality relations of ${\cal Y}_{Llm}$'s, for the given sets of functions $\big\{ \Phi_{Llm}^\tau(x) \big\}$ (the normalizable ones), we have :
\begin{eqnarray}\label{phi inner pro}
\langle \Phi_{L_1l_1m_1}^\tau(x) , \Phi_{L_2l_2m_2}^\tau(x) \rangle = \delta^{}_{L_1L_2} \delta^{}_{l_1l_2} \delta^{}_{m_1m_2} \|\Phi_{L_1l_1m_1}^\tau(x)\|^2\,,
\end{eqnarray}
where, by admitting the form (\ref{ufuods}) and utilizing Eqs. (\ref{1}), (\ref{2}), and (\ref{3}), we obtain:
\begin{eqnarray}\label{norm phi}
\|\Phi_{Llm}^\tau(x)\|^2 &=& \pi R^2 \; 2^{2-2L} \; e^{-\pi \mbox{Im}(\tau)} \; \Big| \frac{\Gamma(L-\tau)}{\Gamma(-\tau)} \Big|^2 \nonumber\\
&& \hspace{1cm} \times \Bigg[ \mbox{Re} \Bigg( \Gamma^\ast\Big( \frac{L-\tau+1}{2} \Big) \; \Gamma\Big( \frac{L-\tau}{2} \Big) \; \Gamma^\ast\Big( \frac{L+\tau+4}{2} \Big) \; \Gamma\Big( \frac{L+\tau+3}{2} \Big) \Bigg) \Bigg]^{-1}\,.
\end{eqnarray}
For real values of $\tau$, by employing the \emph{Legendre duplication formula}\footnote{$\Gamma(z) \; \Gamma(z+1/2) = 2^{1-2z} \sqrt{\pi} \; \Gamma(2z)$.}, Eq. (\ref{norm phi}) simplifies to:
\begin{eqnarray}\label{norm phi'}
\|\Phi_{Llm}^\tau(x)\|^2 = 2^3 R^2 \; \frac{\Gamma(L-\tau)}{\big( \Gamma(-\tau) \big)^2 \; \Gamma(L+\tau+3)}\,.
\end{eqnarray}
According to Eqs. (\ref{norm phi}) and (\ref{norm phi'}), it is manifest that the sets of functions $\big\{ \Phi_{Llm}^\tau(x) \big\}$, with $-3<\mbox{Re} (\tau) < 0$, which is the case for the dS$_4$ scalar principal and complementary series, are normalizable. Hence, considering this fact along with the properties listed below Eqs. (\ref{ufuods}) and (\ref{ufuods alternative}), we argue that the functions $\Phi_{Llm}^\tau(x)$, in the respective ranges of $\tau$ corresponding to the dS$_4$ scalar principal and complementary series, are proper candidates to generate the (respective) carrier Hilbert spaces of the representations. For the scalar discrete series, characterized by $\mbox{Re} (\tau) \leqslant -3$ ($\tau = -p-2$, with $p=1,2,...$), however, the situation is more delicate. To make the point clear, we first need to consider the alternative form (\ref{ufuods alternative}) of the functions $\Phi_{Llm}^\tau(x)$, for which, the range $\mbox{Re} (\tau) \leqslant -3$ is allowed. In this case, the involved hypergeometric functions ${}_2F^{}_1\big(-p, L-p+1;L+2;-e^{-2 \mathrm{i} \rho}\big)$, with $p=1,2,...$, reduce to polynomials of degree $p$ (see appendix \ref{App some}). Accordingly, for the scalar discrete series ($\tau = -p-2$), the corresponding set of functions $\big\{ \Phi_{Llm}^\tau(x) \big\}$ is divided into two parts:
\begin{itemize}
\item{The set of $\big\{ \Phi_{Llm}^\tau(x) \big\}$, with $L < p$ (say $\big\{ \Phi_{L_{< p}lm}^\tau(x) \big\}$), which, with respect to the Klein-Gordon inner product (\ref{Kl-Go inpro}), is of null norm and clearly, considering the allowed ranges of $L,l$, and $m$, is of $p(p+1)(2p+1)/6$ dimension.}
\item{The set of regular normalizable functions $\big\{ \Phi_{Llm}^\tau(x) \big\}$, with $L \geqslant p$ (say $\big\{ \Phi_{L_{\geqslant p}lm}^\tau(x) \big\}$).}
\end{itemize}
Note that the invariant null-norm subspace $\big\{ \Phi_{L_{< p}lm}^\tau(x) \big\}$ can be interpreted as a space of ``gauge" states, which carries the irreducible (nonunitary) finite-dimensional representations of the dS$_4$ group. Regarding the notations introduced in appendix \ref{App Lie algebra B2}, these representations are characterized by ($n_1=0, n_2=p-1$). They are indeed \emph{Weyl equivalent}\footnote{Again, if two representations are Weyl equivalent, then they share same Casimir eigenvalue.} to the discrete series UIR's $\Pi_{p,0}$.

Let us summarize the above results. In the allowed ranges of parameters, the normalized eigenfunctions (denoted here by $\widetilde{\Phi}_{Llm}^\tau(x)$'s), as basis elements of the carrier Hilbert spaces of the dS$_4$ scalar UIR's, read \cite{Chernikov}:
\begin{eqnarray}
\widetilde{\Phi}_{Llm}^\tau(x) = N_L^\tau \; \mathrm{i}^{L-\tau} \; e^{- \mathrm{i} (L+\tau +3)\rho} \; (\cos\rho)^{\tau +3} \; {}_2F^{}_1\big(\tau +2, L+\tau +3;L+2;-e^{-2 \mathrm{i} \rho}\big) \; {\cal Y}_{Llm}(\textbf{u})\,,
\end{eqnarray}
where, with respect to Eq. (\ref{norm phi}), the normalization factors $N_L^\tau$ are:
\begin{eqnarray}
N_L^\tau &=& \frac{1}{R\sqrt{\pi}} \; 2^{L+\tau+2} \; e^{\frac{\pi}{2} \mbox{Im}(\tau)} \; \frac{\Gamma(L-\tau)}{|\Gamma(L-\tau)|} \frac{|\Gamma(-\tau)|}{\Gamma(-\tau)} \frac{1}{(L+1)!} \nonumber\\
&& \times \sqrt{\mbox{Re} \Bigg( \Gamma^\ast\Big( \frac{L-\tau+1}{2} \Big) \; \Gamma\Big( \frac{L-\tau}{2} \Big) \; \Gamma^\ast\Big( \frac{L+\tau+4}{2} \Big) \; \Gamma\Big( \frac{L+\tau+3}{2} \Big) \Bigg)} \; .
\end{eqnarray}
For real values of $\tau$, which is the case for the complementary and discrete series (the normalizable ones), the normalization factors $N_L^\tau$ simplify to:
\begin{eqnarray}
N_L^\tau = \frac{2^{\tau+\frac{3}{2}}}{R} \; \frac{\sqrt{\Gamma(L-\tau) \; \Gamma(L+\tau+3)}}{(L+1)!}\,.
\end{eqnarray}
In particular, for the scalar discrete series, determined by $\tau = -p-2$ ($p=1,2,...$), the orthonormal system turns into ($L \geqslant p$):
\begin{eqnarray}\label{normalized discrete}
\widetilde{\Phi}_{L_{\geqslant p}lm}^\tau(x) = N_{L_{\geqslant p}}^\tau \; \mathrm{i}^{L+p+2} \; e^{- \mathrm{i} (L-p+1)\rho} \; (\cos\rho)^{-p+1} \; {}_2F^{}_1\big(-p, L-p+1;L+2;-e^{-2 \mathrm{i} \rho}\big) \; {\cal Y}_{Llm}(\textbf{u})\,,
\end{eqnarray}
with:
\begin{eqnarray}
N_{L_{\geqslant p}}^\tau = \frac{2^{-p-\frac{1}{2}}}{R} \; \frac{\sqrt{\Gamma(L+p+2) \; \Gamma(L-p+1)}}{(L+1)!}\,.
\end{eqnarray}
As already mentioned, for the scalar discrete series with $p \geqslant 2$, these functions are singular at the limits $\rho = \pm \pi/2$.

Now, for the scalar principal and complementary series, we can rewrite the expansion formula (\ref{nexdspw}), representing the dS$_4$ plane waves as generating functions, in terms of the orthonormal sets $\big\{ \widetilde{\Phi}_{Llm}^\tau(x) \big\}$ (in the respective ranges of $\tau$) as:
\begin{eqnarray}\label{mmmmmmmm}
\Big(\frac{x\cdot\xi}{R}\Big)^\tau = 2\pi^2 \sum_{Llm} \|\Phi_{Llm}^\tau(x)\| \; \widetilde{\Phi}_{Llm}^\tau(x) \; (\xi^0)^\tau \; {\cal Y}_{Llm}^\ast(\textbf{v})\,,
\end{eqnarray}
where $\|\Phi_{Llm}^\tau(x)\|$'s are determined by Eq. (\ref{norm phi}), for the principal series, and by Eq. (\ref{norm phi'}), for the complementary series. For the scalar discrete series, the expansion needs to be split into two sectors:
\begin{eqnarray}\label{nnnnnnnnn}
\Big(\frac{x\cdot\xi}{R}\Big)^{\tau=-p-2} &=& 2\pi^2 \sum_{L=0}^{p-1}\sum_{lm} {\Phi}_{L_{< p}lm}^\tau(x) \; (\xi^0)^{-p-2} \; {\cal Y}_{Llm}^\ast(\textbf{v})\nonumber\\
&& + 2\pi^2 \sum_{L=p}^{\infty}\sum_{lm} \|{\Phi}_{L_{\geqslant p}lm}^\tau(x)\| \; \widetilde{\Phi}_{L_{\geqslant p}lm}^\tau(x) \; (\xi^0)^{-p-2} \; {\cal Y}_{Llm}^\ast(\textbf{v})\,, \;\;\;\;\;\;\; p=1,2,... \;,
\end{eqnarray}
where $\|{\Phi}_{L_{\geqslant p}lm}^\tau(x)\|$'s are determined by Eq. (\ref{norm phi'}), with $\tau=-p-2$. Accordingly, we have the ``spherical" modes in dS$_4$ spacetime in terms of the dS$_4$ plane waves.

\subsubsection{Usual solutions by separation of variables} \label{Subsubsec separation}
Let us now show that the three sets of functions $\big\{\Phi_{Llm}^\tau\big\}$ (or $\big\{\widetilde{\Phi}_{Llm}^\tau\big\}$), corresponding to the three series of the dS$_4$ scalar representations, can be actually obtained by directly solving the scalar-wave equations (\ref{Wave Eq. scalar ambient}). As already pointed out in subsection \ref{Subsec Link to intrinsic coordinates}, the scalar Casimir operator is proportional to the Laplace-Beltrami operator in dS$_4$ spacetime (i.e., $Q^{(1)}_0 = -R^2 \Box_R$), which, with respect to the conformal coordinates $x=x(\rho,\textbf{u})$ (see Eq. (\ref{cbgicoo})), takes the form:
\begin{eqnarray}
\Box_R = \frac{1}{\sqrt{g}} \partial_\nu \sqrt{g} g^{\nu\mu} \partial_\mu = \frac{1}{R^2} \cos^4\rho \frac{\partial}{\partial\rho} \Big( \cos^{-2}\rho \frac{\partial}{\partial\rho} \Big) - \frac{1}{R^2} \cos^2\rho \; \Delta_3\,,
\end{eqnarray}
with:
\begin{eqnarray}
\Delta_3 = \frac{\partial^2}{\partial\psi^2} + 2\cot\psi \; \frac{\partial}{\partial\psi} + \frac{1}{\sin^2\psi} \frac{\partial^2}{\partial\theta^2} + \cot\theta \; \frac{1}{\sin^2 \psi} \frac{\partial}{\partial\theta} + \frac{1}{\sin^2\psi \sin^2\theta} \frac{\partial^2}{\partial\phi^2}\,,
\end{eqnarray}
where the latter is the Laplace operator on $\mathbb{S}^3$ (note that here we have used the parametrization given in Eq. (\ref{z set}), for which $\mathbb{S}^3 \ni \textbf{u} = \textbf{u}(\psi,\theta,\phi)$, with $0 \leqslant \psi , \theta \leqslant \pi$ and $0 \leqslant \phi < 2\pi$). Solutions to the scalar-wave equations (\ref{Wave Eq. scalar ambient}), in the respective ranges of $\tau$, then can be found by separation of variables, $\Phi_{Llm}^\tau(x) = A(\rho) B(\textbf{u})$, based upon which we get \cite{Chernikov,Kirsten}:
\begin{eqnarray}\label{angular part}
\big( \Delta_3 + C \big) B(\textbf{u}) = 0\,,
\end{eqnarray}
\begin{eqnarray}\label{rho dependent part}
\Big( \cos^4\rho \frac{\mathrm{d}}{\mathrm{d} \rho} \; \cos^{-2}\rho \frac{\mathrm{d}}{\mathrm{d} \rho} +C\cos^2\rho - \tau(\tau+3) \Big) A(\rho)= 0\,.
\end{eqnarray}

The relevant solutions to the angular part (\ref{angular part}), with $C=L(L+2)$ and $L \in \mathbb{N}$, are the usual hyperspherical harmonics on $\mathbb{S}^3$, namely, $B(\textbf{u}) = {\cal{Y}}_{Llm}(\textbf{u})$. On the other hand, for the $\rho$-dependent part and for a given $\tau$, the solutions (corresponding to the Euclidean vacuum) are \cite{Kirsten}:
\begin{eqnarray}
A(\rho) \equiv A_{\lambda L} (\rho) = \chi^{}_L \; (\cos\rho)^{\frac{3}{2}} \Big( P_{L+\frac{1}{2}}^\lambda (\sin\rho) - \frac{2 \mathrm{i}}{\pi} Q_{L+\frac{1}{2}}^\lambda (\sin\rho) \Big)\,,
\end{eqnarray}
where $P_n^\lambda$ and $Q_n^\lambda$ are the Legendre functions on the cut, with $\lambda \equiv \lambda(\tau) = \pm (\tau + \frac{3}{2})$, and $\chi^{}_L$ is:
\begin{eqnarray}
\chi^{}_L = \frac{\sqrt{\pi}}{2R} \Bigg( \frac{\Gamma(L-\lambda+\frac{3}{2})}{\Gamma(L+\lambda+\frac{3}{2})} \Bigg)^{\frac{1}{2}}.
\end{eqnarray}
Here, it must be underlined that the above relations provide us with three sets of solutions $\big\{ \Phi_{Llm}^\tau(x) = A_{\lambda L} (\rho) {\cal{Y}}_{Llm}(\textbf{u}) \big\}$ in the allowed ranges of $\tau$ associated with the three series of the dS$_4$ scalar representations, except for the case of discrete series ($\tau=-p-2$), with $L < p$, for which the above formulas break down. These families of solutions obey the following orthogonality prescription:
\begin{eqnarray}
\langle \Phi_{L^\prime l^\prime m^\prime}^\tau , \Phi_{Llm}^\tau \rangle = \delta^{}_{LL^\prime} \delta^{}_{ll^\prime} \delta^{}_{mm^\prime}\,, \;\;\;\;\;\;\; \langle \Phi_{L^\prime l^\prime m^\prime}^\tau , \big(\Phi_{Llm}^\tau\big)^\ast \rangle = 0\,.
\end{eqnarray}

Now, to make apparent the link between the $\rho$-dependent part of the solutions $A_{\lambda L} (\rho)$ and its counterpart in Eq. (\ref{ufuods}) or equivalently in Eq. (\ref{ufuods alternative}), given in terms of the hypergeometric functions, one can either directly expand the Legendre functions in their arguments or solve the differential Eq. (\ref{rho dependent part}), by changing the variables $\rho \mapsto t \; \big( = r^2 \big) = -e^{-2 \mathrm{i} \rho}$:
\begin{eqnarray}\label{rho dependent part'}
\Big( t^2 (1-t)^2 \; \frac{\mathrm{d}^2}{\mathrm{d} t^2} + 2t(1-t) \; \frac{\mathrm{d}}{\mathrm{d} t} - \frac{1}{4} L(L+2)(1-t)^2 - t\tau(\tau+3) \Big) A(t) = 0\,,\;\;\;\;\;\;\; A_{\lambda L} (\rho) \equiv A(t)\,.
\end{eqnarray}
Below, we proceed with the second approach.

\emph{Frobenius solutions to Eq. (\ref{rho dependent part'}) in the neighborhood of $t=0$.}\footnote{At first glance, it might be argued that, with respect to the definition $t = -e^{-2 \mathrm{i} \rho}$, the neighborhood of $t=0$ is not reachable at all (since $|t|=1$). But, the very point to be noticed here is that the whole above construction is performed when we gave a negative imaginary part to the angle $\rho \rightarrow \rho - \mathrm{i} \epsilon$ (see the argument subsequent to Eq. (\ref{dsplwgege})). Therefore, the definition $t = -e^{-2 \mathrm{i} \rho}$ should actually be read as $t = -e^{-2 \mathrm{i} \rho} e^{-2\epsilon}$. In this sense, the neighborhood of $t=0$ is quite reachable by varying the value of $\epsilon \; (>0)$.} For the solutions of the form $A(t) = t^c \sum_{n\geqslant 0} a_nt^n$, the Frobenius indicial equation admits two sets of solutions, respectively, associated with the case $c=c_1= L/2$ and the case $c=c_2=-(L+2)/2$. The former set of solutions, in the neighborhood of $t=0$, reads as:
\begin{eqnarray}\label{6.4}
A(t) = A_{c_1}(t) = t^{\frac{L}{2}} (1-t)^{\tau+3} \; ^{}_2F^{}_1\big( \tau+2, L+\tau+3; L+2; t \big)\,,
\end{eqnarray}
or alternatively, by making use of the Euler's transformation (see Eq. (\ref{Euler's transformation})), as:
\begin{eqnarray}\label{6.4'}
A(t) = A_{c_1}(t) = t^{\frac{L}{2}} (1-t)^{-\tau} \; ^{}_2F^{}_1\big( L-\tau, -\tau-1; L+2; t \big)\,.
\end{eqnarray}
By substituting $t = -e^{-2 \mathrm{i} \rho}$, one can easily check that the above alternative forms of solutions respectively give the $\rho$-dependent part of the solutions (\ref{ufuods alternative}) and (\ref{ufuods}), up to constant factors.

On the other hand, since $c_1 - c_2 = L+1 \in \mathbb{N}$, for the latter set of solutions associated with $c=c_2=-(L+2)/2$, we face a degenerate case, for which linearly independent solutions read:
\begin{eqnarray}\label{6.5}
A(t) = A_{c_2}(t) = \big(\log(t)\big)A_{c_1}(t) + \sum_{n=-L-1}^{\infty} b_n t^{n+\frac{L}{2}}\,,
\end{eqnarray}
where the coefficients $b_n$ are recurrently specified by substituting the above solutions into Eq. (\ref{rho dependent part'}). Note that the second set of solutions is singular at $t=0$, due to the logarithmic term. This set of solutions is relevant to the scalar discrete series case, for which one has to deal with the finite dimensional space of null-norm solutions (in this regard, we would like to draw the attention of readers to subsection \ref{Subsec mc}, where we discuss this matter for the simplest case ($p=1$)).

The above sets of solutions respectively possess the Klein-Gordon norms:
\begin{eqnarray}\label{6.6}
\| A_{c_1} \|^2 = 2^3 R^2 \; \frac{\big((L+1)!\big)^2}{\Gamma(L-\tau) \; \Gamma(L+\tau+3)}\,,
\end{eqnarray}
which correspond to Eq. (\ref{norm phi}), and (for real values of $b_n$'s):
\begin{eqnarray}\label{6.7}
\| A_{c_2} \|^2 = \pi^2 \| A_{c_1} \|^2 + u R^2 \Bigg( \frac{\sqrt{\pi} \; 2^{2-L} (L+1)!}{\Gamma(\frac{L-\tau+1}{2}) \; \Gamma(\frac{L+\tau+4}{2})} + 4(-1)^L v \Bigg)\,,
\end{eqnarray}
where:
\begin{eqnarray}
u = \sum_{n=-L-1}^\infty (-1)^n b_n\,, \;\;\;\;\;\;\; v = \sum_{n=-L-1}^\infty (-1)^n (n+L/2) b_n\,.
\end{eqnarray}
In the discrete series case ($\tau=-p-2$), with $L < p$, we conjecture that the Klein-Gordon norm (\ref{6.7}) vanishes, as the norm (\ref{6.6}) does.

\emph{Frobenius solutions to Eq. (\ref{rho dependent part'}) in the neighborhood of $t=1$, associated with the discrete series} ($\tau=-p-2$). For the solutions of the form $A(t) = (1-t)^d \sum_{n\geqslant 0} c_n(1-t)^n$, the Frobenius indicial equation has two sets of solutions, respectively, associated with $d=d_1=-\tau=p+2$ and $d=d_2=\tau+3=1-p$. The second case represents the $\rho$-dependent part ($t = -e^{-2 \mathrm{i} \rho}$) of the (respective) solutions (\ref{ufuods}) or (\ref{ufuods alternative}). Solutions associated with the first case, in the neighborhood of $t=1$, are given by:
\begin{eqnarray}
A(t) = A_{d_1}(t) = t^{\frac{L}{2}} (1-t)^{p+2} \; ^{}_2F^{}_1\big(L+p+2, p+1; L+2; 1-t \big)\,,
\end{eqnarray}
or equivalently, with respect to the Euler's transformation (see Eq. (\ref{Euler's transformation})), by:
\begin{eqnarray}
A(t) = A_{d_1}(t) = t^{\frac{L}{2}-2p-1} (1-t)^{p+2} \; ^{}_2F^{}_1\big(-p, L-p+1; L+2; 1-t \big)\,.
\end{eqnarray}
Again, because $d_1-d_2=2p+1 \in \mathbb{N}$, one has to deal with a degenerate case in the context of the second set of solutions. Linearly independent solutions then are:
\begin{eqnarray}
A(t) = A_{d_2}(t) = \big(\log(1-t)\big)A_{d_1}(t) + \sum_{n=-2p-1}^{\infty} e_n (1-t)^{n+p+2}\,,
\end{eqnarray}
where the coefficients $e_n$ are recurrently specified by substituting the above solutions into Eq. (\ref{rho dependent part'}), while the latter admits the change of variables $t \mapsto 1-t$. Here, again, the second set of solutions is singular, due to singularity of the logarithmic term at $t=1$.

\subsubsection{DS$_4$ UIR's: spacetime realization versus $\mathbb{S}^3$ realization} \label{Subsubsec versus}
This subsubsection is devoted to a more detailed discussion concerning the Fourier transform (\ref{Fourier principal}). This transform actually intertwines two different realizations of the dS$_4$ scalar principal series UIR's, namely the spacetime and $\mathbb{S}^3$ realizations, while the dS$_4$ plane waves $(x\cdot\xi/R)^\tau$ serve as the (Fourier) kernel for passing from one realization to the other. To see the point, while we have in mind the arguments given in subsubsection \ref{Subsubsec Normalized eigen}, it is sufficient to recall from subsection \ref{Subsec dS4 Principal scalar} that the space spanned by ${\cal Y}_{Llm}$'s on $\mathbb{S}^3$ carries the ten essentially self-adjoint infinitesimal operators associated with the scalar principal series UIR's of the dS$_4$ group Sp$(2,2)$ (the $\mathbb{S}^3$ realization).

Here, for all three series of the dS$_4$ scalar UIR's, we introduce the following kernel:
\begin{eqnarray}
\underline{M}_R \times \mathbb{S}^3 \ni (x, \textbf{v}) \; \mapsto \; K (x, \textbf{v}) \equiv \frac{R^{-1}}{2\pi^2} \; 2^{-\frac{3}{2}} \; e^{\frac{\pi}{2}\mbox{Im}(\tau)} \; |\Gamma(-\tau)| \; (\xi^0)^{-\tau} \; \Big(\frac{x\cdot\xi}{R}\Big)^{\tau},
\end{eqnarray}
where we remind that $\xi = (\xi^0,\xi^0\textbf{v}) \in C^+$, with $\xi^0 >0$ and $\textbf{v}\in \mathbb{S}^3$. This kernel for the scalar principal series representations (with $\tau = -3/2 - \mathrm{i} \nu$ and $\nu\in\mathbb{R}$) expands as:
\begin{eqnarray}
K^{\mbox{\small{ps}}}_{} (x, \textbf{v}) = \sum_{Llm} \widetilde{\Phi}_{Llm}^{\tau}(x) \; \Big({\cal{Y}}_{Llm}^{\tau} (\textbf{v})\Big)^\ast.
\end{eqnarray}
To get the above identity, we have used Eqs. (\ref{mmmmmmmm}) and (\ref{norm phi}), and then, after substituting $\tau = -3/2 - \mathrm{i}\nu$ and considering the fact that $\Gamma^\ast(z)=\Gamma(z^\ast)$, we have applied the Legendre duplication formula. On the other hand, for the scalar complementary series representations (with $\tau= -3/2 - \nu$ and $\nu\in\mathbb{R}$, while $0<|\nu|<3/2$), taking Eq. (\ref{norm phi'}) into account, the expansion of this kernel turns into:
\begin{eqnarray}
K^{\mbox{\small{cs}}}_{} (x, \textbf{v}) = \sum_{Llm} \widetilde{\Phi}_{Llm}^{\tau}(x) \; \Big(\widetilde{{\cal{Y}}}_{Llm}^{\tau} (\textbf{v})\Big)^\ast,
\end{eqnarray}
where $\widetilde{{\cal{Y}}}_{Llm}^\tau (\textbf{v}) = \sqrt{\frac{\Gamma(L-\tau)}{\Gamma(L+\tau+3)}} \; {\cal{Y}}_{Llm}(\textbf{v})$. For the scalar discrete series representations (with $\tau=-p-2$ and $p=1,2,...$), one needs to consider Eqs. (\ref{nnnnnnnnn}) and (\ref{norm phi'}), based upon which:
\begin{eqnarray}
K^{\mbox{\small{ds}}}_{} (x, \textbf{v}) = R^{-1} 2^{-\frac{3}{2}} (p+1)! \sum_{L=0}^{p-1}\sum_{lm} {\Phi}_{L_{< p}lm}^{\tau}(x)\; \Big({\cal{Y}}_{Llm}^{\tau} (\textbf{v})\Big)^\ast + \sum_{L=p}^{\infty}\sum_{lm} \widetilde{\Phi}_{L_{\geqslant p}lm}^{\tau}(x) \; \Big(\widetilde{{\cal{Y}}}_{Llm}^{\tau} (\textbf{v})\Big)^\ast,
\end{eqnarray}
where  $\widetilde{{\cal{Y}}}_{Llm}^{\tau=-p-2} (\textbf{v}) = \sqrt{\frac{(L+p+1)!}{(L-p)!}} \; {\cal{Y}}_{Llm}(\textbf{v})$. We recall from subsections \ref{Subsec complementary dS4} and \ref{Subsec discrete dS4} that $\widetilde{{\cal{Y}}}_{Llm}^\tau (\textbf{v})$'s, respectively, with $\tau= -3/2 - \nu$ and $\tau=-p-2$ ($L\geqslant p$), generate the $\mathbb{S}^3$ realization of the carrier Hilbert spaces of the dS$_4$ scalar complementary and discrete series representations.

Accordingly, for all three series of the dS$_4$ scalar UIR's, we have the two transforms which, in the respective allowed ranges of $\tau$, connect the ``wave functions" $\widetilde{\Phi}_{Llm}^{\tau}(x)$ (square integrable with respect to the Klein-Gordon inner product (\ref{Kl-Go inpro})) with the respective elements $f(\textbf{v})$ in the Hilbert space $L^2_{\mathbb{C}}(\mathbb{S}^3)$, for the scalar principal series, in $L^2_{\mathbb{C}}(\mathbb{S}^3 \times \mathbb{S}^3)$, for the scalar complementary series, and again in $L^2_{\mathbb{C}}(\mathbb{S}^3)$, for the scalar discrete series. These two transforms, which are inverse to each other, read:
\begin{eqnarray}\label{Fourier principal new}
\widetilde{\Phi}_{Llm}^{\tau}(x) = \langle K^\ast_{} (x, \cdot) , f \rangle^{}_{L^2}\,,
\end{eqnarray}
\begin{eqnarray}
f(\textbf{v}) = \langle K_{} (\cdot, \textbf{v}) , \widetilde{\Phi}_{Llm}^{\tau} \rangle^{}_{KG}\,,
\end{eqnarray}
where $\langle \cdot , \cdot \rangle_{L^2}$ and $\langle \cdot , \cdot \rangle^{}_{KG}$ respectively refer to the $L^2(\mathbb{S}^3)$ inner product (see Eq. (\ref{norm scalar prin dS4})) and the Klein-Gordon inner product (\ref{Kl-Go inpro}). One can easily see that, for the principal case, the Fourier transform (\ref{Fourier principal new}) precisely recovers the transform (\ref{Fourier principal}).

\subsection{DS$_4$ plane waves and the zero-curvature limit}\label{Subsec flat limit}
In this subsection, following the lines sketched in Ref. \cite{Garidi zero curvature limit}, we aim to study the behavior of the dS$_4$ scalar principal waves under vanishing curvature.\footnote{Recall that, among all the dS$_4$ UIR's, those admitting a meaningful null-curvature limit merely involve the principal series representations; they contract explicitly to the Poincar\'{e} massive UIR's. In this sense, they are also called dS$_4$ massive representations (see section \ref{Sec contraction}).} We show that, at the flat (Minkowskian) limit, as far as the analyticity domain is chosen properly, they precisely coincide with the usual \emph{positive-frequency}, Minkowskian plane waves of a particle with mass $m$.

According to the discussions given in subsection \ref{Subsec tube}, the dS$_4$ single-valued, global plane waves can be obtained as the boundary values, in the distribution sense, of analytic continuation to the forward (${\cal{T}}^+$) or backward (${\cal{T}}^-$) tubes of the plane-wave solutions to the (relevant) wave equations. Here, for the sake of reasoning, we stick to the waves analytic in the backward tube ${\cal{T}}^-$. Then, for the dS$_4$ scalar principal waves, which are of interest in the current discussion, we have:
\begin{eqnarray}\label{flat limit wave0}
\phi_{\nu,\xi}^{-}(x) = c \; \Big[ \vartheta\Big(\frac{x\cdot\xi}{R}\Big) + \vartheta\Big(-\frac{x\cdot\xi}{R}\Big) e^{- \mathrm{i}\pi\tau} \Big] \; \Big|\frac{x\cdot\xi}{R}\Big|^\tau,
\end{eqnarray}
where $\xi \in C^+$, $\tau=-\frac{3}{2} - \mathrm{i}\nu$ ($\nu\in\mathbb{R}$), and the real-valued constant $c \equiv c_\nu$ is:\footnote{This point will be explicitly discussed later (see Eq. (\ref{c_nu}) and its relevant discussions).}
\begin{eqnarray}\label{ooooo}
c_\nu = \sqrt{ \frac{R^{-2} \; (\nu^2 + 1/4)}{2(2\pi)^3 \; (1+e^{-2\pi\nu}) \; m^2} } \;,
\end{eqnarray}
in which $m$ stands for the Poincar\'{e}-Minkowski mass. Note that, in the sequel (as in section \ref{Sec contraction}), we consider a relation between the representation parameter $\nu$ and $m$ as $\nu = mR$ ($c=1=\hbar$).

Technically, to compute the flat limit ($R\rightarrow\infty$) of the waves (\ref{flat limit wave0}), one needs to consider an area around a given point $x \in \underline{M}_R$, in which all the distances are negligible in comparison with $R$. Regarding the homogeneity of the dS$_4$ hyperboloid $\underline{M}_R$ under the action of Sp$(2,2)$, one can take, for instance, the point $x^4 = R$ and $x^\mu = 0$, with $\mu = 0,1,2,3$.\footnote{From physical point of view, the only physical entity visible to a local observer on the dS$_4$ hyperboloid $\underline{M}_R$ is the gravitational acceleration, naively speaking, the radius of curvature $R$, which is the same all over the hyperbolid. In this sense, the observer can never understand where is he/she exactly located on $\underline{M}_R$. Therefore, for the sake of reasoning and without losing the generality, we can consider any point on $\underline{M}_R$.} Letting $R$ go to infinity, dS$_4$ spacetime in the neighborhood of this point admits its tangent plane, that is, the $1+3$-dimensional Minkowski spacetime. The coordinates $x^\mu$ in this neighborhood are approximated as (see subsection \ref{sec space-time-Lorentz dS4}, and the relations (\ref{coor Min dS4}) and (\ref{coor Min dS4'})):
\begin{eqnarray} \label{contraction coordinates}
x^\mu = x_\circ^\mu + {\cal{O}}(R^{-1})\,, \;\;\;\;\;\;\; x^4 = R + {\cal{O}}(1)\,.
\end{eqnarray}

Now, we can deal with the flat limit of the waves (\ref{flat limit wave0}). In the first step, under the limit $R \rightarrow \infty$, for which $e^{- \mathrm{i} \pi\tau} \approx 0$ (recall that $\tau = -\frac{3}{2} - \mathrm{i} mR$), these waves reduce to:
\begin{eqnarray}\label{xi<0}
\lim_{R\rightarrow\infty} \phi_{\nu,\xi}^{-}(x) \approx \lim_{R\rightarrow\infty} c_\nu \Big[ \vartheta\Big(\frac{x\cdot\xi}{R}\Big) \Big] \; \Big|\frac{x\cdot\xi}{R}\Big|^\tau.
\end{eqnarray}
Subsequently, taking Eq. (\ref{ooooo}) and the parametrization (\ref{contraction coordinates}) into account, we obtain:
\begin{eqnarray}
\lim_{R\rightarrow\infty} \phi_{\nu,\xi}^{-}(x) \approx \frac{1}{\sqrt{2(2\pi)^3}} \; \vartheta(-\xi^4) \; \Bigg[|\xi^4|^\tau \Big( 1 + \frac{\xi_\mu x_\circ^\mu}{R|\xi^4|} \Big)^{-\frac{3}{2}- \mathrm{i} mR}\Bigg]_{R\rightarrow\infty}.
\end{eqnarray}
This limit solely exists for $\xi^4 < 0$, due to the Heaviside function $\vartheta(-\xi^4)$, and for $|\xi^4|=1$, due to the term $|\xi^4|^\tau$. Therefore, we consider the orbital basis $\daleth^-_4$ on $C^+$ given in Eq. (\ref{orbit gamma4}), based upon which we can write $\xi^- \; \big( \in \daleth^-_4 \big) = \Big( \frac{k^0}{m} = \sqrt{\frac{{\vec{k}}^2}{m^2} +1}, \frac{\vec{k}}{m}, - 1 \Big)$, where $(k^0,\vec{k})$ is the four-momentum of a Minkowskian particle with mass $m$. With this choice of orbital basis for $\xi\in C^+$, the above limiting process clearly reveals that, under vanishing curvature, the dS$_4$ principal waves $\phi_{\nu,\xi}^{-}(x)$ meet the usual \emph{positive-frequency}, Minkowskain plane waves of a particle with mass $m$:
\begin{eqnarray}\label{wave flat limit}
\lim_{R\rightarrow\infty} \phi_{\nu,\xi}^{-}(x) =
\left \{ \begin{array}{rl} \frac{1}{\sqrt{2(2\pi)^3}} \; e^{- \mathrm{i} k \cdot x^{}_\circ}, \;\;\;\;\;\;\; {\mbox{for}} \;\; x\cdot\xi > 0\,,\vspace{2mm}\\
\vspace{2mm} 0\,, \hspace{2.2cm}\;\;\;\; {\mbox{for}} \;\; x\cdot\xi < 0\,,\end{array}\right.
\end{eqnarray}
where $k \cdot x^{}_\circ \equiv \eta^{}_{\mu\nu}k^\mu x_\circ^\nu$, with $\mu,\nu = 0,1,2,3$, and $\eta^{}_{\mu\nu} = \mbox{diag}(1,-1,-1,-1)$.

So far, we have shown that under vanishing curvature, thanks to the analyticity requirement at the origin of the term $e^{- \mathrm{i} \pi\tau}$, the modes with $x\cdot\xi < 0$, responsible for the appearance of negative energy modes at the flat limit, are exponentially damped, while the ones with $x\cdot\xi > 0$ meet the legitimate (positive energy) Minkowskian on-shell modes of a particle with mass $m$. Regarding this remarkable result, two critical points must be underlined here. First, this limiting process clearly does not depend on the point we choose. Second, the above result does not mean that the energy concept can be defined globally in dS$_4$ (generally, dS) spacetime. Actually, applying Bogoliubov transformations on the given modes $\phi_{\nu,\xi}^-$ may result in the appearance of conjugate modes $(\phi_{\nu,\xi}^-)^{\ast}$ which their flat limit at some point $x^\prime$ is of negative energy, as soon as the point $x^\prime$ verifies $x^\prime\cdot\xi > 0$ (one can easily check the latter point by repeating the above process for $(\phi_{\nu,\xi}^-)^{\ast}$).

As a final remark, we would like to recall that, on a purely group-theoretical level based upon an \emph{ad hoc} process of contraction, the dS$_4$ principal UIR's contract (at the zero-curvature limit) towards a direct sum of the massive Poincar\'{e} UIR's with positive and negative energy (see Eq. (\ref{massive contraction})). This feature could cause confusion in understanding the physical content of dS$_4$ (generally, dS) relativity under vanishing curvature, strictly speaking, under the Poincar\'{e} contraction of the representations, because it somehow suggests that from the point of view of a local (tangent) Minkowskian observer the curvature is in some sense responsible for the appearance of negative energy modes in the theory. The above result, however, as we will discuss in the next section, lifts this ambiguity and allows for the implementation of group representation theory, in terms of the dS$_4$ (generally, dS) plane waves, to achieve a promising QFT formulation of dS$_4$ (generally, dS) elementary systems (in the Wigner sense) in such a way that, under vanishing curvature, the whole QFT construction meets the ordinary flat one.

\setcounter{equation}{0} \section{QFT in dS$_4$ spacetime}\label{Sec QFT}
In this section, following the seminal works by Bros et al. \cite{Bros 2point func,GazeauPRL}, we are going to present a consistent QFT reading of (free) elementary systems in dS$_4$ spacetime formulated based on a set of fundamental principles, which closely parallel the Wightman axioms for Minkowskian fields, while the usual spectral condition of ``positivity of the energy" is replaced by a certain geometric KMS condition. The latter is equivalent to an exact thermal manifestation of the associated ``vacuum" states. Technically, as already mentioned, the whole quantization process is carried out in terms of the global dS$_4$ plane waves (introduced in the previous section); thanks to the existence of the related (global) dS$_4$-Fourier calculus, one can implement many significant notions like wave propagation, ``particle states", etc. in the context of dS$_4$ QFT quite analogous to its Minkowskian counterpart.

Here and before going into the details, let us once again point out that in this section, in order to keep the argument straight, we merely stick to the case of dS$_4$ scalar quantum fields $\hat{\phi}(x)$, while, technically, we adopt the usage that is (often tacitly) adopted in the vast majority of QFT textbooks to call ``quantization" a Hilbertian Fock space realization of the field algebra (see, for instance, Ref. \cite{Streater}).

\subsection{Local dS$_4$ scalar fields: generalized free fields}\label{Subsec gen free fields}
The starting point, to get an elaborate formulation of the theoretical framework that is meant to be presented here, is to consider the \emph{Borchers-Uhlmann algebra} $\mathfrak{B}$ \cite{Borchers} of terminating sequences of \emph{test-functions} $f=\big(f_0,f_1(x_1), \; ... \;, f_n(x_1,\;...\;,x_n), \; ... \;, 0, 0, \;...\; \big)$ on $\underline{M}_R$, where $f_0\in\mathbb{C}$ and $f_n\in \mathfrak{D}_n$ ($n\geqslant 1$), $\mathfrak{D}_n$ being the space of infinitely differentiable functions with compact support on the cartesian product of $n$ copies of $\underline{M}_R$; $\mathfrak{D}_n \equiv \mathfrak{D}_n\big([\underline{M}_R]^n\big)$. [Note that here the subscript `$n$', as a dynamical variable, refers to the number of particles appeared in the theory; the total particle number is finite, no matter how large: $\sum n_i^{} <\infty$.] In this $\star$-algebra, the product and the involution are respectively given by:
\begin{eqnarray}
(fh)_n = \sum_{\substack{s,t \in \mathbb{N}\\ s+t=n }} f_s \otimes h_t\,,
\end{eqnarray}
\begin{eqnarray}
f \;\mapsto\; f^\ast = \big(f^\ast_0,f^\ast_1(x_1), \; ... \;,f^\ast_n(x_n,\;...\;,x_1), \; ... \;, 0, 0, \;...\; \big)\,,
\end{eqnarray}
where $(f_s \otimes h_t)(x_1,\;...\;,x_{s+t})=f_s(x_1,\;...\;,x_s)h_t(x_{s+1},\;...\;,x_{s+t})$. This algebra is supposed to be equipped with a representation $\underline{U}(\underline{g})$ of the dS$_4$ group Sp$(2,2)$ as follows:
\begin{eqnarray}
\underline{U}(\underline{g}) f = \Big(f_0,\big(\underline{U}(\underline{g})f_{1}\big)(x_1), \; ... \;, \big(\underline{U}(\underline{g}) f_{n}\big)(x_1,\; ... \;,x_n), \;... \;, 0, 0, \;...\; \Big)\,,
\end{eqnarray}
where, besides $f_0$ which is invariant under this action, for $f_{n}$ ($n\geqslant 1$) in a natural way we have:
\begin{eqnarray}
\big(\underline{U}(\underline{g}) f_{n}\big)(x_1,\; ... \;,x_n) = f_{n}\big(\underline{g}^{-1}\diamond x_1, \; ... \;,\underline{g}^{-1}\diamond x_n\big)\,.
\end{eqnarray}

A local QFT then is defined by a continuous \emph{linear functional} ${\cal{W}}(f)$ on $\mathfrak{B}$, that is, a sequence $\big\{ {\cal{W}}_n(f_n) \in \mathfrak{D}^\prime_n, \; n \in \mathbb{N} \big\}$, in which ${\cal{W}}_n(f_n)$'s are distributions (Wightman $n$-point functions):
\begin{eqnarray}
{\cal{W}}_0(f_0) = 1\,, \;\;\;\;\;\;\;\mbox{and} \;\;\;\;\;\;\; \Big\{ {\cal{W}}_n(f_{n}) = \int {\cal{W}}_n(x_1,\;...\;,x_n) f_n(x_1,\;...\;,x_n) \; \mathrm{d}\mu(x_1)\; ...\; \mathrm{d}\mu(x_n), \; n\geqslant 1 \Big\}\,,
\end{eqnarray}
where $\mathrm{d}\mu(x)$ is the invariant measure on $\underline{M}_R$. [The algebra $\mathfrak{B}$ actually allows to elegantly collect the Wightman $n$-point functions to one linear functional ${\cal{W}}(f)$.] In this context, the following conditions must be verified:
\begin{itemize}
\item{\emph{Covariance}. Each ${\cal{W}}_n$ is invariant under the dS$_4$ group action, i.e., ${\cal{W}}_n\big( \underline{U}(\underline{g}) f_{n}\big) = {\cal{W}}_n(f_{n})$, for all $\underline{g} \in \mathrm{Sp}(2,2)$. This is of course equivalent to the invariance of the functional ${\cal{W}}$ itself, i.e., ${\cal{W}}\big(\underline{U}(\underline{g}) f\big) = {\cal{W}}(f)$, for all $\underline{g} \in \mathrm{Sp}(2,2)$.}
\item{\emph{Locality}. There is a ``locality" ideal ${\cal{I}}_{loc}$ in $\mathfrak{B}$, which is defined in the same way as the Minkowskian case (but with respect to the dS$_4$ spacelike separation; see section \ref{Sec dS4 causal structure}), in which the functional ${\cal{W}}$ vanishes (i.e., ${\cal{W}}(f)= 0$, for all $f \in {\cal{I}}_{loc}$).}
\item{\emph{Positivity}. For each $f \in \mathfrak{B}$, characterized by $f_0\in\mathbb{C}$, $f_1\in \mathfrak{D}_1(\underline{M}_R)$, \;...\;, $f_n\in \mathfrak{D}_n\big([\underline{M}_R]^n\big)$:
     \begin{eqnarray}
     \sum_{s,t=0}^n {\cal{W}}_{s+t}( f^\ast_s \otimes f^{}_t ) \geqslant 0\,,
     \end{eqnarray}
     which is equivalent to the positivity of the functional ${\cal{W}}$ itself (i.e., ${\cal{W}}(f^\ast f)\geqslant 0$, for all $f \in \mathfrak{B}$). Note that this assumption should be relaxed to deal with dS$_4$ gauge QFT's \cite{Strocchi}.}
\end{itemize}
Here, we must underline that, at this initial level of generality, it is not necessary to assume any wave equation for ${\cal{W}}_n$.

Now, quite analogous to the Minkowskian case, the \emph{Gel'fand-Naimark-Segal (GNS) construction} (see, for instance, Ref. \cite{Haag}) assures that, given a state ${\cal{W}}(f)$ on $\mathfrak{B}$, one can find a Hilbertian Fock space ${\mathscr{H}}$, a UIR $\underline{U}(\underline{g}) \mapsto \underline{\mathscr{U}}(\underline{g})$ of the dS$_4$ group,\footnote{By $\underline{U}(\underline{g}) \mapsto \underline{\mathscr{U}}(\underline{g})$ is meant the extension of the natural representation $\underline{U}$ of the dS$_4$ group on a Hilbert space ${\cal{H}}$ to the corresponding Fock space ${\mathscr{H}}$, where $\underline{\mathscr{U}}$ denotes the extended representation.} a vacuum vector $\Omega \in {\mathscr{H}}$ invariant under $\underline{\mathscr{U}}$, and finally an operator-valued distribution $\hat\phi$, for which the sequence of its Wightman functions reads:
\begin{eqnarray}
\Big\{ {\cal{W}}_n(x_1,\;...\;,x_n) = \langle\Omega, \hat\phi(x_1) \; ... \; \hat\phi(x_n) \Omega\rangle, \; n\in\mathbb{N} \Big\}\,.
\end{eqnarray}
Moreover, the GNS construction provides us with the vector-valued distribution $\hat{\Phi}_n$ in such a way that:
\begin{eqnarray}
\hat{\Phi}_n(f_n) = \int \Big( \hat\phi(x_1) \; ... \; \hat\phi(x_n)\Omega\Big) f_n(x_1,\;...\;,x_n) \; \mathrm{d}\mu(x_1) \;...\; \mathrm{d}\mu(x_n)\,.
\end{eqnarray}
Note that $\hat{\Phi}_n$ is actually an (unbounded) operator acting on ${\mathscr{H}}$, defined on some dense domain of ${\mathscr{H}}$. The GNS construction also gives a representation $f \mapsto \hat\Phi(f)=\big\{ \hat{\Phi}_n(f_n), \; n \in \mathbb{N} \big\}$ of $\mathfrak{B}$, which contains the basic field $\hat\phi(x)$ as a special case:
\begin{eqnarray}
\hat\phi(f_1) \equiv \hat\Phi\big( (0,f_1,0,\;...\;) \big) = \int \hat\phi(x) f_1(x) \; \mathrm{d}\mu(x)\,.
\end{eqnarray}
The operator $\hat\phi(f_1)$ (that is, a basic field $\hat\phi(x)$ \emph{smeared out} with a test function $f_1$) represents a physical operation performed on the system within the spacetime region determined by the support of $f_1$. Roughly speaking, the argument $x$ of a basic field has direct physical significance. It marks the point where $\hat\phi(x)$ applied to a state produces a change. In this sense, using the basic fields, one can associate with each open region ${\cal{O}}$ in $\underline{M}_R$ a polynomial algebra $P({\cal{O}})$ of operators on Fock space, that is, the algebra generated by all $\hat\Phi(f_0,f_1,\;...\;,f_n,\;...\;)$, the fields smeared out with test functions $f_n(x_1,\;...\;,x_n)$ possessing their support in the region ${\cal{O}}$; $\mbox{supp} f_n(x_1,\;...\;,x_n) \subset {\cal{O}}^{n}$, for $n\geqslant 1$. The set $P({\cal{O}}) \Omega$ is actually a dense subset of ${\mathscr{H}}$. Here, we must underline that our interpretation of the theory is indeed based on the assertion that the elements of this subalgebra of $\hat\Phi(\mathfrak{B})$ can be interpreted as representing physical operations performable within ${\cal{O}}$. This suggests that the net of algebras $P({\cal{O}})$ constitutes the intrinsic mathematical description of the theory. [To get more mathematical details, concerning the above discussions, we encourage readers to refer to Ref. \cite{Haag}.]

At this stage, focusing on a (generalized) free scalar field, we make the above construction more explicit. On the free field level, all the truncated $n$-point functions ${\cal{W}}^{\mbox{\small{tr}}}_n(x_1,\;...\;,x_n)$, with $n \neq 2$, vanish. The quantum theory therefore is completely encoded in the two-point function ${\cal{W}}={\cal{W}}_2(x,x^\prime)$, that is, a distribution on $\underline{M}_R\times \underline{M}_R$ verifying the aforementioned requirements of covariance, locality, and positivity, which imply that:
\begin{itemize}
\item{\emph{Covariance}. For all $\underline{g} \in \mathrm{Sp}(2,2)$, ${\cal{W}}(\underline{g}\diamond x,\underline{g}\diamond x^\prime) = {\cal{W}}(x,x^\prime)$.}
\item{\emph{Locality}. For every spacelike separated pair $(x,x^\prime) \in \underline{M}_R$, namely, $(x-x^\prime)^2<0$ (see section \ref{Sec dS4 causal structure}), ${\cal{W}}(x,x^\prime) = {\cal{W}}(x^\prime,x)$.}
\item{\emph{Positivity}. For all $f_1 \in \mathfrak{D}_1(\underline{M}_R)$:
     \begin{eqnarray} \label{Positivity condition}
     \int_{\underline{M}_R\times \underline{M}_R} {\cal{W}}(x,x^\prime) f_1^\ast(x) f^{}_1(x^\prime) \; \mathrm{d}\mu(x) \mathrm{d}\mu(x^\prime) \geqslant 0\,.
     \end{eqnarray}}
\end{itemize}
The GNS triplet $({\mathscr{H}},\hat\phi,\Omega)$, corresponding to the functional ${\cal{W}}={\cal{W}}_2(x,x^\prime)$, then can be explicitly constructed. Actually, this triplet indicates the Fock representation of a generalized free scalar field $\hat\phi$ verifying the following commutation relations:
\begin{eqnarray}\label{commu}
\big[\hat\phi(f_1),\hat\phi(k^{}_1)\big] = C(f^{}_1,k^{}_1) \times \mathbbm{1} = \int C(x,x^\prime) f_1(x) k^{}_1(x^\prime) \; \mathrm{d}\mu(x) \mathrm{d}\mu(x^\prime) \times \mathbbm{1}\,,
\end{eqnarray}
for all $f_1,k_1 \in \mathfrak{D}_1$, while the commutator function is given by $C(x,x^\prime) = {\cal{W}}(x,x^\prime) - {\cal{W}}(x^\prime,x)$. The associated Hilbertian Fock space representation ${\mathscr{H}}$ of the field algebra is defined by the Hilbertian sum ${\cal{H}}_0 \oplus \big[\oplus_n S ({\cal{H}}_1)^{\otimes n}\big]$, in which $S$ stands for the symmetrization operation, ${\cal{H}}_0 = \big\{ \lambda \Omega \; ; \; \lambda \in \mathbb{C} \big\}$, and the one-particle sector ${\cal{H}}_1$ is defined in such a way that:
\begin{itemize}
\item{A regular element $\dot{h}_1 \in {\cal{H}}_1$ is given by a class of functions $h_1(x) \in \mathfrak{D}_1(\underline{M}_R)$ modulo the functions $k_1$, for which $\int_{\underline{M}_R\times \underline{M}_R} {\cal{W}}(x,x^\prime) k_1^\ast(x) k^{}_1(x^\prime) \; \mathrm{d}\mu(x) \mathrm{d}\mu(x^\prime) = 0$.}
\item{The associated norm, for such an element $\dot{h}_1$, reads:
     \begin{eqnarray}\label{norm Hilbert}
     \langle \dot{h}_1 , \dot{h}_1 \rangle = \int_{\underline{M}_R\times \underline{M}_R} {\cal{W}}(x,x^\prime) h_1^\ast(x) h^{}_1(x^\prime) \; \mathrm{d}\mu(x) \mathrm{d}\mu(x^\prime) \geqslant 0\,.
     \end{eqnarray}}
\item{Eventually, by completion of the space of regular elements, with respect to the above norm, the full Hilbert space ${\cal{H}}_1$ is defined.}
\end{itemize}
Note that a similar procedure for the regular elements $\dot{h}_n \in {\cal{H}}_n = S ({\cal{H}}_1)^{\otimes n}$ is performed.

Each field operator $\hat\phi(f_1)$ can be decomposed into two parts, i.e., ``creation" and ``annihilation" parts: $\hat\phi(f_1) = a^\dagger_{}(f_1) + a(f_1)$. The actions of $a^\dagger_{}(f_1)$ and $a(f_1)$, respectively, on the dense subset of ``regular elements" of the form $\dot{h} = \big( \dot{h}_0, \dot{h}_1, \;...\;, \dot{h}_n, \;...\;,0,0, \; ...\; \big)$ are given by:
\begin{eqnarray}\label{creation}
\Big( a^\dagger_{}(f_1) \dot{h} \Big)_n (x_1,\;...\;, x_n) = \frac{1}{\sqrt{n}} \sum_{i=1}^n f_1(x_i) \; \dot{h}_{n-1}(x_1,\;...\;,\breve{x}_i,\;...\;, x_n)\,,
\end{eqnarray}
\begin{eqnarray}\label{annihilation}
\Big( a(f_1) \dot{h} \Big)_n (x_1,\;...\;, x_n) = \sqrt{n+1} \int_{\underline{M}_R\times \underline{M}_R} f_1(x) {\cal{W}}(x,x^\prime) \; \dot{h}_{n+1}(x^\prime,x_1, \;...\; , x_n) \; \mathrm{d}\mu(x) \mathrm{d}\mu(x^\prime)\,,
\end{eqnarray}
where $\breve{x}_i$ means that this term is omitted. One can easily show that the above formulas give the commutation rules (\ref{commu}).

In summary, within the above framework, on one hand, due to the locality condition of ${\cal{W}}(x,x^\prime)$, the antisymmetric bidistribution $C(x,x^\prime)$ on $\underline{M}_R$ vanishes coherently with the notion of locality inherent to $\underline{M}_R$, i.e., $C(x,x^\prime) = 0$, for every spacelike separated pair $(x,x^\prime) \in \underline{M}_R$. This clearly compels the scalar field $\hat\phi(x)$ to verify the requirement of local commutativity (since $\big[\hat\phi(x),\hat\phi(x^\prime)\big] = C(x,x^\prime)\times\mathbbm{1}$). On the other hand, the covariance requirement of the two-point function necessitates the dS$_4$ covariance of the field operator $\hat\phi(x)$, while the representation $\underline{\mathscr{U}}(\underline{g})$ of the dS$_4$ group in $\mathscr{H}$ is unitary with respect to the same requirement and to the given norm in (\ref{norm Hilbert}).

\subsubsection{Discussion: weak spectral condition}
So far, the requirements of covariance, locality, and positive definiteness are literally carried over from the Minkowskian case to the dS$_4$ one. In dS$_4$ (generally, dS) spacetime, however, no literal or unique analogue of the usual spectral condition of ``positivity of the energy" exists, even worse, it is impossible to define the notion of ``energy" at all; no matter what generator of the dS$_4$ (generally, dS) group is taken into account, the associated Killing vector field, though perhaps timelike in some region of the spacetime, is spacelike in some other region (see subsection \ref{Subsec Energy operator}). Because of this ambiguity, one encounters many inequivalent QFT's (or in other words, the phenomenon of nonuniqueness of the vacuum state) for any single dS$_4$ (generally, dS) field model. As a matter of fact, in the absence of a canonical choice of a time coordinate, based upon which one can classify modes as being positive or negative frequency, the appeared QFT's are mostly relevant to specific choices of time coordinates, which yield associated frequence splittings.

Of course, if one sticks to the free field level, technically there is a possible way out to circumvent the absence of a true spectral condition in dS$_4$ (generally, dS) QFT and to single out a distinguished vacuum state for linear fields. It is indeed a well-established fact that for a wide class of spacetimes with bifurcate Killing horizons, including dS$_4$ (generally, dS) spacetime, the \emph{Hadamard requirement} selects a distinguished vacuum state for linear fields (see Refs. \cite{Haag,Allen,Kay} and references therein). The Hadamard requirement actually postulates that two-point functions of linear fields, for instance, Klein-Gordon fields on curved spacetime, at short distances should behave same as their Minkowskian free-field counterparts. Through this postulate, the selected vacuum state for dS$_4$ (generally, dS) linear fields coincides with the so-called Euclidean \cite{Gibbons} or Bunch-Davies \cite{Bunch} vacuum state. Nevertheless, if one desires to get involved with general interacting fields, the too special character of the Hadamard requirement (which confines it to the free field level) necessitates one to seek another explanation of the existence of preferred vacuum states in the global structure of dS$_4$ (generally, dS) spacetime.

In view of the above considerations, in this section, following the sound arguments given in Refs. \cite{Bros 2point func,GazeauPRL} by Bros et al., we are going to present a rigorous mathematical framework, based on analyticity requirements of $n$-point functions, which sheds a new light on the \emph{preferred representations} of dS$_4$ (generally, dS) QFT (see Ref. \cite{Kay} and references therein) and on the way they solve the problem of the absence of a true spectral condition plaguing QFT in dS$_4$ (generally, dS) spacetime. Of course, in this context, the Hadamard requirement still remains of great significance and indicates the necessity for two-point functions to be the boundary values of analytic functions ``from the good side" (the well-known $i\epsilon$-rule.)

In order to prepare the ground, we recall from Minkowskian QFT the well-known fact that the spectral condition is equivalent to \emph{specific} analyticity properties of the Wightman $n$-point functions ${\cal{W}}^{}_{\circ n}\big( (x^{}_{\circ})^{}_1,\;...\;, (x^{}_{\circ})^{}_n \big)$ \cite{Streater}, arising from the Laplace transform theorem in ${\mathbb{C}}^{4n}$. [Again, the subscript `$\circ$' marks the entities defined in $1+3$-dimensional Minkowski spacetime ${\mathbb{R}}^4$ or in its complex version ${\mathbb{C}}^4$.] Technically, these analyticity properties necessitate that, for each $n$ ($n > 1$), the distribution ${\cal{W}}^{}_{\circ n}\big( (x^{}_{\circ})^{}_1,\;...\;, (x^{}_{\circ})^{}_n \big)$ is the boundary value of an analytic function $W^{}_{\circ n}\big( (z^{}_{\circ})^{}_1,\;...\;, (z^{}_{\circ})^{}_n \big)$ given in the tube:
\begin{eqnarray}
T_{\circ}^{+(n)} = \Big\{ \big( (z^{}_{\circ})^{}_1,\;...\;, (z^{}_{\circ})^{}_n \big) \; &;& \; (z^{}_{\circ})^{}_k = (x^{}_{\circ})^{}_k + \mathrm{i} (y^{}_{\circ})^{}_k \in {\mathbb{C}}^4,\; 1\leqslant k \leqslant n \nonumber\\
\; &;& \; (y^{}_{\circ})^{}_{j+1} - (y^{}_{\circ})^{}_j \in \interior{\underline{V}}_{\circ}^+, \; 1\leqslant j \leqslant n-1 \Big\}\,,
\end{eqnarray}
where the domain $\interior{\underline{V}}_{\circ}^+$ stems from the causal structure in ${\mathbb{R}}^4$ (the definition of $\interior{\underline{V}}_{\circ}^+$ can be simply understood from the corresponding one given in section \ref{Sec dS4 causal structure} for ${\mathbb{R}}^5$). To see the points lying behind the above argument, it would be convenient to briefly review what happens, for instance, in the case of the free (massive) Klein-Gordon scalar field in Minkowski spacetime. Considering the associated Fourier representation, its two-point function ${\cal{W}}^{}_{\circ} = {\cal{W}}^{}_{\circ 2}(x_{\circ},x_{\circ}^\prime)$ reads:
\begin{eqnarray}\label{qqq}
{\cal{W}}^{}_{\circ}(x_{\circ},x_{\circ}^\prime) = \frac{1}{2(2\pi)^2} \int e^{- \mathrm{i} k\cdot x^{}_{\circ}} \; e^{ \mathrm{i} k \cdot x_{\circ}^\prime} \; \vartheta(k^0) \; \delta\big((k)^2 - m^2\big) \; \mathrm{d}^4 k\,,
\end{eqnarray}
where $\vartheta$ refers to the Heaviside function. Note that the measure $\mathrm{d}\mu = \vartheta(k^0) \; \delta\big((k)^2 - m^2\big) \; \mathrm{d}^4 k$ is considered to solve the corresponding Klein-Gordon wave equation and to fulfill the spectral condition \cite{Streater}. According to the spectral condition, or by direct inspection from the convergence properties of the integral at the right-hand side of the above equation, one can easily check that the distribution ${\cal{W}}^{}_{\circ}(x^{}_{\circ},x_{\circ}^\prime)$ appears as the boundary value of a function $W^{}_{\circ}(z^{}_{\circ},z_{\circ}^\prime)$ holomorphic in the domain $T_{\circ}^{+(2)}$, while the boundary value is taken from $T_{\circ}^{+(2)}$.

In the case of dS$_4$ spacetime (embedded in ${\mathbb{R}}^5$), a natural substitute to the above analyticity properties is to requiring the given ${\cal{W}}_n(x_1,\;...\;,x_n)$, for each $n>1$, being the boundary value (in the distribution sense) of a function $W_n(z_1,\;...\;,z_n)$ holomorphic in:
\begin{eqnarray}
{\cal{T}}^{+(n)} = T^{+(n)} \cap \big[\underline{M}_R^{(\mathbb{C})}\big]^n = \Big\{ ( z^{}_1, \;...\;, z^{}_n ) \; &;& \; z^{}_k = x^{}_k + \mathrm{i} y^{}_k \in \underline{M}_R^{(\mathbb{C})}, \; 1\leqslant k \leqslant n \nonumber\\
\; &;& \; y^{}_{j+1} - y^{}_j \in \interior{\underline{V}}^+, \; 1\leqslant j \leqslant n-1 \Big\}\,,
\end{eqnarray}
where $T^{+(n)}$ is the tube associated with $\mathbb{C}^{5n}$ and, again, $\interior{\underline{V}}^+$ is the domain resulting from the causal structure on $\underline{M}_R$ (see section \ref{Sec dS4 causal structure}). As a matter of fact, it has been shown, by Bros et al. \cite{Bros 2point func,Bros 1998}, that ${\cal{T}}^{+(n)}$ is a domain of $\big[\underline{M}_R^{(\mathbb{C})}\big]^n$ and a tuboid above $\big[\underline{M}_R \big]^n$ such that the concept of ``distribution boundary value of a holomorphic function from this domain" remains meaningful. In this sense, it is legitimate to supplement the postulates of covariance, locality, and positivity (pointed out above) by imposing:
\begin{itemize}
\item{\emph{Weak spectral condition}. The distribution ${\cal{W}}_n(x_1,\;...\;,x_n)$, for each $n>1$, is obtained by taking the boundary value of a function $W_n(z_1,\;...\;,z_n)$ holomorphic in the subdomain ${\cal{T}}^{+(n)}$ of $\big[\underline{M}_R^{(\mathbb{C})}\big]^n$.}
\end{itemize}
For a general dS$_4$ two-point function, the above postulate of \emph{normal analyticity} explicitly reads:
\begin{itemize}
\item{The two-point function ${\cal{W}} = {\cal{W}}_2(x,x^\prime) = \langle\Omega, \hat{\phi}(x) \hat{\phi}(x^\prime) \Omega\rangle$ is the boundary value (in the sense of distribution) of a function $W(z,z^\prime)$ analytic in the tuboid domain:
    \begin{eqnarray}
    {\cal{T}}^{+(2)}_{} = \Big\{ (z,z^\prime) \; ; \; z\in {\cal{T}}_{}^-, \; z^\prime \in {\cal{T}}_{}^+ \Big\}\,.
    \end{eqnarray}}
\end{itemize}
[We recall from subsection \ref{Subsec tube} that ${\cal{T}}_{}^{\pm} = T_{}^{\pm} \cap \underline{M}_R^{(\mathbb{C})}$ are respectively the forward and backward tubes of $\underline{M}_R^{(\mathbb{C})}$.]

In the coming subsection, following Refs. \cite{Bros 2point func,GazeauPRL} and by further restricting our attention to the free dS$_4$ Klein-Gordon scalar fields, we will examine the above instruction for a QFT reading of dS$_4$ elementary systems, by presenting the corresponding two-point functions possessing the requirements of covariance, locality, positive definiteness, and normal analyticity. Again, the whole process will be operated in terms of the (related) global dS$_4$ plane-wave type solutions (see the previous section).\footnote{To see relevant arguments for higher spin fields in dS$_4$ spacetime, one can refer to Refs. \cite{Massive 1,Massive 2,Massive 3/2,Massive/Massless 1/2,Massless 1,BehrooziTakook}.} Technically, the genuine dS$_4$-Fourier calculus arising from the latter allows for a spectral analysis of the two-point functions very similar to the Minkowskian case. In this context, we will show that the above set of postulates is entirely encoded in the following \emph{maximal analyticity properties}\footnote{By maximal type is meant that, with respect to the above set of postulates and without imposing very specific conditions (for instance, local commutativity at timelike separation), one cannot enlarge the (physical sheet) holomorphy domain of the corresponding two-point functions.} of the two-point functions $W(z,z^\prime)$:
\begin{itemize}
\item{$W(z,z^\prime)$'s can be analytically continued in the \emph{cut-domain}:
      \begin{eqnarray}\label{cut-domain}
      \Delta = \underline{M}^{(\mathbb{C})}_R \times \underline{M}^{(\mathbb{C})}_R \backslash \Gamma_{}^{(\mathbb{C})}\,,
      \end{eqnarray}
      where \emph{the cut} $\Gamma_{}^{(\mathbb{C})}$ is the set $\big\{ (z,z^\prime) \;;\; z\in\underline{M}^{(\mathbb{C})}_R \times \underline{M}^{(\mathbb{C})}_R, \; (z-z^\prime)^2 = \varrho, \; \forall \varrho \geqslant 0\big\}$.}
\item{$W(z,z^\prime)$'s verify, in the cut-domain $\Delta$, the complex covariance requirement:
     \begin{eqnarray}\label{covariance requirement}
     W(\underline{g}\diamond z,\underline{g}\diamond z^\prime) = W(z,z^\prime)\,,
     \end{eqnarray}
     for all $\underline{g}$ in the complexified dS$_4$ group $\mathrm{Sp}(2,2)_{}^{(\mathbb{C})}$.}
\item{The \emph{permuted Wightman two-point functions} ${\cal{W}}(x^\prime,x) = \langle\Omega, \hat\phi(x^\prime) \hat\phi(x) \Omega\rangle$ are the boundary values of $W(z,z^\prime)$'s from the domain ${\cal{T}}^{-(2)}_{} = \big\{ (z,z^\prime) \; ; \; z\in {\cal{T}}_{}^+, \; z^\prime \in {\cal{T}}_{}^- \big\}$.\footnote{Note that this postulate can also be trivially rephrased as: the permuted Wightman two-point functions ${\cal{W}}(x^\prime,x)$ are obtained by taking the boundary values of $W(z^\prime,z)$'s from the domain ${\cal{T}}^{+(2)}_{} = \big\{ (z,z^\prime) \; ; \; z\in {\cal{T}}_{}^-, \; z^\prime \in {\cal{T}}_{}^+ \big\}$. Nevertheless, the former statement is of more interest here, since through it, each Wightman two-point function ${\cal{W}}(x,x^\prime)$ and the corresponding permuted one ${\cal{W}}(x^\prime,x)$ can be seen as \emph{two different realizations}, strictly speaking, the boundary values from two different domains ${\cal{T}}^{+(2)}_{}$ and ${\cal{T}}^{-(2)}_{}$ (respectively), of same (relevant) analytic two-point function $W(z,z^\prime)$.} [For the sake of simplicity, from now on, the two-point functions $W(z,z^\prime)$, which are analytic in the tuboid domain ${\cal{T}}^{-(2)}_{}$, are denoted by $W^\prime(z,z^\prime)$, and correspondingly, ${\cal{W}}(x^\prime,x)$ by ${\cal{W}}^\prime(x,x^\prime)$.]}
\end{itemize}
In this maximal analytic framework, the given two-point functions specify the dS$_4$ scalar free fields in a \emph{preferred representation}, which is interestingly characterized by a well-established KMS condition defined in proper domains of $\underline{M}_R$ relevant to geodesic observers. This feature remarkably provides us with a simple geometrical interpretation to the Hawking's thermal effects in dS$_4$ Universe \cite{Gibbons,Kay}. As a matter of fact, the KMS condition, as far as maximal analyticity requirements on the temporal geodesics remain verified, represents the ``natural substitute" to the usual spectral condition (here, the Minkowskian linear geodesics are replaced by hyperbolas). It follows that the selected vacuum representation for the fields, in spite of its thermal features, would be the exact counterpart of its relevant Minkowski vacuum representation (the latter appears as the null-curvature limit of the former).

The case of interacting fields of course is more elaborate, since maximal analyticity properties cannot be expected to hold for $n$-point functions ($n > 2$). In this sense, again one is left with the task of finding a proper general setting for dS$_4$ (generally, dS) QFT. However, the above axiomatic approach paves the road for formulating such a theory of interacting fields in dS$_4$ (generally, dS) spacetime (see Refs. \cite{Bros 2point func,GazeauPRL}).

\subsection{(Analytic) Wightman two-point functions for the dS$_4$ (principal and complementary) Klein-Gordon scalar fields} \label{Subsec 2point function}
In the beginning, let us point out that, in this subsection, we merely study the dS$_4$ principal and complementary Klein-Gordon scalar fields. The study of the dS$_4$ discrete scalar case is postponed to the next subsection.

\subsubsection{Plane-wave analysis of the two-point functions}
We begin our discussion with the dS$_4$ principal Klein-Gordon scalar field. With respect to the analytic plane-wave type solutions given in the previous section, we consider the following integral representation of the corresponding two-point function \cite{Bros 2point func,GazeauPRL}:
\begin{eqnarray}\label{2-point kernel}
W_\nu = W(z,z^\prime) = c^{2}_\nu \int_\daleth \Big( \frac{z\cdot\xi}{R} \Big)^{-\frac{3}{2} - \mathrm{i}\nu} \Big( \frac{\xi \cdot z^\prime}{R} \Big)^{-\frac{3}{2} + \mathrm{i}\nu} \; \mathrm{d}\mu^{}_\daleth(\xi)\,, \;\;\;\;\;\;\; \nu\in\mathbb{R}\,,
\end{eqnarray}
where: (i) $z,z^\prime \in \underline{M}_R^{(\mathbb{C})}$, respectively, belong to the backward and forward tubes of $\underline{M}_R^{(\mathbb{C})}$, i.e., $z\in {\cal{T}}_{}^-, \; z^\prime \in {\cal{T}}_{}^+$, (ii) the normalization factor $c^{2}_\nu$ is a positive constant, which will be determined later by applying the local Hadamard condition, and finally (iii) the integration is performed along any orbital basis $\daleth$ of the future null-cone $C^+$ (see subsection \ref{Subsec orbial basis}), while $\mathrm{d}\mu^{}_\daleth$ represents the natural $C^+$ invariant measure on $\daleth$ induced from the $\mathbb{R}^5$ Lebesgue measure. By construction, it is evident that the two-point function $W_\nu (z,z^\prime)$, with respect to both variables $z$ and $z^\prime$, is a solution to the complex version of the dS$_4$ (principal) Klein-Gordon equation (\ref{Wave Eq. scalar ambient}) (with $\tau=-3/2- \mathrm{i} \nu$ and $\nu\in\mathbb{R}$), which is analytic in the tuboid domain ${\cal{T}}^{+(2)}_{}$.

Regarding the integral representation (\ref{2-point kernel}), first of all one must notice that it defines the same analytic two-point function for all $\daleth \subset C^+$, namely, $W^{}_\nu = W^{(\daleth)}_\nu$. In other words, the value of the integral (\ref{2-point kernel}), for given $(z,z^\prime) \in {\cal{T}}^{+(2)}_{}$, is independent of the choice of orbital basis $\daleth$. This point actually stems from the fact that the corresponding integrand appears as the restriction to $\daleth$ of a \emph{closed differential form}\footnote{A closed form is a differential form $\alpha$ whose exterior derivative is zero, $\mathrm{d}\alpha=0$.} \cite{Bros 2point func}. Moreover, for given $(z,z^\prime) \in {\cal{T}}^{+(2)}_{}$, the integrability of (\ref{2-point kernel}) at infinity on noncompact bases $\daleth$ of the type relevant to the stabilizer subgroup $\underline{{\cal{S}}}(x_e^{}) \sim \mathrm{SO}_0(1,3)$ (see subsection \ref{Subsec orbial basis}) is ensured by the homogeneity properties of the integrand, which is clearly a homogeneous function of $\xi$ (on $C^+$) with degree of homogeneity $-3$. To see the point, we can consider, for instance, the basis $\daleth=\daleth^{}_4 \; \big( = \daleth^+_4 \cup \daleth^-_4 \big)$ given in Eq. (\ref{orbit gamma4}). Then, we can write $\xi^\pm_{} \; \big( \in \daleth^\pm_4 \big) = \Big( \frac{k^0}{m} = \sqrt{\frac{{\vec{k}}^2}{m^2} +1}, \frac{\vec{k}}{m}, \pm 1 \Big)$, while the associated invariant measure is $\mathrm{d}\mu^{}_{\daleth_4} = \mathrm{d} \vec{k} / k^0$ (again, $(k^0,\vec{k})$ stands for the four-momentum of a Minkowskian particle with mass $m$). With respect to these variables, the two-point function (\ref{2-point kernel}) is obtained by the following integral:
\begin{eqnarray}
W_\nu (z,z^\prime) = c^{2}_\nu \sum_{l=+,-} \; \int \Big( \frac{z\cdot\xi^l_{}(\vec{k})}{R} \Big)^{-\frac{3}{2} - \mathrm{i}\nu} \Big( \frac{\xi^l_{}(\vec{k}) \cdot z^\prime}{R} \Big)^{-\frac{3}{2} + \mathrm{i}\nu} \; \frac{\mathrm{d} \vec{k}}{k^0}\,,
\end{eqnarray}
which is absolutely convergent.

Now, we show that the corresponding Wightman two-point function ${\cal{W}}_\nu(x,x^\prime)$, which is characterized by taking the boundary value of $W_\nu(z,z^\prime)$ from the tuboid domain ${\cal{T}}^{+(2)}_{}$, verifies the positivity requirement (\ref{Positivity condition}). To do this, with each test function $f_1(x) \in \mathfrak{D}(\underline{M}_R)$ we associate the following expression:
\begin{eqnarray}
\int_{\underline{M}_R \times \underline{M}_R} W_\nu (z, z^\prime) \; f_1^\ast(x) f^{}_1(x^\prime) \; \mathrm{d}\mu(x) \mathrm{d}\mu(x^\prime)\,,
\end{eqnarray}
where $z = x+ \mathrm{i} y \in {\cal{T}}^-$ ($y\in \interior{\underline{V}}^-$) and $z^\prime = x^\prime+ \mathrm{i} y^\prime \in {\cal{T}}^+$ ($y^\prime\in \interior{\underline{V}}^+$). After substituting the kernel (\ref{2-point kernel}) into the above expression and then utilizing the Fourier transform (\ref{Fourier x.xi}) (when $y,y^\prime\rightarrow 0$), we explicitly get the positivity requirement (\ref{Positivity condition}):
\begin{eqnarray}\label{Positivity condition'}
\int_{\underline{M}_R \times \underline{M}_R} {\cal{W}}_\nu (x, x^\prime) \; f_1^\ast(x) f^{}_1(x^\prime) \; \mathrm{d}\mu(x) \mathrm{d}\mu(x^\prime) = \int_{\daleth} \; \mathrm{d}\mu^{}_\daleth(\xi) \Bigg[ \hspace{6cm} \nonumber\\
\times \int_{\underline{M}_R} \Bigg( \underbrace{ c^{}_\nu \Big[ \vartheta\Big(\frac{x\cdot\xi}{R}\Big) + \vartheta\Big(-\frac{x\cdot\xi}{R}\Big) e^{- \mathrm{i} \pi(-\frac{3}{2} - \mathrm{i}\nu)} \Big] \; \Big|\frac{x\cdot\xi}{R}\Big|^{-\frac{3}{2} - \mathrm{i}\nu}}_{\equiv \phi^{-}_{\nu,\xi}(x)} \; f_1^\ast(x) \Bigg) \; \mathrm{d}\mu(x) \nonumber\\
\times \int_{\underline{M}_R} \Bigg( \underbrace{ c^{}_\nu \Big[ \vartheta\Big(\frac{x^\prime\cdot\xi}{R}\Big) + \vartheta\Big(-\frac{x^\prime\cdot\xi}{R}\Big) e^{\mathrm{i}\pi(-\frac{3}{2} + \mathrm{i}\nu)} \Big] \; \Big|\frac{x^\prime\cdot\xi}{R}\Big|^{-\frac{3}{2} + \mathrm{i}\nu}}_{\equiv \phi^{-\ast}_{\nu,\xi}(x^\prime)} \; f^{}_1(x^\prime) \Bigg) \; \mathrm{d}\mu(x^\prime) \Bigg] \nonumber\\
= \int_{\daleth} \big| \phi^{-}_{\nu,\xi}(f^\ast_1) \big|^2 \; \mathrm{d}\mu^{}_\daleth(\xi) \geqslant 0\,. \hspace{3.8cm}
\end{eqnarray}
Note that the hermiticity of the Wightman two-point function ${\cal{W}}_\nu$ is a byproduct of the property of positive definiteness. It can also be realized by considering the boundary values from ${\cal{T}}^{+(2)}$ of the identity $W_\nu(z^\prime,z) = \big(W_\nu(z^\ast,z^{\prime\ast})\big)^\ast$.

At the next step, we examine the covariance property of the two-point function $W_\nu(z,z^\prime)$. We begin by pointing out this fact that, by definition, the forward and backward tubes ${\cal{T}}^\pm = T_{}^{\pm} \cap \underline{M}_R^{(\mathbb{C})}$ are invariant under the action of the real dS$_4$ group Sp$(2,2)$ (as $T^\pm$ and $\underline{M}_R^{(\mathbb{C})}$ are). Therefore, for all $\underline{g} \in \mathrm{Sp}(2,2)$, the following expression still remains meaningful:
\begin{eqnarray}\label{2-point kernel cov}
W_\nu(\underline{g}\diamond z,\underline{g}\diamond z^\prime) = c^{2}_\nu \int_\daleth \Big( \frac{(\underline{g}\diamond z)\cdot\xi}{R} \Big)^{-\frac{3}{2} - \mathrm{i}\nu} \Big( \frac{\xi \cdot (\underline{g}\diamond z^\prime)}{R} \Big)^{-\frac{3}{2} + \mathrm{i}\nu} \; \mathrm{d}\mu^{}_\daleth(\xi)\,.
\end{eqnarray}
Having this point in mind, we first show that, for all $\underline{g} \in \mathrm{Sp}(2,2)$, the following identity holds:
\begin{eqnarray}\label{2-point kernel cov}
W_\nu(\underline{g}\diamond z,\underline{g}\diamond z^\prime) = W_\nu(z,z^\prime)\,.
\end{eqnarray}
Let $\underline{{\cal{S}}}(x^{}_e)$ denote a subgroup of Sp$(2,2)$, which stabilizes a unit vector $x^{}_e$ ($|(x_e^{})^2| =1$) in ${\mathbb{R}}^5$. Let $\daleth$ denote the corresponding orbital basis, invariant under $\underline{{\cal{S}}}(x^{}_e)$, based upon which the integral representation of the two-point function (\ref{2-point kernel}) is written. Regarding the invariance of the measure $\mathrm{d}\mu^{}_\daleth$ and of $\daleth$ under the given subgroup $\underline{{\cal{S}}}(x^{}_e)$, it is evident that the identity (\ref{2-point kernel cov}) is true for any $\underline{g} \in \underline{{\cal{S}}}(x^{}_e) \; \big( \subset \mathrm{Sp}(2,2) \big)$.\footnote{Note that, under the action of the dS$_4$ group Sp$(2,2)$, the dS$_4$ plane waves transform in such a way that:
\begin{eqnarray}
\left( \frac{(\underline{g}\diamond z)\cdot\xi}{R} \right)^\tau = \left( \frac{z\cdot(\underline{g}^{-1}\diamond\xi)}{R} \right)^\tau, \;\;\;\;\;\;\; \mbox{for all}\;\; \underline{g} \in \mathrm{Sp}(2,2)\,. \nonumber
\end{eqnarray}} On the other hand, from the space-time-Lorentz decomposition of Sp$(2,2)$ (see subsection \ref{sec space-time-Lorentz dS4}), we know that any transformation of the Sp$(2,2)$ group may be viewed as the composition of a ``space translation", a ``time translation", and a ``Lorentz boost", i.e., as the composition of transformations belonging to the subgroups $\underline{{\cal{S}}}(x^{}_e)$. Accordingly, by associating with each of these dS$_4$ subgroups $\underline{{\cal{S}}}(x^{}_e)$ the corresponding orbital basis $\daleth$ (invariant under the respective subgroup $\underline{{\cal{S}}}(x^{}_e)$), while we have in mind the above explanation along with the fact that $W^{}_\nu = W^{(\daleth)}_\nu$ for all $\daleth\subset C^+$, one can easily show that the identity (\ref{2-point kernel cov}) holds true for all transformations of the real dS$_4$ group Sp$(2,2)$. Now, to examine the complex covariance property of the two-point function $W_\nu(z,z^\prime)$ (see Eq. (\ref{covariance requirement})), we need to extend Eq. (\ref{2-point kernel cov}) to the complexified dS$_4$ group, $\underline{g} \in \mathrm{Sp}(2,2)^{(\mathbb{C})}$. This task is accomplished by analytic continuation in the group variables. This means that the two-point function $W_\nu(z,z^\prime)$ can be analytically continued in:
\begin{eqnarray}
{\cal{T}}^{+(2)}_{ext} = \Big\{ (z,z^\prime) \in \underline{M}_R^{(\mathbb{C})} \times \underline{M}_R^{(\mathbb{C})} \; ;\;  z=\underline{g}\diamond \bar{z}, \; z^\prime = \underline{g} \diamond \bar{z}^\prime, \; (\bar{z}, \bar{z}^\prime) \in {\cal{T}}^{+(2)},\; \underline{g} \in \mathrm{Sp}(2,2)^{(\mathbb{C})} \Big\}\,,
\end{eqnarray}
which precisely coincides with the cut-domain $\Delta = \big\{ (z,z^\prime) \in \underline{M}^{(\mathbb{C})}_R \times \underline{M}^{(\mathbb{C})}_R \; ; \; (z-z^\prime)^2 < 0\big\}$ given in Eq. (\ref{cut-domain}); the last point appears as a byproduct of the study of the extended tube in two vector variables $T^{+(2)}_{ext}$ in ${\mathbb{C}}^5$ (see, for instance, Ref. \cite{Kallen}). The complex covariance requirement (\ref{covariance requirement}), in $\Delta$, then is fulfilled as a direct result of the identity (\ref{2-point kernel cov}). Note that, proceeding as above, one can similarly show that the two-point function ${W}_\nu^\prime (z,z^\prime)$, which is analytic in the tuboid ${\cal{T}}^{-(2)}$, can also be analytically continued in $\Delta$, in which ${W}_\nu^\prime (z,z^\prime)$ verifies the complex covariance requirement ${W}_\nu^\prime (\underline{g}\diamond z,\underline{g}\diamond z^\prime) = {W}_\nu^\prime (z,z^\prime)$, for all $\underline{g} \in \mathrm{Sp}(2,2)^{(\mathbb{C})}$.

Taking into account, on one hand, the fact that $W_\nu(z,z^\prime)$ is analytic in ${\cal{T}}^{+(2)}$ and, on the other hand, the complex covariance property of $W_\nu(z,z^\prime)$ (more specifically, the transitivity of the $\mathrm{Sp}(2,2)^{(\mathbb{C})}$ group on $\underline{M}_R^{(\mathbb{C})}$), one can easily show that the two-point function $W_\nu(z,z^\prime)$ extends to an \emph{invariant perikernel} \cite{Bros peri,Bros peri''} on $\underline{M}_R^{(\mathbb{C})}$ (with analyticity domain $\Delta$). This practically means that $W_\nu(z,z^\prime)$ is a function of the (pseudo-)distance between the two points $z$ and $z^\prime$ on $\underline{M}_R^{(\mathbb{C})}$, that is, the single (complex) dS$_4$-invariant variable $(z-z^\prime)^2 = -2R^2 - 2z\cdot z^\prime$. This property, which interestingly allows for an explicit calculation by fixing one of the two points $(z,z^\prime)$, will serve below as the starting point to prove the locality requirement of ${\cal{W}}_\nu (x, x^\prime)$.

Technically, let us set the points $(z(\lambda),z^\prime) \in {\cal{T}}^{+(2)}$ in such a way that $z(\lambda)\cdot z^\prime/R^2 = \cosh\lambda$, where $z(\lambda) = \big( - \mathrm{i} R\cosh\lambda, - \mathrm{i} R\sinh\lambda,0,0,0 \big)$ and $z^\prime = \big( \mathrm{i} R,0,0,0,0 \big)$, with $\lambda \geqslant 0$; note that $( z(\lambda)-z^\prime )^2<0$. At these given points, by choosing to integrate on the spherical basis $\daleth_0$ of the future null-cone $C^+$,\footnote{For the sake of reasoning, the spherical basis $\daleth_0 = \big\{ \xi=(\xi^0,\boldsymbol{\xi}) \in C^+ ;\; \xi^0=1\; (|\boldsymbol{\xi}|^2=1) \big\}$ (see subsection \ref{Subsec orbial basis})) is described by the polar coordinates $(\psi,\theta,\varphi)$:
\begin{eqnarray}
\xi^1 &=& \cos\psi\,,\nonumber\\
\xi^2 &=& \sin\psi \cos\theta\,,\nonumber\\
\xi^3 &=& \sin\psi \sin\theta \sin\varphi\,,\nonumber\\
\xi^4 &=& \sin\psi \sin\theta \cos\varphi\,,\nonumber
\end{eqnarray}
with $0\leqslant \psi,\theta \leqslant \pi$ and $0\leqslant \varphi \leqslant 2\pi$. The corresponding invariant measure $\mathrm{d}\mu^{}_{\daleth^{}_0}$ is chosen to be $m^2$ times the rotation-invariant measure on ${\mathbb{S}}^3$, i.e., $\mathrm{d}\mu^{}_{\daleth^{}_0} = m^2\sin^2\psi \sin\theta \; \mathrm{d}\psi \mathrm{d}\theta \mathrm{d}\varphi$ (see appendix \ref{App UIR's SU(2)}).} the integral representation (\ref{2-point kernel}) of the two-point function results in:
\begin{eqnarray}\label{perikernel}
W_\nu \big( z(\lambda),z^\prime \big) = C_\nu \; P^{(5)}_{-\frac{3}{2}- \mathrm{i} \nu} \Big( \frac{z(\lambda)\cdot z^\prime}{R^2} \Big)\,,
\end{eqnarray}
where $C_\nu= {2\pi^2 m^2 e^{-\pi\nu} c^{2}_\nu}$ and:\footnote{Here, we use the fact that the functions $\cosh\lambda$ and $\sinh\lambda$ are even and odd, respectively.}
\begin{eqnarray}\label{generalized leg. func.}
P^{(5)}_{-\frac{3}{2}- \mathrm{i} \nu} \Big( \frac{z(\lambda)\cdot z^\prime}{R^2} \Big) = P^{(5)}_{-\frac{3}{2}- \mathrm{i} \nu} ( \cosh\lambda ) = \frac{2}{\pi} \int_0^\pi \big( \cosh\lambda + \sinh\lambda\cos\psi \big)^{-\frac{3}{2} - \mathrm{i}\nu} \sin^2\psi \; \mathrm{d}\psi\,,
\end{eqnarray}
is the \emph{generalized Legendre function}\footnote{Strictly speaking, in the sense given in appendix \ref{App some}, we actually refer to the integral representation (\ref{generalized leg. func.}) as Legendre function by abuse of notations.} of the first kind (which is proportional to the Gegenbauer function of the first kind $C^{\frac{3}{2}}_{-\frac{3}{2}- \mathrm{i} \nu} (\cosh\lambda)$ \cite{Bateman}; see also appendix \ref{App some}). Equation (\ref{perikernel}) explicitly reveals that the two-point function $W_\nu \big( z(\lambda),z^\prime \big)$ is real valued for all $\lambda$, since:
\begin{eqnarray}\label{p+ = p-}
P^{(5)}_{-\frac{3}{2}- \mathrm{i} \nu} \Big( \frac{z(\lambda)\cdot z^\prime}{R^2} \Big) = P^{(5)}_{-\frac{3}{2}+ \mathrm{i} \nu} \Big( \frac{z(\lambda)\cdot z^\prime}{R^2} \Big)\,.
\end{eqnarray}
Accordingly, we have:
\begin{eqnarray}\label{real-valued peri}
W_\nu \big( z(\lambda),z^\prime \big) = \big( W_\nu \big( z(\lambda),z^\prime \big) \big)^\ast = {W}^\prime_\nu \big( z^\ast(\lambda), z^{\prime\ast} \big)\,.
\end{eqnarray}
Here, we draw attention to the fact that each real spacelike separated pair $\big( x(\lambda),x^\prime \big)$ ($( x(\lambda)-x^\prime )^2<0$), where $x(\lambda)= (-R\sinh\lambda,0,0,0,-R\cosh\lambda)$ and $x^\prime \; \big( = x^{}_\odot \big) = (0,0,0,0,R)$, belongs to the same orbit of $\mathrm{Sp}(2,2)^{(\mathbb{C})}$ as the pairs $\big( z(\lambda),z^\prime \big)$ and $\big( z^\ast(\lambda), z^{\prime\ast} \big)$ do (since $x(\lambda)\cdot x^\prime/R^2 = \cosh\lambda$). In this sense, and also according to the complex covariance property of the two-point function and its permuted counterpart, the identity (\ref{real-valued peri}), for the given $W^{}_\nu$ and $W^\prime_\nu$, can be interpreted as the locality identity ${\cal{W}}^{}_\nu \big( x(\lambda),x^\prime \big) = {\cal{W}}^\prime_\nu \big( x(\lambda) , x^\prime \big) \; \big( = {\cal{W}}_\nu \big( x^\prime , x(\lambda) \big) \big)$ at any such pair. Then, the locality of the construction is proved.

So far, concerning the dS$_4$ principal Klein-Gordon scalar field, we have shown that the corresponding two-point vacuum expectation value of the field, enjoying the maximal analyticity properties, is given by Eq. (\ref{perikernel}); in turn, these maximal analyticity properties, as we have discussed above, completely encode the aforementioned Wightman axioms. Here, the only task, that is left to do, is to determine the constant factor $c^{2}_\nu$. It is unambiguously fixed by considering the canonical commutation relations or the local Hadamard behavior (see, for instance, Ref. \cite{Haag,Kay}) of the corresponding coefficient of the dominant term (that is, the value of the associated quantity in the Minkowskian case). Actually, regarding the fact that the dS$_4$ and Minkowski distances, at short distances (in a tangent plane), are asymptotically equal, one expects that the two-point function (\ref{perikernel}) fulfills the Hadamard requirement, based upon which the constant factor $c^{2}_\nu$ can be determined by considering the canonical normalization of its Minkowskian counterpart, which is equivalent to imposing the canonical commutation relations.

At \emph{coinciding points}, i.e., $\big\{ (z, z^\prime)\in \underline{M}_R^{(\mathbb{C})}\times \underline{M}_R^{(\mathbb{C})}\;;\; z=z^\prime\big\}$, the singular behavior of ${\cal{W}}_\nu$ is determined by the behavior of $P^{(5)}_{-\frac{3}{2}- \mathrm{i} \nu}$ in the vicinity of its singular point ${z\cdot z^\prime}/{R^2} = -1$ \cite{Bros 2point func}:
\begin{eqnarray}\label{Hadamard dS}
W_\nu (z,z^\prime) \approx \frac{2 \; C^{}_\nu}{\Gamma(\frac{3}{2}- \mathrm{i} \nu) \; \Gamma(\frac{3}{2}+ \mathrm{i} \nu)} \Big( 1 + \frac{z\cdot z^\prime}{R^2} \Big)^{-1},
\end{eqnarray}
where $1 + {z\cdot z^\prime}/{R^2} = -{(z-z^\prime)^2}/{2R^2}$. On the other hand, for the Minkowskian Klein-Gordon two-point function, the associated dominant term takes the form:
\begin{eqnarray}\label{Hadamard Minkowski}
- \mathrm{i} D^{(-)}(z^{}_\circ - z^\prime_\circ) = \frac{1}{4\pi^2} \; \frac{-1}{(z^{}_\circ - z_\circ^\prime)^2}\,.
\end{eqnarray}
Accordingly, by comparing the coefficients in Eqs. (\ref{Hadamard dS}) and (\ref{Hadamard Minkowski}), one obtains:
\begin{eqnarray}\label{C_nu}
C^{}_\nu = \frac{\Gamma(\frac{3}{2}- \mathrm{i} \nu) \; \Gamma(\frac{3}{2}+ \mathrm{i} \nu)}{2^4 \pi^2 R^2}\,,
\end{eqnarray}
and correspondingly:
\begin{eqnarray}\label{c_nu}
c^{2}_\nu = \frac{C^{}_\nu}{2 \pi^2 m^2 e^{-\pi\nu}} = \frac{R^{-2} \; (\nu^2 + \frac{1}{4}) }{2 (2\pi)^3 \; \big(1+e^{-2\pi\nu}\big) \; m^2}\,,
\end{eqnarray}
where, to get the above result, we have used the identity $\Gamma(\frac{3}{2}- \mathrm{i} \nu) \; \Gamma(\frac{3}{2}+ \mathrm{i} \nu) = (\frac{1}{2}- \mathrm{i} \nu)(\frac{1}{2}+ \mathrm{i} \nu) \; {\pi}/{\cosh\pi\nu} $. Note that, considering Eqs. (\ref{p+ = p-}) and (\ref{C_nu}), the given two-point function $W_\nu(z,z^\prime) = C^{}_\nu P^{(5)}_{-\frac{3}{2}- \mathrm{i} \nu} \big( \frac{z\cdot z^\prime}{R^2} \big)$ verifies the identity $W_\nu(z,z^\prime) = W_{-\nu}(z,z^\prime)$.

Eventually, the Wightman two-point function for the dS$_4$ principal Klein-Gordon scalar field is obtained by taking the boundary value (in the distribution sense) of Eq. (\ref{2-point kernel}) from the domain ${\cal{T}}^{+(2)}$:
\begin{eqnarray}\label{Wightman 2-point}
{\cal{W}}_\nu (x, x^\prime) &=& c^{2}_\nu \int_\daleth \Big[ \vartheta\Big(\frac{x\cdot\xi}{R}\Big) + \vartheta\Big(-\frac{x\cdot\xi}{R}\Big) e^{- \mathrm{i} \pi(-\frac{3}{2} - \mathrm{i}\nu)} \Big] \; \Big|\frac{x\cdot\xi}{R}\Big|^{-\frac{3}{2} - \mathrm{i}\nu}\nonumber\\
&& \hspace{0.6cm} \times \Big[ \vartheta\Big(\frac{x^\prime\cdot\xi}{R}\Big) + \vartheta\Big(-\frac{x^\prime\cdot\xi}{R}\Big) e^{\mathrm{i} \pi(-\frac{3}{2} + \mathrm{i} \nu)} \Big] \; \Big|\frac{x^\prime\cdot\xi}{R}\Big|^{-\frac{3}{2} + \mathrm{i} \nu} \; \mathrm{d}\mu^{}_\daleth(\xi)\,.
\end{eqnarray}
Here, concerning the above formula, the following points must be underlined:
\begin{itemize}
\item{It remarkably presents a factorization of the Wightman two-point function, in terms of the global plane waves on $\underline{M}_R$, which is quite analogous to the associated Fourier representation for the two-point function of the Minkowski Klein-Gordon scalar field with mass $m$ (see Eq. (\ref{qqq})). Actually, according to the relations given in subsection \ref{Subsec flat limit}, the latter can be simply achieved as the null-curvature limit of the above expression.}
\item{The permuted Wightman two-point function ${\cal{W}}^\prime_\nu(x,x^\prime)$ would be the boundary value of ${W}^\prime_\nu(z,z^\prime)$ from the domain ${\cal{T}}^{-(2)}_{}$. This allows for the explicit construction of the corresponding commutator and the Green functions.}
\item{Considering the instruction given in subsection \ref{Subsec gen free fields}, the one-particle Hilbert space ${\cal{H}}_1$ of the theory can be realized by $L^2(\daleth,\mathrm{d}\mu^{}_\daleth)$, that is, the space of complex-valued functions $\phi^-_{\nu,\xi}(f_1)$ of the variable $\xi$ running in the orbital basis $\daleth \subset C^+$ and square integrable with respect to the measure $\mathrm{d}\mu^{}_\daleth$ (see Eq. (\ref{Positivity condition'})).\footnote{The point that must be noticed here is that, for orbital basis of noncompact type, this statement is valid only in the distribution sense.} [Again, the full Hilbertian Fock space ${\mathscr{H}}$ of the theory is given by the Hilbertian sum ${\cal{H}}_0 \oplus \big[\oplus_n S ({\cal{H}}_1)^{\otimes n}\big]$.] As a matter of fact, any $\dot{f}_1 \in {\cal{H}}_1$ corresponds to a function $f^{}_1(x)$ which, with respect to the Fourier transform (\ref{Fourier x.xi}), is determined by:
    \begin{eqnarray}
    f^{}_1(x) = \int_\daleth \phi^-_{\nu,\xi}(x) \; \phi^-_{\nu,\xi}(f_1) \; \mathrm{d}\mu^{}_\daleth(\xi)\,, \;\;\;\;\;\;\; \phi^-_{\nu,\xi}(f_1) \in L^2(\daleth,\mathrm{d}\mu^{}_\daleth)\,,
    \end{eqnarray}
    where the definitions of $\phi^-_{\nu,\xi}(x)$ and $\phi^-_{\nu,\xi}(f_1)$ have been already given in Eq. (\ref{Positivity condition'}). This manifestation of the elements of the one-particle Hilbert space ${\cal{H}}_1$ interestingly allows us to control the zero-curvature limit of the corresponding representations, i.e., the scalar principal series UIR's. To see the point, proceeding as subsection \ref{Subsec flat limit}, we choose $x^{}_\odot=(0,0,0,0,R)$ as the point of dS$_4$ spacetime in the neighborhood of which the flat limit is going to take, and $\daleth^{}_4 = \daleth^-_4 \cup \daleth^+_4$, given in Eq. (\ref{orbit gamma4}), as the orbital basis on which the integration is going to accomplish; regarding the latter, we can write $\xi^\pm \; \big( \in \daleth^{\pm}_4 \big) = \Big( \frac{k^0}{m} = \sqrt{\frac{{\vec{k}}^2}{m^2} +1}, \frac{\vec{k}}{m}, \pm 1 \Big)$, where $(k^0,\vec{k})$ denotes the four-momentum of a Minkowskian particle with mass $m$ (see Eq. (\ref{orbit gamma4'})). Considering the decomposition of the orbital basis, the (principal) Hilbert space ${\cal{H}}_1$ can be decomposed into two parts ${\cal{H}}^{}_1 = {\cal{H}}^+_1 \oplus {\cal{H}}^-_1$:
    \begin{eqnarray}
    f^{}_1(x) = \int_{\daleth^-_4; \xi^4<0} \phi^-_{\nu,\xi}(x) \;\; \phi^-_{\nu,\xi}(f_1) \;\; \mathrm{d}\mu^{}_{\daleth^-_4}(\xi) \;+\; \int_{\daleth^+_4; \xi^4>0} \phi^-_{\nu,\xi}(x) \;\; \phi^-_{\nu,\xi}(f_1) \;\; \mathrm{d}\mu^{}_{\daleth^+_4}(\xi)\,.
    \end{eqnarray}
    At the flat limit, the second integral vanishes (see Eq. (\ref{xi<0}) and its subsequent discussions) and one is left with:
    \begin{eqnarray}
    \lim_{R\rightarrow\infty} f^{}_1(x) = f^{}_1(x^{}_\circ) = \int \frac{e^{- \mathrm{i} k\cdot x^{}_\circ}}{\sqrt{2(2\pi)^3}} \; \phi^-_{k}(f_1) \; \frac{\mathrm{d}\vec{k}}{k^0}\,,
    \end{eqnarray}
    where $\phi^-_{k}(f_1)$ stands for the zero-curvature limit of $\phi^-_{\nu,\xi}(f_1)$ and, again, $x^{}_\circ = ( x^{0}_\circ, \vec{x}^{}_\circ)$ for the coordinates of the tangent plane at $x^{}_\odot$, that is, the $1+3$-dimensional Minkowski spacetime given by Eq. (\ref{contraction coordinates}). As is evident, taking the global dS$_4$ plane waves and their corresponding Fourier calculus into account, the Poincar\'{e} contraction of the principal series UIR's of the dS$_4$ group, carried by the Hilbert space ${\cal{H}}^{}_1$, merely leads to the irreducible massive representations of the Poincar\'{e} group with exclusively positive energy, as far as the analyticity domain has been chosen properly.}
\item{Finally, we remark that the maximal analytic framework, leading to the Wightman two-point function (\ref{Wightman 2-point}), entails the possibility of going to the \emph{Euclidean sphere ${\mathbb{S}}^4$ of ``imaginary times"} ($x^0= \mathrm{i} y^0$)\footnote{Let $\acute{T}^\pm = {\mathbb{R}}^5 + \mathrm{i} \underline{{V}}^\pm$ and ${\cal{\acute{T}}}^\pm = \acute{T}^\pm \cap M_R^{(\mathbb{C})}$ (for the definition of $\underline{{V}}^\pm$, see section \ref{Sec dS4 causal structure}). Quite similar to $\acute{T}^+ \cup \acute{T}^-$, which contains the ``Euclidean subspace" $\mathbb{E}^5 = \big\{ z=(\mathrm{i} y^0,x^1,x^2,x^3,x^4) \; ; \; (y^0,x^1,x^2,x^3,x^4) \in \mathbb{R}^5 \big\}$ of the complex Minkowski spacetime $\mathbb{C}^5$, it can be simply shown that ${\cal{\acute{T}}}^+ \cup {\cal{\acute{T}}}^-$ contains the sphere ${\mathbb{S}}^4 = \big\{ z=(\mathrm{i} y^0,x^1,x^2,x^3,x^4) \; ; \; (y^0)^2 + (x^1)^2 + (x^2)^2 + (x^3)^2 + (x^4)^2 = R^2 \big\}$. The latter is called ``Euclidean sphere" of $\underline{M}_R^{(\mathbb{C})}$.} by analytic continuation; actually, by confining $W_\nu(z,z^\prime)$ to the Euclidean sphere and then getting the Schwinger function $S_\nu(z,z^\prime)$.\footnote{This is of course permitted by the fact that ${\mathbb{S}}^4 \times {\mathbb{S}}^4$, minus the set of coinciding points $z=z^\prime$, constitutes a subset of the cut-domain $\Delta$.} This feature interestingly allows for the identification of our axiomatic approach with the pioneering Euclidean formulation of the preferred vacuum states introduced by Gibbons and Hawking \cite{Gibbons} (and of course it identifies with the formulation yielded by the Hadamard requirement as well). In this regard, we would like to emphasize that properties of analytic continuation in all the variables must be respected as the basis of every treatment of dS$_4$ (generally, dS) QFT's in the framework of the functional integral on the Euclidean sphere; without the proper analyticity properties, one cannot implement the results concluded by Euclidean approaches in the real dS$_4$ (generally, dS) spacetime. This contains the constructive approach to dS$_4$ (generally, dS) QFT \cite{Figari} or the application that the latter may get in Minkowskian constructive QFT, for which the (constant) radius of curvature $R$ of the dS$_4$ hyperboloid $\underline{M}_R$ appears as a natural infrared cutoff (the Euclidean space turns into a compact one!).

    In connection with the Minkowskian-limit criterion (discussed in the previous item), we would like to supplement the current topic by noting that our QFT construction also explicitly reveals that the Euclidean vacuum has to be preferred, if one desires to get the physically meaningful Minkowskian QFT, under vanishing curvature.}
\end{itemize}

Now, and before getting involved with the physical interpretation of the maximal analyticity properties of the given two-point function, we would like to point out that the two-point function corresponding to the dS$_4$ complementary Klein-Gordon scalar field is simply obtained, in the allowed ranges of $\nu$, by replacing $\nu \mapsto - \mathrm{i} \nu$ in the integral representation (\ref{2-point kernel}):
\begin{eqnarray}\label{2-point kernel cpmplementary}
W_{- \mathrm{i} \nu} (z,z^\prime) = c^{2}_{- \mathrm{i} \nu} \int_\daleth \Big( \frac{z\cdot\xi}{R} \Big)^{-\frac{3}{2} - \nu} \Big( \frac{\xi \cdot z^\prime}{R} \Big)^{-\frac{3}{2} + \nu} \; \mathrm{d}\mu^{}_\daleth(\xi)\,, \;\;\;\;\;\; \nu\in\mathbb{R},\;\; 0<|\nu|<\frac{3}{2}\,,
\end{eqnarray}
where, again, $(z,z^\prime) \in {\cal{T}}^{+(2)}$, $\daleth$ stands for any orbital basis of the future null-cone $C^+$, and $c^{2}_{- \mathrm{i} \nu}$ can be fixed by applying the Hadamard condition. By proceeding as before, one can show that $W_{- \mathrm{i} \nu} (z,z^\prime)$ satisfies all the requirements of positivity, locality, covariance, and normal analyticity. The corresponding Wightman two-point function ${\cal{W}}_{- \mathrm{i} \nu} (x,x^\prime)$ then would be the boundary value of $W_{- \mathrm{i} \nu} (z,z^\prime)$ from the domain ${\cal{T}}^{+(2)}$ (in the same way as Eq. (\ref{Wightman 2-point})).

\subsubsection{Maximal analyticity and KMS condition}
\begin{figure}[H]
    \begin{center}
    \includegraphics[height=.4\textheight]{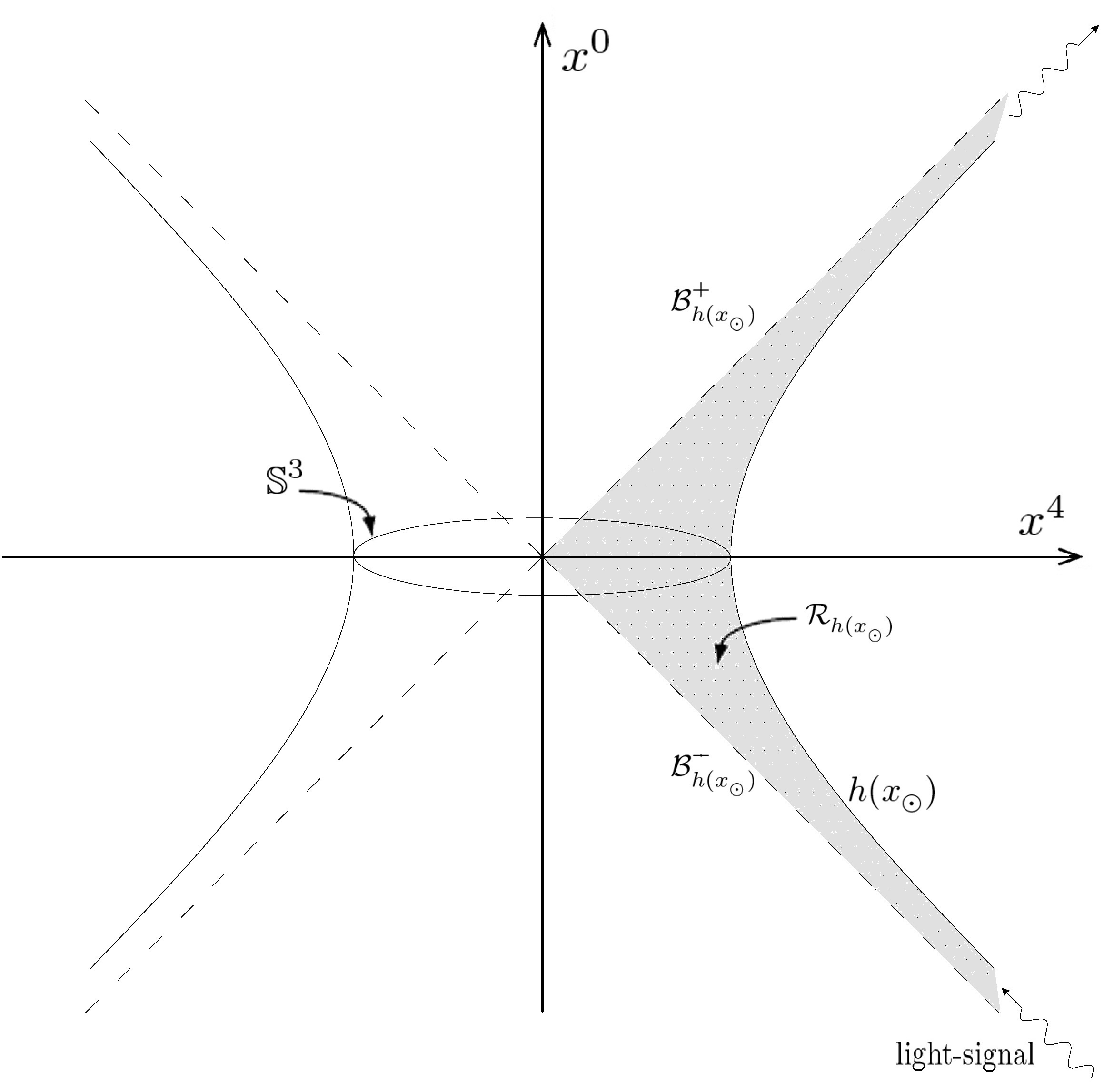}
    \end{center}
    \caption{Causal domains associated with an observer moving on the geodesic $h(x^{}_\odot)$.}
    \label{FIG. KMS}
\end{figure}

We here discuss the physical interpretation of the maximal analyticity properties of $W_\nu(z,z^\prime)$. In order to present this rather elaborate material, as in Ref. \cite{Gibbons}, we adopt the viewpoint of an observer moving on the geodesic $h(x^{}_\odot)$ of the point $x^{}_\odot=(0,0,0,0,R)$ lying at the $(x^0,x^4)$-plane (see FIG. \ref{FIG. KMS}):
\begin{eqnarray}\label{geodesic}
h(x^{}_\odot) = \Big\{ x=x(t) \; ; \; x^0=R\sinh\frac{t}{R},\; \underline{\vec{x}} \equiv (x^1,x^2,x^3)=0, \; x^4= R\cosh\frac{t}{R} \Big\}\,,
\end{eqnarray}
where $t\in\mathbb{R}$. According to the arguments given in section \ref{Sec dS4 causal structure}, all events $x=(x^0,x^1,x^2,x^3,x^4)\in\underline{M}_R$, which can be connected with the observer by the reception of light-signals from the very beginning at $x^\prime\equiv x(t\rightarrow-\infty)$, are determined by the following inequality:
\begin{eqnarray}
x\cdot x^\prime \leqslant -R^2 \;\;\;\Rightarrow \;\;\; \frac{- e^{{-t}/{R}} (x^0+x^4) }{2} \Big|_{t\rightarrow-\infty} \leqslant -R\,. \nonumber
\end{eqnarray}
which merely holds for $(x^0+x^4)>0$. Similarly, all events $x\in\underline{M}_R$, which can ultimately be connected with the observer at $x^\prime\equiv x(t\rightarrow+\infty)$ by the emission of light-signals, are characterized by the requirement $(x^0-x^4)<0$ verifying:
\begin{eqnarray}
x\cdot x^\prime \leqslant -R^2 \;\;\;\Rightarrow \;\;\; \frac{e^{{t}/{R}} (x^0-x^4) }{2} \Big|_{t\rightarrow+\infty} \leqslant -R\,. \nonumber
\end{eqnarray}
The intersection of the above regions, determining all events of $\underline{M}_R$ which can be connected with the observer by the reception and the emission of light-signals, is denoted here by ${\cal{R}}^{}_{h(x^{}_\odot)}$. It explicitly reads:
\begin{eqnarray}
{\cal{R}}^{}_{h(x^{}_\odot)} = \Big\{ x \in \underline{M}_R \; ; \; x^4 > |x^0| \Big\}\,.
\end{eqnarray}
This domain is bordered by:
\begin{eqnarray}
{\cal{B}}^\pm_{h(x^{}_\odot)} = \Big\{ x \in \underline{M}_R \; ; \; x^0 = \pm x^4, \; x^4>0 \Big\}\,,
\end{eqnarray}
where ${\cal{B}}^\pm_{h(x^{}_\odot)}$ are respectively called ``future"/``past" event horizons of the observer sitting on geodesic $h(x^{}_\odot)$.

Now, by interpreting the parameter $t$ (which appears in Eq. (\ref{geodesic})) as the proper time $\boldsymbol{\tau}$ of the observer with the geodesic $h(x^{}_\odot)$, we refer to \emph{the time-translation group corresponding to $h(x^{}_\odot)$}, denoted here by $\underline{T}_{h(x^{}_\odot)}$, as a one-parameter subgroup of the dS$_4$ group ($\underline{T}_{h(x^{}_\odot)}\sim \mathrm{SO}_0(1,1)$; see subsection \ref{sec space-time-Lorentz dS4}). Under the action of $\underline{T}_{h(x^{}_\odot)}$, the region ${\cal{R}}^{}_{h(x^{}_\odot)}$ is foliated by hyperbolic trajectories $h_{\underline{\vec{x}}}(x^{}_\odot)$ parallel to the geodesic $h(x^{}_\odot) = h_{\underline{\vec{0}}}(x^{}_\odot)$. To see the point, let $x=x(\boldsymbol{\tau},\underline{\vec{x}})$ be an arbitrary point in ${\cal{R}}^{}_{h(x^{}_\odot)}$:
\begin{eqnarray}\label{coordinate KMS}
x(\boldsymbol{\tau},\underline{\vec{x}}) =
\left \{ \begin{array}{rl} x^0 &= \sqrt{R^2-\underline{\vec{x}}} \; \sinh\frac{\boldsymbol{\tau}}{R}\,, \vspace{2mm}\\
\vspace{2mm} (x^1,x^2,x^3) &= \underline{\vec{x}}\,,\\
\vspace{2mm} x^4 &= \sqrt{R^2-\underline{\vec{x}}} \; \cosh\frac{\boldsymbol{\tau}}{R}\,, \end{array}\right. \;\;\;\;\;\;\;\;\; \boldsymbol{\tau}\in\mathbb{R}\,,\;\; (\underline{\vec{x}})^2=(x^1)^2+(x^2)^2+(x^3)^2 < R^2\,.
\end{eqnarray}
For $t\in\mathbb{R}$, the action of $\underline{T}_{h(x^{}_\odot)}(t)$ on $x(\tau,\underline{\vec{x}})$, defining a group of isometric automorphisms of the domain ${\cal{R}}^{}_{h(x^{}_\odot)}$, is given by:
\begin{eqnarray}\label{time-translation KMS}
\underline{T}_{h(x^{}_\odot)}(t) \diamond x(\boldsymbol{\tau},\underline{\vec{x}}) = x(t+\boldsymbol{\tau},\underline{\vec{x}}) \equiv x^t_{}\,.
\end{eqnarray}
The corresponding orbits (i.e., $h_{\underline{\vec{x}}}(x^{}_\odot)$'s) clearly represent all branches of hyperbolas of the domain ${\cal{R}}^{}_{h(x^{}_\odot)}$, in two-dimensional plane sections, parallel to the $(x^0,x^4)$-plane. [In this regard, to see a general discussion, we refer readers to Ref. \cite{Kay}.] Here, it is worth noting that, in the given set of orbits of $\underline{T}_{h(x^{}_\odot)}$, the only orbit which represents a geodesic of $\underline{M}_R$ is actually $h(x^{}_\odot)$. In this sense, the interpretation of the group $\underline{T}_{h(x^{}_\odot)}$ as the time translation is merely relevant for either the observers moving on $h(x^{}_\odot)$ or in a vicinity of $h(x^{}_\odot)$, which is \emph{small} in comparison with the radius $R$ of the dS$_4$ hyperboloid $\underline{M}_R$.

From Eq. (\ref{coordinate KMS}), one can easily see that the complexified orbits of $\underline{T}_{h(x^{}_\odot)}$, i.e., the complex hyperbolas $h^{(\mathbb{C})}_{\underline{\vec{x}}}(x^{}_\odot) = \big\{ z^t_{} \equiv z(t+\boldsymbol{\tau},\underline{\vec{x}}),\; t\in\mathbb{C} \big\}$, possess $2 \mathrm{i} \pi R$ periodicity in $t$ (since $z(t+\boldsymbol{\tau},\underline{\vec{x}}) = z(t+\boldsymbol{\tau}+2 \mathrm{i} \pi R,\underline{\vec{x}})$), and that all their nonreal points belong to ${\cal{T}}^\pm$. Now, let $\langle\Omega, \hat{\phi}_\nu(x) \hat{\phi}_\nu(x^{\prime t}) \Omega\rangle = {\cal{W}}_\nu(x,x^{\prime t})$ and $\langle\Omega, \hat{\phi}^{}_\nu(x^{\prime t}) \hat{\phi}_\nu(x) \Omega\rangle = {\cal{W}}^\prime_\nu(x,x^{\prime t}) \; \big( ={\cal{W}}^{}_\nu(x^{\prime t},x) \big)$ be the time-translated correlation functions of two arbitrary events $x$ and $x^\prime$ in ${\cal{R}}^{}_{h(x^{}_\odot)}$. Considering the above, the maximal analyticity properties of $W_\nu(z,z^\prime)$ entail that $W_\nu(x,z^{\prime t})$ characterizes a $2 \mathrm{i} \pi R$-periodic analytic function of $t$ with the domain (periodic cut plane):
\begin{eqnarray}
{\mathbb{C}}_{x,x^\prime}^{\mbox{\tiny{cut}}} = \Big\{ t \in \mathbb{C} \;;\; \mbox{Im} (t) \neq 2n\pi R,\; n\in\mathbb{Z} \Big\} \bigcup \Big\{ t \;;\; t - 2 \mathrm{i} n\pi R \in I_{x,x^\prime},\; n\in\mathbb{Z} \Big\}\,,
\end{eqnarray}
where, for every spacelike separated points $x,x^\prime \in {\cal{R}}^{}_{h(x^{}_\odot)}$ (i.e., $(x-x^{\prime})^2<0$), $I_{x,x^\prime}$ denotes a finite (real) interval on which $x$ and $x^{\prime t}$ remain spacelike separated, or in other words, $(x-x^{\prime t})^2$ remains negative (this interval trivially includes the origin in the real $t$-axis). One can also check that the boundary values of $W_\nu(x,z^{\prime t})$ on $\mathbb{R}$ (in the distribution sense for the variable $t$ and each given values of $x,x^\prime$) coincide with the aforementioned correlation functions (the jumps across the cuts being the advanced and retarded commutators):
\begin{eqnarray}
\lim_{\epsilon\rightarrow 0^+} W_\nu(x,z^{\prime t + \mathrm{i} \epsilon}) &=& {\cal{W}}_\nu(x,x^{\prime t})\,, \nonumber\\
\lim_{\epsilon\rightarrow 0^+} W_\nu(x,z^{\prime t - \mathrm{i} \epsilon}) &=& {\cal{W}}_\nu(x^{\prime t},x)\,.
\end{eqnarray}
These properties explicitly imply that $W_\nu(x,z^{\prime t})$ is analytic in the strip $\mathfrak{S} = \big\{ t \in \mathbb{C} \;;\; 0 < \mbox{Im} (t) < 2\pi R \big\}$,\footnote{For each $t$ in the strip $\mathfrak{S}_+ = \big\{ t \;;\; 0 < \mbox{Im} (t) < \pi R \big\}$ (respectively, $\mathfrak{S}_- = \big\{ t \;;\; \pi R < \mbox{Im} (t) < 2\pi R \big\}$), the associated point $z(t+\boldsymbol{\tau},\underline{\vec{x}})$ is located in the domain ${\cal{T}}^+$ (respectively, ${\cal{T}}^-$) of $\underline{M}^{(\mathbb{C})}_R$.} and verifies the following condition:
\begin{eqnarray}\label{KMS condition}
\lim_{\epsilon\rightarrow 0^+} W_\nu(x,z^{\prime t + 2 \mathrm{i} \pi R - \mathrm{i} \epsilon}) = {\cal{W}}_\nu(x^{\prime t},x)\,, \;\;\;\;\;\;\; x,x^\prime \in {\cal{R}}^{}_{h(x^{}_\odot)}\,,
\end{eqnarray}
which is called KMS condition, after Kubo \cite{Kubo} and Martin and Schwinger \cite{Martin}. This condition reveals that $W_\nu(x,x^{\prime t})$ represents a finite-temperature equilibrium (two-point) correlation function at temperature $1/2\pi R$.

Here, it is worth noting that, as a byproduct of the arguments leading to the KMS condition (\ref{KMS condition}), we get an additional property, associating the domain ${\cal{R}}^{}_{h(x^{}_\odot)}$ by analytic continuation with its antipodal ${\cal{R}}^{}_{h(-x^{}_\odot)} = \big\{ x=(x^0,\underline{\vec{x}},x^4) \in \underline{M}_R, \; -x \equiv (-x^0,\underline{\vec{x}},-x^4) \in {\cal{R}}^{}_{h(x^{}_\odot)} \big\}$ (to see the point, it is sufficient to consider $\mbox{Im}(t)=\pi R$). Note that the natural time variable relevant to an observer moving on the geodesic $h(-x^{}_\odot)$ (antipodal to $h(x^{}_\odot)$) would be equal to $-t$.\footnote{This is because that the associated time-translation group $\underline{T}_{h(-x^{}_\odot)}$ (which is obtained from $\underline{T}_{h(x^{}_\odot)}$, for instance, by a conjugation of the form $\underline{T}_{h(-x^{}_\odot)} = r \underline{T}_{h(x^{}_\odot)} r^{-1}$, $r$ being a rotation of angle $\pi$ in a plane orthogonal to the $x^0$-axis) verifies $\underline{T}_{h(-x^{}_\odot)} (t) = \underline{T}_{h(x^{}_\odot)} (-t)$. This feature is indeed another manifestation of the fact that dS$_4$ (generally, dS) spacetime has no globally timelike Killing vector.} This feature, as far as the correlation functions are concerned, necessitates that the KMS analyticity strip is replaced by $\big\{ t \;;\; -2\pi R < \mbox{Im}(t) < 0 \big\}$, and correspondingly, Eq. (\ref{KMS condition}) by:
\begin{eqnarray}\label{KMS condition'}
\lim_{\epsilon\rightarrow 0^+} W_\nu(x,z^{\prime t - 2 \mathrm{i} \pi R + \mathrm{i}\epsilon}) = {\cal{W}}_\nu(x^{\prime t},x)\,, \;\;\;\;\;\;\; x,x^\prime \in {\cal{R}}^{}_{h(-x^{}_\odot)}\,.
\end{eqnarray}

The ``energy" operator $\mathfrak{E}^{}_{h(x^{}_\odot)}$, corresponding to the geodesic ${h(x^{}_\odot)}$, is eventually defined by considering the \emph{spectral decomposition} (see, for instance, Ref. \cite{Zeidler}) of the unitary representations $\big\{ \underline{\mathscr{U}}_{h(x^{}_\odot)}(t) \;;\; t\in\mathbb{R} \big\}$ of the time-translation group $\underline{T}_{h(x^{}_\odot)}$ in the Hilbertian Fock space ${\mathscr{H}}$ of the theory, namely:
\begin{eqnarray}
\underline{\mathscr{U}}_{h(x^{}_\odot)}(t) = \int_{-\infty}^{\infty} e^{\mathrm{i} \omega t} \; \mathrm{d} E_{h(x^{}_\odot)}(\omega)\,,
\end{eqnarray}
which yields (in a given dense subspace of ${\mathscr{H}}$) the unbounded ``energy" operator as follows:
\begin{eqnarray}
\mathfrak{E}^{}_{h(x^{}_\odot)} = \int_{-\infty}^{\infty} \omega \; \mathrm{d} E_{h(x^{}_\odot)}(\omega)\,.
\end{eqnarray}
Accordingly, one can easily see that, due to the KMS condition (\ref{KMS condition}), energy measurements carried out by an observer at rest at the origin $(x^{}_\odot)$ on states localized in the region ${\cal{R}}_{h(x^{}_\odot)}$ are exponentially damped by a factor $\exp(-2\pi R \omega)$ in the range of negative energies. At the null-curvature limit ($R\rightarrow\infty$), this factor, eliminating all negative energies, remarkably entails that the theory recovers the usual spectral condition of ``positivity of the energy". Note that in the antipodal case ${h(-x^{}_\odot)}$, having this fact in mind that the corresponding time and energy variables are respectively equal to $-t$ and $-\omega$ and proceeding as above, one can show that the construction, possessing the respective KMS condition (\ref{KMS condition'}), recovers the usual spectral condition, as well. [This result once again shows that how the \emph{ad hoc} process of contraction based on group representation theory equipped with the analyticity prerequisite in the complexified dS$_4$ (generally, dS) manifold controls in a very suggestive way the null-curvature limit of dS$_4$ (generally, dS) QFT to its Minkowskian counterpart (in the respective dimensions).]

At the end, we would like to bring up in passing an interesting question: Could the aforementioned antipodal asymmetry explain the matter-antimatter asymmetry problem in our Universe? Matter in our dS$_4$ side might be viewed as antimatter from the antipodal perspective? The answer to this question is of course far beyond the scope of this paper and certainly needs a tremendous amount of work that we leave it to further investigation. Yet, this question in itself well exemplifies the very point lying at the heart of the content of this study, that is, how respecting the whole dS$_4$ (generally, dS) symmetry as a fundamental symmetry of the nature may pave the way for better understanding the Universe.

\subsection{Minimally coupled scalar field as an illustration of a Krein structure}\label{Subsec mc}
We now turn our attention to the dS$_4$ discrete Klein-Gordon scalar fields corresponding to the ``lowest limit" of this series, i.e., the representations $\Pi_{p,0}$ ($p=1,2,...$). These representations of the dS$_4$ discrete series have no physically meaningful Minkowskian limit (see section \ref{Sec contraction}), but still, in the context of a consistent QFT reading of dS$_4$ elementary systems, their corresponding quantum fields are perfectly legitimate to be studied.

Let us begin by recalling from subsection \ref{Subsec generating} that, for a given $p=1,2,...$, the Hilbert space carrying the representation $\Pi_{p,0}$ admits an invariant $p(p+1)(2p+1)/6$-dimensional \emph{null-norm} subspace (with respect to the Klein-Gordon inner product). This null-norm subspace, interpreted as a ``gauge" states space, carries the dS$_4$ irreducible (nonunitary) finite-dimensional representation ($n_1=0,n_2=p-1$) (in the notations given in appendix \ref{App Lie algebra B2}), which is Weyl equivalent to the given UIR $\Pi_{p,0}$. Due to this nonsquare-integrability feature of the representations $\Pi_{p,0}$, quantization of the associated fields yet is not known entirely, with the exception of that corresponding to the lowest case $\Pi_{p=1,0}$, namely, the so-called dS$_4$ minimally coupled scalar field. Concerning the latter in turn, as we will discuss in the current subsection, its nonsquare-integrability feature prohibits the implementation of any quantization scheme based on two-point functions. In this sense, we have to treat the dS$_4$ minimally coupled scalar field in a different manner from the one applied to the other dS$_4$ scalar fields (associated with the principal and complementary UIR's). Presenting a consistent QFT formulation of the minimally coupled scalar field is our task in the current subsection.

\subsubsection{``Zero-mode" problem}
Here, like subsection \ref{Subsec generating}, for the sake of reasoning, we invoke the system of conformal coordinates $x=x(\rho,\textbf{u})$, where $-\pi /2< \rho <\pi /2$ and $\textbf{u} \in \mathbb{S}^3$ (see Eq. (\ref{cbgicoo})).\footnote{Note that, according to the discussions given in subsection \ref{Subsec generating}, among all the scalar fields associated with the discrete series UIR's $\Pi_{p,0}$, the minimally coupled scalar field associated with $\Pi_{p=1,0}$ is the only one for which the use of the conformal coordinates yields no singularity.} Then, adapting the mathematical structure introduced in subsection \ref{Subsec generating}, the \emph{normalizable modes}\footnote{Note that in the current subsection, borrowing the notation introduced in subsection \ref{Subsec generating}, we distinguish the \emph{normalizable} modes (with respect to the Klein-Gordon inner product (\ref{Kl-Go inpro})) by putting the `$\;{\widetilde{}}\;$' symbol on them.}, relevant to the representation $\Pi_{p=1,0}$, read:\footnote{Here, we use Eq. (\ref{normalized discrete}) along with the identity $^{}_2F^{}_1 (-1,L;L+2;-e^{-2 \mathrm{i} \rho}) = 1+\frac{L}{L+2} e^{-2 \mathrm{i} \rho}$.}
\begin{eqnarray}
\underbrace{\widetilde{\Phi}^{\tau=-3}_{L_{\geqslant p=1}lm}(x) \;=\; \widetilde{\Phi}^{}_{L_{\geqslant p=1}lm}(x)}_{\substack{{\mbox{for simplicity of presentation,}}\\{\mbox{we drop the superscript $\tau=-3$}}}} = \frac{Le^{- \mathrm{i} (L+2)\rho} + (L+2)e^{- \mathrm{i} L\rho}}{2 R \; \sqrt{2L(L+1)(L+2)}} \; {\cal{Y}}_{Llm}(\textbf{u})\,.
\end{eqnarray}
Note that for the $L=0$ (constant) mode $\Phi^{}_{0,0,0}$, the normalization constant breaks down. As a matter of fact, adapting the discussions given in subsection \ref{Subsec generating} to the minimally coupled case $\Pi_{p=1,0}$, this mode belongs to the corresponding one-dimensional (null-norm) ``gauge" states space $\big\{\Phi^{}_{L_{< p=1}lm} = \Phi^{}_{0,0,0}\big\}$. In this regard, it is perhaps worthwhile recalling that the Lagrangian:
\begin{eqnarray}\label{hhhhhhhhhh}
\mathfrak{L} = \sqrt{|g|} \; g^{\mu\nu}_{} \big(\partial_\mu \Phi\big) \big(\partial_\nu \Phi\big)\,,
\end{eqnarray}
of the free dS$_4$ minimally coupled scalar field possesses the (global) gauge-like symmetry $\Phi\rightarrow \Phi + \varrho$, where $\varrho$ is a constant function. This makes clear in what sense we call the appeared one-dimensional null-norm subspace of constant functions $\big\{\Phi^{}_{0,0,0}\big\}$ the space of ``gauge" states.

Trivially, the space generated by $\big\{\widetilde{\Phi}^{}_{L_{\geqslant 1}lm}\big\}$ does not form a complete set of modes for the dS$_4$ minimally coupled scalar field. Moreover, this set of modes is not invariant under the dS$_4$ group action. Considering the form of the dS$_4$ infinitesimal generators $M_{AB}$ ($A,B=0,1,2,3,4$) in the conformal coordinates, given in appendix \ref{App infinitesimal}, the latter point can be easily seen, for instance, by the following action \cite{Gupta 2000}:
\begin{eqnarray}\label{action 1}
\big( M_{03} + \mathrm{i} M_{04} \big) \widetilde{\Phi}^{}_{1,0,0} = - \mathrm{i} \frac{4}{\sqrt{6}} \widetilde{\Phi}^{}_{2,1,0} + \widetilde{\Phi}^{}_{2,0,0} + \frac{3}{4\pi R \sqrt{6}}\,,
\end{eqnarray}
where the invariance of the given set of normalizable modes $\big\{\widetilde{\Phi}^{}_{L_{\geqslant 1}lm}\big\}$ is clearly broken, because of the last constant term. Therefore, any application of canonical quantization to $\big\{\widetilde{\Phi}^{}_{L_{\geqslant 1}lm}\big\}$ results in a noncovariant quantum field. Of course, constant functions constitute a part of solutions to the dS$_4$ minimally coupled field equation $Q^{(1)}_0 \Phi = -R^2 \Box_R \Phi = 0$ (for $\tau= -3$ \emph{or} $\tau= 0$ as it is seen in Eq. (\ref{Wave Eq. scalar ambient})). Then, to get rid of this problem, one is naturally led to consider the space constructed over $\big\{\widetilde{\Phi}^{}_{L_{\geqslant 1}lm}\big\}$ and over a constant function as well; the latter, interpreted as a gauge state, is denoted here by ${\Phi}^{}_{g} \; \big( \in \big\{ \Phi^{}_{0,0,0}\big\} \big)$. The obtained space in this way is invariant under the action of the dS$_4$ group and, as will be shown in subsubsection \ref{Subsubsec Gupa-B}, constitutes the physical states space. However, this space as an inner-product space, equipped with the Klein-Gordon inner product (\ref{Kl-Go inpro}), is degenerate; the gauge state ${\Phi}^{}_{g}$ is orthogonal to the entire set of states including itself. Because of this degeneracy, any application of canonical quantization to this states space once again leads to a noncovariant field \cite{deBievre'}.

Technically, for $L=0$, the minimally coupled scalar field equation, strictly speaking, Eq. (\ref{rho dependent part}) when $\tau=-3$, can be directly solved, by proceeding as Eqs. (\ref{6.4}) and (\ref{6.5}). On this basis, one gets two independent solutions (containing the aforementioned constant function ${\Phi}^{}_{g}$) \cite{Gupta 2000}:
\begin{eqnarray}\label{Zero mode}
{\Phi}^{}_{g} = \frac{1}{2\pi R}\,, \;\;\;\;\;\;\; {\Phi}^{}_{s} = \frac{- \mathrm{i}}{2\pi R} \big( \rho + \frac{1}{2} \sin 2\rho \big)\,.
\end{eqnarray}
[The gauge state ${\Phi}^{}_{g}$ is particularly interesting since it is the essence of the cosmological constant. This point was more or less noticed by Kallosh in Ref. \cite{Kallosh}.] Note that these solutions have null norm, i.e., $\langle {\Phi}^{}_{g} , {\Phi}^{}_{g} \rangle = 0 = \langle {\Phi}^{}_{s} , {\Phi}^{}_{s} \rangle$, $\langle \cdot , \cdot \rangle$ being the Klein-Gordon inner product (\ref{Kl-Go inpro}), and that the the constant factors are adjusted in such a way to get $\langle {\Phi}^{}_{g} , {\Phi}^{}_{s} \rangle = 1$. Then, defining $\widetilde{\Phi}^{}_{0,0,0} = {\Phi}^{}_{g}+{\Phi}^{}_{s}/2$, we obtain the \emph{true (normalizable) zero mode}; $\langle \widetilde{\Phi}^{}_{0,0,0} , \widetilde{\Phi}^{}_{0,0,0} \rangle = 1$. This mode supplements the set $\big\{\widetilde{\Phi}^{}_{L_{\geqslant 1}lm}\big\}$ in the sense that the obtained set $\big\{\widetilde{\Phi}^{}_{L_{\geqslant 0}lm}\big\}$ is complete and of strictly positive norm. However, dS$_4$ invariance once again is violated for the space constructed over this new set; for example, we have \cite{Gupta 2000}:
\begin{eqnarray}\label{action 2}
\big( M_{03} + \mathrm{i} M_{04} \big) \widetilde{\Phi}^{}_{0,0,0} &=& \big( M_{03} + \mathrm{i} M_{04} \big) {\Phi}^{}_{s} \nonumber \\
&=& - \mathrm{i} \frac{\sqrt{6}}{4} \Big( \widetilde{\Phi}^{}_{1,0,0} + \big(\widetilde{\Phi}^{}_{1,0,0}\big)^\ast \Big) - \frac{\sqrt{6}}{4} \Big( \widetilde{\Phi}^{}_{1,1,0} + \big(\widetilde{\Phi}^{}_{1,0,0}\big)^\ast \Big)\,.
\end{eqnarray}
These arguments explicitly reveal that, in the sense of the standard Hilbert space quantization (with strictly positive norm modes), one cannot obtain a covariant QFT formulation of the dS$_4$ minimally coupled scalar field; the appearance of negative norm modes is indeed the price that must be paid. This problem, known in the literature under the name of ``zero-mode" problem, was first put forward by Allen in Ref. \cite{Allen}, and during recent three decades, it has been subject to scrutiny in a number of works (see, for instance, Refs. \cite{Allen/Folacci,Kirsten,Tolley,Gupta 2000,deBievre'}).

Here, it is worth noting that, among all the dS$_4$ infinitesimal generators $M_{AB}$, only the four generators $M_{0\texttt{B}}$ ($\texttt{B}=1,2,3,4$) are responsible for the dS$_4$ symmetry breaking of the space of states $\big\{\widetilde{\Phi}^{}_{L_{\geqslant 0}lm}\big\}$. Actually, the other six generators $M_{\texttt{A} \texttt{B}}$ ($\texttt{A},\texttt{B}=1,2,3,4$), corresponding to the compact $SO(4)$ subgroup, preserve dS$_4$ invariance and allow for (in the usual sense) a $SO(4)$-covariant QFT formulation of the dS$_4$ minimally coupled scalar field. Of course, considering $SO(4)$ covariance instead of full dS$_4$ covariance (in other words, spontaneous symmetry breaking) for the formulation, although is of some interest in the context of quantum cosmology, is not relevant to the aim of this paper, that is, presenting a consistent QFT reading of elementary systems in dS$_4$ spacetime.

Nevertheless, in Refs. \cite{Gupta 2000,deBievre'} the authors, employing group representation theory along with a proper adaptation (Krein spaces) of the Wightman-G\"{a}rding axioms for massless fields (the Gupta-Bleuler scheme), have remarkably introduced a consistent way out of this problem, which is of great interest in the context of our study. Considering the fact that the classical free minimally coupled scalar field is, in addition to dS$_4$, also gauge covariant, they have shown that a rather straightforward application of the Gupta-Bleuler formalism allows one to prevent the symmetry breaking altogether in such a way that the minimally coupled scalar quantized field transforms correctly under the dS$_4$ and the gauge transformations, and acts on a states space containing a vacuum invariant under all of them. Note that this appealing result is not in contradiction with the Allen no-go theorem \cite{Allen} (presented above), since the given quantum field in Refs. \cite{Gupta 2000,deBievre'} enjoys a \emph{Krein structure} instead of a Hilbertian one. Below, following Refs. \cite{Gupta 2000,deBievre'}, we will discuss this (Krein-)Gupta-Bleuler quantization scheme in detail.

\subsubsection{(Krein-)Gupta-Bleuler triplet}\label{Subsubsec Gupa-B}
Note that from now on, in two steps, we simplify the notations used above again. First, here, we define the family of indices $K = \big\{ k=(L,l,m)\in \mathbb{N}\times\mathbb{N}\times\mathbb{Z} \;;\; L\neq 0,\; 0\leqslant l \leqslant L,\; -l\leqslant m \leqslant l \big\}$ for the positive norm modes excluding the $L=0$ mode, and $K^\prime=K\cup \big\{ 0 \big\}$ for the whole set of positive norm modes.

\begin{figure}[H]
    \begin{center}
    \includegraphics[height=.25\textheight]{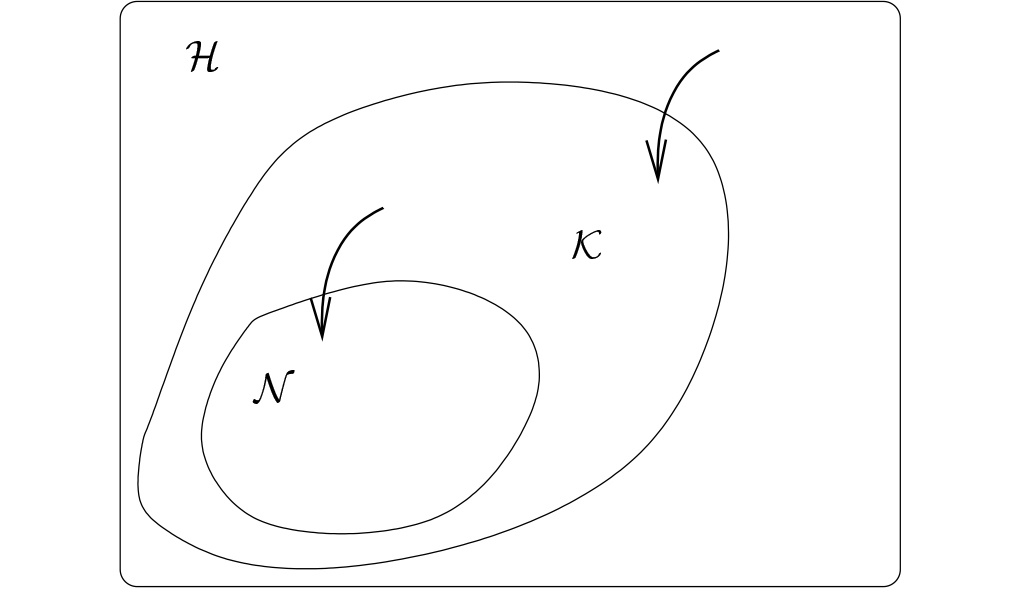}
    \end{center}
    \caption{Gupta-Bleuler type structure lying behind the states space of the dS$_4$ minimally coupled scalar field.}
    \label{FIG. GB 1}
\end{figure}

Now, we elaborate the Gupta-Bleuler type structure lying behind the states space of the dS$_4$ minimally coupled scalar field:
\begin{itemize}
\item{Let ${\cal{H}}$ denote the complete, nondegenerate, and invariant space (say the \emph{total space}) of states, which is equipped with the Klein-Gordon inner product (\ref{Kl-Go inpro}). This space, which is realized by completion of the space of regular elements $\big\{ \widetilde{\Phi}^{}_{k} \;;\; k\in K^\prime \big\}$ by applying the action of the dS$_4$ group (see, for instance, Eqs. (\ref{action 1}) and (\ref{action 2})), reads:
    \begin{eqnarray}\label{pppppppp}
    {\cal{H}} = {\cal{H}}^{}_+ \oplus {\cal{H}}^\ast_+\,,
    \end{eqnarray}
    where:
    \begin{eqnarray}
    {\cal{H}}^{}_+ = \Big\{ c_0^{} \widetilde{\Phi}^{}_{0} + \sum_{k\in K} c^{}_k \widetilde{\Phi}^{}_{k} \;;\; c^{}_0,c^{}_k\in\mathbb{C},\; \sum_{k\in K} |c^{}_k|^2 < \infty \Big\}\,,
    \end{eqnarray}
    stands for a Hilbert space and ${\cal{H}}^\ast_+$ for an anti-Hilbert space (a space with definite negative inner product):
    \begin{eqnarray}
    \langle \widetilde{\Phi}^{}_{k} , \widetilde{\Phi}^{}_{k} \rangle = \langle \widetilde{\Phi}^{}_{0} , \widetilde{\Phi}^{}_{0} \rangle = 1\,, \;\;\;\;\;\;\;\mbox{and}\;\;\;\;\;\;\; \langle (\widetilde{\Phi}^{}_{k})^\ast , (\widetilde{\Phi}^{}_{k})^\ast \rangle = \langle (\widetilde{\Phi}^{}_{0})^\ast , (\widetilde{\Phi}^{}_{0})^\ast \rangle = -1\,,
    \end{eqnarray}
    for all $k\in K$. This shows that the total space of states is a Krein space (an indefinite inner product space). Once again, we must underline that neither ${\cal{H}}^{}_+$ nor ${\cal{H}}^\ast_+$ carries a representation of the dS$_4$ group, and therefore, the decomposition of the total space ${\cal{H}}$ is not covariant (though, it is $SO(4)$ covariant!).}
\item{Considering the arguments given in the previous subsubsection, the total space ${\cal{H}}$ contains the (one-particle) physical states space ${\cal{K}}$, with the following definition, as a closed subspace:
    \begin{eqnarray}
    {\cal{K}} = \Big\{ c_g^{} {\Phi}^{}_{g} + \sum_{k\in K} c^{}_k \widetilde{\Phi}^{}_{k} \;;\; c_g^{},c^{}_k\in\mathbb{C},\; \sum_{k\in K} |c^{}_k|^2 < \infty \Big\}\,.
    \end{eqnarray}
    This subspace is a degenerate (semi-definite) inner product space:
    \begin{eqnarray}
    \langle \widetilde{\Phi}^{}_{k} , \widetilde{\Phi}^{}_{k} \rangle = 1\,, \;\;\;\;\;\;\; \langle \widetilde{\Phi}^{}_{k} , {\Phi}^{}_{g} \rangle = \langle {\Phi}^{}_{g} , {\Phi}^{}_{g} \rangle = 0\,.
    \end{eqnarray}
    for all $k\in K$.}
\item{The subspace ${\cal{K}}$ in turn admits an invariant one-dimensional (gauge) subspace ${\cal{N}}$, which is constructed over ${\Phi}^{}_{g}$. The latter, as already mentioned, is orthogonal to every element in ${\cal{K}}$ including itself.}
\end{itemize}
These three invariant spaces of states all together form the (Krein-)Gupta-Bleuler triplet (see FIG. \ref{FIG. GB 1}):
\begin{eqnarray}
{\cal{N}} \subset {\cal{K}} \subset {\cal{H}}\,,
\end{eqnarray}
which carries the indecomposable structure for the UIR $\Pi_{p=1,0}$ appearing in the case of the dS$_4$ minimally coupled scalar field:
\begin{eqnarray}\label{ffffffffffffff}
\underbrace{\Pi_{1,0}}_{{\cal{H}}/{\cal{K}}=\big\{ c^{}_s {\Phi}^{}_{s} + ({\cal{K}}/{\cal{N}})^\ast_{} \big\}}\;\;\;\longmapsto\;\;\;\underbrace{\Pi_{1,0}}_{{\cal{K}}/{\cal{N}}=\big\{ \sum_{k\in K} c^{}_k \widetilde{\Phi}^{}_{k} \big\}}
\;\;\;\longmapsto\;\;\;\underbrace{\Upsilon_0}_{{\cal{N}}=\big\{ c^{}_g {\Phi}^{}_{g} \big\}}
\end{eqnarray}
where, again, $c_s^{},c_k^{},c_t^{}\in\mathbb{C}$, while $\sum_{k\in K} |c^{}_k|^2 < \infty$, the arrows `$\longmapsto$' show the \emph{leaks under the group action}\footnote{To see the point, it is sufficient to recall Eqs. (\ref{action 1}) and (\ref{action 2}).}, and finally $\Upsilon_0$ stands for the dS$_4$ trivial UIR, on which both dS$_4$ Casimir operators vanish. Note that the trivial representation $\Upsilon_0$ is naturally carried by the space of constant functions, which here plays the role of gauge states space.

Regarding the above structure, we must underline that, although, the subspace ${\cal{K}}$ is referred to as the (one-particle) physical states space, strictly speaking, physical states are those that are defined up to a constant function (gauge state), and the physical space is the one that is characterized by the quotient space ${\cal{K}}/{\cal{N}}$, i.e., the central part of the above indecomposable representation carrying the UIR $\Pi_{p=1,0}$. Another point that must be noticed here is that, in this construction, every positive norm state is not physical (in other words, does not belong to ${\cal{K}}$). For example, despite the positivity of the norm of $\widetilde{\Phi}^{}_{0}$, it is not a physical state ($\widetilde{\Phi}^{}_{0} \notin {\cal{K}}$), since, as already shown, its transformation under the action of the group entails states of negative norm violating unitarity.

\subsubsection{Quantum field}
Since the distinction between the normalizable and non-normalizable modes has been well established above, the `$\;{\widetilde{}}\;$' symbol no longer needs to be kept over the normalizable ones. Therefore, at the second step of simplifying our conventions, from now on (by abuse of notation), we adopt a unified symbol for all the modes as $\phi^{}_k \equiv \widetilde{\Phi}^{}_{k}$, for $k\in K^\prime = K \cup \big\{0\big\}$, $\phi^{}_g \equiv {\Phi}^{}_{g}$, and $\phi^{}_s \equiv {\Phi}^{}_{s}$.

Here, having the above arguments in mind and employing a new representation of the canonical commutation relations, we present a covariant QFT reading \emph{\`{a} la} Gupta-Bleuler of the dS$_4$ minimally coupled scalar field. At a glance, the corresponding quantum field in the usual sense would be a distribution, with values which are operators on the Fock space ${\mathscr{H}}$ constructed over the total (Krein) space ${\cal{H}}$,\footnote{To see the details of the construction of Fock spaces on Krein spaces, one can refer to Ref. \cite{Mintchev}.} while the expectation values of the corresponding observables, as usual in a Gupta-Bleuler construction, would be calculated merely with respect to physical states. The latter are obtained from the Fock vacuum by creation of the elements of the one-particle physical states space ${\cal{K}}$. Technically, the actions of the creation $a^\dagger_{}(\phi)$ and annihilation $a(\phi)$ operators on the dense subset of ``regular elements" of the form $\dot{h} = \big( \dot{h}_0, \dot{h}_1, \;...\; , \dot{h}_n, \;...\; , \dot{h}_N,0,0, \;...\; \big) \in {\mathscr{H}}$ respectively read:
\begin{eqnarray}\label{creation'mc}
\Big( a^\dagger_{}(\phi) \dot{h}_n \Big)(x_1, \;...\; , x_n) = \frac{1}{\sqrt{n}} \sum_{i=1}^n \phi(x_i) \; \dot{h}_{n-1}(x_1, \;...\; ,\breve{x}_i, \;...\; , x_n)\,,
\end{eqnarray}
\begin{eqnarray}\label{annihilation'mc}
\Big( a(\phi) \dot{h}_n \Big)(x_1, \;...\; , x_n) = \mathrm{i} R^2 \sqrt{n+1} \int_{\rho=0} \phi^\ast(\rho,\textbf{u}) \overset{\leftrightarrow}{\partial}_\rho \; \dot{h}_{n+1}\big((\rho,\textbf{u}),x_1, \;...\; , x_n\big) \; \mathrm{d}\mu(\textbf{u})\,,
\end{eqnarray}
where $\mathrm{d}\mu(\textbf{u})$ refers to the invariant measure on $\mathbb{S}^3$ and, again, $\breve{x}_i$ means that this term is omitted.

It can be easily shown that the given creation and annihilation operators verify the usual commutation relations:
\begin{eqnarray}\label{commutations}
[a(\phi),a(\phi^\prime)] = 0\,, \;\;\;\;\;\;\; [a^\dagger_{}(\phi),a^\dagger_{}(\phi^\prime)] = 0\,, \;\;\;\;\;\;\; [a(\phi),a^\dagger_{}(\phi^\prime)] = \langle \phi , \phi^\prime \rangle = 1\,,
\end{eqnarray}
and also:
\begin{eqnarray}
\underline{\mathscr{U}}(\underline{g}) a(\phi) \underline{\mathscr{U}}^{\ast}(\underline{g}) = a\big( \underline{U}(\underline{g}) \phi \big)\,, \;\;\;\;\;\;\; \underline{\mathscr{U}}(\underline{g}) a^\dagger_{}(\phi) \underline{\mathscr{U}}^{\ast}(\underline{g}) = a^\dagger_{}\big( \underline{U}(\underline{g}) \phi \big)\,,
\end{eqnarray}
where $\langle\cdot,\cdot\rangle$ here stands for the Klein-Gordon inner product and, again, $\underline{U}(\underline{g})$ and $\underline{\mathscr{U}}(\underline{g})$, respectively, for the natural representation of the dS$_4$ group on the total (Krein) space ${\cal{H}}$ and its extension to the Fock space ${\mathscr{H}}$.

Denoting the annihilators of the modes $\phi^{}_k$, $\phi^{}_0$, $\phi^{\ast}_0$, and $\phi^{\ast}_k$ ($k\in K$), respectively, by $a^{}_k\equiv a(\phi^{}_k)$, $a^{}_0\equiv a(\phi^{}_0)$, $b^{}_0\equiv a(\phi^{\ast}_0)$, and $b^{}_k\equiv a(\phi^{\ast}_k)$, the (unsmeared) field operator reads:
\begin{eqnarray}\label{field operator mc}
\hat{\phi}(x) &=& \sum_{k} \Big( \phi^{}_k(x) a^{}_k + \phi^{\ast}_k(x) a^{\dagger}_k \Big) - \sum_{k} \Big( \phi^{\ast}_k(x) b^{}_k + \phi^{}_k(x) b^{\dagger}_k \Big) \nonumber\\
&& + \Big( \phi^{}_0(x) a^{}_0 + \phi^{\ast}_0(x) a^{\dagger}_0 \Big) - \Big( \phi^{\ast}_0(x) b^{}_0 + \phi^{}_0(x) b^{\dagger}_0 \Big)\,, \;\;\;\;\;\;\; k\in K\,,
\end{eqnarray}
where the nonvanishing commutation relations between the operators, for all $k \in K^\prime = K \cup \big\{0\big\}$, are:
\begin{eqnarray}
[a^{}_k,a^{\dagger}_k] = 1\,, \;\;\;\;\;\;\; [b^{}_k,b^{\dagger}_k] = -1\,.
\end{eqnarray}
Note that: (i) Regarding the commutation relations given in (\ref{commutations}), the fact that $\langle \phi^{\ast}_k , \phi^{\ast}_k \rangle = -1$ leads to the minus sign in $[b^{}_k,b^{\dagger}_k] = -1$. (ii) It is obvious that the field operator $\hat{\phi}(x)$, which appears as the sum of an operator and its conjugate, is real.

For later use, it is also convenient to reexpress the field operator $\hat{\phi}(x)$ in the following form:
\begin{eqnarray}\label{field operator mc'}
\hat{\phi}(x) &=& \sum_{k} \Big( \phi^{}_k(x) a^{}_k + \phi^{\ast}_k(x) a^{\dagger}_k \Big) - \sum_{k} \Big( \phi^{\ast}_k(x) b^{}_k + \phi^{}_k(x) b^{\dagger}_k \Big) \nonumber\\
&& + \phi^{}_g(x)\big( a^{}_s + a^{\dagger}_s \big) + \phi^{}_s(x) \big( a^{}_g - a^{\dagger}_g \big)\,,\;\;\;\;\;\;\; k\in K\,,
\end{eqnarray}
where $a^{}_s\equiv a(\phi^{}_s)$ and $a^{}_g\equiv a(\phi^{}_g)$; note that $\phi^{\ast}_g=\phi^{}_g$ and $\phi^{\ast}_s=-\phi^{}_s$.

Now, we show that the introduced field operator $\hat{\phi}(x)$ is covariant (in the strong sense) and local. We begin with the covariance property implying that:
\begin{eqnarray}
\underline{\mathscr{U}}(\underline{g}) \hat{\phi}(x) \underline{\mathscr{U}}^{-1}(\underline{g}) = \hat{\phi}(\underline{g} \diamond x)\,,
\end{eqnarray}
for all $\underline{g} \in \mathrm{Sp}(2,2)$. To clarify this point, following the general instruction presented in subsection \ref{Subsec gen free fields}, we smear the field operator $\hat{\phi}(x)$ with a real test function $f_1 \in \mathfrak{D}(\underline{M}_R)$:
\begin{eqnarray}
\hat{\phi}(f_1) = \int \hat{\phi}(x) f_1(x) \; \mathrm{d}\mu(x) &=& \sum_{k} \int \phi^{}_k(x) f_1(x) \; \mathrm{d}\mu(x) \; A^{}_k + \sum_{k} \int \phi^\ast_k(x) f_1(x) \; \mathrm{d}\mu(x) \; A^{\dagger}_k \nonumber\\
&=& \sum_{k} \langle\phi^{\ast}_k(x) , f_1(x) \rangle^{}_{L^2} \; A^{}_k + \sum_{k} \langle\phi^{}_k(x) , f_1(x) \rangle^{}_{L^2} \; A^{\dagger}_k\,,\;\;\;\;\;\;\; k\in K^\prime = K \cup \big\{0\big\}\,,\;\;\;\;\;
\end{eqnarray}
where $A^{}_k \equiv a^{}_k - b^{\dagger}_k$, $\mathrm{d} \mu(x)$ is the invariant measure, and $\langle \cdot , \cdot \rangle^{}_{L^2}$ designates the $L^2(\underline{M}_R)$ inner product:
\begin{eqnarray}\label{L2 product}
\langle g,h \rangle^{}_{L^2} = \int g^{\ast}(x) h(x) \; \mathrm{d}\mu(x)\,.
\end{eqnarray}
[In order to distinguish between the above $L^2(\underline{M}_R)$ inner product and the Klein-Gordon one (which is also used in the current discussion), we make precise our notation by adding proper subscripts $\langle \cdot , \cdot \rangle^{}_{L^2}$ and $\langle \cdot , \cdot \rangle^{}_{KG}$, respectively.] Note that, the operators $A^{}_k=A(\phi^{}_k)$ and $A^{\dagger}_k=A^\dagger(\phi^{}_k)$ are, respectively, anti-linear (or conjugate-linear) and linear in the argument $\phi^{}_k$ (this point can be easily checked by considering the definitions of $A^{}_k$ and $A^{\dagger}_k$ along with Eqs. (\ref{creation'mc}) and (\ref{annihilation'mc})). Hence, we can rewrite the smeared field operator $\hat{\phi}(f_1)$ in the following form:
\begin{eqnarray}
\hat{\phi}(f_1) &=& \sum_{k} \langle\phi^{\ast}_k(x) , f_1(x) \rangle^{}_{L^2} \; A(\phi^{}_k) + \sum_{k} \langle\phi^{}_k(x) , f_1(x) \rangle^{}_{L^2} \; A^{\dagger}(\phi^{}_k)\,,\nonumber\\
&=& A \Big( \sum_{k} \langle\phi^{}_k(x) , f_1(x) \rangle^{}_{L^2} \; \phi^{}_k \Big) +  A^{\dagger} \Big( \sum_{k} \langle\phi^{}_k(x) , f_1(x) \rangle^{}_{L^2} \; \phi^{}_k \Big)\,,\;\;\;\;\;\;\; k\in K^\prime = K \cup \big\{0\big\}\,.
\end{eqnarray}
Defining:
\begin{eqnarray}
p(f_1) \equiv \sum_{k} \langle\phi^{}_k(x) , f_1(x) \rangle^{}_{L^2} \; \phi^{}_k\,,\;\;\;\;\;\;\; k\in K^\prime = K \cup \big\{0\big\}\,,
\end{eqnarray}
which is a vector-valued distribution taking values in the total space ${\cal{H}}$ generated by the modes, we obtain:
\begin{eqnarray}\label{field smeard}
\hat{\phi}(f_1) = A \big( p(f_1) \big) + A^{\dagger} \big( p(f_1) \big)\,.
\end{eqnarray}
Note that the role of $p$ is to associate with each test function $f_1 \in \mathfrak{D}(\underline{M}_R)$ an element of the total space ${\cal{H}}$ ($p(f_1) \in {\cal{H}}$), thus we can consider the associated annihilation and creation operators. $p(f_1)$ is indeed the \emph{unique} vector in ${\cal{H}}$ such that, for any $\psi \in {\cal{H}}$, we have:
\begin{eqnarray}\label{smeared}
\langle p(f_1) , \psi \rangle^{}_{KG} =  \langle f_1 , \psi \rangle^{}_{L^2}\,.
\end{eqnarray}
From the above equation (which its both sides are invariant under isometries) and the nondegeneracy property of the Klein-Gordon inner product on the total space ${\cal{H}}$, on one hand and on the other hand, the invariance of ${\cal{H}}$ itself under the action of the dS$_4$ group, we obtain:
\begin{eqnarray}\label{tttttttttt}
\langle \underline{U}(\underline{g}) p(f_1) \;,\; \psi \rangle^{}_{KG} &=& \langle p(f_1) \;,\; \underline{U}^{-1}(\underline{g})\psi \rangle^{}_{KG} \nonumber\\
&=& \langle f_1 \;,\; \underline{U}^{-1}(\underline{g})\psi \rangle^{}_{L^2} \nonumber\\
&=& \langle \underline{U}(\underline{g})f_1 \;,\; \psi \rangle^{}_{L^2} \nonumber\\
&=& \langle p( \underline{U}(\underline{g}) f_1) \;,\; \psi \rangle^{}_{KG}\,.
\end{eqnarray}
Then, for the extension of the representation $\underline{U}(\underline{g})$ to its counterpart $\underline{\mathscr{U}}(\underline{g})$ on the Fock space ${\mathscr{H}}$ (strictly speaking, on the set of finite length elements of ${\mathscr{H}}$), we get:
\begin{eqnarray}\label{eeeeeeeeeeeeee}
\underline{\mathscr{U}}(\underline{g}) \hat\phi(f_1) \underline{\mathscr{U}}^{-1}(\underline{g}) &=& A \big( \underline{U}(\underline{g}) p(f_1) \big) +  A^{\dagger} \big( \underline{U}(\underline{g}) p(f_1) \big) \nonumber\\
&=& A \big( p(\underline{U}(\underline{g}) f_1) \big) +  A^{\dagger} \big( p(\underline{U}(\underline{g}) f_1) \big) \nonumber\\
&=& \hat\phi\big( \underline{U}(\underline{g}) f_1)\,.
\end{eqnarray}
These identities explicitly reveal that $p(f_1)$, $\hat{\phi}(f_1)$, and, as we will show below, $\hat{\phi}(x)$ transform correctly under the dS$_4$ group action (they are covariant!).

In the distribution sense, one can simply obtain the corresponding unsmeared form of Eq. (\ref{smeared}), which reads $\langle p(x) , \psi \rangle^{}_{KG} = \psi(x)$, for all $\psi \in {\cal{H}}$. The covariant vector-valued distribution $p$, verifying the field equation, now can be expanded in the basis, and in the unsmeared form, as follows:
\begin{eqnarray}\label{unsmeared}
p(x) &=& \sum_{k} \phi^{\ast}_k(x) \phi^{}_k - \sum_{k} \phi^{}_k(x) \phi^{\ast}_k + \phi^{\ast}_0(x) \phi^{}_0 - \phi^{}_0(x) \phi^{\ast}_0
\nonumber\\
&=& \sum_{k} \phi^{\ast}_k(x) \phi^{}_k - \sum_{k} \phi^{}_k(x) \phi^{\ast}_k + \phi^{}_g(x) \phi^{}_s - \phi^{}_s(x) \phi^{}_g\,,\;\;\;\;\;\;\; k\in K\,.
\end{eqnarray}
Then, considering Eq. (\ref{field operator mc}) (or equivalently (\ref{field operator mc'})) along with (\ref{unsmeared}), one can simply show that:
\begin{eqnarray}\label{field unsmeard}
\hat{\phi}(x) = A \big( p(x) \big) +  A^{\dagger} \big( p(x) \big)\,,
\end{eqnarray}
which is the unsmeared form of (\ref{field smeard}). Then, following the identities (\ref{tttttttttt}) and (\ref{eeeeeeeeeeeeee}), the covariance property of the quantum field $\hat{\phi}(x)$ is also justified. The point to be made here is that the definition of the quantum field, as is obvious from the above, does not depend on the modes but on the (dS$_4$-invariant) total space ${\cal{H}}$ they span. Frankly speaking, the modes are merely a tool for calculation. We will come back to this important point later in the last paragraph of this subsection.

In order to check the locality of the field, we calculate $\texttt{W}$, the kernel of $p$, defined formally by:
\begin{eqnarray}
p(f)(x^\prime) = \int \texttt{W}(x^\prime,x) f(x) \; \mathrm{d}\mu(x)\,,
\end{eqnarray}
where $\mathrm{d}\mu(x)$ is the invariant measure. Then, having Eq. (\ref{smeared}) in mind, for $f^{}_1,h^{}_1\in \mathfrak{D}(\underline{M}_R)$, we get:
\begin{eqnarray}
\langle p(f^{}_1) , p(h^{}_1) \rangle^{}_{KG} &=& \langle f^{}_1 , p(h^{}_1) \rangle^{}_{L^2}\nonumber\\
&=& \iint f^\ast_1(x^\prime) \texttt{W}(x^\prime,x) h^{}_1(x) \; \mathrm{d}\mu(x) \mathrm{d}\mu(x^\prime)\,,
\end{eqnarray}
which in the unsmeared form reads:
\begin{eqnarray}
\texttt{W}(x^\prime,x) = \langle p(x^\prime) , p(x) \rangle^{}_{KG}\,.
\end{eqnarray}
Substituting Eq. (\ref{unsmeared}) into the above equation, we obtain:
\begin{eqnarray}\label{commutator}
\texttt{W}(x^\prime,x) &=& \sum_{k} \phi^{\ast}_k(x) \phi^{}_k(x^\prime) - \sum_{k} \phi^{}_k(x) \phi^{\ast}_k(x^\prime) + \phi^{\ast}_0(x) \phi^{}_0(x^\prime) - \phi^{}_0(x) \phi^{\ast}_0(x^\prime) \nonumber\\
&=& \sum_{k} \phi^{\ast}_k(x) \phi^{}_k(x^\prime) - \sum_{k} \phi^{}_k(x) \phi^{\ast}_k(x^\prime) + \frac{1}{2\pi R} \big( \phi^{}_s (x^\prime) - \phi^{}_s(x) \big)\,,\;\;\;\;\;\;\; k\in K\,,
\end{eqnarray}
where we have used the fact that $\phi^{}_g (x)=\phi^{}_g (x^\prime)=1/2\pi R$. The above formula explicitly reveals that $\texttt{W}(x^\prime,x)=- \mathrm{i} \widetilde{G}(x,x^\prime)$, where $- \mathrm{i} \widetilde{G}$ is the natural commutator.\footnote{We remind that dS$_4$ (generally, dS) spacetime is globally hyperbolic, therefore the so-called commutator $\widetilde{G} = G^{\mbox{\tiny{adv}}} - G^{\mbox{\tiny{ret}}}$ is uniquely defined \cite{Isham'}; the propagators $G^{\mbox{\tiny{adv}}}$ and $G^{\mbox{\tiny{ret}}}$ are defined with respect to $\Box_x G^{\mbox{\tiny{adv}}}(x,x^\prime) = \Box_x G^{\mbox{\tiny{ret}}}(x,x^\prime) = - \delta(x,x^\prime)$ and, for a given $x^\prime$, the support in $x$ of $G^{\mbox{\tiny{adv}}}$ (respectively, $G^{\mbox{\tiny{ret}}}$) lies in the past (respectively, future) cone of $x^\prime$. This commutator $\widetilde{G}(x,x^\prime)$ is equal to $+\frac{1}{2}$ for $x$ in the future cone of $x^\prime$, $-\frac{1}{2}$ for $x$ in the past cone of $x^\prime$, and $0$ elsewhere.} Thus, the vector-valued distribution $p$ is just the kernel of the natural commutator.

Having the commutation relations given in (\ref{commutations}) along with Eq. (\ref{field unsmeard}) in mind, now the locality of the field can be explicitly seen through the following relation:
\begin{eqnarray}
[\hat{\phi}(x^\prime), \hat{\phi}(x)] = 2 \langle p(x^\prime) , p(x) \rangle^{}_{KG} = -2 \mathrm{i} \widetilde{G}(x,x^\prime)\,,
\end{eqnarray}
since $\widetilde{G}(x,x^\prime)$ vanishes when $x$ and $x^\prime$ are spacelike separated.

In summary, so far, we have introduced a quantization of the dS$_4$ minimally coupled scalar field satisfying the Wightman axioms. Of course, the appearance of some nonphysical states in this QFT construction, as the price that must be paid to assure the full covariance of the theory, is unavoidable. Now, we must encounter the definition of the observables of the theory to show that these nonphysical states make no trouble, like the appearance of negative energies, for the theory. This is our duty in the coming discussion.

\subsubsection{Stress tensor}
First of all, we point out that the aforementioned classical gauge transformation $\phi(x) \rightarrow \phi(x) + \varrho$ ($\phi \equiv \Phi$ and $\varrho$ being a constant function), on the quantum level, can be implemented by \cite{deBievre'}:
\begin{eqnarray}\label{gauge transform}
V(\varrho) = \exp\big( -\pi\varrho R ( a^\dagger_g - a^{}_g ) \big)\,,
\end{eqnarray}
based upon which, we have:
\begin{eqnarray}
V(-\varrho) \hat{\phi}(x) V(\varrho) = \hat{\phi}(x) + \varrho \mathbbm{1}\,.
\end{eqnarray}

Now, we get involved with the definition of the (second-quantized) physical space, denoted here by ${\mathscr{K}}$. This space is generated from the Fock vacuum $\Omega$ by creating members of ${\cal{K}}$ (i.e., the set of one-particle physical states):
\begin{eqnarray}\label{phys states}
\dot{h} \; \big( \in {\mathscr{K}} \big) = \big(a^\dagger_g\big)^{n_0} \big(a^\dagger_{k_1}\big)^{n_1} ...\; \big(a^\dagger_{k_l}\big)^{n_l} \Omega\,,
\end{eqnarray}
where the indices $k_1,\;...\;,k_l \in K$. The subspace of ${\mathscr{K}}$, constituted by those states orthogonal to every element in ${\mathscr{K}}$ including themselves, is denoted by $\bar{\mathscr{N}}$:
\begin{eqnarray}
\dot{h}\in \bar{\mathscr{N}} \;\;\; \mbox{if} \;\;\; \dot{h} \in {\mathscr{K}} \;\;\; \mbox{and} \;\;\; \langle \dot{h},\dot{f}\rangle=0 \;\;\; \mbox{for all} \;\; \dot{f} \in {\mathscr{K}}\,.
\end{eqnarray}
[In the above equation, $\langle\cdot,\cdot\rangle$ stands for the Klein-Gordon inner product.]

\begin{figure}[H]
    \begin{center}
    \includegraphics[height=.25\textheight]{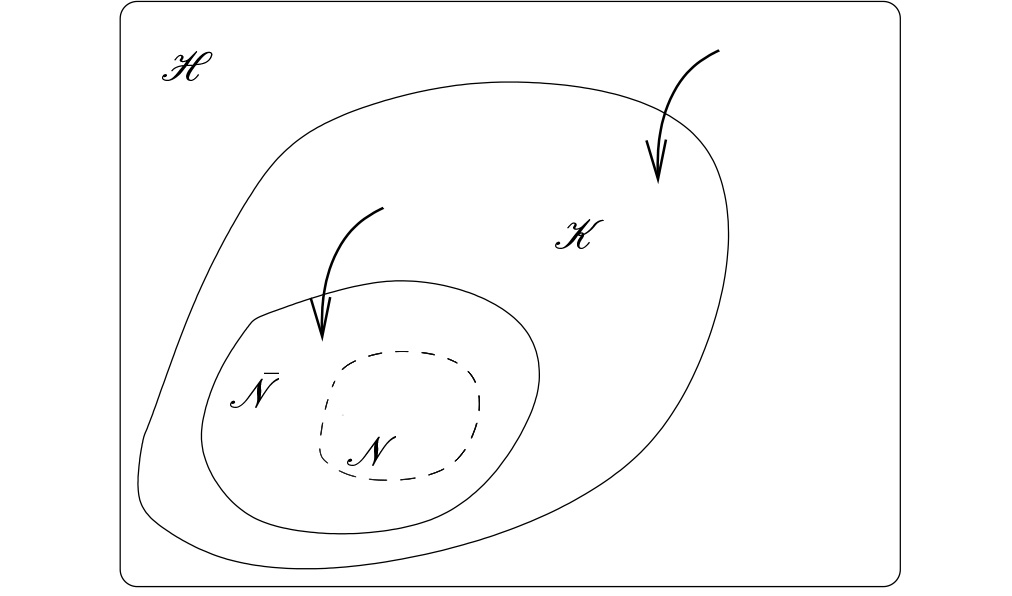}
    \end{center}
    \caption{Second-quantized Gupta-Bleuler structure lying behind the (Fock) states space of the dS$_4$ minimally coupled scalar field.}
    \label{FIG. GB 2}
\end{figure}

Note that the operator $a^{}_g$, when is restricted to the physical states space ${\mathscr{K}}$, is the null operator; for any $\dot{h}_{\breve{g}} \propto \big( a^\dagger_{k_1} \big)^{n_1} ...\; \big( a^\dagger_{k_l} \big)^{n_l} \; \Omega \in {\mathscr{K}}/\bar{\mathscr{N}}$, the state $a^\dagger_g \dot{h}_{\breve{g}}$ belongs to $\bar{\mathscr{N}}$. This implies that, for any $\dot{h}\in {\mathscr{K}}$ and any real $\varrho$, the states $\dot{h}$ and ${V}(\varrho)\dot{h}$ (see Eq. (\ref{gauge transform})) are equal up to an element belonging to $\bar{\mathscr{N}}$. In this sense, we refer to $\bar{\mathscr{N}}$ as the second quantized space of global gauge states. [The above statement therefore can be rephrased as: consistently, two physical states belonging to ${\mathscr{K}}$ are viewed as physically equivalent when they are distinguished by a global gauge state, and under a gauge transformation a physical state goes into an equivalent one.] Accordingly, the corresponding second-quantized Gupta-Bleuler triplet, manifestly invariant under the dS$_4$ group action, reads (see FIG. \ref{FIG. GB 2}):
\begin{eqnarray}
\bar{\mathscr{N}} \subset {\mathscr{K}} \subset {\mathscr{H}}\,.
\end{eqnarray}
Again, ${\mathscr{H}}$ refers to the Fock space constructed over the total (Krein) space ${\cal{H}}$.

We here argue that the associated (Krein-)Fock vacuum state $\Omega$ is unique (strictly speaking, \emph{quasi-unique}) and normalizable. We begin with the first property (the second one will be justified as a byproduct of our next discussion). Let ${\mathscr{N}}$ denote the space of dS$_4$-invariant states of ${\mathscr{H}}$, which technically can be realized by the space generated from the vacuum by applying $a^\dagger_g$. This space, as a subspace of $\bar{\mathscr{N}}$, is trivially of infinite dimension. Then, it seems that the (Krein-)Fock vacuum $\Omega$ is not the unique dS$_4$-invariant state. However, regarding the above statements, one can simply verify that all the states belonging to ${\mathscr{N}}$ are physically equivalent to a member of the one-dimensional space spanned by the vacuum state. Therefore, the (Krein-)Fock vacuum state $\Omega$ is indeed unique, up to physical equivalence (quasi-uniqueness).

We now turn to the duty of defining the observables of the theory. As usual in a Gupta-Bleuler structure, observables are defined with respect to the property that they do not ``see" the gauge states. An observable $\hat{O}$ therefore would be a symmetric operator on ${\mathscr{H}}$ in such a way that, when $\dot{h}$ and $\dot{h}^\prime$ are equivalent physical states (i.e., members of ${\mathscr{K}}$ such that $\dot{h} - \dot{h}^\prime \in \bar{\mathscr{N}}$), it must verify:
\begin{eqnarray}
\langle \dot{h} , \hat{O} \; \dot{h} \rangle = \langle \dot{h}^\prime , \hat{O} \; \dot{h}^\prime\rangle\,.
\end{eqnarray}
Accordingly, it is manifest that the field operator $\hat{\phi}(x)$ itself is not an observable, due to the appearance of $a^\dagger_s$ and $a^{}_s$ (see Eq. (\ref{field operator mc'})) which do not commute with $a^{}_g$ and $a^\dagger_g$, respectively (they ``see" the gauge states); to get the point, one needs to consider Eq. (\ref{Zero mode}) and its subsequent arguments along with (\ref{commutations}). The operators $\partial_\mu\hat{\phi}(x)$, however, no longer carry those terms (containing $a^\dagger_s$ and $a^{}_s$), and hence, the physically interesting observables, such as the stress tensor $T^{}_{\mu\nu}$, can be constructed in terms of them.

The above argument clearly reveals that why the approach through two-point functions is not relevant for the dS$_4$ minimally coupled scalar field. As a matter of fact, due to the gauge dependency of the field operator, different two-point functions, which appear in this context, are gauge dependent as well; with the exception of $\texttt{W}$ (see Eq. (\ref{commutator})), which is defined independently of the field and which, being a commutator, is gauge invariant. For instance, the symmetric two-point function $G^{(1)}(x,x^\prime)= \frac{1}{2} \langle\Omega, \hat{\phi}(x) \hat{\phi}(x^\prime) + \hat{\phi}(x^\prime) \hat{\phi}(x) \;\Omega\rangle$ is not expected to have great meaning in this context and a straightforward calculation shows that it vanishes. This result of course is nothing but another manifestation of the Allen no-go theorem \cite{Allen}.

Now, let us study the behavior of the stress tensor $T^{}_{\mu\nu}$ which, as already mentioned, is an observable in our construction. We show that the ``negative frequency part" (with respect to the conformal time) of the quantum field provides an automatic and covariant renormalization for the theory, allowing for a trivial calculation of the mean values of the components of $T^{}_{\mu\nu}$. We begin with the classical form of the corresponding stress tensor:
\begin{eqnarray}
T^{}_{\mu\nu} = \partial^{}_\mu\phi \; \partial^{}_\nu\phi - \frac{1}{2} g^{}_{\mu\nu} g^{\rho\sigma}_{} \partial^{}_\rho\phi \; \partial^{}_\sigma\phi\,,
\end{eqnarray}
where $\mu,\nu=0,1,2,3$ stem from the conformal coordinates that we have used in our construction. We consider the excited physical states as:
\begin{eqnarray}
\dot{h}_{\breve{g}} \; \big( \in {\mathscr{K}}/\bar{\mathscr{N}} \big) = \frac{1}{\sqrt{n^{}_1 ! \; ... \; n^{}_l !}} \; \big( a^\dagger_{k_1} \big)^{n_1} ...\; \big( a^\dagger_{k_l} \big)^{n_l} \; \Omega\,.
\end{eqnarray}
To compute the mean values $\langle\dot{h}_{\breve{g}} , T^{}_{\mu\nu} \; \dot{h}_{\breve{g}}\rangle$, we start with $\langle\dot{h}_{\breve{g}} , \partial^{}_\mu\hat{\phi}(x) \partial^{}_\nu \hat{\phi}(x) \; \dot{h}_{\breve{g}}\rangle$. Technically, the terms containing $a^{}_{g}$, $a^{}_{s}$, $a^{\dagger}_{g}$, and $a^{\dagger}_{s}$ do not contribute in the calculations, since, on one hand, the terms including $a^{}_{s}$ and $a^{\dagger}_{s}$ disappear due to the derivation and, on the other hand, the operators $a^{}_{g}$ and $a^{\dagger}_{g}$ commute with all the remaining operators including themselves, and hence, the associated terms neutralize each other in the calculations. Then, we get:
\begin{eqnarray}
\langle\dot{h}_{\breve{g}} , \partial^{}_\mu\hat{\phi}(x) \partial^{}_\nu \hat{\phi}(x) \; \dot{h}_{\breve{g}}\rangle = \sum_{k\in K} \big( \partial^{}_\mu\phi^{}_{k}(x) \partial^{}_\nu \phi^\ast_{k}(x) - \partial^{}_\mu\phi^\ast_{k}(x) \partial^{}_\nu \phi^{}_{k}(x) \big) + 2\sum_{i=1}^l n_i \mbox{Re} \big( \partial^{}_\mu\phi^\ast_{k_i}(x) \partial^{}_\nu \phi^{}_{k_i}(x) \big)\,.
\end{eqnarray}
On the right-hand side, the first and third terms are exactly those that appear through the usual (Hilbertian) calculations, while the first one containing infinite terms must be renormalized. In the above Krein QFT construction, however, thanks to the appearance of the unusual second term yielded by the terms of the field operator containing $b^{}_k$ and $b^{\dagger}_k$, we automatically obtain:
\begin{eqnarray}
\langle\dot{h}_{\breve{g}} , \partial^{}_\mu\hat{\phi}(x) \partial^{}_\nu \hat{\phi}(x) \; \dot{h}_{\breve{g}}\rangle = 2\sum_{i=1}^l n_i \mbox{Re} \big( \partial^{}_\mu\phi^\ast_{k_i}(x) \partial^{}_\nu \phi^{}_{k_i}(x) \big)\,.
\end{eqnarray}
Extending the above result to the whole expression $\langle\dot{h}_{\breve{g}} , T^{}_{\mu\nu} \; \dot{h}_{\breve{g}}\rangle$ makes apparent the aforementioned automatic renormalization of the mean values of the stress tensor in the context of Krein QFT reading of the dS$_4$ minimally coupled scalar field. Remarkably, the ``positiveness of the energy", for any physical state $\dot{h}_{\breve{g}}$, is a direct result of this automatic renormalization procedure:
\begin{eqnarray}
\langle\dot{h}_{\breve{g}} , T^{}_{00} \; \dot{h}_{\breve{g}}\rangle \geqslant 0\,.
\end{eqnarray}
The above quantity vanishes if and only if $\dot{h}_{\breve{g}}=\Omega$ (the vacuum is normalizable!). Note that evaluating the above quantity with respect to the other states, which do not belong to the physical states space ${\mathscr{K}}$, may lead to negative values for the energy operator, but obviously this does not raise any concern since these states are not physical. Therefore, on the free field level, the appearance of nonphysical states in the Krein QFT formulation of the dS$_4$ minimally coupled scalar field literally causes no trouble for the theory. Nevertheless, on the interacting level, the situation is different, and the appearance of some virtual particles, analogous to QED in the presence of charges (see, for instance, Ref. \cite{Mandl}), is naturally expected.

Concerning the above automatic renormalization procedure of the stress tensor, we must underline that it remarkably verifies the so-called Wald axioms:
\begin{itemize}
\item{The causality and covariance of the procedure are trivially assured by construction.}
\item{It yields the formal results for the physical states.}
\item{The foundation of the above computations is the identity $[b^{}_k,b^\dagger_k]=-1$, based upon which we get:
     \begin{eqnarray}
     a^{}_k a^{\dagger}_k + a^{\dagger}_k a^{}_k + b^{}_k b^{\dagger}_k + b^{\dagger}_k b^{}_k = 2a^{\dagger}_k a^{}_k + 2 b^{\dagger}_k b^{}_k\,.
     \end{eqnarray}
     Therefore, the above automatic renormalization of the stress tensor is equivalent to reordering when one deals with the physical states (on which $b^{}_k$ vanishes).}
\end{itemize}
To see more on this automatic renormalization procedure and its physical consequences, readers are referred to Refs. \cite{CCP-Krein,Casimir-Krein}.

At the end, we would like to discuss Bogoliubov transformations in the above context. First of all, we draw attention to the fact that the introduced field operator for the minimally coupled scalar field acts on the Fock space ${\mathscr{H}}$ constructed over the total space of states ${\cal{H}}$, which contains all the corresponding positive and negative norm states (see Eq. (\ref{pppppppp})). Then, trivially, under Bogoliubov transformations, any transformed element $\phi_k^\prime= c^{}_k\phi_k^{} - d^{}_k \phi_k^{\ast}$, with $|c^{}_k|^2 - |d^{}_k|^2 = 1$, still belongs to ${\cal{H}}$. In this sense, the given field operator $\hat\phi$ and consequently the corresponding (Krein-)Fock vacuum $\Omega$ are by construction independent of any Bogoliubov transformation; quite contrary to the usual canonical quantization methods in which the definition of such notions depends crucially on the choice of the modes (often being relevant to a particular choice of time coordinate), here their definitions depend only on the total space of states ${\cal{H}}$. Nevertheless, this argument does not mean that Bogoliubov transformations, which are merely simple changes of physical states space, are no longer valid in this construction. Frankly speaking, under Bogoliubov transformations, we have several possibilities for the physical states space, spanned by $\phi_k^\prime$'s, and correspondingly, several possibilities for the physical part $\hat\phi_{\text{\tiny{phys}}}^\prime$ of the quantum field $\hat\phi$, while we have only one field and one vacuum, which are invariant under Bogoliubov transformations. [For more detailed discussions, see Ref. \cite{Garidi2005}.]

%%%%%%%%%%%%%%%%%%%%%%%%%%%%%%%%%%%%%%%%%%%%%%%%%%%%%%%%%%%%%%%%%%%%%%%%%%%%%%%%%%%%%%%%%%%%%%%%%%%%%%%%%%%%%%%%%%%%%%%%
%%%%%%%%%%%%%%%%%%%%%%%%%%%%%%%%%%%%%%%%%%%%%%%%%%%%%%%%%%%%%%%%%%%%%%%%%%%%%%%%%%%%%%%%%%%%%%%%%%%%%%%%%%%%%%%%%%%%%%%%
%%%%%%%%%%%%%%%%%%%%%%%%%%%%%%%%%%%%%%%%%%%%%%%%%%%%%%%%%%%%%%%%%%%%%%%%%%%%%%%%%%%%%%%%%%%%%%%%%%%%%%%%%%%%%%%%%%%%%%%%
%%%%%%%%%%%%%%%%%%%%%%%%%%%%%%%%%%%%%%%%%%%%%%%%%%%%%%%%%%%%%%%%%%%%%%%%%%%%%%%%%%%%%%%%%%%%%%%%%%%%%%%%%%%%%%%%%%%%%%%%
%%%%%%%%%%%%%%%%%%%%%%%%%%%%%%%%%%%%%%%%%%%%%%%%%%%%%%%%%%%%%%%%%%%%%%%%%%%%%%%%%%%%%%%%%%%%%%%%%%%%%%%%%%%%%%%%%%%%%%%%
%%%%%%%%%%%%%%%%%%%%%%%%%%%%%%%%%%%%%%%%%%%%%%%%%%%%%%%%%%%%%%%%%%%%%%%%%%%%%%%%%%%%%%%%%%%%%%%%%%%%%%%%%%%%%%%%%%%%%%%%
%%%%%%%%%%%%%%%%%%%%%%%%%%%%%%%%%%%%%%%%%%%%%%%%%%%%%%%%%%%%%%%%%%%%%%%%%%%%%%%%%%%%%%%%%%%%%%%%%%%%%%%%%%%%%%%%%%%%%%%%

\part{Notion of mass in (A)dS$_4$ relativity}\label{Part mass}
As already mentioned, in dS$_4$ (generally, dS) relativity, some of our most familiar notions (in Poincar\'{e} relativity) like time, rest mass, energy, momentum, spin, etc. may disappear or at least need significant modifications. As a matter of fact, in dS$_4$ (generally, dS) spacetime, granted that no globally timelike Killing vector exists, neither time nor energy can be globally defined. Nevertheless, so far, we have discussed that the physical interpretation of the global analyticity requirements in the complexified dS$_4$ manifold can serve as a natural substitute to the usual spectral condition of ``positivity of the energy" in constructing dS$_4$ QFT's. Moreover, giving explicit arguments based on symmetry considerations in part \ref{Part II}, we have shown that the concept of spin in dS$_4$ relativity can be made precise, as well. Now, pursuing our aim towards a consistent formulation of dS$_4$ elementary systems, we are left with the task of finding a suitable interpretation of the notion of mass in dS$_4$ relativity. This is the matter of our discussion in the present part. [Of course, to compare the results and provide guidelines for future investigations, we also discuss the notion of mass (or at rest energy) in AdS$_4$ relativity.] Again, the premise of our arguments rests on the symmetry considerations in the sense given by Wigner and the mutually deformation/contraction of the UIR's of the relativistic groups (A)dS$_4$ and Poincar\'{e} towards each other.

Note that, in this part again, $c$ (the speed of light) and $\hbar$ (the Planck constant) are no longer normalized to unity. Together with $R$, the radius of curvature of the (A)dS$_4$ hyperboloid, they provide dimensionally independent quantities, which are used to specify the natural unit of ``mass" $\hbar/cR$.

\setcounter{equation}{0} \section{Discussion/reminder: mass and symmetries}
In flat Minkowski spacetime, the notion of (rest) mass originates from the ubiquitous law of energy conservation (regarded as a consequence of the Poincar\'{e} symmetry) and the hypothesis of the existence of elementary systems in Nature (in an asymptotic sense) \cite{Wigner1939,Newton/Wigner}. As a matter of fact, a (free) quantum elementary system in flat Minkowski spacetime (in the sense given by Wigner) forms a UIR space for the Poincar\'{e} relativity group and, according to the Wigner classification (see appendix \ref{App Com UIR's Poincare}), each Poincar\'{e} UIR is completely characterized in terms of two invariant parameters, namely, the (rest) mass $m$ and the spin $s$.

On the other hand, endowing spacetimes with a certain curvature is the only way to deform the Poincar\'{e} group. Such a deformation leads to the dS$_4$ and AdS$_4$ relativity groups \cite{BacryLevi} (see also Ref. \cite{Levy-Nahas}). Accordingly, on the representation level, the massive UIR's of the dS$_4$ \cite{Dixmier, Takahashi'} and the AdS$_4$ \cite{Angelopoulos,Evans67} groups are respected as those that are realized by deformations of the Poincar\'{e} massive UIR's with positive energy ${\cal P}^>_{s,m}$. The dS$_4$ and AdS$_4$ massive representations respectively belong to the dS$_4$ principal series, denoted here by $\underline{U}^{\mbox{\small{ps}}}_{s,\nu}$, with $\nu\in\mathbb{R}$ and $s\in\mathbb{N}/2$, and to the AdS$_4$ discrete series, denoted by $D_{s,\varsigma}$, with $s\in\mathbb{N}/2$ and $\varsigma + n>s+1$, $n\in \N$, with possible extension to all real $\varsigma>s+1$ when dealing with the universal covering of AdS$_4$. In this group-theoretical context, the dS$_4$-invariant parameter $\nu$ and the AdS$_4$-invariant parameter $\varsigma$, as real dimensionless parameters, are considered as replacements for the Minkowskian rest mass $m$ in dS$_4$ and AdS$_4$ relativities, respectively. [For more details, one can refer to section \ref{Sec contraction}.]

Considering the above, there is however an irreconcilable difference between the dS$_4$-invariant parameter $\nu$ and the AdS$_4$-invariant parameter $\varsigma$, which the existence/nonexistence of a lower bound for the ``energy spectrum" lies at its heart. In the AdS$_4$ case, the parameter $\varsigma$ is in fact the lowest value of the discrete spectrum of the generator of ``time" rotations (see section \ref{Sec contraction}, where we have denoted this generator by $L_{50}$). As a result, an unambiguous meaning of a rest energy (mass) in AdS$_4$ relativity can be carried by the AdS$_4$-invariant parameter $\varsigma$ (in the energy AdS$_4$ units $\hbar/cR$):
\begin{eqnarray}\label{adsreste}
E_{\text{\tiny{AdS$_4$}}}^{\;\text{\tiny{rest}}} \equiv \frac{\hbar c \varsigma}{R}\,,\;\;\;\;\;\;\; \varsigma >s+1\,, \;\;\;\mbox{and}\;\;\; s\in\mathbb{N}/2\,.
\end{eqnarray}
Hence, the physical notion of ``energy at rest" survives when deforming the Poincar\'{e} group towards the AdS$_4$ one.

By contrast, the situation is radically different in the dS$_4$ (generally, dS) case, due to the indefinite spectrum of ``energy", common to all UIR's of this group. Indeed, in the dS$_4$ (generally, dS) case, the spectrum of the generator of ``time" hyperbolic rotations covers the whole real line (see section \ref{Sec contraction}). This entails an ambiguous notion of rest energy (mass) in dS$_4$ (generally, dS) relativity. Of course, despite this ambiguity, it is still legitimate to allocate an ``energy" dimension to the dS$_4$-invariant parameter $\nu$ (in the ``energy" dS$_4$ units $\hbar/cR$):
\begin{eqnarray}\label{dsscalee}
E^{}_{\text{\tiny{dS$_4$}}} \equiv \frac{\hbar c \nu}{R}\,,\;\;\;\;\;\;\; \nu \in \mathbb{R}\,.
\end{eqnarray}

Nevertheless, in $2003$, a consistent and univocal definition of mass in dS$_4$ (generally, dS) relativity has been put forward by Garidi \cite{Garidimass}\footnote{Readers should notice that this seminal paper, although never published, has been well acknowledged in the literature.}, which precisely gives sense to terms like ``massive" and ``massless" fields in dS$_4$ (generally, dS) relativity according to their Minkowskian counterparts, yielded by the group contraction procedures. In the coming sections, we will elaborate this definition. We will also compare this definition with other mass formulas introduced within the dS$_4$ context, and will show that it enjoys the advantage to encompass all such formulas.

\setcounter{equation}{0} \section{Garidi mass: definition} \label{Sec Garidi mass}
The Garidi mass formula is given by \cite{Garidimass}:
\begin{eqnarray}\label{conmassf}
\mathfrak{M}_{\text{\tiny{dS$_4$}}}^2 = \frac{\hbar^2}{c^2R^2}\Big( \langle Q^{(1)}\rangle^{}_{\text{\tiny{dS$_4$}}} - \langle Q^{(1)}_{p=q=s}\rangle^{}_{\text{\tiny{dS$_4$}}}\Big)\,,
\end{eqnarray}
where: (i) $\langle Q^{(1)}\rangle^{}_{\text{\tiny{dS$_4$}}}$ stands for the eigenvalues of the dS$_4$ quadratic Casimir operator (see section \ref{Sec Dixmier}) associated with those dS$_4$ UIR's which are \emph{meaningful from the Minkowskian point of view} (see section \ref{Sec contraction}), namely: the whole principal series representations, characterized by $\underline{U}^{\mbox{\small{ps}}}_{s,\nu}$, with $s\in\mathbb{N}/2$ and $\nu\in\mathbb{R}$; the only physical representation (in the sense of contraction/extension to the Poincar\'{e}/conformal UIR's) of the complementary series, characterized by $\underline{U}^{\mbox{\small{cs}}}_{s,\nu}$, with $s=0$ and $\nu = \textstyle\frac{1}{2}$; and finally the representations lying at the lower limit of the discrete series, characterized by $\Pi^\pm_{p,q}$, with $p=q=s\in\mathbb{N}/2$. (ii) The eigenvalues $\langle Q^{(1)}_{p=q=s}\rangle^{}_{\text{\tiny{dS$_4$}}}$ correspond to the dS$_4$ UIR's $\Pi^\pm_{p=s,q=s}$ in the discrete series, the so-called dS$_4$ massless UIR's (see section \ref{Sec contraction}):
\begin{eqnarray}\label{eigdisse}
\langle Q^{(1)}_{p=q=s}\rangle^{}_{\text{\tiny{dS$_4$}}} = -2(s^2-1)\,.
\end{eqnarray}

Before going any further, few points regarding the above mass definition must be underlined:
\begin{itemize}
\item{The mass formula (\ref{conmassf}) is natural in the sense that when, proceeding as in section \ref{Sec wave Eqs.}, the field equations for the dS$_4$ principal (massive) fields (for instance, of integer spin $s$) are rewritten in terms of the Laplace-Beltrami operator $\square_R$, we obtain:\footnote{Here, for the sake of simplicity, we assume that the rank and the spin of the fields $\Psi^{(r)}$ are equal ($r=s$).}
    \begin{eqnarray}
    \Big(Q^{(1)} - \langle Q^{(1)}\rangle^{}_{\text{\tiny{dS$_4$}}}\Big)\Psi^{(r=s)}(x) = 0 \;\;\;\;\;\;\; \Rightarrow \;\;\;\;\;\;\; \Big(\square_R + R^{-2}s(s+1) + R^{-2}\langle Q^{(1)}\rangle^{}_{\text{\tiny{dS$_4$}}}\Big)\Psi^{(s)}(x) = 0\,,
    \end{eqnarray}
    or, according to Eqs. (\ref{conmassf}) and (\ref{eigdisse}):
    \begin{eqnarray}
    \Big( \frac{\hbar^2}{c^2}\big(\square_R + R^{-2}\big[2-s(s-1)\big]\big) + \mathfrak{M}^2_{\text{\tiny{dS$_4$}}}\Big)\Psi^{(s)}(x) = 0\,.
    \end{eqnarray}}
\item{Since the minimum values of $\langle Q^{(1)}\rangle_{\mbox{\tiny{dS}}_4}$ occur at $p=q=s\in\mathbb{N}/2$, which is the case for the lower limit of the discrete series UIR's (the dS$_4$ massless representations), the mass formula (\ref{conmassf}) assures that for every dS$_4$ UIR, \emph{meaningful from the point of view of a Minkowskian observer}, we have ${\mathfrak{M}}^2_{\text{\tiny{dS$_4$}}}\geqslant 0$. To see the point, let us examine all such UIR's one-by-one:
    \begin{itemize}
    \item{For the dS$_4$ massive cases, associated with the principal series representations $\underline{U}^{\mbox{\small{ps}}}_{s,\nu}$, the Casimir eigenvalues are given by Eq. (\ref{Casiprise}), and hence, the Garidi mass takes the form:
        \begin{eqnarray}\label{Garidi mass prin}
        \mathfrak{M}_{\mbox{\tiny{dS}}_4} = \frac{\hbar}{cR}\sqrt{\nu^2 + \Big(s-\frac{1}{2}\Big)^2} = \frac{\hbar|\nu|}{cR}\sqrt{1+\frac{\big(s-\frac{1}{2}\big)^2}{\nu^2}} \;\; > 0\,.
        \end{eqnarray}
        [Note that $\sqrt{\nu^2}=|\nu|$, for $\nu\in\mathbb{R}$.\footnote{Here, one must also notice that, given a specific value of $s$, the principal series representations corresponding to the parameters $\nu$ and $-\nu$ are equivalent (they have same Casimir eigenvalues).} Therefore, the indefiniteness of the dS$_4$-invariant parameter $\nu$ no longer leaks to the definition of Garidi mass $\mathfrak{M}_{\mbox{\tiny{dS}}_4}$.] Moreover, considering the Poincar\'{e}-Minkowski mass as $m=\hbar|\nu|/cR$ and quite similar to the Poincar\'{e} contraction limit (i.e., letting $R$ and $\nu$ tend to infinity, while $m=\hbar|\nu|/cR$ remains unchanged (see section \ref{Sec contraction})), we get:
        \begin{eqnarray}\label{asymp dS}
        \mathfrak{M}_{\mbox{\tiny{dS}}_4} \; \longrightarrow \; m\sqrt{1+\frac{\big(s-\frac{1}{2}\big)^2}{\nu^2}} \;\; \big(\; > 0 \big) =
        \left \{ \begin{array}{rl} m \;\;\;\;\;\;\;\;\;\;\;\;\;\;\;\;\;\;\;\;\;\;\;\; &\mbox{for}\;\;\; s=1/2 \,,\\
        \\ m + {\cal O}(1/\nu) \;\;\;\;\;\;\;\; &\mbox{for}\;\;\; s\neq 1/2\,, \end{array}\right.
        \end{eqnarray}
        where, borrowing the notation used in sections \ref{Sec Group contraction dS4} and \ref{Sec contraction}, the arrow `$\longrightarrow$' stands for the contraction (type) limit under vanishing curvature. [Regarding Eqs. (\ref{Garidi mass prin}) and (\ref{asymp dS}), we also would like to draw attention to the specific symmetric place characterized by the case of spin $s=1/2$, with respect to the scalar $s=0$ and the boson $s=1$ cases.]}
    \item{For the dS$_4$ massless cases, as discussed in section \ref{Sec contraction}, two distinguished cases are involved: the discrete series representations $\Pi^\pm_{p,q}$, with $p=q=s\in\mathbb{N}/2$, for which $\langle Q^{(1)}\rangle^{}_{\mbox{\tiny{dS}}_4} = \langle Q^{(1)}_{p=q=s}\rangle^{}_{\mbox{\tiny{dS}}_4}$, and the complementary series representation $\underline{U}^{\mbox{cs}}_{s,\nu}$, with $s=0$ and $\nu = \textstyle\frac{1}{2}$, for which $\langle Q^{(1)}\rangle^{}_{\mbox{\tiny{dS}}_4} = 2 = \langle Q^{(1)}_{p=q=0}\rangle^{}_{\mbox{\tiny{dS}}_4}$ (see Eqs. (\ref{Casimir dis}) and (\ref{Compleeee}), respectively). Obviously, according to the mass definition (\ref{conmassf}), for both dS$_4$ massless cases, we have $\mathfrak{M}_{\text{\tiny{dS$_4$}}}=0$.}
    \end{itemize}}
\item{In the case of a dS$_4$ UIR with \emph{no} meaningful Minkowskian interpretation, one can still apply the Garidi mass formula (\ref{conmassf}), but without referring to a Minkowskian meaning.}
\end{itemize}

The counterpart of the Garidi mass formula (\ref{conmassf}) in AdS$_4$ relativity reads \cite{GazNo2008, GazNo2011}:
\begin{eqnarray}\label{AdS mass Formula}
\mathfrak{M}^2_{\mbox{\tiny{AdS}}_4} = \frac{\hbar^2}{c^2R^2}\Big( \langle Q^{(1)} \rangle^{}_{\mbox{\tiny{AdS}}_4} - \langle Q^{(1)}_{\varsigma = s+1}\rangle^{}_{\mbox{\tiny{AdS}}_4} \Big)\,,
\end{eqnarray}
where, similarly, $\langle Q^{(1)}\rangle^{}_{\text{\tiny{AdS$_4$}}}$ refers to the eigenvalues of the AdS$_4$ quadratic Casimir operator corresponding to those AdS$_4$ UIR's which are \emph{meaningful from the Minkowskian point of view} (see section \ref{Sec contraction}). Considering Eq. (\ref{Caeigena}), it is easy to check that, for all AdS$_4$ massless cases (see section \ref{Sec contraction}), we have $\mathfrak{M}_{\mbox{\tiny{AdS}}_4}=0$ and, for the AdS$_4$ massive cases, i.e, the UIR's $D_{s, \varsigma}$ with $s\in\mathbb{N}/2$, $\varsigma > s+1$ for $s\neq 0$, and $\varsigma > 2$ for $s= 0$, we have:
\begin{eqnarray}\label{AdS mass discrete}
\mathfrak{M}_{\mbox{\tiny{AdS}}_4} = \frac{\hbar}{cR} \sqrt{\Big( \varsigma - \frac{3}{2} \Big)^2 - \Big(s - \frac{1}{2}\Big)^2} = \frac{\hbar}{cR}\Big(\varsigma - \frac{3}{2}\Big)\sqrt{1-\Big(\frac{s-1/2}{\varsigma -3/2}\Big)^2} \; >0\,.
\end{eqnarray}
The case $\varsigma = 2$ with $s= 0$ is special since it is the unique one for which $\mathfrak{M}_{\mbox{\tiny{AdS}}_4} = 0$ whereas it does correspond to a massive and not massless representation. Moreover, considering the Poincar\'{e}-Minkowski mass as $m=\hbar \varsigma/cR$ and similar to the Poincar\'{e} contraction limit, we obtain:
\begin{eqnarray}\label{asymp AdS}
\mathfrak{M}_{\mbox{\tiny{AdS}}_4} \; \longrightarrow \; m \sqrt{1-\Big(\frac{s-1/2}{\varsigma -3/2}\Big)^2} \; >0\,.
\end{eqnarray}

\setcounter{equation}{0} \section{Garidi mass: a more elaborate discussion}\label{Sec Precision (A)dS mass}
So far, based on a robust group-theoretical approach, we have presented an explicit definition of mass (Garidi mass) in dS$_4$ and AdS$_4$ relativities. In particular, the asymptotic relations between the Minkowskian rest mass $m$ and its (possible) dS$_4$ and AdS$_4$ counterparts have been given (see the contraction (type) formulas (\ref{asymp dS}) and (\ref{asymp AdS}) for the dS$_4$ and AdS$_4$ cases, respectively). Here, we discuss the above results more elaborately.

\subsection{DS$_4$ case}
We begin by recalling the fact that dS$_4$ spacetime is the maximally symmetric solution to the vacuum Einstein's equations with positive cosmological constant $\Lambda$. The latter is connected to the dS$_4$ scalar (Ricci) curvature, which is equal to $4\Lambda$. By taking into account the astrophysical data coming from type Ia supernovae \cite{Perlmutter}, the dS$_4$ radius of curvature is given by $R = \sqrt{3/\Lambda} = c/H$, where $c$ is the speed of light and $H$ is the Hubble constant. The recent estimated value of the Hubble constant is:
\begin{eqnarray}
H\equiv H_0 = 2.5 \times 10^{-18}\;s^{-1}\,.
\end{eqnarray}

Now, according to the formula (\ref{Garidi mass prin}), we can give the abstract (real) dimensionless parameter $\nu$, labeling the dS$_4$ massive UIR's, the status of a physical quantity in terms of other measurable physical quantities:
\begin{eqnarray}\label{nu contraction}
|\nu| = \frac{mcR}{\hbar}=\frac{mc^2}{\hbar H}\,,
\end{eqnarray}
where, again, $m$ is a Minkowskian (rest) mass. Below, we argue that this identity, established based on a pure group representation approach, is by far restrictive.

Considering dS$_4$ spacetime as a perturbation of the flat Minkowski background, it would be typical to define the following dimensionless physical (in the Minkowskian sense) quantity:
\begin{eqnarray}\label{99999999999}
\vartheta \equiv \vartheta_m = \frac{\hbar}{mcR}=\frac{\hbar\sqrt{\Lambda}}{\sqrt{3}mc} = \frac{\hbar H}{mc^2} = \frac{m^{}_H}{m}\,,
\end{eqnarray}
where $m^{}_H=\hbar H/c^2$ is a ``Hubble mass".\footnote{It is worth noting that the parameter $\vartheta$ can also be viewed as $\vartheta =\lambda_{cmp}/R$, the ratio of the (reduced) Compton wavelength $\lambda_{cmp}=\hbar/mc$ of a Minkowskian object with $m>0$ considered at the limit with the universal length $R$ (yielded by the dS$_4$ geometry).} Clearly, for $m=m^{}_H$, we get $\vartheta=1$. For some known masses $m$ and the recent estimated value of the Hubble constant $H_0$, the estimated values of the parameter $\vartheta$ are given in TABLE \ref{Table. mass range}. Accordingly, one can easily see that this parameter is totally negligible for all known massive elementary particles. In other words, the infinitesimal current value of $\Lambda$ does not allow for any practical dS$_4$ effect on the level of high energy physics experiments (no dS$_4$ effect is perceptible in LHC experiments!). However, if $\vartheta$ gets closer to $1$, namely, when one deals either with theories based on large values of $\Lambda$ or with theories giving an infinitely small masses to photons, gravitons, or other massless gauge fields, adopting the dS$_4$ point of view (or $\Lambda$ effect) becomes unavoidable. This is the case, for instance, in the standard inflation scenario.

\begin{table}
\begin{center}
\item[]\begin{tabular}[c]{|l|c|c|}\hline
$\Big|$  Mass $m$ & $\vartheta^{}_m \approx $ \\ \hline
$\Big|$  Hubble mass & 1 \\ \hline
$\Big|$  up. lim. neutrino mass $m_{\nu}$ & $0.165 \times 10^{-32}$ \\ \hline
$\Big|$  electron mass $m_e$ & $0.3 \times 10^{-37}$ \\ \hline
$\Big|$  proton mass $m_p$ & $0.17  \times 10^{-41}$ \\ \hline
$\Big|$  $W^{\pm}$  boson mass & $0.2 \times 10^{-43}$ \\ \hline
$\Big|$  Planck mass $M_{Pl}$ & $0.135 \times 10^{-60}$ \\ \hline
\end{tabular}
\end{center}
\caption{Estimated values of the parameter $\vartheta$ for some known masses $m$ and the recent estimated value of the Hubble constant $H_0$.}
\label{Table. mass range}
\end{table}

Now, considering a Minkowskian (rest) mass $m$ and the fundamental length $R$, nothing prevents us to deal with the dS$_4$ UIR parameter $\nu$, assigned to ``physics" in a constant-curvature spacetime, as a meromorphic function of $\vartheta$; $\nu=\nu(\vartheta)$. Then, in a certain neighborhood of $\vartheta=0$ (i.e, the convergence interval $\vartheta\in (0,\vartheta_{max})$), we have the following Laurent expansion:
\begin{eqnarray}\label{Laurentnu}
\nu = \nu(\vartheta) = \frac{1}{\vartheta} + e_0 + e_1\vartheta + \;...\; + e_n\vartheta^n + \;... \;,
\end{eqnarray}
where $e_n$'s, denoting the expansion coefficients, are pure numbers to be specified.\footnote{Here, one must notice that the above expansion formula does not alter the contraction procedure from the point of view of a local Minkowskian observer, except the fact that it allows us to determine the values of $\nu$ (either positive or negative) with respect to values of $\vartheta$ in a (positive) neighborhood of $\vartheta = 0$.} Note that by multiplying the above equation by $\vartheta$ and letting $\vartheta$ tend to zero, one recovers asymptotically Eq. (\ref{nu contraction}). Actually, the Garidi mass formula appears as a perfect example of this expansion. The point can be easily seen, in the case of the principal (massive) series, by adjusting: \begin{eqnarray}\label{LaurentmassdS}
|\nu| \equiv |\nu(\vartheta)| = \sqrt{\frac{1}{\vartheta^2}-\Big(s-\frac{1}{2}\Big)^2} = \frac{1}{\vartheta}-\Big(s-\frac{1}{2}\Big)^2\Big(\frac{\vartheta}{2}+{\cal O}(\vartheta^2)\Big)\,, \;\;\;\;\;\;\; \vartheta\in (0,1/|s-1/2|]\,,
\end{eqnarray}
in the corresponding mass formula (\ref{Garidi mass prin}). Considering the expansion (\ref{LaurentmassdS}) along with the formula (\ref{asymp dS}), for the ``energy" scale of a dS$_4$ massive elementary system (introduced in Eq. (\ref{dsscalee})), we also obtain:
\begin{eqnarray}
E^{}_{\mbox{\tiny{dS}}_4} = \frac{\hbar c}{R} \; \nu(\vartheta) =
\left \{ \begin{array}{rl} \mathfrak{M}_{\text{\tiny{dS$_4$}}}c^2 \; \longrightarrow \; mc^2 \;\;\;\;\;\;\;\;\;\;\;\;\;\;\;\;\;\;\;\;\;\;\;\;\;\;\;\;\;\;\;\;\; &\mbox{for}\;\;\; s=1/2 \,,\\
\\
\mathfrak{M}_{\text{\tiny{dS$_4$}}}c^2+{\cal O}(\vartheta) \; \longrightarrow \; mc^2 + {\cal O}(\vartheta) \;\;\;\;\;\;\;\; &\mbox{for}\;\;\; s\neq 1/2\,, \end{array}\right.
\end{eqnarray}
where, again, the arrow `$\longrightarrow$' stands for the contraction (type) limit under vanishing curvature.

\subsection{AdS$_4$ case}
Similarly, for the AdS$_4$ discrete (massive) cases, we also consider the following Laurent expansion of $\varsigma$ in a given neighborhood of $\vartheta =0$ ($\vartheta\in (0,\vartheta_{max})$):
\begin{eqnarray}\label{Laurentads}
\varsigma =\varsigma(\vartheta) =\frac{1}{\vartheta} + a_0 + a_1\vartheta + \;...\; + a_n\vartheta^n + \;... \;,
\end{eqnarray}
where, again, the coefficients $a_n$ are pure numbers to be specified. Note that $\lim_{\vartheta\rightarrow 0}\varsigma(\vartheta)\vartheta =1$ recovers asymptotically the identity $m=\hbar \varsigma/cR$. As a matter of fact, the AdS$_4$ mass formula appears as a perfect example of the above expansion formula. To see the point, it is sufficient to substitute the following expansion into the AdS$_4$ mass formula (\ref{AdS mass discrete}) (associated with the AdS$_4$ massive cases):
\begin{eqnarray}\label{Garidireads}
\varsigma(\vartheta) = \frac{3}{2} + \sqrt{\frac{1}{\vartheta^2}+\Big(s-\frac{1}{2}\Big)^2} = \frac{1}{\vartheta} + \frac{3}{2} + \Big(s-\frac{1}{2}\Big)^2 \Big(\frac{\vartheta}{2}+{\cal O}(\vartheta^2)\Big)\,, \;\;\;\;\;\;\; \vartheta\in (0,1/|s-1/2|]\,.
\end{eqnarray}
By comparing the above expansion with the generic one (\ref{Laurentads}), we obtain $a_0 = 3/2$. Moreover, considering the expansion (\ref{Garidireads}) along with the relation (\ref{asymp AdS}), for the AdS$_4$ massive cases, the AdS$_4$ rest energy given in Eq. (\ref{adsreste}) takes the form:
\begin{eqnarray}\label{resteads}
E_{\mbox{\tiny{AdS}}_4}^{\;\mbox{\tiny{rest}}} = \frac{\hbar c}{R} \;  \varsigma(\vartheta) =
\left \{ \begin{array}{rl} \mathfrak{M}_{\mbox{\tiny{AdS$_4$}}}c^2 + \frac{3}{2}\hbar\omega \; \longrightarrow \; mc^2 + \frac{3}{2}\hbar\omega \;\;\;\;\;\;\;\;\;\;\;\;\;\;\;\;\;\;\;\;\;\;\;\;\;\;\;\;\;\;\;\;\; &\mbox{for}\;\;\; s=1/2 \,,\\
\\
\mathfrak{M}_{\mbox{\tiny{AdS$_4$}}}c^2 + \frac{3}{2}\hbar\omega + {\cal O}(\vartheta) \; \longrightarrow \; mc^2 + \frac{3}{2}\hbar\omega + {\cal O}(\vartheta) \;\;\;\;\;\;\;\; & \mbox{for} \;\;\; s\neq 1/2\,. \end{array}\right.
\end{eqnarray}
Again, the arrow `$\longrightarrow$' refers to the contraction (type) limit under vanishing curvature, and $\omega = c/R$. The above relation reveals that, at the first order in the curvature ($\propto1/R$), an AdS$_4$ massive elementary system asymptotically appears as the sum of a relativistic free particle and a quantum harmonic oscillator, respectively, with the rest energy $mc^2$ and zero-point energy $\textstyle\frac{3}{2}\hbar\omega$ (in this regard, see also appendix \ref{App Oscillator}). This is indeed an interesting feature of AdS$_4$ (generally, AdS) relativity, which contains a universal pure vibration energy besides the matter one; the coming subsection is devoted to a brief discussion on this phenomenon and its possible relevance to the current existence of dark matter. The situation for dS$_4$ (generally, dS) relativity, however, is less tractable. The point is well exemplified by the absence of any constant term in Eq. (\ref{LaurentmassdS}).

At the end, for later use, it is useful to give the massless counterpart of Eq. (\ref{resteads}), as well; proceeding as before, by multiplying Eq. (\ref{Garidireads}) by the factor $\hbar\omega$ ($\omega = c/R$), while we have in mind the identity ${\hbar\omega}/{\vartheta} \approx \hbar\omega\varsigma =mc^2$ which vanishes in the massless cases, we have:
\begin{eqnarray}\label{Garidimassless}
E_{\mbox{\tiny{AdS}}_4}^{\;\mbox{\tiny{rest}}} = \hbar\omega (s+1) \,.
\end{eqnarray}

\subsection{Discussion: dark matter as a relic AdS$_4$ curvature energy (?)}
As mentioned above, Eq. (\ref{resteads}) reveals that, at the first order in the curvature ($\propto1/R$), the rest energy of an AdS$_4$ elementary system consists of two sectors, i.e., a ``visible" sector $mc^2$, like in flat Minkowski spacetime, and a ``dark" sector $\textstyle\frac{3}{2}\hbar\omega$, which is reminiscent of the zero-point energy of a quantum (three-dimensional) isotropic harmonic oscillator. In this subsection, taking this fact into account and following the proposal put forward in recent works \cite{Gazeau2021,Gazeau2020} by one of us, we discuss a conjectural interpretation of the origin of \emph{dark matter}\footnote{\label{foot97}We here give some (observational) facts about dark matter. According to the recent Planck analysis \cite{Planck Collaboration} of the power spectrum of the cosmic microwave background (CMB), cold dark (nonbaryonic) matter constitutes $27\%$ of our Universe, while the remaining parts are $5\%$ baryonic matter and $68\%$ dark energy. Dark matter is technically observed by its gravitational effect on luminous (baryonic matter) \cite{Baudis}; its mass halo and the total stellar mass are coupled via a function varying smoothly with mass (see, for instance, Ref. \cite{Behroozi} and references therein), with some possible (and of course notable) exception(s) as recently pointed out in Refs. \cite{Van Dokkum1,Van Dokkum2,Van Dokkum3}. Unfortunately, so far, all attempts in theoretical physics to give a satisfactory explanation of the origin of dark matter have failed. As a matter of fact, all hypothetical particle models (such as WIMP, Axions, Neutrinos, etc.) have failed in direct or indirect detection tests. On the other hand, alternative theories (such as MOND), negating the existence of dark matter as a physical entity, have also failed in the explanation of clusters and the observed pattern in the CMB. [For an overview on the current status of dark matter, including experimental evidence and theoretical motivations, readers are referred to Ref. \cite{Bertone report}.]} and its possible relevance to the ``dark" sector of the rest energy of AdS$_4$ elementary systems. At a glance, according to this proposal, dark matter is nothing but a pure QCD effect, strictly speaking, a gluonic \emph{Bose-Einstein condensate}\footnote{In condensed matter physics, a Bose-Einstein condensate is a state of matter that is typically formed when a gas of bosons at low densities is cooled to temperatures very close to absolute zero.}, which is a relic of the so-called quark epoch of the Universe; the quark epoch began approximately $10^{-12}$ seconds after the Big Bang, when the Universe with temperature $T>10^{12}\mathrm{K}$ was filled with \emph{quark-gluon plasma (QGP)}\footnote{For experimental evidence (RHIC, LHC) in support of QGP, as a super-liquid which is produced in ultra-relativistic heavy ion collisions, see Refs. \cite{Pasechnik,Andronic,Rafelski}. The experimental data show that quarks, anti-quarks, and gluons flow independently in this super-liquid.}, and ended when the Universe was about $10^{-6}$ seconds old, when the temperature had fallen sufficiently to allow the quarks from the preceding quark epoch to bind together into hadrons.\footnote{The following period, when quarks became confined within hadrons, lasting from $10^{-6}\mathrm{s}$ to $1\mathrm{s}$ with temperature $T > 10^{10}\mathrm{K}$, is known as the hadron epoch.} This proposal actually supports the fundamental idea that, aside from the violation of the matter/antimatter symmetry verifying the Sakharov's requirements \cite{Sakharov}, the reconciliation of particle physics and cosmology does not necessitate the recourse to any \emph{ad hoc} fields, particles or hidden variables. The very point that lies at the heart of this proposal and relates it to our group-theoretical reading of AdS$_4$ elementary systems is rooted in the fact that a QCD vacuum energy density, because of trace anomaly, results in a Lorentz-invariant negative-valued contribution to the cosmological constant (see, for instance, Ref. \cite{Pasechnik} and references therein). In this sense, it is naturally expected that the aforementioned primordial QCD world-matter had experienced an effective geometric environment analogous to the AdS$_4$ phase; the quark epoch, which immediately follows the dS$_4$ inflationary phase, may be viewed as an ``AdS$_4$ bounce" (AdS$_4$ phase as anti-inflation!). [To see more arguments on the latter point, see Ref. \cite{Gazeau2021} and references therein.]

\subsubsection{The assumptions}
The aforementioned proposal technically rests on three basic assumptions:
\begin{itemize}
\item{First, the nature of dark matter, which is observed as an energy more or less localized in halos surrounding baryonic matter in galaxies and galaxy clusters, is assumed to be relevant to some effective AdS$_4$ curvature. Then, for instance, for spin $1/2$ elementary systems $\chi$, according to Eq. (\ref{resteads}), we have:
    \begin{eqnarray}
    E_{\mbox{\tiny{AdS}}_4}^{\;\mbox{\tiny{rest}}} \; = \; \underbrace{\mathfrak{M}_{\mbox{\tiny{AdS$_4$}}}(\chi) c^2}_{\mbox{visible}} \;\;\; + \underbrace{\;\frac{3}{2}\hbar\omega\;}_{\mbox{dark}\;\big(\equiv E_{\mbox{\tiny{DM}}}(\chi)\big)}, \;\;\;\;\;\;\; E_{\mbox{\tiny{DM}}}(\chi) = r(\chi) \;  \mathfrak{M}_{\mbox{\tiny{AdS$_4$}}}(\chi) c^2\,,
    \end{eqnarray}
    where the ratio $r(\chi) \equiv E_{\mbox{\tiny{DM}}}(\chi) / \mathfrak{M}_{\mbox{\tiny{AdS$_4$}}}(\chi)c^2$ is expected to reflect \emph{to some extent} the ratio of dark matter to visible matter. Here, by ``to some extent" is meant that this individual ratio should be clearly larger than $1$, to be compatible with the estimated total ratio dark matter/visible matter, that is, $27/5 = 5.4$ (see footnote \ref{foot97}), and of course it should not be too large, to remain compatible with most of the observed ratios halo mass/stellar mass. Generally, a galaxy with a stellar mass $\approx 2\times 10^8 M_\bullet$ ($M_\bullet$ being the solar mass) should possess a dark matter mass $\approx 6\times 10^{10} M_\bullet$ (see, for instance, Ref. \cite{Behroozi}, and for notable exceptions Ref. \cite{Van Dokkum3}).}

\item{Second, the appearance of dark matter is assumed to occur over a period in which, in the sense of validity of the equipartition theorem applied to the quantum-oscillator-like energy spectrum of AdS$_4$ elementary systems, the temperature(s) $T_\chi$ was (were) compatible with a phase of entities $\chi$:
    \begin{eqnarray}
    k_B T_\chi \; \approx \; \hbar \omega \; \approx \; \frac{2 r(\chi)}{3}\; \mathfrak{M}_{\mbox{\tiny{AdS$_4$}}}(\chi) c^2\,,
    \end{eqnarray}
    where $k_B$ refers to the Boltzmann constant. Note that, in agreement with the above assumptions and while we restrict our attention merely to spin $1/2$ elementary systems, the most probable candidates for $\chi$ are the stable light quarks $u$ and $d$ when, at the end of the quark epoch, the QGP had experienced the phase transition, namely, hadronization. Currently, the hadronization temperature for light quarks \cite{Andronic} is estimated to be $T_{cf}=156.5 \pm 1.5 \mathrm{MeV} \approx 1.8\times 10^{12} \mathrm{K}$ (``chemical freeze-out temperature"). Hence, for $\mathfrak{M}_{\mbox{\tiny{AdS$_4$}}} \approx m$ (see Eq. (\ref{asymp AdS})), we get:
    \begin{eqnarray}\label{jjjjjjjjjjj}
    T_\chi \; \approx \; 1.8\times 10^{12} \mathrm{K} \; \approx \; \frac{\hbar\omega}{k_B} \; \approx \; \frac{2 r(\chi)}{3} \; \frac{m(\chi) c^2}{k_B}\,.
    \end{eqnarray}
    Then, for the quarks $u$ (with $m(u)\approx2.2 \; \mathrm{MeV}/c^2$) and $d$ (with $m(d)\approx4.7 \; \mathrm{MeV}/c^2$), the estimated values of $r(\chi=u,d)$ would be:
    \begin{eqnarray}
    r(u) \approx 108\,, \;\;\;\;\;\;\; r(d) \approx 49\,.
    \end{eqnarray}
    Note that these values seem quite reasonable according to the first assumption. Furthermore, the value of $T_\chi$, through Eq. (\ref{jjjjjjjjjjj}), determines the corresponding AdS$_4$ radius curvature $R=c/\omega$ and its lifetime $\textbf{t}\equiv1/\omega$:
    \begin{eqnarray}
    k_B T_\chi \; \approx \; \hbar\omega \;\;\;\;\;\;\; \Rightarrow \;\;\;\;\;\;\; R= 8 \;\mathrm{fm}\,, \;\;\;\;\; \textbf{t} = 2.7\times 10^{-23}\mathrm{s}\,.
    \end{eqnarray}
    This AdS$_4$ length scale $R = 8$ fm is to be compared with the typical distance scales in the context of QGP, which exceed the size of the largest atomic nuclei and the low typical momentum scale (in the $Pb$ case, $R_{Pb} = 5.3788 \;\mathrm{fm}$).}

\item{Third, at the critical hadronization point, the pure AdS$_4$ curvature energy (the ``dark" sector) decouples from the rest mass energy and abides as a free component of the Universe along its following periods.}
\end{itemize}

Note that, besides the supplementary mass granted to the (anti-)quarks living at the quark epoch by an AdS$_4$ environment (as briefly pointed out above), the existence of dark matter may also be originated from the gluonic component of the respective QGP. Concerning the latter case, Eq. (\ref{Garidimassless}) reveals that the rest energy of a gluon, strictly speaking, a spin-$1$ massless boson, living at the quark epoch in an AdS$_4$ effective background is purely ``dark", and is precisely equal to $2\hbar\omega$. Quite similar to the above, according to the equipartition $k_B T_{cf}\approx \hbar\omega$, we qualitatively get (see Eq. (\ref{jjjjjjjjjjj})):
\begin{eqnarray}
\frac{2\hbar\omega}{c^2} \approx \frac{2 \times 2r(u)}{3} \times m(u) = 144 \times m(u) \approx 317 \; \mathrm{MeV}/c^2\,.
\end{eqnarray}
This gluonic effective mass is approximately $4/3$ times the effective mass that quarks and anti-quarks acquired in that QGP-AdS$_4$ environment.

\subsubsection{Sketching a parallel between dark matter and CMB}
Now, it is interesting to sketch a parallel between dark matter and CMB, because the latter is considered as the emergence of the photon decoupling, exactly when photons began to move freely through space instead of constantly being scattered by electrons and protons in plasma. As a matter of fact, one may argue that a (substantial) part of the gluonic component of the QGP living at the quark epoch freely survives after hadronization, within an effective AdS$_4$ environment. This gluonic component is assumed to be an assembly of a large number, say, $N_G$, of non-interacting entities, strictly speaking, of decoupled gluonic colorless systems (not individual free gluons), which form a grand canonical Bose-Einstein ensemble. As discussed in Ref. \cite{Gazeau2021}, the simplest purely gluonic system, susceptible to constitute a Bose-Einstein condensate is a di-gluon. The di-gluons can actually be named ``dark matter quasi-particles", which, with an (effective) AdS$_4$ rest mass equal to ${\hbar}/{cR}$, constitute a Bose-Einstein condensate, since their Compton wavelength and average relative distances are equal to the AdS$_4$ radius of curvature $R$. [Note that the aforementioned (effective) AdS$_4$ rest mass ${\hbar}/{cR}$, with respect to Eq. (\ref{Garidimassless}), is associated with a massless scalar ($s=0$) particle in AdS$_4$ relativity.]

Here, one must notice that the described ``effective dark Universe" above, which is assumed to explain the cosmological standard model with a quantum vacuum or a ground state, is not exactly an ``empty spacetime" in which just some \emph{objects} would move. It is actually a medium which, according to the G\"ursey terminology \cite{Gursey63}, involves \emph{scintillation events},\footnote{The mass scintillation mechanism assumed by G\"ursey is analogous to the Bondi's steady state cosmology \cite{Bondi48} based upon which the creation, at constant density of matter-energy, induces the expansion of the Universe.} i.e., events each consisting in the virtual creation and subsequently, a short time later, annihilation of a particle-antiparticle pair. Note that, the created particle-antiparticle being bosons (respectively, fermions), the corresponding event contributes to the AdS$_4$ (respectively, dS$_4$) world matter, as a kind of a \emph{gluonic Bose-Einstein condensate} (respectively, \emph{baryonic Fermi-Dirac sea}), at the interior (respectively, exterior) of the Hubble horizon \cite{Tannoudji2019}.

Accordingly, we take into account an assembly of these $N_{G}$ dark matter di-gluons, which are practically quasi-particles in an effective AdS$_4$ environment with individual energies $E_n= E_{\mbox{\tiny{AdS}}_4}^{\;\mbox{\tiny{rest}}} + n\hbar\omega$, where the term $E_{\mbox{\tiny{AdS}}_4}^{\;\mbox{\tiny{rest}}}$ corresponding to $m=m_G$ (being zero or negligible) is given by Eq. (\ref{Garidimassless}) at $s=0$, and with degeneracy $g_n= (n+1)(n+3)/2$ \cite{Fronsdal1975}; in this sense, these residue di-gluons behave quite similar to isotropic harmonic oscillators in three-space. These entities are supposed to form a grand canonical Bose-Einstein ensemble, with the chemical potential $\mu$ determined by the condition that the sum over all occupation probabilities at temperature $T$ results in \cite{Grossmann95,Mullin97}:
\begin{equation}\label{NGBEC}
N_G = \sum_{n=0}^{\infty} \frac{g_n}{\exp\left(\frac{\hbar\omega}{k_B T}\left(n+\nu_0 -\mu\right)\right) -1}\,, \;\;\;\;\;\;\; \nu_0 \equiv \frac{E_{\mbox{\tiny{AdS}}_4}^{\;\mbox{\tiny{rest}}}}{\hbar\omega} \,.
\end{equation}
This is a very large number, and therefore, it is expected that this Bose-Einstein gas condensates at temperature:
\begin{equation}\label{TBEC}
T_c \approx \frac{\hbar\omega}{k_B} \left(\frac{N_G}{\zeta(3)}\right)^{1/3}\approx 1.18\times 10^{-3}\times \sqrt{|\Lambda_{\mbox{\tiny{AdS}}_4}|}\,N_G^{1/3}\,,
\end{equation}
to turn into the presently observed dark matter. Note that: (i) For the involved Riemann function, we have $\zeta(3)\approx1.2$. (ii) The above is the standard formula for all isotropic \emph{harmonic traps}\footnote{The term ``harmonic trap", that used here, may cause confusion. As a matter of fact, there is no actual harmonic trap here, but rather a harmonic spectrum on the quantum level which is originated from the AdS$_4$ geometry.} (see, for instance, Ref. \cite{Grossmann95}). In support of this model, with respect to the ultra-cold atoms physics, it has been shown in Ref. \cite{Tannoudji-Lalo} that Bose-Einstein condensation not only can occur in non-condensed matter but also in gas, and that this phenomenon is not rooted in \emph{interactions} but instead in the \emph{correlations implied by quantum statistics.}

Of course, we do not know exactly at which stage after the quark epoch the aforementioned gluonic-Bose Einstein condensation had occurred, but still it is interesting to examine whether Eq. \eqref{TBEC} gives reasonable orders of magnitude by setting $T_c$ equal to the current CMB temperature, that is, $T_c= 2.78$K, and $\vert\Lambda_{\mbox{\tiny{AdS}}_4}\vert \approx \frac{5.5}{6.5}\times \frac{11}{24}\times\Lambda_{\mbox{\tiny{dS}}_4} = 0.39\times \Lambda_{\mbox{\tiny{dS}}_4}$, where $\Lambda_{\mbox{\tiny{dS}}_4}\equiv$ current cosmological constant $\Lambda= 1.1\times10^{-52}\; \mathrm{m}^{-2}$. On this basis, from Eq. \eqref{TBEC}, we obtain the estimated number of di-gluons in the condensate as:
\begin{equation}\label{NGcurrent}
N_G\approx 5\times 10^{88}\,.
\end{equation}
This result already seems reasonable, because the number of gluons are about $10^{9}$ times that of baryons which is estimated to be about $10^{80}$. Keeping this number in Eq. \eqref{TBEC} gives us an estimation of the scaling factor:
\begin{equation}\label{scaleTL}
\sigma_c \equiv \frac{\Lambda_{\mbox{\tiny{dS}}_4}}{T^2_c}\approx 1.85\times10^6\times N_G^{-2/3}\approx 1.36\times10^{-53}\,.
\end{equation}
Therefore, taking $T_c$ to be of the order of ``Matter-dominated era" temperature, that is, $10^4$K, results in the following value for the corresponding $\Lambda_{\mbox{\tiny{dS}}_4}$:
\begin{equation}
\Lambda_{\mbox{\tiny{dS}}_4} \approx 1.36\times 10^{-44}\,\mathrm{m}^{-2}\,.
\end{equation}

\setcounter{equation}{0} \section{Garidi mass: examples and applications}
In this section, following Refs. \cite{Garidimass,GazNo2011} and taking some known examples into account, we discuss how the dS$_4$ Garidi mass formula (\ref{conmassf}), possessing a rich group-theoretical content, simplifies the debate over the notion of mass in dS$_4$ relativity. As the first example, we consider a class of gauge-invariant fields known in the literature under the name of ``partially massless" fields, which have been widely studied by Deser et al. in a series of papers \cite{Deser1983,Deser1984,Deser2001,Deser2001',Deser2001'',Deser2001'''}. Then, we encounter the question of graviton mass in dS$_4$ spacetime, as has been discussed by Novello et al. in Ref. \cite{Novello2003}.

Note that, for the sake of comparison, in this section we consider the units $c=1=\hbar$ and also set $R=H^{-1}$ (since the references cited above, that we are going to compare our result with them, have done so!).

\subsection{``Partially massless" fields}
First of all, we would like to point out that, throughout this paper, the spin-$s$ fields which are referred to as (strictly) massive, admitting no gauge invariance, possess $2s+1$ degrees of freedom. On the other hand, for what we call (strictly) massless fields, the degrees of freedom, due to the appearance of gauge invariance, reduce to $2$ (i.e., the helicities $\pm s$). There are yet intermediate fields, for which gauge invariance allows ``intermediate sets of higher helicities" between the $2s+1$ degrees of freedom of the (strictly) massive fields and the $2$ helicities of the (strictly) massless ones. These fields are relevant to a novel gauge invariance, which was first discovered by Deser et al. in Ref. \cite{Deser1983} in the case of spin-$2$ fields. They are usually called partially massless fields, due to their light-cone propagation properties \cite{Deser2001'''}.

Here, considering the above categories, we evaluate (positive) forbidden ranges of mass for dS$_4$ fields, within our group-theoretical construction, when the fields masses are identified by the Garidi definition (see Eq. (\ref{conmassf})). The very point to be noticed here is that all these three categories of dS$_4$ fields can be entirely characterized in the context of dS$_4$ group representation theory through the invariant parameters $(p,q)$ (in the Dixmier notations; see section \ref{Sec Dixmier}), or equivalently, through the (invariant) Garidi mass and spin parameters $(\mathfrak{M}^{2}_{\mbox{\tiny{dS}}^{}_4},s)$. Considering the latter parameters, it simply follows that, for a given dS$_4$ field with spin $s$, (positive) forbidden ranges of mass are those that trace back to gaps between two unitary representations. As already pointed out, we also would like to compare our results with those that have been already given by Deser et al. in Refs. \cite{Deser1983,Deser1984,Deser2001,Deser2001',Deser2001'',Deser2001'''}. Technically, they have shown that, for a dS$_4$ field with $s>1$, the plane defined by the mass (squared) parameter, say Deser mass ${\mathfrak{M}}^{\prime \;2}_{\mbox{\tiny{dS}}^{}_4}$, and the cosmological constant $\Lambda = 3H^2$ is divided into different unitary and nonunitary areas, which are distinguished by lines of the aforementioned gauge invariant fields. Accordingly, they have argued that the physically irrelevant (positive) mass ranges for the field are those that correspond to the nonunitary areas.

We begin with integer spin dS$_4$ fields, while, to keep the argument straight, we stick to the fields with spins up to $s=3$:
\begin{itemize}
\item{According to Deser et al., for the first two integer cases, i.e., the scalar ($s=0$) and vector ($s=1$) dS$_4$ fields, no new gauge invariance and no (positive) forbidden range of mass exist; the allowed ranges of mass for both cases are determined by ${\mathfrak{M}}^{\prime \;2}_{\mbox{\tiny{dS}}^{}_4}\geqslant 0$. This result completely agrees with the one obtained through our group-theoretical construction, well described by the Garidi mass formula (\ref{conmassf}). As a matter of fact, by $\mathfrak{M}^{2}_{\mbox{\tiny{dS}}^{}_4}\geqslant 0$, for both scalar and vector cases, one can continuously cover the unitary regions, as is shown in FIG. \ref{FIG. mass 0,1}.
    \begin{figure}[H]
    \begin{center}
    \includegraphics[height=.12\textheight]{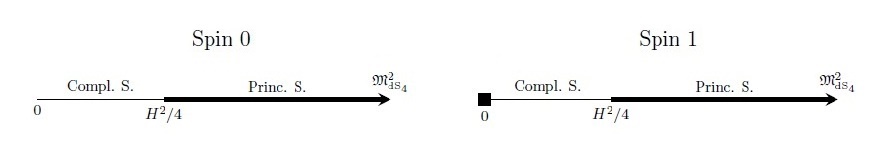}
    \end{center}
    \caption{Mass ranges for the scalar ($s=0$) and vector ($s=1$) dS$_4$ fields; the terms `Compl. S.' and `Princ. S.' are respectively abbreviation for the `complementary series' and `principal series', while the `black squares' refer to the discrete series members, with integer spins.}
    \label{FIG. mass 0,1}
    \end{figure}}
\item{According to Deser et al. (see also Ref. \cite{Higuchi} by Higuchi), in the spin-$2$ case, a partially massless gauge field appears. It is identified by ${\mathfrak{M}}^{\prime \;2}_{\mbox{\tiny{dS}}^{}_4} = 2\Lambda/3 = 2H^2$, while the (strictly) massless one is given by ${\mathfrak{M}}^{\prime \;2}_{\mbox{\tiny{dS}}^{}_4} = 0$. Adopting the Garidi mass formula (\ref{conmassf}), one can simply show that these fields are associated with two specific representations in the discrete series UIR's of the dS$_4$ group, respectively, characterized by $(p=2,q=1)$ or equivalently by $(\mathfrak{M}^{2}_{\mbox{\tiny{dS}}^{}_4}=2H^2,s=2)$, with $4$ degrees of freedom, and by $(p=2,q=2)$ or equivalently by $(\mathfrak{M}^{2}_{\mbox{\tiny{dS}}^{}_4}=0,s=2)$, with $2$ degrees of freedom. The predicted (positive) forbidden mass range in both (say Deser and Garidi) approaches exactly coincides, and based on our group-theoretical terminology, it refers to the gap between the aforementioned unitary representations, as is shown in FIG. \ref{FIG. mass 2}.
    \begin{figure}[H]
    \begin{center}
    \includegraphics[height=.22\textheight]{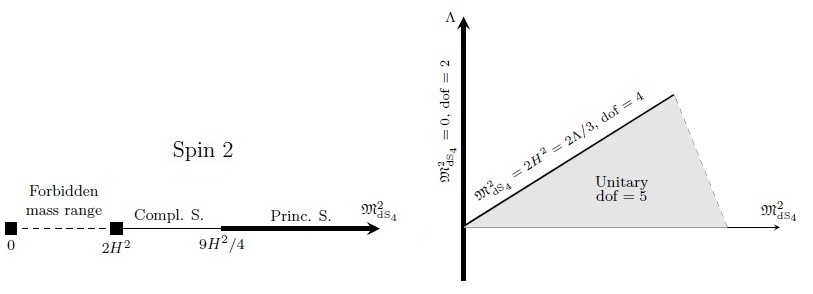}
    \end{center}
    \caption{Mass ranges for spin-2 dS$_4$ fields and phase diagram (showing partially massless lines); the terms `dof' is an abbreviation for the `degrees of freedom'.}
    \label{FIG. mass 2}
    \end{figure}}
\item{According to Deser et al., in the spin-$3$ case, two partially massless gauge fields with ${\mathfrak{M}}^{\prime \;2}_{\mbox{\tiny{dS}}^{}_4} = 2\Lambda = 6H^2$ and ${\mathfrak{M}}^{\prime \;2}_{\mbox{\tiny{dS}}^{}_4} = 4\Lambda/3 = 4H^2$ appear, while the (strictly) massless one is given by ${\mathfrak{M}}^{\prime \;2}_{\mbox{\tiny{dS}}^{}_4} = 0$. Again, with respect to the Garidi mass formula (\ref{conmassf}), one can easily check that they correspond to specific representations in the discrete series UIR's, respectively, characterized by $(p=3,q=1)$ or equivalently by $(\mathfrak{M}^{2}_{\mbox{\tiny{dS}}^{}_4}=6H^2,s=3)$, with $6$ degrees of freedom, by $(p=3,q=2)$ or equivalently by $(\mathfrak{M}^{2}_{\mbox{\tiny{dS}}^{}_4}=4H^2,s=3)$, with $4$ degrees of freedom, and finally by $(p=3,q=3)$ or equivalently by $(\mathfrak{M}^{2}_{\mbox{\tiny{dS}}^{}_4}=0,s=3)$, with $2$ degrees of freedom. And, again, the (positive) forbidden mass ranges in both approaches are exactly the same, and correspond to the gaps between the above unitary representations, as is shown in FIG. \ref{FIG. mass 3}.
    \begin{figure}[H]
    \begin{center}
    \includegraphics[height=.22\textheight]{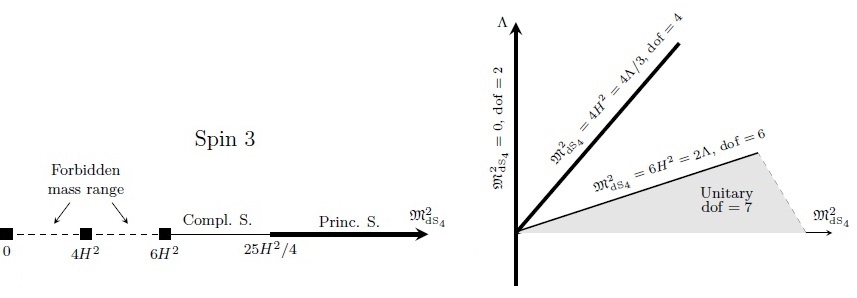}
    \end{center}
    \caption{Mass ranges for spin-3 dS$_4$ fields and phase diagram (showing partially massless lines).}
    \label{FIG. mass 3}
    \end{figure}}
\end{itemize}

Now, we deal with half-integer spin dS$_4$ fields, which are controversial. Again, we stick to the cases with $s\leqslant 3$:
\begin{itemize}
\item{According to Deser et al., for the spin-$1/2$ case (like the scalar and vector cases), no (positive) forbidden mass range exists. This result agrees with what we obtain from our group-theoretical construction through the Garidi mass formula. The situation, however, is different in the spin-$3/2$ and spin-$5/2$ cases.}
\item{According to Deser et al., for the spin-$3/2$ and spin-$5/2$ dS$_4$ fields, FIG. \ref{FIG. mass d 3,5/2} holds. The corresponding (strictly) massless fields, with $2$ degrees of freedom, are respectively determined by ${\mathfrak{M}}^{\prime \;2}_{\mbox{\tiny{dS}}^{}_4} = - \Lambda/3$ (in the spin-$3/2$ case) and by ${\mathfrak{M}}^{\prime \;2}_{\mbox{\tiny{dS}}^{}_4} = -4\Lambda/3$ (in the spin-$5/2$ case). On this basis, since in the dS$_4$ case $\Lambda>0$, the Deser mass formula for both cases lead to negative values; the same result holds for the partially massless case, with $4$ degrees of freedom, in the spin-$5/2$ case, i.e., ${\mathfrak{M}}^{\prime \;2}_{\mbox{\tiny{dS}}^{}_4} = - \Lambda/3 < 0$. In this sense, Deser et al. have argued that these cases belong to AdS$_4$ relativity (for which $\Lambda<0$); no (strictly) massless or partially massless fields with $s=3/2$ and $s=5/2$ (and correspondingly with higher half-integer spins) in dS$_4$ relativity exists! This result is clearly in contradiction with ours. Actually, in our group-theoretical construction, the dS$_4$ (strictly) massless integer and half-integer spin fields all (except the scalar one which belongs to the complementary series) belong to the lower limit of the discrete series UIR's, characterized by $\Pi^\pm_{p=s,q=s}$, with $s\in \mathbb{N}/2$ (see section \ref{Sec contraction}), and all have meaningful dS$_4$ Garidi masses, which are always equal to zero (see section \ref{Sec Garidi mass}).\footnote{Moreover, we must emphasize again that the only group-theoretical consistent formulation for the AdS$_4$ mass is exactly the one given in Eq. (\ref{AdS mass Formula}) \cite{GazNo2008, GazNo2011}.} For instance, according to the Garidi mass formula, for the aforementioned spin-$3/2$ and spin-$5/2$ dS$_4$ cases, FIGs. \ref{FIG. mass g 3/2} and \ref{FIG. mass g 5/2} hold.
    \begin{figure}[H]
    \begin{center}
    \includegraphics[height=.29\textheight]{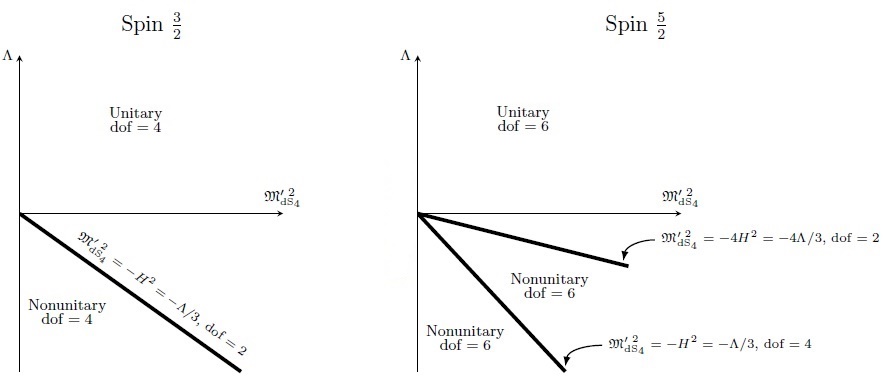}
    \end{center}
    \caption{Phase diagrams (showing partially massless lines) for spin-$3/2$ and spin-$5/2$ dS$_4$ fields, according to the Deser et al. arguments.}
    \label{FIG. mass d 3,5/2}
    \end{figure}

    \begin{figure}[H]
    \begin{center}
    \includegraphics[height=.17\textheight]{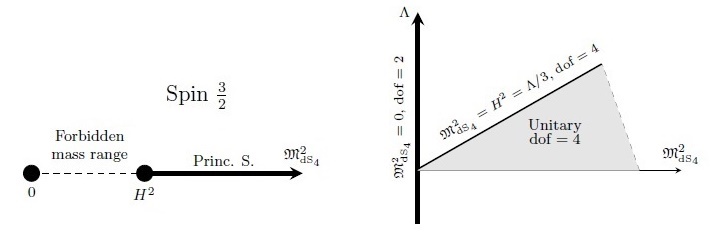}
    \end{center}
    \caption{Mass ranges for spin-$3/2$ dS$_4$ fields and phase diagram (showing partially massless lines), according to the Garidi mass formula; the `black circles' denote the discrete series members, with half-integer spins. Note that the case ${\mathfrak{M}}^{2}_{\mbox{\tiny{dS}}^{}_4} = H^2$, which is ``contiguous" to the principal (massive) series, does not determine a partially massless field.}
    \label{FIG. mass g 3/2}
    \end{figure}

    \begin{figure}[H]
    \begin{center}
    \includegraphics[height=.23\textheight]{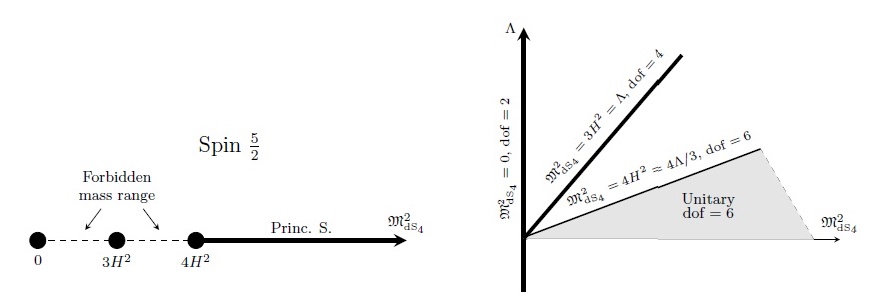}
    \end{center}
    \caption{Mass ranges for spin-$5/2$ dS$_4$ fields and phase diagram (showing partially massless lines), according to the Garidi mass formula. Note that the case ${\mathfrak{M}}^{2}_{\mbox{\tiny{dS}}^{}_4} = 4H^2$, which is ``contiguous" to the principal (massive) series, does not determine a partially massless field.}
    \label{FIG. mass g 5/2}
    \end{figure}}
\end{itemize}

Although, this comparison has been done merely for dS$_4$ fields with spins up to $s=3$, the result is quite clear. The Garidi mass formula, for integer spin dS$_4$ fields, covers the results that have been already given by Deser et al. in Refs. \cite{Deser1983,Deser1984,Deser2001,Deser2001',Deser2001'',Deser2001'''}. In the case of half-integer spin fields, with $s\geqslant 3/2$, however, contradictions come to fore. We believe that these contradictions stem from an improper choice of the mass parameter by Deser at al., since it associates negative values of mass with the known dS$_4$ (strictly) massless unitary representations $\Pi^\pm_{p=s,q=s}$, with $s= 3/2,5/2, \;...\;$, which are meaningful from the point of view of a Minkowskian observer (see section \ref{Sec contraction}). Again, here, it would be convenient to recall from section \ref{Sec Garidi mass} that adopting the Garidi mass formula (\ref{conmassf}), for all dS$_4$ UIR's meaningful from the point of view of a Minkowskian observer (including the aforementioned (strictly) massless UIR's), we have always $\mathfrak{M}^{2}_{\mbox{\tiny{dS}}^{}_4}\geqslant 0$. In this regard, to articulate more clearly our group-theoretical point of view on the very question of the existence of (strictly) massless and partially massless fields in dS$_4$ relativity, we explicitly specify in FIG. \ref{FIG. mass p,q} the allowed ranges of the invariant parameters $(p,q)$ for the dS$_4$ representations with $p=s \leqslant 3$ (in the Dixmier notations; see section \ref{Sec Dixmier}). On this basis, one can recognize all the dS$_4$ (strictly) massless and partially massless fields, as members of the discrete series representations (again, except the scalar massless field which belongs to the complementary series), in terms of their degrees of freedom; contrary to the dS$_4$ (strictly) massless fields, which are meaningful from the Minkowskian point of view (see section \ref{Sec contraction}), the partially massless ones are associated with those discrete series UIR's (i.e., $\Pi^{}_{p,q}$, with $0<q\neq p$) which have no meaning from the viewpoint of a tangent Minkowskian observer.
\begin{figure}[H]
\begin{center}
\includegraphics[height=.31\textheight]{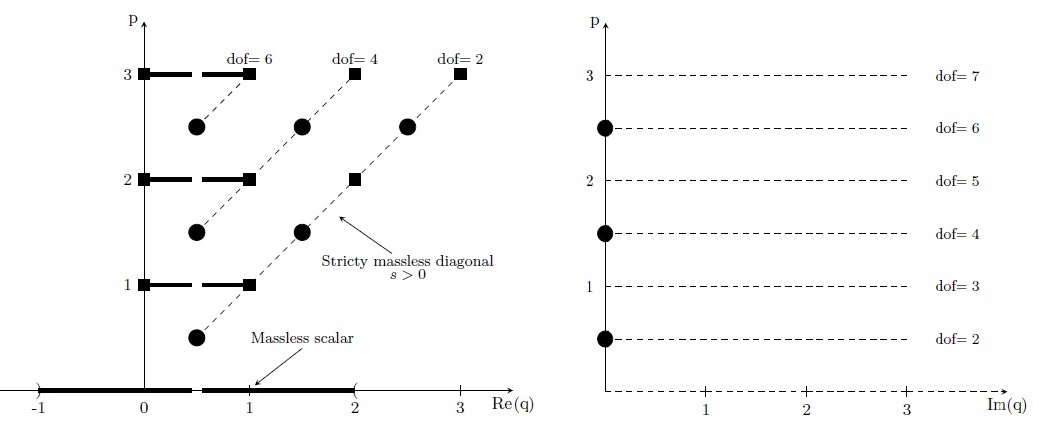}
\end{center}
\caption{Allowed ranges of the invariant parameters $(p,q)$ for the dS$_4$ representations, with $p=s \leqslant 3$. The marked points on the diagonal lines, characterizing the discrete series members, refer to the dS$_4$ (strictly) massless and partially massless fields with (half-)integer spins (with the exception of those with $\mbox{Re}(q) = 1/2$ and $p=3/2,5/2,...$, which are not (strictly) massless, and are ``contiguous" to the principal series).}
\label{FIG. mass p,q}
\end{figure}

\subsection{The question of Graviton ``mass"}
Generally, to describe the graviton field in dS$_4$ spacetime, two different procedures can be followed. One can either begin from the flat Minkowski background and then turn on gravity to get the dS$_4$ structure, or directly begin from the geometry of dS$_4$ spacetime itself. Roughly speaking, the former procedure is based on the following equation:
\begin{eqnarray}\label{EGR1}
R_{\mu\nu} - \frac{1}{2}Rg_{\mu\nu} = - \kappa T_{\mu\nu}\;,
\end{eqnarray}
for which the fundamental state containing the maximum number of symmetries is the Minkowskian geometry, while the latter procedure explicitly includes $\Lambda$ as:\footnote{Note that, here, by $\Lambda$ is meant some sort of bare cosmological constant, and not exactly the observed one which should contain modifications stemmed from the matter fields fluctuations (either classical or quantum fluctuations).}
\begin{eqnarray}\label{EGR2}
R_{\mu\nu} - \frac{1}{2}Rg_{\mu\nu} + \Lambda g_{\mu\nu} = -\kappa T_{\mu\nu}\;,
\end{eqnarray}
for which the fundamental state containing the maximum number of symmetries is the (A)dS$_4$ geometry. Below, following the lines sketched in Refs. \cite{NoveNe2002,Novello2003,GazNo2011}, we will briefly review both procedures, respectively. Then, we will interpret the results within our group-theoretical construction, i.e., by considering the described graviton field as a dS$_4$ elementary system associated with a UIR of the dS$_4$ group. We will show that how our group-theoretical approach, through the Garidi mass formula, simplifies the debate over the question of ``mass" of graviton in dS$_4$ spacetime.

\subsubsection{Passage from flat Minkowski spacetime to a curved background}
The passage of the spin-$2$ field equation from flat Minkowski spacetime to a general curved Riemannian manifold entails some ambiguities, which technically originate from the presence of second order derivatives of the rank-$2$ symmetric tensor ${\Psi}_{\mu\nu}\equiv{\Psi}^{(2)}_{\mu\nu}$ ($\mu,\nu=0,1,2,3$) in the so-called Einstein frame (see, for instance, Refs. \cite{Araser1, Araser2}). All these ambiguities, however, can be lifted, if one takes the Fierz frame representation into account \cite{NoveNe2002, Novello2003}. The latter deals with the three-index tensor:
\begin{eqnarray}
F_{\lambda\mu\nu} = \frac{1}{2} \big(\partial_\mu {\Psi}_{\nu\lambda} - \partial_\lambda {\Psi}_{\nu\mu} + \eta_{\mu\nu} F_\lambda - \eta_{\lambda\nu}F_\mu\big)\,,\nonumber
\end{eqnarray}
where $F_\lambda = \eta^{\mu\nu}F_{\lambda\mu\nu} = \partial_{\lambda}{\Psi} - \partial_\mu {\Psi}_{\lambda}^\mu$ (all the indices take values $0,1,2,3$). There results a unique form of minimal coupling, free of ambiguities. In this context, from ${\Psi}_{\mu\nu}$, two auxiliary fields $\bold{G}_{\mu\nu}^{(a)}$ and $\bold{G}_{\mu\nu}^{(b)}$ are defined:
\begin{eqnarray}\label{Gazgm20}
\bold{G}_{\mu\nu}^{(a)} = \frac{1}{2}\Big(\Box {\Psi}_{\mu\nu} + \nabla_\mu \nabla_\nu {\Psi}^\prime -\underbrace{\big(\nabla_\mu \nabla^\lambda {\Psi}_{\nu\lambda} + \nabla_\nu \nabla^\lambda {\Psi}_{\mu\lambda}\big)} - {g}_{\mu\nu} \big(\Box {\Psi}^\prime - \nabla_\lambda \nabla_\rho {\Psi}^{\lambda\rho}\big)\Big)\,,
\end{eqnarray}
\begin{eqnarray}\label{Gazgm21}
\bold{G}_{\mu\nu}^{(b)} = \frac{1}{2}\Big(\Box {\Psi}_{\mu\nu} + \nabla_\mu \nabla_\nu {\Psi}^\prime -\underbrace{\big(\nabla^\lambda \nabla_\mu {\Psi}_{\nu\lambda} + \nabla^\lambda \nabla_\nu {\Psi}_{\mu\lambda}\big)} - {g}_{\mu\nu} \big(\Box {\Psi}^\prime - \nabla_\lambda \nabla_\rho {\Psi}^{\lambda\rho}\big)\Big)\,,
\end{eqnarray}
where ${\Psi}^\prime = {g}_{\mu\nu}{\Psi}^{\mu\nu}$. Note that these entities are the same except the order of the second derivative in the specified terms. The equation of motion, free of ambiguities, involves:
\begin{eqnarray}\label{Gazgm22}
\hat{\bold{G}}_{\mu\nu} \equiv \frac{1}{2}\big(\bold{G}_{\mu\nu}^{(a)} + \bold{G}_{\mu\nu}^{(b)}\big)\,,
\end{eqnarray}
and precisely reads:
\begin{eqnarray}\label{Gazgm23}
\hat{\bold{G}}_{\mu\nu} + \frac{1}{2} m^2({\Psi}_{\mu\nu}-{g}_{\mu\nu}{\Psi}^\prime)=0\,.
\end{eqnarray}

Note that the field ${\Psi}_{\mu\nu}$ possesses $5$ degrees of freedom. In the flat Minkowski background, the degrees of freedom for the massless field ${\Psi}_{\mu\nu}$ reduce to $2$. This is a direct result of the invariance of the action under the gauge transformation:
\begin{eqnarray}
{\Psi}_{\mu\nu} \rightarrow {\Psi}_{\mu\nu} + \delta{\Psi}_{\mu\nu}\,,
\end{eqnarray}
in which
\begin{eqnarray}\label{Gazgm25}
\delta{\Psi}_{\mu\nu} = \partial_\nu \xi_\mu + \partial_\mu \xi_\nu \,.
\end{eqnarray}

Now, let us restrict our attention to the special case of dS$_4$ background, for which a similar analysis as above holds. The corresponding action, in the Fierz-Pauli frame, reads:
\begin{eqnarray}\label{Gazgm26}
S = \frac{1}{4}\int \sqrt{-{g}} \bigg[ F_{\lambda\rho\mu}F^{\lambda\rho\mu} - F_\mu F^\mu - \frac{1}{2} \mathfrak{M}_{\mbox{\tiny{dS}}^{}_4}^{\prime\prime \;2}\big({\Psi}_{\mu\nu}{\Psi}^{\mu\nu} - {\Psi}^{\prime\;2}\big)\bigg] \mathrm{d}^4x\,.
\end{eqnarray}
With the dS$_4$ counterpart of the transformation (\ref{Gazgm25}):
\begin{eqnarray}\label{gtrans}
\delta{\Psi}_{\mu\nu} = \nabla_\nu \xi_\mu + \nabla_\mu \xi_\nu \,,
\end{eqnarray}
the action (\ref{Gazgm26}) turns into:\footnote{Note that, in an arbitrary curved Riemannian background, the action is not invariant under such a transformation.}
\begin{eqnarray}
\delta S = \frac{1}{2}\int \sqrt{-g} \big(Z^\mu - \mathfrak{M}_{\mbox{\tiny{dS}}^{}_4}^{\prime\prime\;2} F^\mu \big) \xi_\mu \; \mathrm{d}^4x\,,
\end{eqnarray}
where $Z^\mu$, defined as $Z^\mu\equiv 2\nabla_\nu \hat{\textbf{G}}^{\mu\nu}$, takes the value $Z^\mu = -\textstyle\frac{2}{3}\Lambda F^\mu$. For the particular case that the mass parameter $\mathfrak{M}_{\mbox{\tiny{dS}}^{}_4}^{\prime\prime\;2}$ takes the value:
\begin{eqnarray}\label{Gazgm28}
\mathfrak{M}_{\mbox{\tiny{dS}}^{}_4}^{\prime\prime \;2} = -\frac{2}{3}\Lambda \,,
\end{eqnarray}
the above action admits a gauge invariance, based upon which the degrees of freedom of the field reduce to $2$ (the strictly massless spin-$2$, say the graviton, field), while for any other values of $\mathfrak{M}_{\mbox{\tiny{dS}}^{}_4}^{\prime\prime \;2}$, the degrees of freedom remain $5$. At first glance, this result seems quite surprising, because the gauge invariance of the action occurs for the dS$_4$ field ${\Psi}_{\mu\nu}$ with a nonvanishing $\mathfrak{M}_{\mbox{\tiny{dS}}^{}_4}^{\prime\prime \;2}$. Moreover, since in the dS$_4$ case $\Lambda>0$, it yields a negative value for the parameter $\mathfrak{M}_{\mbox{\tiny{dS}}^{}_4}^{\prime\prime \;2}$ relevant to the graviton field in dS$_4$ spacetime (though it is positive in the AdS$_4$ case, since $\Lambda<0$); a massless graviton in dS$_4$ spacetime appears as a tachyonic particle in the Minkowskian sense (!). Before getting involved with the physical interpretation of this result (obtained by Novello et al. in Ref. \cite{Novello2003}), let us revisit the above argument leading to the fine-tuning mass identity (\ref{Gazgm28}) in the context of the second procedure, i.e., when the starting point is the geometry of dS$_4$ spacetime itself.

\subsubsection{Einstein spaces}
By definition, Einstein spaces are described by geometries verifying Eq. (\ref{EGR2}) without matter:
\begin{eqnarray}\label{Gazgm29}
R_{\mu\nu} - \Lambda g_{\mu\nu} = 0\,.
\end{eqnarray}
Such spaces contain the dS$_4$ and AdS$_4$ geometries as special cases. In this context, the Riemannian curvature $R_{\lambda\rho\mu\nu}$ is given by:
\begin{eqnarray}
R_{\lambda\rho\mu\nu} = W_{\lambda\rho\mu\nu} + \frac{\Lambda}{3}g_{\lambda\rho\mu\nu}\,,
\end{eqnarray}
where $W_{\lambda\rho\mu\nu}$ is the Weyl conformal tensor and $g_{\lambda\rho\mu\nu}\equiv g_{\lambda\mu}g_{\rho\nu} - g_{\lambda\nu}g_{\rho\mu}$. One can also rewrite Eq. (\ref{Gazgm29}) as:
\begin{eqnarray}
G_{\mu\nu} + \Lambda g_{\mu\nu} = 0\,,
\end{eqnarray}
where $G_{\mu\nu}$ is the Einstein tensor. Perturbation of this equation around the dS$_4$ solution, for which the Weyl tensor vanishes, yields:
\begin{eqnarray}\label{Gazgm32}
\delta G_{\mu\nu} + \Lambda h_{\mu\nu} = 0\,,
\end{eqnarray}
where $\delta G_{\mu\nu}$ and $\delta g_{\mu\nu}\equiv h_{\mu\nu}$ respectively denote the perturbation of the Einstein tensor and the metric. On this basis, a direct computation leads to:
\begin{eqnarray}\label{Gazgm36}
\hat{G}_{\mu\nu} - \frac{\Lambda}{3}\big( h_{\mu\nu} - h^\prime g_{\mu\nu}\big) =0\,,
\end{eqnarray}
or equivalently, to:
\begin{eqnarray}\label{linfieq}
\big( \Box_H + 2H^{2}\big) h_{\mu\nu} - g_{\mu\nu}\big( \Box_H - H^{2}\big)h^\prime - \big( \nabla_\mu\nabla\cdot h_\nu + \nabla_\nu \nabla\cdot h_\mu \big) + g_{\mu\nu} \nabla^\lambda \nabla^\rho h_{\lambda\rho} + \nabla_\mu \nabla_\nu h^\prime = 0\,,
\end{eqnarray}
where $h^\prime=h_{\mu\nu}g^{\mu\nu}$ and $\hat{G}_{\mu\nu}$ is the dS$_4$ counterpart of Eq. (\ref{Gazgm22}). Note that the above equation is invariant under the gauge transformation (\ref{gtrans}).

A comparison of Eq. (\ref{Gazgm36}) with (\ref{Gazgm23}), quite compatible with the previous result (see Eq. (\ref{Gazgm28})), shows that the fine-tuning mass identity for the graviton field in dS$_4$ spacetime is:
\begin{eqnarray}\label{Gazgm37}
m^2 = -\frac{2}{3}\Lambda \equiv \mathfrak{M}_{\mbox{\tiny{dS}}^{}_4}^{\prime\prime \; 2}\,.
\end{eqnarray}
Again, the above result can be interpreted as a graviton mass in the Minkowskian sense for the AdS$_4$ case (since $\Lambda<0$), while, for the dS$_4$ case ($\Lambda>0$), it involves a tachyonic interpretation (in the Minkowskian sense!) of the existence of the graviton which in turn creates serious interpretative difficulties. Respecting the very approach adopted in this paper, there is yet another interpretation of the above result to be presented in the coming subsubsection. Here, just to prepare the ground, we recall from the beginning of this subsection that, in order to describe the graviton field in (A)dS$_4$ spacetime, one may either begin from a Minkowskian background and then turn on gravity to get the (A)dS$_4$ structure, or begin directly from the (A)dS$_4$ geometry itself. Concerning the latter case, one must notice that giving the bare cosmological constant $\Lambda$ the status of a fundamental constant (as much small as it can be) has a significant impact on our interpretation of basic physical quantities such as mass. As a matter of fact, for a non-vanishing $\Lambda$, the Minkowskian geometry is no longer a solution to the corresponding Einstein's equations and, in this sense, becomes physically irrelevant; in this case, the absence of gravitation is shown by the (A)dS$_4$ geometry. Then, trivially, for a non-vanishing $\Lambda$, the use of ordinary Poincar\'{e} relativity will no longer be justified, and needs to be replaced by (A)dS$_4$ relativity (as we have discussed throughout this paper). In this sense, the Garidi definition of mass in (A)dS$_4$ relativity once again comes to fore.

\subsubsection{Discussion: Garidi interpretation}
Considering the above and for the sake of argument, we convert the field equation (\ref{linfieq}) into its counterpart written in terms of ambient space notations (see the instruction given in section \ref{Sec wave Eqs.}):
\begin{eqnarray}\label{ambicount}
\bar\partial^2\Psi - H^{2}\Sigma_2 x\partial\cdot\Psi - \Sigma_2 \bar\partial\partial\cdot\Psi + \frac{1}{2}\Sigma_2 \bar\partial\bar\partial\Psi^\prime  + \frac{1}{2}H^{2}\Sigma_2 x\bar\partial\Psi^\prime =0\,,
\end{eqnarray}
Pursuing this path, to make the group-theoretical content of the theory explicit, we also express the above field equation in terms of the quadratic Casimir operator of the dS$_4$ group:
\begin{eqnarray}\label{Garmass6.8}
\big( Q^{(1)}_2 + 6 \big) \Psi(x) + D_2\partial_2\cdot\Psi(x) = 0\,,
\end{eqnarray}
where $D_2 \Psi= H^{-2}\Sigma_2 \big(\bar{\partial} - H^{2}x\big)\Psi$ is the generalized gradient and $\partial_2\cdot\Psi = \partial\cdot\Psi - H^{2}x\Psi^\prime - \frac{1}{2}\bar{\partial}\Psi^\prime$ is the generalized divergence; the subscript `2' refers to the fact that the carrier space is constituted by second rank tensors. Note that the above equation is derivable from the following Lagrangian density:
\begin{eqnarray}\label{Lagden}
\mathfrak{L} = -\frac{1}{2x^2}\Psi\cdot\cdot(Q_2+6)\Psi+\frac{1}{2}(\partial_2\cdot\Psi)^2\,,
\end{eqnarray}
where the symbol `$\cdot\cdot$' stands for total contraction. This Lagrangian density is invariant under $\Psi\rightarrow \Psi + D_2\zeta$ ($\zeta$ being an arbitrary vector field), which is the ambient space counterpart of (\ref{gtrans}).

Now, a comparison of Eq. (\ref{Garmass6.8}) with the eigenvalue equation (\ref{Wave Eq. Gen}) reveals that the corresponding physical subspace (which can be associated with a dS$_4$ UIR) is constituted by the solutions to:
\begin{eqnarray}
\big( Q^{(1)}_2 + 6 \big)\Psi(x) = 0\,.
\end{eqnarray}
This subspace carries the discrete series UIR $\Pi^\pm_{p=2,q=2}$ (since $\langle Q_2^{(1)}\rangle = -6$), which means that the field equation corresponding to the fine-tuning mass identity $\mathfrak{M}_{\mbox{\tiny{dS}}^{}_4}^{\prime\prime\;2}=-2\Lambda/3$ and satisfying Eq. (\ref{Gazgm36}) characterizes a usual dS$_4$ (strictly) massless elementary system, with the Garidi mass $\mathfrak{M}^{2}_{\mbox{\tiny{dS}}^{}_4}=0$, for which, due to the gauge invariance, the degrees of freedom reduce to $2$ (i.e., helicities $\pm 2$).

In summary, considering all the above, if we analyze the graviton field in dS$_4$ spacetime in the sense of Poincar\'{e} relativity the result ($\mathfrak{M}_{\mbox{\tiny{dS}}^{}_4}^{\prime\prime\;2}$) shows it as a particle which is not massless, strictly speaking, the result yields a misinterpretation of the graviton mass in dS$_4$ spacetime ($\mathfrak{M}_{\mbox{\tiny{dS}}^{}_4}^{\prime\prime\;2}<0$), while the same analyze in the context of dS$_4$ relativity (through the Garidi mass formula $\mathfrak{M}^{2}_{\mbox{\tiny{dS}}^{}_4}$) indicates the graviton as a massless spin-$2$ particle, with no acausal propagation. The relation between the results obtained by these two approaches, i.e., the relation between $\mathfrak{M}_{\mbox{\tiny{dS}}^{}_4}^{\prime\prime\;2}$ and its Garidi counterpart $\mathfrak{M}^{2}_{\mbox{\tiny{dS}}^{}_4}$, reads as:
\begin{eqnarray}\label{Gazgm66}
\mathfrak{M}_{\mbox{\tiny{dS}}^{}_4}^{\prime\prime\;2} + \frac{2}{3}\Lambda = \mathfrak{M}^{2}_{\mbox{\tiny{dS}}^{}_4} = 0\,.
\end{eqnarray}
As a final remark, we would like to insist on the fact that the current observational data point towards a small but non-vanishing positive cosmological constant, which suggests that our Universe might presently be in a dS$_4$ phase. In consistency with this fact, we propose to reevaluate the above equation which establishes the connection between the bare cosmological term $\Lambda$ and a ``mass" attributed to the graviton.

\setcounter{equation}{0} \section*{Acknowledgements}
Hamed Pejhan is supported by the Bulgarian Ministry of Education and Science, Scientific Programme ‘Enhancing the Research Capacity in Mathematical Sciences (PIKOM)’, No. DO1-67/05.05.2022.

%%%%%%%%%%%%%%%%%%%%%%%%%%%%%%%%%%%%%%%%%%%%%%%%%%%%%%%%%%%%%%%%%%%%%%%%%%%%%%%%%%%%%%%%%%%%%%%%%%%%%%%%%%%%%%%%%%%%%%%%
%%%%%%%%%%%%%%%%%%%%%%%%%%%%%%%%%%%%%%%%%%%%%%%%%%%%%%%%%%%%%%%%%%%%%%%%%%%%%%%%%%%%%%%%%%%%%%%%%%%%%%%%%%%%%%%%%%%%%%%%
%%%%%%%%%%%%%%%%%%%%%%%%%%%%%%%%%%%%%%%%%%%%%%%%%%%%%%%%%%%%%%%%%%%%%%%%%%%%%%%%%%%%%%%%%%%%%%%%%%%%%%%%%%%%%%%%%%%%%%%%
%%%%%%%%%%%%%%%%%%%%%%%%%%%%%%%%%%%%%%%%%%%%%%%%%%%%%%%%%%%%%%%%%%%%%%%%%%%%%%%%%%%%%%%%%%%%%%%%%%%%%%%%%%%%%%%%%%%%%%%%
%%%%%%%%%%%%%%%%%%%%%%%%%%%%%%%%%%%%%%%%%%%%%%%%%%%%%%%%%%%%%%%%%%%%%%%%%%%%%%%%%%%%%%%%%%%%%%%%%%%%%%%%%%%%%%%%%%%%%%%%
%%%%%%%%%%%%%%%%%%%%%%%%%%%%%%%%%%%%%%%%%%%%%%%%%%%%%%%%%%%%%%%%%%%%%%%%%%%%%%%%%%%%%%%%%%%%%%%%%%%%%%%%%%%%%%%%%%%%%%%%
%%%%%%%%%%%%%%%%%%%%%%%%%%%%%%%%%%%%%%%%%%%%%%%%%%%%%%%%%%%%%%%%%%%%%%%%%%%%%%%%%%%%%%%%%%%%%%%%%%%%%%%%%%%%%%%%%%%%%%%%

\part{Appendices}

\begin{appendix}

\setcounter{equation}{0} \section{DS$_2$ Killing vectors}\label{Killing}
In three-dimensional Minkowski $\mathbb{R}^3$, preferred reference frames (called \emph{inertial} frames) exist. An inertial observer can always choose coordinates (strictly speaking, global inertial coordinates) in such a way that the infinitesimal interval between given points $p(x)$ and $p^\prime(x + \mathrm{d} x)$ in $\mathbb{R}^3$ takes the form:
\begin{eqnarray}
\mathrm{d} s^2 = \eta^{}_{ab} \mathrm{d} x^a \mathrm{d} x^b,
\end{eqnarray}
where $a,b=0,1,2$ and $\eta^{}_{ab}=(1,-1,-1)$ is the metric tensor in the inertial frame $S(x)$. The interval between $p$ and $p^\prime$, of course, does not depend on the choice of reference frame, therefore, the transition to another reference frame $S^\prime(x^\prime)$ implies that:
\begin{eqnarray}
\eta^{}_{ab} \mathrm{d} x^a \mathrm{d} x^b = g^\prime_{ab}(x^\prime) \mathrm{d} x^{\prime a} \mathrm{d} x^{\prime b},
\end{eqnarray}
where $g^\prime_{ab}$ is the metric in $S^\prime$. The latter would be inertial if there exist coordinates $x^\prime$ for which we get $g^\prime_{ab} = \eta^{}_{ab}$, or, in other words, if the transition $S\rightarrow S^\prime$ does not change the \emph{form}\footnote{The form variation of a field $F(x)$, defined by $\delta F(x) \equiv F^\prime(x) - F(x)$, must be distinguished from its total variation, given by $\Delta F(x) \equiv F^\prime(x^\prime) - F(x)$. When $x^\prime - x = \zeta$ is infinitesimally small, we have: $\Delta F(x) \approx \delta F(x) + \zeta^a \partial_a F(x)$.} of the metric. Note that such coordinate transformations $x\rightarrow x^\prime$, which do not change the form of the metric, define the symmetry group of $\mathbb{R}^3$. Among all such transformations, in the context of this paper, we are interested in those, which leave invariant the following quadratic form as well:
\begin{eqnarray}\label{Killing symmetry dS2}
(x)^2 = \eta^{}_{ab} x^a x^b.
\end{eqnarray}
These transformations define the dS$_2$ symmetry group.

In order to obtain the explicit form of the infinitesimal dS$_2$ group transformations, we consider the following generic form of coordinate transformations in $\mathbb{R}^3$:
\begin{eqnarray}\label{101010}
x^a \;\rightarrow\; x^{\prime a} = x^a + \zeta^a(x)\,,
\end{eqnarray}
when $\zeta^a(x)$ is infinitesimally small. The expansion of $\zeta^a(x)$ in a power series reads:
\begin{eqnarray}\label{141414}
\zeta^a(x) = \omega^{a}_{} + \omega^a_b x^b + \omega^a_{bc} x^b x^c + \; ...\,,
\end{eqnarray}
with constant parameters $\omega^{a}_{}=0$, $\omega^a_b,\; \omega^a_{bc}$, and $...$ . The vanishing parameter $\omega^{a}_{}=0$ is trivially issued from the condition (\ref{Killing symmetry dS2}). To fix the other coefficients, one must notice that the transformations (\ref{101010}) relate the metrics $\eta^{}_{ab}$ and $g^\prime_{ab}$ by:
\begin{eqnarray}
g^\prime_{ab}(x^\prime) = \frac{\partial x^d}{\partial x^{\prime a}} \frac{\partial x^c}{\partial x^{\prime b}} \eta^{}_{dc} \approx \eta^{}_{ab} - (\partial_a \zeta_{b} + \partial_b \zeta_{a})\,.
\end{eqnarray}
Form invariance of the metric is now expressed by the following \emph{Killing equation}:
\begin{eqnarray}\label{121212}
\delta \eta^{}_{ab} \equiv g^\prime_{ab}(x) - \eta^{}_{ab} \approx - (\partial_a \zeta_{b} + \partial_b \zeta_{a}) = 0\,.
\end{eqnarray}
The requirement (\ref{121212}), considering the expansion formula (\ref{141414}), implies that $\omega^{ab} + \omega^{ba} =0$, and that the remaining parameters $\omega^a_{bc}=...=0$. Then, the infinitesimal, global dS$_2$ transformations would be, of linear type, defined in terms of three constant parameters $\omega^{ab}$ (recall that $\omega^{ab} = - \omega^{ba}$, while $a,b=0,1,2$):
\begin{eqnarray}\label{131313}
\zeta^a(x) = \omega^a_b x^b.
\end{eqnarray}
Taking the above identity into account, one can easily check that the coordinate transformations (\ref{101010}) fulfill the condition (\ref{Killing symmetry dS2}).

Now, we consider a generic field $\phi(x)$, which is scalar with respect to the transformations (\ref{131313}), that is, $\phi^\prime(x^\prime)=\phi(x)$. The change of form of $\phi$ then reads:
\begin{eqnarray}
\delta \phi(x) = (x^a - x^{\prime a}) \partial_a \phi(x) &=& - \omega^a_b x^b \partial_a \phi(x)\nonumber\\
&\equiv& \frac{1}{2} \omega^{ab} K_{ab} \phi(x)\,,
\end{eqnarray}
where the generators, say the Killing vectors, $K_{ab}$ are:
\begin{eqnarray}\label{Killing Eq}
K_{ab} = x_a \partial_b - x_b \partial_a\,.
\end{eqnarray}
They obey the Lie algebra:
\begin{eqnarray}
[K_{12}, K_{20}] = - K_{10}\,,\;\;\;\;\;\;\; [K_{12}, K_{10}] = K_{20}\,,\;\;\;\;\;\;\; [K_{20}, K_{10}] = K_{12}\,,
\end{eqnarray}
or equivalently:
\begin{eqnarray}
[K_{ab},K_{cd}] = - \big( \eta^{}_{ac} {K_{bd}} + \eta^{}_{bd} {K_{ac}} - \eta^{}_{ad} {K_{bc}} - \eta^{}_{bc} {K_{ad}} \big)\,,
\end{eqnarray}
which exhibits the $\mathfrak{su}(1,1)\sim \mathfrak{so}(1,2)$ algebra. According to the space-time-Lorentz factorization of the dS$_2$ group (given in subsection \ref{sec space-time-Lorentz}), one can easily show that $K_{12}$, which is of compact type, stands for the space translations, while $K_{10}$ and $K_{20}$, which both are of noncompact type, refer to the time translations and Lorentz boosts, respectively.

To see more on the above topics, readers are referred to Ref. \cite{Blagojevic}.

\setcounter{equation}{0} \section{$2\times 2$-quaternionic matrices}\label{App quat}
The set of quaternions is defined by:
\begin{eqnarray}\label{quat def}
\mathbb{H} = \Big\{ \textbf{x} \; \equiv \underbrace{(x^4,\; \vec{x})}_{\mbox{\small{scalar-vector notations}}} = \;\;\;\;\;\;\;\underbrace{x^4 \textbf{1} + x^1 {\textbf{e}}^{}_1 + x^2 {\textbf{e}}^{}_2 + x^3 {\textbf{e}}^{}_3}_{\mbox{\small{Euclidean metric notations}}} \;\; ; \;\; x^4,x^1,x^2,x^3 \in \mathbb{R} \Big\}\,,
\end{eqnarray}
where $\{\textbf{1}, {\textbf{e}}^{}_k\}$, with $k=1,2,3$, is a basis such that:
\begin{eqnarray}
&{\textbf{e}}^{}_1 {\textbf{e}}^{}_2 = - {\textbf{e}}^{}_2 {\textbf{e}}^{}_1 = {\textbf{e}}^{}_3\,,&\nonumber\\
&{\textbf{e}}^{}_2 {\textbf{e}}^{}_3 = - {\textbf{e}}^{}_3 {\textbf{e}}^{}_2 = {\textbf{e}}^{}_1\,,&\nonumber\\
&{\textbf{e}}^{}_3 {\textbf{e}}^{}_1 = - {\textbf{e}}^{}_1 {\textbf{e}}^{}_3 = {\textbf{e}}^{}_2\,,&\nonumber\\
&{\textbf{e}}^{}_1 {\textbf{e}}^{}_1 = {\textbf{e}}^{}_2 {\textbf{e}}^{}_2 = {\textbf{e}}^{}_3 {\textbf{e}}^{}_3 = -\textbf{1}\,,&\nonumber\\
& \textbf{1} {\textbf{e}}^{}_k = {\textbf{e}}^{}_k \textbf{1} = {\textbf{e}}^{}_k\,.&
\end{eqnarray}
The multiplication law of quaternions then reads:
\begin{eqnarray} \label{quat times}
{\textbf{x}}{\textbf{x}}^\prime = \Big(x^4 x^{\prime 4} - \vec{x}\cdot\vec{x}^{\prime},\; x^4 \vec{x}^{\prime} + x^{\prime 4}\vec{x} + \vec{x}\times\vec{x}^{\prime}\Big) &=& \big( x^4 x^{\prime 4} - x^1 x^{\prime 1} - x^2 x^{\prime 2} - x^3 x^{\prime 3}\big) \textbf{1}\nonumber\\
&& + \big( x^4 x^{\prime 1} + x^{\prime 4} x^1 + x^2 x^{\prime 3} - x^3 x^{\prime 2} \big) {\textbf{e}}^{}_1 \nonumber\\
&& + \big( x^4 x^{\prime 2} + x^{\prime 4} x^2 + x^3 x^{\prime 1} - x^1 x^{\prime 3} \big) {\textbf{e}}^{}_2 \nonumber\\
&& + \big( x^4 x^{\prime 3} + x^{\prime 4} x^3 + x^1 x^{\prime 2} - x^2 x^{\prime 1} \big) {\textbf{e}}^{}_3\,,
\end{eqnarray}
where $\vec{x}\cdot\vec{x}^{\prime}$ and $\vec{x}\times\vec{x}^{\prime}$ respectively refer to the Euclidean inner product and the cross product in $\mathbb{R}^3$. It is manifest that generally ${\textbf{x}}{\textbf{x}}^\prime \neq {\textbf{x}}^\prime {\textbf{x}}$, unless $\vec{x}\times\vec{x}^{\prime} = 0$. Moreover, the (quaternionic) conjugate of $\textbf{x}$ is given by ${\textbf{x}}^{\scriptscriptstyle\bigstar} = (x^4,\; -\vec{x})$, while $(\textbf{x}\textbf{x}^\prime)^{\scriptscriptstyle\bigstar} = \textbf{x}^{\prime{\scriptscriptstyle\bigstar}} \textbf{x}^{\scriptscriptstyle\bigstar}$, the squared norm by $|\textbf{x}|^2 = \textbf{x} {\textbf{x}}^{\scriptscriptstyle\bigstar} = (x^4)^2 + (x^1)^2 + (x^2)^2 + (x^3)^2 $, and the inverse of a nonzero quaternion by $\textbf{x}^{-1} = {\textbf{x}}^{\scriptscriptstyle\bigstar} / |\textbf{x}|^2$. Another useful identity is:
\begin{eqnarray}\label{exp(quat)}
\exp(\textbf{x}) = \exp(x^4) \Bigg( \cos(\sqrt{(x^1)^2 + (x^2)^2 + (x^3)^2}) + \frac{x^1 {\textbf{e}}^{}_1 + x^2 {\textbf{e}}^{}_2 + x^3 {\textbf{e}}^{}_3}{\sqrt{(x^1)^2 + (x^2)^2 + (x^3)^2}} \sin(\sqrt{(x^1)^2 + (x^2)^2 + (x^3)^2}) \Bigg)\,,
\end{eqnarray}
for $(x^1)^2 + (x^2)^2 + (x^3)^2 \neq 0$.

The basis $\{\textbf{1}, {\textbf{e}}^{}_k\}$ also reveals the natural one-to-one correspondence between $\mathbb{H}$ and $\mathbb{R}^4$ (by abuse of notation, $\mathbb{H}\sim \mathbb{R}^4$) given by the following mapping:
\begin{eqnarray}
\mathbb{H} \ni \textbf{x}=x^4 \textbf{1} +  x^1 {\textbf{e}}^{}_1 + x^2 {\textbf{e}}^{}_2 + x^3 {\textbf{e}}^{}_3 \; \mapsto \; x=(x^1,x^2,x^3,x^4) \in \mathbb{R}^4\,.
\end{eqnarray}
This mapping clearly does not change the norm $|\textbf{x}|^2 = (x^4)^2 + (x^1)^2 + (x^2)^2 + (x^3)^2$, due to the orthonormality of the basis $\{\textbf{1}, {\textbf{e}}^{}_k\}$.

A realization of the basis $\{\textbf{1}, {\textbf{e}}^{}_k\}$ can be achieved by allocating $2\times 2$-matrix representations in such a way that $\textbf{1} \equiv \mathbbm{1}_2$ and ${\textbf{e}}^{}_k \equiv (-1)^{k+1} \mathrm{i} \sigma_k$, where $k=1,2,3$ and $\sigma_k$'s are the Pauli matrices. On this basis, the matrix realization of any $\textbf{x} \in \mathbb{H}$ explicitly reads:
\begin{eqnarray}\label{quat}
\textbf{x} = x^4 \textbf{1} +x^k {\textbf{e}}^{}_k =
\begin{pmatrix}
x^4 + \mathrm{i} x^3 & \mathrm{i} x^1 - x^2\\
\mathrm{i} x^1 + x^2 & x^4 - \mathrm{i} x^3
\end{pmatrix}\,.
\end{eqnarray}
In this realization, the conjugate of the quaternion $\textbf{x}$ is given by the conjugate transpose of the corresponding matrix. Taking this point into account, one can check that the matrix realization (\ref{quat}) of a quaternion $\textbf{x} \in \mathbb{H}$ verifies the aforementioned identities. Moreover, one can show that $\mbox{det} (\textbf{x}) = |\textbf{x}|^2$ and $|\textbf{x}\textbf{x}^\prime|^2 = |\textbf{x}|^2 |\textbf{x}^\prime|^2$ (since $\mbox{det} (\textbf{x}\textbf{x}^\prime) = \mbox{det}(\textbf{x}) \; \mbox{det}(\textbf{x}^\prime) $). Accordingly, for any $\textbf{x} \in \mathbb{H}$, we have:\footnote{In this manuscript, readers must distinguish between the symbol `$^{\scriptscriptstyle\bigstar}$', referring to the quaternionic conjugate, and the symbol `$^\ast$', which refers to the complex conjugate of numbers belonging to $\mathbb{C}$.}
\begin{eqnarray}\label{SU2-quat}
\textbf{x} = |\textbf{x}|
\begin{pmatrix}
\alpha & \beta \\
-{\beta^\ast} & {\alpha^\ast}
\end{pmatrix}\,, \;\;\;\;\;\;\; |\alpha|^2 + |\beta|^2=1 \;\;\; (\mbox{with}\; \alpha,\beta \in \mathbb{C})\,,
\end{eqnarray}
which reveals that the quaternion field, as a multiplicative group, is $\mathbb{H} \sim \mathbb{R}_+\times \mathrm{SU}(2)$. [To see more about the $\mathrm{SU}(2)$ group, one can refer to appendix \ref{App UIR's SU(2)}.]

\setcounter{equation}{0} \section{Wigner classification of the Poincar\'{e} group UIR's}\label{App Com UIR's Poincare}
In this appendix, we briefly recall the Wigner classification of the Poincar\'{e} group UIR's \cite{Wigner1939,Newton/Wigner}, accomplished with respect to the eigenvalues of two Casimir operators of the Poincar\'{e} group. The corresponding quadratic Casimir (or Klein-Gordon) operator takes the form $Q_P^{(1)} = \hat{p}^2 \; \big(\equiv \hat{p}^\mu \hat{p}_\mu\big) = (\hat{p}^0)^2 - (\vec{\hat{p}})^{\;2}$, where $\hat{p}^\mu = (\hat{p}^0,\vec{\hat{p}})$ refers to the four-momentum operator. [Note that here the metric tensor is $\eta^{}_{\mu\nu} = \mbox{diag}(1,-1,-1,-1)$, with $\mu,\nu = 0,1,2,3$.] The quartic Casimir operator, on the other hand, is $Q_P^{(2)} = W^2 \; \big(\equiv W^\mu W_\mu\big)$, with the Pauli-Lubanski operators defined by $W_\mu = \frac{1}{2} {\cal{E}}_{\mu\nu\rho\sigma} \hat{j}^{\nu\rho}\hat{p}^\sigma$, $\hat{j}^{\nu\rho}$ being the relativistic angular momentum tensor operators and ${\cal{E}}_{\mu\nu\rho\sigma}$ the four-dimensional totally antisymmetric Levi-Civita symbol. In the units $c=1=\hbar$, these two Casimir operators respectively possess eigenvalues $\langle Q_P^{(1)} \rangle = m^2$, with $m\geqslant 0$, and in the nonzero mass cases (i.e., $m>0$), $\langle Q_P^{(2)} \rangle = -m^2 s(s+1)$, with $s\in\mathbb{N}/2$. This group-theoretical structure actually makes clear the notion of mass based on spacetime symmetries, in the context of Einstein-Poincar\'{e} relativity.

\begin{table}
\begin{center}
\item[]\begin{tabular}[c]{|l|c|c|}\hline
$\Big|$ First Casimir or squared mass & Point $\hat{p}^{\mu}$ of the mass hyperboloid, invariant under the little group action & Little group\\ \hline
$\Big|$ (a) $\hat{p}^2 = m^2 >0\, , \hat{p}^0 > 0$ & $(m, 0, 0, 0)$& $SO(3)$ \\ \hline
$\Big|$ (b) $\hat{p}^2 = m^2 >0 \, , \hat{p}^0 < 0$ & $(-m, 0, 0, 0)$& $SO(3)$ \\ \hline
$\Big|$ (c) $\hat{p}^2 = 0\, , \hat{p}^0 > 0$ & $(\kappa, \kappa, 0, 0)$& $ISO(2)$ \\ \hline
$\Big|$ (d) $\hat{p}^2 =  0\, , \hat{p}^0 < 0$ & $(-\kappa, \kappa, 0, 0)$& $ISO(2)$ \\ \hline
$\Big|$ (e) $\hat{p}^2 = N^2 >0$ & $(0, N, 0, 0)$& SO$(1,2)$ \\ \hline
$\Big|$ (f) $\hat{p}^{\mu} = 0$ & $(0, 0, 0, 0)$& $SO(1,3)$ \\ \hline
\end{tabular}
\end{center}
\caption{Wigner classification of the Poincar\'e UIR's, according to the mass operator and the little group UIR's.}
\label{Table. Wigner}
\end{table}

According to the Wigner classification of the Poincar\'{e} group UIR's, with respect to the mass operator $Q_P^{(1)}$ and the little (or stabilizer) group UIR's given in TABLE \ref{Table. Wigner}, it is commonly accepted that the physical cases are restricted to:
\begin{itemize}
\item{The case (a), characterizing the massive UIR's with positive energy, ${\cal{P}}^{>}_{s,m}$.}
\item{The case (c), characterizing the massless UIR's with positive energy, ${\cal{P}}^{>}_{s,0}$.}
\item{The case (f), characterizing the vacuum.}
\end{itemize}
For an overview of the Poincar\'{e} group and its representations, one can refer to Refs. \cite{S.T. Ali,S.T. Ali'}.

\setcounter{equation}{0} \section{Some useful relations concerning the $\mathbb{S}^3$ hyperspherical harmonics, Gegenbauer and Legendre polynomials, and hypergeometric functions}\label{App some}
Note that, in this appendix, we just point out those relations which are used in the context of our paper. For more details, one can refer to Refs. \cite{Hua,Magnus,Gradshteyn,JAvery}.

\subsection{Hyperspherical harmonics on $\mathbb{S}^3$}
The hyperspherical harmonics on the unit-sphere $\mathbb{S}^3$ are defined by:
\begin{eqnarray}\label{YLlm}
{\cal{Y}}_{Llm}\big(\textbf{z}(\psi,\theta,\phi)\big) = l! \; 2^{l+1} \Bigg(\frac{(L+1)(L-l)!}{2\pi (L+l+1)!}\Bigg)^{\frac{1}{2}} (\sin\psi)^l \; C_{L-l}^{l+1}(\cos\psi) \; Y_{lm}(\theta,\phi)\,,
\end{eqnarray}
where: (i) $(L,l,m) \in \mathbb{N} \times \mathbb{N} \times \mathbb{Z}$, with $0\leqslant l \leqslant L$ and $-l\leqslant m \leqslant l$, (ii) the quaternion $\textbf{z} \equiv ({z}^4, \vec{z}) \in \mathrm{SU}(2) \sim \mathbb{S}^3$ admits the following parametrization:
\begin{eqnarray}
{z}^4 &=& \cos\psi\,,\nonumber\\
{z}^1 &=& \sin\psi \sin\theta \cos\phi\,,\nonumber\\
{z}^2 &=& \sin\psi \sin\theta \sin\phi\,,\nonumber\\
{z}^3 &=& \sin\psi \cos\theta,
\end{eqnarray}
with $0 \leqslant \psi , \theta \leqslant \pi$ and $0 \leqslant \phi < 2\pi$, (iii) $C_{L-l}^{l+1}$'s are the Gegenbauer polynomials:
\begin{eqnarray}
C_{L-l}^{l+1}(\cos\psi) = \frac{1}{l!} \sum^{\lfloor \frac{L-l}{2} \rfloor}_{k=0} \frac{(-1)^k \; (L-k)!}{k! \; (L-l-2k)!} \; (2\cos\psi)^{L-l-2k}\,,
\end{eqnarray}
(iv) $Y_{lm}$'s are the ordinary spherical harmonics:
\begin{eqnarray}
Y_{lm}(\theta,\phi) = \Bigg( \frac{(2l+1)(l-m)!}{4\pi(l+m)!} \Bigg)^{\frac{1}{2}} \; P_l^m(\cos\theta) \; e^{\mathrm{i} m\phi}\,,
\end{eqnarray}
while $P_l^m$'s stand for the associated Legendre functions:
\begin{eqnarray}\label{AssLP}
P_l^m(\cos\theta) = \frac{(-1)^{l+m}}{2^l \; l!} \; (\sin \theta)^m \; \Big( \frac{\partial}{\partial\cos\theta} \Big)^{l+m} \; (\sin\theta)^{2l}\,.
\end{eqnarray}
Below, we will provide more details on the Gegenbauer and Legendre polynomials.

Note that, with the above choice of constant factors, the set of hyperspherical harmonics fulfills the following orthogonality (and normalization) conditions:
\begin{eqnarray}\label{orthhyha}
\int_{\mathbb{S}^3} {\cal{Y}}^\ast_{Llm}(\textbf{z}) {\cal{Y}}^{}_{L^\prime l^\prime m^\prime}(\textbf{z}) \; \mathrm{d}\mu(\textbf{z}) = \delta^{}_{LL^\prime} \delta^{}_{ll^\prime} \delta^{}_{mm^\prime}\,,
\end{eqnarray}
where $\mathrm{d}\mu(\textbf{z})= \sin^2\psi\sin\theta \;\mathrm{d}\psi \mathrm{d}\theta \mathrm{d}\phi$ is the invariant measure on $\mathbb{S}^3$ (see appendix \ref{App UIR's SU(2)}).

\subsection{Gegenbauer polynomials}
Generally, the Gegenbauer polynomials $C_n^\lambda(x)$ are defined by the formula:
\begin{eqnarray}\label{Gegenbpo}
C_n^\lambda (x) = \sum_{k=0}^{\lfloor \frac{n}{2}\rfloor} (-1)^k \frac{\Gamma(n+\lambda -k)}{k! \; (n-2k)! \; \Gamma(\lambda)}(2x)^{n-2k}\,, \;\;\;\;\;\;\; \lambda > -\frac{1}{2}\,.
\end{eqnarray}
The variable $x$ is real and between $-1$ and $+1$. It can be shown that the $C_n^\lambda(x)$'s are the coefficients of $t^n$ in the power-series expansion of the function:
\begin{eqnarray}\label{G1}
(1 + t^2 - 2xt)^{-\lambda} = \sum_{n=0}^\infty C_n^\lambda(x) \; t^n\,, \;\;\;\;\;\;\; |t| < 1\,.
\end{eqnarray}
Another useful expansion formula is:
\begin{eqnarray}\label{G2}
C_n^\lambda(x) = \frac{1}{\Gamma(\lambda) \; \Gamma(\lambda-1)} \sum_{k=0}^{\lfloor \frac{n}{2}\rfloor} c_k \; C_{n-2k}^1(x)\,,\;\;\;\;\;\;\; c_k = \frac{(n-2k+1) \; \Gamma(k+\lambda-1) \; \Gamma(\lambda+n-k)}{k! \; \Gamma(n-k+2)}\,.
\end{eqnarray}

The Gegenbauer polynomials obey the following recurrence relations:
\begin{eqnarray}
n C_n^\lambda (x) &=& 2\lambda \big(x C_{n-1}^{\lambda+1}(x) - C_{n-2}^{\lambda+1} (x)\big)\,,\\
n C_n^\lambda (x) &=& (2\lambda+n-1) x C_{n-1}^\lambda (x) - 2\lambda (1-x^2) C_{n-2}^{\lambda+1} (x)\,,\\
(2\lambda+n) C_n^\lambda (x) &=& 2\lambda \big(C_n^{\lambda+1} (x) - x C_{n-1}^{\lambda+1} (x)\big)\,,\\
(n+2) C_{n+2}^\lambda (x) &=& 2(\lambda+n+1) x C_{n+1}^\lambda (x) - (2\lambda+n) C_n^\lambda (x)\,,\\
\frac{\mathrm{d}^m}{\mathrm{d} x^m} C_n^\lambda (x) &=& 2^m \lambda (\lambda+1)(\lambda+2) \;...\; (\lambda+m-1) C_{n-m}^{\lambda+m} (x)\,.
\end{eqnarray}

Note the role of the particular rotationally invariant $C_L^1 (\textbf{z}_1 \cdot \textbf{z}_2)$ as generating function of the $\mathbb{S}^3$ hyperspherical harmonics:
\begin{eqnarray}\label{G3}
C_L^1 (\textbf{z}_1 \cdot \textbf{z}_2) = \frac{2\pi^2}{L+1} \sum_{lm} {\cal{Y}}_{Llm}(\textbf{z}_1) {\cal{Y}}_{Llm}^\ast(\textbf{z}_2)\,, \;\;\;\;\;\;\; \textbf{z}_1, \textbf{z}_2 \in \mathbb{S}^3\,.
\end{eqnarray}

The Gegenbauer polynomials can also be expressed as the Legendre functions of the first kind:
\begin{eqnarray}\label{G4}
C^\lambda_\alpha (r) = \pi^{\frac{1}{2}} \; 2^{\frac{1}{2}-\lambda} \; \Gamma(\alpha+2\lambda) \; \big( \Gamma(\alpha+1)\Gamma(\lambda) \big)^{-1} \; (r^2-1)^{\frac{1}{4} - \frac{1}{2}\lambda} \; P^{(\frac{1}{2}-\lambda)}_{\alpha+\lambda-\frac{1}{2}}(r) \,,
\end{eqnarray}
where an integral representation of the Legendre functions of the first kind reads as:
\begin{eqnarray}\label{G5}
\Gamma\big(\frac{1}{2}-\mu\big) \; (r^2-1)^{\frac{1}{2}\mu} \; \pi^{\frac{1}{2}} \; 2^{-\mu} \; P^{(\mu)}_\nu(r) = \int_{0}^{\pi} \big( r+(r^2-1)^{\frac{1}{2}} \cos t \big)^{\nu+\mu} (\sin t)^{-2\mu} \;dt \,,
\end{eqnarray}
with $\mbox{Re} (\mu) < \frac{1}{2}$. Above, the argument $r$ is a point in the complex plane under exclusion of points on the real axis between $+1$ and $-\infty$.

\subsection{Associated Legendre polynomials}
The associated Legendre polynomials $P_l^m(x)$ fulfill the following recurrence relations:
\begin{eqnarray}
x P_l^m (x) - P_{l+1}^m (x) &=& (l+m)(1-x^2)^{\frac{1}{2}} P_l^{m-1} (x)\,,\\
P_{l-1}^m (x) - P_{l+1}^m (x) &=& (2l+1)(1-x^2)^{\frac{1}{2}} P_l^{m-1} (x)\,,\\
P_{l-1}^m (x) - xP_{l}^m (x) &=& (l-m+1)(1-x^2)^{\frac{1}{2}} P_l^{m-1} (x)\,,\\
(1-x^2)^{\frac{1}{2}} P_l^{m+1} (x) &=& (l-m) x P_{l}^m (x) - (l+m) P_{l-1}^m (x)\,,\\
(1-x^2) \frac{\mathrm{d}}{\mathrm{d} x} P_{l}^m (x) &=& (l+1) x P_l^{m} (x) - (l-m+1) P_{l+1}^{m} (x)\nonumber\\
&=& -l x P_l^{m} (x) + (l+m) P_{l-1}^{m} (x)\,,\\
(2l+1) x P_{l}^m (x) &=& (l-m+1) P_{l+1}^{m} (x) + (l+m) P_{l-1}^{m} (x)\,,\\
2 (1-x^2)^{\frac{1}{2}} \frac{\mathrm{d}}{\mathrm{d} x} P_{l}^m (x) &=& (l+m) (l-m+1) P_{l}^{m-1} (x) - P_l^{m+1} (x)\,,\\
- \frac{2m}{(1-x^2)^{\frac{1}{2}}} P_{l+1}^m (x) &=& (l+m) (l+m+1) P_{l}^{m-1} (x) + P_l^{m+1} (x)\,,\\
(1-x^2)^{\frac{1}{2}} P_l^{m+1} (x) &=& (l-m+1) P_{l+1}^{m} (x) - (l+m+1) x P_{l}^{m} (x)\,,\\
(l-m) (l+m+1) P_{l}^{m} (x) &=& - P_l^{m+2} (x) - 2(m+1) x (1-x^2)^{-\frac{1}{2}} P_l^{m+1} (x)\,,\\
(2l+1) (1-x^2)^{\frac{1}{2}} P_l^{m+1} (x) &=& (l-m) (l-m+1) P_{l+1}^{m} (x) - (l+m) (l+m+1) P_{l-1}^{m} (x)\,.
\end{eqnarray}
Again, the variable $x$ is real and between $-1$ and $+1$.

\subsection{Hypergeometric functions}
The hypergeometric functions ${}_2F^{}_1 (a,b;c;r)$ are defined by:
\begin{eqnarray}
{}_2F^{}_1 (a,b;c;r) &=& \frac{\Gamma(c)}{\Gamma(a) \; \Gamma(b)} \sum_{n=0}^\infty \frac{\Gamma(a+n) \; \Gamma(b+n)}{\Gamma(c+n)} \; \frac{r^n}{n!} \nonumber\\
&=& 1 + \frac{a\; b}{c} \; \frac{r}{1!} + \frac{a(a+1)\; b(b+1)}{c(c+1)} \; \frac{r^2}{2!} + ...\,, \;\;\;\;\;\;\; c\neq 0,-1,-2,-3,... \;.
\end{eqnarray}
This series possesses the following properties:
\begin{itemize}
\item{It converges in the open unit disk $|r| < 1$.}
\item{Its behavior on the circle of convergence $|r| =1$ is given by:
     \begin{itemize}
     \item{divergent for $\mbox{Re} (a+b-c) \geqslant 1$,}
     \item{absolutely convergent for $\mbox{Re} (a+b-c) < 0$,}
     \item{conditionally convergent for $0 \leqslant \mbox{Re} (a+b-c) < 1$, the point $r =1$ being excluded.}
     \end{itemize}}
\item{It reduces to a polynomial of degree $n$ in $r$, when $a$ or $b$ is a negative integer $-n$ ($n = 0,1,2,3, ...$).}
\end{itemize}

The Euler's transformation for the hypergeometric functions reads:
\begin{eqnarray}\label{Euler's transformation}
{}_2F^{}_1 (a,b;c;r) = (1-r)^{c-a-b} \; {}_2F^{}_1(c-a,c-b;c;r)\,.
\end{eqnarray}
Other useful relations are:
\begin{eqnarray}\label{limit}
^{}_2F^{}_1(a,b;c;1) = \frac{\Gamma(c) \; \Gamma(c-a-b)}{\Gamma(c-a) \; \Gamma(c-b)}\,, \;\;\;\;\; \mbox{Re} (a+b-c) < 0\,, \;\; \mbox{and} \;\; c\neq 0,-1,-2,...\,,
\end{eqnarray}
\begin{eqnarray}\label{1}
\frac{\mathrm{d}}{\mathrm{d} r} \; ^{}_2F^{}_1(a,b;c;r) = \frac{ab}{c} \; ^{}_2F^{}_1(a+1,b+1;c+1;r)\,,
\end{eqnarray}
\begin{eqnarray}\label{2}
^{}_2F^{}_1(a,b;1+a-b;-1) = 2^{-a} \; \sqrt{\pi} \; \frac{\Gamma(1+a-b)}{\Gamma\big(1+\frac{1}{2}a-b\big) \; \Gamma\big( \frac{1+a}{2} \big)}\,, \;\;\;\;\; 1+a-b \neq 0,-1,-2,...\,,
\end{eqnarray}
\begin{eqnarray}\label{3}
^{}_2F^{}_1(a,b;2+a-b;-1) &=& 2^{-a} \; \sqrt{\pi} \; (b-1)^{-1} \; \Gamma(2+a-b) \\
&& \times \Bigg[ \frac{1}{\Gamma\big( \frac{1}{2}a \big) \; \Gamma\big( \frac{3}{2}+\frac{1}{2}a-b \big)} - \frac{1}{\Gamma\big( \frac{1+a}{2} \big) \; \Gamma\big( 1+ \frac{1}{2}a-b \big)} \Bigg]\,,\;\;\; 2+a-b \neq 0,-1,-2,... \;.\nonumber
\end{eqnarray}

\setcounter{equation}{0} \section{$\mathrm{SU}(2)$ group and its representations}\label{App UIR's SU(2)}

\subsection{$\mathrm{SU}(2)$ group}
The special unitary group $\mathrm{SU}(2)$ consists of unimodular unitary matrices of the second order:
\begin{eqnarray}
\mathrm{SU}(2) = \Bigg\{ g=\begin{pmatrix} \alpha & \beta\\ -\beta^\ast & \alpha^\ast \end{pmatrix}, \;\; \alpha,\beta\in\mathbb{C}, \;\; \mbox{det} (g) = |\alpha|^2 + |\beta|^2 = 1 \Bigg\}\,.
\end{eqnarray}
[This group is indeed a subgroup of the special linear group SL$(2,\mathbb{C})$, that is, the group of all $2\times 2$-complex matrices of determinant unity.] By defining $\alpha = x^4 + \mathrm{i} x^3$ and $\beta = \mathrm{i} x^1 - x^2$, where $x^{\texttt{A}} \in \mathbb{R}$ ($\texttt{A}=1,2,3,4$), we obtain an alternative expression for the unimodular condition $|\alpha|^2 + |\beta|^2 = 1$ as:
\begin{eqnarray}\label{eq3S3}
(x^1)^2 + (x^2)^2 + (x^3)^2 + (x^4)^2 = 1\,,
\end{eqnarray}
which is the equation of the three-sphere $\mathbb{S}^3$ of unit radius in $\mathbb{R}^4$, with center at the origin. On the other hand, one can associate with each point $x=(x^1,x^2,x^3,x^4) \in \mathbb{S}^3$ the matrix $h(x)$ defined by:
\begin{eqnarray}\label{su2itu}
h(x) = \begin{pmatrix}
x^4+ \mathrm{i} x^3 & \mathrm{i} x^1-x^2\\
\mathrm{i} x^1+x^2 & x^4- \mathrm{i} x^3
\end{pmatrix}\,,
\end{eqnarray}
with the property $\mbox{det}\big(h(x)\big) = (x^1)^2 + (x^2)^2 + (x^3)^2 + (x^4)^2 =1$ (then, $h(x) \in \mathrm{SU}(2)$). Therefore, as a topological space, the $\mathrm{SU}(2)$ group is homeomorphic to the unit-sphere $\mathbb{S}^3$. In other words, the underlying manifold of $\mathrm{SU}(2)$ is $\mathbb{S}^3$, which shows that $\mathrm{SU}(2)$ is compact, connected and simply connected.

Considering the above, for any $g\in \mathrm{SU}(2)$ and $x\in\mathbb{S}^3$, a left action defined by $h^\prime (x) = gh(x)$ induces a linear transformation $x^\prime \equiv g\diamond x$ on $\mathbb{S}^3$ in such a way that $\mbox{det} \big(h^\prime(x)\big) = \mbox{det} \big(h(x)\big) = 1$. This means that the induced linear transformation $g\diamond x$ preserves the form (\ref{eq3S3}). Consequently, the left action of $\mathrm{SU}(2)$ on $\mathbb{S}^3$ represents orthogonal transformations belonging to $O(4)$. More precisely, since the $\mathrm{SU}(2)$ group is connected, the orthogonal transformation $g\diamond x$ belongs to $SO(4)$. Similarly, a right action $h(x)g$ also induces a rotation of $\mathbb{R}^4$ associated with $SO(4)$. The sphere $\mathbb{S}^3$ is therefore invariant under either left or right action of $\mathrm{SU}(2)$.

In the present context it is useful to remind the quaternionic realization of the $\mathrm{SU}(2)$ group, which has been previously given in (\ref{SU2-quat}). To do this, we identify $\mathbb{R}^4$ with the field of quaternions $\mathbb{H}$ by the mapping:
\begin{eqnarray}
x=(x^1,x^2,x^3,x^4)\; \mapsto \;\textbf{x} = x^4\textbf{1} + x^k{\textbf{e}}^{}_k\,, \;\;\;\;\; k=1,2,3\,,
\end{eqnarray}
where the quaternionic basis is $\big\{ \textbf{1} \equiv \mathbbm{1}_2, \; {\textbf{e}}^{}_k \equiv (-1)^{k+1} \mathrm{i} \sigma_k, \; k=1,2,3 \big\}$, $\sigma_k$'s being the Pauli matrices. Then, the aforementioned mapping $h$ represents an isomorphism from the multiplicative group of quaternions with norm $1$ (since $\mbox{det} \big(h(x)\big)=1$) onto $\mathrm{SU}(2)$ (in this regard, see also Eq. (\ref{quat})). In this sense, the $\mathrm{SU}(2)$ group can also be realized by:
\begin{eqnarray}
\mathrm{SU}(2)\sim \big\{ \textbf{x}\in\mathbb{H} \; ;\; |\textbf{x}|^2 =1 \big\}\,.
\end{eqnarray}

Regarding the content of our paper, we also would like to study the action of $\mathrm{SU}(2)\times \mathrm{SU}(2)$ on $\mathbb{H}$. Technically, this action is defined by:
\begin{eqnarray}\label{444444444}
\Phi (\textbf{u},\textbf{v}) : \textbf{x} \;\mapsto\; \textbf{u}\textbf{x}\textbf{v}^{-1}\,,
\end{eqnarray}
where $\textbf{u},\textbf{v},\textbf{x}\in\mathbb{H}$, with $|\textbf{u}| = |\textbf{v}| =1$ (in other words, $\textbf{u},\textbf{v}\in \mathrm{SU}(2)$). Since $\mbox{det}\big(\textbf{u}\textbf{x}\textbf{v}^{-1}\big) = \mbox{det}(\textbf{x})$ (that is, $|\textbf{u}\textbf{x}\textbf{v}^{-1}| = |\textbf{x}|$; see appendix \ref{App quat}), the action $\Phi (\textbf{u},\textbf{v})$ represents a rotation in $\mathbb{H}\sim \mathbb{R}^4$ belonging to $O(4)$. More precisely, the fact that $\mathrm{SU}(2)\times \mathrm{SU}(2)$ is connected implies that this action is an element of $SO(4)$. It is also a matter of simple calculations to show that all elements of $SO(4)$, as rotations of $\mathbb{H}\sim\mathbb{R}^4$ around the origin, can be presented by the maps of the form (\ref{444444444}) (see Ref. \cite{Savage}). This reveals a surjective group homomorphism between $\mathrm{SU}(2)\times \mathrm{SU}(2)$ and $SO(4)$. The kernel of this homomorphism is $\big\{\pm(\textbf{1},\textbf{1})\big\}$. To see the point, let $(\textbf{u},\textbf{v}) \in \mbox{ker}(\Phi)$, namely, for each $\textbf{x} \in \mathbb{H}$, we have $\textbf{u}\textbf{x}\textbf{v}^{-1} = \textbf{x}$. Considering $\textbf{x}=\textbf{1}$, it follows that $\textbf{u}=\textbf{v}$, and that $\textbf{u}$ is in the \emph{center}\footnote{The center of a group $G$ is the set of its elements that commute with every element of $G$.} of $\mathbb{H}$, which is equal to $\mathbb{R} \textbf{1}$. On the other hand, since $|\textbf{u}| = |\textbf{v}| =1$, we have $\textbf{u}=\textbf{v}=\pm \textbf{1}$. Then:
\begin{eqnarray}
SO(4)\sim \big(\mathrm{SU}(2)\times \mathrm{SU}(2)\big)/\big\{\pm(\textbf{1},\textbf{1})\big\}\,.
\end{eqnarray}

We end our brief introduction to the $\mathrm{SU}(2)$ group by recalling that the matrices $J_k = \frac{1}{2}(-1)^{k+1} \mathrm{i} \sigma_k$ ($k=1,2,3$) form a basis for its Lie algebra denoted by $\mathfrak{su}(2)$; any element $J \in \mathfrak{su}(2)$ can be written uniquely as:
\begin{eqnarray}
J = t_1J_1 + t_2J_2 + t_3J_3 =\frac{1}{2}
\begin{pmatrix}
\mathrm{i} t_3 & \mathrm{i} t_1-t_2\\
\mathrm{i} t_1+t_2 & - \mathrm{i} t_3
\end{pmatrix}\,, \;\;\;\;\;\;\; t_1,t_2,t_3\in\mathbb{R}\,.
\end{eqnarray}
Note that the matrices $J_1$, $J_2$, and $J_3$ obey the commutation rules:
\begin{eqnarray}
[J_i,J_j] = {{\cal E}_{ij}}^{k} \; J_k\,,
\end{eqnarray}
where $i,j,k = 1,2,3$ and ${{\cal E}_{ij}}^{k}$ is the three-dimensional totally antisymmetric Levi-Civita symbol. The corresponding Casimir operator is:
\begin{eqnarray}
J_1^2 + J_2^2 + J_3^2\,,
\end{eqnarray}
which commutes with all $J_k$'s ($k = 1,2,3$).

\subsection{Haar measure}
Note that on any Lie group $G$ two types of invariant measures (Haar measures) can be defined, by requiring invariance against right translations \cite{Ruhl}:
\begin{eqnarray}
\int_{G} f(g g^\prime) \; \mathrm{d}\mu^{}_r(g) = \int_{G} f(g) \; \mathrm{d}\mu^{}_r(g)\,,
\end{eqnarray}
or against left translations:
\begin{eqnarray}
\int_{G} f(g^\prime g) \; \mathrm{d}\mu^{}_l(g) = \int_{G} f(g) \; \mathrm{d}\mu^{}_l(g)\,,
\end{eqnarray}
for all continuous functions $f(g)$ with compact support on $G$ ($g,g^\prime \in G$). These measures are unique up to a constant factor. For a certain class of groups, including all compact Lie groups (like $\mathrm{SU}(2)$), both measures coincide $\mathrm{d}\mu^{}_r = \mathrm{d}\mu^{}_l = \mathrm{d}\mu$ \cite{Ruhl}.

A Haar measure on $\mathrm{SU}(2)$ is actually a measure on the unit-sphere $\mathbb{S}^3$ invariant under $SO(4)$. To find such a measure, we consider the bicomplex angular coordinates ($\omega, \psi_1, \psi_2$) on $\mathbb{S}^3$ (embedded in $\mathbb{R}^4\sim\mathbb{H}$), for which the following parametrization reads:
\begin{eqnarray}\label{biancopa}
x^4+ \mathrm{i} x^3 = \cos\omega e^{\mathrm{i} \psi_1}\,, \;\;\;\;\;\;\; x^1+ \mathrm{i} x^2 = \sin\omega e^{\mathrm{i} \psi_2}\,,
\end{eqnarray}
with $0\leqslant \omega\leqslant\frac{\pi}{2}$ and $0\leqslant\psi_1 ,\psi_2 < 2\pi$. Then, an element of $\mathrm{SU}(2)$ takes the form (see Eq. (\ref{su2itu})):
\begin{eqnarray}
g(\omega,\psi_1,\psi_2) =
\begin{pmatrix}
\cos\omega e^{\mathrm{i} \psi_1} & \mathrm{i} \sin\omega e^{\mathrm{i} \psi_2}\\
\mathrm{i} \sin\omega e^{- \mathrm{i} \psi_2} & \cos\omega e^{- \mathrm{i} \psi_1}
\end{pmatrix}\,.
\end{eqnarray}
Let $\Theta$ denote the map $(\omega,\psi_1,\psi_2) \mapsto x=(x^1,x^2,x^3,x^4)$. According to the relations given in (\ref{biancopa}), we have:
\begin{eqnarray}
x^1 = \sin\omega\cos\psi_2 \;\; &\Rightarrow& \;\; \mathrm{d} x^1 = \cos\omega\cos\psi_2 \; \mathrm{d}\omega - \sin\omega\sin\psi_2 \; \mathrm{d}\psi_2\,,\nonumber\\
x^2 = \sin\omega\sin\psi_2 \;\; &\Rightarrow& \;\; \mathrm{d} x^2 = \cos\omega\sin\psi_2 \; \mathrm{d}\omega + \sin\omega\cos\psi_2 \; \mathrm{d}\psi_2\,,\nonumber\\
x^3 = \cos\omega\sin\psi_1 \;\; &\Rightarrow& \;\; \mathrm{d} x^3 = -\sin\omega\sin\psi_1 \; \mathrm{d}\omega + \cos\omega\cos\psi_1 \; \mathrm{d}\psi_1\,,\nonumber\\
x^4 = \cos\omega\cos\psi_1 \;\; &\Rightarrow& \;\; \mathrm{d} x^4 = -\sin\omega\cos\psi_1 \; \mathrm{d}\omega - \cos\omega\sin\psi_1 \; \mathrm{d}\psi_1\,.
\end{eqnarray}
We can also rewrite the above equations as:
\begin{eqnarray}
\begin{pmatrix}
\mathrm{d} x^1\\ \mathrm{d} x^2\\ \mathrm{d} x^3\\ \mathrm{d} x^4
\end{pmatrix} &=&
\begin{pmatrix}
\cos\omega\cos\psi_2 \\ \cos\omega\sin\psi_2 \\ -\sin\omega\sin\psi_1 \\ -\sin\omega\cos\psi_1
\end{pmatrix} \mathrm{d}\omega +
\begin{pmatrix}
0\\ 0\\ \cos\psi_1 \\ -\sin\psi_1
\end{pmatrix} \cos\omega \; \mathrm{d}\psi_1 +
\begin{pmatrix}
-\sin\psi_2 \\ \cos\psi_2 \\ 0\\ 0
\end{pmatrix} \sin\omega \; \mathrm{d}\psi_2\nonumber\\
&\equiv & \frac{\partial\Theta}{\partial\omega}\; \mathrm{d}\omega + \frac{\partial\Theta}{\partial\psi_1}\; \mathrm{d}\psi_1 + \frac{\partial\Theta}{\partial\psi_2}\; \mathrm{d}\psi_2\,.\nonumber
\end{eqnarray}
Note that the vector given by $x=\Theta(\omega,\psi_1,\psi_2)$ and the three columns on the right-hand side are unit vectors and orthogonal. They indeed constitute an orthonormal basis, which is direct (this can be easily checked for $\omega = \psi_1 = \psi_2 = 0$). One can show that:
\begin{eqnarray}
\mbox{det}\Bigg(\Theta, \frac{\partial\Theta}{\partial\omega}, \frac{\partial\Theta}{\partial\psi_1}, \frac{\partial\Theta}{\partial\psi_2}\Bigg) = \sin\omega \cos\omega\,.
\end{eqnarray}
The invariant (normalized) measure of $\mathrm{SU}(2)$ therefore takes the form:
\begin{eqnarray}
\mathrm{d}\mu(g) = \frac{1}{4\pi^2}\sin 2\omega \; \mathrm{d}\omega \mathrm{d}\psi_1 \mathrm{d}\psi_2\,, \;\;\;\;\;\;\; \int_{\mathrm{SU}(2)} \; \mathrm{d}\mu(g) = 1\,,
\end{eqnarray}and correspondingly, the invariant integral for an integrable function $f(g)$ on $\mathrm{SU}(2)$ is:

\begin{eqnarray}
\int_{\mathrm{SU}(2)} f(g) \; \mathrm{d}\mu(g) = \frac{1}{4\pi^2} \int_{0}^{\frac{\pi}{2}}\int_{0}^{2\pi}\int_{0}^{2\pi}f(\omega,\psi_1,\psi_2) \; \sin 2\omega \; \mathrm{d}\omega \mathrm{d}\psi_1 \mathrm{d}\psi_2\,.
\end{eqnarray}

In this paper, we also use the standard polar coordinates ($\psi, \theta, \phi$) to parameterize the elements of $\mathrm{SU}(2)\sim\mathbb{S}^3$:
\begin{eqnarray}
x^1 &=& \sin\psi\sin\theta\cos\phi\,,\nonumber\\
x^2 &=& \sin\psi\sin\theta\sin\phi\,,\nonumber\\
x^3 &=& \sin\psi\cos\theta\,,\nonumber\\
x^4 &=& \cos\psi\,,
\end{eqnarray}
with $0\leqslant \psi , \theta \leqslant \pi$ and $0\leqslant \phi \leqslant 2\pi$. Along the lines sketched above, the corresponding invariant (normalized) measure of $\mathrm{SU}(2)$ is:
\begin{eqnarray}\label{Polar measure}
\mathrm{d}\mu(g) = \frac{1}{2\pi^2}\sin^2\psi\sin\theta \;\mathrm{d}\psi \mathrm{d}\theta \mathrm{d}\phi\,.
\end{eqnarray}

To see more on the above topic, one can refer to Refs. \cite{Ruhl,Faraut,Inomata}.

\subsection{Irreducible representations}
Representations of the $\mathrm{SU}(2)$ group can be obtained as a special case of the representations of SL$(2,\mathbb{C})$; the latter acts on the two-dimensional complex linear space ${\mathbb{C}}^2$ of all complex vectors $z= \begin{pmatrix} z_1 \\ z_2 \end{pmatrix}$, where $z_1,z_2\in \mathbb{C}$. Accordingly, following the lines sketched in Ref. \cite{Talman} by Talman, we consider the space $V_j$ of all homogeneous polynomials of degree $2j$ spanned by monomials:
\begin{eqnarray}
f_j^{m}(z) = \frac{z_1^{j+m} \; z_2^{j-m}}{\sqrt{(j+m)!\;(j-m)!}}\,,
\end{eqnarray}
where $j$ is a nonnegative integer or half-integer and $-j \leqslant m \leqslant j$. Every UIR of the $\mathrm{SU}(2)$ group can be constructed on $V_j$, and is characterized by a fixed value of $j$; clearly, there are $(2j+1)$ possible values for the index $m$, and hence, the generated representation is of $(2j+1)$ dimensions.

Since the space $V_j$ is of finite dimension, an explicit matrix representation of $g \in \mathrm{SU}(2)$ can be obtained by solving:
\begin{eqnarray}
f_j^{m_2}(g^{-1} \diamond z) &=& \frac{\big(\cos\omega\;e^{- \mathrm{i} \psi_1}z_1- \mathrm{i} \sin\omega\;e^{\mathrm{i} \psi_2}z_2\big)^{j+m_2} \big(- \mathrm{i} \sin\omega\;e^{- \mathrm{i} \psi_2}z_1 + \cos\omega\;e^{\mathrm{i} \psi_1}z_2\big)^{j-m_2}}{ \sqrt{(j+m_2)!\;(j-m_2)!} }\nonumber\\
&=& \sum_{m_1=-j}^{j} {\cal{D}}^j_{m_1m_2}(g)f_j^{m_1}(z)\,,
\end{eqnarray}
where, in the above, we have employed the aforementioned bicomplex angular coordinates ($\omega, \psi_1, \psi_2$). This expression provides us with a generating function for the matrix elements ${\cal{D}}^j_{m_1m_2}(g)$. The desired elements read:
\begin{eqnarray}
{\cal{D}}^j_{m_1m_2}(g) &=& (-1)^{m_1-m_2} \sqrt{(j+m_1)! \; (j-m_1)! \; (j+m_2)! \; (j-m_2)!} \nonumber\\
&& \times \sum_t \frac{(x^4+ \mathrm{i} x^3)^{j-m_2-t}}{(j-m_2-t)!} \; \frac{(x^4- \mathrm{i} x^3)^{j+m_1-t}}{(j+m_1-t)!} \; \frac{(-x^2+ \mathrm{i} x^1)^{t+m_2-m_1}}{(t+m_2-m_1)!} \; \frac{(x^2+ \mathrm{i} x^1)^t}{t!}\,,
\end{eqnarray}
where the values of $t$ to be summed over are those for which the arguments of the factorial functions remain nonnegative, namely, the values compatible with $m_1-m_2\leqslant t \leqslant j+m_1$ and $0\leqslant t \leqslant j-m_2$. Note that the above representations are unitary, irreducible, and exhaust the $\mathrm{SU}(2)$ irreducible representations \cite{Talman}.

It is also worth noting that, under complex conjugation, these UIR's transform as:
\begin{eqnarray}
\big({\cal{D}}^j_{m_1m_2}(g)\big)^\ast = (-1)^{m_2-m_1}{\cal{D}}^j_{-m_1,-m_2}(g)\,,
\end{eqnarray}
while their transpose are:
\begin{eqnarray}
{\cal{D}}^j_{m_2m_1}(g) = \big( {\cal{D}}^j_{m_1m_2}(g^{-1}) \big)^\ast = (-1)^{m_1-m_2}{\cal{D}}^j_{-m_1,-m_2}(g^{-1})\,.
\end{eqnarray}
The matrix elements ${\cal{D}}^j_{m_1m_2}(g)$ verify the following orthogonality relations:
\begin{eqnarray}
\int_{\mathrm{SU}(2)} {\cal{D}}^j_{m_1m_2}(g)\big( {\cal{D}}^{j^\prime}_{m_1^\prime m_2^\prime}(g) \big)^\ast \; \mathrm{d}\mu (g) = \frac{2\pi^2}{2j+1}\delta^{}_{jj^\prime} \delta^{}_{m_1m_1^\prime} \delta^{}_{m_2m_2^\prime}\,,
\end{eqnarray}
where $\mathrm{d}\mu(g) = \sin^2\psi\sin\theta \;\mathrm{d}\psi \mathrm{d}\theta \mathrm{d}\phi$ (see Eq. (\ref{Polar measure})). Then, the normalized counterparts of ${\cal{D}}^j_{m_1m_2}(g)$'s are:
\begin{eqnarray}
\widetilde{{\cal{D}}}^j_{m_1m_2}(g) = \sqrt{\frac{2j+1}{2\pi^2}} \; {\cal{D}}^j_{m_1m_2}(g)\,.
\end{eqnarray}

Regarding the reduction of the tensor product of two UIR's of $\mathrm{SU}(2)$, the following equivalent formulas, involving the so-called $3-j$ symbols (proportional to the Clebsch-Gordan coefficients), read:
\begin{eqnarray}
{\cal{D}}^j_{m_1m_2}(g) {\cal{D}}^{j^\prime}_{m_1^\prime m_2^\prime}(g) &=& \sum_{j^{\prime\prime} m_1^{\prime\prime} m_2^{\prime\prime}} (2j^{\prime\prime} +1)
\begin{pmatrix}
j & j^\prime & j^{\prime\prime} \\
m_1 & m_1^\prime & m_1^{\prime\prime}
\end{pmatrix}
\begin{pmatrix}
j & j^\prime & j^{\prime\prime} \\
m_2 & m_2^\prime & m_2^{\prime\prime}
\end{pmatrix}
\big( {\cal{D}}^{j^{\prime\prime}}_{m_1^{\prime\prime} m_2^{\prime\prime}}(g) \big)^\ast \nonumber\\
&=&
\sum_{j^{\prime\prime} m_1^{\prime\prime} m_2^{\prime\prime}} (2j^{\prime\prime} +1) (-1)^{m_1^{\prime\prime}-m_2^{\prime\prime}}
\begin{pmatrix}
j & j^\prime & j^{\prime\prime} \\
m_1 & m_1^\prime & -m_1^{\prime\prime}
\end{pmatrix}
\begin{pmatrix}
j & j^\prime & j^{\prime\prime} \\
m_2 & m_2^\prime & -m_2^{\prime\prime}
\end{pmatrix}
{\cal{D}}^{j^{\prime\prime}}_{m_1^{\prime\prime} m_2^{\prime\prime}}(g)\,,
\end{eqnarray}
\begin{eqnarray}
\int_{\mathrm{SU}(2)}{\cal{D}}^j_{m_1m_2}(g) \; {\cal{D}}^{j^\prime}_{m_1^\prime m_2^\prime}(g) \; {\cal{D}}^{j^{\prime\prime}}_{m_1^{\prime\prime} m_2^{\prime\prime}}(g) \; \mathrm{d}\mu(g) = 2\pi^2
\begin{pmatrix}
j & j^\prime & j^{\prime\prime} \\
m_1 & m_1^\prime & m_1^{\prime\prime}
\end{pmatrix}
\begin{pmatrix}
j & j^\prime & j^{\prime\prime} \\
m_2 & m_2^\prime & m_2^{\prime\prime}
\end{pmatrix},
\end{eqnarray}
where one of the multiple expressions of the $3-j$ symbols, in the convention that they are all real, is:
\begin{eqnarray}
\begin{pmatrix}
j & j^\prime & j^{\prime\prime} \\
m & m^\prime & m^{\prime\prime}
\end{pmatrix} = (-1)^{j - j^\prime - m^{\prime\prime}}
\sqrt{\frac{ (j + j^\prime - j^{\prime\prime})! \; (j - j^\prime + j^{\prime\prime})! \; (-j + j^\prime + j^{\prime\prime})! }{(1 + j + j^\prime + j^{\prime\prime})!}} \hspace{4cm}\nonumber\\
\hspace{1cm}\times \sum_t (-1)^t \frac{ \sqrt{(j+m)! \; (j-m)! \; (j^\prime+m^\prime)! \; (j^\prime-m^\prime)! \; (j^{\prime\prime}+m^{\prime\prime})! \; (j^{\prime\prime}-m^{\prime\prime})!} }{t! \; (j^\prime+m^\prime-t)! \; (j-m-t)! \; (j^{\prime\prime}-j^\prime+m+t)! \; (j^{\prime\prime}-j-m^\prime+t)! \; (j+j^\prime-j^{\prime\prime}-t)! }\,.
\end{eqnarray}
Again, the values of $t$ are those for which the arguments of the factorial functions remain nonnegative.

Finally, the relations between the hyperspherical harmonics ${\cal{Y}}_{Llm}(\textbf{u})$ (see Eq. (\ref{YLlm})) and ${\cal{D}}^j_{m_1m_2}(\textbf{u})$, for $\textbf{u} \in \mathrm{SU}(2)$ and $L=2j$, are given by:
\begin{eqnarray}
{\cal{Y}}_{Llm}(\textbf{u}) &=& \sqrt{\frac{L+1}{2\pi^2}} \;\; \mathrm{i}^l \sum_{m_1,m_2} \sqrt{2l+1} \; (-1)^{j-m_2}
\begin{pmatrix}
j & j & l \\
m_1 & -m_2 & m
\end{pmatrix}
{\cal{D}}^j_{m_1m_2}(\textbf{u})\,,\\
{\cal{D}}^j_{m_1m_2}(\textbf{u}) &=& \sqrt{\frac{2\pi^2}{2j+1}} \; \sum_{l,m} \sqrt{2l+1} \; (-1)^{L+l+2m_2}
\begin{pmatrix}
j & j & l \\
m_1 & -m_2 & m
\end{pmatrix}
{\cal{Y}}_{Llm}(\textbf{u})\,.
\end{eqnarray}

\setcounter{equation}{0} \section{Expansions of kernels on $\mathbb{S}^3 \times \mathbb{R}_+ \times \mathbb{S}^3$}\label{App kernel}
Considering Eqs. (\ref{G2}) and (\ref{G3}), an adaptation of the generating function for the Gegenbauer polynomials (\ref{G1}) leads to the following formula:
\begin{eqnarray}\label{B36}
|\textbf{z}_1 - \rho\textbf{z}_2|^{-2\lambda} &=& (1 + \rho^2 - 2\rho \textbf{z}_1 \cdot \textbf{z}_2)^{-\lambda}\nonumber\\
&=& \sum_{L\geqslant0} (L+1) \; \rho^L \; {\cal{P}}_L^\lambda(\rho^2) \; C_L^1(\textbf{z}_1 \cdot \textbf{z}_2)\nonumber\\
&=& 2\pi^2 \sum_{L\geqslant0,l,m} \rho^L \; {\cal{P}}_L^\lambda(\rho^2) \; {\cal{Y}}_{Llm}(\textbf{z}_1) {\cal{Y}}_{Llm}^\ast(\textbf{z}_2)\,, \;\;\;\;\;\;\; \textbf{z}_1, \textbf{z}_2 \in \mathbb{S}^3,
\end{eqnarray}
where:
\begin{eqnarray}
{\cal{P}}_L^\lambda(\rho^2) = \frac{1}{(L+1)!} \frac{\Gamma(\lambda+L)}{\Gamma(\lambda)} \; ^{}_2F^{}_1 (L+\lambda,\lambda-1;L+2;\rho^2)\,.
\end{eqnarray}
This formula is valid for $|\rho| < 1$ in the sense of functions and for $|\rho| = 1$ in the distribution sense. In the latter sense, the expansion (\ref{B36}) explicitly reduces to (see Eq. (\ref{limit})):
\begin{eqnarray}\label{kernel}
|\textbf{z}_1 - \textbf{z}_2|^{-2\lambda} &=& \sum_{L\geqslant0} \frac{\Gamma(L+\lambda) \; \Gamma(3-2\lambda)}{\Gamma(\lambda) \; \Gamma(2-\lambda) \; \Gamma(L-\lambda+3)} \; (L+1) \; C_L^1(\textbf{z}_1 \cdot \textbf{z}_2)\nonumber\\
&=& 2\pi^2 \sum_{L\geqslant0,l,m} \frac{\Gamma(L+\lambda) \; \Gamma(3-2\lambda)}{\Gamma(\lambda) \; \Gamma(2-\lambda) \; \Gamma(L-\lambda+3)} {\cal{Y}}_{Llm}(\textbf{z}_1) {\cal{Y}}_{Llm}^\ast(\textbf{z}_2)\,.
\end{eqnarray}
From the above identity, one can easily obtain Eq. (\ref{Kernel complementary}).

On the other hand, utilizing the identity $\Gamma(n+1) = n\Gamma(n)$, which implies that:
\begin{eqnarray}
\Gamma(3-2\lambda) &=& (2-2\lambda) \; \Gamma(2-2\lambda)\,,\nonumber\\
\Gamma(2-\lambda) &=& (1-\lambda) \; \left( (-1)^L \frac{\Gamma(\lambda+L)}{\Gamma(\lambda)}\right) \; \Gamma(1-\lambda-L)\,,
\end{eqnarray}
we get:
\begin{eqnarray}
|\textbf{z}_1 - \textbf{z}_2|^{-2\lambda} &=& 2\sum_{L=0}^\infty (-1)^L \frac{\Gamma(2-2\lambda)}{\Gamma(1-\lambda-L) \; \Gamma(L-\lambda+3)} \; (L+1) \; C_L^1(\textbf{z}_1 \cdot \textbf{z}_2)\nonumber\\
&=& 4\pi^2 \sum_{L=0}^\infty \sum_{l,m} (-1)^L \frac{\Gamma(2-2\lambda)}{\Gamma(1-\lambda-L) \; \Gamma(L-\lambda+3)} \; {\cal{Y}}_{Llm}(\textbf{z}_1) {\cal{Y}}_{Llm}^\ast(\textbf{z}_2)\,.
\end{eqnarray}
By taking the derivative of the above expression with respect to $\lambda$ and then letting $\lambda \mapsto 1-p$ ($p=1,2,...$), while we consider the Laurent expansion of the function $\psi(z) \equiv \frac{\mathrm{d} \Gamma(z) / \mathrm{d} z}{\Gamma(z)}$ near $z=-n$:
\begin{eqnarray}
\psi(z) = - \frac{1}{z+n} + \psi(n+1) + \sum_{s=2}^\infty A_s(z+n)^{-s}\,, \;\;\;\;\;\;\; A_s = (-1)^s\zeta(s) + \sum_{t=1}^n t^{-s}\,,
\end{eqnarray}
where $\zeta(s)$ is the Riemann Zeta Function (see, for instance, Ref. \cite{Magnus}), we obtain:
\begin{eqnarray}\label{hasan}
|\textbf{z}_1 - \textbf{z}_2|^{2(p-1)} \log |\textbf{z}_1 - \textbf{z}_2|^{-2} &=& 2(2p-1)! \; \Bigg( \sum_{L=0}^{p-1} (-1)^L \frac{c_{p,L}}{(p-1-L)! \; (p+1+L)!} \nonumber\\
&& - (-1)^p \sum_{L=p}^\infty \frac{(L-p)!}{(p+2+L)! \; (p+1+L)!} \Bigg) \; (L+1) \; C_L^1(\textbf{z}_1 \cdot \textbf{z}_2)\nonumber\\
&=& 4\pi^2 (2p-1)! \; \Bigg( \sum_{L=0}^{p-1} \sum_{l,m} (-1)^L \frac{c_{p,L}}{(p-1-L)! \; (p+1+L)!} \nonumber\\
&& - (-1)^p \sum_{L=p}^\infty \sum_{l,m} \frac{(L-p)!}{(p+2+L)! \; (p+1+L)!} \Bigg) \; {\cal{Y}}_{Llm}(\textbf{z}_1) {\cal{Y}}_{Llm}^\ast(\textbf{z}_2)\,,
\end{eqnarray}
where the coefficients $c_{p,L}$ are:
\begin{eqnarray}
\mbox{if}\;\; 0\leqslant L \leqslant p-2 \;\; \Rightarrow \;\; c_{p,L} = -\sum_{s=p-L}^{p+L+1} \frac{1}{s}\,, \;\;\;\;\;\;\; \mbox{and if}\;\; L = p-1 \;\; \Rightarrow \;\; c_{p,L} = \frac{1}{2p} -\sum_{s=1}^{2p-1} \frac{1}{s}\,.
\end{eqnarray}
The identity (\ref{hasan}) serves in subsection \ref{Subsec discrete dS4}.

\setcounter{equation}{0} \section{DS$_4$ complex Lie algebra $\mathfrak{sp}(2,2)^{\mathbb{(C)}}$}\label{App Lie algebra B2}
In this appendix, we present the irreducible (nonunitary!) finite-dimensional representations of the dS$_4$ group, which entail to deal with the complex Lie algebra $\mathfrak{sp}(2,2)^{\mathbb{(C)}}$. As a complex Lie algebra, $\mathfrak{sp}(2,2)^{\mathbb{(C)}}$ is realized by \emph{complexification} of its real counterpart $\mathfrak{sp}(2,2)$ (see Eq. (\ref{algebra dS4})). [The latter is a real, ten-dimensional vector space equipped with an antisymmetric Lie bracket $[.,.]$ verifying the Jacobi identity and linearity.] By complexification of $\mathfrak{sp}(2,2)$, we mean complexifying it as a vector space (i.e., extending its parameter space to complex numbers), and then, extending the corresponding Lie bracket by linearity. Clearly, the complex dimension of $\mathfrak{sp}(2,2)^{\mathbb{(C)}}$ is still $10$, but its real dimension is $2\times 10$. Proceeding as above, while we have in mind Eq. (\ref{algebra dS4}), the generic element of $\mathfrak{sp}(2,2)^{\mathbb{(C)}}$ is given by:
\begin{eqnarray}\label{complsp}
\mathfrak{sp}(2,2)^{\mathbb{(C)}} \ni \underline{X}^{\mathbb{(C)}} = \begin{pmatrix} \vec{\textbf{n}}_{}^{(l,\mathbb{C})} & \textbf{d}_{}^{\mathbb{(C)}} \\ (\textbf{d}_{}^{\mathbb{(C)}})^{\scriptscriptstyle\bigstar} & \vec{\textbf{n}}_{}^{(r,\mathbb{C})} \end{pmatrix},
\end{eqnarray}
where $\textbf{d}_{}^{\mathbb{(C)}}, \vec{\textbf{n}}_{}^{(l,\mathbb{C})}, \vec{\textbf{n}}_{}^{(r,\mathbb{C})} \in \mathbb{H}_{}^{\mathbb{(C})}$ ($\vec{\textbf{n}}_{}^{(l,\mathbb{C})}$ and $\vec{\textbf{n}}_{}^{(r,\mathbb{C})}$ are indeed (complex) pure vector quaternions).

Note that: (i) From now on, for the sake of simplicity, we drop the indices `${\mathbb{C}}$' from the elements of $\mathfrak{sp}(2,2)^{\mathbb{(C)}}$ and $\mathbb{H}^{\mathbb{(C)}}$. (ii) For the algebra of the complex quaternions $\mathbb{H}^{\mathbb{(C)}}$ (which can be realized by its isomorphism to $2\times 2$ matrices over $\mathbb{C}$), we have:\footnote{For the sake of comparison, it is perhaps worthwhile noting that, in the literature, the quaternionic and complex conjugates of a complex quaternion, say, $\textbf{z}$, are usually denoted by $\widetilde{\textbf{z}}$ and $\overline{\textbf{z}}$; the Hermitain conjugate then is naturally denoted by $\overline{\widetilde{\textbf{z}}}$.}
\begin{eqnarray}
\textbf{z},{\textbf{z}^\prime} \in \mathbb{H}^{\mathbb{(C)}} \;\; &\Rightarrow& \;\; \textbf{z}{\textbf{z}^\prime} = (z^4 z^{\prime 4} - \vec{z}\cdot\vec{z}^\prime, z^4\vec{z}^\prime + z^{\prime 4} \vec{z} + \vec{z}\times\vec{z}^\prime)\,,\nonumber\\
\;\; &\Rightarrow& \;\; \textbf{z}^{\scriptscriptstyle\bigstar} = (z^4,-\vec{z})\,,\;\;\;\;\;\;\; \mbox{quaternionic conjugate}\,,\nonumber\\
\;\; &\Rightarrow& \;\; \textbf{z}^\ast = (z^{4\ast},{\vec{z}}^{\;\ast})\,,\;\;\;\;\;\;\; \mbox{complex conjugate}\,,\nonumber\\
\;\; &\Rightarrow& \;\; \big(\textbf{z}^{\scriptscriptstyle\bigstar}\big)^\ast = (z^{4\ast},-{\vec{z}}^{\;\ast})\,,\;\;\;\;\;\;\; \mbox{Hermitain conjugate}\,,\nonumber\\
\;\; &\Rightarrow& \;\; |\textbf{z}|^2 = \textbf{zz}^{\scriptscriptstyle\bigstar} = (z^4)^2 + (z^1)^2 + (z^2)^2 + (z^3)^2\,, \nonumber\\
\;\; &\Rightarrow& \;\; \textbf{z}^{-1} = \textbf{z}^{\scriptscriptstyle\bigstar} /|\textbf{z}|^2\,, \;\;\;\;\;\;\; (\textbf{z}\neq 0)\,.\nonumber
\end{eqnarray}
It can be easily checked that, contrary to the real quaternion algebra $\mathbb{H}$ (see appendix \ref{App quat}), the product of two complex quaternions can be zero, while neither is zero.

\subsection{Cartan-Weyl basis}
We begin with the following generators of $\mathfrak{sp}(2,2)^{\mathbb{(C)}}$:
\begin{eqnarray}\label{basis cartan}
\underline{H}_1 &\equiv &  \frac{1}{2}\begin{pmatrix} \textbf{0} & \textbf{1} \\ \textbf{1} & \textbf{0} \end{pmatrix},\nonumber\\
\underline{H}_2 &\equiv & \frac{\mathrm{i}}{2} \begin{pmatrix} {\textbf{e}}^{}_3 & \textbf{0} \\ \textbf{0} & {\textbf{e}}^{}_3 \end{pmatrix},\nonumber\\
\underline{X}_{1} &\equiv & \frac{1}{2}\begin{pmatrix} {\textbf{e}}^{}_3 & -{\textbf{e}}^{}_3 \\ {\textbf{e}}^{}_3 & -{\textbf{e}}^{}_3 \end{pmatrix},\nonumber\\
\underline{X}_{2} &\equiv & \frac{1}{4} \begin{pmatrix} {\textbf{e}}^{}_1+\mathrm{i}{\textbf{e}}^{}_2 & {\textbf{e}}^{}_1+ \mathrm{i} {\textbf{e}}^{}_2 \\ -{\textbf{e}}^{}_1- \mathrm{i} {\textbf{e}}^{}_2 & -{\textbf{e}}^{}_1- \mathrm{i} {\textbf{e}}^{}_2 \end{pmatrix},\nonumber\\
\underline{X}_{3} &\equiv & \frac{1}{2}\begin{pmatrix} {\textbf{e}}^{}_1+ \mathrm{i} {\textbf{e}}^{}_2 & \textbf{0} \\ \textbf{0} & {\textbf{e}}^{}_1+ \mathrm{i} {\textbf{e}}^{}_2 \end{pmatrix},\nonumber\\
\underline{X}_{4} &\equiv & \frac{1}{4}\begin{pmatrix} {\textbf{e}}^{}_1+ \mathrm{i} {\textbf{e}}^{}_2 & -{\textbf{e}}^{}_1- \mathrm{i} {\textbf{e}}^{}_2 \\ {\textbf{e}}^{}_1+ \mathrm{i} {\textbf{e}}^{}_2 & -{\textbf{e}}^{}_1- \mathrm{i} {\textbf{e}}^{}_2 \end{pmatrix},\nonumber\\
\underline{X}_{5} &\equiv & \frac{1}{2}\begin{pmatrix} {\textbf{e}}^{}_3 & {\textbf{e}}^{}_3 \\ -{\textbf{e}}^{}_3 & -{\textbf{e}}^{}_3 \end{pmatrix},\nonumber\\
\underline{X}_{6} &\equiv & \frac{1}{4} \begin{pmatrix} {\textbf{e}}^{}_1- \mathrm{i} {\textbf{e}}^{}_2 & -{\textbf{e}}^{}_1+ \mathrm{i} {\textbf{e}}^{}_2 \\ {\textbf{e}}^{}_1- \mathrm{i} {\textbf{e}}^{}_2 & -{\textbf{e}}^{}_1+ \mathrm{i} {\textbf{e}}^{}_2 \end{pmatrix},\nonumber\\
\underline{X}_{7} &\equiv & \frac{1}{2}\begin{pmatrix} {\textbf{e}}^{}_1- \mathrm{i} {\textbf{e}}^{}_2 & \textbf{0} \\ \textbf{0} & {\textbf{e}}^{}_1- \mathrm{i} {\textbf{e}}^{}_2 \end{pmatrix},\nonumber\\
\underline{X}_{8} &\equiv & \frac{1}{4}\begin{pmatrix} {\textbf{e}}^{}_1- \mathrm{i} {\textbf{e}}^{}_2 & {\textbf{e}}^{}_1- \mathrm{i} {\textbf{e}}^{}_2\\ - {\textbf{e}}^{}_1 + \mathrm{i} {\textbf{e}}^{}_2 & - {\textbf{e}}^{}_1 + \mathrm{i} {\textbf{e}}^{}_2 \end{pmatrix},
\end{eqnarray}
where here $\big\{ \textbf{1}, \textbf{e}^{}_k \big\}$, with $k=1,2,3$, stands for the complex quaternionic basis. Computing the commutation relations between the above generators, it is quite straightforward to check that the Lie algebra $\mathfrak{sp}(2,2)^{\mathbb{(C)}}$ is \emph{simple}\footnote{A simple Lie algebra, by definition, is a Lie algebra that is non-\emph{abelian} and contains no nonzero \emph{proper ideals}. A direct sum of simple Lie algebras is called a semi-simple Lie algebra (in this regard, see also footnote \ref{foot108}). [Note that: (i) An algebra in which all members commute is called abelian. (ii) An ideal is a special kind of subalgebra. Let $\mathfrak{g}$ be a Lie algebra. If $\mathfrak{g}^{}_{i} \subset \mathfrak{g}$ is an ideal, and $X \in \mathfrak{g}^{}_{i}$ and $Y$ is any element of $\mathfrak{g}$, then $[X,Y] \in \mathfrak{g}_{i}$. Finally, a proper ideal is an ideal that is neither equal to $\{0\}$ nor to $\mathfrak{g}$ itself, which are two obvious ideals of $\mathfrak{g}$.] Moreover, it worth noting that a finite-dimensional simple complex Lie algebra is isomorphic to either of the following: $\mathfrak{sl}(n)^{\mathbb{(C)}}, \mathfrak{so}(n)^{\mathbb{(C)}},\mathfrak{sp}(2n)^{\mathbb{(C)}}$ (classical Lie algebras) or one of the five exceptional Lie algebras. In our case, we have $\mathfrak{sp}(4)^{\mathbb{(C)}} \sim \mathfrak{sp}(2,2)^{\mathbb{(C)}}$.}.

Here, it must be underlined that the above generators have been chosen in such a way that one subset of them, constituted by $\mathfrak{h} = \mbox{span}_{\mathbb{C}}\{ \underline{H}_1, \underline{H}_2 \}$, generates the maximal abelian subalgebra of $\mathfrak{sp}(2,2)^{\mathbb{(C)}}$ ($[\underline{H}_1,\underline{H}_2]=0$), while the other elements of the basis can be viewed as eigenvectors of the \emph{adjoint representation}\footnote{Recall that with each element of a given Lie algebra $\mathfrak{g}$ one can associate a linear transformation $\mbox{ad} : \mathfrak{g} \rightarrow \texttt{End} \; \mathfrak{g}$ (endomorphisms of $\mathfrak{g}$) defined, for any ${X},{Y} \in \mathfrak{g}$, by $\mbox{ad}^{}_{X}(Y)=[{X},{Y}]$; following the arguments presented in subsection \ref{Subsec coadjoint dS2-int}, the latter identity can be easily achieved by putting $g=\exp(rX)$, with $r \in (-\delta,\delta)$ and $\delta> 0$, in the adjoint action (\ref{Ad_g mat}) and taking the derivatives at $r=0$. Utilizing the Jacobi identity, it is a simple task to show that this linear transformation preserves the commutation relations of the algebra, namely, if $[{X}, {Y}] = {Z}$ then we have $[\mbox{ad}^{}_{X}, \mbox{ad}^{}_{Y}] = \mbox{ad}^{}_{Z}$. In this sense, the operator $\mbox{ad}^{}_{X}$ provides a representation, called adjoint representation, of ${X}$, for all ${X} \in \mathfrak{g}$. In the context of adjoint representations, the Lie algebra itself serves as the vector space on which the representations are defined; let $\mathfrak{g}$ be of $n$ dimensions, then these representations would be of $n$ dimensions, namely, if we set up a particular basis, the operators $\mbox{ad}^{}_{X}$ can be shown as $n\times n$ matrices.} of either $\underline{H}_1$ or $\underline{H}_2$, i.e., the members of the aforementioned subalgebra $\mathfrak{h}$ (called Cartan subalgebra). More precisely, for $\underline{H} \equiv a\underline{H}_1 + b\underline{H}_2 \in \mathfrak{h}$, with $a,b \in \mathbb{C}$, we have:
\begin{eqnarray}
\mbox{ad}^{}_{\underline{H}} (\underline{X}_{1}) &=& [\underline{H},\underline{X}_{1}] = a\underline{X}_{1} \equiv \alpha_{1} \underline{X}_{1}\,,\nonumber\\
\mbox{ad}^{}_{\underline{H}} (\underline{X}_{2}) &=& [\underline{H},\underline{X}_{2}] = (-a+b)\underline{X}_{2} \equiv \alpha_{2} \underline{X}_{2}\,, \nonumber\\
\mbox{ad}^{}_{\underline{H}} (\underline{X}_{3}) &=& [\underline{H},\underline{X}_{3}] = b\underline{X}_{3} \equiv \alpha_{3} \underline{X}_{3}\,, \nonumber\\
\mbox{ad}^{}_{\underline{H}} (\underline{X}_{4}) &=& [\underline{H},\underline{X}_{4}] = (a+b)\underline{X}_{4} \equiv \alpha_{4} \underline{X}_{4}\,, \nonumber\\
\mbox{ad}^{}_{\underline{H}} (\underline{X}_{5}) &=& [\underline{H},\underline{X}_{5}] = -a\underline{X}_{5} \equiv \alpha_{5} \underline{X}_{5}\,, \nonumber\\
\mbox{ad}^{}_{\underline{H}} (\underline{X}_{6}) &=& [\underline{H},\underline{X}_{6}] = (a-b)\underline{X}_{6} \equiv \alpha_{6} \underline{X}_{6}\,, \nonumber\\
\mbox{ad}^{}_{\underline{H}} (\underline{X}_{7}) &=& [\underline{H},\underline{X}_{7}] = -b\underline{X}_{7} \equiv \alpha_{7} \underline{X}_{7}\,, \nonumber\\
\mbox{ad}^{}_{\underline{H}} (\underline{X}_{8}) &=& [\underline{H},\underline{X}_{8}] = (-a-b)\underline{X}_{8} \equiv \alpha_{8} \underline{X}_{8}\,. \nonumber
\end{eqnarray}
With respect to the terminology of construction of the \emph{Cartan-Weyl basis} for semi-simple Lie algebras (see, for instance, Refs. \cite{Cahn,Fuchs}), the eigenvalues $\alpha_{i}$ ($i=1,\;...\;,8$), strictly speaking, $\alpha_{i}(\underline{H})$, are called \emph{roots} of the basis $\underline{X}_{i}$; $\alpha_{i}$'s are actually some complex numbers which depend linearly on $\underline{H} \in \mathfrak{h}$, in this sense that they are linear functions $\mathfrak{h}\rightarrow \mathbb{C}$, i.e., elements of the vector space ${\mathfrak{h}}_{}^{\circledast}$ dual to $\mathfrak{h}$. Correspondingly, the generators $\underline{X}_{i}$ are called \emph{root vectors} of $\mathfrak{sp}(2,2)^{\mathbb{(C)}}$. The set of all roots of $\mathfrak{sp}(2,2)^{\mathbb{(C)}}$, denoted here by $\Delta = \big\{ \alpha_i \; ;\; i=1,2,\;...\;,8 \big\}$, is called \emph{root system} of $\mathfrak{sp}(2,2)^{\mathbb{(C)}}$. Accordingly, the given basis (\ref{basis cartan}) can be summarized as:
\begin{eqnarray}\label{basis cartan'}
\Big\{ \underline{H}_i \; ;\; i=1,2\Big\} \bigcup \Big\{ \underline{X}_\alpha \; ;\; \alpha\in\Delta \Big\}\,,
\end{eqnarray}
which is called Cartan-Weyl basis of $\mathfrak{sp}(2,2)^{\mathbb{(C)}}$. Note that, from a physical point of view, the dimension of the Cartan subalgebra, determining the rank of the algebra (for instance, in our case, `the rank of $\mathfrak{sp}(2,2)^{\mathbb{(C)}}$' $ = \mbox{dim}(\mathfrak{h} ) = 2$), reveals the maximum number of quantum numbers which can be used to label (at least partially) the states of a physical system possessing that symmetry algebra \cite{Fuchs}.

Considering the Jacobi identity, for all $\underline{H} \in \mathfrak{h}$ and root vectors $\underline{X}_{\alpha_i}$ and $\underline{X}_{\alpha_j}$ ($\alpha_i, \alpha_j \in \Delta$), we have:
\begin{eqnarray}
\big[H , [\underline{X}_{\alpha_i}, \underline{X}_{\alpha_j}]\big] = \big(\alpha_i(\underline{H}) + \alpha_j(\underline{H})\big)[\underline{X}_{\alpha_i}, \underline{X}_{\alpha_j}]\,,
\end{eqnarray}
based upon which, the following results hold:
\begin{itemize}
\item{If $\alpha_i + \alpha_j \in \Delta$, then $[\underline{X}_{\alpha_i}, \underline{X}_{\alpha_j}] = N_{\alpha_i \alpha_j} \underline{X}_{\alpha_i+\alpha_j}$, with $N_{\alpha_i \alpha_j}\in\mathbb{C}$.}
\item{If $\alpha_i + \alpha_j =0$, then $[\underline{X}_{\alpha_i}, \underline{X}_{-\alpha_i}] = \underline{H}_{\alpha_i} \in \mathfrak{h}$.}
\item{If $\alpha_i + \alpha_j  \neq 0$ and $\alpha_i + \alpha_j \notin \Delta$, then $[\underline{X}_{\alpha_i}, \underline{X}_{\alpha_j}] = 0$.}
\end{itemize}

\subsection{Geometrical picture}
A fundamental step in the analysis of semi-simple Lie algebras is to establish a geometrical picture of the algebra, developed in terms of its root system. In this subsection, following the general instruction given for semi-simple Lie algebras (see, for instance, Refs. \cite{Fuchs,Cahn}), we aim to briefly introduce such a geometrical picture for the $\mathfrak{sp}(2,2)^{\mathbb{(C)}}$ algebra. Accordingly, we first need to define something resembling a scalar product for the elements of the Lie algebra itself.

In the context of our study, admitting the Cartan-Weyl basis (\ref{basis cartan'}), there is a scalar product associated with the algebra, but it is not defined on the algebra itself, but rather on the space containing the roots. Technically, this scalar product is issued from the \emph{Killing form}\footnote{\label{foot108}Recall that, for a given Lie algebra $\mathfrak{g}$, one can define $ K({X},{Y}) = \mbox{tr} \big(\mbox{ad}_X \mbox{ad}_Y\big)$, with ${X},{Y} \in \mathfrak{g}$, called Killing form of $\mathfrak{g}$, which is bilinear and symmetric, $K({X},{Y})=K({Y},{X})$. [Technically, to evaluate $ K({X},{Y})$, first of all one needs to choose a basis for $\mathfrak{g}$, say $X_1,X_2,...$ . Then, one can calculate for each $X_j$, the quantity $\big[X,[Y,X_j]\big]$ and express the result in terms of the $X_i$'s. The coefficient of $X_j$ would be the contribution to the trace. We recall that the trace is independent of the choice of basis.] Here, it is worth noting that there is an intimate relationship between the Killing form and the notion of semi-simpleness of Lie algebras. Actually, a Lie algebra is called semi-simple if and only if the associated Killing form is nondegenerate (Cartan's criterion), in other words, if and only if there exists no element $X\neq 0$ in the algebra obeying $K(X,Y)=0$, for all $Y$ in the algebra.} of $\mathfrak{sp}(2,2)^{\mathbb{(C)}}$ in the following manner. Since the associated Killing form is nondegenerate ($\mathfrak{sp}(2,2)^{\mathbb{(C)}}$ is simple), one can associate with any root $\alpha \in \Delta$ an element $\underline{H}_\alpha \in \mathfrak{h}$, up to a normalization constant, such that:
\begin{eqnarray}\label{Killingfo}
\alpha(\underline{H}) = K(\underline{H}_\alpha, \underline{H})\,,
\end{eqnarray}
for every $\underline{H}\in \mathfrak{h}$. Now, with the help of the elements $\underline{H}_\alpha$, obtained through the above formula, one can define a (nondegenerate) inner product on the space containing the roots:
\begin{eqnarray}\label{equivinaK}
\langle \alpha_i ; \alpha_j \rangle \equiv K(\underline{H}_{\alpha_i} , \underline{H}_{\alpha_j})\,,
\end{eqnarray}
for all roots $\alpha_i, \alpha_j \in \Delta$.

For reasons that will be clarified soon, we here deal with the elements $\underline{H}_{\alpha_1},\underline{H}_{\alpha_2} \in \mathfrak{h}$, respectively, associated with the roots $\alpha_1(\underline{H}) = \alpha_1(a \underline{H}_1+b \underline{H}_2) = a$ and $\alpha_2 (\underline{H}) = \alpha_2(a \underline{H}_1+b \underline{H}_2) = -a+b$, with $a,b \in \mathbb{C}$. Trivially, since $\underline{H}_{\alpha_1}$ and $\underline{H}_{\alpha_2}$ lie in $\mathfrak{h}$, they can be expressed as $\underline{H}_{\alpha_i} = c_i\underline{H}_1 + d_i\underline{H}_2$, with $c_i,d_i \in \mathbb{C}$. Therefore, we have:
\begin{eqnarray}
\alpha_1(a \underline{H}_1) = a \;\;\;\;\; = \;\;\;\;\; &K(\underline{H}_{\alpha_1},a \underline{H}_1) = 6a c_1& \;\;\;\;\; \Rightarrow \;\;\;\;\; c_1 = \textstyle\frac{1}{6}\,,\nonumber\\
\alpha_1(b \underline{H}_2) = 0 \;\;\;\;\; = \;\;\;\;\; &K(\underline{H}_{\alpha_1},b \underline{H}_2) = 6b d_1& \;\;\;\;\; \Rightarrow \;\;\;\;\; d_1=0\,,
\end{eqnarray}
and:
\begin{eqnarray}
\alpha_2(a \underline{H}_1) = -a \;\;\;\;\; = \;\;\;\;\; &K(\underline{H}_{\alpha_2},a \underline{H}_1) = 6a c_2& \;\;\;\;\; \Rightarrow \;\;\;\;\; c_2 = -\textstyle\frac{1}{6}\,,\nonumber\\
\alpha_2(b \underline{H}_2) = b \;\;\;\;\; = \;\;\;\;\; &K(\underline{H}_{\alpha_2},b \underline{H}_2) = 6b d_2& \;\;\;\;\; \Rightarrow \;\;\;\;\; d_2 = \textstyle\frac{1}{6}\,,
\end{eqnarray}
Then, the elements of $\mathfrak{h}$ corresponding to the roots $\alpha_1$ and $\alpha_2$ respectively read:
\begin{eqnarray}
\underline{H}_{\alpha_1} = \textstyle\frac{1}{6} \underline{H}_1\,, \;\;\;\;\;\;\; \mbox{and} \;\;\;\;\;\;\; \underline{H}_{\alpha_2} = -\textstyle\frac{1}{6} \underline{H}_1+\frac{1}{6} \underline{H}_2\,.
\end{eqnarray}
Accordingly, having Eq. (\ref{equivinaK}) in mind, we obtain:
\begin{eqnarray}\label{inroots13}
\langle \alpha_1 ; \alpha_1 \rangle = \textstyle\frac{1}{6}\,, \;\;\;\;\;\;\; \langle \alpha_1 ; \alpha_2 \rangle = -\frac{1}{6}\,, \;\;\;\;\;\;\; \langle \alpha_2 ; \alpha_2 \rangle = \frac{1}{3}\,.
\end{eqnarray}

We now turn back to the root system. The set of roots $\Delta$ (given above) can be split into a subset $\Delta^{}_+ = \big\{ \alpha \in \Delta \;;\; \alpha>0 \big\} = \big\{ \alpha_1,\alpha_2,\alpha_3,\alpha_4\big \}$ of \emph{positive roots} and, since except for $-\alpha$ no other multiple of $\alpha \in \Delta$ is a root, the corresponding subset of \emph{negative roots} $\Delta^{}_- = \big\{ -\alpha \;;\; \alpha \in \Delta^{}_+ \big\} = \big\{ \alpha_5,\alpha_6,\alpha_7,\alpha_8 \big\}$. Hence:
\begin{eqnarray}
\Big\{\underline{X}_\alpha \; ; \; \alpha \in \Delta \Big\} = \Big\{\underline{X}_\alpha \; ; \; \alpha>0 \Big\} \bigcup \Big\{\underline{X}_{-\alpha} \; ; \; \alpha>0 \Big\}\,.
\end{eqnarray}
By definition, a positive root which cannot be obtained as a linear combination of other positive roots with positive coefficients is called a \emph{simple root}. Some important properties of simple roots are: (i) The number of simple roots in a given semi-simple Lie algebra is exactly the rank of the algebra. (ii) The difference of two simple roots is not a root at all. (iii) Simple roots are linearly independent and provide a basis for the root space, which can span the whole root space. In this sense and with respect to our chosen root system, the simple roots are $\big\{ \alpha_1, \alpha_2 \big\}$, for which one can easily see that $\alpha_1 - \alpha_2 \notin \Delta$, and that the other roots can be written in terms of them as $k_\alpha^1 \alpha_1 + k_\alpha^2 \alpha_2 \equiv (k_\alpha^1, k_\alpha^2)$:
\begin{eqnarray}
\alpha_1 = (1,0)\,, \;\;\;\; && \;\;\;\; \alpha_5 = (-1,0)\,,\nonumber\\
\alpha_2 = (0,1)\,, \;\;\;\; && \;\;\;\; \alpha_6 = (0,-1)\,, \nonumber\\
\alpha_3 = (1,1)\,, \;\;\;\; && \;\;\;\; \alpha_7 = (-1,-1)\,, \nonumber\\
\alpha_4 = (2,1)\,, \;\;\;\; && \;\;\;\; \alpha_8 = (-2,-1)\,,\nonumber
\end{eqnarray}
where, as is evident, every positive root has been written as a positive sum of the simple roots. Finally, from the given simple roots, we can form the \emph{Cartan matrix}\footnote{Given a semi-simple Lie algebra, the Cartan matrix, summarizing the structure of the algebra entirely, is constructed over the simple roots by:
\begin{eqnarray}
A_{ij} = 2\frac{\langle \alpha_i ; \alpha_j\rangle}{\langle \alpha_j ; \alpha_j\rangle}\,,\nonumber
\end{eqnarray}
where $\alpha_i$ ($i=1,\;...\;,r$) are the simple roots ($r$ being the rank of the algebra). Clearly, the diagonal elements of this matrix are all equal to two. This matrix is not necessarily symmetric, but if $A_{ij}\neq 0$, then $A_{ji}\neq 0$. Since the scalar product of two different simple roots is nonpositive, the off-diagonal elements can be only $0,-1,-2,$ and $-3$. See more details, in Refs. \cite{Fuchs,Cahn}.} of $\mathfrak{sp}(2,2)^{\mathbb{(C)}}$:
\begin{eqnarray}
A = \begin{pmatrix} 2 & -1 \\ -2 & 2 \end{pmatrix}.
\end{eqnarray}

\subsection{Irreducible representations}
In this subsection, again following the instruction given in Refs. \cite{Fuchs,Cahn}, we are going to present a finite-dimensional irreducible representation of $\mathfrak{sp}(2,2)^{\mathbb{(C)}}$, denoted here by $T_\texttt{ir}$, that is, a mapping of the elements of $\mathfrak{sp}(2,2)^{\mathbb{(C)}}$ into linear operators:
\begin{eqnarray}
\underline{X}_\alpha \;\mapsto\; T_\texttt{ir}(\underline{X}_\alpha)\,,\;\;\;\;\; \alpha \in \Delta\,, \;\;\;\;\;\;\; \mbox{and} \;\;\;\;\;\;\; \underline{H}_i \mapsto T_\texttt{ir}(\underline{H}_i)\,,\;\;\;\;\; i=1,2\,,\nonumber
\end{eqnarray}
which preserve the commutation relations of the Lie algebra; $T_\texttt{ir}\big( [\underline{X},\underline{Y}] \big) = [T_\texttt{ir}(\underline{X}),T_\texttt{ir}(\underline{Y})]$, for all $\underline{X},\underline{Y} \in \mathfrak{sp}(2,2)^{\mathbb{(C)}}$.

Let $V$ denote the vector space in which the linear operators $T_\texttt{ir}(\underline{X}_\alpha)$ and $T_\texttt{ir}(\underline{H}_i)$ act. Since the $\underline{H}_i$'s commute, so $T_\texttt{ir}(\underline{H}_i)$'s do, we can select a basis for $V$ in such a way that the $T_\texttt{ir}(\underline{H_i})$'s are diagonal simultaneously. Therefore, we can write:
\begin{eqnarray}
T_\texttt{ir}(H)\ket{\mu} = \mu(H)\ket{\mu}\,,
\end{eqnarray}
where $\mu$, called a \emph{weight}, is a member of the dual space ${\mathfrak{h}}_{}^{\circledast}$ just as the roots are, and $\ket{\mu}\in V$ is called a \emph{weight vector}. The action of $T_\texttt{ir}(\underline{X}_\alpha)$, with $\alpha \in \Delta^{}_+$, on $\ket{\mu}$ would be a weight vector with weight $\mu+\alpha$ unless $T_\texttt{ir}(\underline{X}_\alpha)\ket{\mu} = 0$. This point can be easily seen through the following identity:
\begin{eqnarray}
T_\texttt{ir}(\underline{H})T_\texttt{ir}(\underline{X}_\alpha)\ket{\mu} &=& \Big(T_\texttt{ir}(\underline{X}_\alpha)T_\texttt{ir}(\underline{H}) + \alpha(\underline{H})T_\texttt{ir}(\underline{X}_\alpha)\Big) \ket{\mu} \nonumber\\
&=& \Big(\mu(\underline{H}) + \alpha(\underline{H})\Big) T_\texttt{ir}(\underline{X}_\alpha) \ket{\mu}\,,\nonumber
\end{eqnarray}
where we have used the identity $[\underline{H},\underline{X}_\alpha] = \alpha(\underline{H})\underline{X}_\alpha$, for all $\underline{H} \in \mathfrak{h}$ and $\alpha \in \Delta^{}_+$, along with the fact that the linear operators $T_\texttt{ir}$ preserve the commutation relations of the Lie algebra. Accordingly, one can refer to $T_\texttt{ir}(\underline{X}_\alpha)$'s as raising operators, and correspondingly, to $T_\texttt{ir}(\underline{X}_{-\alpha})$'s as lowering operators.

A finite-dimensional irreducible representation must possess a highest weight, denoted here by $\Lambda$, such that:
\begin{eqnarray}
T_\texttt{ir}(\underline{X}_\alpha) \ket{\Lambda} = 0\,, \;\;\;\;\;\;\; \forall \alpha\in\Delta^{}_+\,.
\end{eqnarray}
Considering the basis of simple roots $\big\{ \alpha_1, \alpha_2 \big\}$, the highest weight $\Lambda$ is specified in terms of two nonnegative integers, i.e., the so-called ``Dynkin coefficients":
\begin{eqnarray}\label{Dynkin coe}
n_1 = 2\frac{\langle \Lambda ; \alpha_1 \rangle}{\langle \alpha_1 ; \alpha_1 \rangle}\,,\;\;\;\;\;\;\; \mbox{and} \;\;\;\;\;\;\; n_2 = 2\frac{\langle \Lambda ; \alpha_2 \rangle}{\langle \alpha_2 ; \alpha_2 \rangle}\,.
\end{eqnarray}
Having the coefficients $n_1$ and $n_2$ for the highest weight and with respect to the ``Dynkin diagrams" of the algebra (see, for instance, Refs. \cite{Fuchs,Cahn}), it is easy to determine the full set of weights (in terms of their Dynkin coefficients) in the irreducible representation.

Dimension of this representation is given by the Weyl dimension formula \cite{Fuchs,Cahn}:
\begin{eqnarray}
\mbox{dim}^{}_V = \prod_{\alpha >0} \frac{\sum_{i=1,2} \; k_\alpha^i (n_i+1)\langle \alpha_i ; \alpha_i \rangle}{\sum_{i=1,2} \; k_\alpha^i \langle \alpha_i ; \alpha_i \rangle} = \frac{1}{6}(n_1+1)(n_2+1)(n_1+n_2+2)(n_1+2n_2+3)\,.
\end{eqnarray}
Recall that any positive root $\alpha > 0$ can be written in terms of the simple roots $\alpha_1$ and $\alpha_2$ as $\sum_{i=1,2} k_\alpha^i \alpha_i$; $\alpha_1 = (1,0)$, $\alpha_2= (0,1)$, $\alpha_3 = (1,1)$ and $\alpha_4 = (2,1)$.

Finally, the eigenvalues of the corresponding (quadratic) Casimir operator $Q^{(1)}$ can be found by considering its action on the highest weight vector $\ket{\Lambda}$ \cite{Cahn}:
\begin{eqnarray}
Q^{(1)} \ket{\Lambda} = \langle \Lambda ; \Lambda + 2\delta \rangle^{}_2 \ket{\Lambda}\,,
\end{eqnarray}
where, according to the Dynkin coefficients given in (\ref{Dynkin coe}), the highest weight $\Lambda$ is given by $\Lambda = (n_1 + n_2 , \frac{1}{2}n_1+n_2)$, and $\delta = \frac{1}{2}\sum_{\alpha>0}\alpha = (2 , \frac{3}{2})$. Moreover, we should point out that it is traditional to consider a scalar product which gives the \emph{highest root}\footnote{The height of a root $\alpha$ is determined by the sum $\sum_{i=1}^r k_{\alpha}^i$ of the components of the root $\alpha = \sum_{i=1}^r k_{\alpha}^i\alpha_i$ in the basis of simple roots ($r$ being the rank of the algebra). In this regard, the highest root is a root that its height is larger than that of any other root.} a length squared equal to $2$. We have denoted this scalar product in the above equation by $\langle \; ; \; \rangle^{}_2$. In our case, for the highest root determined by $\alpha_4 = (2 , 1)$, we have $\langle \alpha_4 ; \alpha_4 \rangle = 1/3$, and therefore, the scalar product $\langle \; ; \; \rangle^{}_2$ differs from $\langle \; ; \; \rangle$ by a normalization factor equal to $6$. On this basis, one can easily show that:
\begin{eqnarray}\label{speivoCa}
\langle \Lambda ; \Lambda + 2\delta \rangle^{}_2 = \frac{1}{2} \big( n_1^2 + 2n_2^2 + 2n_1n_2 + 4n_1 +6n_2 \big)\,.
\end{eqnarray}

Comparing the Casimir eigenvalues given in Eq. (\ref{speivoCa}) with those appeared in the context of the dS$_4$ UIR's (see section \ref{Sec Dixmier}), up to their signatures,\footnote{The point to be noticed here is that the signature of the Casimir eigenvalues in the above context varies with respect to the chosen root system. On the other hand, in the context of the dS$_4$ UIR's, this signature depends on the signature of the chosen metric. In this sense, we here merely compare the respective Casimir eigenvalues up to their signatures.} reveals that for the possible solutions:
\begin{eqnarray}\label{pqn121}
n_1 = -2q\,, \;\;\;\;\;\;\; n_2 = p+q-1\,,
\end{eqnarray}
\begin{eqnarray}\label{pqn122}
n_1 = 2q-2\,, \;\;\;\;\;\;\; n_2 = p-q\,,
\end{eqnarray}
\begin{eqnarray}\label{pqn123}
n_1 = q-1\,, \;\;\;\;\;\;\; n_2 = -2p-2\,,
\end{eqnarray}
\begin{eqnarray}\label{pqn124}
n_1 = -2p-2\,, \;\;\;\;\;\;\; n_2 = p-q\,,
\end{eqnarray}
the given irreducible representations $T_\texttt{ir}$ share same Casimir eigenvalues with the dS$_4$ UIR's. However, considering the fact that the Dynkin coefficients are nonnegative integers, only the cases (\ref{pqn121}), with $q=0, p\geqslant 1$, and (\ref{pqn122}), with $p \geqslant q \geqslant 1$, represent possible Weyl equivalence between the finite-dimensional irreducible representations $T_\texttt{ir}$ and the dS$_4$ UIR's, strictly speaking, with respect to the allowed ranges of $p$ and $q$, the dS$_4$ discrete series UIR's $\Pi_{p,q}^\pm$.

\setcounter{equation}{0} \section{DS$_4$ UIR's with a unique extension to the UIR's of the conformal group}\label{App massless UIRs}
Let $L_{\mathpzc{AB}}$'s, with $\mathpzc{A},\mathpzc{B}=0,1,2,3,4,5$, denote the generators of the conformal group SO$_0(2,4)$. These generators are assumed to obey the commutation relations:
\begin{eqnarray}\label{cocore}
\big[L^{}_{\mathpzc{AB}}, L^{}_{\mathpzc{CD}}\big] = - \mathrm{i} (\eta^{}_{\mathpzc{AC}}L^{}_{\mathpzc{BD}}
+\eta^{}_{\mathpzc{BD}}L^{}_{\mathpzc{AC}}
-\eta^{}_{\mathpzc{BC}}L^{}_{\mathpzc{AD}}
-\eta^{}_{\mathpzc{AD}}L^{}_{\mathpzc{BC}})\,,
\end{eqnarray}
where $\eta^{}_{\mathpzc{AB}}= \mbox{diag} (1,-1,-1,-1,-1,1)$. Here, following the lines sketched in
Ref. \cite{Barut}, we merely consider those UIR's of SO$_0(2,4)$ which fulfill the following additional condition (representation relation):\footnote{For a complete classification of the SO$_0(2,4)$ UIR's, which is beyond the scope of this paper, readers are referred to Refs. \cite{TYao67, TYao68, MackTodorov, KnappSpeh, Angeloo83}.}
\begin{eqnarray}\label{coantico}
\big\{ L_{\mathpzc{AB}}, L^{\mathpzc{A}}_{\;\mathpzc{C}} \big\} = -2\lambda\eta^{}_{\mathpzc{BC}}\,,
\end{eqnarray}
where $\lambda$ is a number and $\big\{L_{\mathpzc{AB}}, L^{\mathpzc{A}}_{\;\mathpzc{C}}\big\}$ denotes the anti-commutator of $L_{\mathpzc{AB}}$ and $L^{\mathpzc{A}}_{\;\mathpzc{C}}$. Actually, according to Ref. \cite{Barut}, only for special values of $\lambda$ are there nontrivial SO$_0(2,4)$ UIR's. Below, we briefly review how the possible values of $\lambda$ and correspondingly a complete classification of the respective UIR's can be achieved.

From Eq. (\ref{coantico}), it follows immediately that the quadratic Casimir operator of SO$_0(2,4)$ reads as:
\begin{eqnarray}\label{coCaop}
- \frac{1}{2} L_{\mathpzc{AB}}L^{\mathpzc{AB}} = 3\lambda {\mathbbm{1}}\,,
\end{eqnarray}
such that $\lambda$ can only take real values. Utilizing Eqs. (\ref{cocore}), (\ref{coantico}) and (\ref{coCaop}), it is straightforward to show that the two Casimir operators (\ref{Casimir 2}) and (\ref{Casimir 4}) of \emph{the dS$_4$ subgroup}\footnote{In this appendix, to keep the notations visually compatible, we will consider the dS$_4$ group as SO$_0(1,4)$ instead of its universal covering Sp$(2,2)$. Then, one can manifestly visualize the interesting chain $\mathrm{SO}_0(2,4) \supset \mathrm{SO}_0(1,4) \supset \mathrm{SO}(4) \supset \mathrm{SO}(3)$.} SO$_0(1,4)$ ($\subset\mathrm{SO}_0(2,4)$) respectively take the following forms:
\begin{eqnarray}\label{coseop}
Q^{(1)} = - \frac{1}{2} L_{AB}L^{AB} = 2\lambda {\mathbbm{1}}\,,
\end{eqnarray}
\begin{eqnarray}\label{cofoop}
Q^{(2)} = -W_A W^A &=& \lambda(1-\lambda) {\mathbbm{1}}\,.
\end{eqnarray}
Recall from section \ref{Sec Dixmier} that $L_{AB}$'s, with $A, B = 0,1,2,3,4$, are the generators of SO$_0(1,4)$ and that $W_A = - \frac{1}{8} {\cal{E}}_{\tiny{ABCDE}} L^{BC} L^{DE}$, where ${\cal{E}}_{\tiny{ABCDE}}$ is the five-dimensional totally antisymmetric Levi-Civita symbol. [Note that there are two distinct (but algebraically identical) SO$_0(1,4)$ subgroups in SO$_0(2,4)$, one generated by $\big\{L_{AB}\big\}=\big\{L_{0\texttt{A}}, L_{\texttt{A} \texttt{B}}\big\}$ and the other by $\big\{L_{5\texttt{A}}, L_{\texttt{A} \texttt{B}}\big\}$, where $L_{\texttt{A} \texttt{B}}$'s, with $\texttt{A}, \texttt{B} = 1,2,3,4$, are the generators of the SO$(4)$ subgroup of SO$_0(1,4)$. Above, we have considered the former.] The only generator of the conformal group that lies outside SO$_0(1,4)$ is $\Gamma_0 \equiv L_{50}$, for which, we have:
\begin{eqnarray}\label{gam0o}
\big(\Gamma_0\big)^2 = \frac{1}{2} L_{\texttt{A} \texttt{B}} L^{\texttt{A} \texttt{B}} + \lambda {\mathbbm{1}}\,.
\end{eqnarray}
Note that, to get the above equation, we have used Eqs. (\ref{coCaop}) and (\ref{coseop}).

According to Eqs. (\ref{coseop}) and (\ref{cofoop}), both dS$_4$ quadratic and quartic Casimir operators are constants for representations of the conformal group SO$_0(2,4)$ characterized by (\ref{coantico}). We then naturally suspect that the irreducible representations of SO$_0(2,4)$ will remain irreducible also with respect to SO$_0(1,4)$. Yet, we only know that the irreducible representations into which it reduces must have the same values of the quadratic and quartic Casimir operators. If the irreducible representations remain irreducible with respect to SO$_0(1,4)$, then the SO$(4)$ $\supset$ SO$(3)$ basis of SO$_0(1,4)$ is already a complete basis of the SO$_0(2,4)$ irreducible representation. To examine this very point, we first recall that the UIR's of the SO$(4)$ group, denoted here by $\textbf{D}^{(j_l,j_r)}$, are determined by two numbers $j_l,j_r \in \mathbb{N}/2$, which are related to the values of the SO$(4)$ Casimir operators by:
\begin{eqnarray}\label{so4co1}
\frac{1}{2}L_{\texttt{A} \texttt{B}} L^{\texttt{A} \texttt{B}} = 2\big( j_l (j_l + 1) + j_r (j_r + 1)\big) {\mathbbm{1}}\,,
\end{eqnarray}
\begin{eqnarray}\label{so4co2}
\frac{1}{8} {\cal E}^{\texttt{A} \texttt{B} \texttt{C} \texttt{D}} L_{\texttt{A} \texttt{B}} L_{\texttt{C} \texttt{D}} = \big( j_l (j_l + 1) - j_r (j_r + 1) \big) {\mathbbm{1}}\,.
\end{eqnarray}
From Eqs. (\ref{gam0o}) and (\ref{so4co1}), we directly obtain:
\begin{eqnarray}\label{specgam}
\mbox{spectrum} \; (\Gamma_0)^2 = \lambda + 2\big( j_l (j_l + 1) + j_r (j_r + 1)\big)\,.
\end{eqnarray}
This implies that, besides the SO$_0(2,4)$ generators lying inside SO$_0(1,4)$, the spectrum of $\Gamma_0$ (up to a sign) is also characterized by that of SO$(4)$ in the irreducible representation of SO$_0(2,4)$. Therefore, the SO$(4)$ $\supset$ SO$(3)$ basis of SO$_0(1,4)$ is indeed a complete basis of the SO$_0(2,4)$ irreducible representation.

We are now ready to go through a complete classification of UIR's of SO$_0(2,4)$ characterized by the additional representation relation (\ref{coantico}). Considering all the above, while we compare our UIR's characterized by (\ref{coseop}) and (\ref{cofoop}) with the complete list of the SO$_0(1,4)$ UIR's given by Dixmier (see section \ref{Sec Dixmier}), possible values of $\lambda\in\mathbb{R}$ are:
\begin{eqnarray}\label{povaa}
\lambda = 1 - s^2 \,, \;\;\;\;\;\;\; s=0,\frac{1}{2},1,\frac{3}{2},2,\;... \,,
\end{eqnarray}
where:
\begin{itemize}
\item{The case $s=0$ corresponds to the dS$_4$ UIR ${\underline{U}}^{\mbox{\small{cs}}}_{0,\frac{1}{2}}$, which in turn with respect to $\mathrm{SO}_0(1,4) \supset \mathrm{SO}(4)$ reduces as ${\underline{U}}^{\mbox{\small{cs}}}_{0,\frac{1}{2}}\;\big|_{\mathrm{SO}(4)} = {\textbf{D}}^{(0,0)}$ (see section \ref{Sec Dixmier}).}
\item{Other values of $s$ correspond to the dS UIR's $\Pi^{\pm}_{s,s}$, for which we have $\Pi^{-}_{s,s}\;\big|_{\mathrm{SO}(4)} = {\textbf{D}}^{(s,0)}$ and $\Pi^{+}_{s,s}\;\big|_{\mathrm{SO}(4)} = {\textbf{D}}^{(0,s)}$ (see section \ref{Sec Dixmier}).}
\end{itemize}

The above SO$_0(1,4)$ UIR's not only extend to SO$_0(2,4)$, but they also precisely extend to two inequivalent UIR's of SO$_0(2,4)$. This additional doubling comes to fore when we take into account the sign of $\Gamma_0$. Actually, if we substitute the possible values of $\lambda$ given in Eq. (\ref{povaa}) (and correspondingly, the respective values of $(j_l,j_r)$ given in the above itemization) into (\ref{specgam}), we find:
\begin{eqnarray}
\mbox{spectrum}\; \Gamma_0 \equiv E_0 = \pm (s+1)\,.
\end{eqnarray}
Note that no operator in SO$_0(2,4)$ changes the sign of $(s + 1)$ and consequently the sign of conformal energy $E_0$. Therefore, the sign of $E_0$ is another invariant of the SO$_0(2,4)$ UIR's. Accordingly, in this paper, we denote the SO$_0(2,4)$ UIR's by ${\cal{C}}^{\gtrless}_{E_0,j_l,j_r}$, in which the superscripts `$\gtrless$' refer to the positive/negative sign of conformal energy, respectively.

At the end, we must underline that the conformal extension of the above dS$_4$ UIR's is equivalent to the conformal extension of the Poincar\'{e} massless UIR's. [For the latter point, which is beyond the scope of this paper, readers are referred to Ref. \cite{Mack1977}.] In this sense, the above dS$_4$ UIR's are recognized as dS$_4$ massless representations.

\setcounter{equation}{0} \section{DS$_4$ infinitesimal generators in terms of the conformal coordinates}\label{App infinitesimal}
In this appendix, we present the (orbital part of the) dS$_4$ infinitesimal generators $M_{AB} = - \mathrm{i} \big( x^{}_A \partial^{}_B - x^{}_B \partial^{}_A \big)$, with $A,B=0,1,2,3,4$, in terms of the conformal coordinates $x=x(\rho,\textbf{u})$, where $-\pi /2< \rho <\pi /2$ and $\textbf{u} \in \mathbb{S}^3$ (see Eq. (\ref{cbgicoo})). In this context, the three infinitesimal generators of boosts take the form:
\begin{eqnarray}
M_{01}&=& \mathrm{i} \Big(-\cos\rho\sin\psi\sin\theta\cos\phi\frac{\partial}{\partial\rho} - \sin\rho\cos\psi\sin\theta\cos\phi\frac{\partial}{\partial\psi}
- \frac{\sin\rho\cos\theta\cos\phi}{\sin\psi}\frac{\partial}{\partial\theta} + \frac{\sin\rho\sin\phi}{\sin\psi\sin\theta}\frac{\partial}{\partial\phi} \Big)\,,\\
M_{02}&=& \mathrm{i} \Big(-\cos\rho\sin\psi\sin\theta\sin\phi\frac{\partial}{\partial\rho} - \sin\rho\cos\psi\sin\theta\sin\phi\frac{\partial}{\partial\psi}
- \frac{\sin\rho\cos\theta\sin\phi}{\sin\psi}\frac{\partial}{\partial\theta} - \frac{\sin\rho\cos\phi}{\sin\psi\sin\theta}\frac{\partial}{\partial\phi} \Big)\,,\\
M_{03}&=& \mathrm{i} \Big(-\cos\rho\sin\psi\cos\theta\frac{\partial}{\partial\rho} - \sin\rho\cos\psi\cos\theta\frac{\partial}{\partial\psi}
+ \frac{\sin\rho\sin\theta}{\sin\psi}\frac{\partial}{\partial\theta} \Big)\,.
\end{eqnarray}
On the other hand, the infinitesimal generator of time translation reads:
\begin{eqnarray}
M_{04}= \mathrm{i} \Big(-\cos\rho\cos\psi\frac{\partial}{\partial\rho} + \sin\rho\sin\psi\frac{\partial}{\partial\psi} \Big)\,.
\end{eqnarray}
The other six generators (of space rotations $M_{ki}$ and space translations $M_{4k}$ ($k,i=1,2,3$)), associated with the compact $SO(4)$ subgroup, take the same form as those already given in subsection \ref{Subsec dS4 Principal scalar} (see Eq. (\ref{M_AB})).

Note that the $O(1,4)$-invariant measure on the dS$_4$ hyperboloid is:
\begin{eqnarray}
\mathrm{d}\mu = (\cos\rho)^{-4} \; \mathrm{d}\rho \mathrm{d}\mu(\textbf{u})\,,
\end{eqnarray}
where $\mathrm{d}\mu(\textbf{u})$ refers to the $O(4)$-invariant measure on $\mathbb{S}^3$ (see appendix \ref{App UIR's SU(2)}).

\setcounter{equation}{0} \section{Precision on Eq. (\ref{resteads})}\label{App Oscillator}
In section \ref{Sec Precision (A)dS mass}, we have shown that the AdS$_4$ group SO$_0(2,3)$ represents the relativity group for an elementary system which is a deformation of both a relativistic free particle (with the rest energy $mc^2$) and a harmonic oscillator (with the rest energy $\frac{3}{2}\hbar\omega$, $\omega=\frac{c}{R}$). In this appendix, following Ref. \cite{GazRe1993}, we aim to make this argument more explicit through the contraction process on the group/algebra level, as we have discussed in section \ref{Sec Group contraction dS4}. Of course, in order to avoid technical (but not conceptual) difficulties, we deal with ($1+1$)-dimensional spacetime (AdS$_2$ case). The rest energy of the (relevant) harmonic oscillator then would be $\frac{1}{2}\hbar\omega$ instead of $\frac{3}{2}\hbar\omega$.

We adapt the notion of elementary systems to the one-dimensional harmonic oscillator, while the Newton group \cite{BacryLevi} is considered as the relativity group. The Newton Lie algebra is generated by:
\begin{eqnarray}
H=\frac{p^2}{2m} + \frac{1}{2}m\omega^2 q^2\,, \;\;\;\;\;\;\; P\equiv p\,, \;\;\;\;\;\;\; L\equiv q\,,
\end{eqnarray}
with Poisson brackets:
\begin{eqnarray}\label{PBN}
\big\{H,P\big\}= m\omega^2 L\,, \;\;\;\;\;\;\; \big\{H,L\big\}= -\frac{P}{m}\,, \;\;\;\;\;\;\; \big\{P,L\big\}=1\,.
\end{eqnarray}
We also consider the Lie algebra of the Poincar\'{e} group generated by:
\begin{eqnarray}
H^{}_{\mbox{\tiny{tot}}}= \sqrt{p^2c^2 + m^2c^4}\,, \;\;\;\;\;\;\; P\equiv p\,, \;\;\;\;\;\;\; L\equiv \frac{qH^{}_{\mbox{\tiny{tot}}}}{mc^2}\,,
\end{eqnarray}
where $H^{}_{\mbox{\tiny{tot}}}$ is the energy, $P$ the momentum, and $L$ the observable associated with the (properly re-scaled) Lorentz boost. The corresponding Poisson brackets are given by:
\begin{eqnarray}
\big\{H^{}_{\mbox{\tiny{tot}}},P\big\}= 0\,, \;\;\;\;\;\;\; \big\{H^{}_{\mbox{\tiny{tot}}},L\big\}= -\frac{P}{m}\,, \;\;\;\;\;\;\; \big\{P,L\big\}=-\frac{H^{}_{\mbox{\tiny{tot}}}}{mc^2}\,.
\end{eqnarray}
If we restrict our attention to the kinetic energy instead of the total one, $H=H^{}_{\mbox{\tiny{tot}}}-mc^2$, we get:
\begin{eqnarray}\label{PBP}
\big\{H,P\big\}= 0\,, \;\;\;\;\;\;\; \big\{H,L\big\}= -\frac{P}{m}\,, \;\;\;\;\;\;\; \big\{P,L\big\}=1-\frac{H}{mc^2}\,.
\end{eqnarray}
One can simply show that the Lie algebras (\ref{PBN}) and (\ref{PBP}) contract, respectively, for $\omega\rightarrow 0$ and for $c\rightarrow \infty$, towards the symmetry Lie algebra for the Galilean free particle:
\begin{eqnarray}
\big\{H,P\big\}= 0\,, \;\;\;\;\;\;\; \big\{H,L\big\}= -\frac{P}{m}\,, \;\;\;\;\;\;\; \big\{P,L\big\}=1\,.
\end{eqnarray}

Now, we take into account the Lie algebra for the AdS$_2$ elementary system, namely, for a free particle with mass $\mathfrak{M}_{\mbox{\tiny{AdS}}_2}$ living in AdS$_2$ spacetime:
\begin{eqnarray}
\big\{H,P\big\}= \frac{\mathfrak{M}_{\mbox{\tiny{AdS}}_2} c^2}{R^2}L\,, \;\;\;\;\;\;\; \big\{H,L\big\}= -\frac{P}{\mathfrak{M}_{\mbox{\tiny{AdS}}_2}}\,, \;\;\;\;\;\;\; \big\{P,L\big\}=1-\frac{H}{\mathfrak{M}_{\mbox{\tiny{AdS}}_2}c^2}\,.
\end{eqnarray}
This algebra can be identified to the AdS$_2$ Lie algebra $\mathrm{SO}_0(2,1)$. The Poincar\'{e} and Newton relativities then can be simply achieved from the AdS$_2$ relativity through the contraction process; the limit $R\rightarrow\infty$ results in the Poincar\'{e} algebra ($\mathfrak{M}_{\mbox{\tiny{AdS}}_2} \longrightarrow m$), whereas the limit $c,R\rightarrow\infty$ ($c/R=\omega$ being fixed) results in the Newton algebra. Actually, quite similar to the dS$_4$ case (see section \ref{Sec Group contraction dS4}), we have:
\begin{eqnarray}
\begin{array}{ccccccc}
{\mbox{AdS$_2$ group (algebra)}} & \longrightarrow & {\mbox{Poincar\'{e} group (algebra)}} & \\
\\
\big\downarrow &  & \big\downarrow &\\
\\
{\mbox{Newton group (algebra)}} & \longrightarrow & {\mbox{Galilei group (algebra)}} &
\end{array}
\end{eqnarray}
where, again, the arrows `$\longrightarrow$' stand for the group (algebra) contractions.

\end{appendix}

\end{document}